\documentclass[12pt,chapterprefix=true]{report}

\usepackage{longtable}
\usepackage{verbatimbox}
\usepackage{changepage}
\usepackage{graphicx}
\usepackage{amsmath}
\usepackage{graphicx,epsf,epsfig}
\usepackage{dcolumn}
\usepackage{bm}
\usepackage{ulem}
\usepackage[usenames]{color}
\usepackage{cite}
\usepackage{setspace}
\usepackage{verbatim}
\usepackage{hyperref}
\usepackage{caption}
\usepackage{subcaption}
\usepackage{mathtools}
\usepackage{dsfont} 
\usepackage{amssymb}
\usepackage{graphicx}
\usepackage{dsfont} 
\usepackage{amssymb}
\usepackage{tabularx}
\usepackage{multirow}
\usepackage{makecell}
\usepackage{diagbox}
\usepackage[abs]{overpic}
\usepackage{tocloft,calc}

\allowdisplaybreaks

\hypersetup{
    colorlinks=true,
    linkcolor=blue,
    filecolor=magenta,      
    urlcolor=blue,
    citecolor=blue
}
\renewcommand{\baselinestretch}{2.0} 

\newcommand{\be}[0]{\begin{equation}}
\newcommand{\ee}[0]{\end{equation}}
\newcommand{\ba}[0]{\begin{align}}
\newcommand{\ea}[0]{\end{align}}

\definecolor{red(ncs)}{rgb}{0.77, 0.01, 0.2}
\newcommand{\sci}[2]{#1 \times 10^{#2}}

\newcommand{\NS}{{\mbox{\tiny NS}}}
\newcommand{\BH}{{\mbox{\tiny BH}}}
\newcommand{\R}{{\mbox{\tiny R}}}
\newcommand{\I}{{\mbox{\tiny I}}}
\newcommand{\ISCO}{{\mbox{\tiny ISCO}}}
\newcommand{\TT}{{\mbox{\tiny TT}}}
\newcommand{\EdGB}{{\mbox{\tiny EdGB}}}
\newcommand{\PK}{{\mbox{\tiny PK}}}
\newcommand{\NC}{{\mbox{\tiny NC}}}
\newcommand{\RD}{{\mbox{\tiny RD}}}
\newcommand{\ST}{{\mbox{\tiny ST}}}
\newcommand{\MDR}{{\mbox{\tiny MDR}}}
\newcommand{\ppE}{{\mbox{\tiny ppE}}}
\newcommand{\GR}{{\mbox{\tiny GR}}}
\newcommand{\X}{{\mbox{\tiny X}}}

\newcommand{\GB}{{\mbox{\tiny GB}}}
\newcommand{\EH}{{\mbox{\tiny EH}}}
\newcommand{\dCS}{{\mbox{\tiny dCS}}}
\newcommand{\ppN}{{\mbox{\tiny ppN}}}
\newcommand{\MR}{{\mbox{\tiny MR}}}
\newcommand{\GW}{{\mbox{\tiny GW}}}
\newcommand{\II}{{\mbox{\tiny I}}}
\newcommand{\K}{{\mbox{\tiny K}}}
\newcommand{\A}{{\mbox{\tiny A}}}
\newcommand{\B}{{\mbox{\tiny B}}}
\newcommand{\JP}{{\mbox{\tiny JP}}}
\newcommand{\MD}{{\mbox{\tiny MD}}}
\newcommand{\IMR}{{\mbox{\tiny IMR}}}
\newcommand{\DEF}{{\mbox{\tiny DEF}}}

\definecolor{red(ncs)}{rgb}{0.77, 0.01, 0.2}

\newcommand{\KrangeTab}[2]{\scriptsize#1 MeV $\leq K_0 \leq$ #2 MeV\normalsize}

\newcommand{\KsymrangeTab}[2]{\scriptsize#1 MeV $\leq K_\mathrm{sym,0} \leq$ #2 MeV\normalsize}
\newcommand{\MrangeTab}[2]{\scriptsize#1 MeV $\leq M_0 \leq$ #2 MeV\normalsize}

\newcommand{\Krange}[2]{#1 MeV $\leq K_0 \leq$ #2 MeV}

\newcommand{\Ksymrange}[2]{#1 MeV $\leq K_\mathrm{sym,0} \leq$ #2 MeV}
\newcommand{\Mrange}[2]{#1 MeV $\leq M_0 \leq$ #2 MeV\normalsize}

\usepackage{array}
\newcolumntype{C}[1]{>{\centering\let\newline\\\arraybackslash\hspace{0pt}}m{#1}}

\newcolumntype{C}[1]{>{\centering\arraybackslash}m{#1}}

\def\addsymbol #1: #2#3{$#1$ \> \parbox{5in}{#2 }\\}

\makeatletter
\def\@part[#1]#2{%
    \ifnum \c@secnumdepth >-2\relax
      \refstepcounter{part}%
      \addcontentsline{toc}{part}{\partname~\thepart\hspace{1em}#1}%
    \else
      \addcontentsline{toc}{part}{#1}%
    \fi
    \markboth{}{}%
    {\centering
     \interlinepenalty \@M
     \normalfont
     \ifnum \c@secnumdepth >-2\relax
       \huge\bfseries \partname\nobreakspace\thepart
       \par
       \vskip 20\p@
     \fi
     \Huge \bfseries #2\par}%
    \@endpart}
\makeatother

\begin{document}

\pagenumbering{roman}	
\pagestyle{headings}
\newpage 
\thispagestyle{empty}
\begin{center}

\singlespacing


{\LARGE \textbf{Probing Fundamental Physics with Gravitational Waves}}\\
\vspace{1.5cm}
{\Large Zachary Carson}\\
Salt Lake City, Utah, USA\\
B.S.~Physics, B.S.~Mathematics, University of Utah, 2014\\

\vspace{0.8cm}
A Dissertation presented to the Graduate Faculty \\
of the University of Virginia in Candidacy for the degree of \\
Doctor of Philosophy\\
\vspace{0.8cm}
Department of Physics\\
University of Virginia\\
May 2020\\
\end{center}
\vspace{0.8cm}
\begin{flushright}
\singlespacing
\rule{2.3in}{0.5pt} \\
{\large Professor Kent Yagi \\ \textit{Research Advisor}}\\ 
\end{flushright}

\vspace{0.4cm}
\begin{flushright}
\singlespacing
\rule{2.3in}{0.5pt} \\
\hfill {\large Professor David Nichols}\\
\end{flushright}

\vspace{0.4cm}
\begin{flushright}
\singlespacing
\rule{2.3in}{0.5pt} \\
\hfill {\large Professor Simonetta Liuti}\\
\end{flushright}

\vspace{0.4cm}
\begin{flushright}
\singlespacing
\rule{2.3in}{0.5pt} \\
\hfill {\large Professor Scott Ransom}\\
\end{flushright}




\newpage
\section*{\Huge Abstract}
\addcontentsline{toc}{section}{Abstract} 
\indent
\indent
\begin{spacing}{1.8}
\thispagestyle{plain}
The explosive coalescence of two black holes $1.3$ billion light years away has for the very first time allowed us to peer into the \textit{extreme gravity} region of spacetime surrounding these events.
With these maximally compact objects reaching speeds up to $60\%$ the speed of light, collision events such as these create harsh spacetime environments where the fields are strong, non-linear, and highly dynamical -- a place yet un-probed in human history.
On September 14, 2015, the iconic chirp signal from such an event was registered simultaneously by both of the Laser Interferometer Gravitational-Wave Observatory (LIGO) detectors -- by an unparalleled feat of modern engineering.
Dubbed ``GW150914'', this gravitational wave event paved the way for an entirely new observing window into the universe, providing for the unique opportunity to probe fundamental physics from an entirely new viewpoint.
Since this historic event, the LIGO/Virgo collaboration (LVC) has further identified ten additional gravitational wave signals in its first two observing runs, composed of a myriad of different events.
Important among these new cataloged detections is GW170817, the first detection of gravitational waves from the merger of two neutron stars, giving way to new insight into the supranuclear physics resident within.

This thesis explores this new unique opportunity to harness the information encoded within gravitational waves in regards to their source whence they came, to probe fundamental physics from an entirely new perspective.
Part~\ref{part:ProbingNuclearPhysics} focuses on probing nuclear physics by way of the tidal information encoded within gravitational waves from binary neutron star mergers.
By finding correlations between this tidal
\newpage
\thispagestyle{myheadings}
\markright{Abstract}
\noindent information and fundamental nuclear matter parameters, we find new constraints on the latter with both current and future gravitational wave observations.
Finally, by making use of constraints on the nuclear matter equation of state from GW170817, we develop improved universal relations between neutron star observables, which assist in better parameter estimation for future observations.

Another enticing subject one might consider is the validity of Einstein's general relativity.
While observationally confirmed in every spacetime region reachable over the last century, it has yet to be probed in extreme gravity environments, such as those outside binary black hole mergers.
Part~\ref{part:TestingGR} focuses on testing general relativity from such events by way of the remnants of such spacetime encoded within the gravitational wave signal.
By considering both parameterized tests and by testing the consistency between the inspiral and merger-ringdown signals, we find strong constraints on several alternative theories of gravity with both current and future observations, including the combination of multiple events and with the multi-band detections between both space-based and ground-based detectors.
Finally, we devise a new general spacetime metric which is parameterized beyond the Kerr one that describes black holes in general relativity.
We find corrections to several astrophysical phenomena in the new beyond-Kerr metric which could be observed with future observations by e.g. the Event Horizon Telescope.

\end{spacing}
\newpage


\newpage
\section*{\Huge Acknowledgments}
\addcontentsline{toc}{section}{Acknowledgments} 
\indent
\indent
\begin{spacing}{2}
\thispagestyle{plain}
This thesis and all of its included works would not have been possible without several different people I've met throughout my six year tenure at the University of Virginia.
First and foremost, I would like to give great thanks to my incredible advisor Kent Yagi.
His patient help and guidance has taught me so many things about both gravitational wave physics and how good research is conducted.
His passion for gravitational wave physics is evident in the way that he taught me everything I know about the field, and I greatly appreciate all the work he put into checking my calculations, editing my many papers, and meeting with me often to discuss our research.

I would also like to thank all of my fantastic collaborators and co-workers who have provided me with valuable information that has helped me along with this research.
In particular, the knowledgeable Andrew Steiner has been extremely helpful by always being available to answer my many questions in regards to nuclear physics and neutron stars.
Additionally, I have Nicol\'{a}s Yunes to thank for discussions on his great writing practices.
I also give my thanks to Katerina Chatziioannou and Carl-Johan Haster who taught me many things about the LIGO/Virgo collaboration and the analyses done there.
Finally I thank my many co-workers including Shammi Tahura, Brian Seymour, Alex Saffer, Nan Jiang, and Josef Zimmerman for useful, and illuminating discussions on both their research and my own.

Next I would like to thank several people within the University of Virginia physics department.
Firstly, I would like to thank professor Maxim Bychkov for teaching me what it takes to be an effective physics educator. 
I have taught with him for several years as both a TA and a ``super TA'', and I have become a much better teacher for it.
Next I would like to thank professor Xiaochao Zheng whom I graded for several semesters, for always being a friendly face around the department who would occasionally even bring me delicious home-cooked food!
I would also like to thank my previous advisors Olivier Pfister and Chris Neu who gave me my start in physics research, and for being patient as I discovered my true passion in gravitational wave physics.
Finally I would especially like to thank professors Xiaochao Zheng, Maxim Bychkov, and Stefan Bae{\ss}ler for nominating me for the graduate teaching award, and for all of the hard work they put into completing my amazing profile which ended up winning the award.

\thispagestyle{myheadings}
\markright{Acknowledgements}

Next I would like to thank the members of my thesis committee Kent Yagi, Scott Ransom, David Nichols, and Simonetta Liuti, for taking time from their very busy schedules to read my thesis and attend my defense.
I would also like to thank all of the physics department support staff, especially Peter Cline and Dawn Shifflett, for their kind and supporting help, as well as all of their hard work behind-the-scenes.

Last but not least, I would like to thank my friends and family, without whom this would never have been possible.
I thank my Utahn friends Q Taylor Hunt, Brock Giles, Adam and Brittney Magonigal, and James Duran for always being there to hang out and relax whenever I visited home.
I also thank my close friend Sina Tafti in the physics department for always being there to grab a drink, or to drive me to the airport at 3 AM with no cabs to be found.
Next, I'd like to thank Gen Schmitt, Aaron and Allie Wegner, Ben Godek, Rajveer Nehra, Cristiana Principato, and Chun-Hung Chang for being good friends in Charlottesville, as well as the rest of the physics department graduate student population.
Next I want to thank my incredible parents Maria Addison, and Monty and Melonie Carson for always being proud and supportive of me on the long (and far away from home) doctoral road.
I would also like to thank my fianc\'{e}'s family including Nan Gray, Eric Day, Dena and Todd Spencer, John and Cheryl Price, and Zak and Donna Price and their six amazing kids for all being so caring and making me feel welcome in the family.
Finally, I thank my wife-to-be Kacey Price for your continuous emotional and moral support which helped immensely throughout both the good and the bad of the Ph.~D. journey.
In particular, I thank you for always putting a smile on my face during the stressful times, and for giving me the distractions (primarily in the form of food and board games) that gave me some semblance of work-life-balance that I probably never would have found on my own.

All of the work in this thesis was supported by the NSF (Award PHY-1806776), the Ed Owens Fund, the University of Virginia Physics departmental fellowship, as well as graduate teaching funds from the same department.
\end{spacing}


\newpage
\section*{\Huge List of abbreviations}
\addcontentsline{toc}{section}{List of abbreviations} 
\thispagestyle{plain}
\indent
\indent
\thispagestyle{plain}
\begin{table*}[!h]
\centering
\renewcommand{\arraystretch}{1.17}
\begin{tabular}{C{4cm} | C{12cm}}
\textbf{GR} & General relativity \\
\textbf{GW} & Gravitational wave \\
\textbf{BH} & Black hole \\
\textbf{NS} & Neutron star \\
\textbf{PSR} & Pulsar \\
\textbf{SMBH} & Supermassive black hole \\
\textbf{SMBHB} & Supermassive black hole binary \\
\textbf{EMRI} & Extreme-mass-ratio inspiral \\
\textbf{EoS} & Equation of state \\
\textbf{PN} & post-Newtonian \\
\textbf{ppE} & parameterized post-Einsteinian \\
\textbf{SNR} & Signal-to-noise-ratio \\
\textbf{PhenomD} & Phenomenological waveform model D \\
\textbf{NR} & Numerical relativity \\
\textbf{RD} & Ringdown \\
\textbf{QNM} & Quasinormal mode \\
\textbf{I} & Inspiral \\
\textbf{MR} & Merger-ringdown \\
\textbf{IMR} & Inspiral-merger-ringdown \\
\textbf{ISCO} & Innermost stable circular orbit \\
\textbf{SEP} & Strong equivalence principle \\
\textbf{LI} & Lorentz invariance \\
\textbf{K} & Kerr \\
\textbf{JP} & Johannsen-Psaltis \\
\textbf{MD, mod.~$\Delta$} & Modified~$\Delta$ \\
\textbf{STT} & Scalar tensor theory \\
\textbf{EdGB} & Einstein-dilaton Gauss-Bonnet \\
\textbf{dCS} & dynamical Chern-Simons \\
\textbf{GB} & Gauss-Bonnet \\
\textbf{DEF} & Damour and Esposito-Far$\acute{e}$se \\
\textbf{MO} & Mendes-Ortiz \\
\end{tabular}
\end{table*}

\newpage
\thispagestyle{myheadings}
\markright{List of abbreviations}

\begin{table*}[!ht]
\centering
\renewcommand{\arraystretch}{1.17}
\begin{tabular}{C{4cm} | C{12cm}} 
\textbf{MDR} & Modified dispersion relation \\
\textbf{LIGO} & Laser Interferometer Gravitational Wave Observatory \\
\textbf{O$\bm{i}$} & Observing run ``$i$'' \\
\textbf{LVC} & LIGO/Virgo collaboration \\
\textbf{CE} & Cosmic Explorer \\
\textbf{ET} & Einstein Telescope \\
\textbf{LISA} & Laser Interferometer Space Antenna \\
\textbf{(B-)DECIGO} & Deci-hertz Interferometer Gravitational Wave Observatory (B) \\
\textbf{EHT} & Event Horizon Telescope \\
\textbf{VLBI} & Very long baseline interferometry \\
\textbf{NICER} & Neutron Star Interior Composition ExploreR \\
\textbf{SKA} & Square Kilometre Array \\
\textbf{TT} & Transverse traceless (gauge) \\
\end{tabular}
\end{table*}

\clearpage


\newpage
\section*{\Huge List of symbols}
\addcontentsline{toc}{section}{List of symbols} 
\thispagestyle{plain}
\indent
\indent
\thispagestyle{plain}
\begin{table*}[!h]
\centering
\renewcommand{\arraystretch}{1.17}
\begin{tabular}{C{2cm} | C{14cm}}
$t$ & Time coordinate \\
$r$ & Radial coordinate \\
$\theta$ & Polar coordinate \\
$\phi$ & Azimuthal coordinate \\
$\text{ M}_\odot$ & Solar mass $1.989\times10^{30}\text{ kg}=5\times10^{-6}\text{ s}=1.5\text{ km}$\\
$m$ & Isolated stellar object mass\\
$m_\A$ & $A$-th mass in a stellar binary\\
$M$ & Total mass in a stellar binary $m_1+m_2$\\
$\delta_m$ & Weighted mass difference $(m_1-m_2)/M$\\
$\mu$ & Reduced stellar binary mass $(m_1m_2)/M$ \\
$\eta$ & Symmetric mass ratio $\mu/M$ \\
$q$ & Mass ratio $m_1/m_2$ \\
$\mathcal{M}$ & Chirp mass $\eta^{3/5}M$ \\
$M_f$ & Remnant black hole mass \\
$\Delta M_f$ & Remnant black hole mass difference $M_f^\I-M_f^\MR$ \\
$\bar{M}_f$ & Average remnant black hole mass $(M_f^\I+M_f^\MR)/2$ \\
$\epsilon$ & Weighted remnant black hole mass difference $\Delta M_f/\bar{M}_f$\\
$J_z$ & Rotational angular momentum in the $\hat{z}$-direction \\
$a$ & Isolated stellar object spin parameter $J_z/m$\\
$a_\A$ & $A$-th body spin parameter in a stellar binary\\
$a_{s,a}$ & Symmetric/Anti-symmetric spins $(m_1\chi_1 \pm m_2\chi_2)/2$\\
$\chi$ & Isolated stellar object dimensionless spin $a/m$ \\
$\chi_\A$ & $A$-th body dimensionless spin \\
$\chi_{s,a}$ & Symmetric/Anti-symmetric dimensionless spins $(\chi_1 \pm \chi_2)/2$\\
$\chi_f$ & Remnant black hole spin \\
$\Delta \chi_f$ & Remnant black hole spin difference $\chi_f^\I-\chi_f^\MR$ \\
$\bar{\chi}_f$ & Average remnant black hole spin $(\chi_f^\I+\chi_f^\MR)/2$ \\
$\sigma$ & Weighted remnant black hole spin difference $\Delta \chi_f/\bar{\chi}_f$\\
$Q_{ij}$ & Quadrupole moment \\
$\mathcal{E}_{ij}$ & Tidal field \\
$\lambda$ & Tidal deformability $-Q_{ij}/\mathcal{E}_{ij}$ \\
\end{tabular}
\end{table*}

\newpage
\thispagestyle{myheadings}
\markright{List of symbols}

\begin{table*}[!ht]
\centering
\renewcommand{\arraystretch}{1.17}
\begin{tabular}{C{2cm} | C{14cm}} 
$\lambda_0$ & Tidal deformability at mass $1.4\text{ M}_\odot$ \\
$\Lambda$ & Dimensionless tidal deformability $\lambda/m^5$ \\
$\Lambda_\A$ & $A$-th body Dimensionless tidal deformability in a stellar binary \\
$\Lambda_{s,a}$ & Symmetric/Anti-symmetric tidal deformabilities $(\Lambda_1 \pm \Lambda_2)/2$\\
$\Lambda_m$ & Dimensionless tidal deformability at mass $m$ \\
$\tilde\Lambda$ & Mass-weighted tidal deformability \lbrack Eq.~\eqref{eq:Lt}\rbrack \\
$\delta\tilde\Lambda$ & 6PN tidal deformability correction \lbrack Eq.~\eqref{eq:dLt}\rbrack \\
$R$ & Stellar radius \\
$C$ & Stellar compactness $m/R$ \\
$I$ & Stellar moment of inertia  \\
$\bar{I}$ & Dimensionless stellar moment of inertia $I/m^3$ \\
$Q$ & Stellar quadrupole moment \\
$\bar{Q}$ & Dimensionless stellar quadrupole moment $Q/{m^3\chi^2}$ \\
$n_{p,n}$ & Proton/Neutron number densities \\
$n$ & Total number density $n_n+n_p$ \\
$\delta$ & Isospin symmetry parameter $(n_n-n_p)/n$ \\
$\rho_0$ & Nuclear saturation density $\sim2.3\times10^{17}\text{ kg m}^{-3}$ \\
$e_0$ & Energy per particle at saturation density \\
$K_0$ & Incompressibility at saturation density \lbrack Eq.~\eqref{eq:nuclearParameters}\rbrack\\
$Q_0$ & Third derivative of symmetric matter at saturation density \lbrack Eq.~\eqref{eq:nuclearParameters}\rbrack\\
$M_0$ & Slope of the incompressibility at saturation density $Q_0+12K_0$ \\
$J_0$ & Symmetry energy at saturation density \lbrack Eq.~\eqref{eq:nuclearParameters}\rbrack\\
$L_0$ & Slope of symmetry energy at saturation density \lbrack Eq.~\eqref{eq:nuclearParameters}\rbrack\\
$K_\text{sym,0}$ & Curvature of symmetry energy at saturation density \lbrack Eq.~\eqref{eq:nuclearParameters}\rbrack\\
$E_\text{orb}$ & Specific orbital energy \lbrack Eq.~\eqref{eq:Veff}\rbrack \\
$E_b$ & Binding energy $1-E_\text{orb}$ \\
$L_z$ & Specific orbital angular momentum in the $\hat{z}$-direction \lbrack Eq.~\eqref{eq:Veff}\rbrack \\
$\Omega$ & Orbital angular velocity \lbrack Eq.~\eqref{eq:KeplerianFrequency} \rbrack\\
$f$ & Gravitational wave frequency $\Omega/\pi$ \\
$v$ & Relative velocity of binary constituents $(\pi M f)^{1/3}$ \\
$u$ & Effective relative velocity of binary constituents $(\pi \mathcal{M} f)^{1/3}$ \\
$r_\ISCO$ & Radius of the innermost-stable-circular orbit \lbrack Eq.~\eqref{eq:risco}\rbrack \\
$f_\ISCO$ & Frequency at the innermost-stable-circular orbit $(6^{3/2}\pi M)^{-1}$ \\
$\theta^a$ & Template waveform parameters vector \lbrack Eqs.~\eqref{eq:templateBH},~\eqref{eq:templateNS}\rbrack\\
$D_L$ & Luminosity distance  \\
$z$ & Redshift \lbrack Eq.~\eqref{eq:redshift}\rbrack\\

\end{tabular}
\end{table*}

\begin{table*}[!ht]
\centering
\renewcommand{\arraystretch}{1.17}
\begin{tabular}{C{2cm} | C{14cm}} 
$H_0$ & Hubble constant $70\text{ km s}^{-1}\text{Mpc}^{-1}$~\cite{Aghanim:2018eyx} \\
$\Omega_\Lambda$ & Relative universal vacuum energy density $0.67$~\cite{Aghanim:2018eyx} \\
$\Omega_M$ & Relative universal matter density $0.33$~\cite{Aghanim:2018eyx} \\
$\mathcal{M}_z$ & Redshifted chirp mass $\mathcal{M}(1+z)$ \\
$\mathcal{A}$ &  Sky-averaged gravitational wave amplitude $\mathcal{M}_z^{5/6}/(\sqrt{30}\pi^{2/3}D_L)$\\
$t_c$ & Binary coalescence time \\
$\phi_c$ & Binary coalescence phase \\
$S_n^\A$ & Power spectral noise density for gravitational wave detector $A$ \lbrack Fig.~\ref{fig:sensitivities}\rbrack\\
$(a|b)$ & Detector sensitivity weighted inner product of $a$ and $b$ \lbrack Eq.~\eqref{eq:overlap}\rbrack \\
$\rho$ & Signal-to-noise ratio $\sqrt{(h|h)}$ \\
$\Gamma_{ij}$ & Fisher information matrix $(\partial_i h|\partial_j h)$ \\
$\Sigma_{ij}$ & Covariance matrix $\Gamma_{ij}^{-1}$ \\
$h(t)$ & Time-domain gravitational waveform \\
$\tilde{h}(f)$ & Frequency-domain gravitational waveform \lbrack Eq.~\eqref{eq:PhenomDwaveform}\rbrack\\
$A$ & Non-sky-averaged gravitational wave amplitude in frequency domain \\
$\Psi$ & Gravitational waveform phase in frequency domain \\
$(\beta_\ppE,b)$ & Parameterized post-Einstein phase correction parameters \lbrack Eq.~\eqref{eq:ppe}\rbrack \\
$(\alpha\ppE,a)$ & Parameterized post-Einstein amplitude correction parameters \lbrack Eq.~\eqref{eq:ppe}\rbrack \\
$\alpha_\EdGB$ & Einstein-dilaton Gauss-Bonnet coupling parameter \lbrack Eq.~\eqref{eq:linear-coupling}\rbrack \\
$\zeta_\EdGB$ & Einstein-dilaton Gauss-Bonnet dimensionless coupling $16\pi\alpha_\EdGB^2/M^4$ \\
$\alpha_\dCS$ & Dynamical Chern-Simons coupling parameter \\
$\zeta_\dCS$ & Dynamical Chern-Simons dimensionless coupling $16\pi\alpha_\dCS^2/M^4$ \\
$\epsilon_3$ & Johannsen-Psaltis spacetime metric deviation parameter \lbrack Eq.~\eqref{eq:JPparameter}\rbrack \\
$\beta$ & Modified $\Delta$ spacetime metric deviation parameter \lbrack Eq.~\eqref{eq:Delta_replacement}\rbrack \\
$A_i(r)$ & New spacetime metric deviation functions $\sum\limits_n^\infty\alpha_{in}\frac{m^n}{r^n}$ \\
$f(r)$ & New spacetime metric deviation function $r^2\sum\limits_n^\infty\epsilon{n}\frac{m^n}{r^n}$ \\
$\Gamma_{\alpha\mu\nu}$ & Christoffel symbols $\frac{1}{2}(g_{\alpha\mu,\nu}+g_{\alpha\nu,\mu}-g_{\mu\nu,\alpha})$  \\
$R^\mu{}_{\nu\alpha\beta}$ & Riemann curvature tensor $\Gamma^\mu{}_{\nu\beta,\alpha}-\Gamma^\mu{}_{\nu\alpha,\beta}+\Gamma^\mu{}_{\delta\alpha}\Gamma^\delta{}_{\nu\beta}-\Gamma^\mu{}_{\delta\beta}\Gamma^\delta{}_{\nu\alpha}$  \\
$R_{\mu\nu}$ & Ricci tensor $R^\delta{}_{\mu\delta\nu}$ \\
$R$ & Ricci scalar $R^\mu{}_\mu$ \\
$T_{\mu\nu}$ & Stress-Energy tensor  \\
$\eta_{\alpha\beta}$ & Flat spacetime metric tensor $\text{diag}\lbrack-1,1,1,1\rbrack$\\
$g_{\alpha\beta}^\X$ & Curved spacetime metric tensor \\
$\Sigma$ & Kerr metric parameter $r^2+a^2\cos^2\theta$ \\
$\Delta$ & Kerr metric parameter $r^2+a^2-2mr$ \\
\end{tabular}
\end{table*}

\clearpage


\newpage
\phantomsection
\addcontentsline{toc}{section}{Table of contents} 
\renewcommand{\baselinestretch}{1.0} 
\begin{spacing}{1.2}

\setlength{\cftchapnumwidth}{\widthof{\textbf{Appendix~999~}}}
\makeatletter
\g@addto@macro\appendix{%
   \addtocontents{toc}{%
     \protect\renewcommand{\protect\cftchappresnum}{\appendixname\space}%
   }%
}
\renewcommand\contentsname{Table of contents}
\tableofcontents
\end{spacing}


\newpage
\begin{spacing}{1.195}
\phantomsection
\addcontentsline{toc}{section}{List of figures}
\renewcommand\listfigurename{List of figures}
\listoffigures
\vspace{-1in}
\end{spacing}


\newpage
\begin{spacing}{1.2}
\phantomsection
\renewcommand\listtablename{List of tables}
\addcontentsline{toc}{section}{List of tables}
\listoftables
\end{spacing}


\newpage
\thispagestyle{plain}
\section*{\Huge Preface}
\indent
\indent
\addcontentsline{toc}{section}{Preface}
\begin{spacing}{2}
\thispagestyle{plain}
The work presented in this thesis has all been developed in collaboration with my advisor Kent Yagi as well as with several researchers at other institutions.
All of the work within this document has been accepted for publication in peer-reviewed journals.
Part~\ref{part:ProbingNuclearPhysics} is based on three works all published to \textit{Physical Review D}.
In particular, Chapter~\ref{chap:ConstrainingNuclearParameters} was based on two publications carried out in collaboration with Andrew W. Steiner at the University of Tennessee and Oak Ridge National Laboratory.
Further, Chapter~\ref{chap:EoSInsensitiveRelations} was based on one publication accomplished in collaboration with Katerina Chatziioannou at the Flatiron institute, Carl-Johan Haster at the Massachusetts Institute of Technology, and Nicol\'{a}s Yunes at the University of Illinois at Urbana-Champaign.
Part~\ref{part:TestingGR} is based on eight works published to \textit{Physical Review D} (Chapters~\ref{chap:Parameterized} and~\ref{chap:beyondKerr}), \textit{Classical and Quantum Gravity} (Chapter~\ref{chap:Parameterized}), \textit{Classical and Quantum Gravity Letters} (Chapter~\ref{chap:IMRconsistency}), and \textit{MDPI Proceedings} (Chapters~\ref{chap:Parameterized} and~\ref{chap:IMRconsistency}), and were carried out locally at the University of Virginia by me and Kent Yagi.
In addition, one section of Chapter~\ref{chap:Parameterized} is based on a publication carried out in collaboration with Brian C. Seymour, a former undergraduate at the University of Virginia.

My participation in the above projects has been extensive.
As first author of all of the above papers, most of the analytical calculations have been carried out by me, with double-checks done by several collaborators, especially Kent Yagi.
In addition, in the works presented in Chapter~\ref{chap:ConstrainingNuclearParameters}, Andrew W. Steiner generated the 
\newpage
\thispagestyle{myheadings}
\markright{Preface}
\noindent phenomenological neutron star equations of states used in the analysis.
Also in one section of Chapter~\ref{chap:Parameterized} Brian C. Seymour did all of the calculations based on pulsar timing observations, while I focused on the ones from gravitational waves.
Finally, in Chapter~\ref{chap:EoSInsensitiveRelations}, Carl-Johan Haster provided the spectral equation of state posterior samples by using the same analysis and settings as those used by the LVC with GW170817.
I am grateful to Nicol\'{a}s Yunes for use of his Fisher analysis Mathematica script which several of my projects have utilized with heavy modifications.
I am also grateful to Kent Yagi for use of his \texttt{C} script used for solving the neutron star Tolman Oppenheimer Volkoff equations, which I used extensively in Part~\ref{part:ProbingNuclearPhysics}.
A majority of my calculations have been done in the \texttt{Mathematica} computer algebra system, both locally and with the University of Virginia's High Performance Computing cluster Rivanna.
\end{spacing}
\newpage


\newpage			

\bibliographystyle{hIEEEtran}

\setcounter{page}{1}	
\pagenumbering{arabic}	

\setcounter{secnumdepth}{5}
\renewcommand{\thepart}{\Alph{part}}


\chapter{Introduction}\label{chap:introduction}
\begin{figure}
\begin{center} 
\includegraphics[width=0.7\columnwidth]{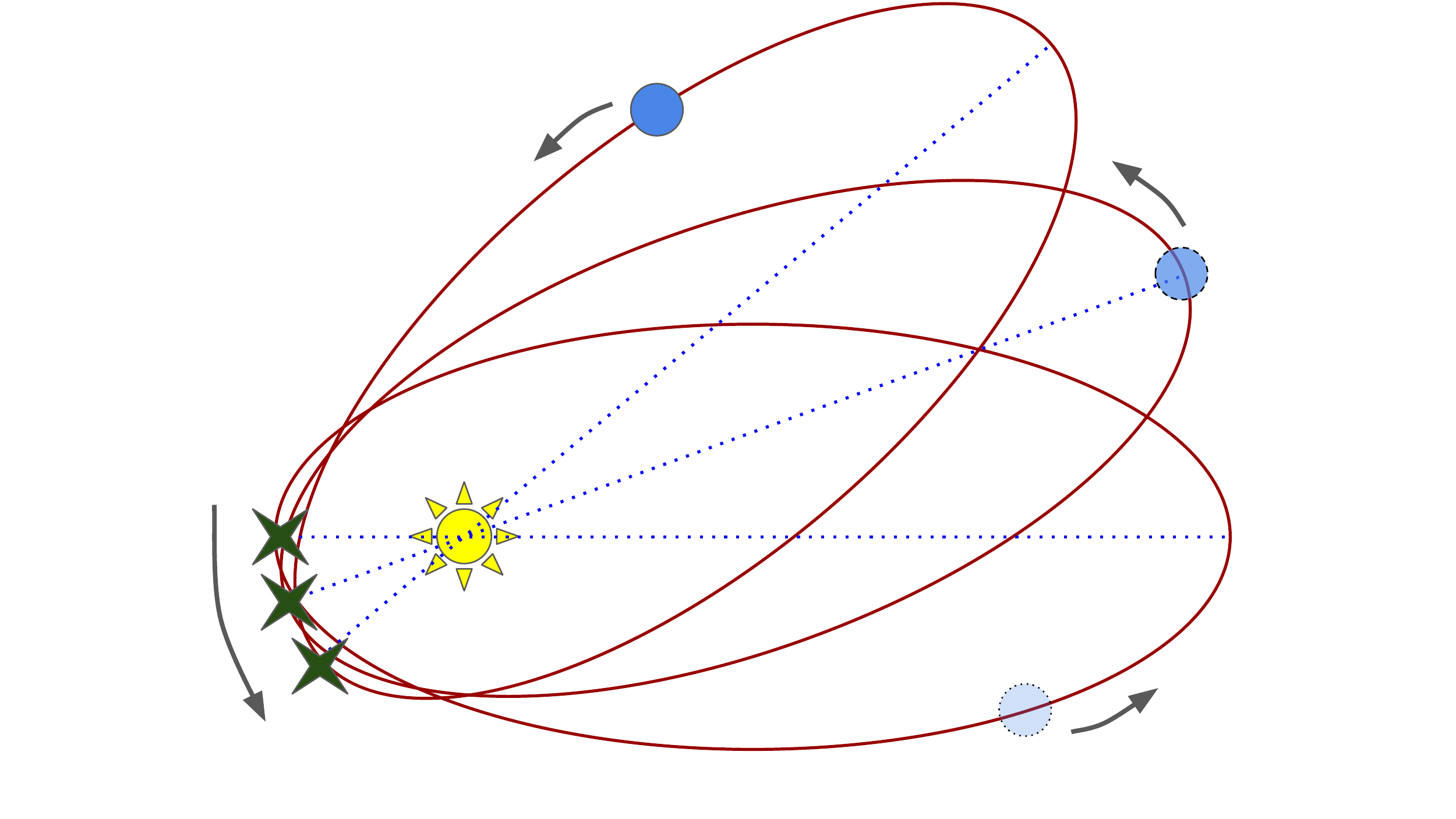}
\end{center}
\caption[Perihelion advance of Mercury]{
Exaggerated demonstration of the advance of Mercury's (blue) perihelion (green star) about the Sun (yellow).}
\label{fig:perihelion}
\end{figure}

\begin{spacing}{2.0}
Before Einstein's famous advent of general relativity (GR) in 1915, the law of universal gravitation put forth by Newton over two hundred years ago in 1687 was accepted as the ultimate expression of gravity.
At that time, while Newton himself was skeptical on this point, his theory was extremely accurate in its predictions of the attractive forces between massive objects.
However, as human ingenuity advanced over the following century, careful astronomical observations proved there to be some minute deviations between Newton's theory and what they found.
Foremost among such observations was the peculiar precession of Mercury's perihelion (as demonstrated in Fig.~\ref{fig:perihelion}), an effect well explained in Newton's gravity by the perturbations of other planets, the oblateness of the sun, and also tidal effects.
However, the observed precession of $574$'' per century did not agree well with the Newtonian calculation of $531$'' per century, an extra $43$'' per century that would later be perfectly accounted for by GR~\cite{RevModPhys.19.361,Park_2017}.
This effect, among others first prompted Einstein to begin work on his historic theory.

Built upon his theory of special relativity in 1905\footnote{Einstein's special relativity taught us that the laws of physics are the same for all observers in any inertial reference frame, and that the speed of light in vacuum is constant for all observers.}, Einstein's full theory of GR elegantly describes the relationships between mass and the curvature of spacetime.
Incredibly so, this complex theory of gravity is described by the simple set of field equation that made Einstein famous:
\begin{equation}
G_{\mu\nu}=\frac{8\pi G}{c^4} T_{\mu\nu}.
\end{equation}
In the above expression, the right hand side describes the energy/matter content of the system via the stress-energy tensor $T_{\mu\nu}$, while the left hand side describes the curvature of spacetime via the Einstein Tensor $G_{\mu\nu}$.
As John Archibald Wheeler aptly put it, ``\textit{Spacetime tells matter how to move; matter tells spacetime how to curve}''~\cite{misner_thorne_wheeler_2008}.
One astounding phenomena that appears out of the new theory of gravity is the presence of gravitational radiation, or \textit{gravitational waves} (GWs), from sources with time-dependent mass quadrupole moments.
Such radiation manifests itself as ripples in (or perturbations of) spacetime itself, traveling across the universe at the speed of light.

In the following century, Einstein's celebrated theory of gravity has made waves in the scientific community as it has perfectly explained every gravitational phenomena observed, and passed every test with flying colors.
The first three ``classical tests'' proposed by Einstein himself consisted of the following three observations: the perihelion precession of Mercury's orbit, the deflection of light about the Sun, and finally the gravitational redshift of light~\cite{Einstein}. 
The first test, the advancement of Mercury's perihelion as shown in Fig.~\ref{fig:perihelion}, can be shown to agree with the carefully observed value of $574$'' per century by including general relativistic effects.
In particular, the planets' deeper descent into the Sun's gravitational well at the perihelion relative to the aphelion causes an additional shift of $43$'' per century for Mercury, perfectly explaining the disagreement.
The second test of the deflection of light about the Sun was famously confirmed by the Eddington experiment on May 29, 1919.
By observing stars near the line-of-sight of the edge of the Sun both at night and during the total solar eclipse, they were amazingly able to observe the predicted $1.75$'' shift of star light predicted by GR.
Finally, in 1925 Walter Sydney Adams initially measured the predicted gravitational redshift\footnote{Gravitational redshift describes the shift of photon wavelengths to longer wavelengths when observed from a point higher in the gravitational potential. For example, light emitted by a star becomes redshifted as it travels out of the gravitational ``well''.} of light radiating from the massive star Sirius-B.
All such experiments have been improved upon significantly in the past century with continued outstanding agreement with Einstein's theory of GR.

Since the historic classic confirmations of Einstein's theory of GR, several more modern tests have been enacted in every spacetime environment available, all still agreeing strongly with GR's predictions.
Here in the local solar system for example, in 1964 Irwin Shapiro demonstrated the time-dilation radar photons experienced (the Shapiro delay) on a round-trip from Mercury and Venus just after being eclipsed by the Sun, an effect that agreed with the predictions of GR to $5\%$~\cite{Shapiro}.
More recent observations with the Cassini probe have narrowed down this agreement to $0.002\%$~\cite{Bertotti:2003rm}.
Later in 1968, Kenneth Nordtvedt proposed a test to probe the equivalence principle, which is a fundamental pillar of GR telling us that the trajectories of freely-falling bodies is independent of their internal structure.
The so-called Nordtvedt effect compares the relative motion between two bodies (e.g. the Earth and moon) which would experience no difference in acceleration in GR towards a third body (e.g. the Sun)~\cite{Nordtvedt_theory}.
This effect has been studied both here on Earth via Lunar Laser Ranging techniques~\cite{Nordtvedt_lunar} which has accurately measured the Earth-Moon distance to within a centimeter, as well as with binary pulsar\footnote{Pulsars are rapidly rotating neutron stars which emit extremely regular radio pulses on each rotation. Such stars have extremely stable orbits which make the pulses very precise and allow for very accurate monitoring of their orbital mechanics.} timing of triple systems~\cite{Stairs_BinaryPulsarTest,Wex_BinaryPulsarTest}.
All such observations to date have all similarly been found to agree with GR~\cite{Will_SolarSystemTest}.
Similarly, further pulsar timing observations have confirmed GR yet again from the accurate measurement of their perihelion precession~\cite{Stairs_BinaryPulsarTest,Wex_BinaryPulsarTest}, an effect much stronger than found here in the solar system.
Several other tests of GR have been enacted including the observations of gravitational lensing of distant objects behind massive galaxies, with table-top experiments of small masses, and even with large-scale cosmological observations as discussed in~\cite{Ferreira_CosmologyTest,Clifton_CosmologyTest,Joyce_CosmologyTest,Koyama_CosmologyTest,Salvatelli_CosmologyTest,Ishak:2018his}.
All such tests to date have proven Einstein to be correct yet again.

\begin{figure}
\begin{center} 
\includegraphics[width=\columnwidth]{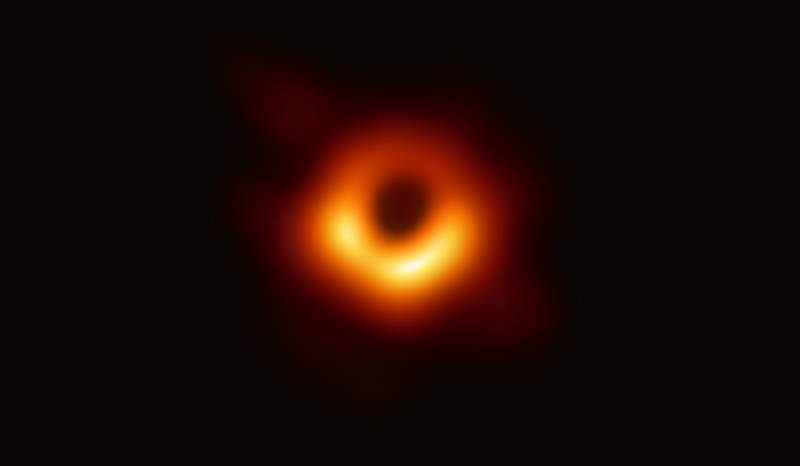}
\end{center}
\caption[The first picture of a black hole]{The very first image of a black hole. This image is of the supermassive black hole resident at the center of M87, and was taken by the EHT. (Taken from Ref.~\cite{1435174}).}
\label{fig:EHT}
\end{figure}

Even more recently, we have had the unique opportunity to probe the spacetime surrounding a supermassive black hole (BH) at the center of the M87 galaxy.
As described by the no-hair theorem, in GR, uncharged, stationary BHs are described by the Kerr spacetime metric which is parameterized only by their mass and spin.
Very recently in 2019, through superior feats of human engineering and scientific collaboration, the Event Horizon Telescope (EHT) effectively made use of an Earth-sized telescope~\cite{doeleman2009imaging} to take an image of the famous supermassive BH M87$^*$~\cite{1435171,1435174,1435175,1435177,1435168} as shown in Fig.~\ref{fig:EHT}.
By combining the observations of seven millimeter and sub-millimeter instruments across the globe (very-long baseline interferometric, or VLBI, array), this incredible feat has once again failed to disprove Einstein's theory of gravity in yet another observation.

\begin{figure}
\begin{center} 
\includegraphics[width=\columnwidth]{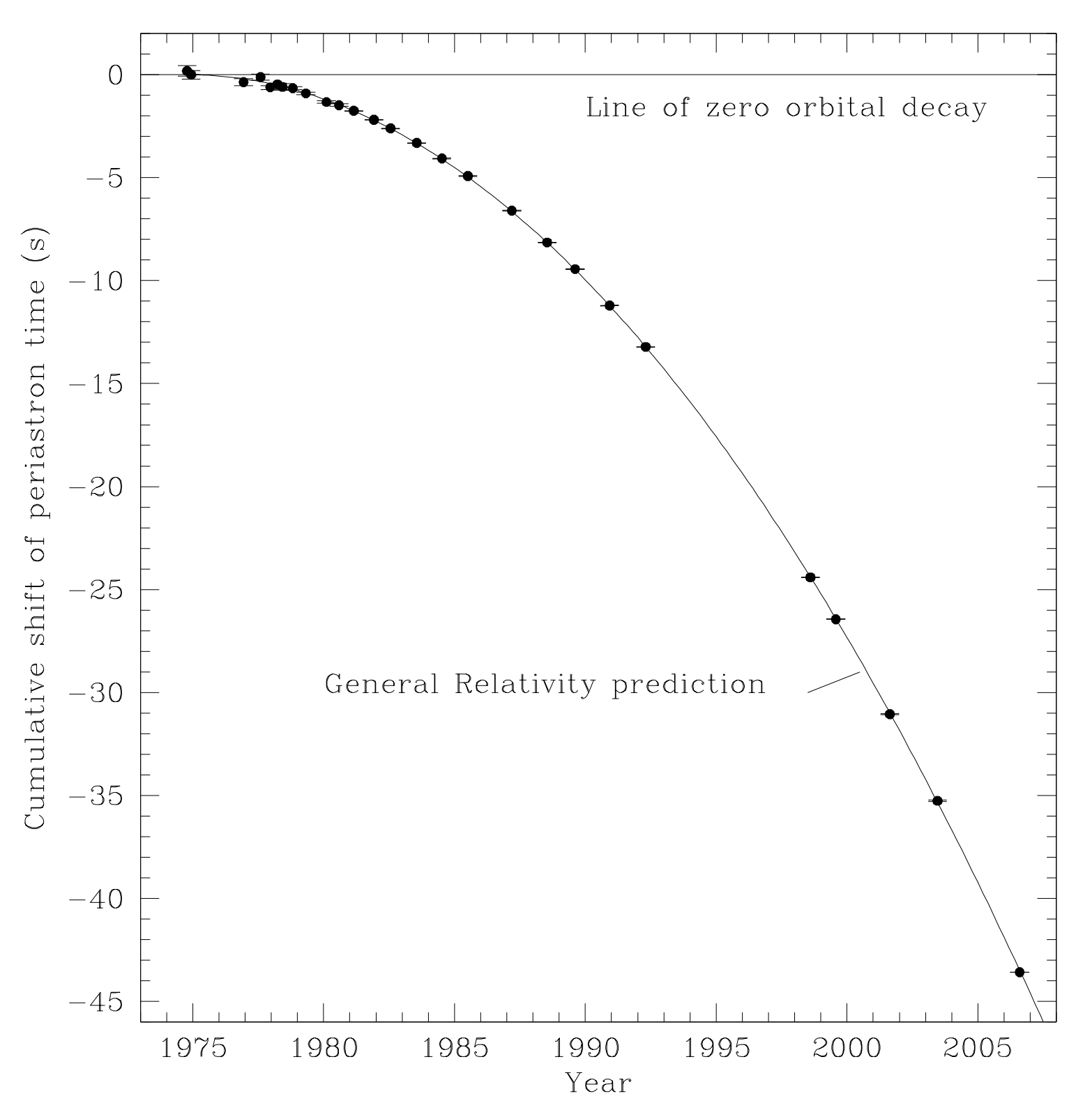}
\end{center}
\caption[Orbital decay of PSR B1913+16 via emission of gravitational waves]{Decay of the orbital period of the Hulse-Taylor binary pulsar system PSR B1913+16~\cite{HulseTaylor} taken with three decades of timing observations (data points with error bars ``mostly too small to see''). The solid line depicts the expected orbital decay from the radiation of GWs, which matches very well. (Taken from Ref.~\cite{Weisberg:2010zz}).}
\label{fig:indirect}
\end{figure}

However, one aspect of Einstein's theory was still missing.
Prior to the year 1974, there had been no convincing proof of one of the most outstanding predictions of Einstein's -- the existence of GWs.
In this same year, Hulse and Taylor discovered what would turn out to be one of the first clearly convincing cases of gravitational radiation in an indirect observation.
Using the $305$ m Arecibo radio observatory, the team at the University of Massachusetts Amherst famously observed the familiar pulses from a rapidly rotating pulsar $6.4$ kpc away, with a period of only 59 ms.
By modeling the originating system, they found it to be in a binary orbit with what turned out to be another NS -- the first, and only discovery of a binary pulsar system to date.
Named PSR B1913+16, or the ``Hulse-Taylor'' binary~\cite{HulseTaylor} after its discoverers, this fascinating system was carefully studied over the following three decades.
The resulting work by Weisburg, Nice, and Taylor over thirty years later offered us the first indirect observation of gravitational radiation~\cite{Weisberg:2010zz}.
As shown in Fig.~\ref{fig:indirect}, we see that the observed decay in the orbital period of the Hulse-Taylor binary matches perfectly with the predictions of GR, which tells us that GWs carry energy and angular momentum away from the system.
This loss results in a slow inspiral between the two objects until they finally lose enough energy to merge together.
This first remarkable indirect observation of GWs became yet another convincing win for Einstein, nearly sixty years past GR's formulation, and in result, it led the duo to receive the Nobel Prize in 1993 ``\textit{for the discovery of a new type of pulsar, a discovery that has opened up new possibilities for the study of gravitation}''~\cite{hulse_taylor}.

\begin{figure}
\begin{center} 
\includegraphics[width=\columnwidth]{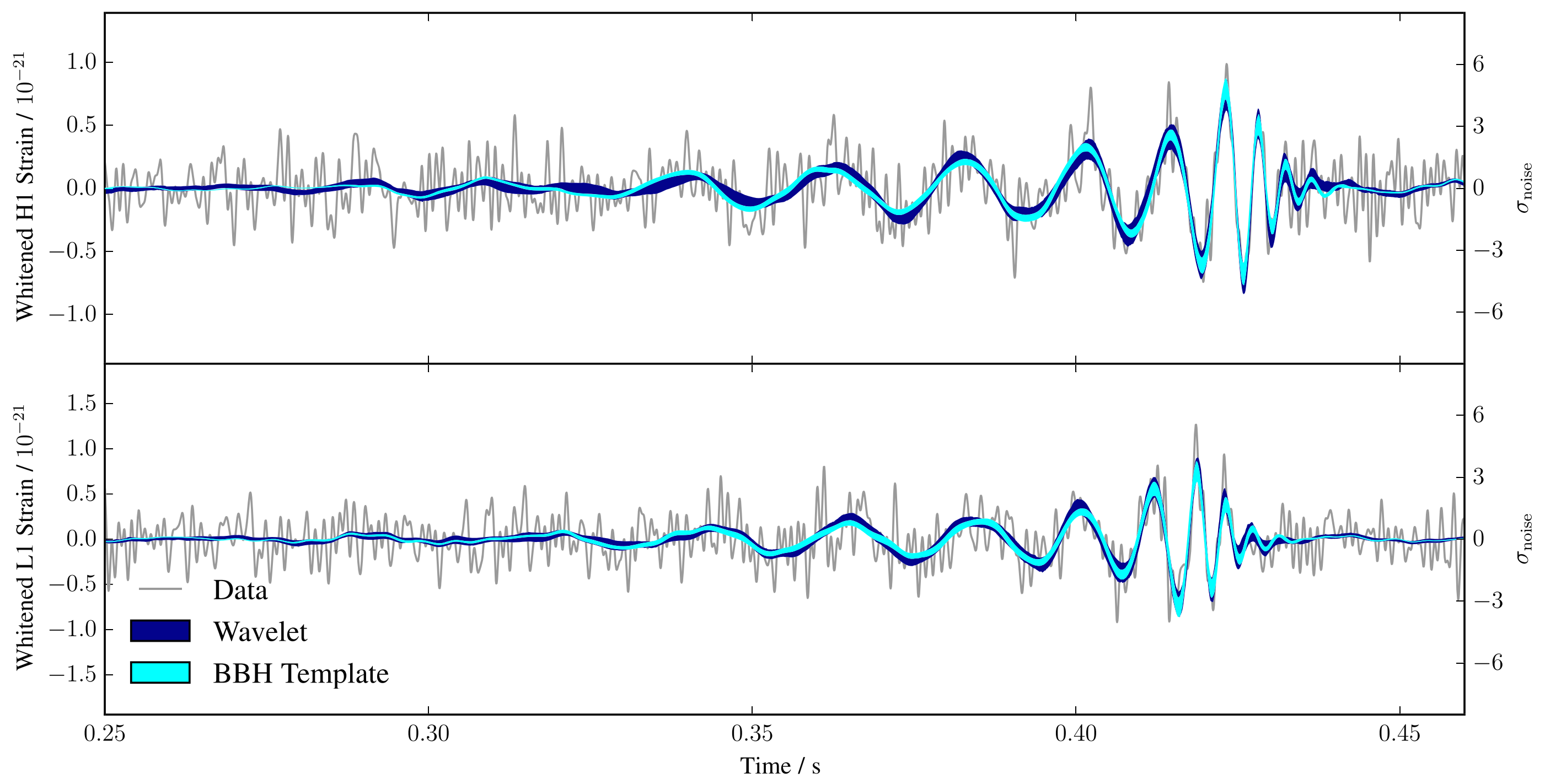}
\end{center}
\caption[GW150914: The first observation of gravitational waves]{Gravitational wave strain signal (gray) from the Hanford (top) and Livingston (bottom) GW interferometers. The overlaid blue curve was obtained from a linear combination of sine-Gaussian wavelets that did not assume any particular morphology, while the cyan curve is a result from the PhenomD waveform model. (Taken from Ref.~\cite{GW150914}).}
\label{fig:GW150914}
\end{figure}

Forty years later on the historic day of September 14, 2015 the LVC for the very first time directly observed GWs from a coalescing pair of BHs~\cite{GW150914}.
Aptly named GW150914, this famous event rocked the world of GW physics, finally confirming one of Einstein's last predictions from his theory of gravity.
Through a true marvel of modern engineering and scientific collaboration, both of the LIGO interferometers located in Hanford Washington and Livingston Louisiana simultaneously observed the characteristic GW chirp signal of a merging compact binary as shown in Fig.~\ref{fig:GW150914}.
Careful Bayesian analysis of this event led the LVC to determine this signal to have originated from a merging pair of BHs with masses of $35.8\text{ M}_\odot$ and $29.1\text{ M}_\odot$, located $410$ Mpc away.
This cataclysmic event reached nearly $60\%$ the speed of light before finally colliding, releasing a total amount of energy equivalent to $3\text{ M}_\odot$~\cite{GW150914}.
This incredibly powerful event released fifty times the energy radiated from all of the stars in the observable universe over a period of only fractions of a second.
The remnant BH was left behind with $62\text{ M}_\odot$, rotating with a dimensionless spin\footnote{The dimensionless spin $\chi$ is defined as the total angular momentum in the $\hat{z}$-direction divided by its mass squared, or $\chi\equiv J_z/m^2$. The maximum value this parameter can take is $1$, else the appearance of naked singularities, which have yet to be observed.} of $67\%$ of its maximum value.
The resulting GWs from both the inspiral, the merger, and the ringdown of this event then traveled for $1.3$ billion years to Earth at the speed of light, to finally be detected with minute deviations of the LIGO interferometer arm lengths.
One typically classifies such perturbations of spacetime by a quantity known as the \textit{strain}, simply given by $\Delta L/L$, where $L$ is the characteristic distance between two test masses, and $\Delta L$ is the resulting change in length between the two given by the passing of a gravitational wave.
As can be seen in Fig.~\ref{fig:GW150914}, such strains are typically on the order of $\sim10^{-3}$ of the radius of a proton for arm lengths of $L\sim\mathcal{O}(1\text{ km})$!
Over the following 4 years, the LVC has further detected nine more GW signals from merging BH binaries~\cite{LIGOScientific:2018mvr}, as well as one detection of GWs from a merging pair of neutron stars (NSs), GW170817~\cite{TheLIGOScientific:2017qsa} in their first two observing runs (O1 and O2).
The current observing run (O3) has further identified over fifty new GW candidates~\cite{gracedb} of binary BH, binary NS, and even BH-NS merger systems.
Among such exciting new candidate events is the second confirmed observation of a binary NS coalescence~\cite{GW190425} with a large total mass of $\sim3.4\text{ M}_\odot$.

The famous GW event GW150914 has opened an entirely new window into observing the universe from an entirely new, unique perspective.
With gravitational interactions being the weakest of all four forces found in our universe, the resulting interaction between GWs and the matter they encounter is extremely limited, and they can pass through many obstacles without reserve.
For example, GWs can easily escape unhindered from the optically thick regions of spacetime that are occupied by high densities of matter, which can not be probed by traditional electromagnetic (EM) observations.
Primary among examples of such situations is in the very early universe prior to the last scattering surface when photons were unable to escape the hot-dense gas that was the universe.
Previously undetectable with EM observations, such times may now be observed via the primordial gravitational radiation which was emitted long before in the period of cosmic inflation.
In addition, GWs allow us to probe, for the very first time in human history, the \textit{extreme gravity} environments of spacetime, where the fields are extremely strong, non-linear, and highly dynamical. 
Prior to this new era of astrophysical observation, we only had access to the weak and static regions of spacetime found in e.g. the solar system, on Earth, with pulsar timing observations, and of large-scale cosmological observations.
This new window into the universe will now allow us to probe several aspects of fundamental physics from an entirely new standpoint.

\begin{figure}
\begin{center} 
\includegraphics[width=0.7\columnwidth]{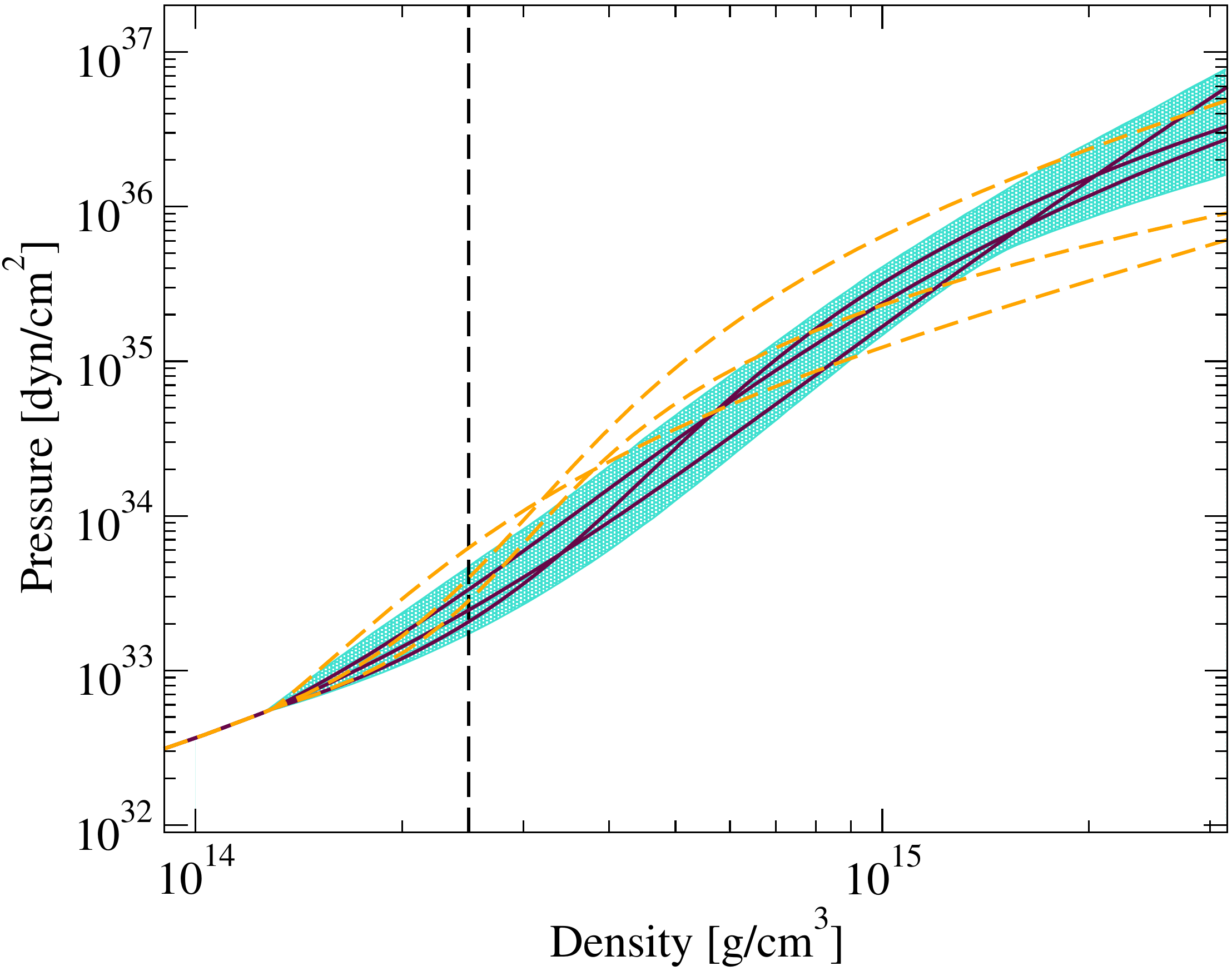}
\end{center}
\caption[The neutron star equation of state]{Six example EoS's of nuclear matter found within NSs. The shaded turquoise region corresponds to the 90\% confidence interval on the EoS from the observation of GWs from GW170817~\cite{LIGO:posterior}. The dashed vertical line corresponds to the nuclear saturation density. The spectral EoSs displayed here were randomly generated for demonstration purposes, as discussed further in Chapter~\ref{chap:EoSInsensitiveRelations}, with the solid curves agreeing with the LVC result, and the dashed ones not.}
\label{fig:introEos}
\end{figure}

Besides testing Einstein's theory of GR in the extreme gravity environments found outside binary BH coalescences, another fascinating subject to explore with our newfound observation of GWs is nuclear physics.
While nuclear matter has been extensively studied here in our local terrestrial environments, the behavior of nuclear matter beyond the saturation density of $\rho_0 \approx 2.5 \times 10^{14} \text{ g/cm}^3$ has been poorly modeled up until now, due to the lack of physical observations.
Closely related to the short-range nature of the strong force~\cite{saturation}, the aforementioned nuclear saturation density dictates the average density of atomic nuclei, regardless of their location on the periodic table.
As a result, without large external forces acting on matter, here on Earth we are mostly limited to nuclear matter observations at or below the saturation density $\rho_0$.
A useful identifying thermodynamic function of nuclear matter is known as the \textit{equation of state} (EoS), which relates the pressure and density of nuclear matter.
See Figure~\ref{fig:introEos} for several examples of the NS EoS.
Until now, terrestrial experiments have probed and constrained the nuclear EoS up to around the saturation density with various experiments studying e.g. heavy-ion, neutron-rich collision interactions~\cite{Li:HeavyIon,Tsang:SymmetryEnergy,Centelles:NeutronSkin,Li:CrossSections,Chen:SymEnergy}.
However, the behavior of the EoS beyond the saturation density is mostly uncertain up until now, and is currently one of the largest unsolved mysteries in both nuclear physics and astrophysics.

Throughout a star's life, the intense inward gravitational pressure creates a hot, dense environment in the core which initiates nuclear fusion, resulting in an outward flow of energy that supports the star against gravitational collapse~\cite{clayton_2007,Laughlin_1997}.
Beginning with the fusion of the lightest element helium into hydrogen, this process continues until the star runs out of fuel, at which point the process will start over again with the fusion of hydrogen into carbon.
The cycle repeats, moving through the heavier elements in the periodic table until a lack of stellar mass is available to re-ignite fusion of the next heavier element.
For the most massive of stars, the process continues through neon, oxygen, silicon, iron, and even nickel; at which point there exists no remaining stable elements for fusion to occur.
At this point, the inward gravitational force overcomes any outward forces present within the star, and the core  collapses inwards as the outer lays are ejected out into space in a supernova event, providing the universe with several of the heavier elements that we see today here on Earth\footnote{As confirmed with GW170817~\cite{TheLIGOScientific:2017qsa} by Kasen \textit{et al} in Ref.~\cite{Kasen:2017sxr}, the heaviest elements found on the periodic table are formed from the kilanova event preceding the coalescence of two NSs.}.
With smaller-mass stars below $10\text{ M}_\odot$, the core collapse is eventually halted by the presence of an electron degeneracy pressure -- a direct outcome of the Pauli exclusion principle restricting fermions to exist in the same state as each other.
The resulting white dwarf is a dense body with masses between $0.15\text{ M}_\odot$ and $1.2\text{ M}_\odot$, and radii between 4,000 km and 16,000 km (comparable to that of Earth).
On the other hand, for giant stars with masses between $10\text{ M}_\odot$ and $29\text{ M}_\odot$, the extreme gravitational force exceeds that of even the electron degeneracy pressure, further collapsing the core down until it is halted yet again by a similar neutron degeneracy pressure, along with strong repulsive nuclear forces.
At this point, the pressures are so extreme that even molecules and individual atoms are torn apart, with the electrons and protons energetic enough to form more neutrons.
The result is an ultra-compact object (NS) composed mostly of neutrons and subatomic particles, with masses between $1\text{ M}_\odot$ and $2\text{ M}_\odot$, and radii around $11-12$ km (comparable to that of a city!).
Within the NS is the existence of \textit{supranuclear matter}, with densities reaching upwards of five times that of the nuclear saturation density - providing us with the excellent opportunity to probe the EoS beyond the atomic densities found here on Earth. 
Finally, for extremely massive stars with greater than $29\text{ M}_\odot$, the inward gravitational pressure is large enough to overcome even the neutron degeneracy pressure and strong repulsions present in the NS, and it collapses down to a single point of singularity masked by an event horizon, known as a BH.

Apart from BHs which are vacuum solutions to Einstein's field equation, NSs exist as the most compact (observed) objects in the universe.
The resulting supranuclear matter present within these extreme objects is vital to the goal of studying the EoS beyond the saturation density, something unreachable here with terrestrial observations.
In particular, the nuclear matter EoS is extremely important because it is intrinsically deterministic of the global NS observables, including the mass, radius, moment of inertia, quadrupole moment, and tidal deformability.
Before the observation of GWs from such objects, several EM observations of NSs gave way to new probes of the EoS.
For example, in the past, x-ray observations of the NS's mass and radius have been used to constrain the EoS via the mass-radius relationship~\cite{guver,ozel-baym-guver,steiner-lattimer-brown,Lattimer2014,Ozel:2016oaf}.
More recently in 2019, the Neutron Star Internal Composition Explorer (NICER) mounted on the International Space Station successfully mapped the NS mass-radius profile via x-ray observations of local hotspots rotating with the NS~\cite{Riley_2019,Miller_2019,Bogdanov_2019,Bogdanov_2019b,Guillot_2019,Raaijmakers_2019}.
Several follow-up investigations have further transformed such probability distributions of the NS radius with uncertainties on the order of $\sim10\%$ into constraints on the EoS~\cite{Raaijmakers_2019,Christian:2019qer,Jiang:2019rcw,Raaijmakers:2019dks}.
Further in the future, the Square Kilometre Array (SKA) projected to be built in Australia and South Africa by $\sim$2027 is expected to make measurements of the NS's moment of inertia via high precision timing of the periastrion advancement, further contributing to constraints on the nuclear EoS.

\begin{figure}
\begin{center} 
\includegraphics[width=.7\columnwidth]{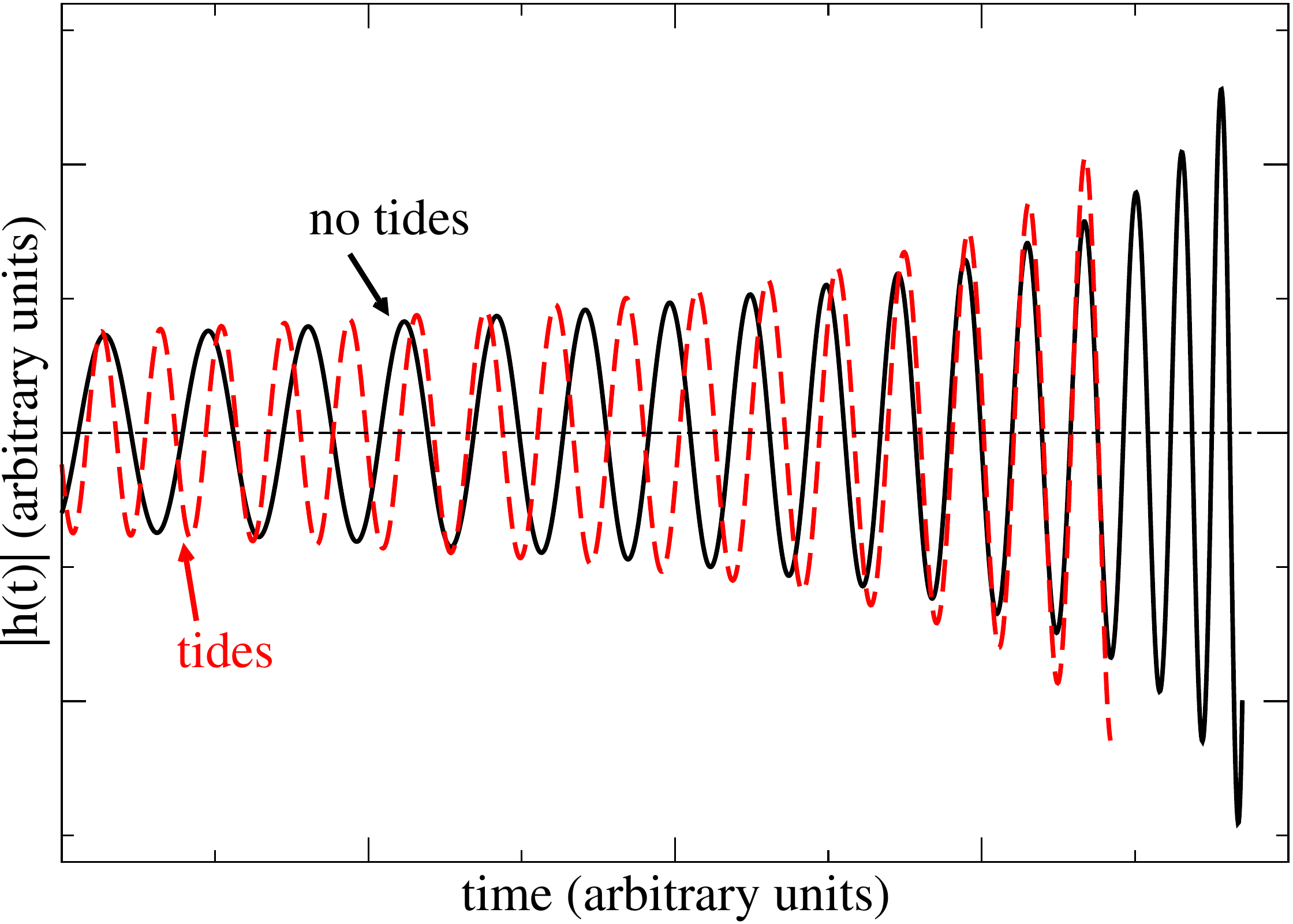}
\end{center}
\caption[Gravitational wave signal with and without tides]{Gravitational wave signal with and without tides imposed on the compact objects. The former occurs for e.g. merging BHs, while the latter occurs for e.g. NSs.}
\label{fig:tides}
\end{figure}

On August 17, 2017 the prospects of observing NSs changed forever with the historic GW detection of a binary NS coalescence, named GW170817~\cite{TheLIGOScientific:2017qsa}.
By measuring the amount of tidal influence on the gravitational waveform (as demonstrated in Fig.~\ref{fig:tides}), the LVC was able to place for the very first time an observational constraint on the NS \textit{tidal deformability}, which characterizes the deformation of nuclear matter in response to an external tidal field.
This observation, not possible with ordinary EM detections, led the LVC to further place yet another constraint on the supranuclear matter EoS~\cite{LIGO:posterior} as shown by the shaded region in Fig.~\ref{fig:introEos}.
Followed up by the energetic gamma-ray burst GRB170817A~\cite{Kienlin2017,Goldstein2017} and several resulting EM observations over the following weeks~\cite{Lipunov2017} from the same source, this event became the poster-child for multi-messenger astronomy.
For the first time ever, we now have the unique ability to combine both EM observations (e.g. the mass, radius, moment of inertia) with GW observations of the tidal deformability to place even further constraints on the nuclear matter EoS.

Another method one can use to probe the NS EoS was first introduced by Yagi and Yunes in 2013~\cite{Yagi2013}, known as the ``I-Love-Q'' universal relations.
In particular, here they found that certain combinations of the dimensionless NS observables I (moment of inertia), Love (tidal deformability), and Q (quadrupole moment) displayed EoS-insensitive properties when plotted amongst each other, with uncertainties below $\sim1\%$ for each independent relationship.
Similarly in 2013, the same authors (and others) found approximately universal relations between the NS tidal deformability and the stellar compactness (the ratio between the mass and radius)~\cite{Yagi:2013bca,Yagi:ILQ,Maselli:2013mva,Yagi:2016bkt}.
Later in 2014, Yagi developed the ``multipole Love'' universal relations~\cite{Yagi:Multipole} between higher-order multipolar moments of the NS tidal deformabilities.
Such universal relations are highly useful in their ability to analytically express NS observables in terms of others in an EoS-independent way.
For example, a determination of one NS observable (say, the dimensionless moment of inertia) could allow an automatic determination of the dimensionless quadrupole moment with relatively small systematic uncertainties using the I-Love-Q universal relations.
This same idea was implemented recently to place EoS-independent constraints on the mass-radius curve using GW data~\cite{Katerina:residuals,LIGO:posterior}.
Finally, in 2015 and 2016 the same authors developed yet another important universal relation used extensively by the LVC, known as the ``Binary Love'' universal relations~\cite{Yagi:2015pkc,Yagi:binLove}.
Within the gravitational waveform for binary NS mergers, the individual tidal deformabilities $\Lambda_{1,2}$ of each NS are highly degenerate with each other, and are very difficult to measure.
Instead, the \textit{mass-weighted tidal deformability} $\tilde\Lambda$ is dominant in the gravitational waveform among all tidal effects present, and is thus what the LVC measured in~\cite{TheLIGOScientific:2017qsa}.
The Binary Love universal relations then show EoS-independence between symmetric and anti-symmetric combinations of the individual tidal deformabilities $\Lambda_{s,a}=(\Lambda_1\pm\Lambda_2)/2$.
With such relations, one can express $\Lambda_s(\Lambda_a)$ (or vice-versa) which allows the mass-weighted tidal deformability to be written as a function of $\Lambda_a$ only.
Finally, the constraint on $\tilde\Lambda$ allows a constraint on $\Lambda_a$, which implies a bound on $\Lambda_s$, which further allows one to infer properties of the individual tidal deformabilities $\Lambda_{1,2}$ with only a small amount of systematic uncertainty.

In this thesis, we first consider how one can harness the information encoded within GWs to probe various aspects of fundamental physics, such as were previously impossible before the first detections by the LVC.
In Part~\ref{part:ProbingNuclearPhysics} we focus on probes of nuclear physics, in particular of the supranuclear matter found only within NSs.
Specifically, as two NSs orbit one another such as in GW170817, the immense gravitational tidal fields from the neighboring stars induces a deformation of each body.
Characterized by the tidal deformability of nuclear matter, this effect is magnified as the NSs approach one another under the emission of gravitational radiation.
This deformation ultimately alters the ensuing inspiral trajectory, a relic which is encoded directly into the GW signal as demonstrated in Fig.~\ref{fig:tides}\footnote{We observe that systems with tides merge sooner than those without. This is because compact objects with tides such as NSs, have an extended structure and thus a non-zero quadrupole moment. This additional moment introduces new gravitational radiation than would be expected for point particles such as BHs.}.
Being vacuum solutions to Einstein's field equations, merging BH binaries do not experience such an effect, thus the tidal deformation present in a GW signal can be used to constrain the tidal deformability of the NSs.
For the first time ever, GWs have allowed us to measure the tidal deformability of compact objects as was done by the LVC in Ref.~\cite{TheLIGOScientific:2017qsa}, something not previously possible with EM observations alone.
By using such constraints, we show how one can further constrain the EoS of supranuclear matter by probing the fundamental nuclear matter parameters which determine it in a model-independent way. 
In particular, we consider constraints on the EoS found from the current observation of GW170817, as well as improvements to be made with future GW detections, including the combination of multiple events.
Finally, we find improvements to several universal relationships between NS observables from the observation of GW170817.
Such new relations are then shown to assist in better parameter estimation of future GW detections.

For the second half of the thesis, we consider how one can test GR using GW observations.
For the first time in history, we have the unique opportunity to probe the extreme gravity spacetime environments present around the coalescence of binary BHs.
While GR has been experimentally and observationally confirmed everywhere else in the spacetime parameter space, one can imagine that this might cease to be true in more exotic environments.
Similar to the accuracy of Newton's theory of gravity before the advent of high-powered telescopes and observations of stronger gravitational effects, GR may yet prove to simply be a subset of a more grand theory of gravity.
While GR may be perfectly accurate in the weak-field or static spacetime environments considered thus far, an alternative theory of gravity could potentially activate in such extreme-gravity spacetimes that we may now probe.
However, despite this we still ask -- why must we continue to test a theory of gravity that continually surprises us with its extreme accuracy of our observations, even in light of the eleven observed GW signals all remaining consistent with GR?
This is because even with all of its successes, there still remains several astrophysical mysteries and open questions, which can not be described by GR alone.
For example, the yet-unexplained unification of GR with quantum mechanics~\cite{Clifton:2011jh,Joyce:2014kja,Famaey:2011kh,Milgrom:2008rv,Jain:2010ka,Koyama:2015vza}, dark matter and the unexpected rotation curves of galaxies~\cite{Famaey:2011kh,Milgrom:DarkMatter,Milgrom:2008rv,Clifton:2011jh,Joyce:2014kja}, dark energy and the late-time accelerated expansion of the universe~\cite{Jain:2010ka,Salvatelli:2016mgy,Koyama:2015vza,Joyce:2014kja}, or even the rapid inflationary period of the early universe~\cite{Joyce:2014kja,Clifton:2011jh,Famaey:2011kh,Koyama:2015vza} are all prevalent examples of unexplained phenomena which could potentially be attributed to an alternative theory of gravity.

\begin{figure}
\begin{center} 
\includegraphics[width=.7\columnwidth]{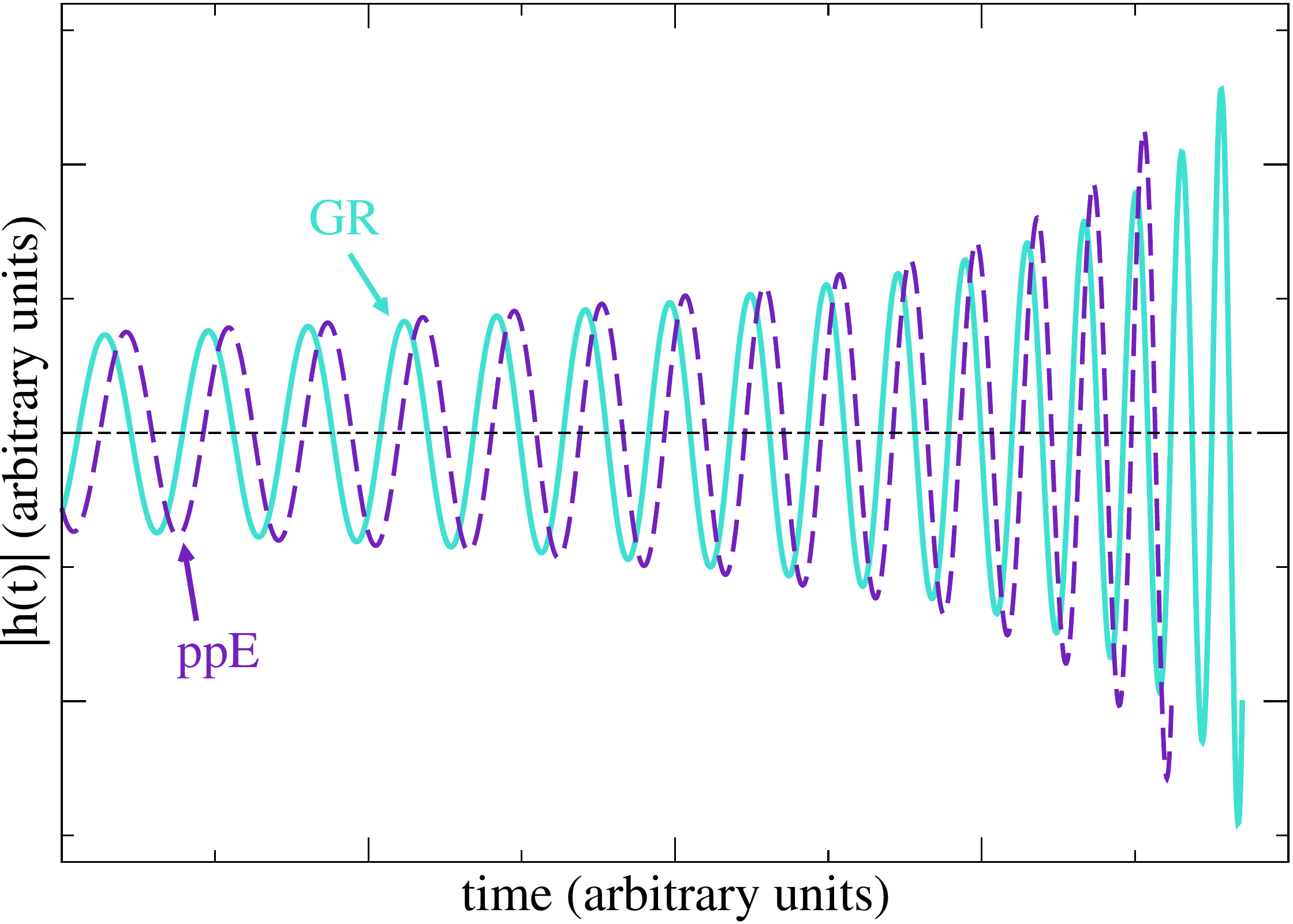}
\end{center}
\caption[Gravitational wave signal with and without ppE corrections]{Gravitational wave signal with and without ppE corrections injected beyond GR. This signal was generated for $-1$PN effects (i.e. $(v/c)^{-7}$) corrections to the the waveform for demonstration purposes.}
\label{fig:GRvsNonGR}
\end{figure}

In Part~\ref{part:TestingGR} we continue to test GR using the observations of GW signals from merging BH (and BH-NS) binaries.
As one can imagine, scientific theories such as GR can never be entirely proven.
However, what we can do is test them in an agnostic way to constrain alternative theories of gravity that work beyond-GR.
To do this, one popular method is to parameterize the gravitational waveform found in GR ($\tilde{h}_\GR(f)=A_\GR e^{i\Psi_\GR}$) in a theory agnostic way that deviates beyond GR.
Known as the \textit{parameterized post-Einsteinian} (ppE) formalism as laid out by Yunes \textit{et al} in Ref.~\cite{Yunes:2009ke}, we introduce general arbitrary corrections to the GR amplitude ($\Delta A$) and phase ($\Delta\Psi$) in the frequency domain as
\begin{equation}
\tilde{h}_\GR(f)=A_\GR(1+\Delta A) e^{i\Psi_\GR+i\Delta\Psi}.
\end{equation}
The arbitrary modifications $\Delta A$ and $\Delta\Psi$\footnote{These corrections depend on the power of velocity relative to the speed of light at which given effects alter the gravitational waveform.} are completely theory-agnostic and can be tested and constrained as so.
As demonstrated in Fig.~\ref{fig:GRvsNonGR}, ppE modifications like so alter the ensuing orbital binary inspiral, which affects the resulting detected gravitational waveform signal.
Typically, modified theories of gravity increase the amount of radiation emitted off the binary system, accelerating the inspiral thus predicting an earlier time of merger than would be predicted by GR alone.
By constraining the sizes of theory-agnostic parameters $\Delta A$ and $\Delta\Psi$, one can then map such constraints to ones on various theory-specific parameters found in proposed alternative theories of gravity.

\begin{figure}
\begin{center} 
\includegraphics[width=1\columnwidth]{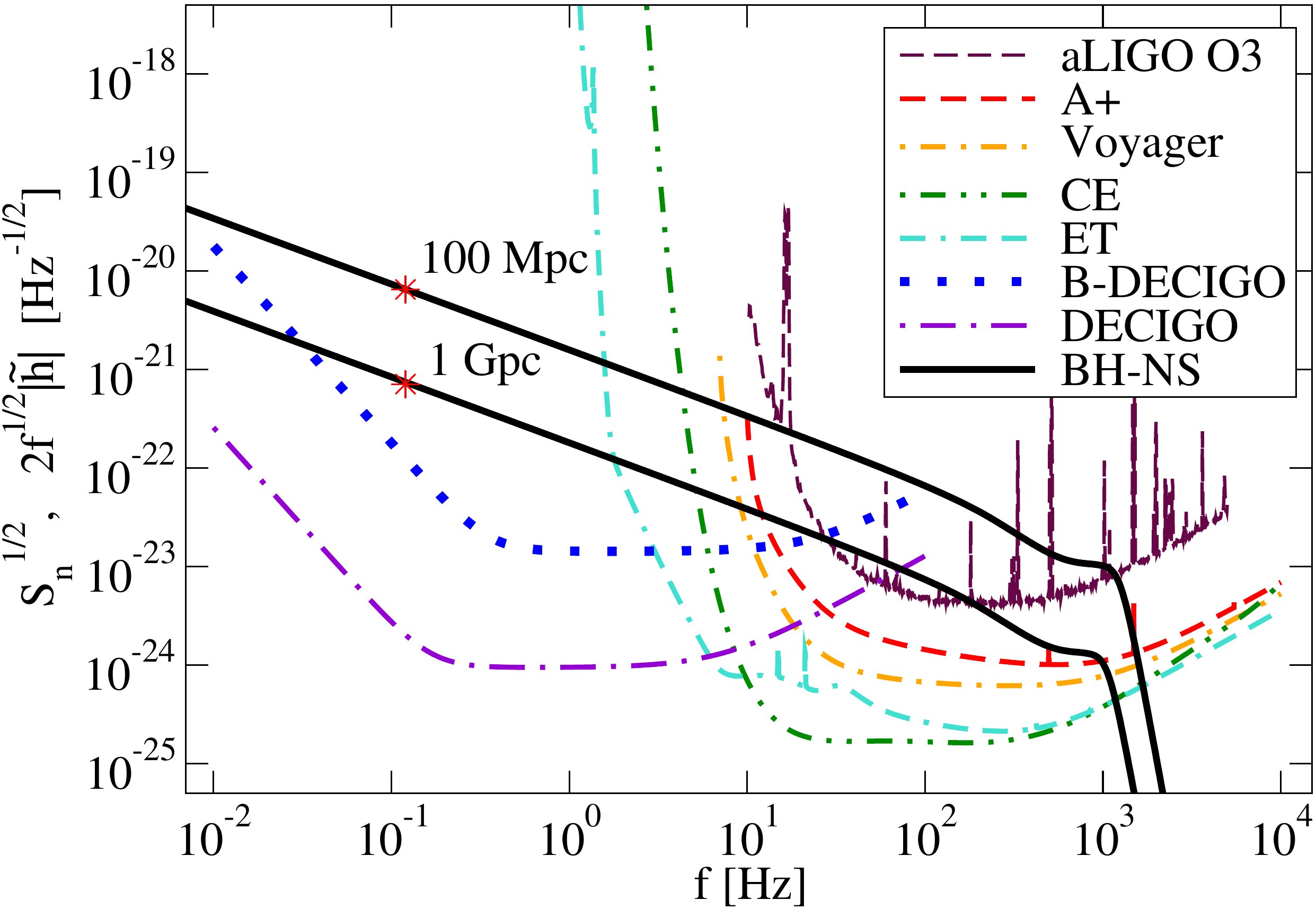}
\end{center}
\caption[Gravitational wave detector sensitivities]{
Spectral noise densities $\sqrt{S_n(f)}$ for both the ground-based (solid) and space-based (dashed) detectors considered in this thesis.
In particular, we consider the ground-based detectors LIGO O2~\cite{aLIGO,O3}, aLIGO~\cite{aLIGO}, A\texttt{+}~\cite{AppData,Ap_Voyager_CE}, Voyager~\cite{VRTCEETData,Ap_Voyager_CE}, CE~\cite{VRTCEETData,Evans:2016mbw,Ap_Voyager_CE}, and ET-D~\cite{ET,Evans:2016mbw,VRTCEETData} as interpolated from publicly available data.
The space-based detectors considered are TianQin~\cite{TianQin}, LISA~\cite{LISA}, B-DECIGO~\cite{B-DECIGO}, and DECIGO~\cite{DECIGO}.
Additionally shown is the characteristic amplitude (dotted) $2\sqrt{f}|\tilde{h}(f)|$ for a GW150914-like event using the IMRPhenomD~\cite{PhenomDI,PhenomDII} gravitational waveform template. The ratio between GW spectrum and signal roughly corresponds to signal-to-noise ratio.
}
\label{fig:sensitivities}
\end{figure}

In this thesis, we first consider two tests of GR: the parameterized tests and the inspiral-merger-ringdown consistency tests.
In the former, we introduce arbitrary ppE modifications to the GW signal to ultimately constrain several alternative theories of gravity, both with current GW observations and future ones, including with future GW interferometers, with the combination of multiple detected events, and even with the multi-band observations between both ground-based and space-based GW detectors.
As displayed in Fig.~\ref{fig:sensitivities}, future GW detectors both on the ground (aLIGO O2~\cite{aLIGO,O3}, aLIGO~\cite{aLIGO}, A\texttt{+}~\cite{AppData,Ap_Voyager_CE}, Voyager~\cite{VRTCEETData,Ap_Voyager_CE}, CE~\cite{VRTCEETData,Evans:2016mbw,Ap_Voyager_CE}, and ET-D~\cite{ET,Evans:2016mbw,VRTCEETData}) as well as ones in space (TianQin~\cite{TianQin}, LISA~\cite{LISA}, B-DECIGO~\cite{B-DECIGO}, and DECIGO~\cite{DECIGO}) promise hefty improvements in sensitivity in both the low-frequency and high-frequency regimes.
In the latter, we find new corrections to not only the inspiral gravitational waveform, but also to the merger-ringdown portion, and to the predictions of the remnant BH mass and spin.
We then test the consistency between the inspiral and merger ringdown signals, finding the magnitude of beyond-GR corrections required to become inconsistent, and thus providing evidence of non-GR behavior present within the observed signal.
Finally, we consider probes of the spacetime environments themselves outside BHs with EM observations.
We first construct a new spacetime metric that is generally parameterized beyond the Kerr one that describes BHs in GR, while still preserving the Kerr symmetries.
In the new spacetime, we then find the resulting corrections to several astrophysical phenomena including the photon rings (stable circular equatorial photon orbits), and the orbital energy, angular momentum, Keplerian and epicyclic frequencies of particle orbits.
Such phenomena could be observed and tested in the future by e.g. the EHT.

\section{Executive summary of results}\label{intro:Summary}
 
Here we provide a brief summary of results for busy readers.
In particular, we begin with a recap of the primary findings of Chapter~\ref{chap:ConstrainingNuclearParameters}, where the observation of binary NS merger event GW170817 allowed us to place constraints on nuclear matter physics both in the present and in the future.
Following this, we summarize Chapter~\ref{chap:EoSInsensitiveRelations} where the results of the same event GW170817 allowed us to create more accurate universal relationships between NS observables.
We then show how such new relationships allow for better determination of such observables with future GW detections.
We then recap Chapters~\ref{chap:Parameterized} and~\ref{chap:IMRconsistency} where we focus on testing GR in the extreme gravity regime of binary BH coalescences both in the present and the future, using both parameterized tests and inspiral-merger-ringdown consistency tests.
Finally, we wrap up with a summary of Chapter~\ref{chap:beyondKerr}, where we focused on alternative spacetime metrics beyond the Kerr one and their resulting corrections to both the gravitational waveform and also several EM astrophysical observables.

\subsection{Constraining nuclear matter parameters}
Here we summarize the important findings of Chapter~\ref{chap:ConstrainingNuclearParameters}.
Dominant among all tidal parameters in the gravitational waveform phase, the mass-weighted tidal deformability $\tilde\Lambda$ was measured stringently by the LVC from binary NS merger GW170817.
First, we find new universal relations between this tidal deformability and nuclear matter parameters $K_0$, $M_0$, or $K_{\text{sym},0}$ for a number of mass ratios allowed by GW170817.
Contrary to previous work, we find low-order nuclear parameters $K_0$ and $M_0$ to have very poor correlations, due to our inclusion of a broad new class of EoSs, taking into account all possible sources of systematic uncertainties.

Additionally, we studied similar universal relations between $\tilde{\Lambda}$ and linear combinations of nuclear parameters, such as $K_0 + \alpha L_0$, $M_0 + \beta L_0$, and $K_{\text{sym},0} + \gamma L_0$.
We found that such relations typically have a stronger correlation than that in the case of individual nuclear parameters.
This is consistent with the findings of Ref.~\cite{Malik2018} on correlations between nuclear parameters and \textit{individual} tidal deformabilities, though the correlations presented here are much lower than that reported in the previous work.
Contrary to Ref.~\cite{Malik2018} where coefficients are chosen such that correlation is maximal, we choose coefficients $\alpha = 2.27$, $\beta = 24.28$, and $\gamma = 0$.
To minimize the propagation of uncertainties from $L_0$, we manually choose $\alpha$ and $\beta$ to be as small as possible, while keeping in mind that the correlation with $\tilde\Lambda$ must be large enough to determine bounds on nuclear parameters.
We arbitrarily choose $\alpha$ and $\beta$ such that correlations are $50\%$ to give one example of the derived bounds.
The parameter $\gamma$ was chosen to be $0$ in order to neglect the additional uncertainty accrued by the addition of $L_0$, possible in this case only due to the high correlations between $K_{\text{sym},0}$ and $\tilde\Lambda$.

\begin{figure*}
\begin{center} 
\includegraphics[width=.45\linewidth]{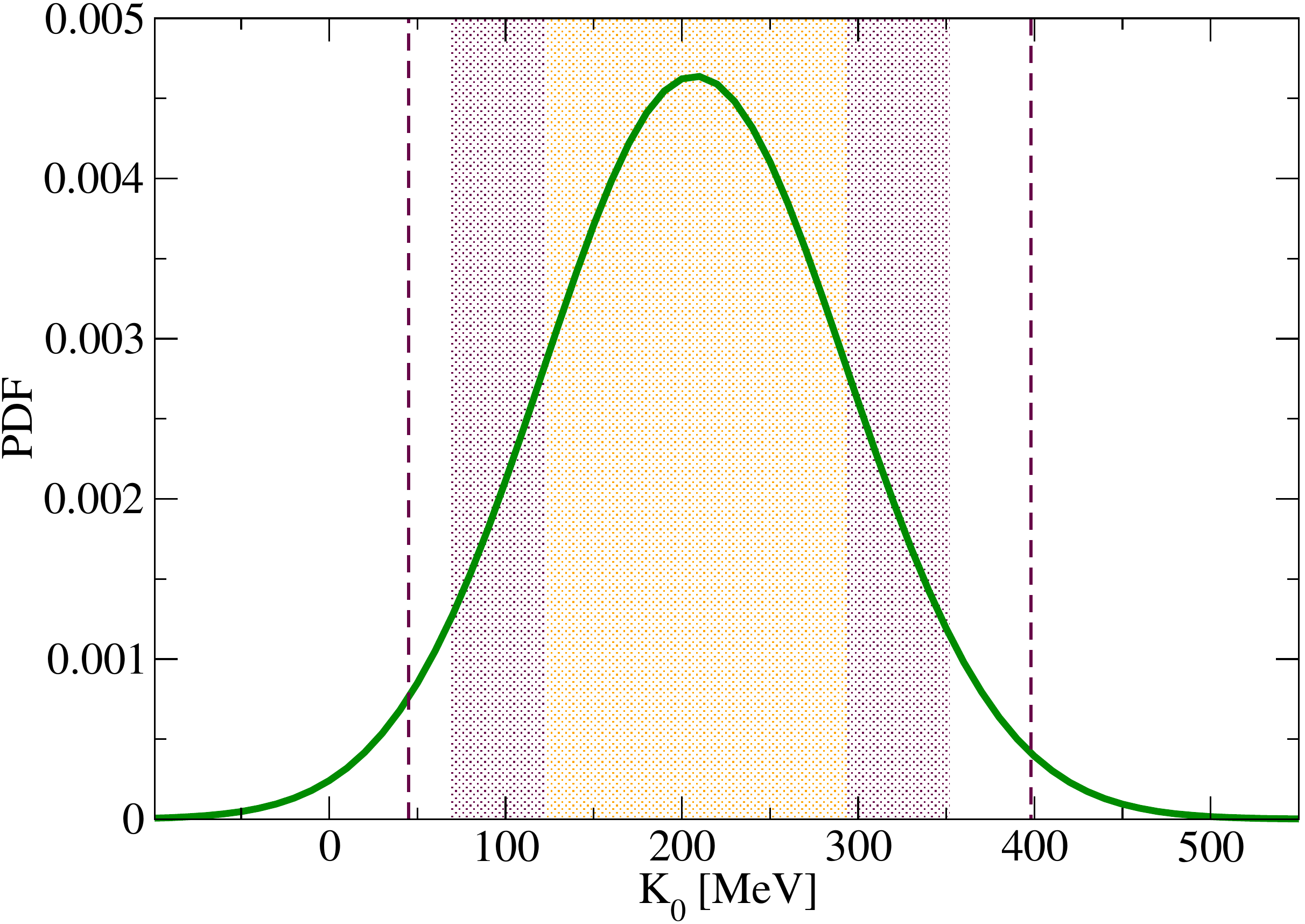}
\includegraphics[width=.45\linewidth]{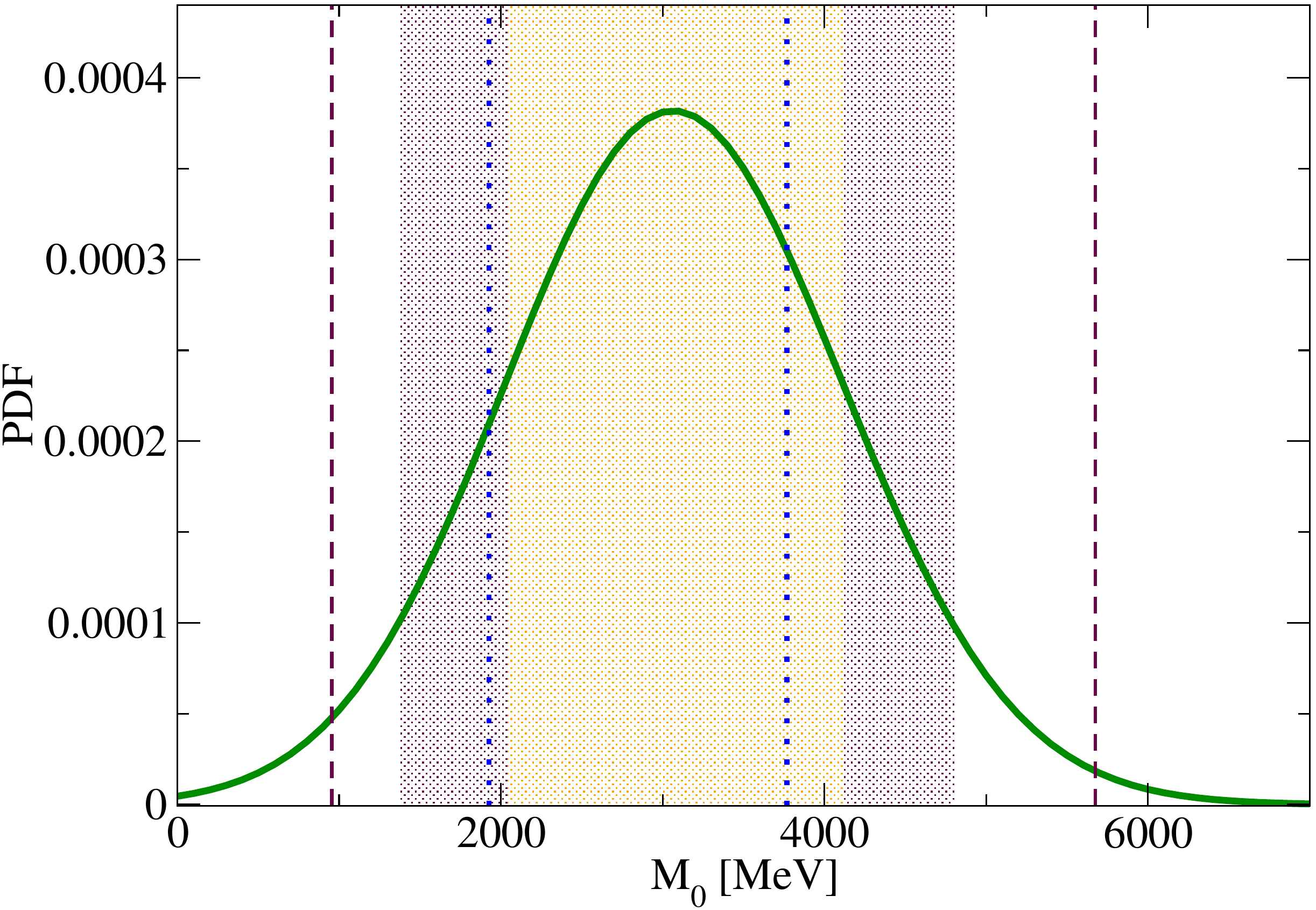}
\includegraphics[width=.45\linewidth]{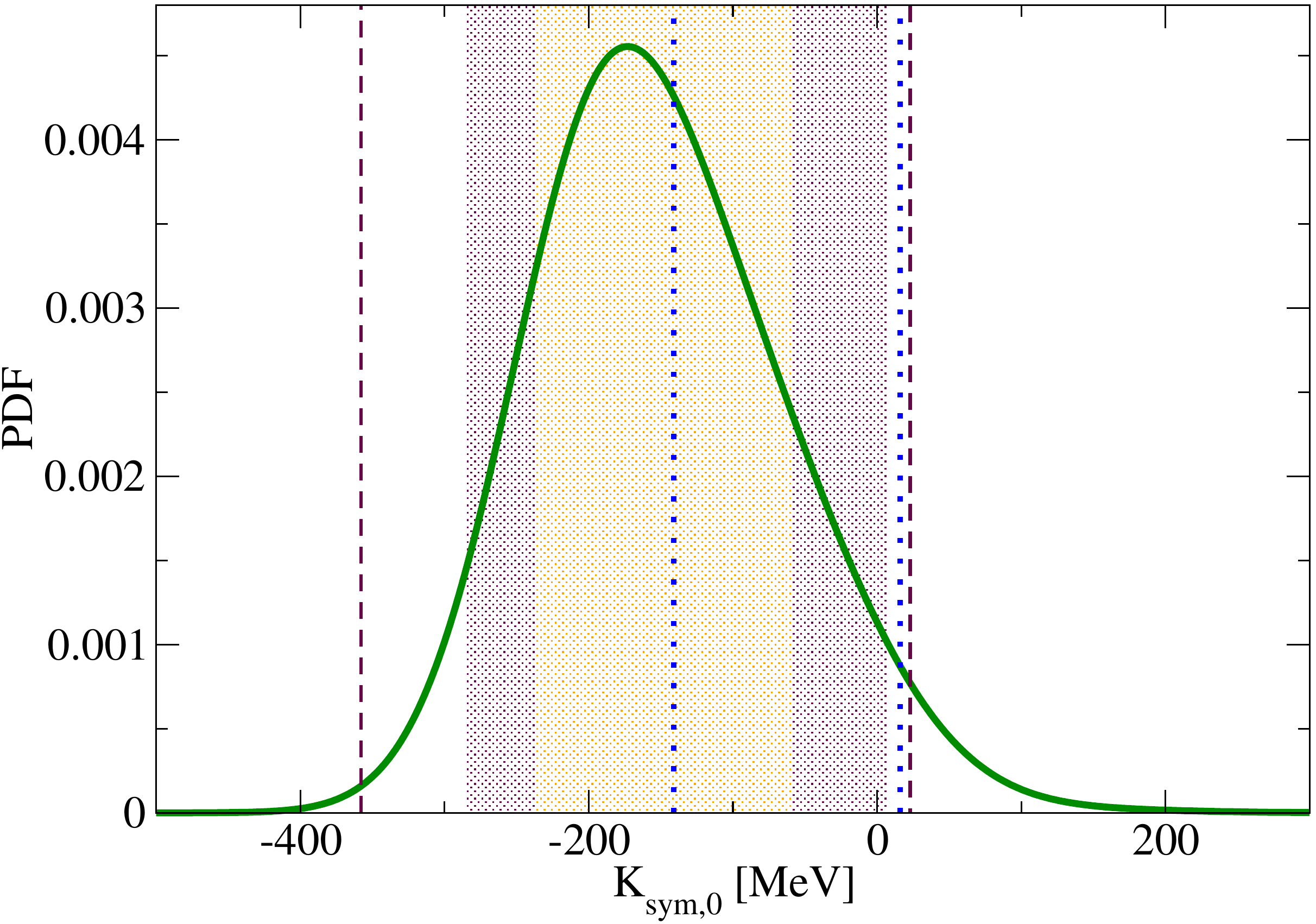}
\end{center}
\caption[Nuclear matter constraints]{Resulting posterior distributions on the nuclear incompressibility $K_0$ and its slope $M_0$, and the curvature of symmetry energy $K_{\text{sym},0}$ derived by integrating over the product of one-dimensional conditional probability distributions ($P(K_0 + \alpha L_0|\tilde{\Lambda})$, $P(M_0+\beta L_0|\tilde{\Lambda})$, and $P(K_{\text{sym},0}|\tilde{\Lambda})$) and $P_{\text{LIGO}}(\tilde{\Lambda})$ in Fig.~\ref{fig:LigoPrior}.
Further, for the linear combinations of $K_0 + \alpha L_0$ and $M_0 + \beta L_0$, one more integration over the probability distribution of $30 \text{ MeV}\leq L_0 \leq 86 \text{ MeV}$ was required to directly find the posterior distributions on $K_0$ and $M_0$. 
Overlayed are the resulting 68\% and 90\% confidence intervals in orange and maroon respectively, as well as the corresponding bounds calculated in Sec.~\ref{sec:linear} shown by dashed maroon vertical lines.
Additionally shown in dotted blue are the corresponding bounds on $M_0$ and $K_{\text{sym},0}$ computed by Ref.~\cite{Malik2018}, using priors of $\tilde{\Lambda} \in \lbrack 70,720 \rbrack$ and $L_0 \in \lbrack 30, 86 \rbrack$ MeV.
Observe how the results for the 90\% confidence intervals obtained in this section are slightly smaller than those found in Sec.~\ref{sec:linear}, indicating that previously the error was slightly overestimated (as the probability distribution on $\tilde\Lambda$ and the covariances between $\tilde\Lambda$ and the nuclear parameters were not properly taken into account). \hyperlink{link:Posteriors}{ \small (Return to the first reference of this figure in the main text.)}
}
\label{fig:Posteriors}
\end{figure*}

Figure~\ref{fig:Posteriors} presents the 90\% confidence interval on nuclear matter parameters after GW170817, based on the universal relations obtained previously.
In the computation of these above bounds, the posterior probability distribution on $\tilde\Lambda$ as derived by the LIGO Collaboration~\cite{Abbott:LTposterior} was used.
In particular, we find such bounds to be \Ksymrange{-285}{7}~\footnote{The constraint on $K_{\text{sym,0}}$ bears a close resemblance to that in Refs.~\cite{Margueron:Ksym,Mondal:Ksym}.} at the $90\%$ confidence interval.
Additionally, we find bounds on $K_0$ and $M_0$ to be \Krange{69}{352} and \Mrange{1371}{4808}.
Such results are much weaker than the ones found in Ref.~\cite{Malik2018}, born from the inclusion of systematic errors from a broader class of EoSs and the scatter uncertainty from EoS variation on universal relations.
These results lead us to conclude that it is important to account for the large systematic errors accrued from a wider range of valid EoSs and EoS variation in the approximate universal relations.

We then expand this work into the future of GW astronomy.
In particular, we attempt to find constraints on the nuclear matter parameters as a function of the binary systems' chirp mass.
The chirp mass $\mathcal{M}$ is the dominant parameter in the gravitational waveform, and thus has the largest potential to be accurately measured.
We begin by finding the correlations between the mass-weighted tidal deformability $\tilde\Lambda$ and various nuclear parameters, and combinations thereof.
We find that the low-order nuclear parameters $K_0$ and $M_0$ observe small correlations with $\tilde\Lambda$, while the correlations for higher-order parameter $K_{\text{sym},0}$ remains high at $\sim 80\%$.
For this reason, we focus on constraints on the curvature $K_{\text{sym},0}$ of the symmetry energy which is one of the most uncertain of nuclear matter parameters around the nuclear saturation density~\cite{Li:2019xxz}.

We compute the posterior probability distribution for the curvature of symmetry energy $K_{\text{sym},0}$, for 22 different values of chirp mass between $0.94 \text{ M}_\odot$ and $1.6 \text{ M}_\odot$.
For each value of chirp mass $\mathcal{M}_i$ considered, we compute the single-event $\tilde\Lambda$ uncertainties using Fisher analysis techniques.
Approximated as a Gaussian prior, the uncertainty in $\tilde\Lambda$ may be used to estimate the posterior probability distribution on $K_{\text{sym},0}$, by multiplication with the one-dimensional conditional probability distribution on $K_{\text{sym},0}$ given an observation of $\tilde\Lambda$, and then integrating over all $\tilde\Lambda$ values.
The process is then repeated for each value of chirp mass $\mathcal{M}_i$, resulting in a relationship between the uncertainties in $K_{\text{sym},0}$ and the chirp mass $\mathcal{M}$.

\begin{figure}
\begin{center} 
\begin{overpic}[width=0.7\columnwidth]{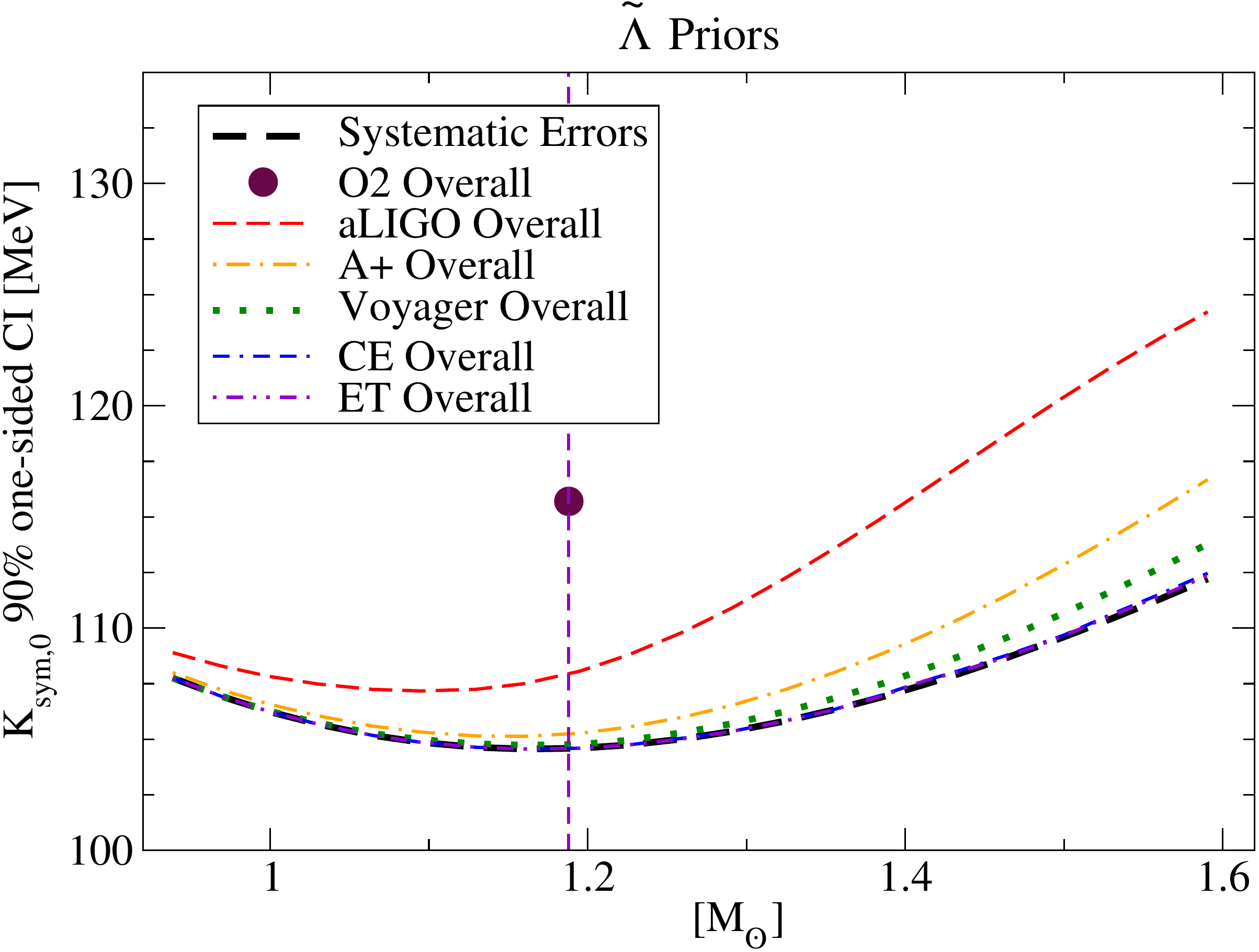}
\put(150,3.5){\small$\mathcal{M}$}
\end{overpic}
\end{center}
\caption[Statistical+systematic errors on $K_{\text{sym},0}$]{
The overall (statistical plus systematic) errors on $K_{\text{sym},0}$ using priors in $\tilde\Lambda$ for single-event measurements, plotted as a function of the binary systems' chirp mass -- applicable to any future binary NS merger. 
We fix the mass ratio as $q=0.9$ consistent with GW170817 and choose the distance and sky location of the binaries such that it gives the signal-to-noise ratio (SNR) of 32.4 for the O2 run, again corresponding to GW170817.
This is repeated for 6 different interferometers (O2, aLIGO, A\texttt{+}, Voyager, CE, ET).
Note that the error on O2 appears as a single point, corresponding to GW170817 - the single event observed on O2.
We also present systematic errors due to scattering in correlations between $K_{\text{sym},0}$ and $\tilde \Lambda$.
Observe how as one improves the detector sensitivity, the statistical errors become subdominant, and the overall errors approach the systematic uncertainties' ``wall".
This indicates the need to further reduce the EoS variation in the scattering that is the origin of systematic uncertainties before stronger constraints on $K_{\text{sym},0}$ can be derived. Additionally shown by the dashed vertical line is the chirp mass $\mathcal{M}=1.188\text{ M}_{\odot}$ corresponding to GW170817. \hyperlink{link:OverallVsSystematic}{ \small (Return to the first reference of this figure in the main text.)}
}
\label{fig:OverallVsSystematic}
\end{figure}

The corresponding one-sided 90\% confidence intervals on $K_{\text{sym},0}$ for single-event detections, along with the calculated systematic errors, are plotted in Fig.~\ref{fig:OverallVsSystematic} as a function of chirp mass for 6 different GW interferometers.
In this figure, we observe that as the detector sensitivity is increased, the statistical errors become subdominant rather quickly.
This in turn forces the overall errors to approach the
systematic error ``wall" at $\sim104$ MeV, caused partially by uncertainties in the EoS at low-density which are less sensitive to neutron star tidal deformabilities.
For this reason, the curves corresponding to the 3rd-generation detectors CE and ET become indistinguishable from the systematic errors -- indicating the necessity to reduce such errors in order for the further constraint of $K_{\text{sym},0}$ to become possible.

Following this, we offer a method to further decrease the statistical errors in the measurement of $K_{\text{sym},0}$.
This is accomplished by repeating the same analysis for the fixed chirp mass of $1.188\text{ M}_\odot$ with the first coefficient, $\lambda_0$, of the Taylor expanded dimensionless tidal deformability $\Lambda \approx \lambda_0+\lambda_1(1-\frac{m}{m_0})$ about $m_0=1.4\text{ M}_{\odot}$\footnote{$m_0$ will remain fixed for the remainder of the chapter, with the exception of Sec.~\ref{sec:systematics}, where we consider the effect of variations in $m_0$.}, rather than the mass-weighted tidal deformability $\tilde\Lambda$.
$\lambda_0$ (or the tidal deformability at $1.4\text{ M}_{\odot}$) is mass-independent, and thus, it is identical for all future GW events.
This way, we can combine the uncertainties for multiple detected events when it becomes applicable for future detectors.
As was observed in Fig.~\ref{fig:OverallVsSystematic}, we found that the uncertainties in $K_{\text{sym},0}$ became dominated by systematics for the single-event analyses on Voyager-era detectors and beyond.
By combining GW170817-like events detected on aLIGO and A\texttt{+}, we find that one can further reduce the statistical errors in $K_{\text{sym},0}$ such that the errors similarly become dominated by systematics.

Finally, we investigate the reduction of systematic errors by adding information about the tidal deformability at various different masses.
We begin by generating a four-dimensional Gaussian probability distribution $P(K_{\text{sym},0},\Lambda_{m_x},\Lambda_{m_y},\Lambda_{m_z})$.
The systematic error on $K_{\text{sym}}$ is obtained by first evaluating $P(K_{\text{sym},0},\Lambda_{m_x},\Lambda_{m_y},\Lambda_{m_z})$ at the fiducial values of $\Lambda_{m_x}$, $\Lambda_{m_y}$, and $\Lambda_{m_z}$ and estimating the 90\% confidence interval.
We then present the systematic errors when $m_z$ is fixed to be $1.5\text{ M}_\odot$, and ($m_x,m_y$) are varied between $1.0\text{ M}_\odot$ and $2.0\text{ M}_\odot$,
observing that the resulting systematic uncertainties are reduced to $\sim74$ MeV for certain combinations of $m_x$ and $m_y$.

\subsection{Equation of state insensitive relations}
Here we recap the primary findings of Chapter~\ref{chap:EoSInsensitiveRelations}.
We begin by investigating the increase in EoS insensitivity due to the constraints placed by GW170817 on the allowed space of EoSs. 
We use the posterior probability distribution on the pressure-density plane~\cite{Carney:2018sdv,LIGO:posterior} obtained from GW170817~\cite{TheLIGOScientific:2017qsa} to inform the set of EoSs that is compatible with this observation. We then generate two large samples of spectral EoSs~\cite{Lindblom:2018rfr}, one in which the EoSs are directly sampled from the posterior probability distribution (the ``constrained" sample) and another in which this constraint is not enforced (the ``unconstrained" sample).
We repeat the analysis done by Yagi and Yunes~\cite{Yagi:binLove,Yagi:ILQ} on both sets of EoSs and find that the EoS-insensitive relations present less EoS variability with the constrained set. 
In particular, the EoS-insensitivity increases by a factor of $\sim 60$\% in the binary Love relations (for stars with mass ratio larger than $0.75$), by a factor of $\sim 70\%$ in the C-Love relations, and by factors of $\sim 50$\% in the I-Love-Q relations. 

With this study at hand, we then carry out additional related studies on EoS-insensitive relations that go beyond the work in~\cite{Yagi:binLove,Yagi:ILQ}. First, we investigate the relation between the NS radius and its tidal deformability, the R-Love relations, for both sets of EoSs, as these are critical in order to place constraints on the radius from a measurement of the Love number.  We use the C-Love relations to construct the R-Love relations and find that the maximum EoS variability drops from $\sim 880$ m in the unconstrained case to $\sim 360$ m in the constrained case. Second, we study the EoS-universality of hybrid stars, which experience strong first-order phase transitions from hadronic to quark matter in the core~\cite{Paschalidis2018,Most:2018eaw,Burgio:2018yix,Montana:2018bkb}. We find that the I-Love-Q and C-Love relations remain EoS-insensitive for these hybrid stars, although the EoS variability increases slightly. However, we also find that the binary Love relations are not EoS-insensitive for a mixed binary with a (massive) hybrid star and a (low-mass) hadronic star, due to the large separation in mass-weighted tidal deformability between the constituent stars.

Last but not least, we study the importance of using the improved EoS-insensitive relations in future GW observations. The use of EoS-insensitive relations introduces systematic uncertainties in parameter estimation because of the intrinsic non-zero EoS-variability in these relations, which one must marginalize over. These uncertainties are currently irrelevant because statistical uncertainties in parameter estimation are much larger with current detectors. But as the detector sensitivity is improved, the signal-to-noise ratio and the number of events that will be detected will increase, therefore decreasing the statistical uncertainties below the systematic ones due to EoS variability. We carry out Fisher analyses to estimate when the statistical uncertainties become comparable to the systematic uncertainties due to EoS variability and find that this occurs for Voyager-class detectors. 

\begin{figure}
\begin{center} 
\includegraphics[width=0.9\columnwidth]{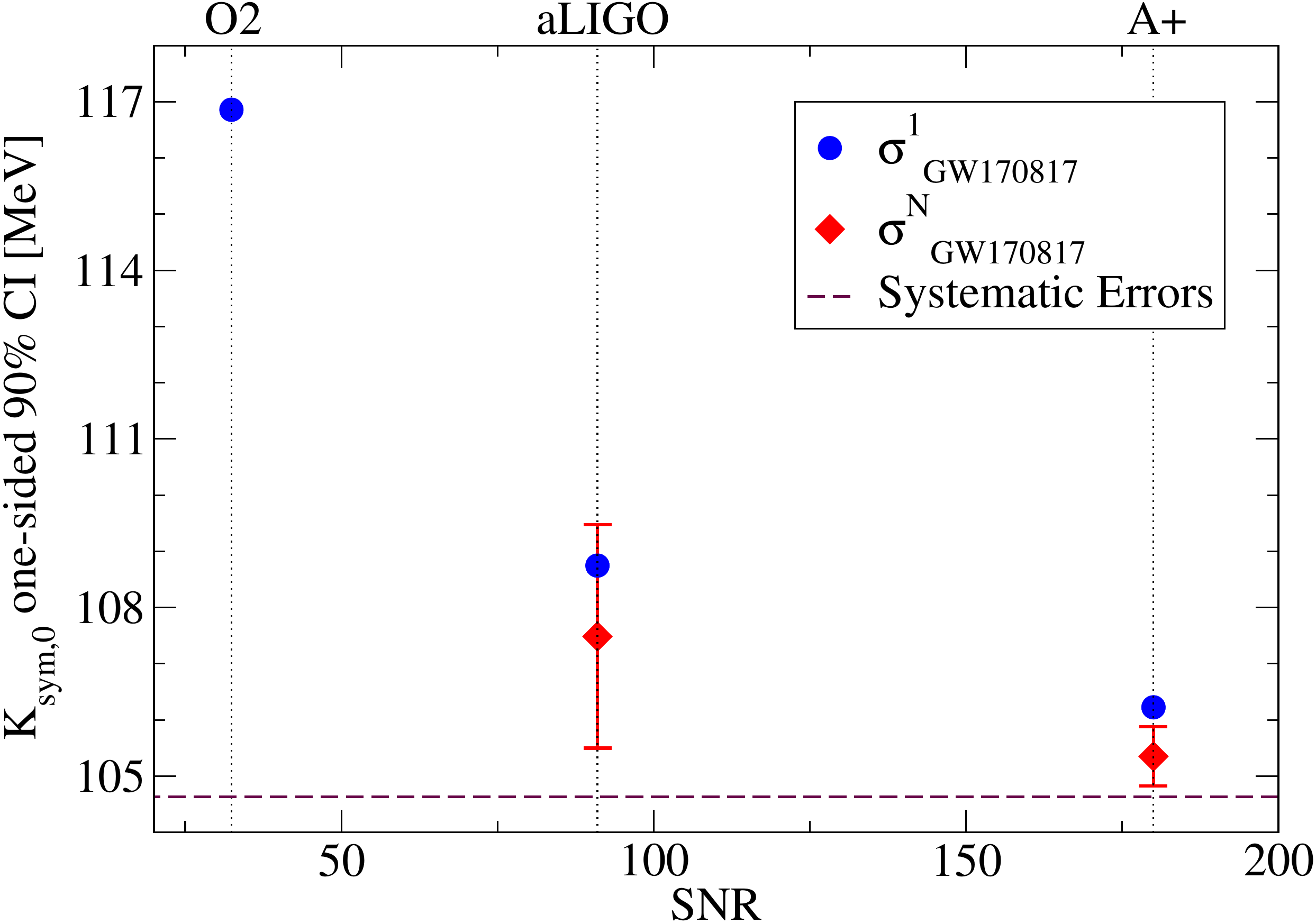}
\end{center}
\caption[GW170818-like event statistical uncertainties on $\lambda_0$]{Fisher-estimated statistical uncertainties on the extraction of $\lambda_{0}$ with interferometer (O2, aLIGO, A\texttt{+}, Voyager, CE, ET-D) as a function of the signal-to-noise-ratio expected in each of these instruments, given a single GW170817 detection (circles). The statistical uncertainties with ET are lower than with CE in spite of a lower signal-to-noise ratio because the former is more sensitive above 300 Hz, where tidal effects matter the most (see Sec.~\ref{sec:futureObservations} for further discussion). We also plot the combined statistical uncertainty given $N$ observations consistent with the NS binary merger rate for a 1 year observation (regions), with the top and bottom edges of the regions corresponding to pessimistic and optimistic merger rates. These statistical uncertainties should be compared to the systematic uncertainty on the extraction of $\lambda_{0}$ due to EoS-variability (horizontal dashed line). The statistical and systematic uncertainties cross for Voyager-class detectors. We confirm this conclusion by repeating the statistical analysis with two different waveform models: PhenomD~\cite{PhenomDI,PhenomDII} plus NRTidal corrections~\cite{Dietrich:2017aum,Samajdar:NRTidal} and PhenomD~\cite{PhenomDI,PhenomDII} plus 6PN tidal corrections~\cite{Vines:2011ud,Wade:tidalCorrections}. \hyperlink{link:stackedFisher}{ \small (Return to the first reference of this figure in the main text.)}}
\label{fig:stackedFisher}
\end{figure} 

Figure~\ref{fig:stackedFisher} shows this result in more detail. We present the (Fisher-estimated) statistical uncertainty in the measurement of $\lambda_{0}$ with various detectors (LIGO O2~\cite{aLIGO}, Advanced LIGO at design sensitivity (aLIGO)~\cite{aLIGO}, LIGO A\texttt{+} (A\texttt{+})\cite{Ap_Voyager_CE}, Voyager~\cite{Ap_Voyager_CE}, the Einstein Telescope (ET)~\cite{ET}, and the Cosmic Explorer (CE)~\cite{Ap_Voyager_CE}) for an event similar to GW170817. The x-axis shows the signal-to-noise ratio for a GW170817 event observed with each of these detectors. We also present the combined statistical uncertainty $\sigma^A_N$ after $N$ detections with each of these instruments, with the top and bottom of the region representing the most optimistic and pessimistic expectation for the number of detections expected from the binary NS merger rate~\cite{Abbott2017}. These statistical uncertainties should be compared to the systematic uncertainty in $\lambda_{0}$ due to EoS-variability improved with the constrained set. 
We find that the statistical and the systematic uncertainties cross for Voyager-class detectors; indicating the need to reduce systematic uncertainties before the Voyager era.

\subsection{Parameterized tests of general relativity}
Now let us preview the main results found in Chapter~\ref{chap:Parameterized}.
We begin by finding constraints on the ppE magnitude parameter $\beta_\ppE$~\cite{Yunes:2009ke}, which describes the strength of a generalized deviation from the GR waveform $A_\GR e^{i\Psi_\GR} \rightarrow A_\GR e^{i(\Psi_\GR+\beta_\ppE u^{b})}$, where $A_\GR$ and $\Psi_\GR$ are the amplitude and phase of gravitational waveforms in the Fourier domain predicted from GR, $u$ is the effective velocity of the binary BH system, and $b$ categorizes the power of velocity at which the modified theory of gravity affects the waveform.
Using Fisher analysis techniques~\cite{Poisson:Fisher,Berti:Fisher,Yagi:2009zm}, we estimate the maximum magnitude $\beta_\ppE$ can take while remaining consistent with the statistical detector noises 
for GW150914-like~\cite{GW150914} events observed on a ground-based detector (CE~\cite{Ap_Voyager_CE}), space-based detectors (LISA~\cite{LISA}~\cite{LISA}, TianQin~\cite{TianQin}, B-DECIGO~\cite{B-DECIGO}, and DECIGO~\cite{DECIGO}), and the multi-band combinations thereof.
We find that, as expected, space-based detectors which are sensitive to low-frequencies (or small relative velocities of binaries) are most proficient at constraining $\beta_\ppE$ at small values of $b$, while ground-based detectors are most proficient at large values of $b$, which is consistent with e.g.~\cite{Chamberlain:2017fjl}.
When combining measurements from both types of detector, we find improvements across all values of $b$.

Following this, constraints on $\beta_\ppE$ can be mapped to the associated parameters of various theories of gravity identified by their value of $b$, as summarized in Table~\ref{tab:theories}.
We see that EdGB, dCS, noncommutative gravity, and massive graviton (both dynamical and propagating effects) theories can provide stronger constraints than the current best bounds found in the literature, displayed in the last column of Table~\ref{tab:theories}. For dCS in particular, multi-band GW observations are crucial in most cases to place meaningful bounds, which are more stringent than the existing bounds by $\sim 7$ orders of magnitude.

\renewcommand{\arraystretch}{1.2}
\begin{table*}
\centering
\resizebox{\textwidth}{!}{%
\begin{tabular}{|c|c|c||c|c|c|c|c||c|}
\hline
&&&&&&&&\\[-1em]
\multirow{2}{*}{Theory} & \multirow{2}{*}{PN ($b$)} & \multirow{2}{*}{Parameter} & \multirow{2}{*}{CE~\cite{Ap_Voyager_CE}} & TianQin~\cite{TianQin} & LISA~\cite{LISA} & B-DECIGO~\cite{B-DECIGO} & DECIGO~\cite{DECIGO} & \multirow{2}{*}{Current}\\
&&&&&&&&\\[-1em]
 &  & & & \textbf{(+CE)} & \textbf{(+CE)} & \textbf{(+CE)} & \textbf{(+CE)} & \\
\hline
\hline
&&&&&&&&\\[-1em]
\multirow{2}{*}{EdGB~\cite{Maeda:2009uy,Yagi_EdGBmap}} &  \multirow{2}{*}{$-1$ ($-7$)} & \multirow{2}{*}{$\sqrt{\alpha_\EdGB}$ \lbrack km\rbrack} & \multirow{2}{*}{$4.2$} & $0.55$ & $0.72$ & $0.29$ & $0.12$ & \multirow{1}{*}{$10^7$~\cite{Amendola_EdGB}}\\
&&&&&&&& $2-6$~\cite{Kanti_EdGB,Pani_EdGB,Yagi_EdGB,Nair_dCSMap,Yamada:2019zrb,Tahura:2019dgr} \\[-1em]
 & & &  & ($\bm{0.30}$) & ($\bm{0.37}$) & ($\bm{0.25}$) & ($\bm{0.11}$) & \\
\hline
&&&&&&&&\\[-1em]
\multirow{2}{*}{dCS~\cite{Jackiw:2003pm,Alexander_cs,Yagi:2012vf,Nair_dCSMap}} &  \multirow{2}{*}{+2 ($-1$)} & \multirow{2}{*}{$\sqrt{\alpha_\dCS}$ \lbrack km\rbrack} & \multirow{2}{*}{--} & -- & -- & -- & $23$ & \multirow{2}{*}{$10^8$~\cite{AliHaimoud_dCS,Yagi_dCS}}\\
&&&&&&&&\\[-1em]
 & & & & ($\bm{31}$) & ($\bm{31}$) & ($\bm{29}$) & ($\bm{18}$) & \\
\hline
&&&&&&&&\\[-1em]
\multirow{2}{*}{Scalar-tensor~\cite{Horbatsch_STcosmo,Jacobson_STcosmo}} &  \multirow{2}{*}{$-1$ ($-7$)} & \multirow{2}{*}{$\dot \phi$ \lbrack sec$^{-1}$\rbrack} & \multirow{2}{*}{$340$} & $6.1$ & $10.8$ & $1.6$ & $0.25$ & \multirow{2}{*}{$10^{-6}$~\cite{Horbatsch_STcosmo}}\\
&&&&&&&&\\[-1em]
 & & &  & ($\bm{1.8}$) & ($\bm{2.7}$) & ($\bm{1.2}$) & ($\bm{0.21}$) & \\
\hline
&&&&&&&&\\[-1em]
\multirow{2}{*}{Noncommutative~\cite{Harikumar:2006xf,Kobakhidze:2016cqh}}  & \multirow{2}{*}{+2 ($-1$)} & \multirow{2}{*}{$\sqrt{\Lambda}$} & \multirow{2}{*}{$0.74$} & $1.5$ & $1.5$ & $0.83$ & $0.57$ & \multirow{2}{*}{3.5~\cite{Kobakhidze:2016cqh}}\\
&&&&&&&&\\[-1em]
 & & &  & ($\bm{0.66}$) & ($\bm{0.66}$) & ($\bm{0.64}$) & ($\bm{0.50}$) & \\
\hline
&&&&&&&&\\[-1em]
\multirow{2}{*}{Varying $G$~\cite{Will_SEP,Yunes_GdotMap,Tahura_GdotMap}} &  \multirow{2}{*}{$-4$ ($-13$)} & \multirow{2}{*}{$\dot G$ \lbrack yr$^{-1}$\rbrack} & \multirow{2}{*}{$990$} & $1.6\times 10^{-7}$ & $2.3\times 10^{-6}$ & $9.4\times 10^{-8}$ & $3.4\times 1.9\times10^{-9}$ & \multirow{1}{*}{$10^{-13}-10^{-12}$}\\
&&&&&&&& \cite{Bambi_Gdot,Copi_Gdot,Manchester_Gdot,Konopliv_Gdot} \\[-1em]
 & & &  & ($\bm{3.0\times 10^{-8}}$) & ($\bm{5.5\times 10^{-7}}$) & ($\bm{7.2\times 10^{-8}}$) & ($\bm{1.5\times 10^{-9}}$) & \\
\hline
&&&&&&&&\\[-1em]
\multirow{2}{*}{Varying $M$~\cite{Yagi_EDmap,Berti_ModifiedReviewSmall}} & \multirow{2}{*}{$-4$ ($-13$)} & \multirow{2}{*}{$\dot{M}$ \lbrack M$_\odot$ yr$^{-1}$\rbrack} & \multirow{2}{*}{$1.9\times 10^4$} & $5.8\times10^{-6}$ & $9.0\times10^{-5}$ & $3.8\times10^{-6}$ & $7.2\times10^{-8}$ & \multirow{2}{*}{--}\\
&&&&&&&&\\[-1em]
 & &  & & ($\bm{1.2\times 10^{-6}}$) & ($\bm{2.0\times 10^{-5}}$) & ($\bm{2.8\times 10^{-6}}$) & ($\bm{5.9\times 10^{-8}}$) & \\
\hline
&&&&&&&&\\[-1em]
\multirow{1}{*}{Massive graviton~\cite{Zhang:2017jze,Finn:2001qi}} &\multirow{2}{*}{$-3$ ($-11$)} & \multirow{2}{*}{$m_g$ \lbrack eV\rbrack} & \multirow{2}{*}{$6.7\times10^{-16}$} & $1.9\times10^{-19}$ & $3.1\times10^{-19}$ & $6.6\times10^{-20}$ & $9.5\times10^{-21}$ & \multirow{1}{*}{$10^{-14}$~\cite{Chung:2018dxe}}\\
 (dynamical) &&&&&&&& $5.2\times10^{-21}$~\cite{Miao:2019nhf} \\[-1em]
 & & & & ($\bm{9.2\times10^{-20}}$) & ($\bm{1.1\times10^{-19}}$) & ($\bm{5.6\times10^{-20}}$) & ($\bm{8.5\times10^{-21}}$) & \\
\hline \hline
&&&&&&&&\\[-1em]
\multirow{1}{*}{Massive graviton~\cite{Mirshekari_MDR}} & \multirow{2}{*}{+1 ($-3$)} & \multirow{2}{*}{$m_g$ \lbrack eV\rbrack} & \multirow{2}{*}{$1.3\times10^{-24}$} & $1.6\times10^{-22}$ & $1.6\times10^{-22}$ & $1.4\times10^{-23}$ & $3.4\times10^{-24}$ & \multirow{2}{*}{$5\times10^{-23}$~\cite{Abbott_IMRcon}}\\
 (propagation)&&&&&&&&\\[-1em]
 & & &  & ($\bm{6.1\times10^{-25}}$) & ($\bm{6.3\times10^{-25}}$) & ($\bm{2.9\times10^{-25}}$) & ($\bm{1.8\times10^{-25}}$) & \\
\hline 
\end{tabular}
}
\caption[Constraints on modified theories of gravity]{
Tabulated list of modified theories of gravity considered in this chapter.
The first column displays the modified theory of gravity in question, the second column indicates the post-Newtonian (PN) order (or ppE exponent $b$) at which the effect enters the gravitational waveform phase, and the third column identifies the appropriate parameters associated with the theory.
The fourth column tabulates the estimated constraints on the above-mentioned theoretical parameter as if a GW150914-like event were detected on ground-based detector CE, while the fifth to eighth columns likewise display the same bound as observed by space-based detectors LISA, TianQin, B-DECIGO, and DECIGO (top), and again for the multi-band GW observations in combination with CE (bottom).
Finally, the last column displays the current constraints on the theoretical parameters as found in the literature.
Entries with a horizontal dash (in the fifth to eighth columns) correspond to bounds on parameters which do not satisfy the small-coupling approximation, indicating that no meaningful bounds may be placed.
All GW bounds are derived from GW generation mechanism bounds, except for the last row, which comes from bounds on the GW propagation correction.
We note that such bounds are obtained via an initial LISA detection exactly four years prior to merger corresponding to the LISA missions lifetime.
This assumption is investigated for validity in Sec.~\ref{sec:results}. \hyperlink{link:theories}{ \small (Return to the first reference of this table in the main text.)}
}\label{tab:theories}
\end{table*}

We then consider the present and future implications on constraining scalar tensor theories, in particular, quasi-Brans-Dicke theory and Einstein-dilaton Gauss-Bonnet (EdGB) gravity from the GW detections of BH-NS coalescences.
Such sources are extremely timely to consider as some of the candidate events in the O3 run by the LVC, such as S190814bv, are likely to be the merger of a black hole and a neutron star~\cite{gracedb,gracedb2}.
In scalar tensor theories, BHs and/or NSs can become scalarized\footnote{in EdGB gravity, BHs become scalarized and NSs don't, while in quasi-Brans-Dicke theories the opposite occurs.}, which gives rise to a ``fifth force'' that depends on the internal structure of the massive objects and violates the strong equivalence principle (SEP). 
Binaries consisting of scalarized astrophysical objects further emit scalar dipole radiation (on top of gravitational quadrupolar radiation in GR), causing the binaries to evolve faster.
Such radiation becomes larger when the difference between the scalar charges of the binary constituents become larger, and thus a mixed binary consisting of one black hole and one neutron star system is ideal for probing such theories~\cite{Berti:spaceFreq,Takahiro,Sagunski:2017nzb,Huang:2018pbu}. 

We consider the single BH-NS detections with future GW detectors, as well as the multi-band detections between both space- and ground-based detectors~\cite{Barausse:2016eii,Nair:2015bga,Nair:2018bxj}, and finally the combination of multiple observations~\cite{Takahiro,Abadie:2010cf} made on future detectors with expanded horizons. 
We find that future observations (especially multi-band ones) can constrain both theories of gravity several orders-of-magnitude stronger than the current bounds.
More importantly, we find that if the candidate events S190426c and S190814bv on the LVC O3 are indeed mixed binaries, we can place strong constraints on the EdGB theory of gravity almost an order-of-magnitude stronger than the current one.

Finally, we probe EdGB gravity in particular with GWs from binary BH mergers by including both inspiral and ringdown corrections to the gravitational waveform.
The former correction is computed using the ppE formalism described above, where the mapping to EdGB gravity is known~\cite{Yagi:2011xp,Yunes_ModifiedPhysics,Tahura:2019dgr}.
The latter corrections are computed with the EdGB corrections to the individual QNM ringing frequency and damping time found in Ref.~\cite{Blazquez-Salcedo:2016enn} (see also~\cite{Blazquez-Salcedo:2017txk,Blazquez-Salcedo:2016yka,Blazquez-Salcedo:2018pxo}). Moreover, we take into account EdGB corrections to the final mass and spin of the remnant BH as a function of the initial masses and spins, which can be estimated from corrections to the orbital energy and angular momentum found in Ref.~\cite{Ayzenberg:2014aka}.

With the EdGB corrections to the inspiral signal as well as the remnant BH mass, spin, and QNMs in hand, we derive current and projected future bounds on the EdGB coupling parameter $\alpha_\EdGB$\footnote{Note that in the EdGB theory of gravity with coupling parameter $\alpha_\EdGB$, GR is recovered in the limit of $\alpha_\EdGB\to0$.}.
As a first step calculation, we adopt the Fisher analysis~\cite{Cutler:Fisher}, which is known to agree well with Bayesian analyses for loud enough signals~\cite{Vallisneri:FisherSNR,Vallisneri:FisherSNR2}, such as GW150914~\cite{Yunes_ModifiedPhysics}.
We first consider four GW events, in order of increasing mass: GW170608~\cite{GW170608}, GW151226~\cite{GW151226}, GW150914~\cite{GW150914}, and GW170729~\cite{GW170729}.
We find that GW events detected during the O1/O2 runs by LVC detectors have varying success on the constraint of $\sqrt{\alpha_\EdGB}$ while varying the type of EdGB corrections introduced to the template waveform (inspiral only, axial or polar QNMs only, or both).
We find that for more massive events, the inclusion of corrections to the merger-ringdown are necessary in order to satisfy the small coupling approximation corresponding to $16\pi\alpha_\EdGB^2/M^4 \ll 1$.
This stresses the need for the inclusion of merger-ringdown corrections to the template waveform, especially for more massive events where such contributions become important.
Further, we find that future GW150914-like events detected by CE~\cite{Ap_Voyager_CE}, LISA~\cite{LISA}, or the multiband combination of the two improve the constraints considerably, going beyond current bounds.

\subsection{Inspiral-merger-ringdown consistency tests of general relativity}
\begin{figure}
\centering
\includegraphics[width=0.7\columnwidth]{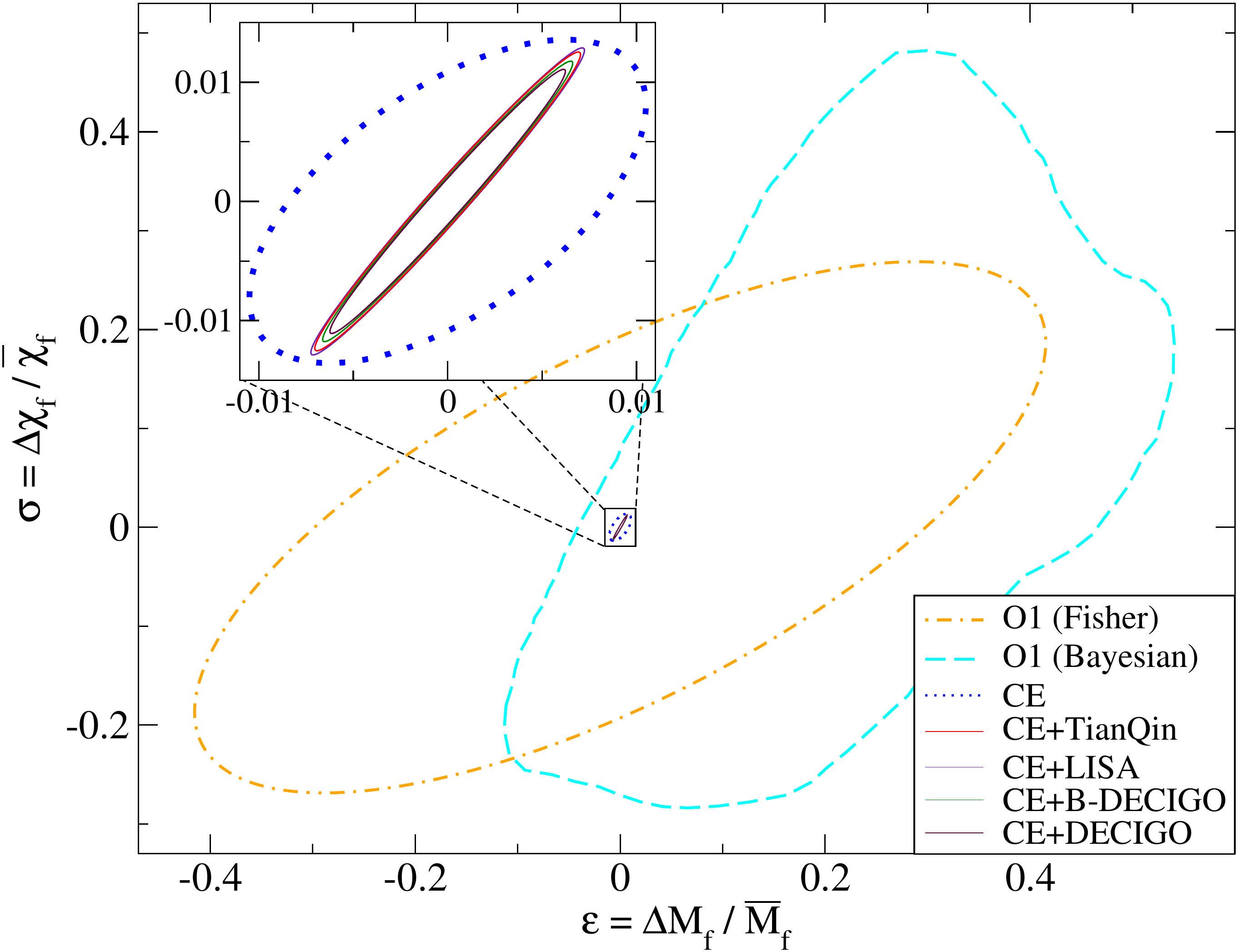}
\caption[IMR consistency test (transformed) for GW150914-like events]{
90\% credible region contours of the transformed probability distributions in the $\epsilon-\sigma$ plane (see Eqs.~\eqref{eq:epsilon}--\eqref{eq:transform}), describing the difference in the remnant mass and spin predictions between the inspiral and merger-ringdown estimate for GW150914-like event using the GR templates. 
Here we display the results for LIGO O1 (Fisher and Bayesian~\cite{Abbott_IMRcon} for comparison), CE, and the multi-band observations of CE and LISA, TianQin, B-DECIGO, and DECIGO.
The areas of such confidence regions show the following: (i) good agreement within 10\% between the Fisher and Bayesian analyses, (ii) three orders-of-magnitude improvement from LIGO O1 to CE, and (iii) up to an additional order-of-magnitude improvement with multi-band observations. \hyperlink{link:IMRDconsistencyTransformed}{ \small (Return to the first reference of this figure in the main text.)}
}\label{fig:IMRDconsistencyTransformed}
\end{figure}

Here we briefly summarize our important findings found in Chapter~\ref{chap:IMRconsistency}.
We begin by offering a new simplified (Fisher-analysis-based), predictive inspiral-merger-ringdown (IMR) consistency test~\cite{Ghosh_IMRcon,Ghosh_IMRcon2,Abbott_IMRcon,Abbott_IMRcon2} for future GW150914-like events.
This is done by computing the Gaussian posterior probability distributions between the remnant BH's mass $M_f$ and spin $\chi_f$ obtained independently from both the inspiral (I) and merger-ringdown (MR) signals.
The consistency between such final mass and spin parameters obtained independently from the inspiral and merger-ringdown signals could then tell one something about the underlying theory of gravity.
In particular, if such predictions disagree with each other to a statistically significant level, evidence of non-GR behavior emergent within the signal can be presented.
Such distributions are then combined into the joint-probability distribution between non-GR parameters $\Delta M_f\equiv M_f^\II-M_f^\MR$ and $\Delta \chi_f\equiv \chi_f^\II-\chi_f^\MR$, with $(\Delta M_f,\Delta \chi_f)|_\GR=(0,0)$ being the GR value.
We estimate the effective \textit{size} of the 90\% confidence region in Fig.~\ref{fig:IMRDconsistencyTransformed}
for ground based detectors, as well as the combination with different space-based detectors.
The areas of such posterior probability distributions can be used to predict the amount of ``wiggle" room a given non-GR theory of gravity will have to become the correct theory of gravity with future detectors, as any separation of the two posteriors could indicate deviations from GR.
We find that the ground-based detector is optimal at measuring the merger-ringdown portion of the signal, the space-based detectors are efficient at measuring the inspiral portion of the signal, and the combination of the two\footnote{Note that space-based detectors can not observe the merger-ringdown phase of GW150914-like events. Thus, signals can \textit{only} be combined for the inspiral portion of the signal.} proves to reduce the posterior sizes by up to an order of magnitude.

We follow this up by applying the IMR consistency tests to probe the specific example theory of EdGB gravity.
In particular, we devise a new method to consider EdGB corrections to not only the inspiral properties of a binary BH coalescence~\cite{Takahiro} using the ppE formalism, but also to the characteristic quasinormal modes (QNMs)~\cite{Blazquez-Salcedo:2016enn} and final properties of the post-merger BH~\cite{Ayzenberg:2014aka} as discussed further in Chapter~\ref{chap:Parameterized}.
We then apply the IMR consistency test with full waveform corrections by steadily increasing the magnitude of EdGB corrections present within the observed GW signal until the inspiral and merger-ringdown portions of the signal disagree to a statistically significant degree.
At this point, we can claim the presence of non-GR behavior in the signal.
We find that the IMR consistency test applied in this way can give comparable bounds to those found from the parameterized tests found in Chapter~\ref{chap:Parameterized}.
Additionally, we find that future observations, especially multi-band ones, can constrain the EdGB theory of gravity by up to an order-of-magnitude stronger than previously found.
To the best of our knowledge, the IMR consistency test has been put into context here for an example modified theory of gravity for the first time, and can indeed be easily applied to other alternative theories of gravity, given the required ingredients.

\subsection{Testing beyond-Kerr spacetimes}
Finally, we provide brief details on our primary findings in Chapter~\ref{chap:beyondKerr}.
We begin by presenting for the very first time a recipe for one to quickly estimate corrections to the inspiral, merger-ringdown, and remnant BH properties given only an arbitrary spacetime metric $g_{\alpha\beta}^\X$.
For demonstration purposes, we apply this to two example beyond-Kerr spacetime metrics JP and mod.-$\Delta$ spacetime metrics $g_{\alpha\beta}^\JP$ and $g_{\alpha\beta}^\MD$, which are each parameterized only by the single parameters $\epsilon_3$ and $\beta$, deviating from the Kerr metric $g_{\alpha\beta}^\K$.
We then demonstrate the power of such corrections by using the IMR consistency test from Chapter~\ref{chap:IMRconsistency} to predict the magnitudes of $\epsilon_3$ and $\beta$ required for one to observe statistically significant deviations from the Kerr result.

Table~\ref{tab:results} presents a summary of the main results found.
Here we compare constraints on the JP and mod.-$\Delta$ deviation parameters $\epsilon_3$ and $\beta$ for each GW event and detector considered in this thesis.
In particular, constraints are obtained using two different methods: (i) using the inspiral merger-ringdown consistency tests of GR in which one compares the inspiral and merger-ringdown signal's predictive power of the remnant BH mass and spin; and (ii) using the parameterized tests of GR, in which the Fisher analysis parameter estimation method is used to estimate the statistical uncertainties on template waveform parameters.
With the current-generation GW detector aLIGO O2, we find comparable constraints on the JP deviation parameter $\epsilon_3$ to those from x-ray observations of BH accretion disks~\cite{Kong:2014wha,Bambi:2015ldr}, found to be $\epsilon_3\lessapprox5$. 
With future space- and ground-based GW observatories, we find constraints several orders of magnitude stronger.
We find that such results from the IMR consistency tests are mostly comparable to those from the parameterized tests as discussed in Chapter~\ref{chap:Parameterized}.
In particular, we find that the extreme-mass-ratio-inspirals observable by future space-based detector LISA~\cite{LISA} can probe such effects by three orders-of-magnitude stronger than the current constraints found in the literature.
We additionally find that the additional use of merger-ringdown and remnant BH property corrections to the waveform can alter results by up to 15\% as compared to using inspiral ones alone, as is commonly done.

\renewcommand{\arraystretch}{1.2}
\begin{table*}
        \centering
        \begin{tabular}{|cc|c|c|c|c|}
             \cline{1-6}
			 && \multicolumn{2}{c|}{}&\multicolumn{2}{c|}{}\\[-1em]
             && \multicolumn{2}{c|}{$\epsilon_3$ (JP~\cite{Johannsen:2011dh})}  & \multicolumn{2}{c|}{$\beta$ (mod.-$\Delta$~\cite{Johannsen:2015mdd,Johannsen:2015pca})}\\
             && \multicolumn{2}{c|}{}&\multicolumn{2}{c|}{}\\[-1em]
             && IMR & Param. &  IMR & Param.\\
             &&&&&\\[-1em]
            \hline
            &&&&&\\[-1em]
            \multirow{2}{*}{O2~\cite{aLIGO}} & GW150914~\cite{GW150914} & (7)$^*$ & (5)$^*$ & (2)$^*$ & (1)$^*$\\
            \cline{2-2}
            & GW170729~\cite{GW170729} & (10)$^*$ & (14)$^*$ & (14)$^*$ & (11)$^*$ \\
            &&&&&\\[-1em]
            \hline
            &&&&&\\[-1em]
            \multirow{2}{*}{CE~\cite{Ap_Voyager_CE}} & GW150914~\cite{GW150914} & 0.05 & 0.05 & 0.05 & 0.02\\
            \cline{2-2}
            & GW170729~\cite{GW170729} & 0.6 & 0.5 & 0.06 & 0.07\\
            &&&&&\\[-1em]
            \hline
             &&&&&\\[-1em]
            \multirow{2}{*}{CE+LISA~\cite{Ap_Voyager_CE,LISA}} & GW150914~\cite{GW150914} & 0.02 & 0.03 & $5\times10^{-3}$ & $4\times10^{-3}$\\
            \cline{2-2}
            & GW170729~\cite{GW170729} & 0.05 & 0.09 & 0.05 & 0.03\\
            &&&&&\\[-1em]
            \hline
			&&&&&\\[-1em]
            \multirow{2}{*}{LISA~\cite{LISA}} & EMRI & $(2\times10^{-3})^\dagger$ & $10^{-3}$ & ($2\times10^{-4}$)$^\dagger$ & $10^{-4}$ \\
            \cline{2-2}
            & SMBHB & 0.02 & 0.01 & $10^{-3}$ & $10^{-3}$\\
            \hline
        \end{tabular}
        \caption[Constraints on $\epsilon_3$ and $\beta$]{
        Summary of results obtained in this chapter for both the Johannsen-Psaltis and modified-$\Delta$ metrics. Here we compare constraints on the deviation parameters $\epsilon_3$ and $\beta$ obtained via the IMR consistency tests of GR (IMR), and the parameterized tests of GR (Param.) for each gravitational wave event and detector considered. In particular, bounds are presented for GW150914-like events ($m_1=36\text{ M}_\odot$, $m_2=29\text{ M}_\odot$), GW170729-like events ($m_1=50.6\text{ M}_\odot$, $m_2=34.4\text{ M}_\odot$), EMRIs ($m_1=10^6\text{ M}_\odot$, $m_2=10\text{ M}_\odot$), and super-massive black hole binaries (SMBHBs,  $m_1=10^6\text{ M}_\odot$, $m_2=5\times10^4\text{ M}_\odot$). Observe that the bounds with the two methods are comparable in all cases presented here. \hyperlink{link:results}{ \small (Return to the first reference of this table in the main text.)}\\
        $*$  \scriptsize{Constraints with the aLIGO O2 detector are not as reliable because they fall beyond the small-deviation approximation made when deriving ppE parameters.}\\
        $\dagger$  \scriptsize{Constraints from EMRIs with IMR consistency tests may not be accurate since the IMRPhenomD waveforms were calibrated to numerical relativity simulations with mass ratios only up to 1:18. In the parameterized test, all such numerical relativity (NR) fits have been removed, and integrations stopped before the merger to avoid such inaccuracies.}
        }\label{tab:results}
\end{table*}

We then follow up this work by designing a more generic, stationary and axisymmetric, asymptotically flat Kerr-like BH metric with separable structure like in the Kerr metric.
Following the important work of Johannsen in~\cite{Johannsen:2015pca}, this is done by first introducing the most generic deviations possible into the contravariant Kerr metric in such a way that the Hamilton-Jacobi separability condition is preserved as in the case of a Kerr spacetime.
The new metric is then simplified by imposing the constraint of asymptotic flatness at spatial infinity. 
Further, we impose constraints consistent with the weak-field solar system tests as Johannsen did in~\cite{Johannsen:2015pca}, however we note that such constraints may not be explicitly valid in the strong-gravity regions surrounding BHs.
We then found that our new metric can be mapped to at least eight known BH solutions~\cite{Papadopoulos:2018nvd,Randall:1999ee,Aliev:2005bi,Jai-akson:2017ldo,Ding:2019mal,Kerr-Sen,Kanti_EdGB,Maeda:2009uy,Sotiriou:2014pfa,Ayzenberg:2014aka,Jackiw:2003pm,Yagi_dCS,Yunes_dcs,Bardeen,Kumar:2019uwi,Pani:2011gy,Kumar:2020hgm}.

\begin{figure*}[!htbp]
\begin{center}
\includegraphics[width=.4\textwidth]{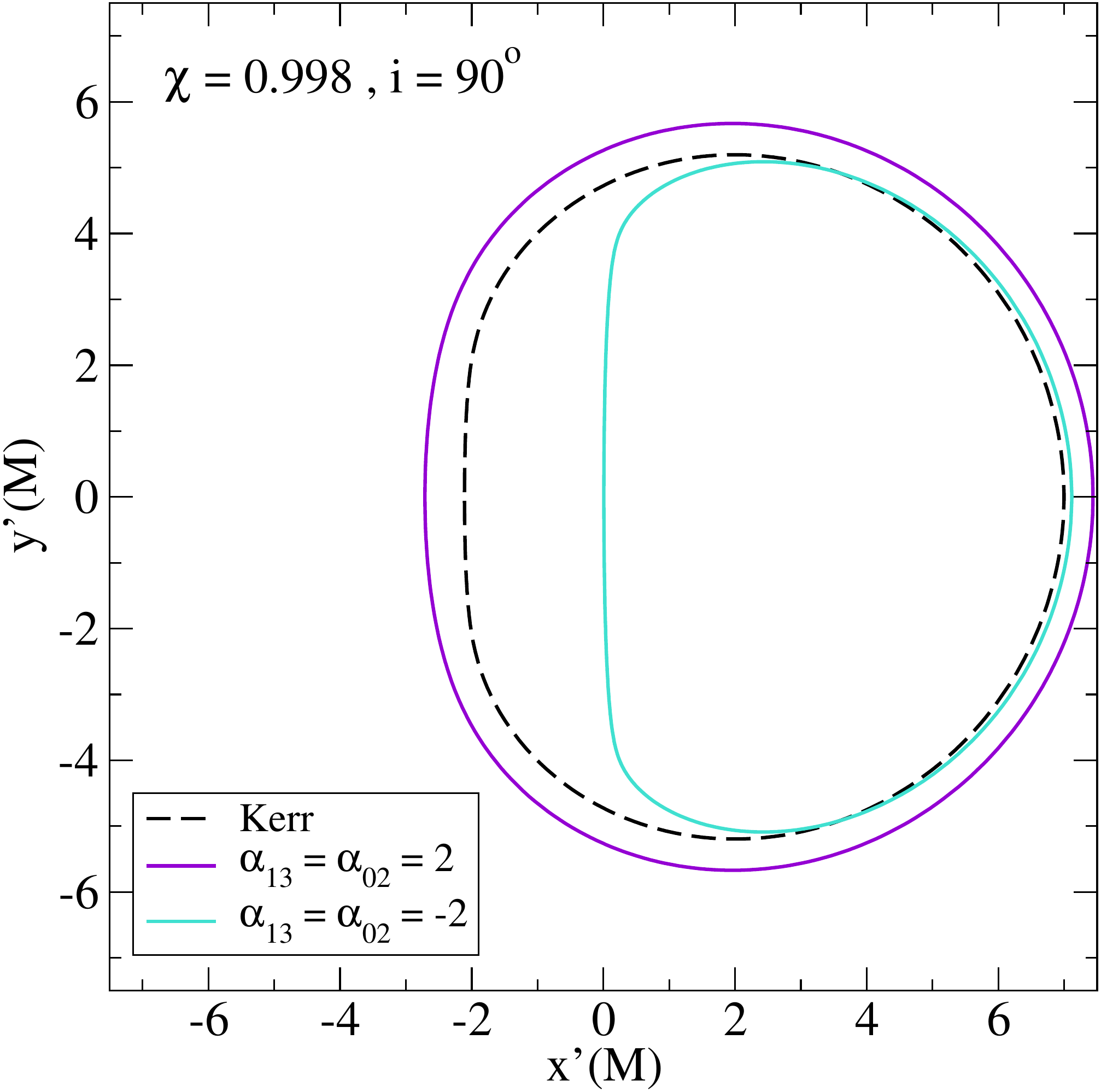}
\includegraphics[width=.4\textwidth]{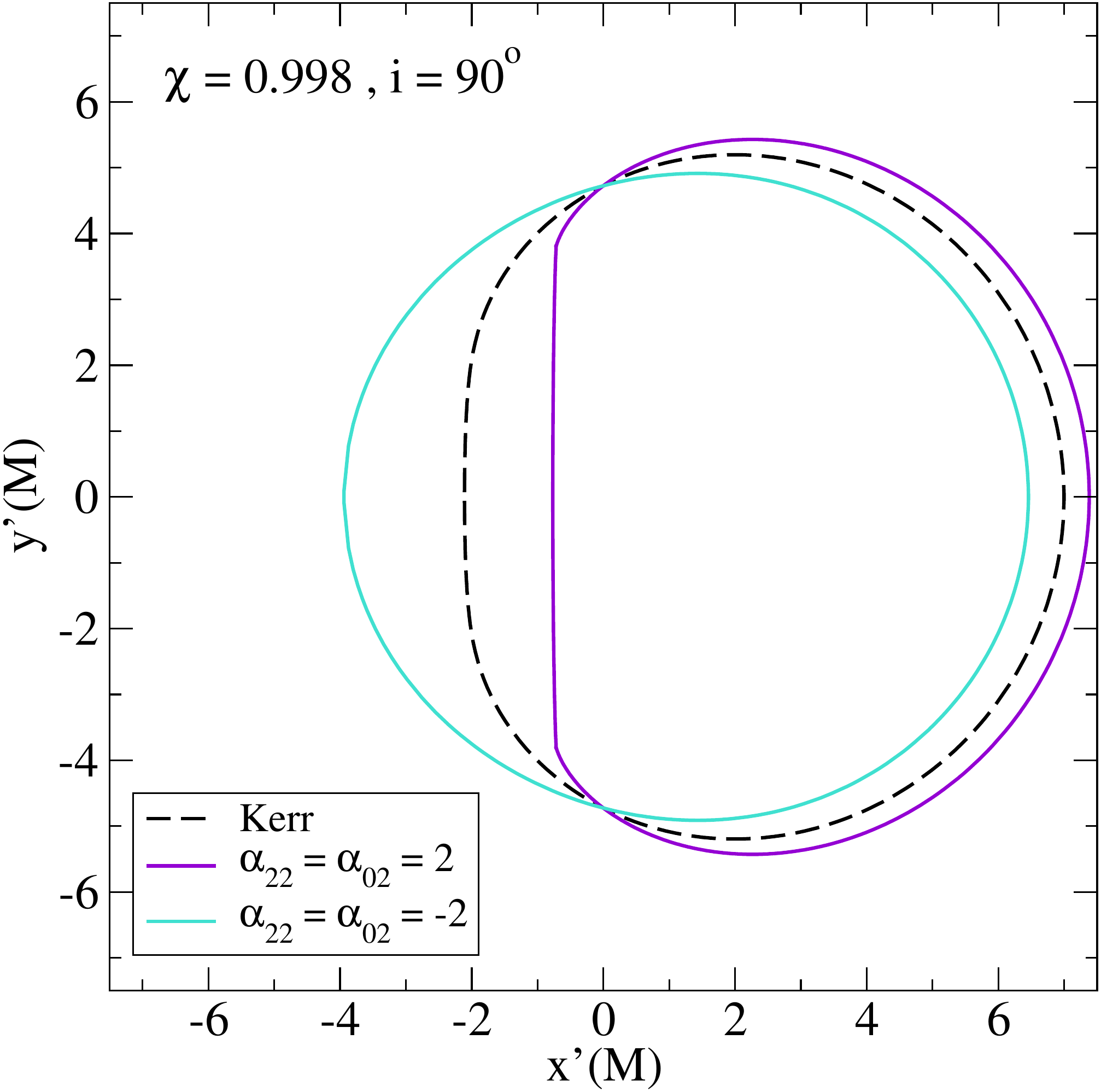}\\
\includegraphics[width=.4\textwidth]{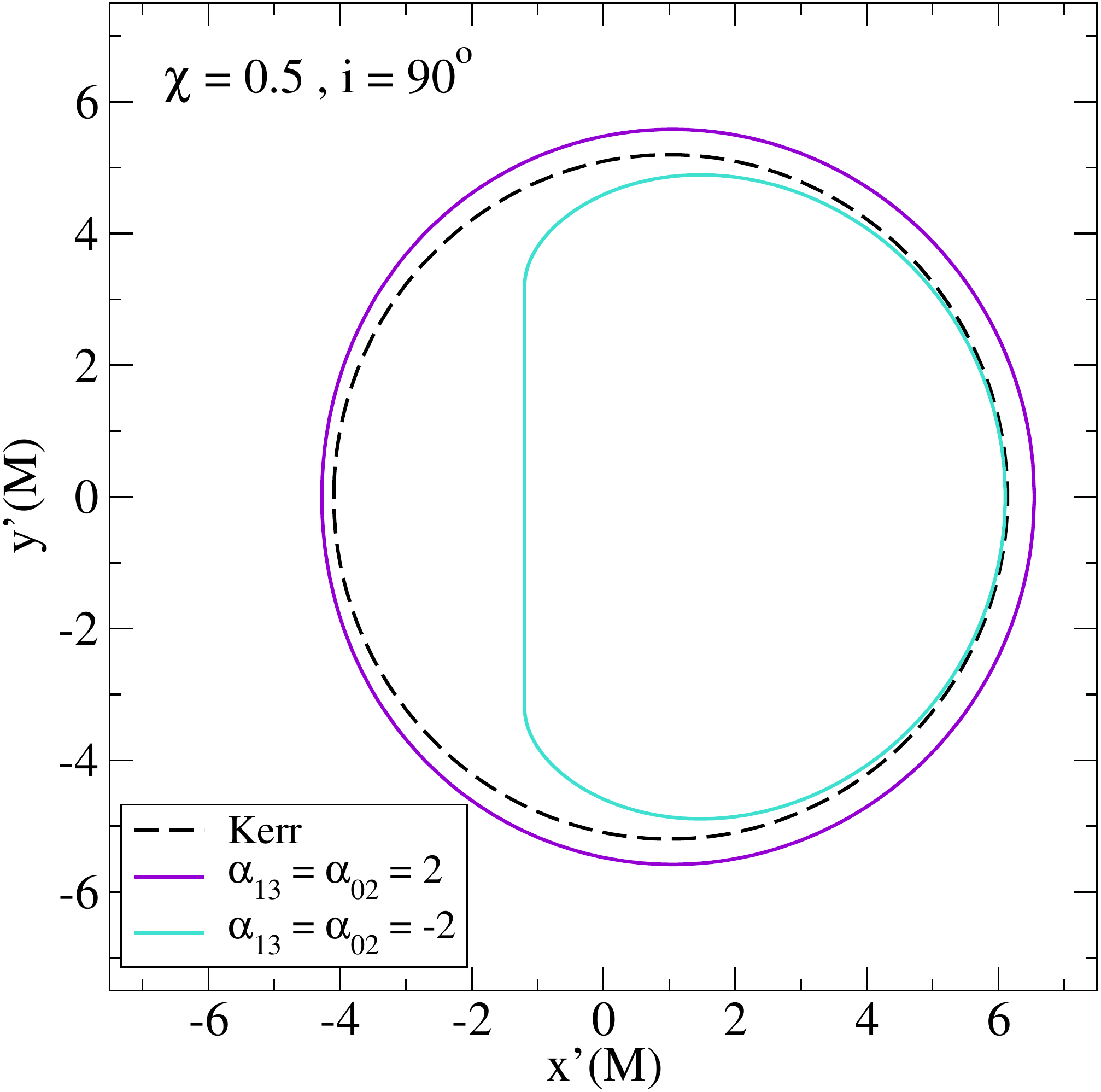}
\includegraphics[width=.4\textwidth]{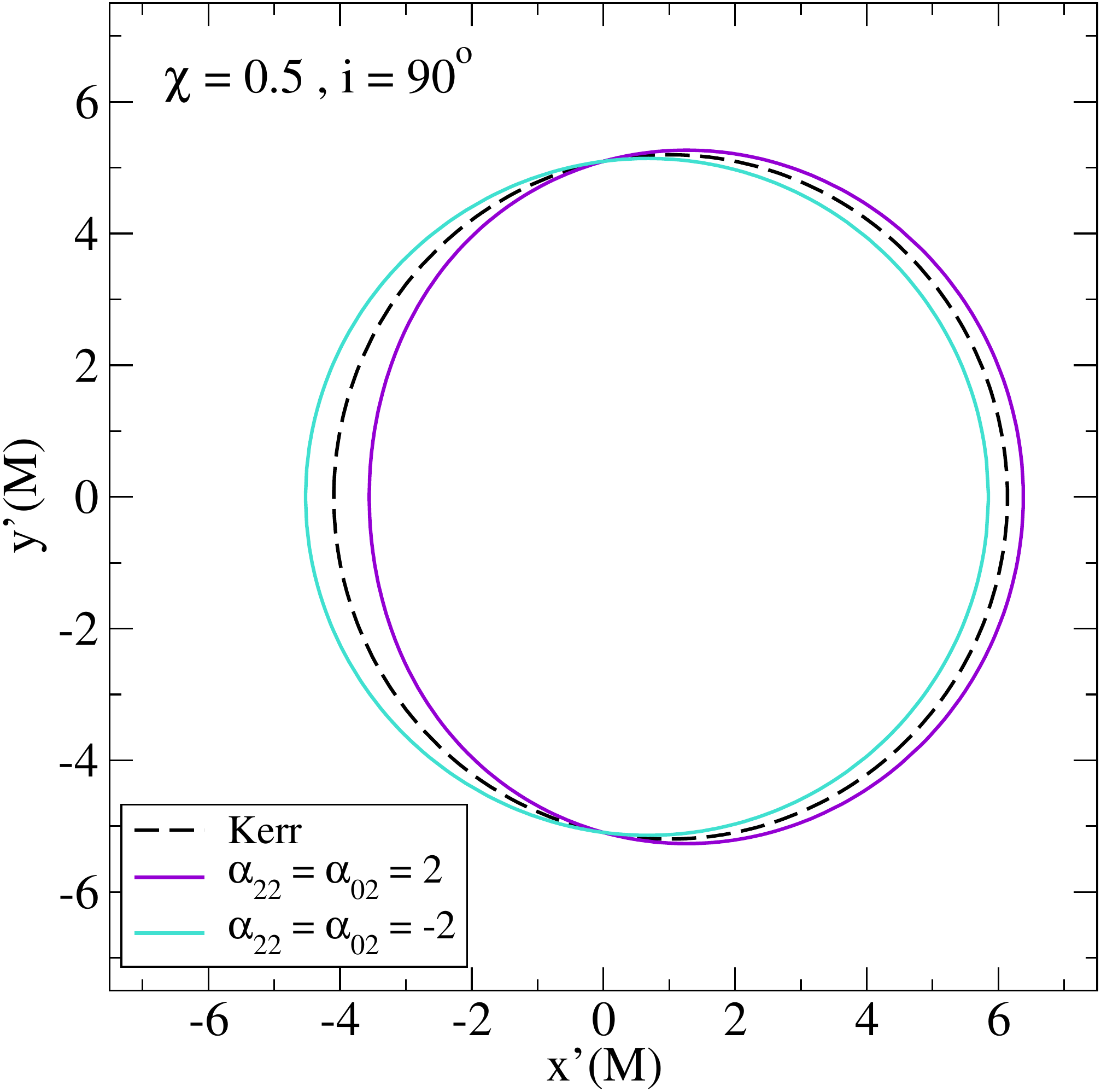}\\
\includegraphics[width=.4\textwidth]{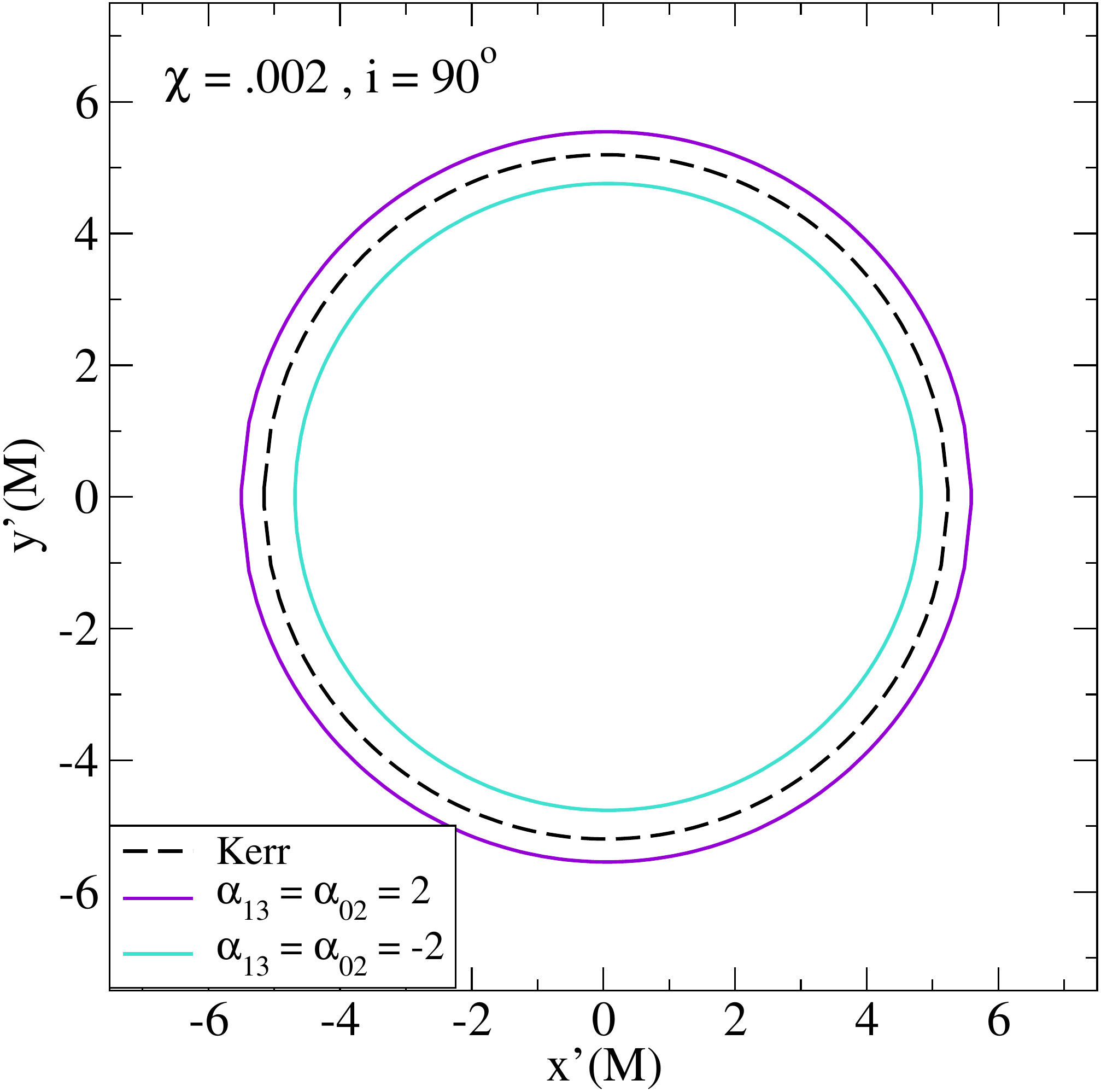}
\includegraphics[width=.4\textwidth]{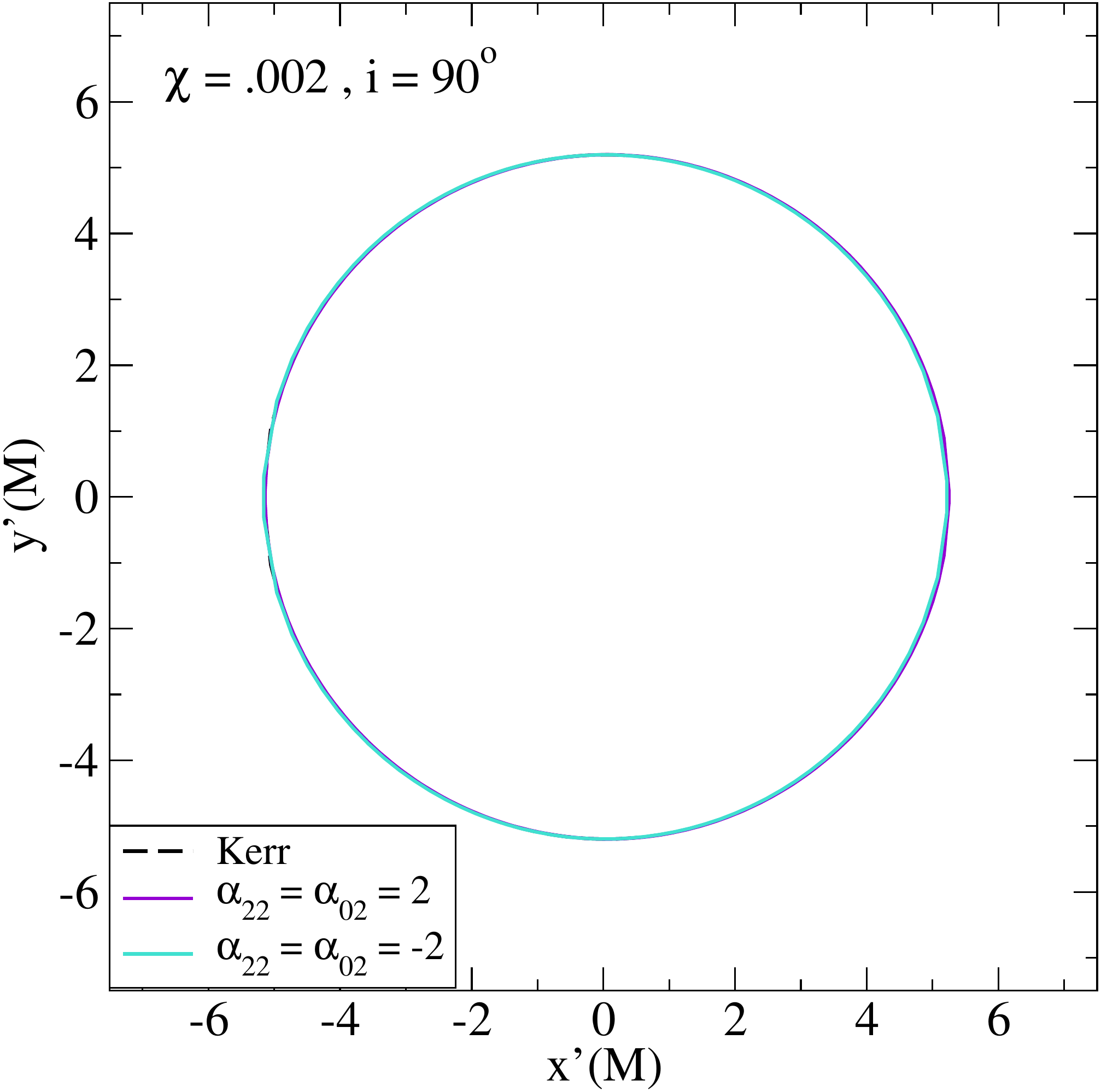}\\
\caption[Photon orbits]{Images of photon rings about a high-spin $\chi=0.998$ (top), medium-spin $\chi=0.5$ (middle), and low-spin $\chi=0$ (bottom) BH, for various non-vanishing GR deformation parameters.
We avoid combinations of such parameters that produce naked singularities, and let $\alpha_{13}=\alpha_{02}=\pm2$ (left) and $\alpha_{22}=\alpha_{02}=\pm 2$ (right).
The inclination is fixed at the extreme case of $i=90^\circ$ in every scenario, for demonstration purposes. \hyperlink{link:BHshadowsNonGR}{ \small (Return to the first reference of this figure in the main text.)}
}\label{fig:BHshadowsNonGR}
\end{center}
\end{figure*}

We then explore several properties of the new spacetime, which could possibly be translated into future EM observables.
We first locate the positions of the event horizon, Killing horizon, and ergosphere, finding that the former two reduce to the Kerr horizons, and the latter depends on just one of the five free functions (unlike the JP and mod.~$\Delta$ metrics that only have one free function or parameter) found in the metric.
Following this, we investigate the orbital properties of circular equatorial particle orbits, finding analytic expressions for the orbital energy and angular momentum, the Keplerian and epicyclic frequencies, and also the location of the ISCO.
We next derive analytic expressions for the photon rings as can be observed by e.g. the EHT, and present plots of the viewing plane as seen by a distant observer at spatial infinity for several parameterizations of the metric.
We then demonstrate the effect each parameterization has on each of the above BH properties in Fig.~\ref{fig:BHshadowsNonGR}, and also investigate the presence of naked singularities emergent for certain parameterizations.

\section{Organization and conventions}\label{intro:organization}
Let us now present the outline of the following thesis, which is based on Refs.~\cite{Zack:nuclearConstraints,Zack:futureNuclearConstraints,Zack:URrelations,Carson_multiBandPRD,Carson_multiBandPRL,Zack:Proceedings,Zack:mixedBinaries,Carson_BumpyPhotonRings,Carson_QNM_PRL,Carson_BumpyQNM,Carson_QNMPRD} by the same authors.
In Chapter~\ref{chap:GWs} we introduce a theoretical background of GWs, starting from their generation from arbitrary sources.
We then focus on circular orbits of compact object binaries as is considered for the remainder of the thesis and derive the leading-order (Newtonian orbit) gravitational waveform and explain the post-Newtonian corrections.
We then introduce the full PhenomD gravitational waveform model with numerical relativity fits that is utilized for the remainder of this thesis, along with a description of modifying the waveform in alternative theories of gravity beyond-GR, or for extended structures such as NSs.
Finally, we introduce the various GW sources considered throughout the thesis.
In Chapter~\ref{chap:ParameterEstimation} we introduce the Fisher analysis method of parameter estimation for both single GW event detections, as well as the combination of several observations.
We also describe the mechanics of a GW interferometer and describe the various current GW detectors, and future third-generation ground-based and space-based GW observatories.
Here, the contents of the thesis is then broken up into two separate parts.

In Part~\ref{part:ProbingNuclearPhysics} we discuss how one can use the detection of GWs from binary NS mergers such as GW170817 to probe nuclear physics.
We begin in Chapter~\ref{chap:ConstrainingNuclearParameters} by introducing the fundamental nuclear matter parameters which determine the EoS of nuclear matter.
We then show how detections of GWs from binary NS mergers both in the present and in the future can constrain such nuclear matter parameters.
In Chapter~\ref{chap:EoSInsensitiveRelations}, we find more accurate universal relations between NS observables using knowledge of GW170817, and show how they can be used for better parameter estimations in future observations.

Next, in Part~\ref{part:TestingGR} we show how one can test GR using the coalescences of binary BHs across the universe.
In Chapter~\ref{chap:Parameterized} we consider several alternative theories of gravity and test them against GR in a parameterized way for both current, and future GW observations including multi-band and multi-signal detections.
Chapter~\ref{chap:IMRconsistency} similarly tests alternative theories of gravity by building corrections to the gravitational waveform in both the inspiral and merger-ringdown components.
We then compare the inspiral and merger-ringdown GW signals for consistency with each other, providing a test of the underlying theory of gravity.
Finally, in Chapter~\ref{chap:beyondKerr} we investigate probes of beyond-Kerr spacetimes surrounding BHs. 
We first consider singly-parameterized beyond-Kerr spacetime metrics and show how we can test them in a similar manner to the previous chapter.
We then build a new general spacetime metric that preserves Kerr symmetries and find corrections to several astrophysical phenomena which could be linked to EM observations.

Finally, in the Appendices, we provide supplementary material to the main thesis.
In Appendix~\ref{app:LTL14} we study the correlations between the weighted tidal deformability $\tilde\Lambda$ and the tidal deformability at $1.4\text{ M}_\odot$, $\lambda_0$.
In Appendix~\ref{app:sansHybrid} we re-compute constraints on nuclear matter parameters as done in Chapter~\ref{chap:ConstrainingNuclearParameters} with the exclusion of the phenomenological equations of state.
In Appendix~\ref{app:hybrid} we consider hybrid stars, which are exotic NSs with quark-matter cores and hadronic-matter crusts.
Appendix~\ref{app:EoS-comparison} demonstrates a comparison to the analysis done in Chapter~\ref{chap:ConstrainingNuclearParameters} without restricting the equations of state to previous experimental knowledge.
In Appendix~\ref{app:multiplicative} we consider the multiplicative combinations of nuclear matter parameters, rather than the linear ones previously utilized.
Next, in Appendix~\ref{app:posteriorExample} we provide a direct demonstration of computing posterior probability distributions on nuclear matter parameters.
We follow this up in Appendix~\ref{app:theoryCompare} where we compare two similar quasi-Brans-Dicke theories of gravity by Mendes-Ortiz and Damour and Esposito-Far$\acute{e}$se.
Appendix~\ref{app:EosSpinCompare} then investigates the role of the NS EoS and spin on the constraint of scalar tensor theories of gravity.
In Appendix~\ref{app:arbitraryCorrections} we display the lengthy expressions for corrections to a remnant BH's mass and spin in an arbitrary spacetime given a metric tensor $g_{\alpha\beta}^\X$.
In Appendix~\ref{app:lowerOrder} we analyze the effects of non-vanishing lower-order metric deviation parameters on direct EM observables.
Finally, in Appendix~\ref{app:nakedSingularities} we discuss the appearance of naked singularities for certain parameterizations of the new beyond-Kerr spacetime. 

Now let us explain some of the common conventions used throughout this thesis.
Unless specifically noted, we make use of the Einstein summation notation, where the use of repeated indices implies the summation over the number of dimensions of the tensors in question.
For example,
\begin{equation}
g_{\mu\nu}dx^\mu dx^\nu\equiv \sum\limits_{\mu,\nu=0,1,2,3}g_{\mu\nu}dx^\mu dx^\nu
\end{equation}
for four-vectors $dx^\mu$.
Additionally in our tensor notation, we make use of both Greek indices which pertain to all four spacetime indices $\mu,\nu,\dots=0,1,2,3$ ($0,1,2,3$ commonly refer to $t,x,y,z$, or  $t,r,\theta,\phi$, etc.), and Latin indices which correspond to only the three spatial indices $i,j,\dots=1,2,3$.
When pertaining to derivatives, we will commonly make use of the shorthand notation $\partial_\X\equiv\frac{\partial}{\partial X}$.
Occasionally, we also utilize the convention that $\dot{F}\equiv \frac{dF}{dt}$ represents a time derivative, while $F'\equiv\frac{dF}{dr}$ represents a radial derivative.
Additional dots and primes correspond to additional consecutive time and radial derivatives.
When considering tensor derivatives, we use the standard ``comma'' notation, in which one takes a partial derivative over indices located after a comma.
For example,
\begin{equation}
T^{\alpha}{}_{\beta,\mu\nu\dots}\equiv\frac{\partial}{\partial x^\mu}\frac{\partial}{\partial x^\nu}\dots T^{\alpha}{}_\beta
\end{equation}
for some arbitrary tensor $T^{\alpha}{}_{\beta}$.
We also make use of the d'Alembert operator, which is simply defined as $\Box\equiv\partial^\mu\partial_\mu$.
Finally, throughout this thesis, we have adopted geometric units of $G=c=1$, unless otherwise stated.
In these useful units, mass, length, time, and energy all have the same unit (usually of length or time).
A useful conversion factor in the geometric unit system is $1\text{ M}_\odot=5\times10^{-6} \text{ s}=1.5\text{ km}$.

\end{spacing}


\chapter{Gravitational waves}\label{chap:GWs}
\begin{adjustwidth}{\parindent}{\parindent}
\begin{spacing}{1.5}

Gravitational waves fill our observable universe, embedded with a rich source of information regarding their creation.
Traveling at the speed of light, such radiation originates from several different astrophysical sources across the universe.
Dominant among these sources are the binary inspirals of compact objects such as black holes and neutron stars, culminating in a violent coalescence event accompanied by a large burst of gravitational radiation.
In particular, tidal deformation effects present in the coalescence of neutron stars are embedded into the gravitational waveform which tells us something about the nuclear physics present within.
On the other hand, features present within the waveform emanating from the inspiral of black holes reflect details about the spacetime surrounding such objects.
In this chapter, we find the gravitational waveform generated from the circular inspiral of compact objects which contains remnants whence they originated.
\end{spacing}
\end{adjustwidth}
\newpage


\section{Generation of gravitational waves}
In this section our main goal is to express the radiation of GWs from various sources across the universe.
For a more thorough description on the generation of GWs, see Refs.~\cite{schutz_2018,carroll_2019}.
In particular, we treat GWs as simply leading-order perturbations to a flat spacetime.
In order for one to characterize the properties of a given spacetime, we introduce the notion of a \textit{metric} to quantify scalar line elements $ds^2$.
Specifically, the line element in an arbitrary spacetime described by a metric $g_{\mu\nu}$ can be written as
\begin{equation}
ds^2=g_{\mu\nu}dx^\mu dx^\nu.
\end{equation}
Here, $g_{\mu\nu}$ is a covariant (0,2)-tensor which takes two vectors $dx^\mu$ and $dx^\nu$ as input to recover the invariant scalar property $ds^2$.
In an arbitrary spacetime, the symmetric metric tensor $g_{\mu\nu}$ can have non-vanishing off-diagonal elements as well as non-unity elements which describe the curvature of the spacetime in question.
In a flat, or Minkowski spacetime, the metric can be written as
\begin{equation}
g_{\mu\nu}=\eta_{\mu\nu} = \begin{pmatrix}
   -1 & 0 & 0 & 0 \\
	0 & 1 & 0 & 0 \\
	0 & 0 & 1 & 0 \\
	0 & 0 & 0 & 1 \end{pmatrix},
\end{equation}
where $\eta_{\mu\nu}$ is known as the \textit{Minkowski metric}.
We then consider GWs as small linear perturbations $h_{\mu\nu}$ to a Minkowski background like so
\begin{equation}\label{eq:linearized}
g_{\mu\nu}=\eta_{\mu\nu}+h_{\mu\nu},
\end{equation}
where we assume $|h_{\mu\nu}| \ll 1$.
This formalism is known as \textit{linearized gravity}.

Now that we have our perturbed spacetime metric on hand, we must solve for their solutions in GR.
To do this, we consider the famous Einstein's equations
\begin{equation}\label{eq:EinsteinEquations}
R_{\mu\nu}-\frac{1}{2}g_{\mu\nu}R=8\pi T_{\mu\nu}
\end{equation}
where $T_{\mu\nu}$ is the stress-energy tensor which describes the matter and energy densities present.
The Ricci tensor $R_{\mu\nu}$ and scalar $R$ can be obtained from the Riemann curvature tensor $R^\mu{}_{\nu\alpha\beta}$ like so
\begin{gather}
R^\mu{}_{\nu\alpha\beta}\equiv \Gamma^\mu{}_{\nu\beta,\alpha}-\Gamma^\mu{}_{\nu\alpha,\beta}+\Gamma^\mu{}_{\delta\alpha}\Gamma^\delta{}_{\nu\beta}-\Gamma^\mu{}_{\delta\beta}\Gamma^\delta{}_{\nu\alpha}\\
R_{\mu\nu}=R^\delta{}_{\mu\delta\nu}\\
R=R^\mu{}_\mu,
\end{gather} 
where the Christoffel symbols $\Gamma_{\alpha\mu\nu}=\frac{1}{2}(g_{\alpha\mu,\nu}+g_{\alpha\nu,\mu}-g_{\mu\nu,\alpha})$ describe the connections between different tangent spaces on a manifold.
We finally substitute our perturbed spacetime metric $g_{\mu\nu}$ from Eq.~\eqref{eq:linearized} into Einstein's equation from Eq.~\eqref{eq:EinsteinEquations} to obtain a wave equation of the form
\begin{equation}\label{eq:LinearizedEE}
\Box \bar{h}_{\mu\nu}=-16\pi T_{\mu\nu}
\end{equation}
to linear order in perturbation $h_{\mu\nu}$.
In the above expression, we have defined $\bar h_{\mu\nu}\equiv h_{\mu\nu}-\frac{1}{2}h\eta_{\mu\nu}$ to be the trace-reversed metric perturbation, where the Lorentz gauge of $\bar{h}^{\mu\nu}{}_{,\nu}=0$ has been assumed.

Following this, assuming we are in vacuum, there exists additional gauge freedom which can be used to further allow $\bar{h}=h^{0i}=0$.
This combined with the Lorentz gauge above results in $h^{00}{}_{,0}=0$, or $\frac{\partial}{\partial t}h^{00}=0$.
This tells us that the $(0,0)$, or $tt$-component of our metric perturbation is non-dynamical.
Time-independent contributions do not affect the observable GWs, which allow us to neglect the component, so $h^{00}=0$.
This set of gauge conditions is known as the transverse-traceless (TT) gauge, which is conveniently summarized as:
\begin{equation}\label{eq:TTgauge}
h_{0\mu}^\TT=h^\TT=h_{ij}^{\TT,j}=0.
\end{equation}

Finally, we are left with two residual degrees of freedom in our perturbation metric $h_{\mu\nu}$.
These are commonly decomposed into two linearly independent polarizations for GWs, known as the $+$ (plus), and $\times$ (cross) modes.
Each polarization mode has amplitudes defined to be $h_+$ and $h_\times$.
For example, GWs propagating in the $+\hat{z}$-direction results in a perturbation metric in the TT gauge to be
\begin{equation}
h^\TT_{\mu\nu} = \begin{pmatrix}
	0 & 0 & 0 & 0 \\
	0 & h_+ & h_\times & 0 \\
	0 & h_\times & -h_+ & 0 \\
	0 & 0 & 0 & 0 \end{pmatrix}.
\end{equation}
We see that GW spacetime perturbations $h^\TT_{\mu\nu}$ are then composed of both plus and cross polarization modes.

\begin{figure}
\begin{center} 
\includegraphics[width=\columnwidth]{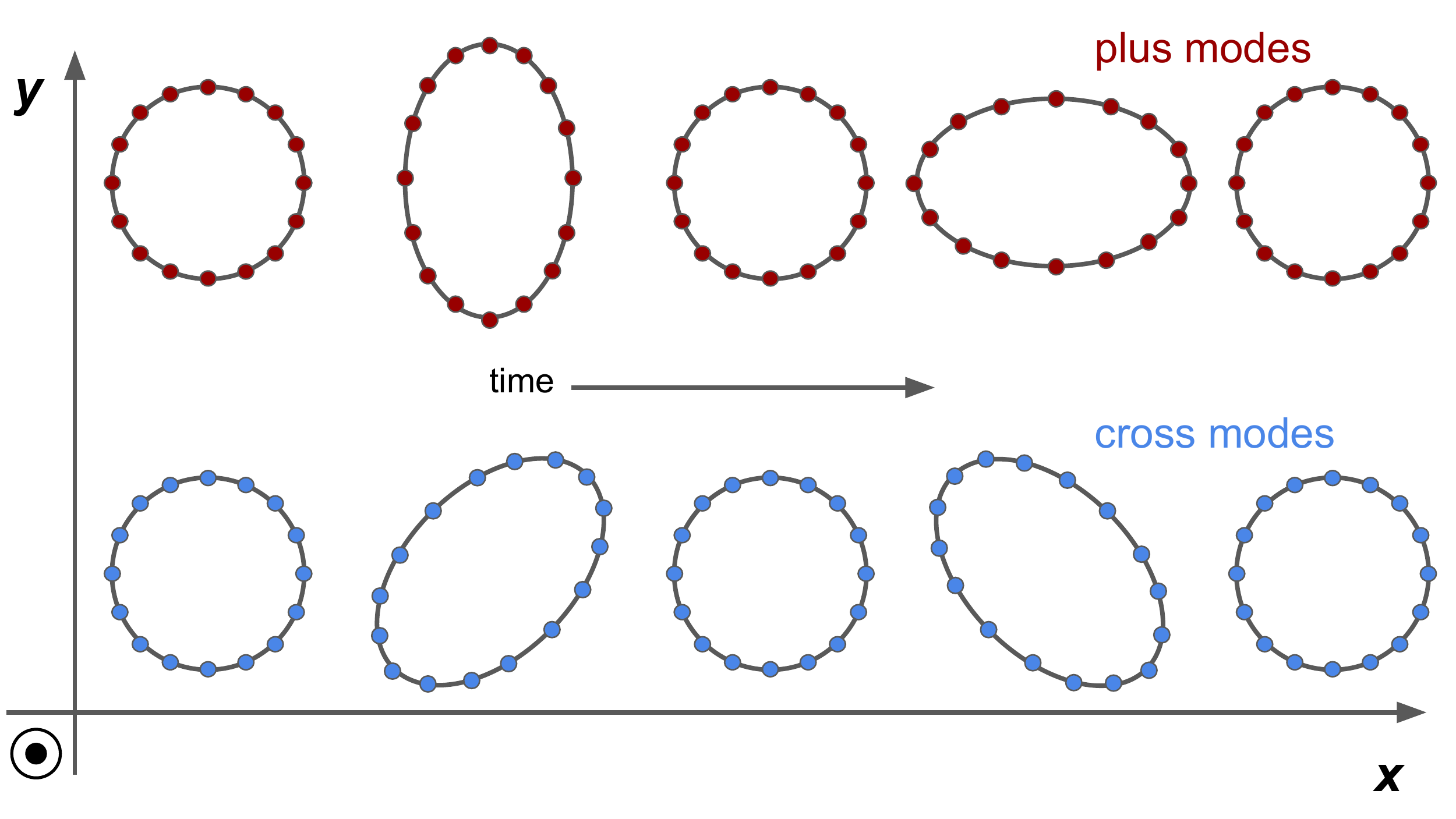}
\end{center}
\caption[GWs on a test ring of particles]{
Reaction of a ring of test particles to the passing of plus-polarized (top, red) and cross-polarized (bottom, blue) GWs traveling in the $+\hat{z}$ direction.}
\label{fig:ringParticles}
\end{figure}

Now we consider the effects of GWs as they pass through matter.
To visualize this, we first consider a simple ring of particles situated in spacetime as indicated by the left-most diagrams in Fig.~\ref{fig:ringParticles}.
We then subject the rings of particles to both plus (top) and cross (bottom) polarized GWs.
As time passes (to the right), we see that rings of particles become deformed in an oscillatory behavior. 
For example, we see that under the influence of plus-polarized GWs, the ring is stretched first in the $\hat{y}$-direction, followed by the $\hat{x}$-direction.
On the other hand, cross-polarized GWs do the same thing but rotated by $45^\circ$.
Such an example demonstrates the resulting stretching and compressing of spacetime under the influence of GW perturbations.
We can see how such effects allow interferometers to observe GWs by their differential stretching/compressing of each orthogonal arm, as discussed further in Sec.~\ref{intro:Detectors}.

Finally, we must discuss the generation of GW amplitudes $h_+$ and $h_\times$ from arbitrary sources with stress-energy tensor $T_{\mu\nu}$.
We begin by solving the Einstein equations found in Eq.~\eqref{eq:LinearizedEE} with use of a retarded Green's function.
The resulting generalized solution for $\bar{h}_{\mu\nu}$ can be found to be
\begin{equation}
\bar{h}_{\mu\nu}(t,\bm{x})=4\int d^3x'\frac{T_{\mu\nu}(t-|\bm{x}-\bm{x}'|,\bm{x}')}{|\bm{x}-\bm{x}'|}
\end{equation}
for source and field vectors $\bm{x'}$ and and $\bm{x}$ respectively.
Now we apply the TT gauge found in Eq.~\eqref{eq:TTgauge} to find 
\begin{equation}
h_{ij}^\TT(t,\bm{x})=4\int d^3x'\frac{T_{ij}^\TT(t-|\bm{x}-\bm{x}'|,\bm{x}')}{|\bm{x}-\bm{x}'|},
\end{equation}
where $T_{ij}^\TT$ is the transverse-traceless part of the stress-energy tensor.
In particular, we are now interested in gravitational waves outside of the GW source, which has approximate size $L$, with the field-points in question having $r=|\bm{x}|\gg L$.
Further, if we assume (usually correctly) that the source velocity is much smaller than the speed of light, we can employ a Taylor formula (reminiscent of the Laplace expansion) to expand our integrand into multipole moments like so
\begin{equation}
\frac{T_{ij}(t-|\bm{x}-\bm{x}'|,\bm{x}')}{|\bm{x}-\bm{x}'|}\to\frac{T_{ij}(t-r,\bm{x}')}{r}+\mathcal{O}\left(\frac{1}{r^2}\right),
\end{equation}
where we have only taken into account the leading order multipole.
This then gives us our metric perturbation in the TT gauge as
\begin{equation}\label{eq:multipole}
h_{ij}^\TT(t,\bm{x})\approx \frac{4}{r}\int d^3x'T_{ij}^\TT(t-r,\bm{x}'),
\end{equation}
where we note that to leading order, gravitational radiation decreases only at a linear rate of $1/r$.

Finally, we consider the conservation of energy in the system.
In particular, this constraint tells us that the stress-energy tensor must obey the expression 
\begin{equation}
T^{\mu\nu}{}_{,\nu}=0.
\end{equation}
By expanding the conservation equation into spatial and temporal components, we arrive at
\begin{gather}
T^{tt}{}_{,t}=-T^{ti}{}_{,i}\\
T^{ti}{}_{,t}=-T^{ij}{}_{,j}
\end{gather}
which allows one to find
\begin{equation}
T^{tt}{}_{,tt}=T^{ij}{}_{,ij}.
\end{equation}
With several more manipulations, one ultimately finds that
\begin{equation}
T^{kl}=\frac{1}{2}(T_{tt}x^kx^l)^{,tt}-2(T_{kl}x^l+T_{il}x^k)^{,i}.
\end{equation}
Inserting the above expression into the leading-order multipolar expansion in Eq.~\eqref{eq:multipole}, one arrives at the desired result:
\begin{equation}
h_{ij}^\TT=\frac{2}{r}\frac{\partial^2}{\partial t^2}\int d^3x' \rho(t-r,\bm{x}')x_i' x_j'\equiv\frac{2}{r} \ddot{M}_{ij}^\TT,
\end{equation}
where $\rho\equiv T_{00}$ is the mass density.
We have defined $M_{ij}$ to be the \textit{mass quadrupole moment}, or
\begin{equation}\label{eq:quadrupole}
M_{ij}(t-r)\equiv\int d^3x \rho(t-r,\bm{x}')x_i'x_j'.
\end{equation}
Finally, we can express the plus- and cross-polarized GW perturbations as
\begin{gather}\label{eq:GWs}
h_+(t,\bm{x})=\frac{1}{r}\left( \ddot{M}_{11}(t-r)-\ddot{M}_{22}(t-r) \right)\\
h_\times(t,\bm{x})=\frac{2}{r}\ddot{M}_{12}(t-r).
\end{gather}
We note that the above expressions are valid only on the $\hat{z}$-axis, while Ref.~\cite{misner_thorne_wheeler_2008} describes the more general projections onto arbitrary axes.
Thus, ultimately we find that the presence of quadrupolar gravitational radiation requires the non-vanishing second-derivatives of the mass quadrupole moment.
For example, this radiation occurs within the orbiting binary of stellar objects which we will generalize to in the following section.


\section{The gravitational waveform}\label{intro:PhenomD}
Now that we have established the GW perturbation $h$ from an arbitrary source with mass quadrupole moment $M_{ij}$, we focus our attentions on our main source of interest in the following thesis - the inspiral merger of binary compact objects.
In particular, we first focus on the leading-order GW contributions for the case of a circular binary.
We then expand to the higher-order corrections to the waveform with expansion parameter $(v/c)^{2}$.
Finally, we present the PhenomD~\cite{PhenomDI,PhenomDII} gravitational waveform which incorporates numerical-relativity (NR) fits to further increase the accuracy of our waveform model.
We follow this up with a description of the effect of extended structures such as NSs in the gravitational waveform.
Finally, we conclude by expanding to beyond GR in the ppE formalism by inserting parameterized corrections to the waveform at arbitrary PN orders. 

\subsection{Leading order quadrupole radiation from circular orbit binaries}\label{intro:PN}
\begin{figure}
\begin{center} 
\includegraphics[width=\columnwidth]{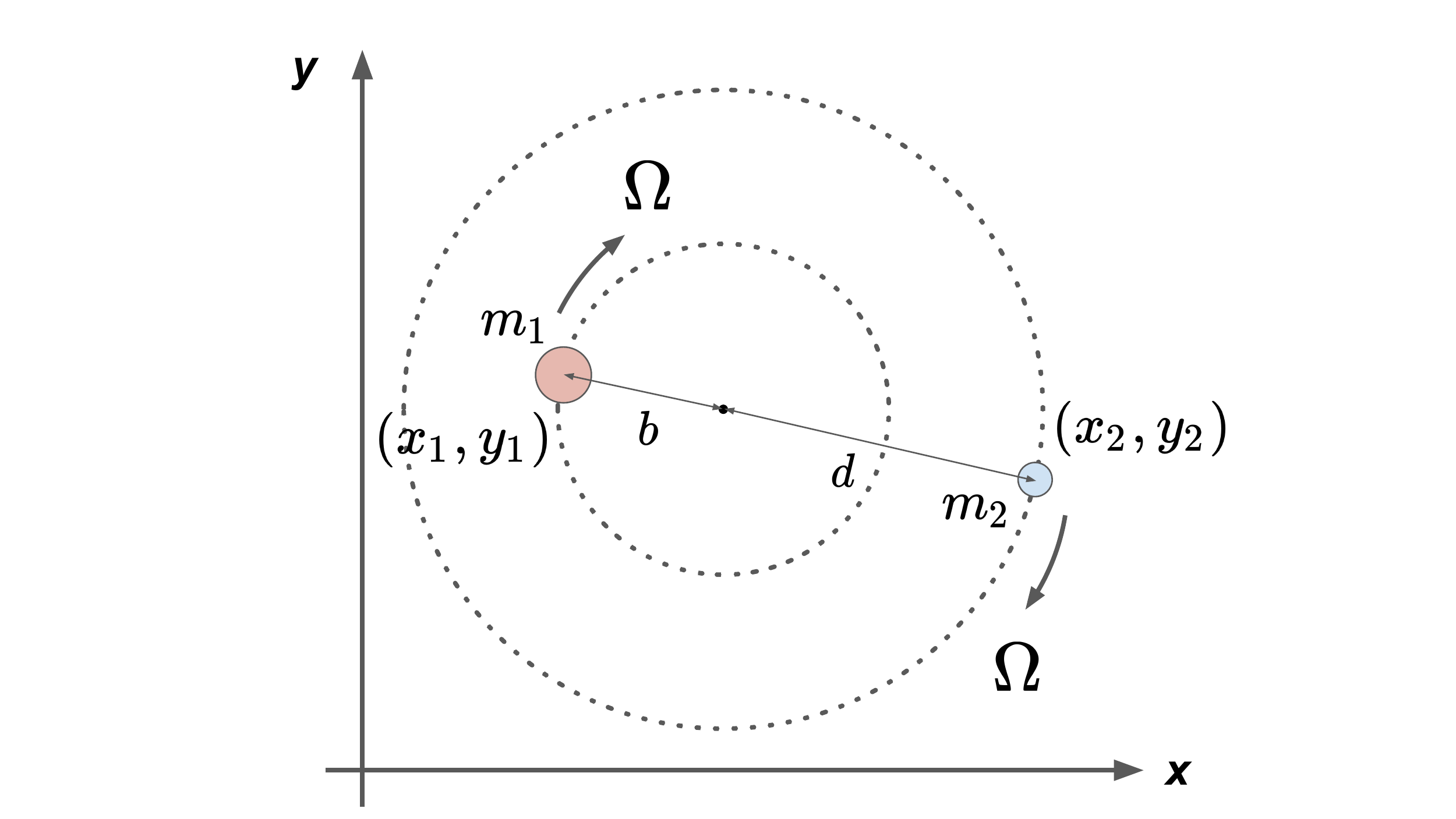}
\end{center}
\caption[Circular binary inspiral]{Diagram of a circular binary orbit with two objects of mass $m_1$ and $m_2$ separated by a total distance $a=b+d$.}
\label{fig:circularOrbit}
\end{figure}

In this section we will focus on the simple case of two stellar bodies in a perfectly circular binary inspiral as shown in Fig.~\ref{fig:circularOrbit}.
As the objects radiate gravitational radiation which removes energy and angular momentum from the system, they slowly inspiral towards each other
At the point that the \textit{innermost-stable-circular-orbit} (ISCO) is reached, the objects finally enter a plunging orbit due to the absence of stable orbits and finally make contact with each other~\cite{Pretorius:2005gq}.
At this point, if the objects are black holes (BHs) with initial masses $m_1$ and $m_2$, a common horizon is formed and a remnant BH is created with total mass $M_f<m_1+m_2$ (due to the energy loss from GWs), which promptly rings down to a relaxed state via the emission of \textit{quasinormal modes} (QNMs).
Similarly, if the bodies are neutron stars (NSs), a hypermassive NS is formed which is either stable (for small-mass NSs), or collapses (promptly with slow rotations, and delayed with fast rotations) into a small-mass BH.

As shown in Fig.~\ref{fig:circularOrbit}, we assume two compact objects with mass $m_1$ and $m_2$ in a circular binary orbit.
We generalize the orbit to the the $x-y$ plane for simplicity, to be located a luminosity distance $D_L$ from Earth.
If the total mass of the system is $M=m_1+m_2$, and the orbital separation is given by $a$, we can find the orbital angular velocity to be $\Omega^2=M/a^3$ to leading order with Kepler's Law.
We can then find the positions of each object $(x_\A^1,x_\A^2)$ in the center of mass frame to be simply
\begin{gather}
(x_1^1,x_1^2)=\left( \frac{m_2}{M}a\cos\left( \Omega t+\frac{\pi}{2} \right) , \frac{m_2}{M}a\sin\left( \Omega t+\frac{\pi}{2} \right) \right)\\
(x_2^1,x_2^2)=\left( - \frac{m_1}{M}a\cos\left( \Omega t+\frac{\pi}{2} \right) , - \frac{m_2}{M}a\sin\left( \Omega t+\frac{\pi}{2} \right) \right).
\end{gather}

Finally, if we assume the compact objects to be point masses, we can simplify the mass quadrupole moment formula found in Eq.~\eqref{eq:quadrupole} with Dirac-delta functions.
The resulting definition of point-particle quadrupole moment is given by
\begin{equation}
M_{ij}=\sum\limits_{\A=1,2}m\A x_\A^ix_\A^j.
\end{equation}
For the case of the circular binary found in Fig.~\ref{fig:circularOrbit}, we find the second derivative of the $(1,1)$-component of the quadrupole moment to be
\begin{align}
\nonumber \ddot{M}_{11}&=\frac{d^2}{dt^2}(m_1x_1^1x_1^2+m_2x_2^1x_2^2)\\
\nonumber&=\eta M a^2\frac{d^2}{dt^2}\cos^2\left( \Omega t+\frac{\pi}{2} \right)\\
&=2\eta M a^2\Omega^2\cos(2\Omega t)
\end{align}
for symmetric mass ratio $\eta\equiv\frac{m_1 m_2}{M^2}$.
Similarly, the $(2,2)$- and $(1,2)$-components can be found as
\begin{align}
\ddot{M}_{22}&=-2\eta M a^2\Omega^2\cos(2\Omega t),\\
\ddot{M}_{12}&=2\eta M a^2\Omega^2\sin(2\Omega t).
\end{align}
We can then substitute the above expressions into the final quadrupole radiation formula in Eq.~\eqref{eq:GWs} to find
\begin{gather}\label{eq:hplushcross}
h_+=\frac{4\eta M}{D_L}a^2\Omega^2\cos(2\Omega t)=\frac{4\eta M^2}{a D_L}\cos(2\Omega t)\\
h_\times=\frac{4\eta M}{D_L}a^2\Omega^2\sin(2\Omega t)=\frac{4\eta M^2}{a D_L}\sin(2\Omega t),
\end{gather}
where once again the above expressions are only valid along the $\hat{z}$-axis, and can be seen more generally in~\cite{misner_thorne_wheeler_2008}.

From the above formulae, we notice a few key points.
First, we find that the amplitude of GWs only falls of as $1/r$, contrary to typical EM radiation, where the flux falls off as $1/r^2$.
In the former case, while the flux of GWs still fall off as $1/r^2$ like for EM waves, we can directly measure the amplitude as a distortion of spacetime, as opposed to EM radiation where we can only measure the flux impacted on a photodetector.
Secondly, we observe that the amplitude of GWs are proportional to the total mass $M^2$ of the binary system.
Finally, we see that the GW magnitude is inversely proportional to the orbital separation $a$.
The latter two points tell us that the more compact an object is (i.e. the ratio of the mass to the size, or quantitatively $C\equiv M/R$ for compactness $C$) the larger magnitude GWs are radiated. 
This is because for objects with smaller compactnesses, the separation distance can approach smaller values while maximizing the total mass.
For this reason, binary BH inspirals with individual compactnesses of $C=1/2$ make ideal candidates for GW sources, as well as tests of extreme gravity.

Next we must find the evolution of the frequency and orbital separation of the binary system to estimate the GW phase.
To do so, we begin by finding the total energy of the system, given by
\begin{equation}
E=\frac{1}{2}m_1v_1^2+\frac{1}{2}m_2v_2^2-\frac{m_1m_2}{a}
\end{equation}
to leading order, where the first two terms are the individual kinetic energies of each object, and the last is the gravitational potential energy between the two.
By taking the velocity of each object to be $v_\A=a\frac{m_\B}{M}\Omega$ with $\Omega^2=M/a^3$, we can reduce the above expression to 
\begin{equation}
E=-\frac{\eta M^2}{2a}.
\end{equation}
By following Refs.~\cite{schutz_2018,carroll_2019}, the gravitational wave luminosity $L_\GW\equiv\frac{dE}{dt}$ emitted from the binary system can be written as
\begin{equation}
\frac{dE}{dt}=-\frac{1}{5}\left\langle \dddot{M}_{ij}\dddot{M}_{ij}-\frac{1}{3}\left(\dddot{M}_{kk}\right)^2\right\rangle.
\end{equation}
By substituting the above expressions for $M_{ij}$ into this formula, we get a total GW luminosity of
\begin{equation}
L_\GW=\frac{32}{5}\frac{\eta^2M^2}{a^5}\equiv -\frac{dE}{dt},
\end{equation}
which is exactly equal to the loss of energy in the system.

Knowing the above expressions for $E$ and $L_\GW$, we next want to find the evolution of the orbital separation distance, or $a(t)$.
We do this via a simple chain-rule expansion
\begin{equation}
\frac{da}{dt}=\frac{da}{dE}\frac{dE}{dt}=\left(\frac{dE}{da}\right)^{-1} L_\GW=-\frac{64\eta M^3}{5a^3}.
\end{equation}
The solution to this simple differential equation with the initial condition of $a(t=t_c)=0$ for coalescence time $t_c$ is then
\begin{equation}\label{eq:orbitalSeparation}
a(t)=4\left\lbrack \frac{1}{5} \eta M^3 (t_c-t)\right\rbrack^{1/4}.
\end{equation}
Finally, we must find the frequency evolution of the binary system $f(t)$.
We start from the definition of the GW frequency $f$
\begin{equation}
f\equiv\frac{\Omega}{\pi}=\frac{1}{\pi}\sqrt{\frac{M}{a^3}}
\end{equation}
using Kepler's Law once again.
Substituting $a(t)$ from Eq.~\eqref{eq:orbitalSeparation} gives us the final frequency evolution of GWs to leading order as
\begin{equation}\label{eq:foft}
f(t)=\frac{5^{3/8}}{8\pi}\frac{1}{\mathcal{M}^{5/8}(t_c-t)^{3/8}},
\end{equation}
for \textit{chirp mass} $\mathcal{M}=M\eta^{3/5}$.

Finally, we estimate the GW phase as shown in Ref.~\cite{Cutler:Fisher}.
This is done by integrating the following expression
\begin{equation}\label{eq:phioft}
\phi(t)=\int2\pi f(t) dt=-2\left( \frac{1}{5\mathcal{M}}(t_c-t) \right)^{5/8}+\phi_c
\end{equation}
for coalescence phase $\phi_c$.
To find the gravitational waveform in the Fourier domain $\tilde{h}(f)$ which we can observe in, we make use of the stationary phase approximation.
By assuming that the binary velocities are much less than the speed of light, the Fourier component $\tilde{h}(f)$ of $h(t)=A(t)\cos\phi(t)$ under the stationary phase approximation can be found to be~\cite{StationaryPhase}
\begin{equation}\label{eq:stationaryPhase}
\tilde{h}(f)\approx\frac{1}{2}A\left(t(f)\right)e^{i\left(2\pi f t(f)-\phi(f)-\pi/4\right)}.
\end{equation}
We begin building this expression by inverting Eq.~\eqref{eq:foft} to find $t(f)=t_c-5\mathcal{M}(8\pi\mathcal{M}f)^{-8/3}$, which can be substituted into Eq.~\eqref{eq:phioft} to find $\phi(f)=\phi_c-2(8\pi\mathcal{M}f)^{-5/3}$.
By substituting these expressions into Eq.~\eqref{eq:stationaryPhase} along with the amplitudes $A\left(t(f)\right)$ from Eq.~\eqref{eq:hplushcross}, we get the final result after some manipulation
\begin{gather}\label{eq:FourierWaveform}
\tilde{h}_+(f)=A f^{-7/6}e^{i\Psi(f)}\\
\tilde{h}_\times(f)=A f^{-7/6}e^{i\Psi(f)+i\pi/2},
\end{gather}
where once again the expressions are valid on the $\hat{z}$-axis, described more generally in~\cite{misner_thorne_wheeler_2008}.
Here we have defined $A\equiv\sqrt{5/24}\pi^{-2/3}\mathcal{M}^{5/6}/D_L$ and $\Psi(f)=2\pi f t_c-\phi_c-\pi/4+\frac{3}{128}(\pi\mathcal{M}f)^{-5/3}$.
See in e.g. Fig.~\ref{fig:sensitivities} for the gravitational waveform characteristic strain amplitude $2\sqrt{f}|\tilde{h}(f)|$ for a GW150914-like event.

The above expressions for the Fourier-space gravitational waveform $\tilde{h}(f)$ have been derived to leading order, which we will call the $0$-th \textit{post-Newtonian} (PN) order term in the waveform.
At the $0$-th PN order, we have assumed a Newtonian circular orbit.
In the PN formalism, we can then provide corrections to this by expanding to higher PN orders proportional to $(v/c)^{2n}$ for $n$-PN corrections relative to the leading order Newtonian contributions as found in Eq.~\eqref{eq:FourierWaveform}.
Here, $v$ is the characteristic relative velocity of the objects in the binary, and is given by $v=(\pi M f)^{1/3}$.
Higher PN-order corrections to the gravitational waveform amplitude and phase can be found in e.g. Refs.~\cite{{PhysRevD.80.084043,Blanchet:2013haa,Bohe:2013cla,Poisson:1997ha,Arun:2008kb,Mikoczi:2005dn}
} for up to $3.5$PN order, by following the same process used above for modified effects.


\subsection{PhenomD gravitational waveform}\label{sec:phenomD}
Accurate models of the gravitational waveform signal are crucial in the \textit{matched filtering} process used to detect such signals, as well as in the extraction of observables from them.
In the matched filtering stage of detection used by the LVC, a potential GW signal detected in the interferometer is then matched to a large bank of gravitational wave templates with a large variety of parameterizations, individually.
If a high correlation between the detected signal and a given template waveform is observed, the detection becomes a candidate event with a quick estimation of its parameters.
At this point, the candidate event is passed along for a more in-depth Bayesian analysis to extract most-likely parameters from the observed signal relative to the assumed template waveform.

\begin{figure}
\begin{center} 
\includegraphics[width=.7\columnwidth]{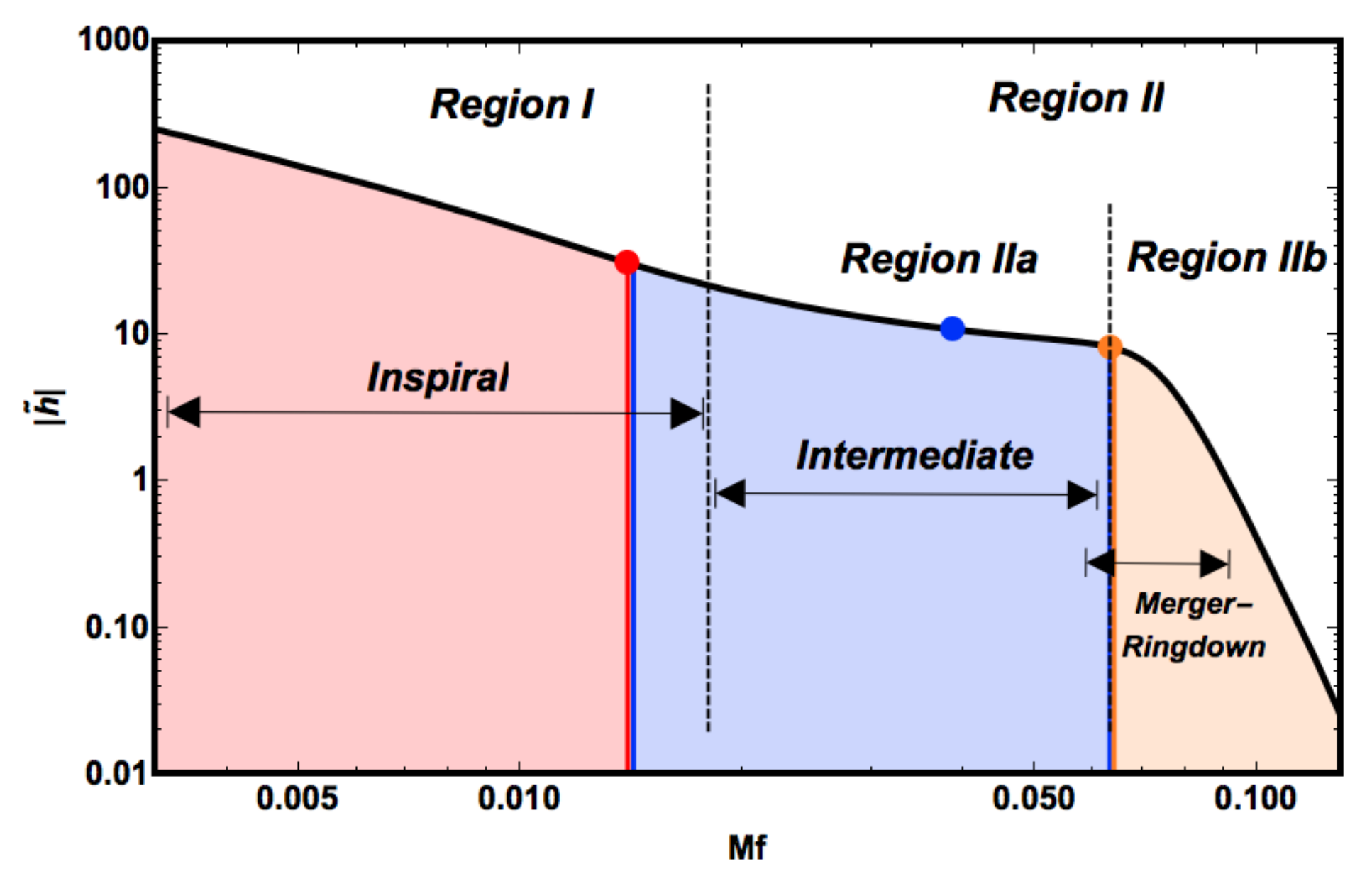}
\end{center}
\caption[PhenomD gravitational waveform]{Amplitude of the PhenomD gravitational waveform $|\tilde{h}(f)|$. Region I ($M f<0.018$) corresponds to the waveform portion with a hybrid of PN terms and NR data, while region II ($M f\geq0.018$) is purely composed of NR data. (Taken from Refs.~\cite{PhenomDI,PhenomDII}.)}
\label{fig:PhenomD}
\end{figure}

Now let us consider further corrections to the gravitational waveform, beyond the PN ones discussed in the previous section.
More specifically, one can build a frequency-domain phenomenological gravitational waveform model beyond the PN formalism in each of the inspiral, merger, and ringdown portions of the signal.
In particular, as shown in Fig.~\ref{fig:PhenomD}, the inspiral portion (Region I) corresponds to low-frequencies when the binary objects are far apart, and slowly-orbiting.
Under the emission of gravitational radiation, the bodies inspiral together with increased frequency until they reach the location of the ISCO (corresponding to a frequency $f_\ISCO=(6^{3/2}\pi M)^{-1}$), where circular orbits cease to exist.
At this point, the compact objects enter a plunging orbit until they make contact and form a remnant NS or BH.
The remnant object then undergoes a ringdown, radiating QNMs to settle down into its final relaxed state.
As shown in Fig.~\ref{fig:PhenomD}, at this point the orbit enters Region II which is broken up into the intermediate and merger-ringdown stages of the coalescence.

For the remainder of this thesis, we consider the sky-averaged ``IMRPhenomD" gravitational waveform template $\tilde{h}(f)$ found in Refs.~\cite{PhenomDI,PhenomDII} for point particles such as BHs.
For the case of NSs, the PhenomD point-particle waveform can be modified for extended structure as discussed in Sec.~\ref{sec:tidal}.
In the PhenomD gravitational waveform, authors Khan and Husa \textit{et al.} first separated both the template waveform amplitude $A$ and phase $\Psi$ into three portions: The inspiral ($A_\I$, $\Psi_\I$), intermediate  ($A_\text{int}$, $\Psi_\text{int}$), and merger-ringdown  ($A_\MR$, $\Psi_\MR$).
Within the former region, the authors built a hybrid model between the PN formalism discussed above in Sec.~\ref{intro:PN}, and additional \textit{numerical relativity} (NR) fits to complex simulations.
In the latter two regions, the authors utilized pure NR data to construct the intermediate and merger-ringdown waveform templates.
The authors calibrated the phenomenological NR fits in each region to 19 different effective-one-body-numerical-relativity\footnote{Effective-one-body (EOB) waveforms map the two-body problem into that of a single particle  moving in an effective metric to non-perturbatively determine the late binary evolution~\cite{EOB}. Effective-one-body-numerical-relativity (EOBNR) waveforms extend this by stitching NR simulation data to extend the waveform into the merger-ringdown phases~\cite{EOBNR}.} waveforms with mass ratios up to 1:18, and BH spins of up to $|\chi|\sim0.85$ (up to $0.98$ for equal-mass systems), for dimensionless BH spin $\chi\equiv a/m$, with $a=J_z/m$ for $\hat{z}$ angular momentum $J_z$.
Next, the independent waveform amplitudes ($A_\I$, $A_\text{int.}$, $A_\MR$) and phases ($\Psi_\I$, $\Psi_\text{int.}$, $\Psi_\MR$) are ``stitched'' together by matching the waveform and its derivative at the boundary frequencies separating each phase, which were determined by maximizing the waveform mismatch with template waveforms.
The resulting full-waveform amplitude ($A_\IMR$) and phase ($\Psi_\IMR$) is now valid for the entire binary coalescence event, where the Fourier-component waveform is given by
\begin{equation}\label{eq:PhenomDwaveform}
\tilde{h}(f)=A_\IMR e^{i\Psi_\IMR}.
\end{equation}
With the final inspiral $+$ merger-ringdown (IMR) PhenomD gravitational waveform template available, the authors then tested it for mismatch against the 19 calibration waveforms in addition to 29 verification hybrids.
The PhenomD waveform model typically exhibited no more than $1\%$ mismatch against all 48 NR waveforms.

Within the non-precessing sky-averaged PhenomD gravitational waveform model, we consider the following template parameters in our analysis.
For the case of binary BHs, we utilize a template parameter vector $\theta^a$ consisting of
\begin{equation}\label{eq:templateBH}
\theta^a_\BH=(\mathcal A,\phi_c,t_c,\mathcal{M}_z,\mathcal{\eta},\chi_s,\chi_a),
\end{equation}
where $\mathcal A=\frac{\mathcal{M}_z^{5/6}}{\sqrt{30}\pi^{2/3}D_L}$ is a normalized, sky-averaged, frequency-independent amplitude factor in the fourier domain, $\mathcal M_z = (1+z) \mathcal M$ is the redshifted chirp mass, and $\chi_{s,a}=\frac{1}{2}(\chi_1 \pm \chi_2)$ are the symmetric and anti-symmetric combinations of individual dimensionless spins $\chi_{1,2}$.
Further, we have averaged over the entire sky to remove all sky-localization parameters in the template.
The above parameters have been specifically chosen for their dominance in the gravitational waveform (for example we use $\mathcal{M}$ and $\eta$ instead of $m_1$ and $m_2$ because the latter two are highly degenerate with each other).

Now that we have our full gravitational waveform in GR with leading-order terms modified by higher-order PN corrections and NR data, we must assume prior knowledge on the parameters entering the template.
Typically, when one (such as the LVC) extracts posterior probability distributions on waveform template parameters $\theta^a$, prior probability distributions are first injected, imparting previous knowledge on such parameters into the analysis.
For a Bayesian analysis such as the one used by the LVC, such prior probability distributions can take any form required, typically uniform in many cases.
For the more computationally inexpensive Fisher analysis we consider in this thesis as described later in Chapter~\ref{chap:ParameterEstimation}, all prior distributions must be Gaussian-distributed.
To approximate such distributions, we pick a root-mean-square value $\sigma_{\theta^a}^{(0)}$ corresponding to previous knowledge, along with a central ``fiducial'' value which allows one to construct a Gaussian prior probability distribution.
For uniformly-distributed parameters with no prior knowledge or constraints, we assume $\sigma_{\theta^\A}^{(0)}=\infty$.
In particular, for this thesis investigation we always impose priors on the coalescence phase of $|\phi_c|\leq\pi$ as a general constraint.
For BHs, we impose a spin prior of $|\chi_{s,a}|<1$, corresponding to the maximum rotation a BH can take before exposing a naked singularity, which is not physically or observationally motivated.
All other parameters $\mathcal{A}$, $t_c$, $\mathcal{M}_z$, and $\eta$ have no prior motivations or constraints and we assume $\sigma_{\theta^\A}^{(0)}=\infty$.
As for fiducial values, we assume $t_c=\phi_c=0$ for simplicity, and directly calculate for the remaining parameters $\mathcal{A}$, $\mathcal{M}$, $\eta$, and $\chi_{s,a}$ from the assumed event intrinsic properties, as considered in a case-by-case basis, and discussed further in Sec.~\ref{sec:sources}.

\subsection{Extended structures in the waveform}\label{intro:tidal}
Now that we have a gravitational waveform with both PN corrections and NR fits for point-particles such as BHs, how do we extend this template to extended structures such as NSs?
In 2007, authors Flanagan and Hinderer first considered this effect on the gravitational waveform, deriving the leading-order quadrupolar corrections first entering the gravitational waveform at 5PN order~\cite{Flanagan2008}.
Later in 2011, Vines \textit{et al} extended this analysis to find the next-leading-order tidal correction entering at 6PN order~\cite{Vines:2011ud}, which was shown again in~\cite{Wade:tidalCorrections}. 
See also Ref.~\cite{tidal7.5} where Damour \textit{et al} derived incomplete tidal corrections to the gravitational waveform phase up to $7.5$PN order in 2012. 
Finally, in 2018 Yagi computed the $\ell$th-order multipole moments for extended structures in the gravitational waveform in Ref.~\cite{Yagi:Multipole}.

Such corrections to the gravitational waveform can ultimately be expressed by finding the conservative and dissipative corrections to the binary system from such tidal effects.
By first finding tidal corrections to Kepler's law $r(\Omega)$, one can find a new expression for the binding energy of the system $E_b$, giving the conservative tidal corrections.
Following this, the quadrupolar radiation formula combined with the modified Kepler's law can give a corrected expression for the radiated GW flux, or $\dot{E}$ of the system in the presence of tides, giving the ensuing dissipative corrections to the system.
Finally, the resulting GW phase in the presence of tidal deformations can be found by following e.g.~\cite{Tichy:1999pv,Tahura:2019dgr,Cutler:Fisher}.
Such corrections enter the gravitational waveform phase at 5PN and 6PN orders respectively as described in Refs.~\cite{Flanagan2008,Vines:2011ud,Yagi:Multipole,Wade:tidalCorrections}.
The modifications to the waveform are characterized by the new tidal parameters $\tilde\Lambda$ (mass-weighted tidal deformability) and $\delta\tilde\Lambda$, which are given by
\begin{align}
\label{eq:Lt}\tilde{\Lambda}&=\frac{8}{13}\left\lbrack (1+7\eta-31\eta^2)(\Lambda_1+\Lambda_2)+\sqrt{1-4\eta}(1+9\eta-11\eta^2)(\Lambda_1-\Lambda_2) \right\rbrack\\
\nonumber\delta\tilde{\Lambda}&=\frac{1}{2}\Bigg\lbrack \sqrt{1-4\eta}(1-\frac{13272}{1319}\eta+\frac{8944}{1319}\eta^2)(\Lambda_1+\Lambda_2)\\
\label{eq:dLt}&\hspace{2cm}+(1-\frac{15910}{1319}\eta+\frac{32850}{1319}\eta^2+\frac{3380}{1319}\eta^3)(\Lambda_1-\Lambda_2) \Bigg\rbrack,
\end{align}
for individual dimensionless tidal deformabilities $\Lambda_\A$, which characterize the NS deformation in response to an external tidal field (i.e. from the neighboring star).
For a more thorough description of the NS tidal deformability see Chapter~\ref{chap:ConstrainingNuclearParameters} of this thesis.

In a binary system, if both compact objects are NSs\footnote{In the case of mixed binaries, we take $\Lambda_1\to0$ (the tidal deformability of a non-rotating BH is zero~\cite{damour-nagar,Binnington:2009bb,Kol:2011vg,Chakrabarti:2013lua,Gurlebeck:2015xpa}) and $\Lambda_2\to\Lambda_\NS$, making $\tilde\Lambda$ and $\delta\tilde\Lambda$ functions of of only $\eta$ and $\Lambda_\NS$, which can then be re-parameterized into $\Lambda_s$. See also Ref.~\cite{Gralla:2017djj} in which it was shown that the standard computation of the tidal deformability relies on the comparison to the BH one and the effective BH tidal deformability may have a small non-zero effect on the gravitational waveform.}, we simply modify the PhenomD waveform model with the 5PN+6PN tidal corrections for extended structures in a binary orbit as found in Refs.~\cite{Flanagan2008,Vines:2011ud,Yagi:Multipole,Wade:tidalCorrections}\footnote{See Ref.~\cite{Thompson:2020nei} for a recent, more accurate phenomenological BH/NS waveform model, named ``IMRPhenomNSBH''. See also Ref.~\cite{Lackey:2013axa}, where a phenomenological BH/NS waveform model was constructed where the phase exceeded the NR results by 30\%. See also Ref.~\cite{Kumar:2016zlj}, which updates the model from Ref.~\cite{Lackey:2013axa} with a more accurate baseline binary BH model, and Ref.~\cite{Pannarale:2015jka} for a BH/NS amplitude model, Ref.~\cite{Hinderer:2016eia} for an effective-one-body model applicable to BH/NS systems, or Ref.~\cite{Barkett:2019tus} for BH/NS models computed with tidal splicing. Finally, refer to Ref.~\cite{Chakravarti:2018uyi} for an analysis on the waveform systematic uncertainties present in such models.}.
The resulting template parameters now consist of
\begin{equation}\label{eq:templateNS}
\theta^a_\NS=(\mathcal A,\phi_c,t_c,\mathcal{M}_z,\mathcal{\eta},\chi_s,\chi_a,\tilde{\Lambda}, \delta\tilde{\Lambda}).
\end{equation}
For NSs, we additionally impose a small-spin prior of $|\chi_{s,a}|<0.05$ motivated by theoretical considerations and pulsar timing observations, which is also used by the LVC in e.g.~\cite{LIGO:posterior,Abbott:2018exr,Abbott2017,TheLIGOScientific:2017qsa,GW190425}.
Finally, for NSs we also impose priors  on the tidal parameters of $0<\tilde\Lambda<3000$ as motivated by the LVC detection of binary NS merger GW170817~\cite{LIGO:posterior,Abbott:2018exr,Abbott2017,TheLIGOScientific:2017qsa}, and $|\delta\tilde\Lambda|<500$ as motivated in~\cite{Wade:tidalCorrections}.

\subsection{Beyond general relativity}\label{intro:beyondGR}
In the previous sections, we have described in detail the gravitational waveform in a universe ruled solely by GR.
How would this change if instead some modified or alternative theory of gravity ran the show?
To date, a large number of alternative theories of gravity have been proposed, several of which could potentially be likely to exist in nature.
Therefore, one must remain as agnostic as possible when testing observed GW signals for any minute hints of beyond-GR behaviors present within.
To do so, we consider the ppE formalism first suggested by Yunes \textit{et al.} in~\cite{Yunes:2009ke}, where one can introduce arbitrary beyond-GR modifications to the waveform template in an agnostic, parameterized way.
In particular, the ppE formalism allows one to inject modifications to the GW template at any given arbitrary PN order.
The ppE formalism is highly advantageous, as it allows one to constrain the effects of any generic modification to GR into one parameter, controlled by the chosen power of velocity, which can then be mapped backwards to many modified theories of gravity which alter the waveform at the same PN order.

Let us now consider the properties of the ppE waveform template.
In Fourier space, a modified template waveform in the ppE formalism can be written generically as
\begin{equation}\label{eq:ppe}
\tilde{h}_{\text{ppE}}(f)=A_\GR(f)(1+\alpha_\ppE u^a)e^{i[\Psi_\GR(f)+\beta_\ppE u^b]},
\end{equation}
where $u=(\pi \mathcal{M} f)^{1/3}$ is the effective relative velocity of the gravitating bodies in a binary.
The ppE parameters $(a,b)$ then characterize the velocity dependence at which non-GR modifications of magnitude $(\alpha_\ppE,\beta_\ppE)$ enter the waveform in the amplitude and phase, respectively.
See Fig.~\ref{fig:GRvsNonGR} for a schematic comparison between the GR and ppE waveforms $h(t)$ in the time-domain for arbitrary corrections $\beta_\ppE$ and $b$.
Observe how there is a noticeable shift in the waveform phase when injecting ppE modifications characterizing beyond-GR behaviors.
The ppE exponents $a$ and $b$ can then be mapped to the familiar PN order $n$ corrections to the gravitational waveform.
In particular, we define the ppE phase corrections to the gravitational waveform as absolute, or $\Psi \to \Psi_\GR + \beta u^b$, where the leading order PN ($0$PN) correction enters the waveform at $(v/c)^{-5}$.
On the other hand, the ppE amplitude corrections are chosen to be relative to the GR components, or $A=A_\GR(1+\alpha u^a)$.
Therefore, the final relationships between $n$-PN order and absolute power of $(v/c)$ can be written as
\begin{gather}\label{eq:PNorder}
a=2n\\
b=2n-5
\end{gather}
for amplitude corrections ($a$) and phase corrections ($b$).

We can consider two different modifications to the gravitational waveform: (a) those that alter the generation of GWs, and (b) those that alter the propagation of GWs.
For modifications to generation mechanisms, the ppE correction can only be included in the inspiral portion of the waveform.
This is because beyond the inspiral phase of the waveform, only NR data is available rather than the PN framework found in the inspiral.
On the other hand, with corrections to propagation mechanisms, we can include modifications to the entire gravitational waveform. This is because the generation of such GWs is irrelevant to their propagation across spacetime.

With a new ppE gravitational waveform template describing arbitrary modifications beyond GR, we can then map constraints on theory-agnostic parameters $\beta_\ppE$ or $\alpha_\ppE$ into theory-specific parameters corresponding to given alternative theories of gravity.
In particular, we carry out Fisher analyses as explained below in Chapter~\ref{chap:ParameterEstimation} assuming that GR is the true theory of gravity found in nature (i.e. choosing the fiducial value of the ppE parameter as $\beta_\ppE=0$ or $\alpha_\ppE=0$) to estimate the statistical errors on $\beta_\ppE$ or $\alpha_\ppE$.
Such statistical errors describe the non-GR ``fuzziness'' one can statistically expect the parameter to obey while still remaining consistent within the detector noise, and can therefore be taken as an upper-bound constraint on $\beta_\ppE$ or $\alpha_\ppE$.
By finally choosing the power $b$ or $a$ corresponding to the modified theory of gravity one wishes to study, the size of the effect ($\beta_\ppE$ or $\alpha_\ppE$) can then be mapped to the corresponding theoretical constants\footnote{Under consideration of propagation of uncertainties when transforming $\sigma_{\beta_\ppE}$ to $\sigma_\epsilon$ for some theory-specific parameter $\epsilon$, all terms containing measurement errors on intrinsic template parameter vanish due to their proportionality with $\beta_\ppE\rightarrow0$. See Ref.~\cite{Perkins:2018tir} for a more in-depth discussion on this topic.}. 
The function mapping the theory-agnostic parameters $\beta_\ppE$ to theory-specific parameters can be found by computing the binding energy (conservative corrections) and the radiated GW flux (dissipative corrections) within the new theory of gravity, which can then be converted to ppE parameters by e.g. following Refs.~\cite{Tahura_GdotMap,Yunes:2016jcc}.
See Chapter~\ref{chap:Parameterized} where we list all of the example theories considered in this thesis in detail, together with the mapping functions between the theory-agnostic $\beta_\ppE$ and theory-specific parameters.
Additionally, see Chapter~\ref{chap:beyondKerr} for example computations of such parameters in two alternative beyond-GR spacetimes, given only an arbitrary spacetime metric $g_{\alpha\beta}^\X$.

\section{Gravitational wave sources}\label{sec:sources}
As previously discussed, gravitational radiation is present whenever there exists a time-varying mass quadrupole moment.
This applies to several phenomena present across the universe, however due to the inherent weakness of gravitational interactions, only a certain few produce spacetime perturbations large enough to be detected here on Earth.
In specific, the coalescence of compact objects produce the largest amplitude of GWs, whether they be BHs or NSs (or a combination thereof).
Other weaker sources of GWs include the rotation of asymmetric NSs (e.g. with $\sim\mathcal{O}(1\text{ cm})$ mountains on their surface), asymmetric supernova explosions, primordial GWs arose from the cosmic inflation of the universe, the inspiral of binary white dwarfs, and many others.
Because such events are not currently detectable with the current LVC equipment, we focus our attention on the inspiral of BHs, NSs, and combinations of the two in the following thesis.

Within the LIGO detector's very first observing period spanning from 2002 to 2010, no gravitational waves were found.
However, within days of operation of the advanced LIGO (aLIGO) detectors\footnote{The redesign of LIGO's interferometers improved the sensitivity by an order-of-magnitude, which corresponds to three orders-of-magnitude more volume of space searched.} first observing run (O1) in 2015, GWs from a pair of merging BHs was observed for the very first time on the historic day of September 14, 2015~\cite{GW150914}.
This event was dubbed ``GW150914'', and it heralded an entirely new era of astrophysics, opening a new observing window into the universe.
Within the first and second observing runs, the LVC further observed 9 GW signals from merging BHs~\cite{LIGOScientific:2018mvr}.
Further, on August 17, 2017 the LVC for the first time observed the GW signal from a distant pair of merging NSs, observing a tidal signature present within the waveform~\cite{TheLIGOScientific:2017qsa}.
This historic event named ``GW170817'' made history by opening the doors to multi-messenger astronomy, due to the follow-up detections of the gamma-ray burst GRB 170817A as well as EM observations everywhere from the radio to the x-ray bands, observed by numerous facilities across the planet.
The third LVC observing run (O3) began on April 1, 2019, and has since detected over fifty BH/BH, NS/NS, and BH/NS candidate events~\cite{gracedb} which could be of possible future interest.

\renewcommand{\arraystretch}{1.2}
\begin{table}
\centering
\resizebox{\linewidth}{!}{%
\addvbuffer[12pt 8pt]{\begin{tabular}{c | c c c c c c}
Event & $m_1$ $(M_\odot)$ & $m_2$ $(M_\odot)$ & $\chi_1$ & $\chi_2$ & $D_L$ (Mpc) & SNR\\
\hline
GW170817~\cite{TheLIGOScientific:2017qsa} (NS/NS) & $1.48$ & $1.27$ & $0.02$ & $-0.02$ & $40$ & $32.4$\\
GW170608~\cite{GW170608} (BH/BH) & $12.0$ & $7.0$ & $0.5$ & $-0.66$  & $340$ & $9.0$\\
GW151226~\cite{GW151226} (BH/BH) & $14.2$ & $7.5$ & $0.5$ & $-0.36$ & $440$ & $13.0$\\
GW150914~\cite{GW150914} (BH/BH) & $35.8$ & $29.1$ & $0.32$ & $-0.44$ & $410$ & $25.1$\\
GW170729~\cite{GW170729} (BH/BH) & $51.0$ & $31.9$ & $0.60$ & $-0.57$ & 2,900 & $10.7$\\
\hline
BHNS (BH/NS) & $1.4$ & varies & varies & varies & varies & varies\\
EMRI (BH/BH) & $10^6$ & $10$ & $0.90$ & $-0.50$ & 3,000 & --\\
SMBHB (BH/BH) & $10^6$ & $5\times10^{4}$ & $0.90$ & $-0.90$ & 3,000 & --\\
\end{tabular}}
}
\caption[List of GW events considered]{(top) List of notable GW events investigated in the following thesis, along with their constituent masses $m_{1,2}$ and dimensionless spins $\chi_{1,2}$, the luminosity distance $D_L$, and finally the detection SNR. Such events include the binary NS merger (GW1780817) as well as the most and least massive binary BH events yet detected (GW170729 and GW170608 respectively), and also the ``golden event'' GW150914.
(bottom) Same as the top but for future events detectable by the LVC and future space-based interferometers. 
Here, many parameters are unknown, but are varied on a case-by-case basis for practical purposes.
}\label{tab:events}
\end{table}

In the following thesis, we consider five notable GW events as if they were observed by current or future GW interferometers.
See Tab.~\ref{tab:events} for a comprehensive list of the following GW events, including their constituent masses $m_{1,2}$, dimensionless spins $\chi_{1,2}$, their luminosity distance $D_L$, and finally the detection signal-to-noise-ratio (SNR).
In particular, we first consider the binary NS merger event GW170817~\cite{TheLIGOScientific:2017qsa}, in which constraints on the tidal deformability allows us to study the properties of supranuclear matter within the star, as discussed in Part~\ref{part:ProbingNuclearPhysics}.
We then focus our attention on four binary BH events as detected by the LVC, useful in testing GR as found in Chapters~\ref{chap:Parameterized} and~\ref{chap:IMRconsistency}.
The first binary BH event considered is the least-massive event detected to date, GW170608~\cite{GW170608} with a total mass of only $19\text{ M}_\odot$.
This is followed by the ``Christmas event'', GW151226~\cite{GW151226}, and then the ``golden event'' GW150914~\cite{GW150914} with large SNR.
Finally, we consider the most massive event detected to date, GW170729~\cite{GW170729} with a total mass of $83.5\text{ M}_\odot$.
Each above event has various useful properties in different situations, including larger/smaller mass ratios, total masses, luminosity distances, and spins.
Such differences prove beneficial when testing different aspects of GR as found in Chapters~\ref{chap:Parameterized} and~\ref{chap:IMRconsistency}.
For each event considered in the following thesis, we scale the detection SNR observed on the O2 GW detector to the actual event SNR as detected by the LVC in order to account for differences between our template waveform and the actual GW event observed.
The same scaling ratio is then applied to future GW detectors for consistency.

Finally, it sometimes proves useful to consider more exotic events which have yet to be observed by the LVC.
In particular, in Chapter~\ref{chap:Parameterized} we consider the merger of a BH and a NS (BHNS), useful for constraining scalar-tensor theories of gravity.
Here we assume a NS of mass $1.4\text{ M}_\odot$ and a BH of varying mass.
Such an event is extremely timely to consider as two of the candidates in the O3 run by the LVC, S190426c and S190814bv, are likely to be such a merger, if they are of astrophysical origin~\cite{gracedb,gracedb2}.
Next we consider the merger of extreme-mass-ratio inspirals (EMRIs) detectable in the future by the space-based GW detector LISA in Chapter~\ref{chap:beyondKerr}.
We consider such events with masses of $10^6\text{ M}_\odot$ and $10\text{ M}_\odot$, and they are highly useful for constraining theories of gravity that depend on the symmetric mass ratio $\eta$.
Finally, in Chapter~\ref{chap:beyondKerr} we consider the extreme inspiral of super-massive black hole binaries (SMBHBs) again detectable by LISA.
Such events with masses of $10^6\text{ M}_\odot$ and $5\times10^4\text{ M}_\odot$ are useful in constraining modified theories of gravity that depend strongly on the total mass.
See Tab.~\ref{tab:events} for more details on the above events yet to be detected by the LVC, including their individual masses $m_{1,2}$, dimensionless spins $\chi_{1,2}$, the luminosity distance $D_L$, and finally the detection SNR.
Because the above events are yet to be detected, in the following thesis we vary several parameters to consider a wide range of possibilities.


\chapter{Parameter estimation}\label{chap:ParameterEstimation}
\begin{adjustwidth}{\parindent}{\parindent}
\begin{spacing}{1.5}
The most reliable, comprehensive method used to extract parameters from a given gravitational wave signal $s=h_t + n$ (the sum of the true gravitational waveform $h_t$ with true parameters $\theta^a_t$ and noise $n$), with known GW template $h$, is through a full \textit{Bayesian analysis}.
In such analyses, one reconstructs the full posterior probability distributions for parameters $\theta^a$, given a signal $s$.
With such a large parameter space, this form of analysis proves to be quite computationally expensive, and infeasible when many samples are required.
Alternatively, in the limit of large signal-to-noise ratio, one can use a \textit{Fisher analysis}, which assumes Gaussian noise and prior distributions.
Such analyses have been shown to agree very well to their Bayesian counterparts for signal-to-noise-ratios of $\mathcal{O}(25)$, corresponding to the GW event GW150914.
Further, the Fisher analysis process is significantly less computationally expensive than a Bayesian one.
\end{spacing}
\end{adjustwidth}
\newpage


\section{The Fisher analysis method}\label{intro:Fisher} 
For large enough SNRs~\cite{Vallisneri:FisherSNR,Vallisneri:FisherSNR2}, a Fisher analysis~\cite{Cutler:Fisher,Poisson:Fisher,Berti:Fisher,Yagi:2009zm} may be used as a reliable approximation to the Bayesian analysis used by e.g. the LVC. 
Assuming that we have a perfect waveform template ($h=h_t$), the signal-to-noise ratio (SNR) is given by the inner product of the waveform with itself, weighted by the spectral noise density $S_n(f)$ of the detector:
\begin{equation}\label{eq:SNR}
\rho\equiv\sqrt{(h|h)},
\end{equation}
where the inner product is defined as
\begin{equation}\label{eq:overlap}
(a|b) \equiv 2 \int^{f_{\text{high}}}_{f_{\text{low}}}\frac{\tilde{a}^*\tilde{b}+\tilde{b}^*\tilde{a}}{S_n(f)}df.
\end{equation}
The limiting frequencies $f_{\text{low}}$ and $f_{\text{high}}$ depend on the specific GW detector configurations, and are chosen on a case-by-case basis as discussed further in Sec.~\ref{intro:Detectors}.

In a Fisher analysis, we make the assumptions that the detector noise is stationary, uncorrelated, and Gaussian.
Following Refs.~\cite{Cutler:Fisher,Yagi:2009zm}, the noise in a GW interferometer follows a probability distribution roughly of the form
\begin{equation}
p(n) \propto \exp \left\lbrack  -\frac{1}{2}(n|n) \right\rbrack.
\end{equation}

Given a successful detection of a GW signal $s$ given by the sum of the noise $n$ in the detector and the true gravitational waveform $h_0(\theta^a)$ with true parameters $\theta^a$, we have~\cite{Vallisneri}
\begin{equation}
s=h_0(\theta^a)+n.
\end{equation}
We can then re-write the detector noise probability distribution as
\begin{align}\label{eq:fisher}
\nonumber p(\theta^a_t|s) &\propto  p^{(0)} \exp \left\lbrack -\frac{1}{2}(s-h_t|s-h_t) \right\rbrack\\
\nonumber & = p^{(0)} \exp \left\lbrack -\frac{1}{2}(s|s)-\frac{1}{2}(h_t|h_t)+(h_t|s) \right\rbrack\\
& \propto p^{(0)} \exp \left\lbrack(h_t|s) -\frac{1}{2}(h_t|h_t)\right\rbrack,
\end{align}
where we let $p^{(0)}$ be the prior distributions on parameters $\theta^a$, to be determined later.
Additionally, the term $(s|s)$ will act as an overall constant and becomes irrelevant because it is purely signal, containing no parameters to be varied.
We next strive to determine the maximally-likely parameters $\hat{\theta}^a$ that most agree with the observed, true parameters $\theta^a$ contained in the true waveform $h_0$.
Such parameters describe our template waveform $h_t$ which is found by expanding a generic waveform $h(t)$ about the true waveform $h_0$ like so~\cite{Vallisneri}
\begin{equation}
h_t=h_0+\partial_i h_0 \Delta\theta^i + \mathcal{O}(\Delta\theta^2),
\end{equation}
for error in our determination of best-fit parameters relative to true parameters $\Delta\theta^i\equiv\theta^i-\hat{\theta}^i$.
Such parameters can be determined by maximizing the distribution in Eq.~\eqref{eq:fisher} with respect to $\theta^a$, resulting in the expression $(\partial_ih_t|s)-(\partial_ih_t|h_t)=0$.
Substituting in the above expression and expanding Eq.~\eqref{eq:fisher} to quadratic order in error $\Delta\theta$ results in the final expression given by
\begin{equation}\label{eq:fisherPDF}
p(\theta^a|s) \propto p^{(0)}_{\theta^a} \exp \left\lbrack -\frac{1}{2} \Gamma_{ij} \Delta \theta^i \Delta \theta^j \right\rbrack
\end{equation}
where the \textit{Fisher information matrix} $\Gamma_{ij}$ is determined to be
\begin{equation}
\Gamma_{ij}\equiv(\partial_i h | \partial_j h),
\end{equation}
at leading order, in the limit of large SNR.

Equation~\eqref{eq:fisherPDF} is reminiscent of the multivariate Gaussian probability distribution about best-fit parameters $\hat{\theta}^i$ with variance-covariance matrix given by $\Sigma_{ij}\Rightarrow\Gamma_{ij}^{-1}$.
In this thesis, we follow the works of~\cite{Cutler:Fisher,Poisson:Fisher,Berti:Fisher} and assume that the prior distributions on parameters $\theta^a$, represented as $p^{(0)}_{\theta^a}$ are distributed as Gaussian, for simplicity.
In reality, we could represent the prior probability distributions $p^{(0)}_{\theta^a}$ with any distribution one wanted.
For example, the LVC commonly utilizes uniform priors on many parameters which is more realistic than the Gaussian ones considered here.
However, such a choice makes the following computations more complicated and strays away from the key purpose of the Fisher analysis: quick, simple estimations on waveform parameters $\theta^a$ for large SNR events. 
As such, we construct prior distributions by translating the (typically non-Gaussian) prior information values into the 68\% confidence interval root-mean-square errors of a Gaussian probability distribution for our priors.
In addition, we remain conservative in the estimation of priors as described previously in Sec.~\ref{sec:phenomD}, with many chosen to be $\sigma=\infty$ (corresponding to uniform unknown distributions).
Then, we can define the effective Fisher matrix to be
\begin{equation}
\tilde \Gamma_{ij} \equiv \Gamma_{ij} +\frac{1}{\left(\sigma^{(0)}_{\theta^i}\right)^2}\delta_{ij},
\end{equation}
where $\sigma^{(0)}_{\theta^i}$ are the prior root-mean-square estimates of parameters $\theta^i$. 
The resulting $1\sigma$ root-mean-square errors on best-fit parameters $\hat{\theta}^i$ can be simply written as 
\begin{equation}
\Delta\theta_i=\sqrt{\tilde \Gamma^{-1}_{ii}}.
\end{equation}
Additionally, the covariances between parameters $\theta^i$ and $\theta^j$ can be simply given as the off-diagonal components, or $\tilde\Gamma_{ij}^{-1}$.
Finally, if one utilizes information from multiple detectors A and B, the resulting Fisher matrix becomes
\begin{equation}
\tilde \Gamma_{ij}^{\text{total}}=\Gamma_{ij}^{\text{A}}+ \Gamma_{ij}^{\text{B}}+\frac{1}{\left(\sigma^{(0)}_{\theta^i}\right)^2}\delta_{ij}.
\end{equation}
In result, the estimated root-mean-square errors on best-fit parameters $\Delta\theta_a$ agree with those from the LVC's Bayesian analysis to within $\sim40\%$ for GW150914-like events with SNRs of $25.1$, as thoroughly investigated in Ref.~\cite{Yunes_ModifiedPhysics}.
Such agreements only strengthen considerably as the SNR increases, becoming especially negligible for future events on third-generation detection with $\rho>1000$, which are mostly considered in the following thesis.

\subsection{Combined uncertainties from multiple detections}\label{intro:Combined}
Now we explain how one can combine statistical uncertainties on certain parameters $\theta^i$ from multiple observed GW events.
This is accomplished with a simple Fisher analysis as described in Sec.~\ref{intro:Fisher}, where we first consider single-event detections observed on GW detectors $A \equiv ($aLIGO O2~\cite{aLIGO,O3}, aLIGO~\cite{aLIGO}, A\texttt{+}~\cite{AppData,Ap_Voyager_CE}, Voyager~\cite{VRTCEETData,Ap_Voyager_CE}, ET~\cite{ET,Evans:2016mbw,VRTCEETData}, CE~\cite{VRTCEETData,Evans:2016mbw,Ap_Voyager_CE}, LISA~\cite{LISA}, TianQin~\cite{TianQin}, B-DECIGO~\cite{B-DECIGO}, DECIGO~\cite{DECIGO}$)$ as described further in Sec.~\ref{intro:Detectors}.
We next simulate a population of $N_A$ events for each interferometer, approximating the number of events detected on interferometer $A$ over an observing period of one year.
We then then combine the statistical uncertainties from $N_A$ events, resulting in an approximation on the overall measurement accuracy of $\theta^i$.
The process used to achieve this is outlined below:

\begin{table}
\centering
\begin{tabular}{|c|c|c|c|}
\hline
\multirow{2}{*}{Detector} & \multicolumn{3}{c|}{Detection Rate [events/yr]} \\
\cline{2-3}\cline{3-4}
& Pessimistic & Realistic & Optimistic\\
\hline
\hline
A\texttt{+}&5&270&5,500\\
Voyager&72&3,600&74,000\\
CE&720&36,000&730,000\\
ET&510&25,000&520,000\\
B-DECIGO&43&2,200&44,000\\
DECIGO&730&37,000&730,000\\
\hline
\end{tabular}
\caption[One year black hole - neutron star detection rates]{Pessimistic, mean, and optimistic 1 year detection rates for $10\text{ M}_\odot-1.4\text{ M}_\odot$ BH-NS binaries assuming a local BH-NS coalescence rate of $\mathcal{R} \in \lbrack 0.6, 610 \rbrack \text{Gpc}^{-3}\text{yr}^{-1}$, with a ``realistic value" of $30\text{ Gpc}^{-3}\text{yr}^{-1}$~\cite{Abadie:2010cf,LIGOScientific:2018mvr}.
}\label{tab:eventRates}
\end{table}

\begin{enumerate}
\item[(i)] Perform a Fisher analysis as outlined in Sec.~\ref{intro:Fisher} using detector sensitivity $S_n^A(f)$, at various values of the redshift $z$.
The redshift of a merger event is computed by solving
\begin{equation}\label{eq:redshift}
D_L=\frac{1+z}{H_0}\int\limits^z_0\frac{dz'}{\sqrt{\Omega_M(1+z')^3+\Omega_\Lambda}},
\end{equation}
where $D_L$ is the luminosity distance to the merger, $H_0 = 70 \text{km s}^{-1}\text{Mpc}^{-1}$ is the local Hubble constant~\cite{Aghanim:2018eyx}, $\Omega_{\Lambda}=0.67$ is the universe's vacuum energy density~\cite{Aghanim:2018eyx}, and $\Omega_M=1-\Omega_{\Lambda}$ is the matter density.
This results in the single-event statistical uncertainty $\sigma^A(z)$ accrued in the extraction of some parameter $\theta^i$ on detector $A$ as a function of the event redshift.

\item[(ii)] Estimate the total number of binary merger events $N_A$ detected (for BH/BH, NS/NS, or BH/NS events) over one observing year\footnote{The population of $N_A$ events follows the probability distribution~\cite{Shutz:SNR,Chen:SNR} given by $f(\rho)=3 \rho_{\text{th}}^3/\rho^4$ with a network SNR threshold of $\rho_{\text{th}}=8$.} on detector $A$ by taking into account the merger rate history throughout all redshift values within detector A's horizon redshift $z_h$\footnote{The horizon redshift is computed to be the redshift where the SNR equals the SNR threshold of the detector.}, as shown by Eq. (10) of Ref.~\cite{Cutler:BNSmerger}:
\begin{equation}
N_A=\Delta \tau_0 \int\limits^{z_{h}}_0 4 \pi \lbrack  a_0r_1(z)\rbrack^2 \mathcal{R} r(z) \frac{d \tau}{dz} dz.
\end{equation}
Here, $a_0r_1(z)$, $\frac{d\tau}{dz}$, and $r(z)$ for our chosen cosmology are given by:
\begin{equation}
a_0r_1(z) = \frac{1}{H_0}\int\limits^z_0 \frac{dz'}{\sqrt{(1-\Omega_{\Lambda})(1+z')^3+\Omega_{\Lambda}}},
\end{equation}
\begin{equation}
\frac{d\tau}{dz} = \frac{1}{H_0} \frac{1}{1+z}\frac{1}{\sqrt{(1-\Omega_{\Lambda})(1+z')^3+\Omega_{\Lambda}}},
\end{equation}
\begin{equation}
r(z) = \left\{
\begin{array}{ll}
      1+2z & z \leq 1 \\
      \frac{3}{4}(5-z) & 1\leq z\leq 5 \\
      0 & z\geq 5\\ 
\end{array}\,.
\right.
\end{equation}
Here we choose an observing period of $\Delta \tau_0 = 1$ year, and calculate the detection rate for the upper, central, and lower limits of the local binary NS coalescence rate density $\mathcal{R}=1540^{+3200}_{-1220} \text{ Gpc}^{-3}\text{yr}^{-1}$~\cite{Abbott2017}, giving the rates $N_A$ shown in i.e. Table~\ref{tab:eventRates} and the second column of Table~\ref{tab:variances}.

\item[(iii)] Compute the combined population standard deviation $\sigma_{N_A}$, taking into account sources at varying redshifts as was done in Eq. (3) of Ref.~\cite{Takahiro}:
\begin{equation}\label{eq:stacking}
\sigma_{N_A}^{-2}=\Delta \tau \int\limits^{z_h}_04 \pi \lbrack a_0 r_1(z)\rbrack^2 \mathcal{R}r(z)\frac{d\tau}{dz}\sigma^A(z)^{-2}dz.
\end{equation}
Here, $\sigma^A(z)$ is the redshift dependence of the root-mean-square error on detector $A$, evaluated in step (i).
This results in the combined statistical uncertainties for the lower, central, and upper limits of the local binary NS coalescence rate density $\mathcal{R}$.
\end{enumerate}

See also Sec.~\ref{sec:varymass} for an alternative analysis in which the BH and NS masses are varied for a more comprehensive combined uncertainty calculation.


\section{Gravitational wave interferometers}\label{intro:Detectors}
Since the successful detection of GWs by the LVC detectors' second observing run in Hanford, Washington and Livingston, Louisiana~\cite{aLIGO} (aLIGO O2), several future interferometers have been proposed and even funded.
By both increasing our sensitivity to high-frequency GWs (with frequencies between $1$ Hz and $10^4$ Hz), and also expanding into the low-frequency regime (with frequencies between $10^{-4}$ Hz and $1$ Hz), we can further probe fundamental physics which might currently be hiding within aLIGO O2's limited capabilities.
In particular, we here discuss the future prospects for GW interferometry, with expeditions both here on the ground, as well as up in space.

\begin{figure}
\begin{center} 
\includegraphics[width=\columnwidth]{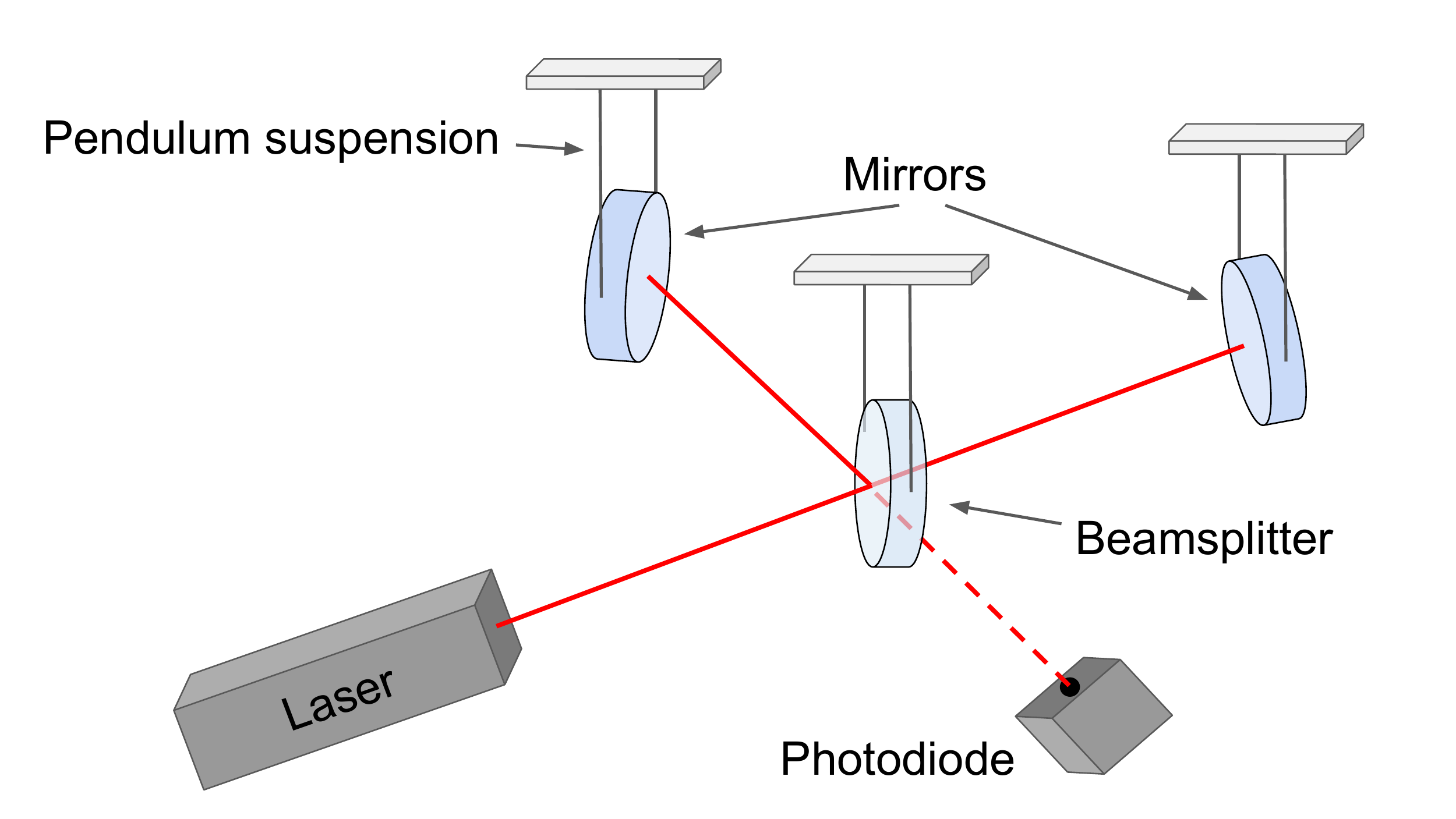}
\end{center}
\caption[Michelson GW interferometer]{
Michelson GW interferometer, similar to the setup found in aLIGO.
}
\label{fig:michelson}
\end{figure}

A GW interferometer detects minute GW signals with use of a Michelson interferometer as shown in Fig.~\ref{fig:michelson}.
In particular, this is done by splitting a laser beam (with a wavelength of 1064 nm in the LIGO detectors) into two orthogonal arms (of length $4$ km in the LIGO detectors).
With the use of partially-reflecting mirrors, a Fabry-P\'{e}rot cavity is formed, allowing the light to be recycled in each arm to effectively increase the path length traveled (280 round-trips for the LIGO detector).
The beams then recombine at the beamsplitter, where interference patterns can be observed by a photodetector on the output arm.
When the arm lengths are exactly fixed, the light completely destructively interferes and as a result, nothing is detected.
Under the presence of a GW, one or both of the cavities will undergo a length change, which results in a slight detuning of the beams relative to one another.
The resulting interference signal can be used to infer the properties of the observed GW signal.

Such GW interferometers can be characterized by their distribution of power over frequency. 
Such a distribution is known as the power spectral density, which for simplicity we refer to as the detector sensitivity, denoted by $S_n(f)$, with units of Hz$^{-1/2}$.
Through rigorous experimental work, the power spectral density for the $4$-km arm length aLIGO O2 detectors have been characterized in~\cite{TheLIGOScientific:2016zmo}.
The sources of noise taken into account consist of, but are not limited to: seismic effects, thermal fluctuations in the mirrors, quantum uncertainty in the light, laser intensity and frequency variations, and several more.
Such effects are reduced in future observing runs until finally the aLIGO design sensitivity is reached~\cite{aLIGO} (corresponding to observing run 5, or O5).

Future upgrades to the aLIGO design sensitivity have been proposed which further improve upon the detector sensitivity in the late 2020's.
In the LIGO A\texttt{+}~\cite{AppData,Ap_Voyager_CE} proposal, upgrades to the mirrors and their suspension, as well as the injection of squeezed light would almost double the design sensitivity.
Finally, a planned upgrade known as LIGO Voyager~\cite{VRTCEETData,Ap_Voyager_CE} will replace the mirrors with more massive, cooled ones, and the laser with one of a longer wavelength to once again double the sensitivity.
Figure~\ref{fig:sensitivities} displays the proposed detector sensitivities $S_n(f)$ for aLIGO and its future upgrades, as well as the future detectors considered in this analysis\footnote{Here we also show the characteristic strain of a GW signal from GW150914, characterized by $2\sqrt{f}|\tilde{h}(f)|$ with the same units as $S_n(f)$. Roughly speaking, the ratio between this and the detector sensitivity corresponds to the SNR of the event.}.
The number of independent interferometers (e.g. 2 for aLIGO) have been accounted for by directly modifying the spectral noise density $S_n(f)$ by $N_{\text{detectors}}^{-1}$. 
Similarly, any detector geometry that is not $90^{\circ}$ obtains an additional factor of $1/\sin{\theta_{\text{arm}}}$ applied to the spectral noise density.
Further, Tab.~\ref{tab:detectors} tabulates useful information for each GW interferometer, such as the arm length, cut-off frequencies, GW150914-like SNR, and the number of independent interferometers.

\begin{table*}
\centering
\resizebox{\textwidth}{!}{%
\begin{tabular}{c|c|c|c|c|c|c}
Detector & Location & GW150914 $f_{\text{low}}$ (Hz) & GW150914 $f_{\text{high}}$ (Hz) & GW150914 SNR & Arm length & Interferometers\\
\hline
aLIGO O2~\cite{aLIGO,O3} & Ground & {23} & {4000} & $25.1$ & $4$ km & 2\\
aLIGO~\cite{aLIGO} & Ground & {23} & {4000} & $67.1$ & $4$ km & 2\\
A\texttt{+}~\cite{AppData,Ap_Voyager_CE} & Ground & {10} & {4000} & $133.0$ & $4$ km & 2\\
Voyager~\cite{VRTCEETData,Ap_Voyager_CE} & Ground & {10} & {4000} & $313.0$ & $4$ km & 2\\
CE~\cite{VRTCEETData,Evans:2016mbw,Ap_Voyager_CE} & Ground & {1} & {4000} & $1.64\times10^3$ & $40$ km & 1\\
ET~\cite{ET,Evans:2016mbw,VRTCEETData} & Ground & {1} & {4000} & $714.0$ & $10$ km & 3\\
LISA~\cite{LISA} & Space & {0.01} & {1} & $9.3$ & 2.5 Gm & 2\\
TianQin~\cite{TianQin} & Space & {0.01} & {1} & 10.7 & 0.173 Gm & 2\\
B-DECIGO~\cite{B-DECIGO} & Space & {0.01} & {100} & $607.0$ & 100 km & 2\\
DECIGO~\cite{DECIGO} & Space & {0.01} & {100} & $1.53 \times 10^4$ & 1,000 km & 8\\
\end{tabular}
}
\caption[Gravitational wave detectors]{
Tabulated information for the ground-based detectors and all 4 space-based detectors considered in our analysis.
The frequency integration ranges $f_{\text{low}}$-$f_{\text{high}}$ are computed using Eqs.~\eqref{eq:groundFreq} and~\eqref{eq:spaceFreq} for the example of GW150914.
The lower ground-based and upper space-based frequency limits correspond to the detector limits $f_{\text{low-cut}}$ and $f_{\text{high-cut}}$, while the upper ground-based and lower space-based limits correspond to an arbitrary value such that the gravitational wave spectrum is sufficiently small compared to the detector sensitivity, and the GW frequency 4 years prior to merger.
The GW150914 SNR is computed via Eq.~\eqref{eq:SNR}.
}\label{tab:detectors}
\end{table*}

Future planned GW detectors to be operational in the late $\sim$2030's, which we refer to as third-generation interferometers, can successfully harbor up to an order-of-magnitude improvement to the aLIGO design sensitivity.
Two such detectors are now being planned, the first of which is known as the Cosmic Explorer~\cite{VRTCEETData,Evans:2016mbw,Ap_Voyager_CE} (CE) being planned by the LVC.
This feat of human engineering will boast $40$-km arm lengths enclosed in ultra-high vacuum tubes with cryogenic technology to allow up to an order-of-magnitude improvement to aLIGO's design sensitivity (or O5).
Further, the proposed Einstein Telescope~\cite{ET,Evans:2016mbw,VRTCEETData} (ET) by several institutions in the European Union consists of a triangular $10$-km arm pattern, effectively creating two independent interferometers.
This planned detector will be built underground to reduce seismic noise, and will have cryogenically cooled vacuum tubes, which will enable it to outperform CE's sensitivity below $10$ Hz, while it is slightly less sensitive beyond $10$ Hz.
Such third generation detector sensitivities and details are additionally shown in Fig.~\ref{fig:sensitivities} and Tab.~\ref{tab:detectors}.

Finally, an entirely new class of GW detectors have been proposed and planned to be operational entirely in space in the $\sim$2030's.
Due to this unique opportunity to increase the arm lengths (as well as eliminate various terrestrial noises), such detectors have sensitivity in the mHz regime, extending from $10^{-4}$ Hz to $1$ Hz.
First, proposed by the European Space Agency is the Laser Interferometer Space Antenna~\cite{LISA} (LISA), which will exist in a trailing orbit behind Earth about the sun, with a triangular arm configuration with lengths of $2.5$ Gm.
Similarly, the Chinese-proposed GW detector TianQin~\cite{TianQin} has similar capabilities to LISA, albeit with more sensitivity above $10^{-2}$ Hz, and slightly less below.
Finally, the Deci-hertz Interferometer Gravitational wave Observatory~\cite{DECIGO} (DECIGO) is a Japanese-proposed GW detector with two $10^4$ km triangular interferometers interlocked to form a star-shape, along with two additional triangular interferometers on independent orbits for improved sky localization, resulting in a total of 8 independent interferometers.
With the shorter arm-lengths, DECIGO and it's less-sensitive counterpart B-DECIGO~\cite{B-DECIGO} bridge the gap between ground-based detectors and LISA/TianQin, operating in a frequency range between $10^{-2}$ Hz and $10^2$ Hz.
Such space-based detector sensitivities and details are also displayed in Fig.~\ref{fig:sensitivities} and Tab.~\ref{tab:detectors}.

Now we must decide the frequency ranges $f_{\text{low}}$ to $f_{\text{high}}$ which are valid in each GW detector, as used in the integration ranges from e.g.  Eq.~\eqref{eq:overlap}.
For all of the ground-based detectors, we choose the upper and lower integration frequencies as
\begin{equation}\label{eq:groundFreq}
f_{\text{low}}^{\text{ground}}=f_{\text{low-cut}},\quad
f_{\text{high}}^{\text{ground}}=f_{\text{small}},
\end{equation}
where the upper limit of $f_{\text{small}}$ is chosen such that the gravitational wave spectrum $2\sqrt{f}|\tilde{h}|$ is sufficiently small compared to the detector sensitivity $S_n$ (for e.g. GW150914-like events this number is chosen to be 4,000 Hz).
The detector lower cut-off frequencies $f_{\text{low-cut}}$ depend on the specific equipment used, and is given by $23$ Hz, $10$ Hz, and $1$ Hz for the aLIGO detector, its upgrades, and the third generation detectors respectively.
Similarly, for space-based detectors, we choose
\begin{equation}\label{eq:spaceFreq}
f_{\text{low}}^{\text{space}}=\text{max}\lbrack 10^{-4}\text{ Hz}, f_{\text{4yrs}} \rbrack, \quad
f_{\text{high}}^{\text{space}}=f_{\text{high-cut}},
\end{equation}
where $10^{-4}\text{ Hz}$ is the default detector cutoff frequency~\cite{Berti:spaceFreq}, and $f_{\text{4yrs}}$ is the frequency $T_\mathrm{obs}$ years prior to the merger.
This frequency is given by~\cite{Berti:spaceFreq}
\begin{equation}\label{eq:f4yr}
f_{T_\mathrm{obs}} = 1.84\times10^{-2} \left( \frac{\mathcal{M}}{28\text{ M}_\odot} \right)^{-5/8} \left( \frac{T_{\text{obs}}}{1\text{ yr}} \right)^{-3/8}\,.
\end{equation}
Finally, $f_{\text{high-cut}}$ is the detector-dependent high cut-off frequency which is found to be $1$ Hz for LISA and TianQin, and $100$ Hz for B-DECIGO and DECIGO.
For demonstrative purposes, $f_{\text{low}}$ and $f_{\text{high}}$ are tabulated in Tab.~\ref{tab:detectors} for GW150914-like events on each GW detector considered in the following thesis.


\part{Probing nuclear physics}\label{part:ProbingNuclearPhysics}
\addtocontents{toc}{\protect\begin{adjustwidth}{0.5cm}{0cm}}


\chapter{Constraining nuclear matter parameters}\label{chap:ConstrainingNuclearParameters}
\begin{adjustwidth}{\parindent}{\parindent}
\begin{spacing}{1.5}
The gravitational wave emission from the merging binary neutron star system GW170817 arrived full of tidal information which can be used to probe the fundamental ultra-dense nuclear physics residing in these stars.
We find strong correlations among neutron star tidal deformabilities and certain combinations of nuclear parameters associated with the equation of state.
These relations are then used to derive bounds on such parameters from GW170817, with several sources of systematic uncertainties taken into account. 
We then extend this analysis into the future of gravitational wave astronomy by finding similar correlations for varying chirp masses, the dominant determining factor in the frequency evolution of the inspiral, such that one can apply the same method to future detections.
We estimate how accurately one can measure nuclear parameters with future gravitational wave interferometers and show how such measurements can be improved by combining multiple events.
\end{spacing}
\end{adjustwidth}
\newpage
\begin{spacing}{2.0}


\section{Introduction}\label{sec:intro}

NSs exist in one of the most extreme states of matter found in the universe\footnote{This chapter is based on the following papers: Z. Carson, A. W. Steiner and K. Yagi, \textit{Constraining nuclear matter parameters with GW170817}, Phys. Rev. D \textbf{99} 043010 (2019) and Z. Carson, A. W. Steiner and K. Yagi, \textit{Future Prospects for Constraining Nuclear Matter Parameters with Gravitational Waves}, Phys. Rev. D \textbf{100}, 023012 (2019)}.
However, the determination of the equation of state (EoS) of ultra-dense matter found exclusively in such compact objects remains to be one of the largest unsolved mysteries in both nuclear physics and astrophysics to date.
The nuclear matter EoS determines many important stellar properties, such as the mass, radius, and tidal deformability, and is vital to the further study of supranuclear matter.
Independent measurements of certain macroscopic NS observables determined by the EoS, such as the mass and radius, can be used to constrain the EoS, as was indeed done in Refs.~\cite{guver,ozel-baym-guver,steiner-lattimer-brown,Lattimer2014,Ozel:2016oaf} via x-ray observations of the mass-radius relationship.
However, such measurements potentially suffer from large systematic errors due to uncertainties in the astrophysical modeling of x-ray bursts.

Recent observations of gravitational waves (GWs) from a merging binary NS system (GW170817~\cite{TheLIGOScientific:2017qsa}) have been used to probe the interior nuclear structure via imprinted tidal effects~\cite{Abbott2018,Abbott:2018exr,Paschalidis2018,Burgio2018,Malik2018,Landry:2018prl}, which offers us a cleaner method of determining the nuclear matter EoS than EM observations of neutron stars.
As the NSs lose energy via GW emission, they inspiral towards each other and become increasingly tidally deformed in response to the companion stars' tidal field.
This deformation is characterized by the \textit{tidal deformability}~\cite{Flanagan2008} of the NS, and is strongly dependent on the underlying EoS.
Further, the mass-weighted combination of such tidal deformabilities from each star is the leading tidal parameter in the gravitational waveform, which has been constrained by the LIGO and VIRGO Collaboration to a 90\% credible bound of $70 \leq \tilde{\Lambda} \leq 720$~\cite{Abbott2018,De:2018uhw}.
Such observations have also been mapped to the NS radius in Refs.~\cite{Annala:2017llu,Abbott:2018exr,Lim:2018bkq,Bauswein:2017vtn,De:2018uhw,Most:2018hfd}.

While all currently-proposed EoSs to date utilize various different
approximations, one way to effectively study them is by measuring the
nuclear parameters which parameterize the EoSs using a model-independent formalism.
One such method for doing this\footnote{Piecewise polytropic constructions~\cite{Read2009,Lackey:2014fwa,Carney:2018sdv} and spectral EoSs~\cite{Lindblom:2010bb,Lindblom:2012zi,Lindblom:2013kra,Lindblom:2018rfr,Abbott:2018exr} similarly parameterize nuclear matter EoSs in a model-independent way. See also~\cite{Landry:2018prl,Kumar:2019xgp} for related works on piecewise unified EoSs.} is to Taylor expand the energy per nucleon of asymmetric nuclear matter about the saturation density of symmetric matter~\cite{Myers:1969zz,Vidana2009}.
The resulting coefficients are known as the ``nuclear matter parameters" and consist of the following: the slope of the symmetry energy $L_0$; the nuclear incompressibility $K_0$; the slope of the incompressibility $M_0$; the curvature of the symmetry energy $K_{\text{sym},0}$; and higher orders, each evaluated at the nuclear saturation density.
Interestingly, approximate universal relations exist among nuclear physics parameters mentioned above and the NS radius at a given mass~\cite{Alam2016} (see e.g.~\cite{Sotani:2013dga,Silva:2016myw} for other universal relations involving nuclear parameters). 
The authors found that while individual nuclear parameters are only weakly correlated with the stellar radius, linear combinations of the form $K_0+\alpha L_0$ and $M_0+\beta L_0$ become highly correlated, where $\alpha$ and $\beta$ are chosen such that the correlation becomes maximum.

Previous important analyses by Alam \textit{et al.}~\cite{Alam2016} found approximately universal relations between the NS radius at a given mass and the nuclear parameters mentioned above (similar work can be found in Refs.~\cite{Sotani:2013dga,Silva:2016myw}).
Further, Malik \textit{et al.}~\cite{Malik2018} found that certain linear combinations of nuclear parameters (such as $K_0+\alpha L_0$ with $\alpha$ chosen to give maximal correlation) gave way to heightened correlations with the individual tidal deformabilities evaluated at a given mass.
By assuming individual masses for GW170817 and taking the approximate universal relations to be exact, the authors utilized prior constraints on the tidal deformability from GW170817~\cite{Abbott2017,Radice2018} and $L_0$~\cite{Abbott2018,Oertel2017,Lattimer2014} to derive new constraints on the nuclear parameters.

This important first-step work of Ref.~\cite{Malik2018} needs to be improved in various ways. 
In this chapter, we propose an extension upon this work by taking into account at least the following five points of interest for current constraints on nuclear parameters.
First, we consider a broader class of EoSs by phenomenologically varying nuclear parameters.
Second, we consider correlations among the mass-weighted tidal deformability (instead of the individual tidal deformabilities) and nuclear parameters for various mass ratios. This allows us to eliminate the need to choose specific NS masses $m_1$ and $m_2$, as was done in Ref.~\cite{Malik2018}.
Third, instead of assuming perfect linear regression between nuclear parameters and $\tilde{\Lambda}$, the uncertainty from scatter (corresponding to the EoS variation in the approximate universal relations) is taken into account, including the covariances among parameters.
Fourth, we use the recent updated posterior distribution of the dominant tidal deformability $\tilde \Lambda$ by LVC~\cite{Abbott2018}.
Finally, we investigate constraints on the incompressibility $K_0$ in addition to its slope $M_0$ and the curvature of symmetry energy $K_{\text{sym},0}$.

We then extend the work into the future of GW astronomy.
While every future merger event will be composed of NSs with varying individual masses which are difficult to measure, we can categorize them by the \textit{chirp mass} $\mathcal{M}$, which is the dominant driving factor in the frequency evolution of the inspiral event given by a certain combination of individual NS masses.
We repeat the analysis as a function of chirp mass, applicable to any future event.
Further, we restrict the set of EoSs to those that obey the nuclear parameter correlations of Ref.~\cite{Tews2017}, and consider the implications of observations using future GW interferometers: Advanced LIGO (aLIGO)~\cite{aLIGO}, LIGO A\texttt{+} (A\texttt{+})~\cite{Ap_Voyager_CE}, Voyager~\cite{Ap_Voyager_CE}, Cosmic Explorer (CE)~\cite{Ap_Voyager_CE}, and Einstein Telescope (ET)~\cite{ET}.
We will consider not only the increased sensitivities from current detectors but also the combined uncertainties from multiple-event detections (relevant for future detectors with expanded horizon volumes). 
 

\section{Background and theory}\label{sec:theory}

In this section we begin with a review on the NS tidal deformability in Sec.~\ref{sec:tidal}, followed up by a review on the NS equation of state and its constituent nuclear parameters in Secs.~\ref{sec:parameters} and~\ref{sec:eos}.


\subsection{Neutron star tidal deformability}\label{sec:tidal}
We begin by reviewing how one can extract internal structure information of NSs via GW measurement.
In the presence of a neighboring tidal field $\mathcal{E}_{ij}$, such as the binary NS system found in GW170817, NSs tidally deform away from sphericity and acquire a non-vanishing quadrupole moment $Q_{ij}$ that is characterized by the \textit{tidal deformability} $\lambda$~\cite{Flanagan2008,hinderer-love,Yagi2013}:
\begin{equation}
Q_{ij} = - \lambda \mathcal{E}_{ij}.
\end{equation}
Such tidal deformability can be made dimensionless as:
\begin{equation}
\Lambda \equiv \frac{\lambda}{m^5},
\end{equation}
with $m$ representing the stellar mass.
The magnitude of $\lambda$ depends strongly on the underlying structure, as well as the fluid nature of the NS~\cite{PhysRevD.98.084010}. Therefore, this parameter has the largest impact on the GW phase, and thus, it is encoded in GW observations of binary NS coalescence. 

Following Refs.~\cite{hinderer-love,damour-nagar,Yagi2013}, the dimensionless tidal deformability $\Lambda$ can be computed by isolating different asymptotic limits of the gravitational potential in the buffer zone $R \ll r \ll \mathcal{L}$ given by
\begin{align}
\Phi (x^i)=&\frac{1+g_{tt}}{2} =-\frac{M}{r} - \frac{3}{2}\frac{Q_{ij}}{r^3} \Bigg(\frac{x^i}{r} \frac{x^j}{r}-\frac{1}{3}\delta_{ij} \Bigg) + \mathcal{O} \Bigg( \frac{\mathcal{L}^4}{r^4} \Bigg)+ \frac{1}{2} \mathcal{E}_{ij} x^i x^j + \mathcal{O} \Bigg( \frac{r^3}{R^3} \Bigg),
\end{align}
where $r = |x^i|$, $\mathcal{L}$ is the length scale of the companion-induced curvature, and $R$ is the stellar radius.
Here, $g_{tt}$ corresponds to the $tt$-component of the full spacetime metric:
\begin{equation}
g_{\alpha\beta}=g_{\alpha\beta}^{(0)}+h_{\alpha\beta},
\end{equation}
constructed via a non-spinning, spherically-symmetric background solution $g_{\alpha\beta}^{(0)}$ perturbed by the tidal deformation with metric components $h_{\alpha\beta}$.

The resulting expression describing $\Lambda$ can be found to be
\begin{align}
\begin{split}
\Lambda =& \frac{16}{15} (1-2\bar{C})^2[2+2\bar{C}(y_\R-1)-y_\R]\{2\bar{C}[6-3y_\R+3\bar{C}(5y_\R-8)]+4\bar{C}^3[13-11y_\R\\
&+\bar{C}(3y_\R-2)+2\bar{C}^2(1+y_\R)]+3(1-2\bar{C})^2[2-y_\R+2\bar{C}(y_\R-1)]\ln{(1-2\bar{C})}\}^{-1}.
\end{split}
\end{align}
Here $\bar{C} \equiv m / R$ is the stellar compactness with $R$ representing the NS radius, and $y_\R \equiv y(R)$ with $y(r) \equiv r h'(r)/h(r)$, where a prime stands for taking a derivative with respect to the radial coordinate $r$. $h$ represents the quadrupolar part of the $(t,t)$ component of the metric perturbation satisfying the following differential equation:
\begin{align}\label{eq:deq} 
\begin{split}
h''+ \Big\{ \frac{2}{r} + \Big\lbrack \frac{2m}{r^2}+4 \pi r (p - \epsilon ) \Big\rbrack e^{\lambda} \Big\} h'+ \Big\{&4 \pi \Big\lbrack 5 \epsilon + 9p + (p+ \epsilon) \frac{d \epsilon}{dp} \Big\rbrack e^{\lambda}\\
&- \frac{6}{r^2}e^{\lambda} - \Big(\frac{d \nu}{dr} \Big)^2 \Big\}h =0,
\end{split}
\end{align}
with background metric coefficients $e^{\nu} = g_{tt}$ and $e^{\lambda} = (1-2m/r)^{-1} = g_{rr}$, while $p$ and $\epsilon$ represent pressure and energy density respectively.

The above differential equation can be solved as follows.
First, one needs to prepare unperturbed background solutions by choosing a specific EoS, or $p(\epsilon)$, and solve a set of Tolman-Oppenheimer-Volkoff (TOV) equations with a chosen central density (or pressure) and appropriate boundary conditions (the exterior metric being the Schwarzschild one). The stellar radius is determined from $p(R)=0$ while the mass is given by $m=m(R) = 4\pi \int_0^R \epsilon(r)\, r^2 dr$.
Having such solutions at hand, one then plugs them into Eq.~\eqref{eq:deq} and solves it with the boundary condition $y(0)=2$~\cite{hinderer-love}. 

In this investigation, we consider the scenario of two NSs orbiting each other in a binary system, such as GW170817.
In this case, each NS individually obtains quadrupole moments from the neighboring tidal field, resulting in two highly-correlated tidal deformabilities $\Lambda_1$ and $\Lambda_2$.
Due to these correlations, individual tidal deformabilities are very difficult to extract from GW observations\footnote{One way to cure this problem is to use universal relations between them~\cite{Yagi:2015pkc,Yagi:2016qmr,De:2018uhw,Zhao:2018nyf}.}.
Typically, it is useful to reparameterize the waveform via independent linear combinations of $\Lambda_1$ and $\Lambda_2$ which enter the gravitational waveform at 5th post-Newtonian (PN) and 6PN orders\footnote{$n$PN order corrections enter the gravitational waveform at relative powers of $(v/c)^{2n}$.} respectively.
The dominant tidal effect in the resulting waveform is known as the \textit{mass-weighted tidal deformability}, and is given by~\cite{Flanagan2008}
\begin{equation}
\tilde{\Lambda} = \frac{16}{13} \frac{(1+12q) \Lambda_1+(12+q)q^4\Lambda_2}{(1+q)^5},
\end{equation}
with mass ratio $q \equiv m_2/m_1$ ($m_1 \geq m_2$).
Here we also define the \textit{chirp mass} of the binary system, which is the primary controlling factor of the merger inspiral defined by
\begin{equation}
\mathcal{M}\equiv\Bigg(\frac{q^3}{1+q}\Bigg)^{1/5}m_1.
\end{equation}
Similarly to $\tilde\Lambda$, this quantity can be measured with much higher accuracy than either of the individual masses $m_1$, $m_2$, or the mass ratio $q$.
For this reason, we consider the binary chirp mass $\mathcal{M}$ to be the dominant dependent variable in this analysis, cataloging our various results as a function of $\mathcal{M}$ for any future GW event.


\subsection{Asymmetric nuclear matter parameters}\label{sec:parameters}

While the NS EoS is not currently known, there are many methods one can use to restrict it using various observations. 
This is because the structure of a NS and many of its observables such as mass, radius, tidal deformability, etc. rely strongly on the underlying EoS of nuclear matter.
For example, GW observations may help constrain the EoS in the pressure-density plane~\cite{LIGO:posterior}.
In this chapter, we show how GW detections can aid in the constraint of various characteristics of the EoS, known as the nuclear matter parameters~\cite{Vidana2009}.

As originally considered in Ref.~\cite{Myers:1969zz} and followed up in i.e. Refs.~\cite{Vidana2009,Alam2016,Malik2018}, we offer a generic method to parameterize NS EoSs.
Our starting point is with the definition of the EoS, given by
\begin{equation}
P(n)=n^2 \frac{\partial}{\partial n} e(n)
\end{equation}
for total nuclear number density $n \equiv n_p + n_n$, energy per nucleon of asymmetric nuclear matter $e$, and proton (neutron) number density $n_p$ ($n_n$).
We then expand the energy per nucleon $e$ with the isospin symmetry parameter $\delta \equiv (n_n-n_p)/n$ about $\delta=0$ (symmetric nuclear matter case) as~\cite{Vidana2009}:
\begin{equation}
e(n,\delta)=e(n,0)+S_2(n)\delta^2+ \mathcal{O}(\delta^4),
\end{equation}
where $e(n,0)$ corresponds to the energy of symmetric nuclear matter, and $S_2(n)$ represents the ``symmetry energy'' of asymmetric nuclear matter at quadratic order in $\delta$.
We note that in the above expansion, odd powers of $\beta$ are missing. 
This is because the quantum chromodynamic isospin symmetry leads one to conclude that even powers of $\delta$ (corresponding to strong nuclear forces) dominate over the odd ones (corresponding to EM forces).
Further, we note that the lower-order terms such as $e(n,0)$ dominate in near symmetric nuclear matter as found in e.g. terrestrial environments, while the higher order terms such as $S_2(n)$ tend to become more important in neutron-rich matter as found in NSs.

We follow this up by further characterizing the above terms $e(n,0)$ and $S_2(n)$ as yet another expansion about the \textit{saturation density} $n_0$ as:
\begin{align}
\begin{split}\label{eq:nuclearParameters}
e(n,0)&=e_0+\frac{K_0}{2} y^2 + \frac{Q_0}{6}y^3 + \mathcal{O}(y^4),\\
S_2(n)&=J_0+L_0 y + \frac{K_\mathrm{sym,0}}{2} y^2 + \mathcal{O}(y^3),
\end{split}
\end{align}
where $y \equiv (n - n_0)/3 n_0$ characterizes the relative difference from the saturation density. 
The above coefficients in each individual expansion are simply known as the ``nuclear matter parameters'' which determine the EoS in a model-independent way.
In particular, the nuclear matter parameters are more specifically known as the energy per particle $e_0$, incompressibility coefficient $K_0$, third derivative of symmetric matter $Q_0$, symmetry energy $J_0$, its slope $L_0$, and its curvature $K_\mathrm{sym,0}$ at saturation density, respectively. Following Refs.~\cite{Alam2014,Malik2018}, we further introduce the slope of the incompressibility to be:
\begin{equation}
M_0 \equiv Q_0 + 12 K_0.
\end{equation}
We once again observe that the lower-order nuclear matter parameters such as $e_0$, $J_0$, and $L_0$ become dominant in e.g. terrestrial observations (see the ``PREX'' and ``CREX'' experiments on neutron distributions in heavy elements such as $^{208}$Pb~\cite{Dong:2015vga,Meucci:2014mha,Horowitz:2013wha,Michaels:2015iea,Tagami:2020shn,Yang:2019pbx,Roca-Maza:2015eza}, which can be related to such lower-order parameters) with near-saturation density observations, while the supranuclear densities present within NSs result in the new opportunity to probe high-order parameters, such as $K_\text{sym,0}$.

In this chapter, we investigate correlations between the various nuclear parameters $L_0$, $K_0$ $M_0$, $K_{\text{sym},0}$ and the mass-weighted average tidal deformability $\tilde{\Lambda}$ in order to derive bounds on nuclear parameters from GW170817.  Bounds on $M_0$ and $K_\mathrm{sym,0}$ have previously been derived in Ref.~\cite{Malik2018} using GW170817, which we revisit in this chapter. Current experiments and astrophysical observations place bounds on $L_0$ as $40\text{ MeV} < L_0 < 62\text{ MeV}$~\cite{Lattimer2013,Lattimer2014,Tews2017}, and $30 \text{ MeV} < L_0 < 86\text{ MeV}$~\cite{Oertel2017}.

\subsection{The supranuclear equation of state}\label{sec:eos}

The structure of a NS and its tidal interactions in a binary system rely heavily on the underlying EoS of nuclear matter. 
Because of this, we employ a wide range of 120 different nuclear models in our analysis. 
These EoSs can be classified into three broad categories: 24 non-relativistic EoSs with Skyrme-type interaction, 9 RMF EoSs, and 88 EoSs derived through phenomenological variation.
Following Ref.~\cite{Read2009}, the high-density core EoSs listed above are all matched to the low-density EoS of Douchin and Haensel~\cite{Douchin:2001sv} at the transition density $\epsilon_{\text{tr}}$ such that the pressures are equivalent.

The EoSs in the first two classes are used also in~\cite{Alam2016,Malik2018}.
The Skyrme models used here are: SKa, SKb~\cite{Koehler1976}, SkI2, Sk13, SkI4, SkI5~\cite{Reinhard1995}, SkI6~\cite{Nazarewicz1996}, Sly230a~\cite{Chabanat1997}, Sly2, Sly9~\cite{Chabanat1995}, Sly4~\cite{Chabanat1998}, SkMP~\cite{Bennour1989}, SkOp~\cite{Reinhard1999}, KDE0V1~\cite{Agrawal2005}, SK255, SK272~\cite{Agrawal2003}, Rs~\cite{Friedrich1986}, BSK20, BSK21~\cite{Goriely2010}, BSK22, BSK23, BSK24, BSK25, BSK26~\cite{Goriely2013}.
On the other hand, the RMF models selected are BSR2, BSR6~\cite{Dhiman2007,Agrawal2010}, GM1~\cite{Glendenning1991}, NL3~\cite{Lalazissis1997}, NL3$\omega \rho$~\cite{Carriere2003}, TM1~\cite{Sugahara1994}, DD2~\cite{Typel2010}, DDH$\delta$~\cite{Gaitanos2004},  DDME2~\cite{Typel1999}.

One of the new EoS classes that we consider is the phenomenological
EoSs (PEs). To construct these EoSs, we followed the
formalism of Ref.~\cite{Margueron2018} by randomly sampling nuclear parameters $J_0$, $K_0$, $L_0$, $Q_0$ and $K_\mathrm{sym,0}$ as found in Table I of the above reference.  Following this, nonphysical EoSs with acausal structure ($v_s > c$), or having decreasing pressure as a function of density were removed.

For the second half of our analysis in regards to future GW observations, we employ a restricted set of the same EoSs as was used previously, taking into account the observed correlations between nuclear parameters $J_0$ and $L_0$.
Starting with the 120 nuclear EoS models described above, we further remove 63 EoS models which do not comply with the allowed regions shown in Fig. 8 of Tews \textit{et al.}~\cite{Tews2017}.
Here, they combined an exclusion region $J_0(L_0)$ with the ``accepted" 95.4\% correlation confidence bands between $J_0$ and $L_0$. See App.~\ref{app:EoS-comparison} for the impact of restricted EoSs on correlations between $K_\mathrm{sym,0}$ and $\tilde \Lambda$.

Taking the shared region between the above two exclusions results in 58 different nuclear EoS models, which can be classified into 3 distinct classes: 13 non-relativistic ``Skyrme-type" EoSs, 5 relativistic-mean-field (RMF) EoSs, and 40 PEs.
The Skyrme-type models remaining consist of: SKa, Sly230a, Sly2, Sly9, Sly4, SkOp, SK255, SK272, BSK20, BSK21, BSK22, BSK24, BSK26.
Further, the RMF models remaining are: BSR2, BSR6, NL3$\omega \rho$, DD2, DDH$\delta$.
All 18 of the above EoSs originate from the minimal set of EoSs used in Refs.~\cite{Alam2016,Malik2018}, now restricted by nuclear matter correlations.

One last class of EoSs indirectly used in this chapter can be found later in Chapter~\ref{chap:EoSInsensitiveRelations}, which we call ``LVC constrained" EoSs in this chapter.
By sampling the full physical EoS parameter space, the LIGO and Virgo Collaboration~\cite{LIGO:posterior,Carney:2018sdv} derived a marginalized 90\% posterior region on the NS pressure as a function of mass density (EoS) from GW170817, as seen in Fig. 2 of~\cite{LIGO:posterior}.
By randomly sampling the EoS posteriors from this analysis, a set of 100 ``constrained" EoSs were obtained, restricted by the GW observation of GW170817. 
While we do not directly utilize these 100 EoSs in the current analysis, we use them to estimate the mean value of the mass-weighted tidal deformability $\tilde\Lambda$ in Sec.~\ref{sec:futureSingle}, seen by Fig.~\ref{fig:meanLt}.

\begin{figure}
\begin{center} 
\includegraphics[width=0.7\columnwidth]{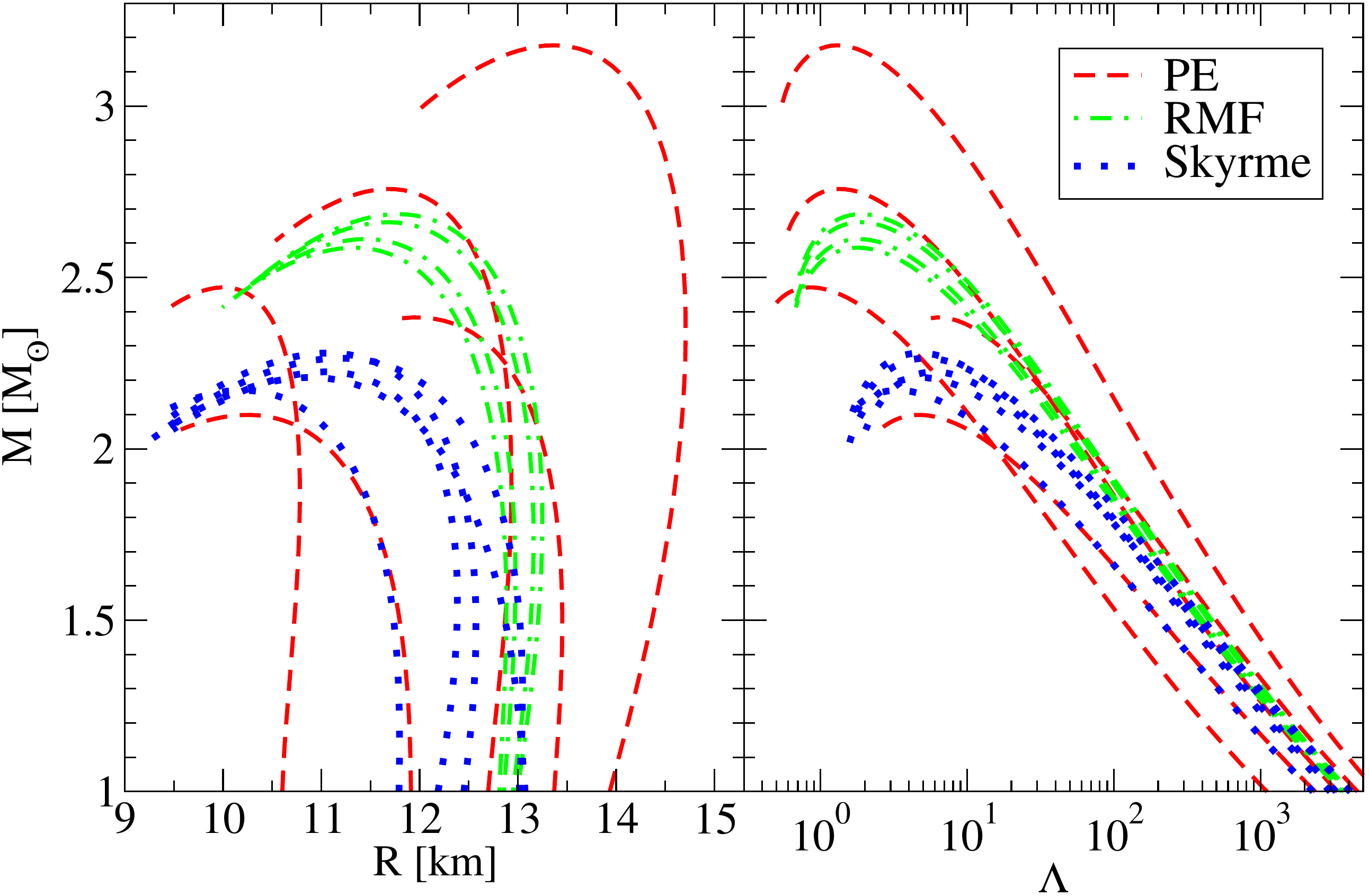}
\end{center}
\caption[Neutron star $M-R-\Lambda$ relationships]{Neutron star mass as a function of radius (left) and tidal deformability $\Lambda$ (right) for a representative set of the EoSs used in our analysis, separated into groups of phenomenological (red dashed), RMF (green dotted-dashed) and Skyrme-type (blue dotted). 
Observe how Skyrme and RMF EoSs follow self-consistent behavior, while PEs see a wide variance in properties such as maximum mass and radius, due to the nature of the random sampling in nuclear parameters.
}
\label{fig:M_RL}
\end{figure} 

Finally, in Fig.~\ref{fig:M_RL} we present the relations among the NS mass, radius and tidal deformability for selected EoSs in different classes mentioned above. Observe that RMF EoSs tend to produce NSs with larger radii and maximum mass than those for Skyrme-types, while the PE ones generate NSs with a wide range of properties.


\section{Correlations between tidal deformability and nuclear parameters}\label{sec:correlations}

In this section, we study correlations among nuclear parameters and tidal deformability, where the latter can be measured from GW observations. 
The amount of correlation between two variables $x$ and $y$ with $N$ data points can be quantified by the Pearson correlation coefficient $C$ defined by:
\begin{equation}\label{eq:correlation}
C(x,y) = \frac{\sigma_{xy}}{\sqrt{\sigma_{xx} \sigma_{yy}}},
\end{equation}
where the covariances $\sigma_{xy}$ are given by:
\begin{equation}\label{eq:covariance}
\sigma_{xy} = \frac{1}{N} \sum\limits^N_{i=0}x_i y_i - \frac{1}{N^2}\Big( \sum\limits^N_{i=0}x_i \Big) \Big( \sum\limits^N_{i=0}y_i \Big).
\end{equation}
$C= \pm 1$ represents absolute (anti-)correlation, while $C= 0$ corresponds to having no correlation. 

\subsection{$\tilde \Lambda$ versus nuclear parameters}

\begin{figure*}
\begin{center} 
\includegraphics[width=\textwidth]{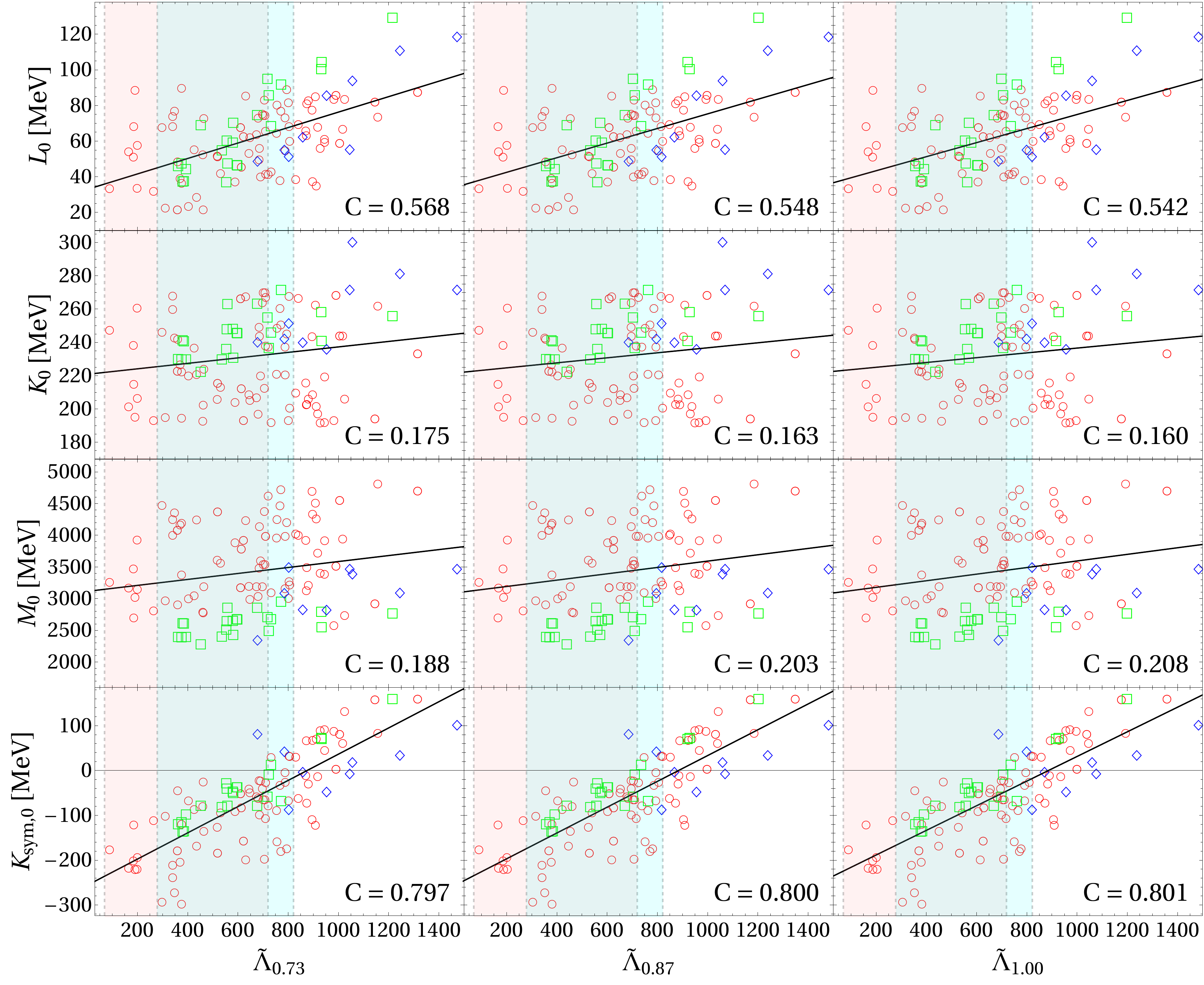}
\end{center}
\caption[Nuclear matter parameters vs. $\tilde\Lambda$]{
Correlations between nuclear parameters $L_0$, $K_0$, $M_0$, $K_{\text{sym},0}$ and the mass-weighted average tidal deformability $\tilde{\Lambda}_q$ for a chirp mass of $\mathcal{M} = 1.188M_\odot$ corresponding to GW170817, using Skyrme EoSs (green square), RMF EoSs (blue diamond), and PEs (red circle). Mass ratios are chosen as $q=0.73$ (left), $0.875$ (middle), and $1.00$ (right) consistent with GW170817.
The shaded cyan and magenta regions represent the measurement constraints on $\tilde{\Lambda}$ from GW170817~\cite{Abbott2018,Coughlin:2018fis}.
The solid black line in each panel represents the best fit line through the data, and the Pearson correlation coefficient $C$ measures the amount of correlation ($C=1$ being the absolute correlation and $C=0$ being no correlation).
}
\label{fig:nuclear}
\end{figure*} 

Reference~\cite{Malik2018} first studied the universal relations between nuclear parameters and the tidal deformability for isolated neutron stars. 
The authors then map this to the GW measurement on $\tilde \Lambda$ by using yet another universal relation between $\tilde \Lambda$ and $\lambda_0$ (the tidal deformability at $1.4\text{ M}_\odot$) for a specific choice of masses in a binary neutron star that is consistent with GW170817. 
However, the mass ratio $q \in \lbrack 0.73, 1.00\rbrack$~\cite{TheLIGOScientific:2017qsa} for this event has not been measured very precisely (the lower bound of this constraint has recently been improved to $0.8$ in Ref.~\cite{Coughlin:2018fis}), and the question arises as to whether such relation holds for various $q$. 
As we show in Appendix~\ref{app:LTL14}, indeed the universal relation is highly insensitive to the choice of $q$. 
This suggests that there are universal relations between nuclear parameters and $\tilde \Lambda$ for a given chirp mass $\mathcal{M}$ which has been measured with high accuracy for GW170817. 
Finding these universal relations is the focus of this section. 
Universal relations involving $\tilde \Lambda$ are, in some sense, practically more useful than those with $\lambda_0$, because the former is a quantity which can be directly measured from GW observations.

Figure~\ref{fig:nuclear} shows the correlations between nuclear parameters ($L_0$, $K_0$, $M_0$, $K_{\text{sym},0}$), and the mass-weighted average tidal deformability $\tilde{\Lambda}_q$ evaluated at mass ratios of $q=0.73$, $0.87$ and $1.00$. 
The linear regression shown in each panel represents the best fit line describing the relation between nuclear parameters and $\tilde{\Lambda}$. 
Observe that $K_0$ and $M_0$ show very poor correlations, resulting from a disconnect between PEs and EoSs found in Ref.~\cite{Malik2018}.
On the other hand, higher order parameter $K_{\text{sym},0}$ sees a fairly strong correlation of $\sim 0.80$.
It is noted that PEs typically have values of $K_0$ that are much lower than those for Skyrme or RMF EoSs, while $M_0$ is much higher, and $L_0$ and $K_{\text{sym},0}$ are very similar. Let us emphasize that we have restricted to physically valid PEs which have increasing pressure, and this is why we do not have PEs with e.g. $M_0 < 2500$ MeV~\footnote{This does \textit{not} mean that Skyrme and RMF EoSs with $M_0 < 2500$ are nonphysical.}.
The above finding indicates a necessity in using a large number of EoSs as nuclear parameters can take on a much wider range of values than considered in~\cite{Malik2018}.
Observe also that the behavior of the scattering and the amount of correlation found in Fig.~\ref{fig:nuclear} is not very sensitive to $q$.  This can also be seen from Fig.~\ref{fig:correlation}, where correlations between various nuclear parameters and $\tilde{\Lambda}$ are plotted as a function of mass ratio q.

\subsection{$\tilde \Lambda$ versus linear combinations of nuclear parameters}\label{sec:lin-comb}

References~\cite{Alam2016,Malik2018} report that correlations among nuclear parameters and NS observables become stronger if one considers certain combinations of the former, which we study here. In Refs.~\cite{Oertel2017,Lattimer2014,Tews2017}, tight constraints on the slope of the symmetry energy $L_0$ were derived. Thus we focus on constraining the incompressibility $K_0$, its slope $M_0$, and the symmetry energies' curvature $K_{\text{sym},0}$, utilizing prior bounds on $L_0$ and $\tilde{\Lambda}$ by considering linear combinations of the form $K_0+\alpha L_0$, $M_0+\beta L_0$, and $K_{\text{sym},0}+\gamma L_0$ with some coefficients $\alpha$, $\beta$ and $\gamma$. In previous literature~\cite{Alam2016,Malik2018}, these coefficients are chosen such that correlations become maximum.

\begin{figure}
\begin{center} 
\includegraphics[width=0.7\columnwidth]{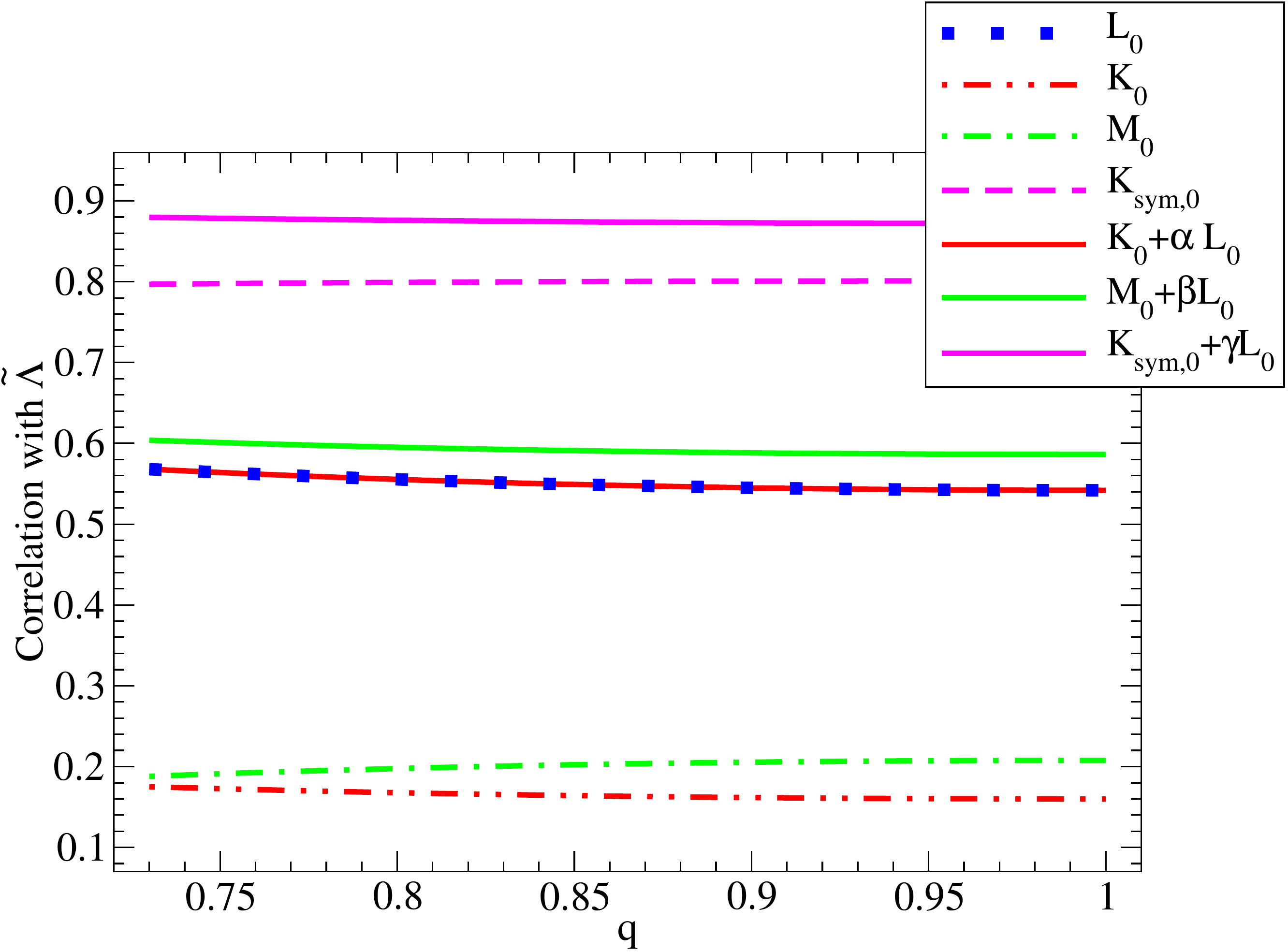}
\end{center}
\caption[Nuclear matter correlations as a function of mass ratio]{Correlations with $\tilde{\Lambda}$ as a function of mass ratio $q$ for $K_0+\alpha L_0$, $M_0+\beta L_0$, and $K_{\text{sym},0}+\gamma L_0$ for $\mathcal{M} = 1.188M_\odot$.
These are much stronger than those involving single nuclear parameters, which is also shown for reference. 
Here we choose $\alpha=2.27$ and $\beta=24.28$ giving 50\% correlations in the universal relations, while we choose $\gamma=2.63$ such that the correlation is maximized (see Sec.~\ref{sec:constraints} for more details).
Observe that correlations do not change significantly with $q$ across a wide range of mass ratios.
}
\label{fig:correlation}
\end{figure} 

Figure~\ref{fig:correlation} presents the correlations between $\tilde{\Lambda}$ and linear combinations of nuclear parameters as a function of mass ratio $q$. 
We found that the values of $\alpha$ and $\beta$ which give maximal correlation are unnecessarily large. For practical purposes, we choose here $\alpha =2.27$ and $\beta = 24.28$, such that a correlation of 50\% in the universal relations is achieved. For $\gamma$, we use $\gamma = 2.63$ which maximizes the correlation, as was done previously (see Sec.~\ref{sec:constraints} for more details).
For reference, we also show correlations involving single nuclear parameters. 
Observe that the former correlations are much stronger than the latter (except for $K_0 + \alpha L_0$ whose correlation is comparable to that of $K_0$) and remain to be strong over the acceptable region of mass ratio.
This implies that our choice of $q$ when calculating bounds on nuclear parameters does not matter significantly.
Therefore, we consider universal relations evaluated at the central mass ratio of $q=0.87$, shown in Fig.~\ref{fig:combinedKsymCorrelations}.
Also notice how linear combinations involving high-order nuclear parameter $K_{\text{sym},0}$ continue to significantly outperform lower-order parameters.

\begin{figure}
\begin{center} 
\includegraphics[trim={0 0.45cm 0 0},clip,width=0.65\columnwidth]{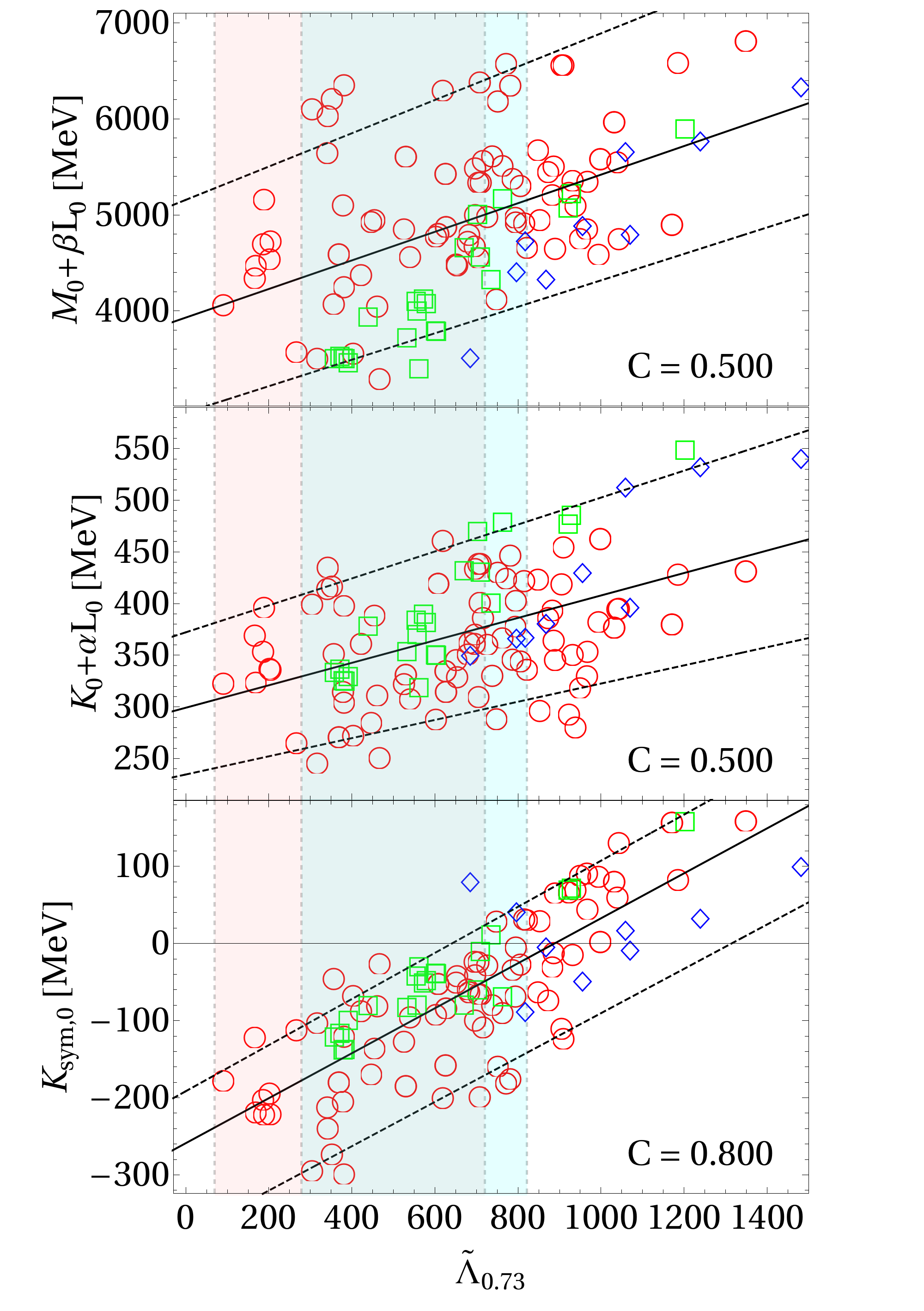}
\end{center}
\caption[Linear combinations of nuclear parameters vs. $\tilde\Lambda$]{
(Top) Correlations between mass-weighted average tidal deformability $\tilde{\Lambda}$ and linear combination of nuclear parameters $M_0$ and $L_0$ for a chirp mass of $\mathcal{M} = 1.188M_\odot$ corresponding to GW170817, using Skyrme (green square), RMF (blue diamond), and PE EoSs (red circle). 
The first two classes were also considered in Ref.~\cite{Malik2018} while the last class is considered here for the first time.
Mass ratio is chosen to be $q=0.87$, consistent with GW170817, though such correlations are insensitive to $q$.
The shaded cyan and magenta regions represent the measurement constraints on $\tilde{\Lambda}$ from GW170817~\cite{Abbott2018,Coughlin:2018fis}.
The solid black line represents the best fit line through the data, while the dashed lines correspond to the lines drawn with 90\% error bars on y-intercept and slope. 
The constant $\beta$ for the linear combination $M_0 + \beta L_0$ is chosen to be $\beta=24.28$ such that the correlation between observables becomes $50\%$.
(middle) Similar to the top panel but for the linear combination of $K_0+\alpha L_0$, with $\alpha=2.27$.
(bottom) Similar to the top panel but for the curvature of symmetry energy $K_{\text{sym},0}$.
}
\label{fig:combinedKsymCorrelations}
\end{figure} 

\subsection{Future prospects}

In the previous section, we described the correlations as a function of the mass ratio $q \equiv m_1/m_2$ for the fixed chirp mass of $1.188\text{ M}_{\odot}$; corresponding to GW170817.
We here supplement this investigation by considering the correlations as a function of varying chirp mass at a fixed mass ratio, applicable to any number of future GW observations.
The left panel of Fig.~\ref{fig:LT} justifies the use of a fixed mass ratio by presenting $\tilde\Lambda$ as a function of $q$ for the various EoS models used in this analysis.
Observe how $\tilde\Lambda$ is insensitive to the choice of $q$. Such feature is absent in the right panel of Fig.~\ref{fig:LT}, where $\tilde\Lambda$ is plotted as a function of chirp mass for fixed $q$.
Thus, for the remainder of this chapter we fix the mass ratio to be $q=0.90$, corresponding to the center of $0.80 \leq q \leq 1.00$ derived in Ref.~\cite{Coughlin:2018fis} for GW170817.

\begin{figure}
\begin{center} 
\begin{overpic}[width=0.8\linewidth]{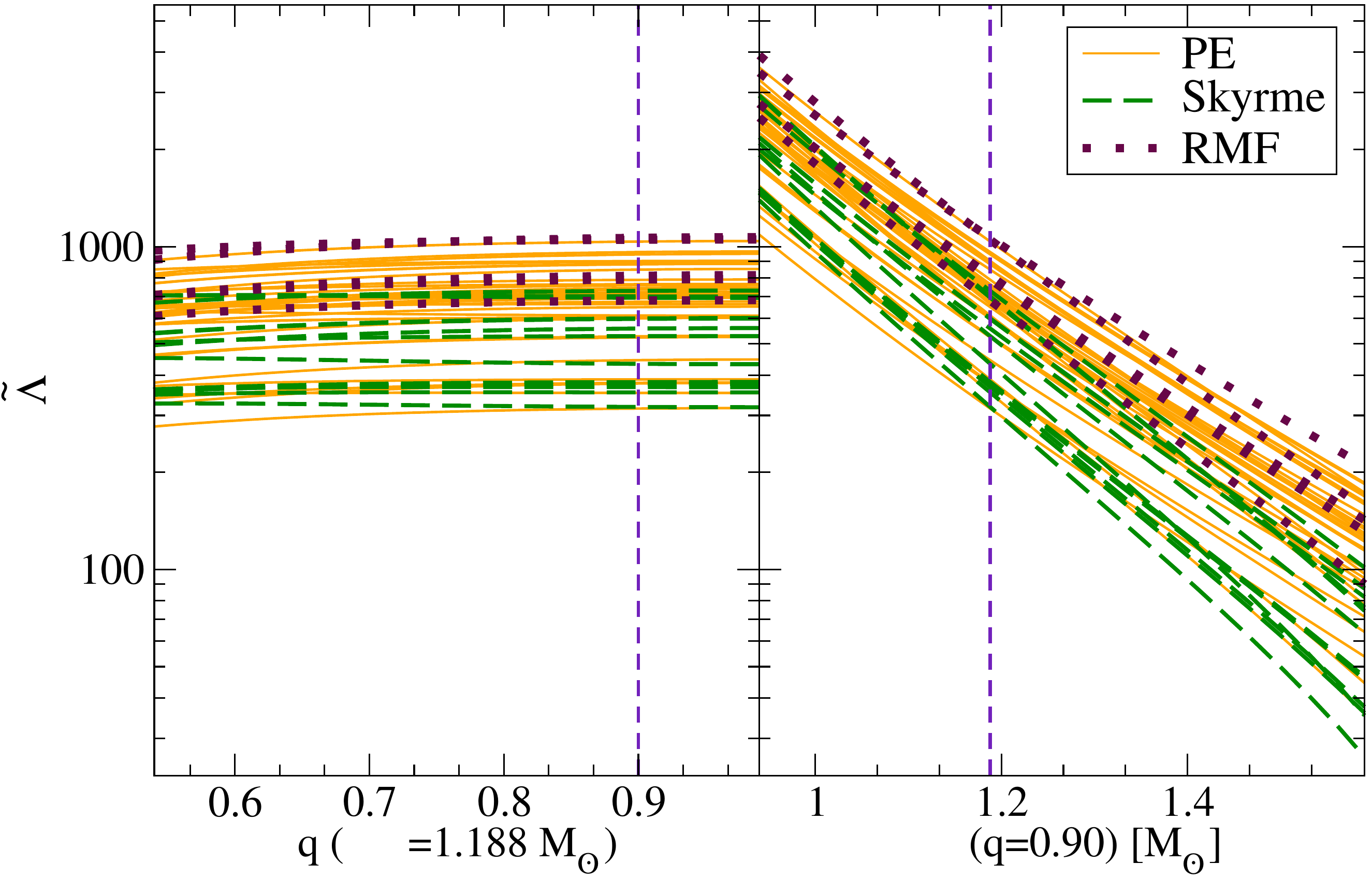}
\put(89,5){\small$\mathcal{M}$}
\put(228,5){\small$\mathcal{M}$}
\end{overpic}
\end{center}
\caption[$\tilde\Lambda$ as a function of chirp mass and mass ratio]{
(left) Mass-weighted tidal deformability $\tilde\Lambda$ for each EoS model used in this analysis as a function of the mass ratio $q$ for a fixed chirp mass of $\mathcal{M}=1.188 \text{ M}_{\odot}$. The vertical dashed line at $q=0.90$ corresponds to GW170817. Observe that $\tilde\Lambda$ is insensitive to the choice of $q$, which justifies our method of keeping the mass ratio fixed. (right) Similar to the left panel but as a function of the chirp mass $\mathcal{M}$ for a fixed mass ratio of $q=0.90$. The vertical dashed line at $\mathcal{M}=1.188 \text{ M}_{\odot}$ corresponds to GW170817.
}
\label{fig:LT}
\end{figure} 

It was shown in Refs.~\cite{Alam2016,Malik2018} that certain linear combinations of nuclear parameters, specifically $K_0+\alpha L_0$, $M_0+\beta L_0$, and $K_{\text{sym},0}+\gamma L_0$, exhibit heightened correlations, allowing one to derive more accurate constraints on the individual nuclear parameters.
However, in this chapter we show that this comes at the expense of additional sources of uncertainty which, if properly accounted for, enlarges the resulting constraints on the nuclear parameters $K_0$, $M_0$, and $K_{\text{sym},0}$.
It was also found that the single nuclear parameters as well as the linear combinations involving $K_0$ and $M_0$ observed poor correlations of $\mathcal{C} \lesssim 0.50$; indicating somewhat unreliable constraints on the nuclear parameters.

Figure~\ref{fig:CorrOfM} similarly shows the above correlations as a function of chirp mass for a fixed mass ratio of $q=0.90$. 
Observe how, similar to what was found in Ref.~\cite{Zack:nuclearConstraints}, the correlations for $K_0$, $M_0$, $K_0+\alpha L_0$, and $M_0+\beta L_0$ are exceedingly poor for all values of chirp mass.
$K_{\text{sym},0}$ on the other hand, remains highly correlated with $\tilde\Lambda$ across the entire range of $\mathcal{M}$.
We also observe how correlations are not improved by much when considering linear combinations between $K_{\text{sym},0}$ and $L_0$.

Could other combinations of nuclear parameters give stronger correlations? To address this question, we further explore new combinations of nuclear parameters in App.~\ref{app:multiplicative}. 
In particular, we consider the ``multiplicative" combinations of $K_0 L_0^{\eta}$, $M_0 L_0^{\nu}$, and $K_{\text{sym},0} L_0^{\mu}$ that is motivated from Refs.~\cite{Sotani:2013dga,Silva:2016myw}. We found that such new combinations do not offer any advantages in terms of correlations and constraints.

For the above reasons, we consider only the curvature of the symmetry energy $K_{\text{sym},0}$ when considering future-obtained constraints, without combinations with other parameters which would otherwise introduce additional uncertainties in the computation of constraints.
For the present considerations on nuclear matter combinations from GW170817, we present bounds on each of $K_0$, $M_0$, and $K_\text{sym,0}$.

\begin{figure}
\begin{center} 
\begin{overpic}[width=0.8\linewidth]{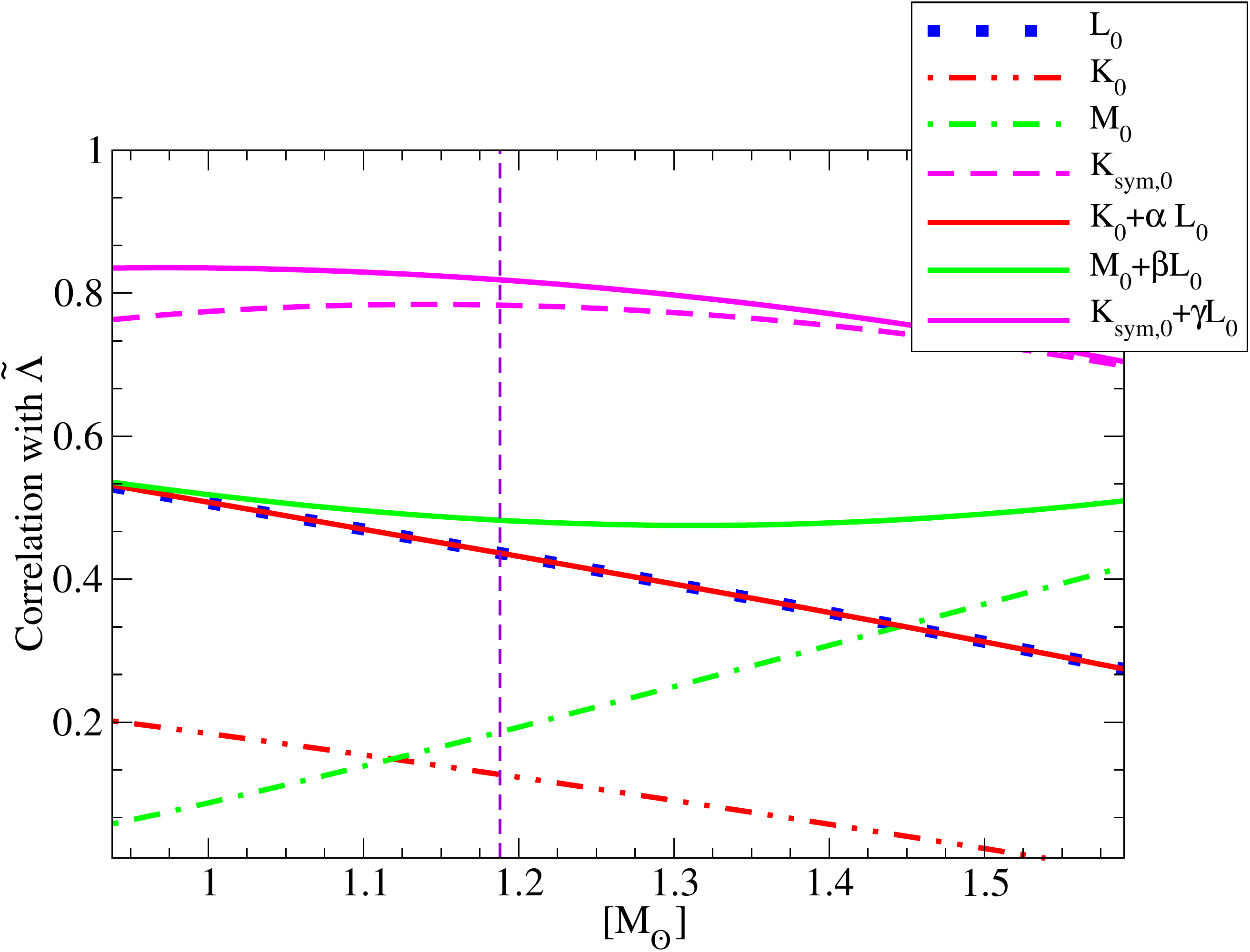}
\put(150,5){\small$\mathcal{M}$}
\end{overpic}
\end{center}
\caption[Correlations between $\tilde\Lambda$ and nuclear parameters as a function of $\mathcal{M}$]{
Correlations between $\tilde\Lambda$ and various nuclear parameters as a function of the chirp mass.
Observe how low-order nuclear parameters $K_0$ and $M_0$ show poor correlations, while high-order parameter $K_{\text{sym},0}$ is highly correlated -- both with and without a linear combination with $L_0$.
Additionally shown by the dashed vertical line is the chirp mass of $1.188\text{ M}_{\odot}$ corresponding to GW170817, studied in detail by Ref.~\cite{Zack:nuclearConstraints}.
}
\label{fig:CorrOfM}
\end{figure} 


\section{Constraints on nuclear matter parameters}\label{sec:constraints}

Here we compute constraints on nuclear matter parameters.
We begin with the current bounds formed by the binary NS merger GW170817.
We follow this up with the future prospects of nuclear matter parameters observed by advanced GW detectors, specifically focusing on the curvature of the symmetry energy $K_{\text{sym},0}$.


\subsection{GW170817}\label{sec:current}

\begin{figure*}
\begin{center} 
\includegraphics[width=.39\textwidth]{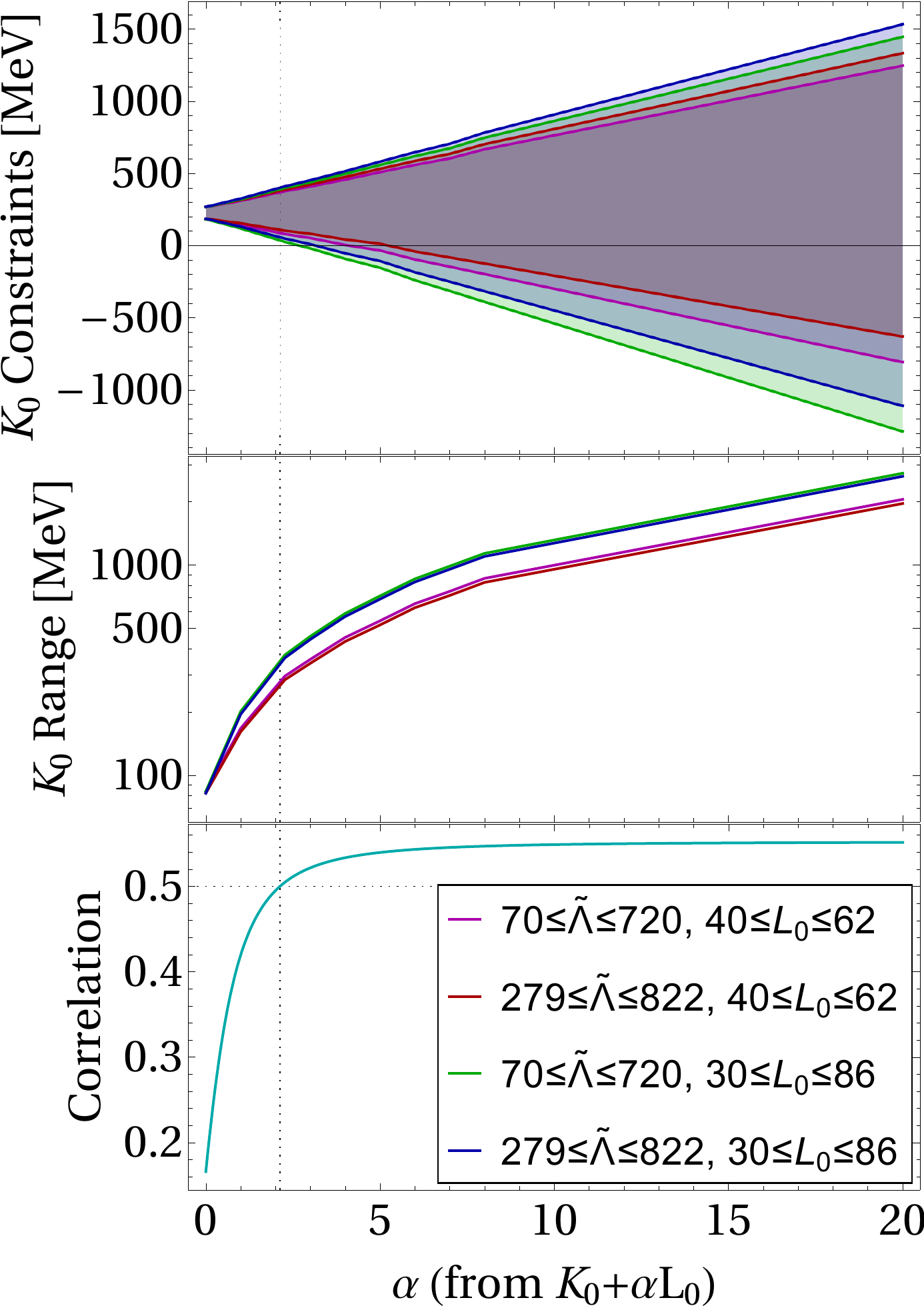}
\includegraphics[width=.39\textwidth]{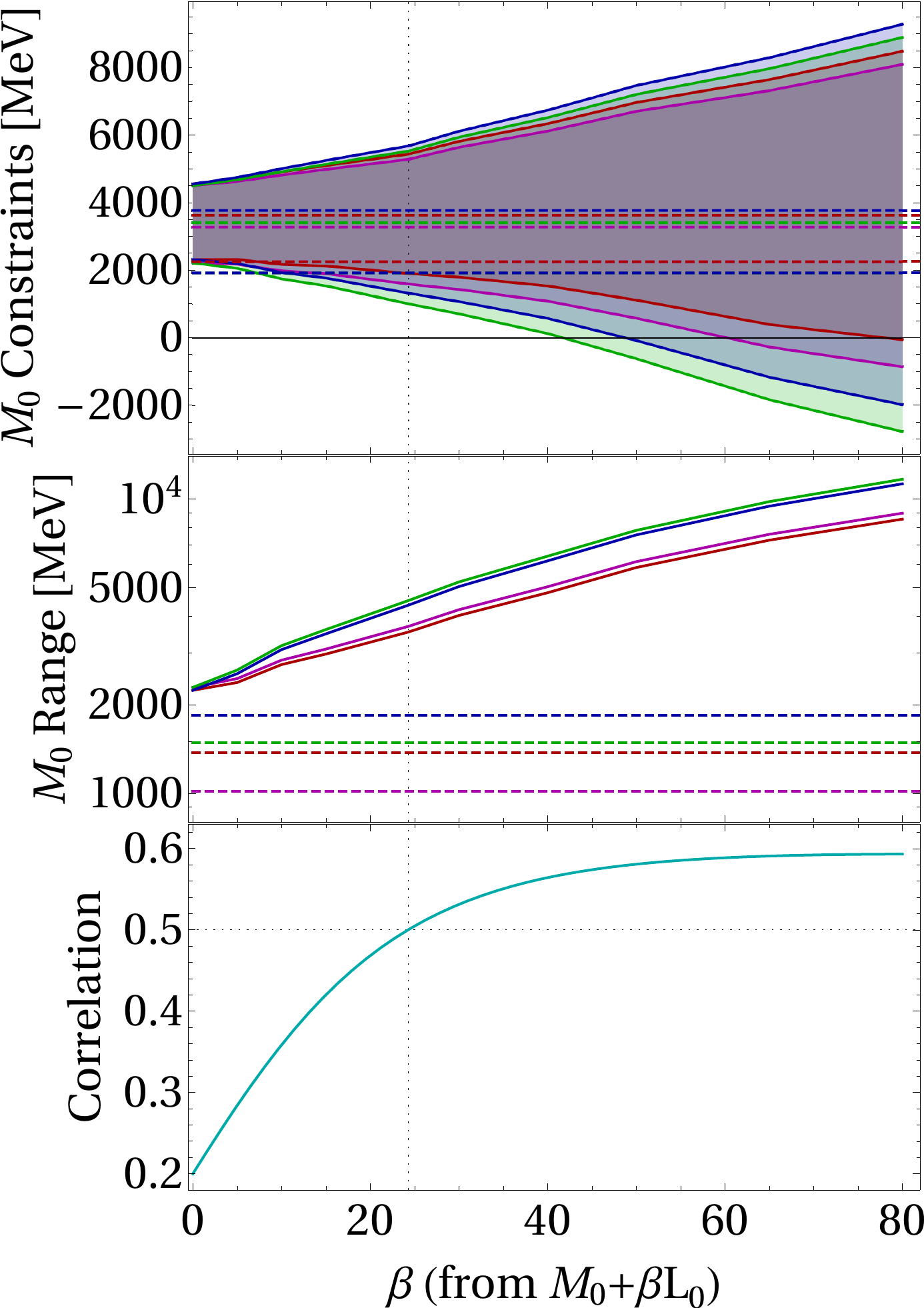}
\includegraphics[width=.39\textwidth]{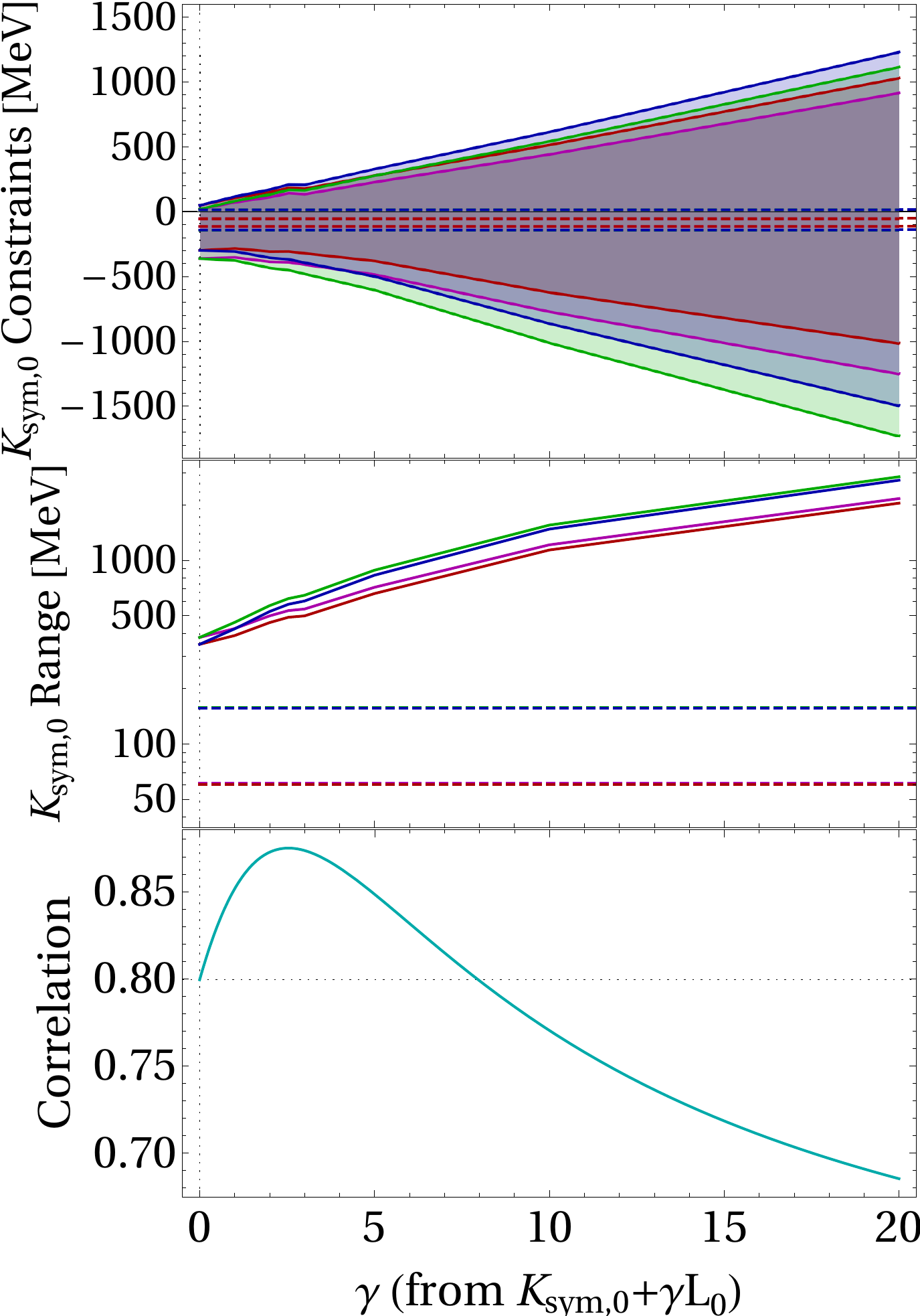}
\end{center}
\caption[Nuclear constraints' dependence on linear combination coefficient]{
Comparisons between nuclear parameter constraints and correlations with $\tilde{\Lambda}$, evaluated at $q=0.87$, as functions of $\alpha$, $\beta$, and $\gamma$. 
(Top) Estimated nuclear parameter constraints for different combinations of priors.
Dashed horizontal lines correspond to bounds derived by~\cite{Malik2018} under similar prior assumptions.
(Middle) Constraint ranges given as the difference between upper and lower limits.
(Bottom) Correlations between $\tilde{\Lambda}$ and linear combinations of nuclear parameters. 
Dotted vertical lines represent chosen values of $\alpha$, $\beta$, and $\gamma$ for deriving final bounds on the nuclear parameters.}
\label{fig:ConstraintsOfAlpha}
\end{figure*}

Let us now use the approximate universal relations among combined nuclear parameters and $\tilde \Lambda$ to derive bounds on the former from the measurement of the latter with GW170817. 
In this section, we detail the process used to estimate nuclear parameter bounds, taking into account the EoS scattering uncertainty.
We offer two alternative methods of accomplishing this.
In Sec.~\ref{sec:linear}, we offer a crude estimation of the constraints by finding linear regressions between the nuclear parameters and $\tilde\Lambda$.
We estimate $90\%$ confidence integrals on such regressions which allows us to predict bounds on nuclear parameters.
The linear regressions provide ready-to-use type results that can easily be implemented as the measurement on $\tilde \Lambda$ from GW170817 are updated.
In Sec.~\ref{sec:posteriors}, we detail a more comprehensive analysis in which we first compute the 2-dimensional probability distribution between the nuclear parameters and $\tilde\Lambda$.
We then combine this with the probability distribution on $\tilde\Lambda$ computed by Ref.~\cite{Abbott:LTposterior} to estimate the posterior distribution on nuclear parameters $K_0$, $M_0$, and $K_{\text{sym},0}$.

\subsubsection{Constraint estimation via linear regressions}\label{sec:linear}
In this simple error analysis, we first construct linear regressions of the form $(a\pm \delta_a^{\pm})\tilde{\Lambda} + (b \pm \delta_b^{\pm})$ on the relations evaluated at the central mass ratio of $q=0.87$ with the ``90\%'' error on the slope and $y$-intercept as follows:

\begin{equation}
\label{eq:regressions1}
\frac{K_0}{\text{MeV}} + \alpha \frac{L_0}{\text{MeV}} = 0.1086^{+0.02172}_{-0.02064}\, \tilde{\Lambda} + 299.1^{+72.97}_{-64.60}\,,
\end{equation}
\vspace{-5mm}
\begin{equation}
\label{eq:regressions2}
\frac{M_0}{\text{MeV}} + \beta \frac{L_0}{\text{MeV}} = 1.488^{+0.2456}_{-0.2038}\, \tilde{\Lambda} + 3929^{+1226}_{-990.2}\,,
\end{equation}
\vspace{-5mm}
\begin{equation}
\label{eq:regressions3}
\frac{K_{\text{sym},0}}{\text{MeV}} + \gamma \frac{L_0}{\text{MeV}} = 0.2915^{+0.007287}_{-0.004080}\, \tilde{\Lambda} - 259.1^{+67.36}_{-118.9}\,.
\end{equation}

The uncertainties on the slope and $y$-intercept, $\delta_a^{\pm}$ and $\delta_b^{\pm}$, are found by varying the upper and lower error bars throughout the parameter space, selecting only combinations of $\delta_a^{\pm}$ and $\delta_b^{\pm}$ which form ``90\% error lines" $(a \pm \delta_a^{\pm})\tilde{\Lambda} + (b \pm \delta_b^{\pm})$ containing 90\% of the data points between them.
Further, we choose the ``best fit'' 90\% error lines by minimizing the residual sum of squares, $\sum_{1=1}^{n}(y_i-f(x_i))^2$, as denoted by the dashed black lines in Fig.~\ref{fig:combinedKsymCorrelations}.
For reference, the covariances $\sigma_{ab}$ from Eq.~\eqref{eq:correlation} between $a$ and $b$ are found to be approximately $0.7274$, $124.5$, and $0.4235$ for Eqs.~\eqref{eq:regressions1}--\eqref{eq:regressions3}, respectively.
Using this method of uncertainty prediction, we find a 90\% confidence interval on the value of $b$ and $a$, allowing us to account for the EoS scatter in the universal relations when deriving bounds on nuclear parameters from GW170817, as we will study next.

Let us now use Eqs.~\eqref{eq:regressions1}--\eqref{eq:regressions3} to derive bounds on $K_0$, $M_0$, and $K_{\text{sym},0}$, as was done in Ref.~\cite{Malik2018}. 
We utilize prior bounds obtained from nuclear experiments and astrophysical observations as $L_0 \in \lbrack 40 , 62 \rbrack$ MeV~\cite{Lattimer2013} and $L_0 \in \lbrack 30 , 86 \rbrack$ MeV~\cite{Oertel2017,Lattimer2014,Tews2017}, as well as tidal deformability ranges of $\tilde{\Lambda} \in \lbrack 70 , 720 \rbrack$~\cite{Abbott2018}  and $\tilde{\Lambda} \in \lbrack 279 , 822 \rbrack$~\cite{Coughlin:2018fis}.
Utilizing the 90\% confidence interval's range on y-intercepts, we find constraints on $K_0$, $M_0$, and $K_{\text{sym},0}$ within priors of $L_0$ and $\tilde{\Lambda}$ such that minimal and maximal values of nuclear parameters are obtained. 
Therefore, 2 constraints on $\tilde{\Lambda}$ and 2 constraints on $L_0$ allow us to derive 4 possible constraints on each nuclear parameter $K_0$, $M_0$, and $K_{\text{sym},0}$. 
This particular method of estimating the probability distribution is conservative by nature, and also takes into account the uncertainty from scatter in our relations.

The top panels of Fig.~\ref{fig:ConstraintsOfAlpha} show comparisons between estimated nuclear parameter limits, while the central panels show constraint ranges (maximum value minus minimum value) as the linear combination coefficient ($\alpha$, $\beta$, or $\gamma$) is increased. The bounds are stronger if the ranges are smaller. For comparison, the bottom panels display the correlation between the nuclear parameter combinations and $\tilde{\Lambda}$. Observe that the bounds become weaker as one increases the coefficients, as we are introducing an additional source of uncertainty from $L_0$. Does this mean that it is always better to set the coefficients to 0 and consider universal relations with individual nuclear parameters? The answer is no because correlations are too small when $\alpha=\beta=0$, as can be seen from the bottom panels of Fig.~\ref{fig:ConstraintsOfAlpha}. If such correlations are too small, the relations can easily be affected by the addition of new EoSs and the bounds derived from these relations become unreliable. 

Therefore, we need to find the balance between having large enough correlations and yet to have reasonable bounds on the nuclear parameters.
Regarding $\alpha$ and $\beta$, notice that bounds on $K_0$ and $M_0$ increase approximately linearly with the coefficients, while correlations with $\tilde{\Lambda}$ quickly asymptote to values of $\sim 0.60$.
Thus we choose $\alpha=2.27$ and $\beta = 24.28$ such that correlations evaluated at central mass ratio $q=0.87$ are an arbitrary value of $C=0.50$, chosen to keep correlations as high as possible, while keeping $\alpha$ and $\beta$ as small as possible to avoid the propagation of uncertainty in $L_0$.
Regarding $\gamma$, because $K_{\text{sym},0}$ starts off with strong correlation at $\gamma=0$, we choose this value to remove any additional uncertainty in $\gamma$ and $L_0$ from our calculations
(Note this can not be done for the cases of $K_0$ and $M_0$ due to weak individual correlations with $\tilde{\Lambda}$).
Observe that the coefficient choices discussed in Ref.~\cite{Malik2018}, to maximize correlations to the level of $0.8$ and beyond is not necessarily applicable to every situation. 
As seen in Fig.~\ref{fig:ConstraintsOfAlpha}, high correlations are unobtainable for linear combinations involving $K_0$ and $M_0$, yielding no bounds under such a selection criteria.
Instead, reducing the threshold to $0.50$ returns constraints as shown below, albeit being less reliable.

Table~\ref{tab:Constraints} summarizes the bounds on the nuclear parameters with these fiducial choices of $\alpha$, $\beta$ and $\gamma$, using both this method of constraint estimation, and the more-comprehensive method described in Sec.~\ref{sec:posteriors}.
The constraints on $M_0$ and $K_{\mathrm{sym},0}$ are additionally visualized in \hypertarget{link:Posteriors}{Fig.~\ref{fig:Posteriors}}.
Notice how our conservative constraints (found by using the largest-range priors on both $L_0$ and $\tilde{\Lambda}$) on the slope of incompressibility and the curvature, \Mrange{955}{5675} and \Ksymrange{-358}{23}, are much weaker than those found in Ref.~\cite{Malik2018} (see Fig.~\ref{fig:Posteriors}), due to the consideration of EoS scatter uncertainty, and of additional PEs with a wider range of nuclear values.
We observe that the constraints derived here on $K_{\text{sym,0}}$ show good agreement with that of Refs.~\cite{Margueron:Ksym,Mondal:Ksym}.
Let us emphasize that the bounds on $K_0$ and $M_0$ should be considered as rough estimates, as the correlation of 0.50 is not very large; thus these bounds are more easily affected by inclusion of yet additional EoSs than the bounds on $K_\mathrm{sym,0}$.

\begin{table*}
\centering
\resizebox{\linewidth}{!}{%
\begin{tabular}{|| C{3.3cm} || C{4.6cm} || C{4.6cm} | C{4.6cm}||@{}m{0pt}@{}} 
\cline{2-3}\cline{3-4}
\multicolumn{1}{c|}{} & Method 2 & \multicolumn{2}{c||}{Method 1}\\
\hline
\diagbox[width=3.7cm]{\hspace*{0pt} \raisebox{2pt}{$L_0$ \small{[MeV]}}}{\raisebox{-5pt}{$\tilde{\Lambda}$} \hspace*{0.5cm}} & $\tilde\Lambda$ Posterior Distribution~\cite{Abbott:LTposterior} & 70--720~\cite{Abbott2018} & 279--822~\cite{Coughlin:2018fis} \\
\hline
\hline
 40--62~\cite{Lattimer2013,Lattimer2014,Tews2017} & \makecell{\\ \KrangeTab{69}{352}  \\ \\ \MrangeTab{1371}{4808} \\ \\ \KsymrangeTab{-285}{7} \\ \\} & \makecell{\\ \KrangeTab{100}{375}  \\ \\ \MrangeTab{1538}{5433} \\ \\ \KsymrangeTab{-358}{23} \\ \\}  & \makecell{\\ \KrangeTab{118}{388} \\ \\ \MrangeTab{1849}{5609} \\ \\ \KsymrangeTab{-298}{54} \\ \\ } \\
 \hline
 30--86~\cite{Oertel2017} & \makecell{\\ \KrangeTab{123}{330}  \\ \\ \MrangeTab{1884}{4635} \\ \\ \KsymrangeTab{-285}{7} \\ \\} & \makecell{\\ \KrangeTab{45}{398}  \\ \\ \MrangeTab{955}{5675} \\ \\ \KsymrangeTab{-358}{23} \\ \\} & \makecell{\\ \KrangeTab{63}{411} \\  \\ \MrangeTab{1266}{5852} \\  \\ \KsymrangeTab{-298}{54}\\ \\ }\\
 \hline
\end{tabular}
}
\caption[Nuclear matter parameter constraints]{
GW170817 constraints on the incompressibility $K_0$ (top row), its slope $M_0$ (middle row), and the symmetry energy curvature $K_{\text{sym},0}$ for 4 different sets of priors on $L_0$~\cite{Lattimer2013,Oertel2017,Lattimer2014,Tews2017}, and $\tilde{\Lambda}$~\cite{Coughlin:2018fis,Abbott2018}.
These quantities are computed using two different methods: (i) a simple linear regression estimation described in Sec.~\ref{sec:linear} (labeled ``Method 1" on the right column), and (ii) a comprehensive computation of the nuclear parameter posterior probability distributions described in Sec.~\ref{sec:posteriors} (labeled ``Method 2" on the left column).
The two methods show moderate agreement, although the first method can be seen to over-estimate the errors -- thus we recommend the use of the more accurate distributions computed in method 2, which properly take into account the covariances between the parameters, as well as utilizes the full posterior distribution on $\tilde\Lambda$ derived by the LIGO Collaboration~\cite{Abbott:LTposterior}.
The bounds on nuclear parameter $M_0$ and $K_{\text{sym},0}$ are weaker but more reliable than those found in~\cite{Malik2018} due to the inclusion of scatter uncertainty in our linear regressions. The bounds on $K_0$ and $M_0$ should be taken as a rough estimate as the correlation in universal relations that were used to derive them are not large, and thus, may be subject to change with inclusion of further EoSs.
}\label{tab:Constraints} 
\end{table*}

\subsubsection{Constraint estimation via LIGO posterior distributions}\label{sec:posteriors}
In this section, we offer a more comprehensive method of estimating nuclear matter constraints than was found in Sec.~\ref{sec:linear}.
Previously, a rough estimate on the nuclear matter constraints was computed by finding linear regressions between $\tilde\Lambda$ and nuclear parameters.
By estimating the $90\%$ errors on these lines, bounds on the nuclear parameters were manually approximated.
In this section, we improve upon this method by (i) properly taking into account the covariance between $\tilde\Lambda$ and nuclear parameters by generating a multivariate probability distribution, and (ii) taking into account the full posterior probability distribution on $\tilde\Lambda$ as derived by the LIGO Collaboration~\cite{Abbott:LTposterior}.

We begin by generating the 2-dimensional probability distribution between $\tilde\Lambda$ and the nuclear parameters, taking into account the specific covariances between them.
For the example of $K_{\text{sym},0}$ the distribution is given by:
\begin{equation}\label{eq:probDist}
P(\tilde{\Lambda},K_{\text{sym},0})=\frac{1}{2\pi\sqrt{|\Sigma|}}e^{-\frac{1}{2}(x-\mu)^T\Sigma^{-1}(x-\mu)},
\end{equation}
where $x$ is the 2-dimensional vector containing $\tilde\Lambda$ and the given nuclear parameter, $\mu$ is the 2-dimensional vector containing the expected values of $x$, and $\Sigma$ is the $2\times2$ covariance matrix defined with elements given by Eq.~\eqref{eq:covariance}.
This distribution is displayed in Fig.~\ref{fig:probDist} for each nuclear parameter.
Notice here the high degree of covariance between the variables used in this analysis - indicative of the importance for using this method of constraint extraction.

\begin{figure*}
\begin{center} 
\includegraphics[width=.45\linewidth]{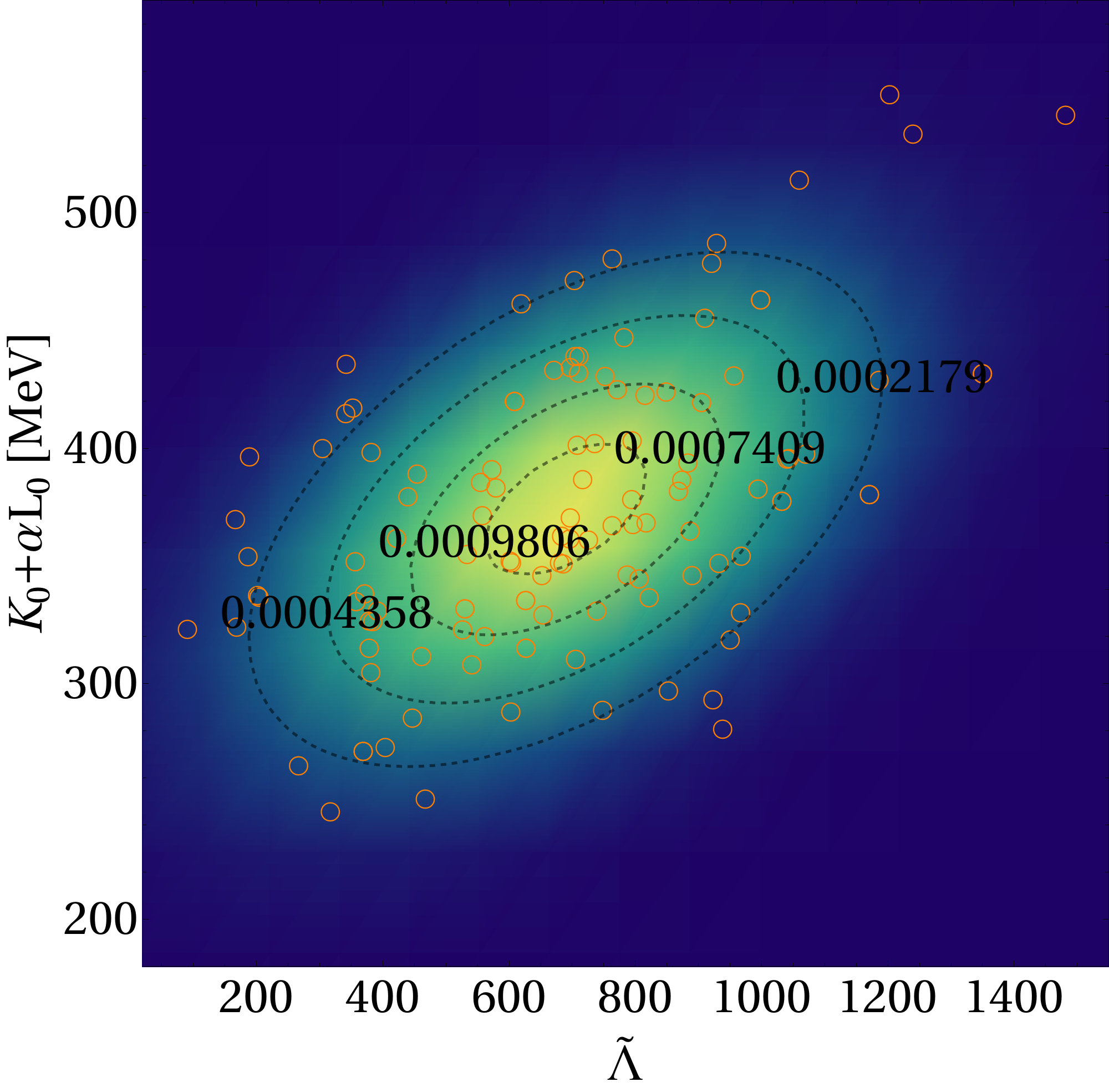}
\includegraphics[width=.46\linewidth]{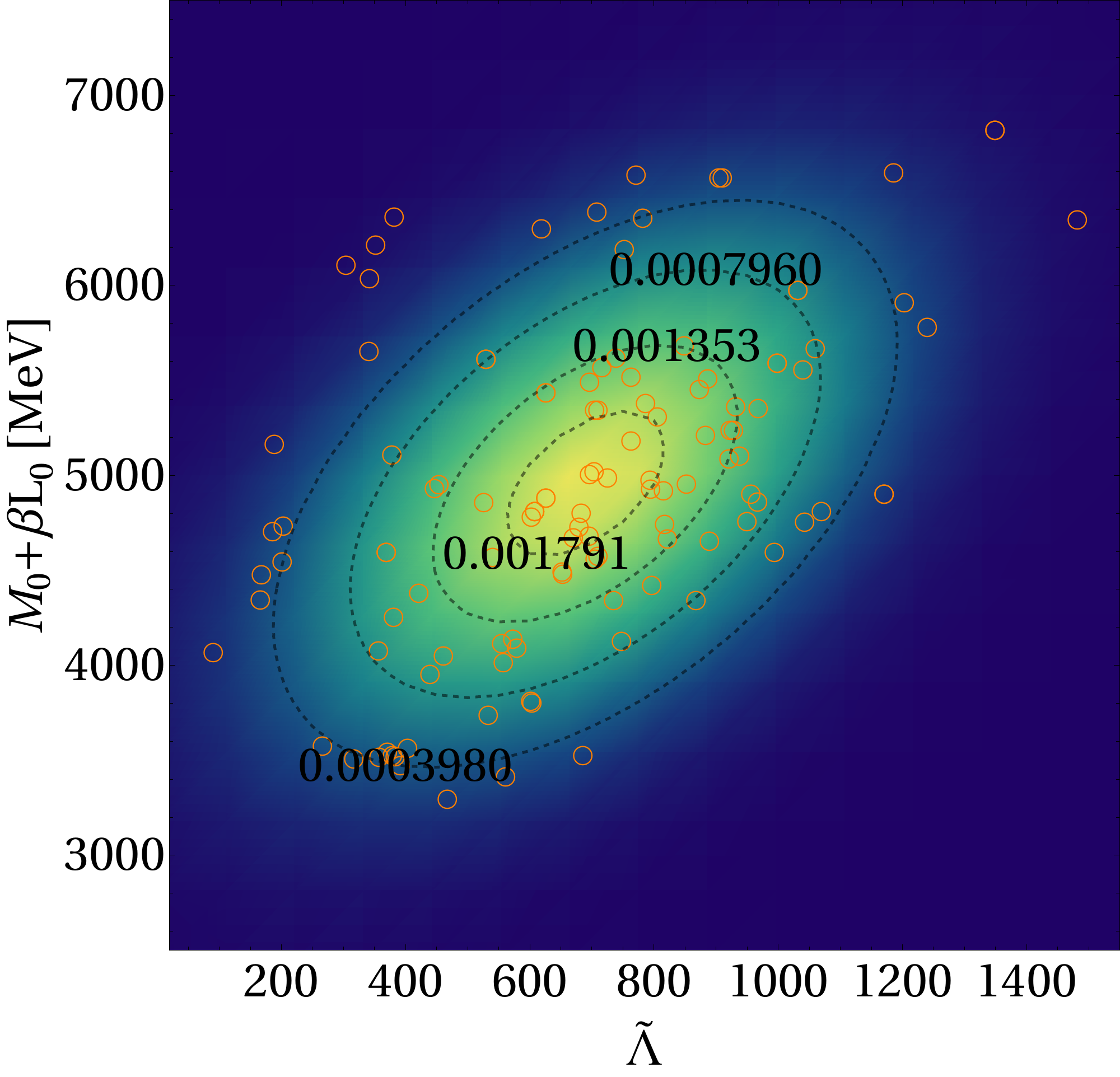}
\includegraphics[width=.45\linewidth]{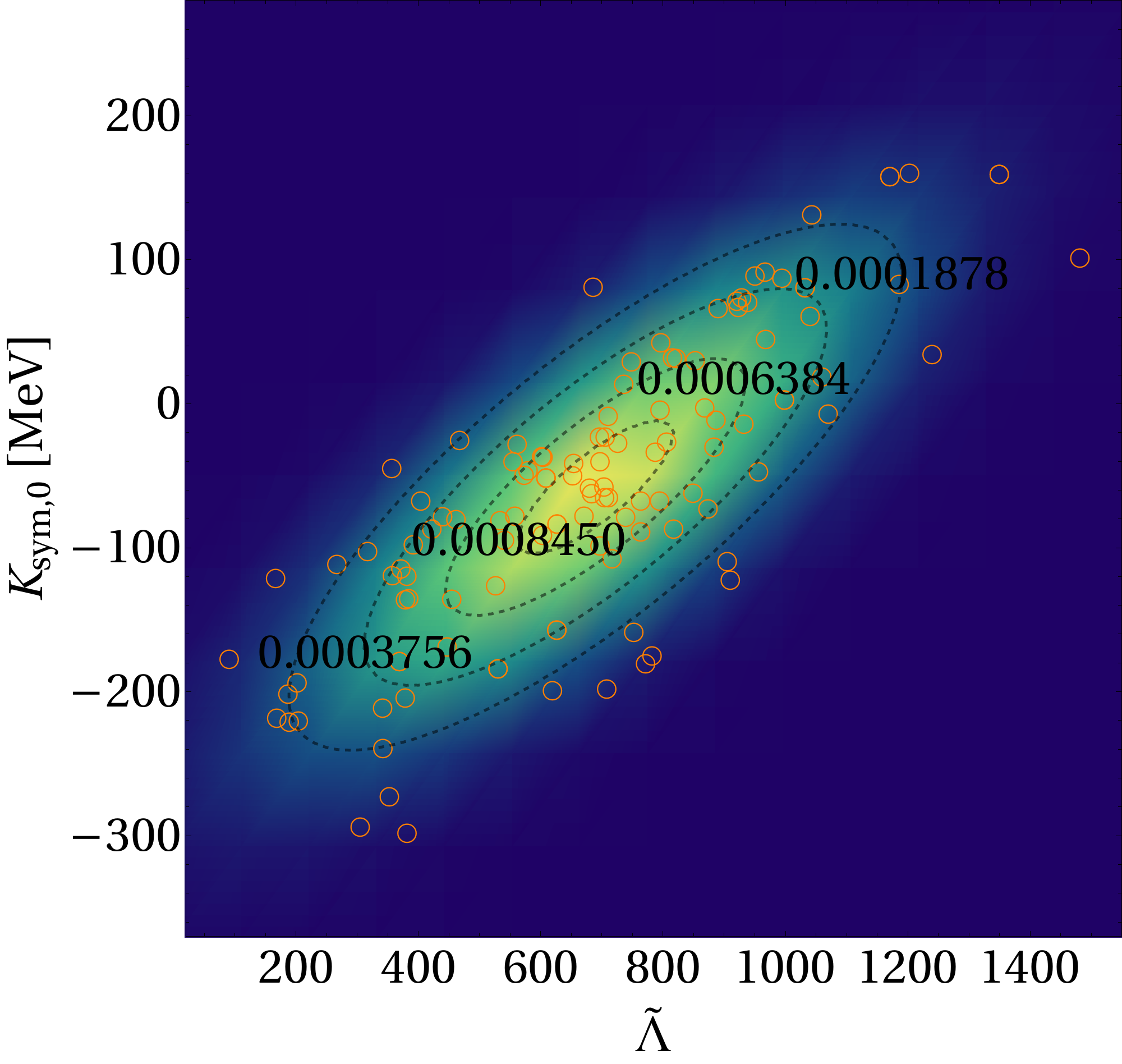}
\end{center}
\caption[$\tilde\Lambda$ - nuclear parameter PDFs]{Two-dimensional normalized probability distributions between $\tilde\Lambda$ and nuclear parameters $K_0+\alpha L_0$ (left), $M_0+\beta L_0$ (right), and $K_{\text{sym},0}$ (center) generated via Eq.~\eqref{eq:probDist}.
Overlayed on the distributions is the set of 120 data points corresponding to each EoS used in this investigation for comparison.
Observe how the multivariate Gaussian distributions indicate high levels of covariance between the variables, indicating the importance of estimating bounds using this method.}
\label{fig:probDist}
\end{figure*}

Following this, we compute the conditional probability distributions on nuclear matter parameters given a tidal observation of $\tilde\Lambda_\text{obs}$.
Following Ref.~\cite{jensen_2007}, the one-dimensional conditional probability distribution on nuclear parameter $Y$ is then given by
\begin{equation}
P(Y|\tilde\Lambda_\text{obs})\sim \mathcal{N}\left(\mu_Y+\frac{\sigma_Y}{\sigma_{\tilde\Lambda_\text{obs}}}C(\tilde\Lambda_\text{obs}-\mu_{{\tilde\Lambda_\text{obs}}}),(1-C^2)\sigma_Y^2\right).
\end{equation}
In the above expression, $\mathcal{N}(\mu,\sigma^2)$ is the normal distribution with mean and variance $\mu$ and $\sigma^2$, and $\mu_\A$ and $\sigma_\A^2$ are the mean and variances of $Y$ and $\tilde\Lambda_\text{obs}$.

\begin{figure}
\begin{center} 
\includegraphics[width=0.7\columnwidth]{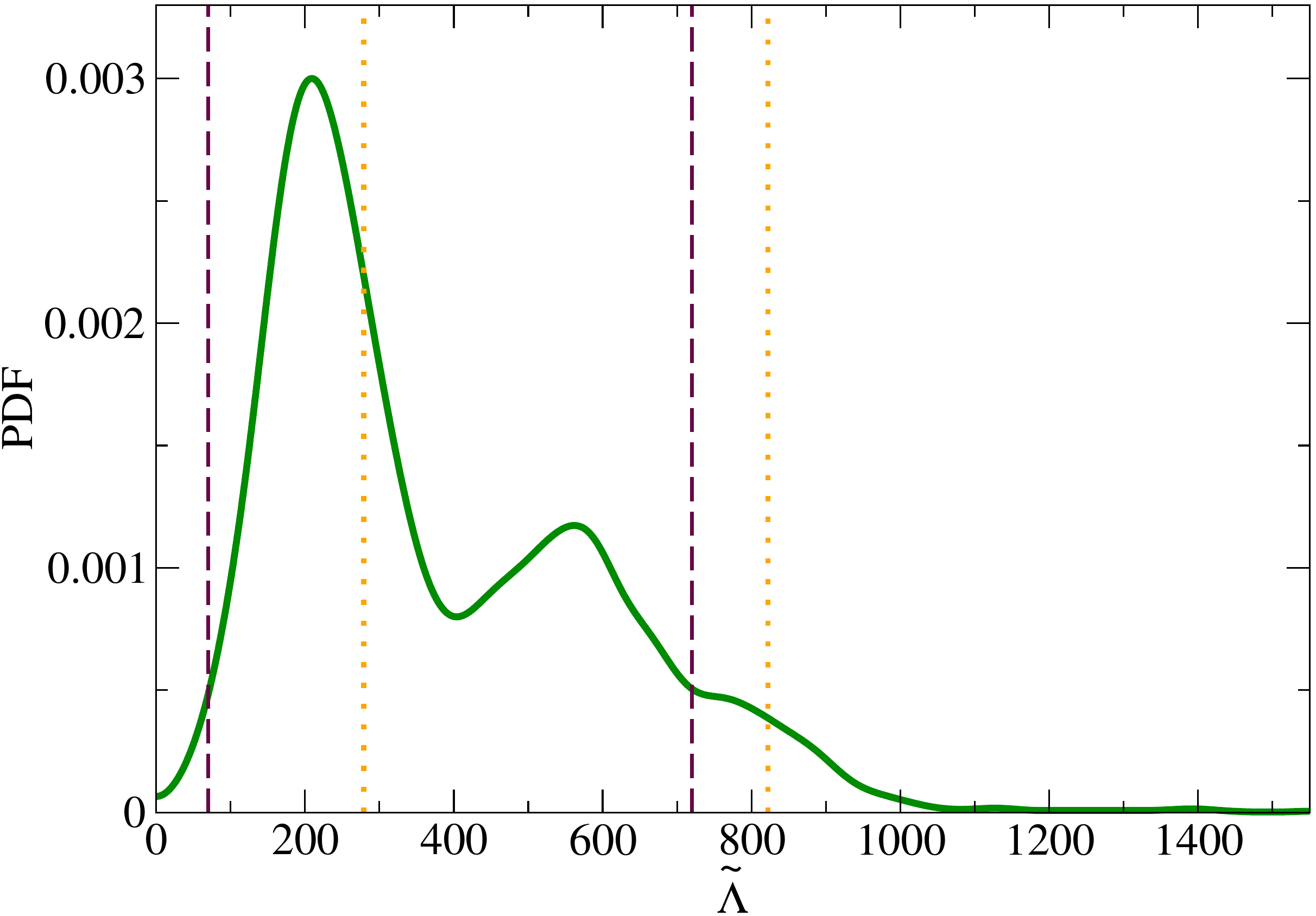}
\end{center}
\caption[$\tilde\Lambda$ posterior PDF]{Posterior probability distribution on $\tilde\Lambda$ as derived by the LIGO Collaboration in Ref.~\cite{Abbott:LTposterior}. We take this as a prior distribution when computing the posteriors on nuclear parameters. Additionally shown are the GW and EM counterpart bounds of $70 \leq \tilde{\Lambda} \leq 720$~\cite{Abbott2018} (dashed maroon) and $279 \leq \tilde{\Lambda} \leq 822$~\cite{Coughlin:2018fis} (dotted orange) for comparison.}
\label{fig:LigoPrior}
\end{figure}

Next, we extract the one-dimensional probability distributions on $K_0+\alpha L_0$, $M_0 + \beta L_0$, and $K_{\text{sym},0}$ by combining the one-dimensional conditional distributions $P(Y|\tilde\Lambda)$ found in Eq.~\eqref{eq:conditional} with the probability distribution $P_{\text{LIGO}}(\tilde\Lambda)$ on $\tilde\Lambda$  derived by the LIGO Collaboration in Ref.~\cite{Abbott:LTposterior} for GW170817, shown in Fig.~\ref{fig:LigoPrior}.
For example, the posterior probability distribution on $K_0+\alpha L_0$ is given by:
\begin{equation}
P(K_0+\alpha L_0)=\int\limits_{-\infty}^{\infty}P(K_0+\alpha L_0|\tilde{\Lambda})P_{\text{LIGO}}(\tilde{\Lambda})d\tilde{\Lambda},
\end{equation}
and similarly for $M_0+\beta L_0$ and $K_{\text{sym},0}$.
Additionally, to find the probability distributions on $K_0$ and $M_0$, we perform one last integration over the prior probability distribution of $L_0$, assumed to be Gaussian with standard deviation $\sigma=\frac{1}{2}(80+36)$ and mean $\mu=\frac{1}{2}(80-36)$~\cite{Oertel2017} (or $\sigma=\frac{1}{2}(62+40)$ and $\mu=\frac{1}{2}(62-40)$~\cite{Lattimer2013,Lattimer2014,Tews2017} for the alternative priors on $L_0$).
For example, the probability distribution on $K_0$ is given by:
\vspace{-.1cm}
\begin{equation}
P(K_0)=\int\limits_{-\infty}^{\infty}P(K_0+\alpha L_0)P(L_0)dL_0,
\end{equation}
with $\alpha=2.27$.

The results of these computations are shown in Fig.~\ref{fig:Posteriors} for the more conservative priors on $L_0$.
We observe that $K_0$, $M_0$, and $K_{\text{sym},0}$ now obey distributions that look like skewed Gaussians centered at $K_0=208^{+86}_{-85} \text{ MeV}$, $M_0=3075^{+1045}_{-1033}\text{ MeV}$, and $K_{\text{sym},0}=-156^{+97}_{-81} \text{ MeV}$ ($68\%$ standard deviations).
This results in  $90\%$ confidence intervals of \Krange{69}{352}, \Mrange{1371}{4808}, and \Ksymrange{-285}{7}.
We tabulate these values for both priors on $L_0$ in Table~\ref{tab:Constraints} for comparison to the simple method described in Sec.~\ref{sec:linear}.
These constraints on the nuclear parameters are comparable to, yet smaller than that found in Sec.~\ref{sec:linear}, although are much more accurate because the covariances between $\tilde\Lambda$ and such nuclear parameters were properly taken into account, as well as considering the true probability distribution on $\tilde\Lambda$ from GW170817 as derived by the LIGO Collaboration.

How much does the addition of PEs affect the bounds on $K_0$, $M_0$, and $K_{\text{sym},0}$? To address this, we repeat our analysis without including these additional EoSs (see Appendix~\ref{app:sansHybrid} for more details). We find that the removal of such EoSs gives strong improvement in both correlations and nuclear constraints for low-order nuclear parameters $K_0$ and $M_0$, and the results are consistent with those in Ref.~\cite{Malik2018}. This further illuminates the need to study a wider variety of EoSs for use in universal relations to properly account for systematic errors.


\subsection{Future prospects}\label{sec:chirpmass}

Now that we have identified the high-correlation behavior of $K_{\text{sym},0}$, we proceed to compute projected bounds on the curvature of the symmetry energy as a function of chirp mass that is applicable to any future event.
Additionally we offer the same analysis repeated for 5 anticipated future detector sensitivities $S_n(f)$ for detectors O2~\cite{aLIGO}, aLIGO~\cite{aLIGO}, A\texttt{+}~\cite{Ap_Voyager_CE}, Voyager~\cite{Ap_Voyager_CE}, ET~\cite{ET} and CE~\cite{Ap_Voyager_CE} (see Fig.~\ref{fig:sensitivities}), which would allow one to compute the corresponding posterior distribution on $K_{\text{sym},0}$ given an events' chirp mass $\mathcal{M}$.


\subsubsection{Single events}\label{sec:futureSingle}

Previously in Sec.~\ref{sec:current}, a posterior distribution on $\tilde\Lambda$ as derived from GW170817, was utilized in order to compute posterior distributions on the nuclear parameters.
In this analysis of future observations however, no such distribution is available.
To remedy this, we approximate the effective ``future" posterior distribution on $\tilde\Lambda$ as a Gaussian probability distribution given by
\begin{equation}\label{eq:LtPrior}
P_A(\tilde\Lambda)= \frac{1}{\sqrt{2\pi\sigma^2_{A}}} e^{-(\tilde\Lambda-\mu_{\tilde\Lambda})^2/2\sigma_{A}^2}
\end{equation}
for detector $A$.
Here, $\mu_{\tilde\Lambda}=\mu_{\tilde\Lambda}(\mathcal{M})$ is computed from the mean value of the ``LVC constrained" EoSs found in Chapter~\ref{chap:EoSInsensitiveRelations} and again described in Sec.~\ref{sec:eos} for each value of chirp mass, as shown by Fig.~\ref{fig:meanLt}.
Further, $\sigma_{A}$ is approximated via simple Fisher analyses described in Sec.~\ref{intro:Fisher}, which estimates the measurement accuracy on $\tilde\Lambda$ under the assumption of detector sensitivity $A$. Figure~\ref{fig:sigmaLt} presents $\sigma_{A}$ for all 6 detectors.

\begin{figure}
\begin{center}
\begin{overpic}[width=0.7\columnwidth]{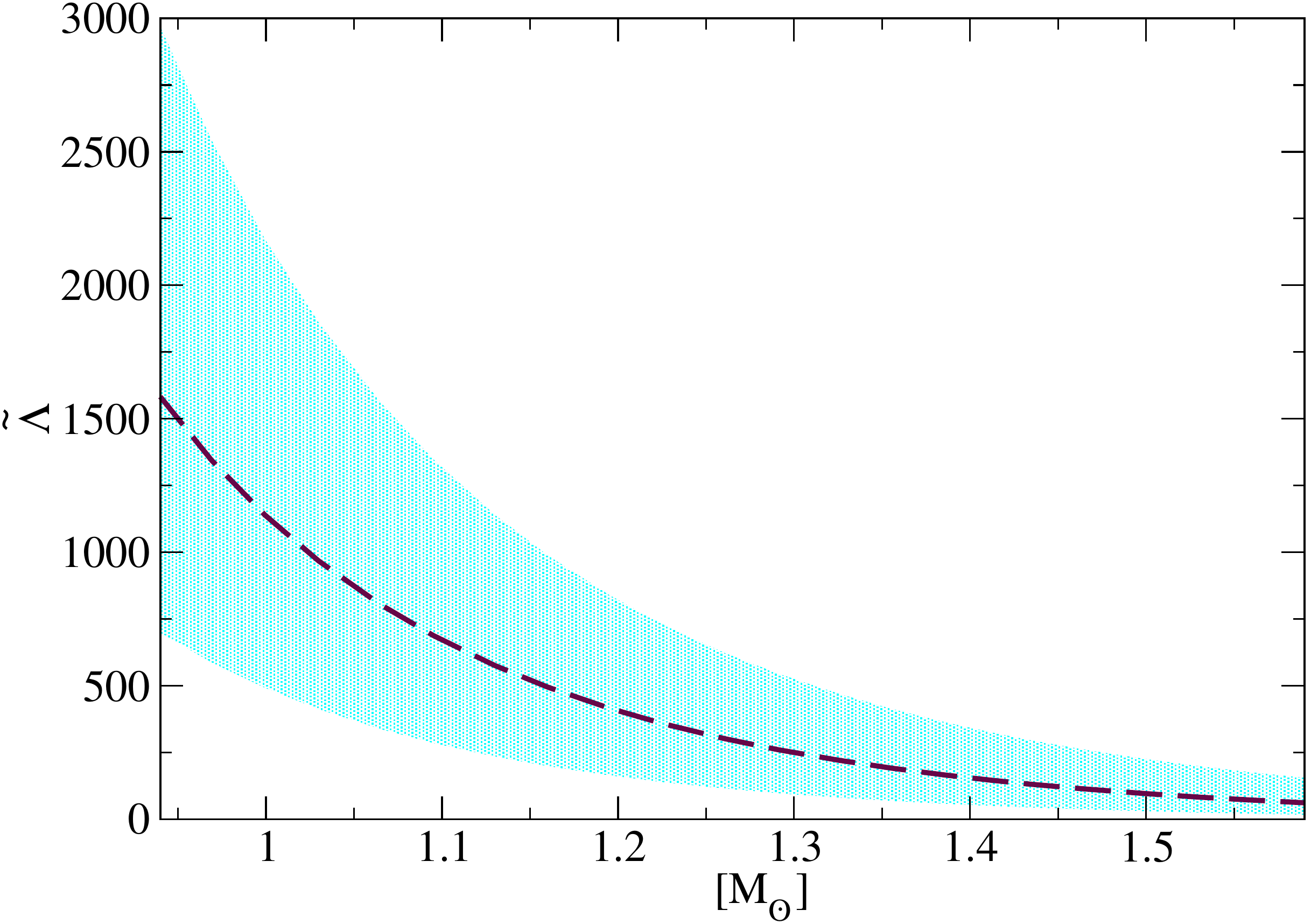}
\put(150,4){\small$\mathcal{M}$}
\end{overpic}
\end{center}
\caption[Mean $\tilde\Lambda$ value as a function of chirp mass]{
Mean value of $\tilde\Lambda$ (dashed maroon curve) as a function of chirp mass $\mathcal{M}$, computed as the mean value of the ``LVC constrained" EoSs (cyan shaded region) from Chapter~\ref{chap:EoSInsensitiveRelations} for each value of chirp mass.
This mean value corresponds to $\mu_{\tilde\Lambda}(\mathcal{M})$ used in the generation of the approximated $\tilde\Lambda$ probability distributions in Eq.~\eqref{eq:LtPrior} needed to compute constraints on $K_{\text{sym},0}$ in Eq.~\eqref{eq:P_Ksym}.
}
\label{fig:meanLt}
\end{figure} 

\begin{figure}
\begin{center} 
\begin{overpic}[width=0.8\columnwidth]{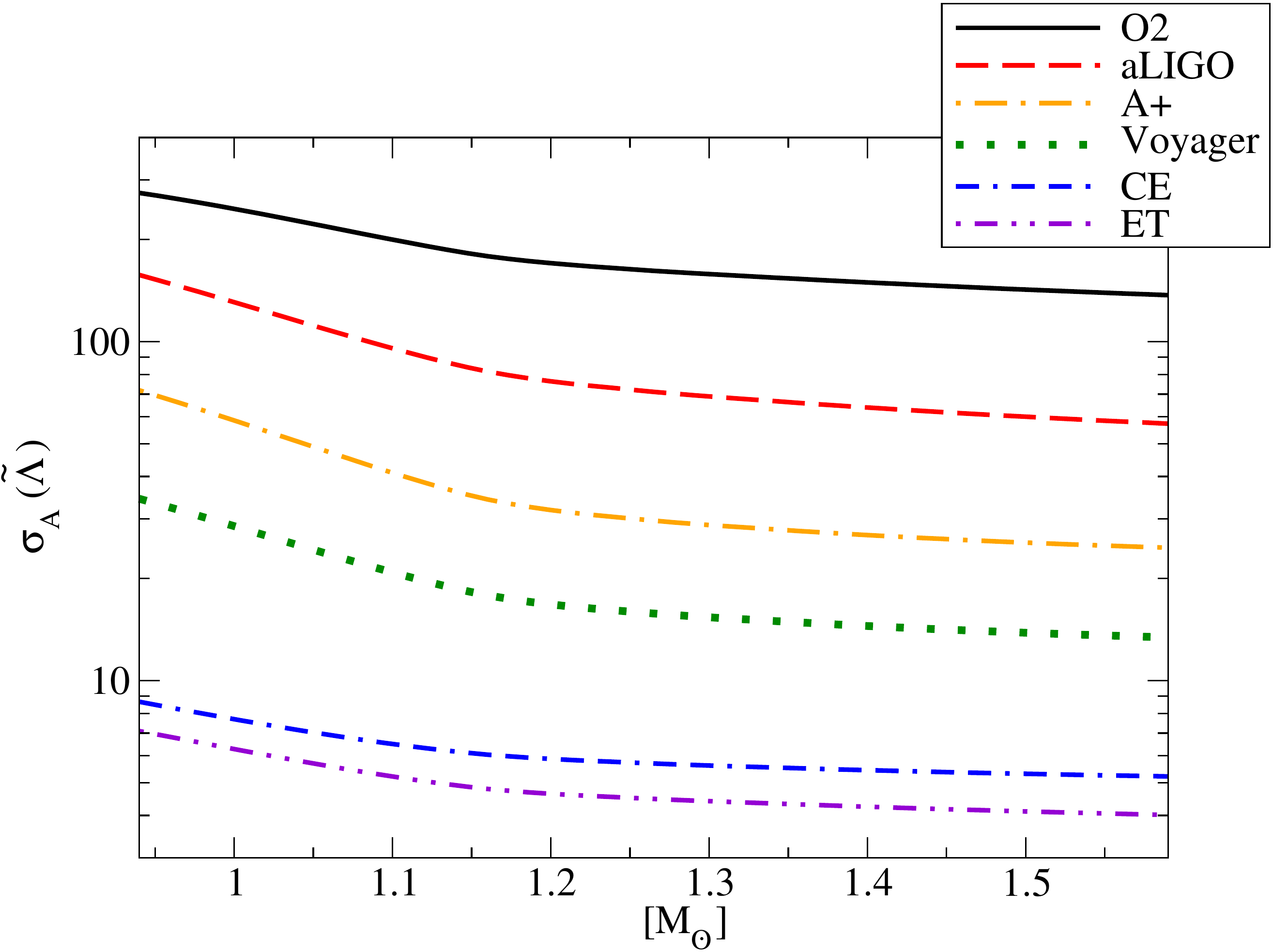}
\put(160,5){\small$\mathcal{M}$}
\end{overpic}
\end{center}
\caption[$\tilde\Lambda$ measurement accuracies]{
Approximate 90\% measurement accuracies of $\tilde\Lambda$ on detectors $A=($O2, aLIGO, A\texttt{+}, Voyager, CE, ET$)$ as a function of chirp mass, computed via simple Fisher analyses.
These correspond to the standard deviations $\sigma_A$ used in the generation of the approximated $\tilde\Lambda$ probability distributions needed to compute constraints on $K_{\text{sym},0}$.
}
\label{fig:sigmaLt}
\end{figure} 

Now we compute the posterior distributions on $K_{\text{sym},0}$ using the Gaussian prior distributions on $\tilde\Lambda$ computed above as a function of chirp mass, for future detectors.
Following the process used previously in in Sec.~\ref{sec:posteriors}, this is accomplished by first generating a two-dimensional Gaussian probability distribution between $K_{\text{sym},0}$ and $\tilde\Lambda$, taking into account the covariances between the two as
\begin{equation}\label{eq:2dPDF}
P(\tilde{\Lambda},K_{\text{sym},0})=\frac{1}{2\pi\sqrt{|\bm{\Sigma}|}}e^{-\frac{1}{2}(\bm{x}-\bm{\mu})^T\bm{\Sigma}^{-1}(\bm{x}-\bm{\mu})}.
\end{equation}
Here $\bm{x}$ and $\bm{\mu}$ are the 2D vectors containing $(\tilde\Lambda, K_{\text{sym},0})$ and their means respectively, and $\bm{\Sigma}$ is the covariance matrix with elements given by Eq.~\eqref{eq:covariance}.

Let us now offer readers the means to fully reproduce the results of the above analysis for any future event by constructing a fit for $\bm \mu$ and $\bm \Sigma$ in terms of chirp mass $\mathcal{M}$. Based on the relations between the former and the latter as shown in Fig.~\ref{fig:CovMusLt}, we create a fit in a logarithmic power expansion as
\begin{equation}
\log{y_i}=a_i+b_i \log{\mathcal{M}}+c_i (\log{\mathcal{M}})^2,
\end{equation}
with $y_i$ being the various parameters $\sqrt{\Sigma_{ab}}$ and $\mu_a$, and fitting coefficients $a_i$, $b_i$, and $c_i$ which are summarized in Table~\ref{tab:CovMusLt}.
Observe how well the fit agrees with the numerical data in Fig.~\ref{fig:CovMusLt}.

\begin{figure}
\begin{center} 
\begin{overpic}[width=0.7\columnwidth]{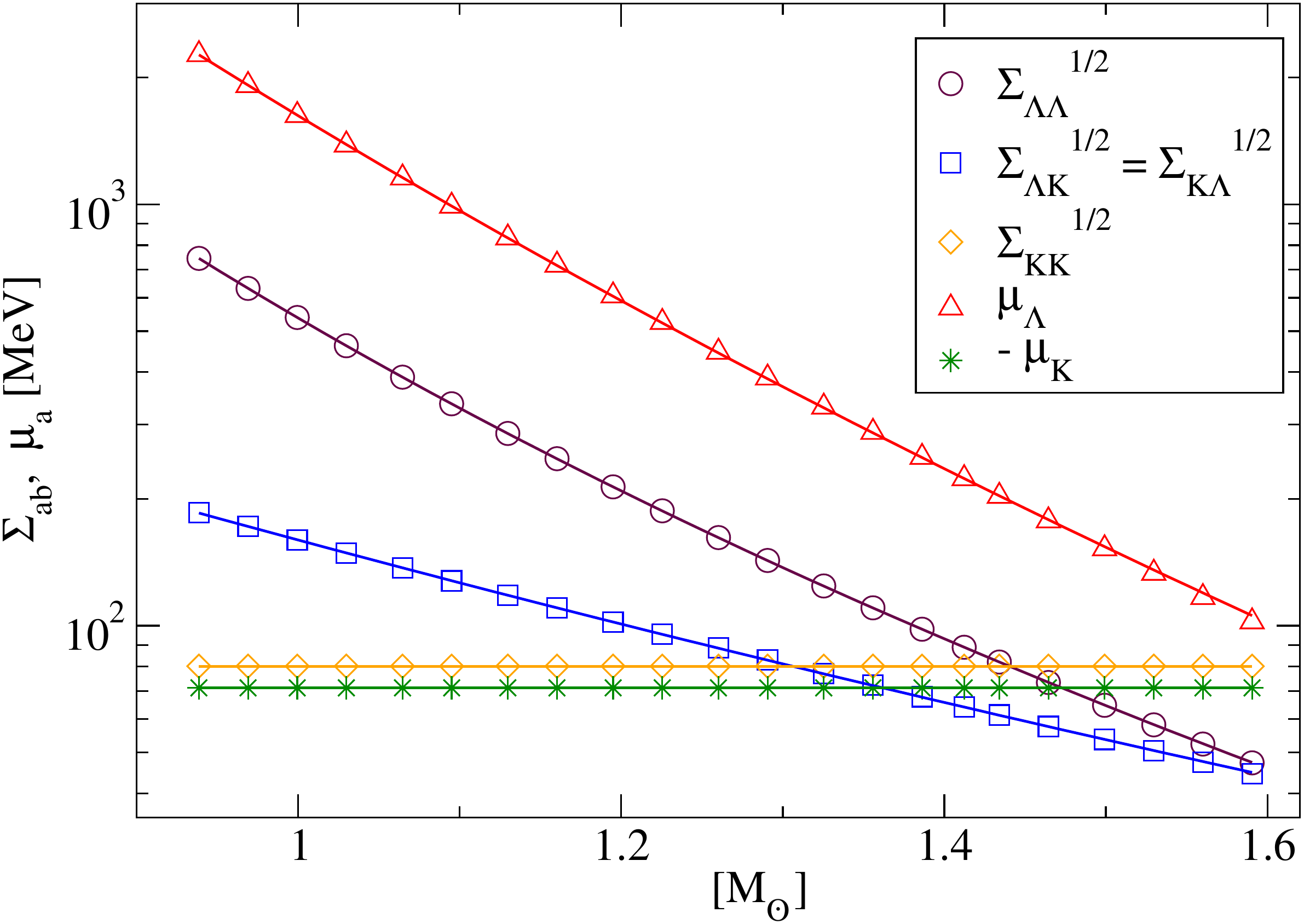}
\put(150,4){\small$\mathcal{M}$}
\end{overpic}
\end{center}
\caption[Covariance matrix elements and means for $\tilde\Lambda$ -- $K_{\text{sym},0}$ PDF]{
Values of $\Sigma_{ab}$ and $\mu_a$ in Eq.~\eqref{eq:2dPDF} along with their respective fits (tabulated in Table~\ref{tab:CovMusLt}), necessary for the full reconstruction of the two-dimensional probability distributions between $\tilde\Lambda$ and $K_{\text{sym},0}$.
}
\label{fig:CovMusLt}
\end{figure} 

\begin{table}
\centering
\begin{tabular}{c || c} 
Parameter & Fitting Function\\
\hline
$\sqrt{\Sigma_{\Lambda\Lambda}}(\mathcal{M})$ [MeV] & $\text{Exp}\lbrack 6.287 - 11.86 \log{\mathcal{M}}-0.9803\log^2\mathcal{M} \rbrack$\\
$\sqrt{\Sigma_{\Lambda\text{K}}}(\mathcal{M})$ [MeV] & $\text{Exp}\lbrack 5.073 - 5.477 \log{\mathcal{M}}-4.103\log^2\mathcal{M} \rbrack$\\
$\sqrt{\Sigma_{\text{KK}}}(\mathcal{M})$ [MeV] & $80.11$\\
$\mu_{\Lambda}(\mathcal{M})$ [MeV] & $\text{Exp}\lbrack 7.394 - 12.27 \log{\mathcal{M}}-6.399\log^2\mathcal{M} \rbrack$\\
$\mu_{\text{K}}(\mathcal{M})$ [MeV] & $-71.6$
\end{tabular}
\caption[Covariance matrix elements and means for $\tilde\Lambda$ -- $K_{\text{sym},0}$ PDF]{
Respective fitting functions for the covariance matrix $\bm \Sigma$ and the mean vector $\bm \mu$ in Eq.~\eqref{eq:2dPDF} necessary for the full reconstruction of the two-dimensional probability distributions between $\tilde\Lambda$ and $K_{\text{sym},0}$.
Here, the values of $\Sigma_{\text{KK}}$ and $\mu_{\text{K}}$ correspond to the variance and mean of $K_{\text{sym},0}$, which are independent of chirp mass, thus require no fitting function.
}\label{tab:CovMusLt} 
\end{table}

Constraints on $K_{\text{sym},0}$ are extracted by first computing the conditional probability distributions on $K_\text{sym,0}$ given a tidal deformability observation of $\tilde\Lambda_\text{obs}$.
By following Ref.~\cite{jensen_2007}, we can generate the one-dimensional conditional probability distribution on $K_\text{sym,0}$ by taking
\begin{equation}\label{eq:conditional}
 P(K_\text{sym,0}|\tilde\Lambda_\text{obs})\sim \mathcal{N}\Bigg(\mu_{K_\text{sym,0}}+\frac{\sigma_{K_\text{sym,0}}}{\sigma_{\tilde\Lambda_\text{obs}}}C(\tilde\Lambda_\text{obs}-\mu_{{\tilde\Lambda_\text{obs}}}),(1-C^2)\sigma_{K_\text{sym,0}}^2\Bigg).
\end{equation}
Above, $\mathcal{N}(\mu,\sigma^2)$ is the normal distribution with mean and variance $\mu$ and $\sigma^2$, while $\mu_\A$ and $\sigma_\A^2$ are the mean and variances of $K_\text{sym,0}$ and $\tilde\Lambda_\text{obs}$.
Finally, we can combine the one-dimensional conditional probability distribution function of Eq.~\eqref{eq:conditional} with the one-dimensional prior distribution on $\tilde\Lambda$ of Eq.~\eqref{eq:LtPrior}.
Marginalizing over $\tilde\Lambda$ results in a posterior probability distribution on $K_{\text{sym},0}$
\begin{equation}
\label{eq:P_Ksym}
P_A(K_{\text{sym},0})=\int\limits^{\infty}_{-\infty}P(K_{\text{sym},0}|\tilde{\Lambda})P_A(\tilde\Lambda) \, d\tilde\Lambda,
\end{equation}
from which 90\% confidence intervals on the curvature of symmetry energy can be extracted.
This process is then repeated for $22$ values of chirp mass $\mathcal{M}$ across its feasible range, and then for each interferometer $A$.
Appendix~\ref{app:posteriorExample} exemplifies this by demonstrating the procedure for one value of chirp mass $\mathcal{M}=1.188 \text{ M}_{\odot}$ on interferometer O2, corresponding to GW170817.
The results found there are compared to those found in Sec.~\ref{sec:current} from GW170817 in order to demonstrate the accuracy of our approximated Gaussian $\tilde\Lambda$ priors, rather than the full posterior distribution found in Ref.~\cite{LIGO:posterior}. We found that we slightly \textit{underestimate} the errors in $K_{\text{sym},0}$ by using this method.

There is one important question to analyze here: how do the statistical errors on $K_{\text{sym},0}$ ($\sigma_A$ in $P_A(\tilde \Lambda)$ given in Eq.~\eqref{eq:LtPrior} that enters in Eq.~\eqref{eq:P_Ksym}) compare to the systematic errors (covariance $\bm \Sigma$ in $P(\tilde{\Lambda},K_{\text{sym},0})$ given in Eq.~\eqref{eq:2dPDF} that also enters in Eq.~\eqref{eq:P_Ksym})?
As more events are observed and the detector sensitivities $S_n(f)$ drop, the statistical errors on the measurement of $K_{\text{sym},0}$ approach zero, and the overall errors limit closer to the systematic error ``wall" introduced from the EoS variation in the universal relations.
We study this effect by first plotting the overall errors on $K_{\text{sym},0}$ as a function of chirp mass, defined to be the one-sided 90\% confidence interval on the posterior distribution of $K_{\text{sym},0}$.
Following this, we define the systematic errors to be the one-sided 90\% confidence interval of $K_{\text{sym},0}$ in the two-dimensional probability distribution evaluated at the central value $\mu_{\tilde\Lambda} (\mathcal M)$ of the $\tilde\Lambda$ prior distribution shown in Fig.~\ref{fig:meanLt}.
Equivalently, the fixed diagonal $K_{\text{sym},0}$ coefficient of the Gaussian argument $\exp \lbrack -\Sigma^{-1}_{\text{KK}}(K_{\text{sym},0}-\langle K_{\text{sym},0} \rangle)^2/2 + \dotsc \rbrack$ shows the systematic errors to be exactly equal to $(\Sigma_{\text{KK}}^{-1})^{-1/2}$.

\hypertarget{link:OverallVsSystematic}{Figure~\ref{fig:OverallVsSystematic}} displays the results of the above described procedure; plotting the (one-sided 90\% confidence interval) overall and systematic errors on the measurement of $K_{\text{sym},0}$ as a function of chirp mass. 
We observe here the presence of a minimum in the uncertainties with respect to the chirp mass - a relic originating from the correlations between $K_{\text{sym},0}$ and $\tilde\Lambda$ seen in Fig.~\ref{fig:CorrOfM}, which similarly observe a maximum at the same chirp mass (and thus minimum EoS variation that generates systematic errors).
We do note, however, that while previous analyses by Refs.~\cite{Alam2016,Malik2018} required high correlations for the computation of constraints\footnote{Refs.~\cite{Alam2016,Malik2018} assumed the relationship between $\lambda_0$ (the tidal deformability at $1.4\text{ M}_{\odot}$) and nuclear parameters to lay exactly on the best-fit line between the two. Thus, high degrees of correlation were absolutely necessary for accuracy on this claim.}, our analysis does not, as all covariances between $\tilde\Lambda$ and $K_{\text{sym},0}$ are taken into account by the two-dimensional probability distribution of Eq.~\eqref{eq:2dPDF}.

Observe also how, as predicted, the statistical errors drop as the more sensitive detectors are analyzed, reducing to almost zero as the overall errors limit to the fixed systematic error ``wall".
The overall errors on the highly-sensitive third generation interferometers CE and ET are indistinguishable from the systematic errors -- indicating that the error budget is highly dominated by systematics at this point.
Once the errors are dominated by systematics, improving detector sensitivities or observing new events will not aid in the further constraint of $K_{\text{sym},0}$.
This indicates the urgent need to reduce the systematic errors found in the EoS-variation of the universal relations for the Voyager-class detectors and beyond\footnote{A similar conclusion is reached later in Chapter~\ref{chap:EoSInsensitiveRelations}, where the detector statistical errors became comparable to the systematic errors from the binary Love universal relations for future detectors Voyager and beyond.}.


\subsubsection{Multiple events}\label{sec:futureMultiple}

The future of GW astronomy will become quite busy in terms of detected events. 
For example, future GW interferometer Cosmic Explorer will be detecting anywhere from $3\times 10^5$ to $4\times 10^6$~\cite{Zack:URrelations} binary NS merger events within its horizon distance per year - a staggering number which will certainly help reduce the statistical errors on tidal measurements.
How does one account for this effect when studying the uncertainties in future, undetected, events?
The dominant tidal parameter in the gravitational waveform, $\tilde\Lambda$, depends strongly on the subsequent masses in the binary system, something difficult to predict beforehand.
Ultimately, this prevents one from combining the uncertainties on $\tilde\Lambda$ for multiple events.

Fortunately, this can be remedied by following in the footstpdf of Chapter~\ref{chap:EoSInsensitiveRelations}, where we reparameterized the gravitational waveform to instead consider the $\lambda_0$ and $\lambda_1$ tidal coefficients, generated by Taylor expanding the tidal deformability $\Lambda$ about the reference mass of $m_0=1.4 \text{ M}_{\odot}$~\cite{delPozzo:TaylorTidal,Yagi:binLove}:\footnote{We note here that a linear truncation of this Taylor series is valid for our purposes. By taking into account an additional quadratic term identified by $\lambda_2$, we found a reduction in measurement accuracy in $\lambda_0$ by only $\leq5\%$, across various detectors and values of chirp mass.}
\begin{equation}
\Lambda \approx \lambda_0+\lambda_1\left( 1-\frac{m}{m_0} \right).
\end{equation}
Here, $\lambda_0 \equiv \Lambda|_{1.4M_\odot}$ and $\lambda_1 \equiv -d\Lambda/d\ln m|_{1.4M_\odot}$ are the dimensionless tidal deformability and its slope at $1.4 \text{ M}_{\odot}$, and they do not depend on the individual NS masses $m$ or any combination thereof (however they do depend on the fiducial mass value $m_0$ chosen).
Therefore, they are identical for every future binary NS merger event, and may be combined in uncertainty.

Similar to the correlation between $\tilde \Lambda$ and $K_\mathrm{sym,0}$, we also find a correlation between $\lambda_0$ and $K_\mathrm{sym,0}$. We constructed a 2D Gaussian distribution $P(\lambda_0,K_{\text{sym},0})$ similar to $P(\tilde{\Lambda},K_{\text{sym},0})$ in Eq.~\eqref{eq:2dPDF} and find
\begin{equation}
 \Sigma^{1/2}_{ab} = 
\begin{pmatrix}
193.6 & 97.10 \\
97.10  & 80.11
\end{pmatrix} \ [\mathrm{MeV}]\,,
\quad
\bm \mu = \begin{pmatrix}
543.2 \\
 -71.164
\end{pmatrix} \ [\mathrm{MeV}]\,,
\end{equation}
for $\bm x = (\lambda_0, K_\mathrm{sym,0})$. 
Notice that both $\bm \Sigma$ and $\bm \mu$ are independent of $\mathcal M$ in this case.

\begin{figure}
\begin{center} 
\includegraphics[width=0.7\columnwidth]{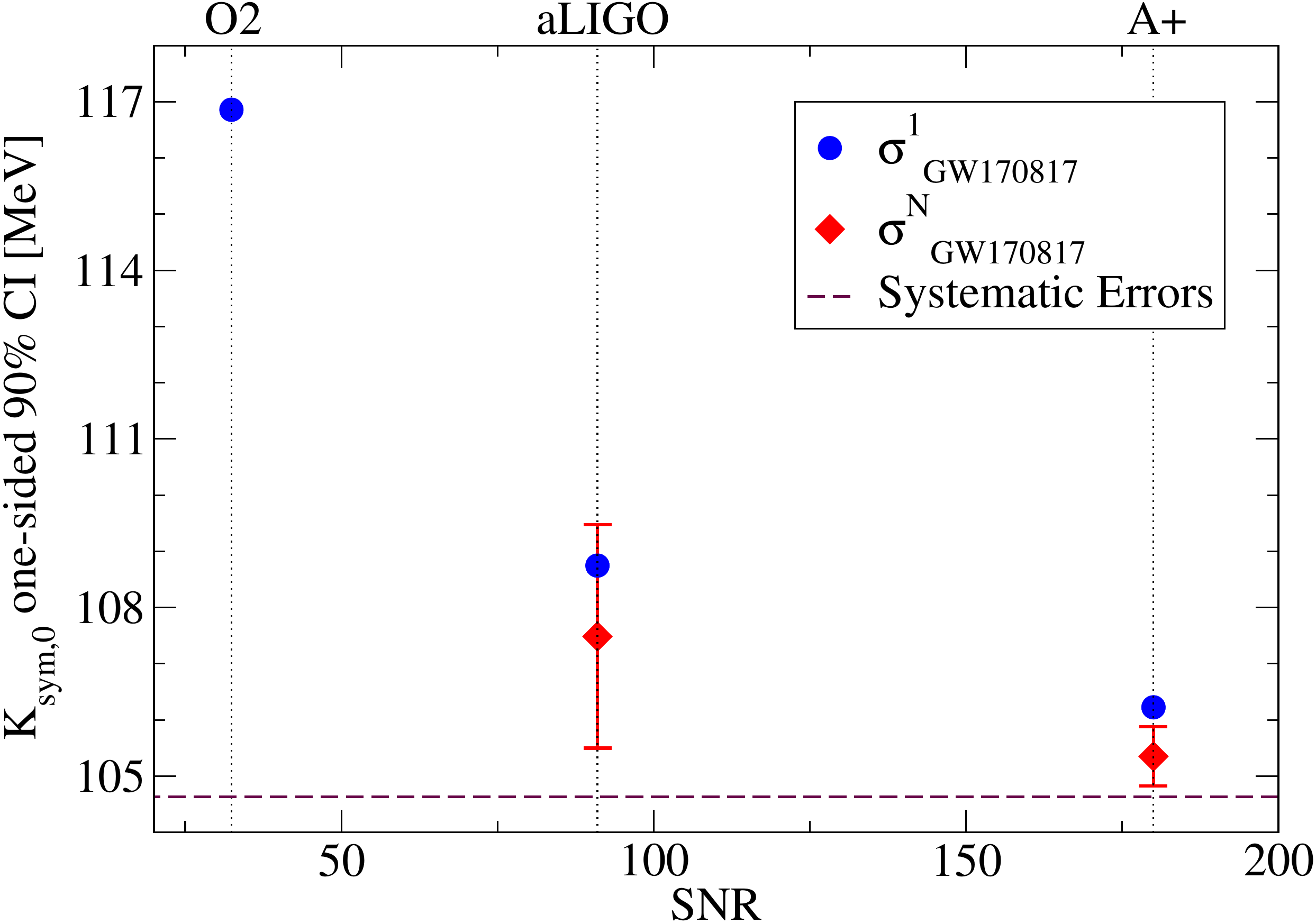}
\end{center}
\caption[Statistical+systematic uncertainties on $K_{\text{sym},0}$]{
The overall errors on $K_{\text{sym},0}$ using priors on the combined $\lambda_0$ (tidal deformability at $1.4\text{ M}_{\odot}$) uncertainty of multiple events (described in Chapter~\ref{chap:ParameterEstimation}), evaluated at the chirp mass $\mathcal{M}=1.188\text{ M}_{\odot}$.
The 90\% uncertainties on $K_{\text{sym},0}$ are shown as a function of the SNR of GW170817 as detected on each interferometer.
$\sigma^1_{\text{GW170817}}$ corresponds to the constraint formed with 1 GW170817-like observation, while $\sigma^N_{\text{GW170817}}$ forms the range bounded by the optimistic and pessimistic local binary NS coalescence rates.
While the single-event analysis of Fig.~\ref{fig:OverallVsSystematic} shows that single detections are nearly saturated by systematic uncertainties for Voyager-class detectors and beyond, here we show the effect stacking events can have on the aLIGO and A\texttt{+} analyses.
We observe that by combining multiple detections, even the aLIGO and A\texttt{+} interferometers approach the systematic error ``wall" (dashed horizontal line) with an optimistic number of detections.
}
\label{fig:stackedFisherNuc}
\end{figure} 

In this section, we repeat the analysis performed in Sec.~\ref{sec:futureSingle} using the combined uncertainties on $\lambda_0$ from $N_A$ unique events with chirp mass $1.188\text{ M}_\odot$, corresponding to the number of observed binary NS mergers within one observing year on detector $A$. We refer to Sec.~\ref{intro:Combined} for details on how to combine information from multiple events.
Fiducial values of $\lambda_0$ and $\lambda_1$ were computed to be the mean values of $\Lambda|_{1.4M_\odot}$ and $-d\Lambda/d\ln m|_{1.4M_\odot}$ from the ``LVC constrained" EoSs.
Figure~\ref{fig:stackedFisherNuc} shows how the combined-event uncertainties on $K_{\text{sym},0}$ for the fixed chirp mass of $1.188\text{ M}_\odot$ further become saturated on the aLIGO, and A\texttt{+} detectors as well.
As is shown in Fig.~\ref{fig:OverallVsSystematic}, the single-event uncertainties on $K_{\text{sym},0}$ become dominated by systematic errors for Voyager-class detectors and beyond, and thus there is not much point in stacking multiple events for these detectors to further reduce statistical errors on $\lambda_0$.

\section{Reducing systematic errors via multidimensional correlations}\label{sec:systematics}

Let us now consider how we can reduce the systematic ``walls'' present in Fig.~\ref{fig:OverallVsSystematic}.
In Sec.~\ref{sec:futureMultiple}, this was computed by evaluating the two-dimensional probability distribution between $K_{\text{sym},0}$ and $\lambda_0$ at the fiducial value of $\lambda_0$, and then finding the 90\% confidence interval of the resulting probability distribution of $K_{\text{sym},0}$ to yield $\sim104$ MeV.
We here construct multidimensional correlations among $K_\mathrm{sym,0}$ and $\Lambda_{m_x}$ at a few different masses $m_x$ (since we expect to detect GWs from binary NSs with different masses with future observations) to see how adding information of the tidal deformability at multiple different masses may help us to reduce the systematic errors on $K_\mathrm{sym,0}$.

Let us begin by using $\Lambda$ at two different masses $m_x$ and $m_y$. This requires us to find a three-dimensional correlation among $K_{\text{sym},0}$, $\Lambda_x (\equiv \Lambda_{m_x})$, and $\Lambda_y (\equiv \Lambda_{m_y})$, and construct a three-dimensional Gaussian distribution $P(K_{\text{sym},0}, \Lambda_x,\Lambda_y)$.
Figure~\ref{fig:3dPlot} shows an example of such a distribution for the case of $m_x=1.3\text{ M}_\odot$ and $m_y=1.6\text{ M}_\odot$.
The systematic error is then computed by evaluating the three-dimensional distribution at the fiducial values of $\Lambda_x$ and $\Lambda_y$, and then evaluating the resulting one-dimensional $K_{\text{sym},0}$ probability distribution at the 90\% confidence interval.

\begin{figure}
\begin{center} 
\includegraphics[width=0.8\columnwidth]{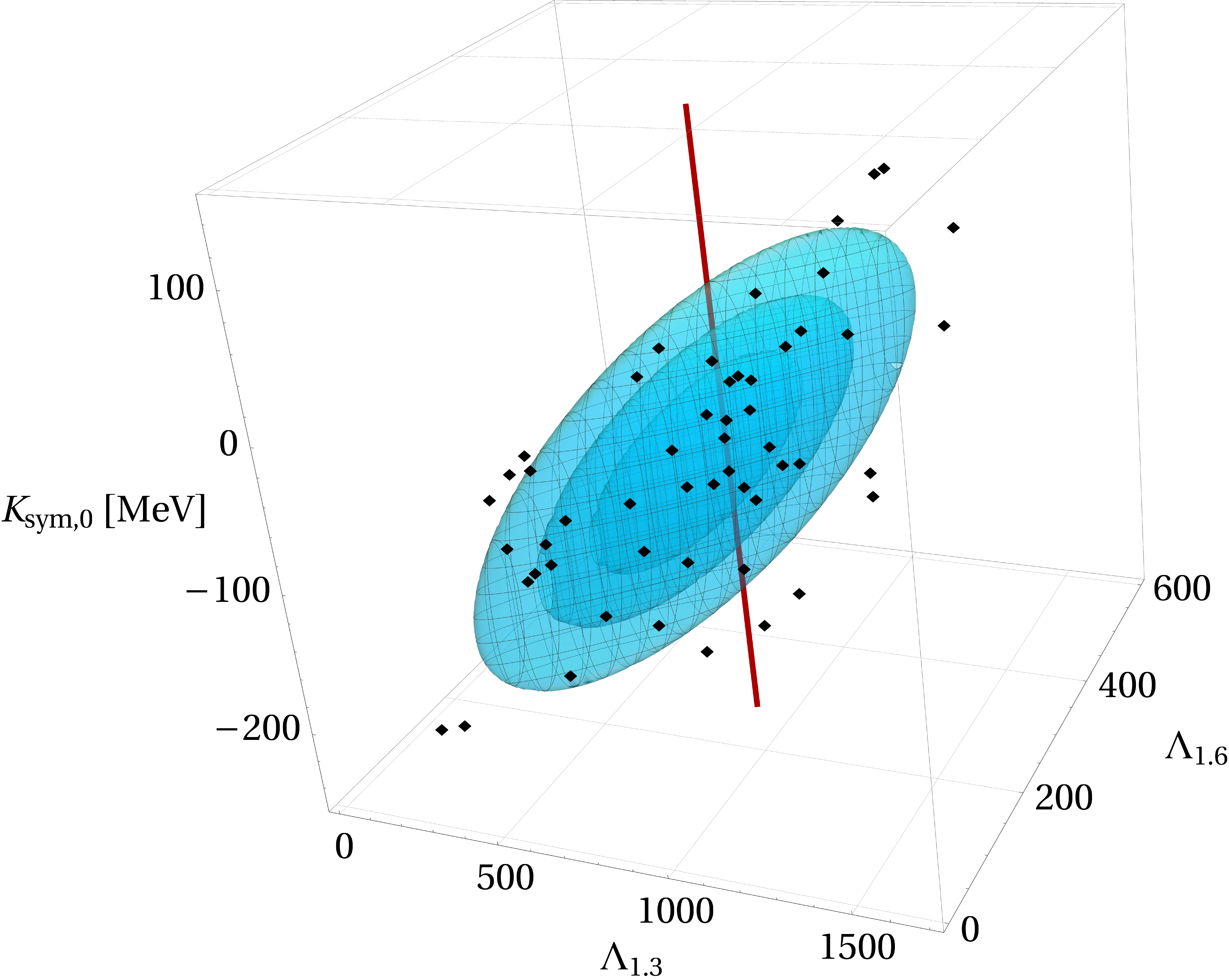}
\end{center}
\caption[Three-dimensional PDF between $\Lambda_{1.3}$, $\Lambda_{1.6}$, and $K_{\text{sym},0}$]{
Example three-dimensional probability distribution between $\Lambda_{1.3}$, $\Lambda_{1.6}$, and $K_{\text{sym},0}$ (blue density contour).
Each EoS yields one point (black dot) in this space.
The resulting systematic errors in $K_{\text{sym},0}$ are computed by evaluating the probability distribution at the fiducial values of $\Lambda_{1.3}=886.8$ and $\Lambda_{1.6}=269.4$ (maroon line), at the 90\% confidence level.
}
\label{fig:3dPlot}
\end{figure} 

Figure~\ref{fig:systematicContour} displays the resulting systematic uncertainties on $K_{\text{sym},0}$ using canonical masses $m_x$ and $m_y$ between $1\text{ M}_\odot$ and $2\text{ M}_\odot$.
Observe that the systematic errors can be reduced by setting both $m_x$ and $m_y$ to be large or small. However, this means that $m_x \approx m_y$, which corresponds to effectively using a two dimensional probability distribution. Thus, in this case, having additional information on $\Lambda$ at a different mass does not help to reduce the systematic errors.

\begin{figure}
\begin{center} 
\includegraphics[width=0.7\columnwidth]{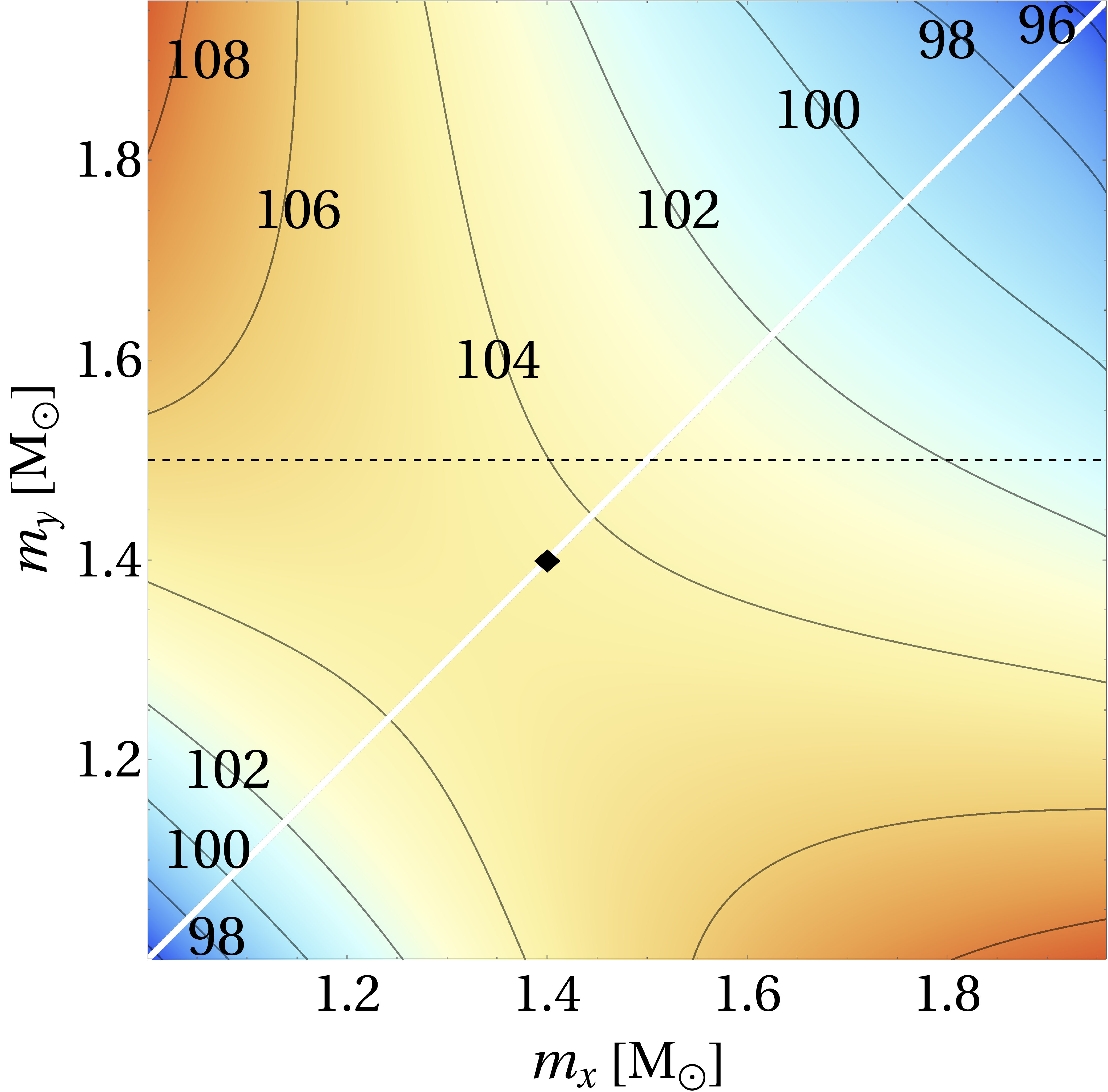}
\end{center}
\caption[Systematic uncertainty contour plot from three-dimensional PDF]{
Similar to Fig.~\ref{fig:systematicContourFixly}, but computed from the three-dimensional probability distribution between $K_{\text{sym},0}$, $\Lambda_x$, and $\Lambda_y$, and evaluated at the fiducial values of $\Lambda_x$ and $\Lambda_y$.
The white diagonal line at $m_x=m_y$ corresponds to the systematic errors obtained from the reduced two-dimensional probability distribution $P(K_{\text{sym},0},\Lambda_x)$.
In particular, the black diamond represents the systematic error obtained with such a function with $\lambda_0$ (the horizontal dashed line of Fig.~\ref{fig:stackedFisherNuc}). 
The systematic errors along the horizontal dashed line at $m_y = 1.5M_\odot$ corresponds to $P(K_{\text{sym},0},\Lambda_x,\Lambda_{1.5})$, which is equivalent to $P(K_{\text{sym},0},\Lambda_x,\Lambda_x,\Lambda_{1.5})$ along the diagonal line in Fig.~\ref{fig:systematicContourFixly}.
Observe that having the information of additional $\Lambda$ values does not help in this case, and what matters is to have $m_x$ and $m_y$ to be both small or large.
}
\label{fig:systematicContour}
\end{figure} 

It may sound strange that adding more pieces of information does not help to reduce the systematic errors. Let us explain why this is the case by comparing the systematic errors at $(m_x,m_y) = (1,2)M_\odot$ and $(m_x,m_y) = (2,2)M_\odot$.
Figure~\ref{fig:contourCases} compares the two-dimensional 90\% contours between $K_{\text{sym},0}$ and $\Lambda_{2.0}$ from two different methods by computing (i) directly the two-dimensional probability distribution from Eq.~\eqref{eq:2dPDF}, and (ii)  the three-dimensional probability distribution between $K_{\text{sym},0}$, $\Lambda_{2.0}$, and $\Lambda_{1.0}$, and then evaluating it at the fiducial value of $\Lambda_{1.0}$.
We observe that while the contour from the first case has a larger area (and value of $|\bm \Sigma|$) as expected due to the use of less information, it becomes distorted such that the systematic uncertainty (along the dashed horizontal line corresponding to the fiducial value of $\Lambda_{2.0}$) becomes smaller than that from the first case.

\begin{figure}
\begin{center} 
\includegraphics[width=0.7\columnwidth]{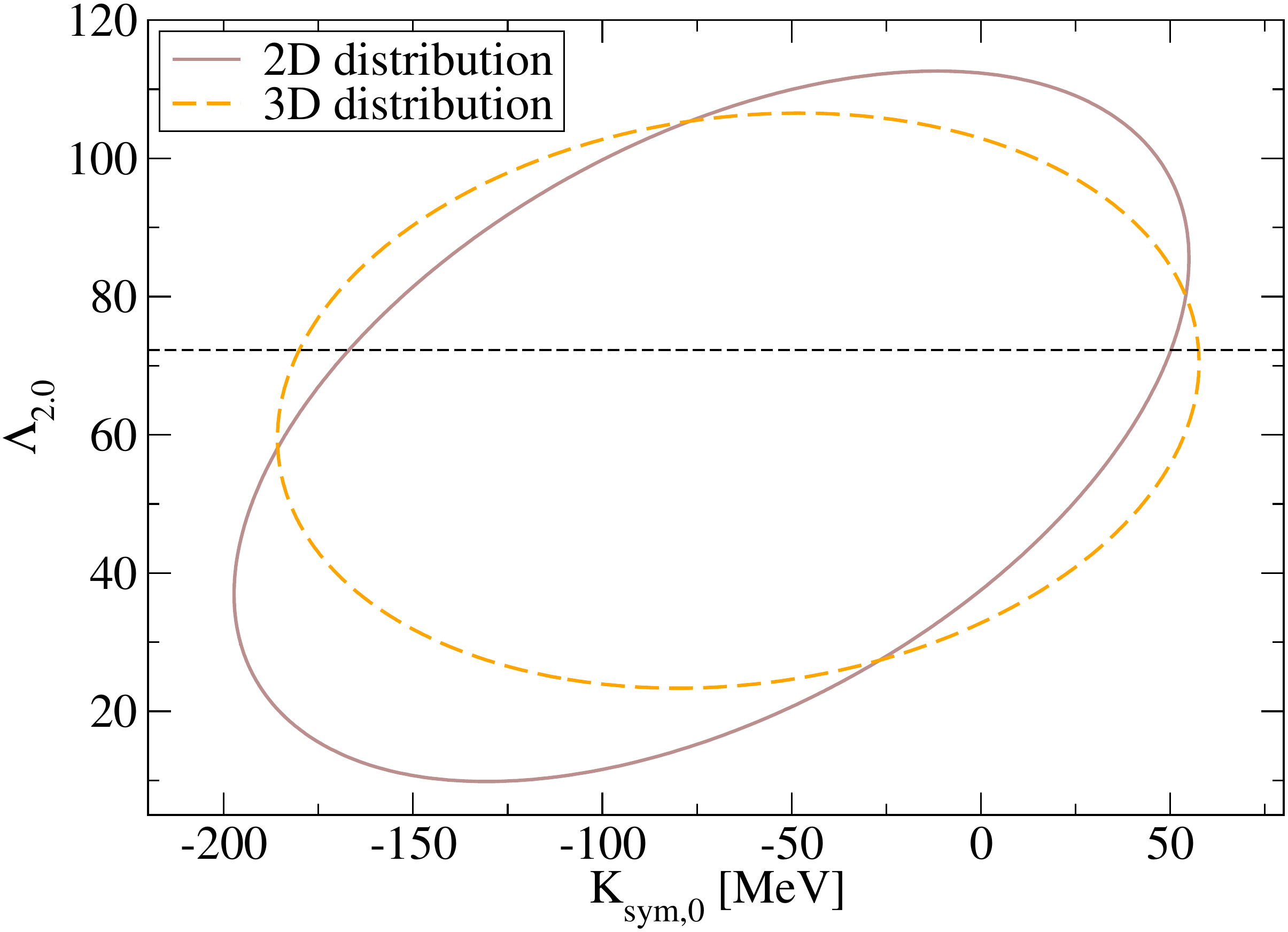}
\end{center}
\caption[$K_{\text{sym},0}$ -- $\Lambda_{2.0}$ PDF computed from two- and three-dimensional PDFs]{90\% confidence interval contours of the two-dimensional probability distribution between $K_{\text{sym},0}$ and $\Lambda_{2.0}$ computed using two different methods:
(i) (brown) the two-dimensional probability distribution between $K_{\text{sym},0}$ and $\Lambda_{2.0}$, and (ii) (dashed orange) the three-dimensional probability distribution between $K_{\text{sym},0}$, $\Lambda_{2.0}$, and $\Lambda_{1.0}$. (We evaluate the latter at the fiducial value of $\Lambda_{1.0}$.)
To compute the systematic errors in $K_{\text{sym},0}$, one would evaluate such contours at the fiducial value of $\Lambda_{2.0}$, denoted by the horizontal line, and finding the 90\% confidence interval of the resulting one-dimensional probability distribution in $K_{\text{sym},0}$.
Observe that although the area of the brown contour is larger than that of the orange, the systematic error on $K_{\text{sym},0}$ from the former is smaller than that of the latter.
}
\label{fig:contourCases}
\end{figure}

Let us now consider using $\Lambda$ at three different masses, $m_x$, $m_y$ and $m_z$. 
This requires us to find a four-dimensional correlation and construct the four-dimensional Gaussian probability distribution $P(K_{\text{sym},0},\Lambda_x,\Lambda_y,\Lambda_z)$.
We fix $m_z=1.5\text{ M}_\odot$, and allow $m_x$ and $m_y$ to vary between $\lbrack 1.0, 2.0 \rbrack\text{ M}_\odot$.
Similar to the process used previously, this probability distribution is evaluated at the fiducial values of $\Lambda_x$, $\Lambda_y$ and $\Lambda_{1.5}$: 
\begin{equation}
P'''(K_{\text{sym},0})=P(K_{\text{sym},0}, \bar{\Lambda}_x,\bar{\Lambda}_y,\bar{\Lambda}_{1.5}).
\end{equation}
The resulting 90\% confidence intervals are presented in Fig.~\ref{fig:systematicContourFixly} for the entire range of $m_x$ and $m_y$ mass values.
We observe that by including information about binaries with large, medium, and small masses together, the systematic errors can be improved drastically, down to $\sim74$ MeV. 
We also see that along the diagonal line of $m_x=m_y$, the four-dimensional probability distribution $P(K_{\text{sym},0},\Lambda_x,\Lambda_x,\Lambda_{1.5})$ reduces to the three-dimensional case $P(K_{\text{sym},0},\Lambda_x,\Lambda_{1.5})$, with uncertainties $\sim104$ MeV approaching that of Fig.~\ref{fig:systematicContour} along the horizontal dashed line. 

\begin{figure}
\begin{center} 
\includegraphics[width=0.7\columnwidth]{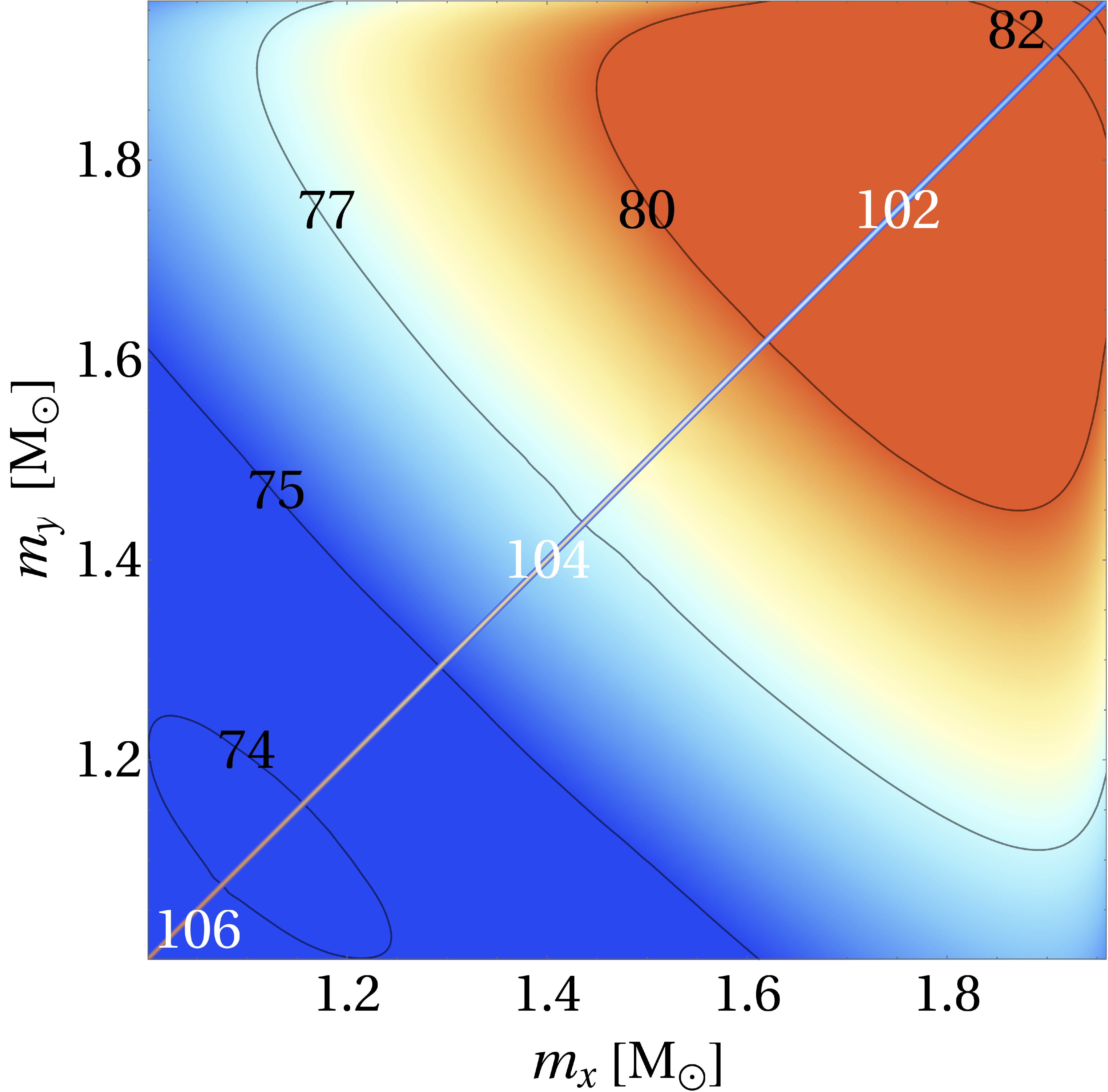}
\end{center}
\caption[$K_{\text{sym},0}$ systematic uncertainty contour plot]{
Contours displaying the systematic errors in $K_{\text{sym},0}$ [MeV] as a function of the masses $m_x$ and $m_y$ used to compute the four-dimensional probability distribution between $K_{\text{sym},0}$, $\Lambda_{1.5}$, $\Lambda_x$, and $\Lambda_y$.
The systematic errors are then computed by evaluating the probability distribution at the fiducial values of $\Lambda_{1.5}$, $\Lambda_x$, and $\Lambda_y$, and then taking the 90\% confidence interval of the resulting distribution in $K_{\text{sym},0}$.
Observe how a large reduction in systematic errors to $\sim74$ MeV can occur by including information about the tidal deformability at 3 different NS masses spread throughout their realistic range.
The diagonal contours at $m_x = m_y$ labeled in white correspond to the systematic errors obtained from the reduced three-dimensional probability distribution $P(K_{\text{sym},0},\Lambda_x,\Lambda_{1.5})$.
}\label{fig:systematicContourFixly}
\end{figure}


\section{Conclusions}\label{sec:conclusion}

The recent GW observation GW170817 coupled with the IR/UV/optical counterpart placed upper and lower bounds on the mass-weighted average tidal deformability $\tilde{\Lambda}$. 
We take advantage of this by selecting a diverse set of NS EoSs encompassing non-relativistic Skyrme-type interactions, RMF interactions and phenomenological variation of nuclear parameter models in order to constrain the nuclear matter parameters which are vital to limiting physically valid EoSs. 
We first found that approximate universal relations exist between linear combinations of nuclear parameters and $\tilde \Lambda$ for all values of mass ratio $q$ allowed from GW170817. 
We next constructed 2-dimensional probability distributions between $\tilde\Lambda$ and such nuclear parameters, converted them into one-dimensional conditional probability distributions on $K_\text{sym,0}$ given observations of $\tilde\Lambda$, and finally combined them with a posterior probability distribution on $\tilde\Lambda$ from LIGO and integrated them over $\tilde \Lambda$ in order to obtain posterior distributions on the nuclear parameters. 
From these posterior distributions, we derived 90\% confidence intervals on the incompressibility $K_0$, its slope $M_0$, and the curvature of symmetry energy $K_{\text{sym},0}$ at saturation density as \Krange{69}{352}, \Mrange{1371}{4808}, and \Ksymrange{-285}{7}.
The bounds on $M_0$ and $K_{\text{sym},0}$ are more conservative and safer to quote than those found in~\cite{Malik2018}. 
In addition, the constraints derived on $K_{\text{sym,0}}$ shows agreement with those in Refs.~\cite{Margueron:Ksym,Mondal:Ksym}.
We also note that bounds on $K_0$ and $M_0$ are less reliable than those on $K_\mathrm{sym,0}$ due to smaller correlations in the universal relations. 

Are there any other ways to further improve the constraints on $K_{\text{sym},0}$ using observed GW events?
One might think that the constraint on $\tilde\Lambda$ with GW170817 may help in this direction.
However, the restriction of data in \textit{only} the $\tilde\Lambda$ dimension does not help as systematic errors are found by evaluating the scattering width in the $K_{\text{sym},0}$ direction.
We continue this analysis into the future of GW astronomy.
In particular, we compute constraints on $K_{\text{sym},0}$ as a function of chirp mass when observed in multidude by future gravitational wave detectors.
We find such constraints to improve down to $\sim104$ MeV, when progress is halted by the presence of systematic mismodeling uncertainties.
Such systematic errors must be reduced for future progress to be made.

Finally, we briefly discuss the possibility of NSs with strong first-order phase transitions from hadronic to quark matter in the core, as described in Ref.~\cite{Paschalidis2018}.
With high enough observed chirp masses $\mathcal{M}$, future binary NS merger events could potentially be composed of one or both \textit{hybrid stars} (HSs) with quark-matter cores. 
The tidal deformabilities and thus, the nuclear parameters, depend on such structure, and could potentially disagree between events with varying chirp masses and combinations of NS/HS~\cite{Montana:2018bkb}.
Thus, significant variations between nuclear parameter measurements with future GW observations with varying chirp masses could potentially present evidence of strong phase transitions at around $2-3$ times the nuclear saturation density. 
If such transitions are present at sufficiently low-densities, then the nuclear matter parameters will be further decoupled from the GW observations and thus our lower limit for the nuclear matter parameter uncertainties will increase.
Alternatively, similar measurements of nuclear parameters could either indicate a pure hadronic matter EoS, or phase transitions occurring at higher nuclear densities.
The structure of such high-density transitions could be probed by the GW post-merger oscillation signal. 
The bounds derived in this chapter are only valid for NSs and may not be valid for hybrid stars (HSs) with quark core and nuclear matter envelope. 
We discuss this point in more detail in Appendix~\ref{app:hybrid}.

As we showed in the previous section, one can use multidimensional correlations to reduce the systematic errors. 
Instead of using tidal deformabilities from different NS masses obtained from GWs alone, one can consider combining information from multi-messenger observations.
For example, pulse profiling techniques by NASA's Neutron star Interior Composition Explorer (NICER) instrument may provide high-precision measurements on the NS radius down to $5\%$~\cite{NICER:nsradius,NICER:nsEoS}. 
Thus, one can construct multidimensional correlations among nuclear parameters, tidal deformabilities and radii of NSs. 
The work along this direction has been done in Ref.~\cite{Zimmerman:2020eho}.

\end{spacing}


\newpage
\chapter{Equation-of-state insensitive relations}\label{chap:EoSInsensitiveRelations}
\begin{adjustwidth}{\parindent}{\parindent}
\begin{spacing}{1.5}
The thermodynamic relation between pressure and density (i.e.~the equation of state) of cold supranuclear matter is critical in describing neutron stars, yet it remains one of the largest uncertainties in nuclear physics. 
The extraction of tidal deformabilities from the gravitational waves emitted in the coalescence of neutron star binaries, such as GW170817, is a promising tool to probe this thermodynamic relation.
Equation-of-state insensitive relations between symmetric and anti-symmetric combinations of individual tidal deformabilities, the so-called  ``binary Love relations", have proven important to infer the radius of neutron stars, and thus constrain the equation of state, from such gravitational waves. 
A similar set of relations between the moment of inertia, the tidal deformability, the quadrupole moment, and the compactness of neutron stars, the so-called ``I-Love-Q" and ``C-Love" relations, allow for future tests of general relativity in the extreme gravity regime. 
But even the most insensitive of such relations still presents some degree of equation-of-state variability that could introduce systematic uncertainties in parameter extraction and in model selection. 
We here reduce this variability by more than $50\%$ by imposing a prior on the allowed set of equations of state, derived from the posteriors generated from the analysis of GW170817.  
\end{spacing}
\end{adjustwidth}
\newpage


\section{Introduction}
\label{sec:intro}

The thermodynamic relation between pressure and density in cold, supranuclear matter, the EoS, remains a largely unsolved problem in nuclear physics\footnote{This chapter is based on the following paper: Z. Carson, K. Chatziioannou, C. Haster, K. Yagi, and N. Yunes \textit{Equation-of-state insensitive relations after GW170817}, Phys. Rev. D \textbf{99} 083016 (2019)}.
The EoS is critical to our understanding of NSs because it determines many NS observables, such as their mass, radius, moment of inertia ($I$), quadrupole moment ($Q$) and tidal deformability (or Love number). Unfortunately, terrestrial experiments can only probe the EoS to around nuclear saturation density ($\rho_0 \approx 2.5 \times 10^{14} \text{ g/cm}^3$)~\cite{Li:HeavyIon,Tsang:SymmetryEnergy,Centelles:NeutronSkin,Li:CrossSections,Chen:SymEnergy}. Although some temperature-dependent heavy-ion collision experiments can probe higher densities~\cite{Danielewicz:2002pu}, astrophysical observations of NSs remain ideal for constraining the EoS of cold and ultra dense, nuclear matter.

Independent measurements of NS observables can be used to constrain the nuclear EoS. For example, EM observations of the mass and radius of certain NSs have been used to place confidence limits in the mass-radius plane, and thus constrain the EoS~\cite{guver,ozel-baym-guver,steiner-lattimer-brown,Lattimer2014,Ozel:2016oaf}. These observations, however, may potentially suffer from large systematic errors~\cite{Miller:2016pom, Miller2013} due to uncertainties in the astrophysical modeling of X-ray bursts. The GWs emitted in the coalescence of NS binaries may be a cleaner probe of nuclear physics. During the early inspiral, the orbital separation is large enough that the tidal fields are negligible; but as the orbital separation decreases due to GW emission, tidal forces grow and the NSs respond by developing deformations determined by their nuclear EoS. 
These deformations source additional multipole radiation as well as affect the orbital trajectory of the binary, thus altering the GWs emitted, encoding within the latter the NS EoS~\cite{hinderer-love,Flanagan2008}.

The GWs emitted by binary NSs in the late inspiral must then depend on the \textit{tidal deformabilities} $\Lambda_1$ and $\Lambda_2$, which control the linear response of the star's quadrupole deformation to the (quadrupole) tidal field of the companion (to leading order in a post-Newtonian expansion~\cite{Blanchet:2013haa})~\cite{Flanagan2008,Vines:2011ud}. These parameters, however, enter the GW waveform model at the same post-Newtonian order, making them degenerate, and thus, very difficult to estimate independently with current GW data~\cite{Wade:tidalCorrections}. Instead, one can extract certain combinations of the tidal deformabilities, such as a certain mass-weighted tidal deformability $\tilde{\Lambda}$~\cite{Favata:2013rwa,Wade:tidalCorrections}, or one can extract the (mass-independent) coefficients $(\lambda_{0},\lambda_{1},\ldots)$ of a Taylor expansion of the tidal deformabilities about some fiducial mass~\cite{Messenger:2011gi,delPozzo:TaylorTidal,Yagi:binLove}. Current detectors are not sensitive enough to accurately measure any of these coefficients, but future detectors will, and the information from multiple events could then be combined, since the Taylor expansion coefficients should be common to all events. 

Lacking enough sensitivity in current GW observations, one is forced to only estimate the mass-weighted tidal deformability, but this prevents the independent extraction of $\Lambda_{1}$ and $\Lambda_{2}$. Yagi and Yunes~\cite{Yagi:2015pkc,Yagi:binLove} proposed a solution to this problem by finding ``approximately universal'' or ``EoS-insensitive'' relations between the symmetric and anti-symmetric combinations of the tidal deformabilities $\Lambda_{s,a}=\frac{1}{2}(\Lambda_1 \pm \Lambda_2)$, the so-called ``binary Love relations." These relations can be used to analytically express $\Lambda_{s}$ in terms of $\Lambda_{a}$ (or vice-versa), making the mass-weighted tidal deformability a function of only $\Lambda_{a}$. A measurement of the mass-weighted tidal deformability then implies a measurement of $\Lambda_{a}$, and through the use of the binary Love relations, also a measurement of $\Lambda_{s}$, which then allows for the inference of the individual tidal deformabilities $\Lambda_{1}$ and $\Lambda_{2}$~\cite{Yagi:2015pkc,Yagi:binLove}. With those at hand, one can further use EoS-insensitive relations between the tidal deformabilities and the compactness, the so-called ``C-Love relations''~\cite{Yagi:2013bca,Yagi:ILQ,Maselli:2013mva,Yagi:2016bkt}, to infer the radii of the NSs, and thus, to place two constraints in the mass-radius plane, one for each star in the binary. This idea was recently implemented for GW170817~\cite{TheLIGOScientific:2017qsa}, allowing EoS-independent constraints on the mass-radius curve using GW data~\cite{Katerina:residuals,LIGO:posterior}.     

EoS-insensitive relations can in fact be used for more than just measuring the nuclear EoS. For years, the theoretical physics community considered the possibility of using measurements of NS properties, such as the mass, the radius and the moment of inertia, to constrain deviations from General Relativity in the strong-field regime. Certain modified theories of gravity, such as scalar tensor theories with spontaneous scalarization~\cite{Damour:1996ke}, Einstein-\AE ther and Ho\v rava gravity~\cite{Eling:2007xh,Yagi:2013ava,Yagi:2013qpa}, dynamical Chern-Simons gravity~\cite{Yunes:2009ch,Yagi:2013mbt,Gupta:2017vsl}, beyond Horndesky theories~\cite{Babichev:2016jom,Sakstein:2016oel}, modify such NS observables, but unfortunately, these modifications are typically degenerate with the nuclear EoS~\cite{Pani:2014jra,Minamitsuji:2016hkk,Maselli:2016gxk}. Yagi and Yunes proposed to solve this problem by finding EoS-insensitive relations between the moment of inertia, the tidal deformability (or Love number) and the quadrupole moment, the so-called ``I-Love-Q'' relations~\cite{Yagi:2013bca,Yagi:ILQ}. Given a measurement of the Love number for a given NS, for example through GW observations, the I-Love-Q relations can be used to infer the moment of inertia or the quadrupole moment. A second independent measurement of either of these two quantities, for example through binary pulsar observations~\cite{Lattimer:2004nj} or observations with the Neutron star Interior Composition ExploreR (NICER)~\cite{Ozel:2015ykl}, then allows an EoS-insensitive test of General Relativity in the strong field regime~\cite{Yagi:2013bca,Yagi:ILQ,Gupta:2017vsl,Doneva:2017jop}. 

The implementation of the EoS-insensitive relations in data analysis has to somehow contend with the fact that these relations are in fact not exactly universal, but rather present different (albeit small) levels of EoS variability. In the case of the binary Love and C-Love relations to infer the radii of NSs with GW170817, the problem is solved by marginalizing over the EoS variability~\cite{Katerina:residuals}. Presumably, this same procedure can be applied in the future when carrying out tests of General Relativity with the I-Love-Q relations, using a combination of binary pulsar, NICER and GW data. But this marginalization procedure may not always be as important; as constraints in the mass-radius plane become more stringent with future GW observations, the allowed space of EoSs will shrink, which in turn must naturally decrease the degree of EoS variability in all EoS-insensitive relations. This is the main focus of this chapter.  


\section{Background and theory}\label{sec:eostheory}

In this section we review how the EoS can be represented analytically
through a spectral decomposition, and how the observation of GW170817
constrains the space of possible EoSs. We then proceed to discuss the EoS-insensitive relations. 
See Sec.~\ref{sec:tidal} from Chapter~\ref{chap:ConstrainingNuclearParameters} for details on the NS tidal deformability $\Lambda$, and how to determine it.

\subsection{Spectral representations of NS EoSs}
\label{sec:eos}

The structure of a NS and its tidal interactions in a binary system rely heavily on the underlying state function (or equation of state - EoS) describing the relationship between the pressure ($p$) and energy density ($\epsilon$) of nuclear matter.
Given that all currently proposed EoSs utilize certain approximations~\cite{Oertel:Review,Baym:Review}, one method to study a wide range of physically realizable EoSs is to parameterize them such that any realistic EoS can be represented with a small number of parameters.
Spectral representations~\cite{Lindblom:2010bb,Lindblom:2012zi,Lindblom:2013kra,Lindblom:2018rfr,Abbott:2018exr} parameterize EoSs by performing spectral expansions on the adiabatic index $\Gamma(p)$\footnote{Another way of parameterizing EoSs is through a piecewise polytropic formulation~\cite{Read2009,Lackey:2014fwa,Carney:2018sdv}.}:
\begin{equation}
\Gamma(x) = \exp{\sum_k^{N}\gamma_k x^k},
\end{equation}
where $x \equiv \log{(p/p_0)}$ for a minimum pressure $p_0$.
The EoS is then determined by an integration of the differential equation:
\begin{equation}
\frac{d \epsilon(p)}{dp}=\frac{\epsilon(p)+p}{p \Gamma(p)}.
\end{equation}
Using this formalism, any valid EoS can be approximated through the choice of $N$ spectral coefficients $\gamma_k$, and we here choose $N=4$, tabulated for several common EoSs in Table 1 of~\cite{Lindblom:2018rfr}.

\begin{figure*}
\begin{center} 
\includegraphics[width=.7\columnwidth]{EoSs.pdf}
\includegraphics[width=.7\columnwidth]{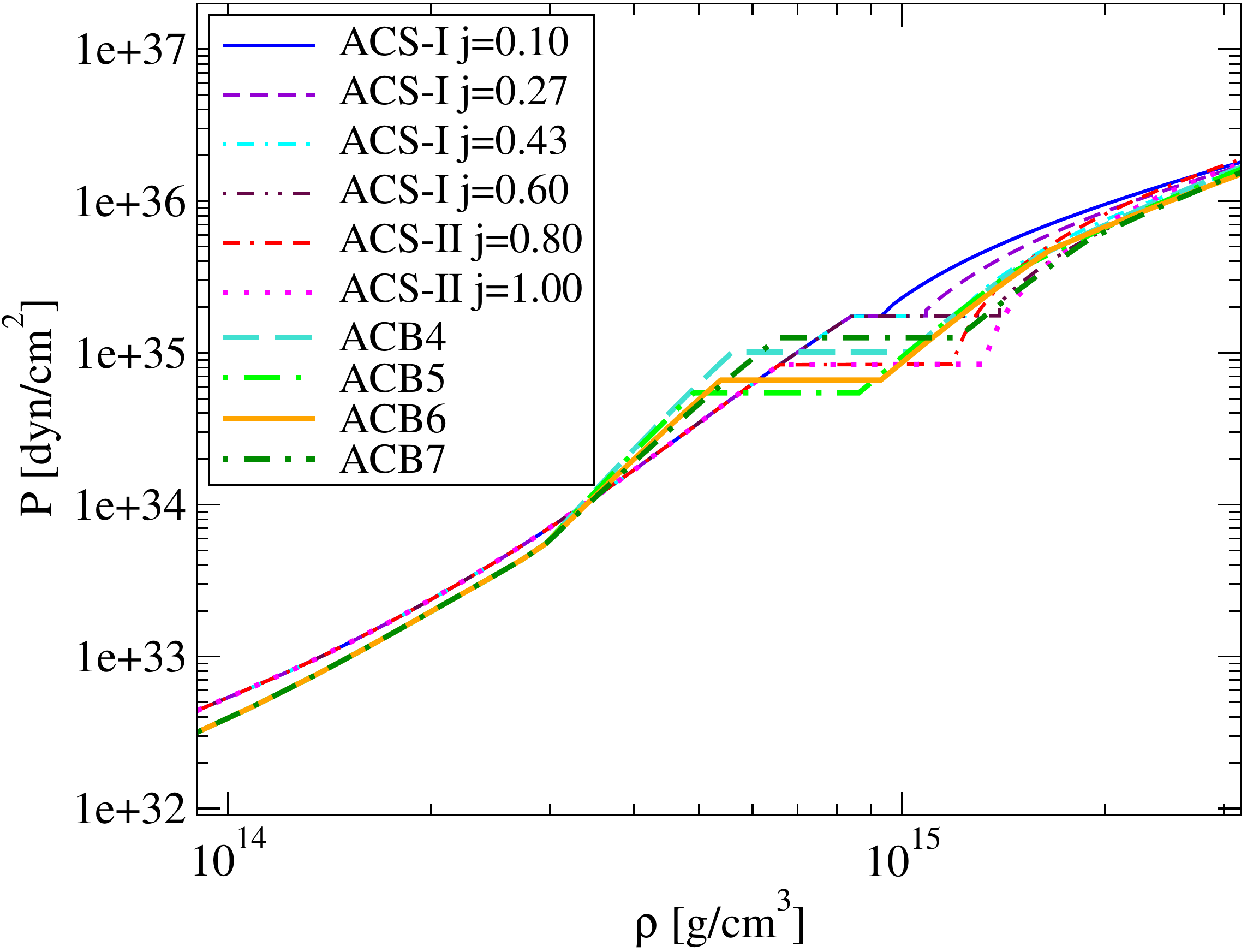}
\end{center}
\caption[Equations of state]{Top: small representative samples of the unconstrained (dotted) and constrained (solid) sets, together with the 90\% marginalized posterior distribution from the observation of GW170817 (cyan shaded region)~\cite{LIGO:posterior}. There is significantly less variability in the constrained set of EoSs due to the requirement that they be consistent with the GW170817 observation. Bottom: EoSs for ACS and ACB hybrid stars~\cite{Paschalidis2018}, each transitioning from a hadronic branch (corresponding to a pure hadronic-matter NS) into a quark-matter branch (quark-matter inner core surrounded by hadronic matter) at various transition pressures $P_{\text{tr}}$.
}
\label{fig:eos}
\end{figure*} 

We here wish to consider EoS-insensitive relations using two sets of EoSs: a ``constrained set'' consistent with the observation of GW170817 and an ``unconstrained set'' that does not impose this prior. In \textit{both} cases, we model the EoS with a piecewise function that equals the low-density crust EoS of SLy\footnote{SLy represents the unified equation of state based on the effective Skyrme Lyon nucleon-nucleon interactions, developed by Douchin and Haensel, 2001~\cite{Douchin:2001sv}.}~\cite{Douchin:2001sv} below half nuclear saturation density $\rho_{\text{stitch}}=1.3 \times 10^{14} \text{ g/cm}^3$~\cite{Read2009}, and equals the spectral decomposition described above outside the crust. For the latter, we restrict the spectral coefficients to the ranges $\gamma_0 \in \lbrack 0.2,2 \rbrack$, $\gamma_1 \in \lbrack -1.6,1.7 \rbrack$, $\gamma_2 \in \lbrack -0.6,0.6 \rbrack$, $\gamma_3 \in \lbrack -0.02,0.02 \rbrack$, and the adiabatic index is further restricted to $\Gamma \in \lbrack 0.6,4.5 \rbrack$~\cite{Lindblom:parameters}. Moreover, we impose the following two restrictions:  (i) causality within 10\%, i.e. that the speed of sound of the fluid be less than the speed of light to 10\% (following the analysis of Ref.~\cite{LIGO:posterior}), and (ii) a high maximum mass, i.e.~that the resulting EoS supports NSs with masses at least as high as $1.97 \text{ M}_{\odot}$, consistent with astrophysical observations~\cite{1.97NS,2.01NS,Zhao:massiveNS}. 

The unconstrained set is then defined by drawing random samples in the spectral coefficients within their allowed prior ranges, and then eliminating any EoS that either leads to an adiabatic index $\Gamma$ outside the allowed range, breaks the causality restriction, or breaks the maximum mass restriction. 
To obtain the constrained set of EoSs we analyze the publicly available data for GW170817 from the Gravitational Wave Open Science Center~\cite{GWOSC,Vallisneri:2014vxa}. 
We use the publicly available software library {\tt LALInference}~\cite{Veitch:2014wba,lalinference_o2} to sample the EoS posterior, similar to Ref~\cite{LIGO:posterior}.
Our analysis uses the same settings and prior choice as those of~\cite{LIGO:posterior,Carney:2018sdv}. 
In both cases, each set consists of 100 members, with a subset of these shown in the top panel of Fig.~\ref{fig:eos}.

In addition, we investigate 10 transitional quark-hadron matter stars, which undergo strong first-order phase transitions at a pressure $P_{\text{tr}}$, leading to the hadronic branch departing into a quark-matter branch at a given transitional mass~\cite{Paschalidis2018,Alford:2017qgh,1971SvA....15..347S,Zdunik:2012dj,Alford:2013aca}. In particular, we focus on the ACS and ACB models (corresponding to speed of sound, and piecewise polytropic representations of the EoS beyond the phase transition, respectively) described in~\cite{Paschalidis2018}, and also shown in the right panel of Fig.~\ref{fig:eos}. These result in two distinct types of NSs, based on their mass: (i) massive ($m \geq \text{ m}_{\text{tr}}$) hybrid stars which have quark-matter inner cores and nuclear matter elsewhere (henceforth, we denote such stars as hybrid stars (HSs)), and (ii) low-mass ($m \leq \text{ m}_{\text{tr}}$) hadronic stars with no internal transition to quark matter (henceforth, we denote these stars as simply NSs).

\subsection{EoS-insensitive relations: previous work}\label{sec:eosInsensitive}
Current GW interferometry is not yet sensitive enough to accurately extract both tidal parameters $\tilde{\Lambda}$ and $\delta\tilde{\Lambda}$.
In a search to remedy this, Yagi and Yunes~\cite{Yagi:binLove} found that symmetric and anti-symmetric combinations of the tidal deformabilities
\begin{equation}
\Lambda_s \equiv \frac{\Lambda_2 + \Lambda_1}{2}, \hspace{6mm} \Lambda_a \equiv \frac{\Lambda_2 - \Lambda_1}{2},
\end{equation}
display EoS-insensitive properties to a high degree, showing EoS variations of at most 20\% for binaries with masses less than $1.7 \text{ M}_{\odot}$ and using a representative sample of 11 EoSs. 
These ``binary Love relations" allow one to analytically break degeneracies between the tidal parameters: one can substitute $\Lambda_{a}=\Lambda_{a}(\Lambda_{s})$ in the GW model, thus completely eliminating $\Lambda_{a}$ from the parameter list (or vice-versa). This is important for two reasons: (i) the new model allows for the more accurate extraction of $\Lambda_{s}$ (or $\Lambda_{a}$ if $\Lambda_{s}$ is eliminated), and (ii) the relations allow for the inference of $\Lambda_{a}$ given a measurement of $\Lambda_{s}$ (or vice-versa), and from this for the inference of the individual tidal deformabilities $\Lambda_{1}$ and $\Lambda_{2}$. A simple Fisher analysis has shown that the binary Love relations improve parameter estimation of $\tilde{\Lambda}$ by up to an order of magnitude~\cite{Yagi:2015pkc,Yagi:binLove}.

Similar EoS-insensitive relations have been found between individual NS observables: the moment of inertia ($I$), the tidal deformability (Love), the quadrupole moment ($Q$), and the compactness ($C$), known as the ``I-Love-Q" and ``C-Love" relations~\cite{Yagi:2013bca,Yagi:ILQ, Maselli:2013mva}.
These relations are EoS-insensitive to better than 1\% and 6\% respectively, and they have important applications in both GW astrophysics~\cite{Kumar:2019xgp} and experimental relativity~\cite{Yagi:2013bca,Yagi:ILQ,Gupta:2017vsl,Doneva:2017jop}. For example, these relations and the measurement of the tidal deformabilities allow for the inference of several other stellar properties, such as the moment of inertia, the  compactness, the spin ($\chi$), and the radius of NSs~\cite{Kumar:2019xgp}.
Analyses such as that of~\cite{Kumar:2019xgp} could benefit from the improvement of such EoS-insensitive relations, which we derive in this chapter. 


\begin{table*}[htb]
\centering
\resizebox{\linewidth}{!}{%
\begin{tabular}{ c  c  | c c c c c c c c c c} 
 \hline
 \hline
 $y$ & $x$ & $\alpha$ & $K_{yx}$ & $a_1$ & $a_2$ & $a_3$ & $a_{4}$ & $a_{5}$ & $b_1$ & $b_2$ & $b_3$ \\
 \hline
  $\bar{I}$ & $\Lambda$ & (--) & (--) & $1.493$ & $0.06410$ & $0.02085$ & $-5.018 \times 10^{-4}$ & $3.16 \times 10^{-7}$ & (--) & (--) & (--)  \\
 $\bar{Q}$ & $\Lambda$ & (--) & (--) & $0.2093$ & $0.07404$ & $0.05382$ & $-5.018 \times 10^{-3}$ & $1.576 \times 10^{-4}$ & (--)  &(--)  & (--) \\ 
  $\bar{I}$ & $\bar{Q}$ & (--) & (--) & $1.383$ & $0.5931$ & $-0.02161$ & $0.04190$ & $-2.968 \times 10^{-3}$ &(--)  &(--) & (--) \\
 \hline 
 $\bar{I}$ & $\Lambda$ & $2/5$ & $0.5313$ & $1.287$ & $0.09888$ & $-2.300$ & (--) & (--) & $-1.347$ & $0.3857$ & $-0.02870$\\
 $\bar{Q}$ & $\Lambda$ & $1/5$ & $3.555$ & $-2.122$ & $2.72$ & $-1.491$ & (--) & (--) & $0.8644$ & $-0.1428$ & $-1.397$\\
 $\bar{I}$ & $\bar{Q}$ & $2$ & $0.008921$ & $10.59$ & $-37.46$ & $43.18$ &(--)  &(--)  & $-2.361$ & $1.967$ & $-0.5678$\\
 $C$ & $\Lambda$ & $-1/5$ & $0.2496$ & $-919.6$ & $330.3$ & $-857.2$ & (--) & (--) & $-383.5$ & $192.5$ & $-811.1$\\
\hline
\hline
\end{tabular}
}
\caption[I-Love-Q and C-Love relations]{
Fit parameters for the I-Love-Q and C-Love relations using the constrained set and the fitting functions in Eq.~\eqref{eq:ILQfit} (top) and in Eq.~\eqref{eq:ILQfitNew} (bottom).
}\label{tab:ILQfitNew}
\end{table*}

\section{EoS-insensitive relations}
\label{sec:universal}

In this section, we repeat and improve the analyses of~\cite{Yagi:binLove,Yagi:ILQ} through the use of the two sets of EoSs (the constrained and the unconstrained set) and through new fitting functions that properly limit to the Newtonian results.   


\subsection{I-Love-Q relations}
\label{sec:ilq}

Here we present our results on the I-Love-Q universality, comparing when possible to Fig.~1 of~\cite{Yagi:ILQ}.
In particular, we consider two distinct classes of NSs: nuclear matter EoSs and hybrid quark-hadron star EoSs as described in Sec.~\ref{sec:eostheory}.
We begin in Sec.~\ref{sec:ilq-nuc} by fitting the new I-Love-Q relations using the constrained set of EoSs.
This is followed in Sec.~\ref{sec:ilq-hyb} by an analysis and discussion of how well the hybrid star EoSs agree with the improved binary Love relations. 

\subsubsection{Nuclear matter stars}
\label{sec:ilq-nuc}

Following~\cite{Yagi:ILQ}, we first fit the data for each EoS-insensitive relation to the following function:
\begin{equation}\label{eq:ILQfit}
\ln{y}=a_{1}+a_{2} \ln{x} + a_{3} (\ln{x})^2 + a_{4} (\ln{x})^3 + a_{5} (\ln{x})^4,
\end{equation}
where $y$ and $x$ correspond to NS observables $\bar{I}$, $\bar{Q}$, and $\Lambda$, and the updated coefficients are given in the top of Table~\ref{tab:ILQfitNew}. This fitting function, however, does not limit properly to the Newtonian results given by~\cite{Yagi:ILQ}:
\begin{equation}\label{eq:Newtonian}
\bar{I}^{\text{N}} = K_{\bar{I}\Lambda}\Lambda^{2/5}, \hspace{3mm} \bar{Q}^{\text{N}} = K_{\bar{Q}\Lambda}\Lambda^{1/5}, \hspace{3mm} \bar{I}^{\text{N}} = K_{\bar{I}\bar{Q}}\bar{Q}^{2},
\end{equation}
where $\Lambda^{-1/5} \sim C$ when $C \ll 1$. One can thus improve the fitting function to 
\begin{equation}\label{eq:ILQfitNew}
y=K_{yx} x^{\alpha} \frac{1+\sum_{i=1}^3 a_i x^{-i/5}}{1+\sum_{i=1}^3 b_i x^{-i/5}},
\end{equation}
where $\alpha$ is either $2/5$, $1/5$, or $2$ for the $\bar{I}-\Lambda$, $\bar{Q}-\Lambda$, and $\bar{I}-\bar{Q}$ relations respectively, and where the new fitting coefficients are presented in the bottom of Table~\ref{tab:ILQfitNew}. While the two fits result in similar $R^2$ values\footnote{$R^2$ is the coefficient of determination, defined as $\sum_i(f_i-\bar{y})^2/\sum_i(y_i-\bar{y})^2$, where $\bar{y}$ is the mean data value, and $f_i$, $y_i$ are the modeled and actual data values} of $\sim 0.999995$, the fit in Eq.~\eqref{eq:ILQfitNew} has the advantage that it properly limits to the Newtonian result as $\Lambda \gg 1$~\cite{Yagi:binLove}.

Figure~\ref{fig:ILQ} shows the improved I-Love-Q relations between the dimensionless moment of inertia $\bar{I} \equiv I/m^3$, the dimensionless quadrupole moment $\bar{Q} \equiv Q/m^3$, and the dimensionless tidal deformability $\Lambda$. The fits are done using the new fitting function in Eq.~\eqref{eq:ILQfitNew} and over either the constrained set or the unconstrained set separately. The bottom panels show the relative fractional difference between the fit and the data for each set of EoSs. The new fits to the constrained set shows considerably more EoS-insensitivity than the fit to the unconstrained set. These results are summarized in Table~\ref{tab:maxVar}, which tabulates the maximum EoS variation in each fit. Clearly then, the EoS-insensitive relations can be made more universal by restricting the EoSs through the use of observations. Such improvements can be beneficial to future studies in GW astrophysics, such as those of~\cite{Kumar:2019xgp}, and in experimental relativity, such as those of~\cite{Gupta:2017vsl}. 

\begin{figure*}[!htbp]
\begin{center} 
\includegraphics[width=.45\textwidth]{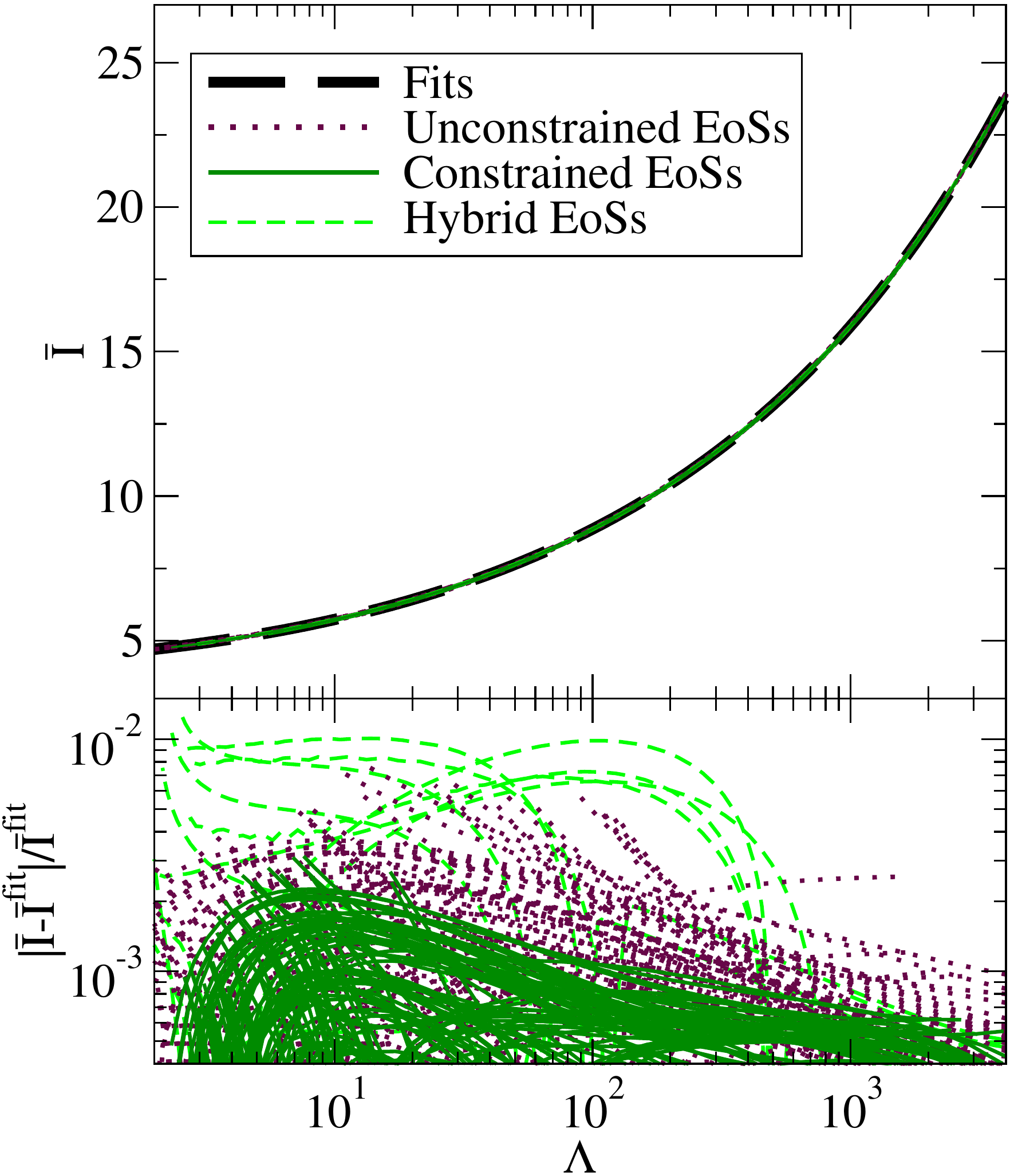}
\includegraphics[width=.45\textwidth]{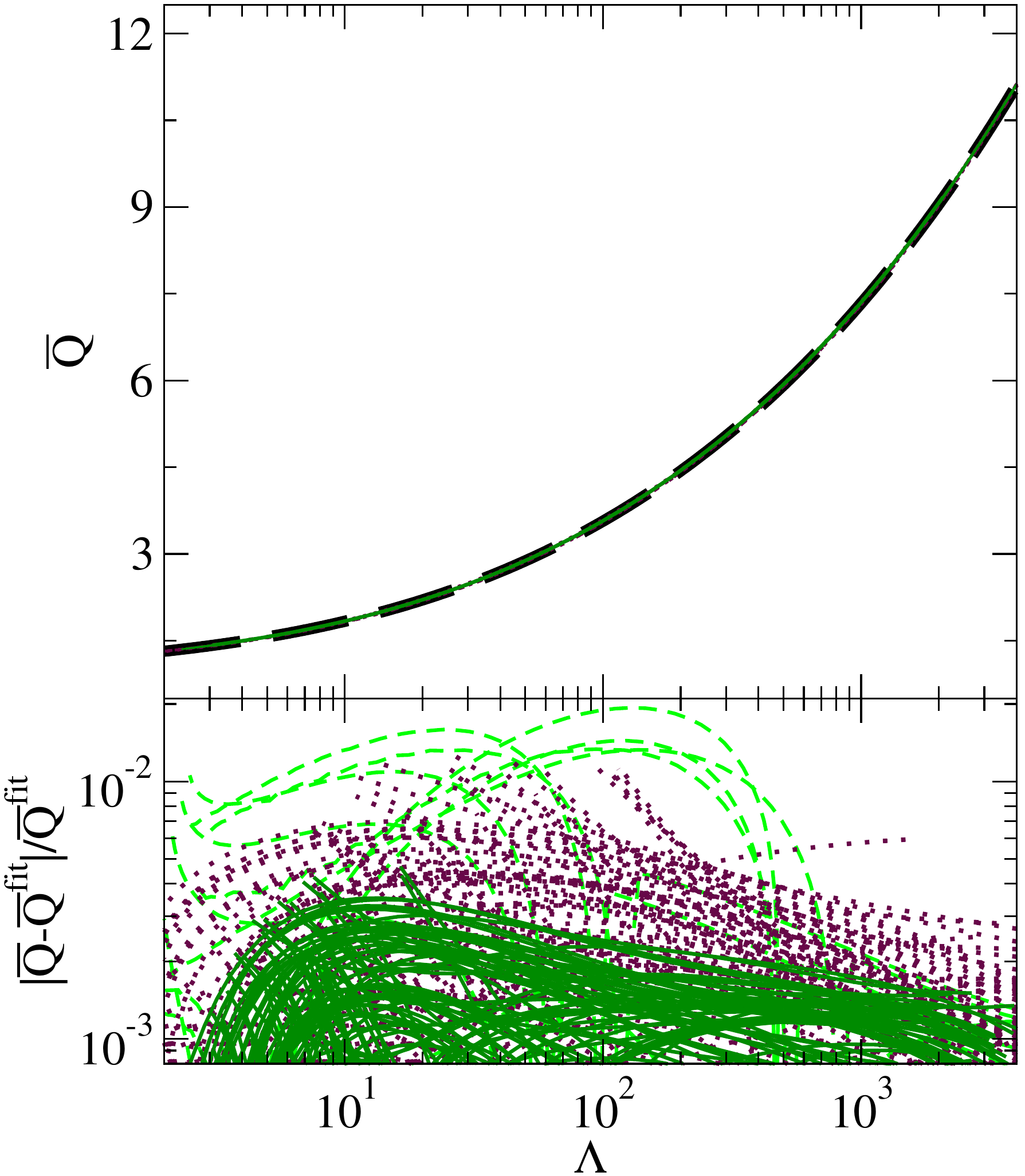}
\includegraphics[width=.45\textwidth]{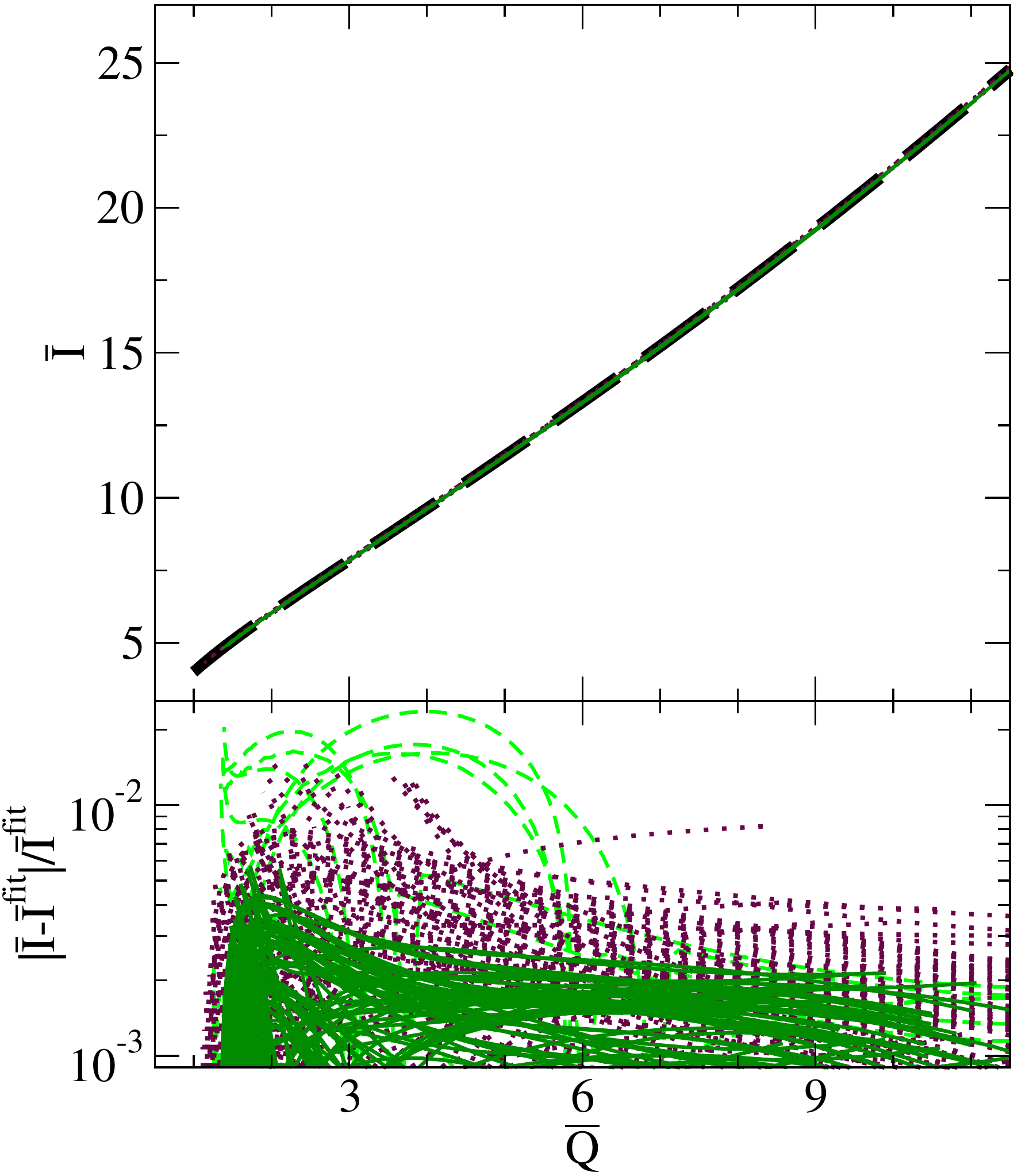}
\end{center}
\caption[I-Love-Q relations]{
Individual I-Love-Q relations $\bar{I}-\Lambda$ (left), $\bar{Q}-\Lambda$ (right), and $\bar{I}-\bar{Q}$ (center), shown for both the constrained EoSs (solid green) and unconstrained EoSs (dotted maroon).
In these figures, the black dashed lines correspond to the fits given by Eq.~\eqref{eq:ILQfitNew}.
The fractional difference from the fits, shown in the bottom panels, is greatly suppressed for the constrained case, compared to both the unconstrained case, and results from previous works~\cite{Yagi:2013bca,Yagi:ILQ}.
The maximal EoS variation from the fits for the unconstrained and constrained sets of EoSs are compared in Table~\ref{tab:maxVar}.
Additionally shown in this figure is the fractional difference from the nuclear matter fits for the 10 hybrid star EoSs (dashed green).
}
\label{fig:ILQ}
\end{figure*}

\begin{table}[htb]
\centering
\begin{tabular}{ c  || c c c } 
 \hline
 \hline
 \textbf{EoS-insensitive} & \multicolumn{3}{c}{\textbf{Maximal EoS Variability}} \\
 \cline{2-4}
 \textbf{Relation} & \multicolumn{1}{c|}{\textit{Previous}} & \multicolumn{1}{c|}{\textit{Unconstrained}} & \textit{Constrained}\\
 \hline
 $\bar{I}-\Lambda$ &  $0.0059$ & $0.0077$ & $0.0031$\\
 $\bar{Q}-\Lambda$ & $0.010$ & $0.013$ & $0.0047$\\
 $\bar{I}-\bar{Q}$ & $0.012$ & $0.015$ & $0.0057$\\
 \hline
 \multirow{2}{*}{$C-\Lambda$} & $0.065$ & $0.072$ & $0.022$\\
 & (--) & ($0.018$) & ($0.0066$)\\
  \hline
 \multirow{2}{*}{$R-\Lambda$} & -- & $0.056$ & $0.022$\\
 & (--) & ($880 \text{ m}$) & ($360 \text{ m}$) \\
 \hline
 $\Lambda_a-\Lambda_s$ & $\sim0.50$ & $0.57$ & $0.21$\\
 $q=0.90$ & (--) & ($190$) & ($37$) \\
 \cline{1-1}
 $\Lambda_a-\Lambda_s$ & $\sim0.20$ & $0.25$ & $0.083$\\
  $q=0.75$ & (--) & ($320$) & ($52$) \\
  \cline{1-1}
 $\Lambda_a-\Lambda_s$ & $\sim0.025$ & $0.038$ & $0.018$\\
  $q=0.50$ & (--) & ($240$) & ($29$) \\
  \cline{1-1}
\hline
\hline
\end{tabular}
\caption[Maximal EoS variation in the universal relations]{Maximum relative and fractional EoS variation in the I-Love-Q, C-Love, R-Love, and binary Love relations using the unconstrained set, the constrained set and variations reported in previous work~\cite{Yagi:ILQ,Yagi:binLove}. The maximum absolute EoS variation is also reported in the C-Love, R-Love and binary Love cases in parentheses. The maximum variation in the constrained set case is better than a factor of two smaller than the variability in the unconstrained case and in previous work. The maximum EoS variation in the unconstrained set is slightly larger than that found in previous work because the former is built from a large random sampling of EoSs. 
}
\label{tab:maxVar}
\end{table}

\subsubsection{Hybrid quark-hadron stars}
\label{sec:ilq-hyb}

Let us now focus on the I-Love-Q relations of hybrid stars, and their compatibility with their nuclear matter counterparts. For concreteness, we consider three different sets of EoSs in the hybrid case:
\begin{enumerate}
\item the complete set of 100 constrained EoSs combined with the 10 hybrid star EoSs,
\item the complete set of 100 constrained EoSs alone,
\item the complete set of 10 hybrid star EoSs alone.
\end{enumerate}
For each of these cases, we compute the I-Love-Q relations, we fit the data to Eq.~\eqref{eq:ILQfitNew} and we compute the relative fractional difference. 

\begin{table}[htb]
\centering
\begin{tabular}{ c  || c c } 
 \hline
 \hline
 \textbf{Fitting} & \multicolumn{2}{c}{\textbf{Maximal EoS Variability}} \\
 \cline{2-3}
 \textbf{Case} &  \multicolumn{1}{c|}{\textit{Constrained}} & \textit{Hybrid}\\
 \hline
 \textit{Combined} &  \multirow{2}{*}{$0.0044$} & \multirow{2}{*}{$0.014$}\\
 \textit{(Case 1)} & &\\
 \cline{1-1}
 \textit{Constrained only} & \multirow{2}{*}{$0.0031$} & \multirow{2}{*}{$0.017$}\\
  \textit{(Case 2)} & &\\
  \cline{1-1}
 \textit{Hybrid only} & \multirow{2}{*}{$0.0084$} & \multirow{2}{*}{$0.010$}\\
  \textit{(Case 3)} & &\\
  \cline{1-1}
\hline
\hline
\end{tabular}
\caption[I-Love maximal EoS variation with different EoS sets]{
Maximum relative and fractional EoS variation in the I-Love relation, fitting to three different sets of data: using the constrained set plus hybrid EoSs, using the constrained set alone, and using only the hybrid EoSs. In all 3 cases the hybrid EoSs are EoS-insensitive to $\sim1$\%, which is a slight decrease in universality relative to the hadronic only EoSs.
}\label{tab:hybridCompare}
\end{table}

The fractional differences of hybrid stars from the fit for the second case (fit to only the constrained EoSs) is shown with dashed green lines in Fig.~\ref{fig:ILQ}, while the maximum EoS variation for the I-Love relation is shown in Table~\ref{tab:hybridCompare} for both the constrained and hybrid star cases. The hybrid star EoSs obey the I-Love-Q relations in each case to better than $\sim1.7$\%, a variability that is slightly higher than that found when using only nuclear matter EoSs in previous works~\cite{Yagi:2013bca,Yagi:ILQ}, and is consistent with~\cite{Paschalidis2018}. The universality cannot be improved much through the introduction of new fits, only bringing the maximal EoS variation down to $\sim1$\% for the fits constructed with only hybrid star EoSs. From this study, we conclude that hybrid star EoSs \textit{do} obey the traditional nuclear I-Love-Q relations computed with nuclear EoS data, albeit with a slight decrease in universality to $\sim1.7$\%.  
We find this decrease in universality to be somewhat consistent with similar works~\cite{Paschalidis2018,Bandyopadhyay2018,PhysRevD.95.101302,Han:2018mtj}.
In such investigations, departures from universality of up to $\sim1\%-2\%$ were found with various hybrid star EoSs, compatible with our result of $\sim1.7$\%.


\begin{figure*}
\begin{center} 
\includegraphics[width=.49\columnwidth]{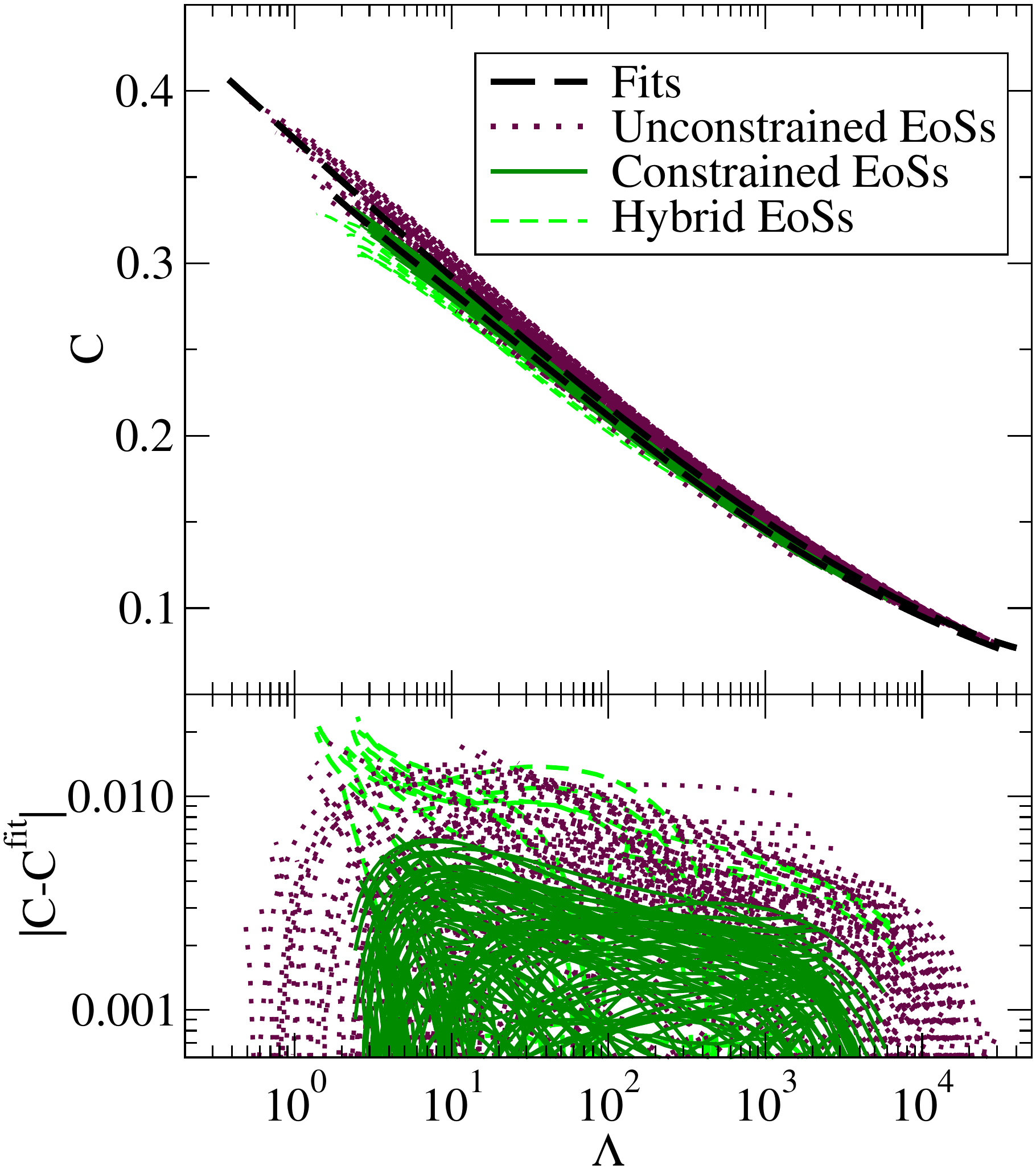}
\includegraphics[width=.47\columnwidth]{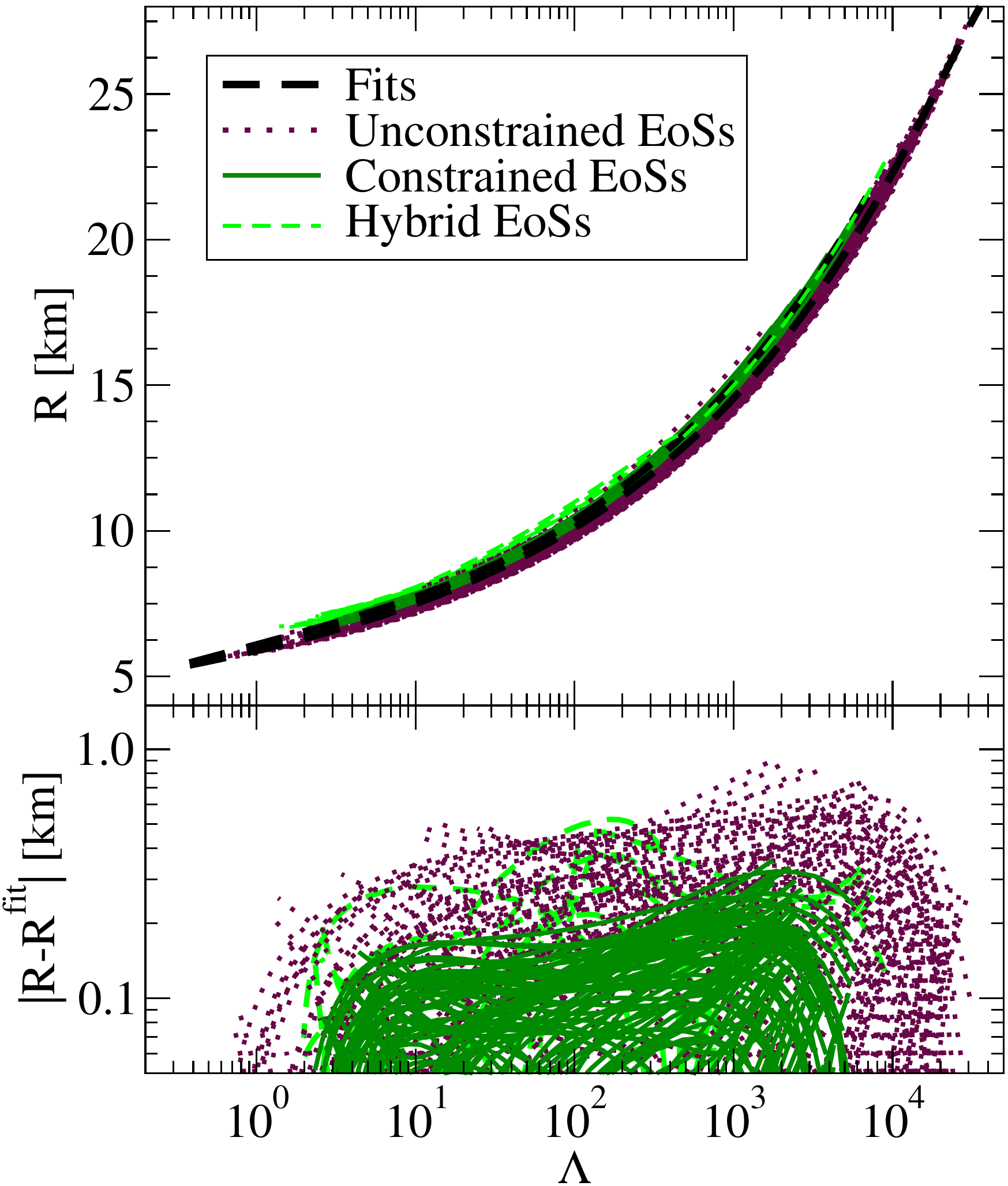}
\end{center}
\caption[C-Love and R-Love universal relations]{
Similar to Fig.~\ref{fig:ILQ} but for the C-Love (left) and the R-Love relations (right). In the top panels, we show two different fits, one for the unconstrained set and one for the constrained set of EoSs. The bottom panels show the \textit{absolute differences} (rather than fractional difference as in Fig.~\ref{fig:ILQ}) from the fit. The absolute difference is suppressed in the constrained set case relative to the unconstrained set, and results from previous work~\cite{Maselli:2013mva}. In the left panel, we also present the corresponding relations in the hybrid star cases (dashed green) for comparison, where, although still EoS-insensitive, the degree of universality decreases.
}
\label{fig:clove}
\end{figure*} 

\subsection{C-Love relations}
\label{sec:clove}

In this subsection we focus on the EoS-insensitive C-Love relations, as introduced in~\cite{Yagi:2013bca,Yagi:ILQ,Maselli:2013mva}, as well as on the R-Love relations, since these play a key role in GW data analysis for the extraction of M-R credible intervals. As in the previous subsection, we consider both nuclear matter EoSs, as well as hybrid quark-hadron star EoSs.

\subsubsection{Nuclear matter stars}
\label{sec:clove-nuc}

Following Ref.~\cite{Maselli:2013mva}, we begin by fitting the data for each set of EoSs to the simple curve
\begin{equation}
C = \sum^2_{k=0} a_k (\ln{\Lambda})^k.
\end{equation}
Doing so, yields $a_0 = 0.3617$, $a_1 = -0.03548$, and $a_2 = 0.0006194$ for the constrained set of EoSs, similar to what was found in~\cite{Maselli:2013mva}. As in the I-Love-Q case, however, the above fitting function does not limit properly to the Newtonian result 
\begin{equation}\label{eq:cloveFit}
C^N=K_{C\Lambda}\Lambda^{-1/5}.
\end{equation}
We thus repeat the fit using Eq.~\eqref{eq:ILQfitNew} as the fitting function, with the new fitting coefficients presented in Table~\ref{tab:ILQfitNew}.

The left panel of Fig.~\ref{fig:clove} shows the C-Love relations for both the constrained and unconstrained sets, along with the corresponding \textit{absolute differences}\footnote{We present absolute differences instead of fractional differences since the former is what matters directly to the GW data analysis.} from the fits (instead of the fractional differences as done back in Fig.~\ref{fig:ILQ}). The fit to the constrained EoSs suppresses the EoS variability compared to the fit to the unconstrained set, as well as that of previous work~\cite{Maselli:2013mva}. The maximal EoS variation is compared between these three cases in Table~\ref{tab:maxVar}.

From the C-Love relations, we can also compute directly the R-Love relation using $R(\Lambda)=m/C(\Lambda)$ for NSs of mass $m$. The right panel of Fig.~\ref{fig:clove} shows the R-Love relations for the constrained and unconstrained sets, with the bottom panel showing the absolute difference of the data and the fits. The C-Love relations allow us to infer the NS radius to better than $\sim 350 \textrm{m} $ in the constrained case, while the error goes up to $1,000 \textrm{m}$ in the unconstrained case, as also tabulated in Table~\ref{tab:maxVar}. The systematic uncertainty in the radius using the constrained fit is thus comparable to the $\sim 140\textrm{m}$ systematic uncertainty in the radius due to the choice of EoS to model the crust~\cite{Gamba:2019kwu}. 
See also Refs.~\cite{Annala:2017llu,Raithel:2018ncd} for related work on the R-Love relations.

\subsubsection{Hybrid quark-hadron stars}
\label{sec:clove-hyb}

Let us now focus on whether the C-Love relations hold for hybrid stars. Much like in Sec.~\ref{sec:ilq-hyb}, we perform 3 separate fits: one to the constrained set plus 10 hybrid EoSs, another to the constrained set only, and a third to the 10 hybrid EoSs only. We then compare the EoS insensitivity in each case.

In the top left panel of Fig.~\ref{fig:clove}, we show the C-Love relations for hybrid stars (dashed green curves), while in the bottom panel, we show the absolute difference of the relations for such stars from the fit constructed only from the constrained set of EoSs (dashed green curves). The C-Love relations for hybrid stars remain EoS-insensitive, but the degree of universality is not as high as in the case of hadronic NSs. Table~\ref{tab:hybridCompareClove} compares the maximal EoS variability for the constrained and hybrid star EoSs fitting to the three data sets described above. As in Sec.~\ref{sec:ilq-hyb}, the maximal EoS variation for hybrid stars fluctuates only slightly ($\sim 4.5\% - 7\%$) in each case. From this, we conclude that hybrid stars do obey the C-Love relation derived for nuclear matter stars, with the caveat that the maximum universality increases to $\sim 7.1\%$.
For completeness, we also show the R-Love relations for hybrid stars in Fig.~\ref{fig:clove}.
The absolute differences from the (constrained EoS) fit are similarly displayed in the bottom panel, showing a maximum EoS variability of $\sim50\text{ m}$, consistent with the uncertainties found in the unconstrained relations.

\begin{table}
\centering
\begin{tabular}{ c  || c c } 
 \hline
 \hline
 \textbf{Fitting} & \multicolumn{2}{c}{\textbf{Maximal EoS Variability}} \\
 \cline{2-3}
 \textbf{Case} &  \multicolumn{1}{c|}{\textit{Constrained}} & \textit{Hybrid}\\
 \hline
 \textit{Combined} &  \multirow{2}{*}{$0.037$} & \multirow{2}{*}{$0.055$}\\
 \textit{(Case 1)} & &\\
 \cline{1-1}
 \textit{Constrained only} & \multirow{2}{*}{$0.022$} & \multirow{2}{*}{$0.072$}\\
  \textit{(Case 2)} & &\\
  \cline{1-1}
 \textit{Hybrid only} & \multirow{2}{*}{$0.058$} & \multirow{2}{*}{$0.045$}\\
  \textit{(Case 3)} & &\\
  \cline{1-1}
\hline
\hline
\end{tabular}
\caption[C-Love maximal EoS variation with different EoS sets]{
Similar to Table~\ref{tab:hybridCompare} but for the C-Love relation.
For all 3 cases, the hybrid EoSs are only universal up to a minimum of $\sim1$\% (fractional difference from the fit), and the constrained EoSs typically outperform the hybrid ones (other than the third case where the hybrid stars show slightly better agreement to the fit). The second case is also shown in Fig.~\ref{fig:clove}.
}\label{tab:hybridCompareClove}
\end{table}


\subsection{Binary love relations}
\label{sec:binary}

\begin{figure}[!htbp]
\begin{center} 
\includegraphics[width=0.5\columnwidth]{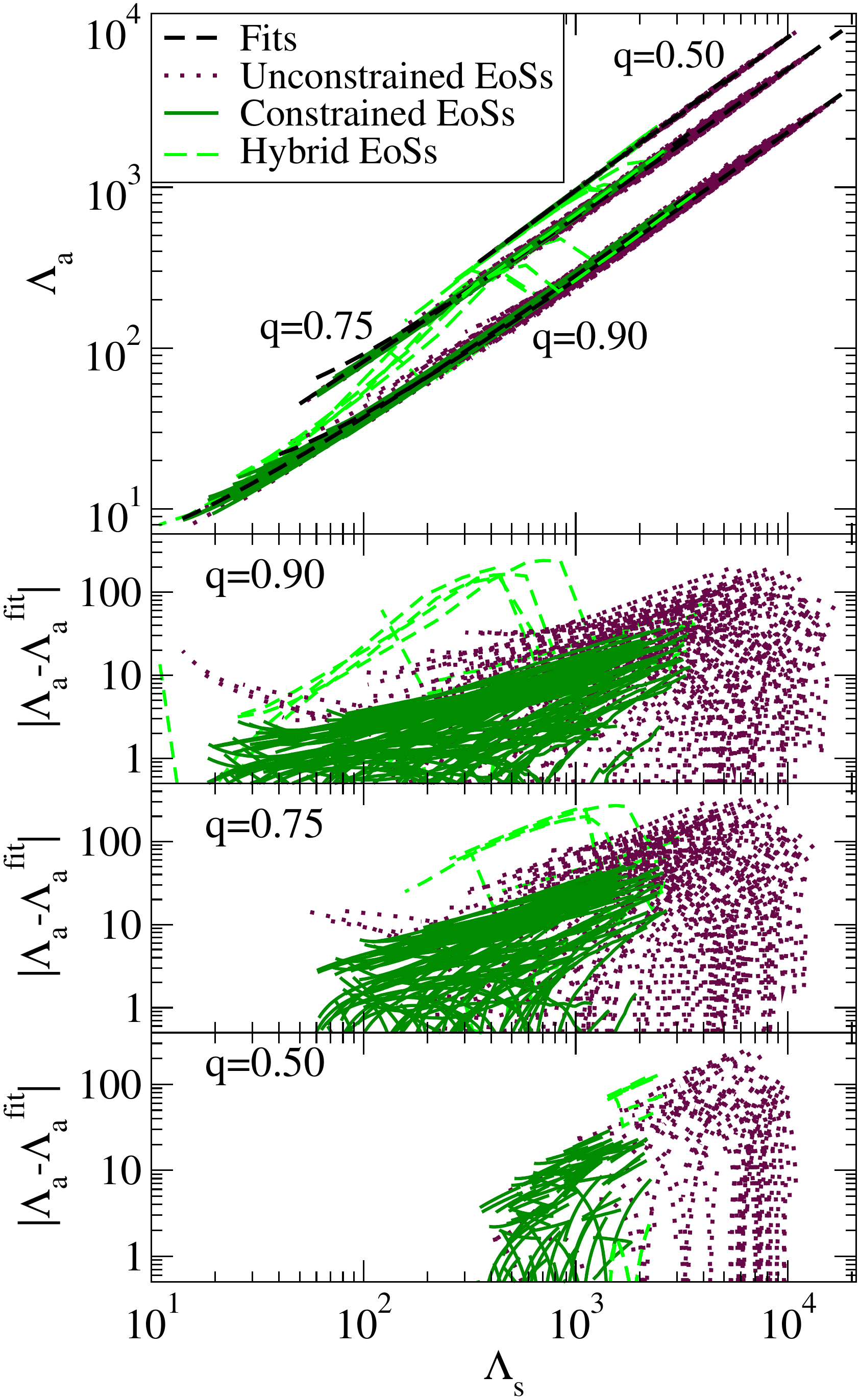}
\end{center}
\caption[Binary Love relations]{Binary Love relations using Eq.~\eqref{eq:binLovefit} fitted to the constrained (dotted maroon) and unconstrained sets (solid green), fixing $q=0.90$, 0.75 and 0.50. The bottom panels show the absolute difference (not the relative difference as in Fig.~\ref{fig:ILQ}) of the fit and the data. The fit to the constrained set shows a reduction in EoS variation relative to both the fit to the unconstrained set and previous work~\cite{Yagi:binLove}. For comparison, we also show the binary Love relations for the 10 hybrid star EoSs (dashed green curves), which present much larger EoS variation.
Observe here how both hadronic and hybrid star EoSs alike seemingly observe increasingly small distinction from the fits as the mass ratio $q$ approaches 0.
While the absolute errors remain somewhat consistent with mass ratio, the fractional differences from the fit approach $0$ as $q\rightarrow0$ as the universal relation becomes exact ($\Lambda_s\rightarrow\Lambda_a$).
}
\label{fig:binLove}
\end{figure} 

Let us now consider the binary Love relations. As in other relations, we consider nuclear matter stars and hybrid stars separately.

\subsubsection{Nuclear matter stars}

Following Ref.~\cite{Yagi:binLove}, we begin by fitting the binary Love relations to the constrained and unconstrained sets using the two-dimensional curve:
\begin{equation}\label{eq:binLovefit}
\Lambda_a=F_n(q) \frac{1+ \sum_{i=1}^3 \sum_{j=1}^2 b_{ij}q^j \Lambda_s^{i/5}}{1 + \sum_{i=1}^3 \sum_{j=1}^2 c_{ij}q^j \Lambda_s^{i/5}} \Lambda_s^{\alpha},
\end{equation}
where $q\equiv m_{2}/m_{1}$ is the mass ratio with $m_2 \leq m_1$, and $F_n(q)$ is the Newtonian-limiting controlling factor, given by
\begin{equation}\label{eq:control}
F_n(q) \equiv \frac{1-q^{10/(3-n)}}{1+q^{10/(3-n)}}.
\end{equation}
The coefficients of the fit to the constrained set are 
\begin{align}
b_{ij} = \begin{bmatrix}
    -14.40 & 14.45   \\
    31.36 & -32.25   \\
    -22.44 & 20.35   
\end{bmatrix}
\quad
c_{ij} = \begin{bmatrix}
    -15.25 & 15.37   \\
    37.33 & -43.20   \\
    -29.93 & 35.18   
\end{bmatrix},
\end{align}
where we set $n = 0.743$ and $\alpha = 1$ as was done in Ref.~\cite{Yagi:binLove}.
Unlike the individual I-Love-Q relations, the binary Love relations depend also on the mass ratio. 

Figure~\ref{fig:binLove} shows the improved binary Love relation for 3 different mass ratios ($q=0.9$, 0.75 and 0.5) for both the constrained and unconstrained sets of EoSs. Once again we find that the constrained set shows a considerable increase in EoS-insensitivity relative to both the unconstrained set and previous works~\cite{Yagi:binLove}. As before, the fit to the unconstrained set shows similar, yet slightly larger EoS variation than that found in previous work~\cite{Yagi:binLove} due to the random sampling used in our analysis. The absolute error between the fit and the data is approximately independent of the mass ratio $q$, with a subtle maximum at $q=0.75$, which mimics the behavior of the controlling factor $F_n(q)$. As before, maximum EoS variation is tabulated in Table~\ref{tab:maxVar} for each value of mass ratio considered here.

We can extract two additional conclusions from Fig.~\ref{fig:binLove}. First, as the mass ratio becomes more extreme (i.e.~away from unity), the ranges of values that $\Lambda_s$ and $\Lambda_a$ can take decrease. This is because as the separation between $m_1$ and $m_2$ increases, the effective number of possible $\Lambda_1(m_1)$-$\Lambda_2(m_2)$ configurations decrease, and thus, the allowed parameter space shrinks. Second, the fit to the constrained set does not extend to as high values of $\Lambda_s$ as the fit to the unconstrained set does. This is because of how the sets of EoSs were generated in our analysis, with the unconstrained set including some EoSs that are stiffer\footnote{Typically, the ``stiffness" of an EoS is determined by the amount of pressure gained given an increase in density. Stiff EoSs have steep pressures and predict larger maximum mass NSs, while soft EoSs have shallow pressures and predict smaller maximum mass NSs.} and others that are softer than those constrained by the GW170817 event. This results in the unconstrained set  containing a larger range of tidal deformabilities $\Lambda_{1,2}$ (from both above and below) than the constrained set, which ultimately leads to a \textit{larger} range of $\Lambda_s=\frac{1}{2}(\Lambda_1+\Lambda_2)$ values.

\subsubsection{Hybrid quark-hadron stars}

Let us now consider the binary Love relations for hybrid stars. As described in Sec.~\ref{sec:eostheory}, transitional quark-hadron matter stars undergo strong first-order phase transitions at a transitional pressure $P_{\text{tr}}$, where the hadronic branch departs into a quark-matter branch at the corresponding transitional mass $m_{\text{tr}}$. These transitions result in two distinct types of NSs, based on their observed mass: (i) massive ($m \geq m_{\text{tr}}$) hybrid stars that have quark-matter inner cores and nuclear matter elsewhere (which recall we denote ``HS"); and (ii) low-mass ($m \leq m_{\text{tr}}$) hadronic stars with no internal transition to quark matter (which recall we denote ``NS"). We therefore expect the binary Love relations for hybrid stars to present behavior identical to their purely hadronic counterparts below the transitional mass (or correspondingly above a transitional $\Lambda_{s,\textrm{tr}}$), and very different behavior for higher masses (or correspondingly for smaller $\Lambda_{s,\textrm{tr}}$). 

This behavior is exactly what we observe in Fig.~\ref{fig:binLove} for the binary Love relations of hybrid stars. Indeed, contrary to the case of EoS-insensitive relations for isolated hadronic stars (I-Love-Q, C-Love, and R-Love) which remain moderately EoS-insensitive for all hybrid stars, the binary Love relations depart from their hadronic counterparts below some transitional $\Lambda_{s}$. This is because at low pressures ($\Lambda_s>\Lambda_{s,\textrm{tr}}$) the binary is purely of NS/NS type (with the EoS for both stars a member of the constrained set), but once the critical pressure is reached ($\Lambda_{s}<\Lambda_{s,\textrm{tr}}$), one or both stars transition into the hybrid branch, resulting in large reductions in tidal deformability, as shown in the left panel of Fig.~\ref{fig:clove}. When one star lies on the hadronic-matter branch while the other is on the quark-matter branch, the difference between tidal deformabilities becomes larger than expected for pure hadronic-matter stars. This results in large deviations in the sums and differences of tidal deformabilities $(\Lambda_2 \pm \Lambda_1)$, disrupting the overall universality and generating the large ``bump" in the binary Love relations for hybrid stars\footnote{This discrepancy is not present in the I-Love-Q and C-Love relations due to their single-star nature.} seen in Fig~\ref{fig:binLove}.

From this analysis, we conclude that binaries containing one or more hybrid stars do \textit{not} satisfy the same binary Love relations as binaries that contain only traditional nuclear matter stars. This does not imply, however, that more sophisticated binary Love relations cannot be constructed that will remain EoS-insensitive and able to model both types of binary systems. Such relations, however, would necessarily have to include (at least) one new parameter that determines the transition between the hadronic and the quark branches, such that the ``bump'' in the binary Love relations can be properly modeled. Work along these lines is outside the scope of this chapter.


\section{Impact on future observations}
\label{sec:observations}

We have now shown that binary NS merger observations can help improve the EoS-insensitive relations, but the question remains: is it worth it?
Current interferometer sensitivities are not yet high enough to accurately constrain the dominant tidal parameter $\tilde{\Lambda}$.
For example, GW170817 was detected by the second LIGO observing run (``O2")~\cite{aLIGO} and Virgo~\cite{TheVirgo:2014hva}, and was able to constrain $\tilde{\Lambda}$ to a $90\%$ credible interval centered at $\mu_{\tilde{\Lambda}}=395$ and with a width of 325~\cite{Abbott2018} (or $\sigma_{\tilde{\Lambda}} \approx 198$).  This corresponds to statistical uncertainties of ${\cal{O}}(80\%)$, which dominates the error budget compared to the small systematic uncertainties picked up by EoS variation in the EoS-insensitive relations.
This implies that currently, the use of improved EoS-insensitive relations will only make a negligible difference on the extraction of tidal parameters.

In this section, we further explore this question and study when in the future the new set of improved relations will become important as current detectors are improved and new ones are built. In Sec.~\ref{sec:marginalization}, we first estimate the systematic uncertainties introduced by using the improved binary Love relations.
In Sec.~\ref{sec:futureObservations}, we estimate the statistical uncertainties on the extraction of tidal parameters, and compare them to the above-mentioned systematic uncertainties.
This is repeated for 5 future detectors, where multiple detections become important, but to combine constraints from multiple events we cannot use the $\tilde\Lambda$ parameterization, as this depends on the masses of the binary constituents. To remedy this, we re-parameterize the waveform in terms of the coefficients $\lambda_0$ (sometimes called $\Lambda_{1.4}$) and $\lambda_1$, obtained from a Taylor expansion of the dimensionless tidal deformability $\Lambda$ about a ``canonical" reference mass $m_0=1.4\text{ M}_{\odot}$~\cite{delPozzo:TaylorTidal,Yagi:binLove}:
\begin{equation}
\Lambda = \lambda_0 + \lambda_1 \left(1-\frac{m}{m_0}\right) + {\cal{O}}\left[\left(1-\frac{m_{0}}{m}\right)^{2}\right].
\end{equation}
The Taylor coefficients $\lambda_0$ and $\lambda_1$ are mass independent, and thus they are identical in all binary NS observations, and their posteriors may be combined.

\subsection{Error marginalization}\label{sec:marginalization}

Although the improved binary Love relation $\Lambda_a^{\text{relation}}(\Lambda_s,q)$ shows a high degree of universality, any residual EoS dependence in the relation could introduce a systematic bias and lead to incorrect inferences about the correct EoS. Let us then discuss a method to marginalize over the residual EoS-dependence of the binary Love relations. The residual is here defined as 
\begin{equation}\label{eq:residual}
r(\Lambda_{s},q) \equiv \Lambda_a^{\text{relation}}(\Lambda_s,q)-\Lambda_a^{\text{true}}\,,
\end{equation}
where $\Lambda_a^{\text{relation}}(\Lambda_s,q)$ is given by the binary Love fit found in Sec.~\ref{sec:binary}, while $\Lambda_a^{\text{true}}$ is the true value predicted by choosing a particular EoS in the constrained set and solving for the tidal deformabilities numerically. 

Following the proposal in~\cite{Katerina:residuals}, one can model the residual EoS-sensitivity by enhancing the binary Love relations through 
\begin{equation}
\Lambda_a^{{\textrm{new relation}}}=\Lambda_a^{\text{relation}}(\Lambda_s,q)+\mathcal{N}(\mu_{r}(\Lambda_s,q),\sigma_{r}(\Lambda_s,q)),
\end{equation}
where $\mathcal{N}(\mu_r,\sigma_r)$ is a normal distribution with mean and variance $\mu_{r}$ and $\sigma_{r}^2$. Let us further assume that the residuals $r$ obey a Gaussian distribution with mean and standard deviation that can be decomposed as
\begin{align}
\mu_{r}(\Lambda_s,q) &=\frac{\mu_{r}(\Lambda_s)+\mu_{r}(q)}{2},\\ 
\sigma_{r}(\Lambda_s,q) &=\sqrt{\sigma_{r}^2(\Lambda_s) + \sigma_{r}^2(q)},
\end{align}
where $\mu_{r}(q)$ and $\mu_{r}(\Lambda_{s})$ are the means marginalized over $\Lambda_{s}$ and $q$ respectively, while $\sigma_{r}(q)$ and $\sigma_{r}(\Lambda_{s})$ are the standard deviations marginalized over $\Lambda_{s}$ and $q$ respectively.

Clearly then, to account for the residual EoS-sensitivity, we must first find the marginalized mean and standard deviation of the residual function, which we accomplish as follows. We first generate $\Lambda_{s}^{\textrm{true}}$ data by sampling through various values of mass ratio $q\in \lbrack0.36,1\rbrack$ and symmetric tidal deformability $\Lambda_{s} \in \lbrack 4,4600 \rbrack$ over all 100 elements of the constrained set of EoSs. For each value of $(q,\Lambda_{s})$ we then compute $\Lambda_{a}^{\textrm{relation}}$ from the binary Love relation, and then compute the residuals through Eq.~\eqref{eq:residual}. We then proceed to marginalize over $q$ by binning the residuals in $\Lambda_{s}$, which returns a distribution of residuals that is only a function of the binned $\Lambda_{s}$, and from which we can compute the mean $\mu_{s}(\Lambda_{s})$ and the standard deviation $\sigma_{r}(\Lambda_{s})$. Repeating this procedure by marginalizing over $\Lambda_{s}$ then allows us to compute the mean $\mu_{s}(q)$ and the standard deviation $\sigma_{r}(q)$. Finally, we fit the means and standard deviations to the functions~\cite{Katerina:residuals}
\begin{align}
\mu_{r}(\Lambda_s) &= \mu_1 \Lambda_s + \mu_2, \label{eq:margFit1}\\ 
\mu_{r}(q) &= \mu_3 q^2 + \mu_4 q + \mu_5, \label{eq:margFit2}\\ 
\sigma_{r}(\Lambda_s) &= \sigma_1 \Lambda_s^{5/2} + \sigma_2 \Lambda_s^{3/2} + \sigma_3 \Lambda_s +  \sigma_4 \Lambda_s^{1/2} + \sigma_5, \label{eq:margFit3}\\ 
\sigma_{r}(q) &= \sigma_6 q^3 + \sigma_7 q^2 + \sigma_8 q + \sigma_9, 
\label{eq:margFit4}
\end{align}
where the fitting parameters $\mu_i$ and $\sigma_i$ are tabulated in Table~\ref{tab:marginalized}. 

\begin{table}
\centering
\addtolength{\tabcolsep}{1pt} 
\begin{tabular}{ c | c || c | c}
\hline 
\noalign{\smallskip}
$\mu_1$ & $3.509 \times 10^{-3}$ & $\sigma_1$ & $-2.074 \times 10^{-7}$\\
$\mu_2$ & $9.351 \times 10^{-1}$ & $\sigma_2$ & $-1.492 \times 10^{-3}$\\
$\mu_3$ & $-18.07$ & $\sigma_3$ & $-4.891 \times 10^{-2}$\\
$\mu_4$ & $27.56$ & $\sigma_4$ & $8.207 \times 10^{-1}$\\
$\mu_5$ & $-10.10$ & $\sigma_5$ & $-1.308$\\
 &  & $\sigma_6$ & $-63.76$\\
 &  & $\sigma_7$ & $11.14$\\
 &  & $\sigma_8$ & $75.25$\\
 &  & $\sigma_8$ & $-23.69$\\
 \noalign{\smallskip}
 \hline
\end{tabular}
\caption[Binary Love $\lambda_a$ relative error fit coefficients]{
Coefficients to the fits given by Eqs.~\eqref{eq:margFit1}-\eqref{eq:margFit4} for the relative error on $\Lambda_a$ in the improved binary Love EoS-insensitive relations presented in this chapter.
}\label{tab:marginalized}
\addtolength{\tabcolsep}{-1pt}
\end{table}

In what follows, however, we will be interested in comparing an estimate of the systematic uncertainties in $\lambda_0$ due to the residual EoS-sensitivity in the binary Love relations to the statistical error in the extraction of this parameter. To estimate the former, we first numerically calculate $\lambda_{0}^{\textrm{true}}$ by sampling again in $(q,\Lambda_{s})$ within the same ranges as before and over all 100 elements of the constrained and the unconstrained sets of EoSs. We then use the binary Love relation to calculate $\lambda_0^{\text{relation}}$ over the $(q,\Lambda_{s})$ sampled points, and from this we compute the residual.  
\begin{equation}
{\cal{R}}(\Lambda_{s},q) = \lambda_0^{\text{relation}}(\Lambda_s,q)-\lambda_0^{\text{true}}.
\end{equation}
Figure~\ref{fig:residuals} shows a histogram of ${\cal{R}}$ for both the constrained and unconstrained set of EoSs. The standard deviations of the two histograms, $\sigma=9.764$ and $\sigma=78.28$, show a large decrease in EoS-variability from the unconstrained to the constrained sets of EoSs. From this distribution, we can additionally find the 90\%, 99\%, and 100\% credible interval on $\lambda_0$, namely $P_{90}=13.19$, $P_{99}=42.38$, and $P_{100}=111.32$ for the constrained EoSs.

Figure~\ref{fig:qLsResiduals} shows the standard deviations of the $\lambda_0$ residuals binned in $q$ (top panel) and $\Lambda_{s}$ (bottom panel), as was done previously for the $\Lambda_a$ residuals. The EoS-variability in $\lambda_0$ is dominated by the region around $q \sim 0.5$ and $\Lambda_{s} \sim 2000$. Henceforth, in order to take into account these high-error regions of parameter space which may get averaged out, we approximate the EoS-variability in the binary Love relations with the 90\% credible interval (rather than the $1\sigma$ uncertainty), which for the $\lambda_{0}$ parameter is $P_{90}=13.19$ (indicated by the dashed green line in Fig.~\ref{fig:qLsResiduals} and the dashed indigo line in \hypertarget{link:stackedFisher}{Fig.~\ref{fig:stackedFisher})}. We can similarly derive a conservative estimate of the systematic uncertainty due to EoS-variability in the binary Love relations in terms of $\tilde\Lambda_{a}$ to find $12.01$, which is what we use later in Fig.~\ref{fig:singleFisherLt}.
Here we note that these results rely on the assumption of a purely hadronic EoS, as the constrained set of EoSs do not contain any hybrid stars.
This assumption was also made in Ref.~\cite{LIGO:posterior} upon the derivation of the posterior probability distribution.

\begin{figure}[!htbp]
\begin{center} 
\includegraphics[width=0.7\columnwidth]{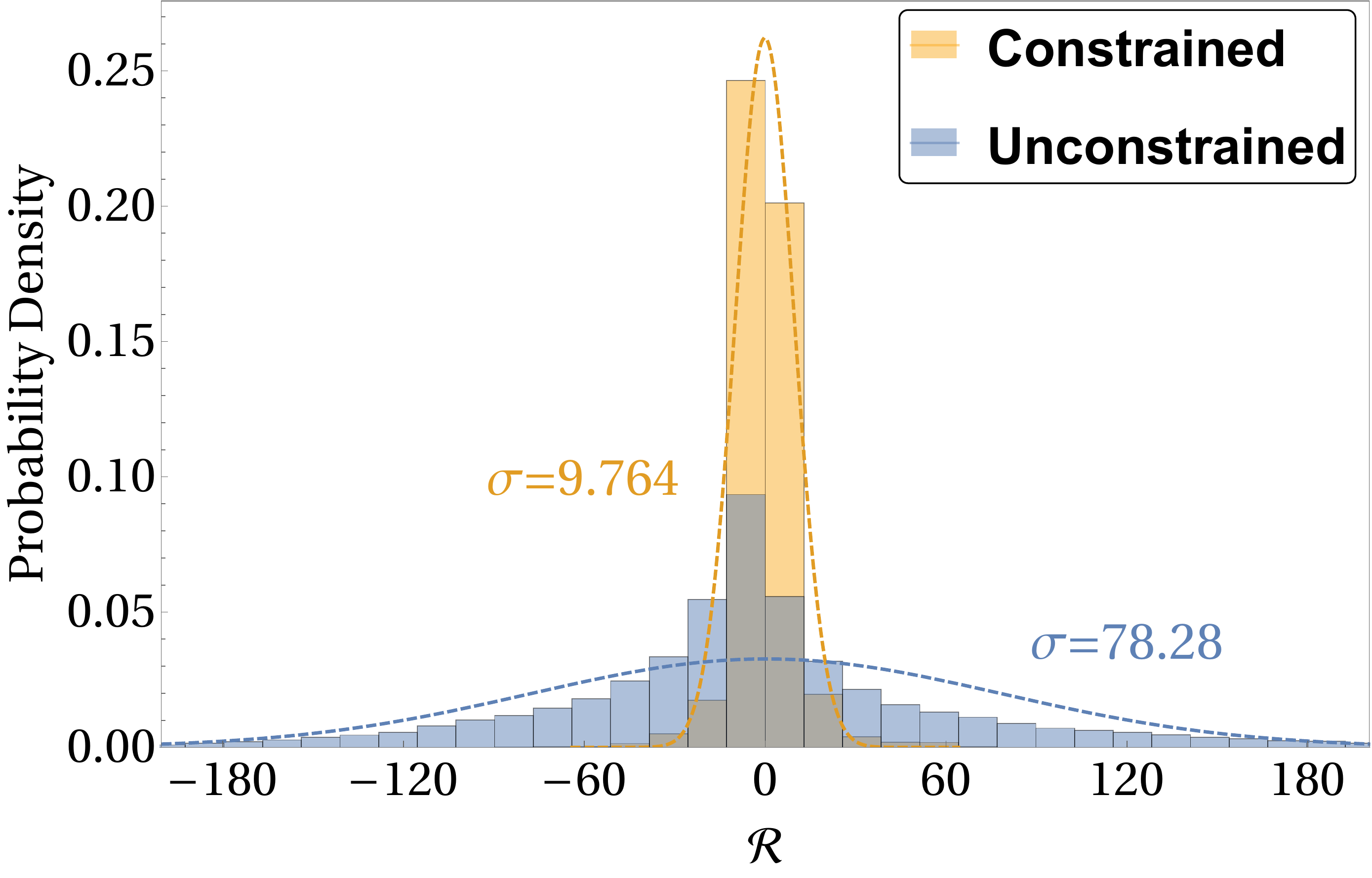}
\end{center}
\caption[$\lambda_0$ residuals from the binary Love relation]{
Residuals on $\lambda_0$ computed as $\mathcal{R}=\lambda_0^{\text{relation}}-\lambda_0^{\text{true}}$ (computed by sampling $q\in \lbrack0.36,1\rbrack$ and $\Lambda_{s} \in \lbrack 4,4600 \rbrack$ for each of the 100 EoS samples) for the binary Love relations modeled in Sec.~\ref{sec:binary} for both constrained and unconstrained sets of EoSs.
We fit these residuals with Gaussian distributions centered at $\mu=-0.1530$ and $\mu=0.2710$ with standard deviations of $\sigma=9.764$ and $\sigma=78.28$ for the constrained and unconstrained sets of EoSs, respectively.
These uncertainties correspond roughly to the systematic uncertainties introduced on the parameter extraction of $\lambda_0$ upon the use of binary Love relations.
However, to take into account the systematic uncertainties found in high-error regions of the parameter space, we instead set the systematic uncertainty to be the 90th percentile, $P_{90}=13.19$.
The systematic uncertainties from using the improved (constrained) binary Love relations are negligible compared to the statistical uncertainties accrued on parameter extraction from GW170817, found to be $\sigma_{\lambda_0}=170.1$.
}
\label{fig:residuals}
\end{figure}

\begin{figure}
\begin{center} 
\begin{overpic}[width=0.7\columnwidth]{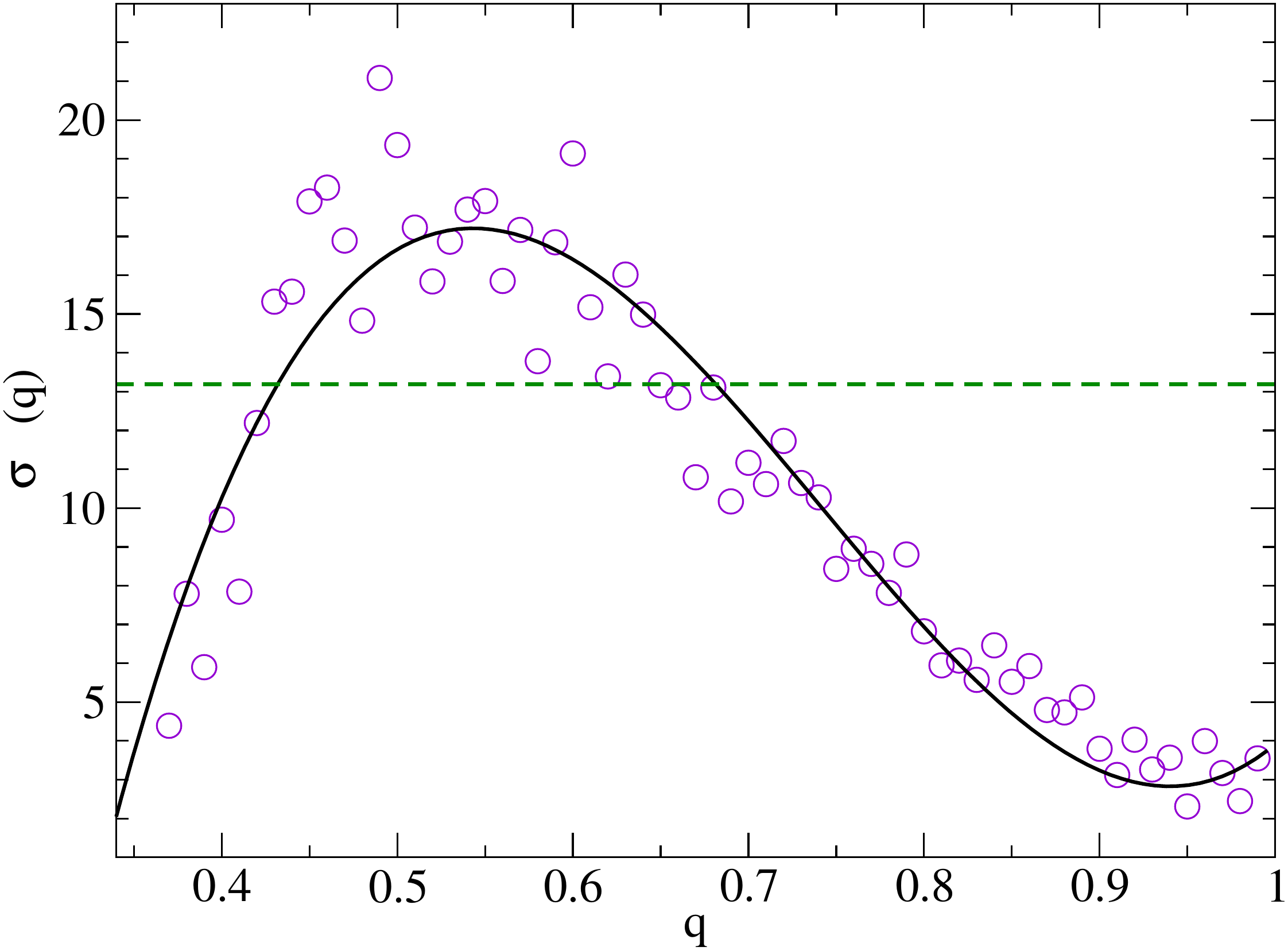}
\put(7,116){\tiny\rotatebox{90}{$\mathcal{R}$}}
\end{overpic}
\begin{overpic}[width=0.7\columnwidth]{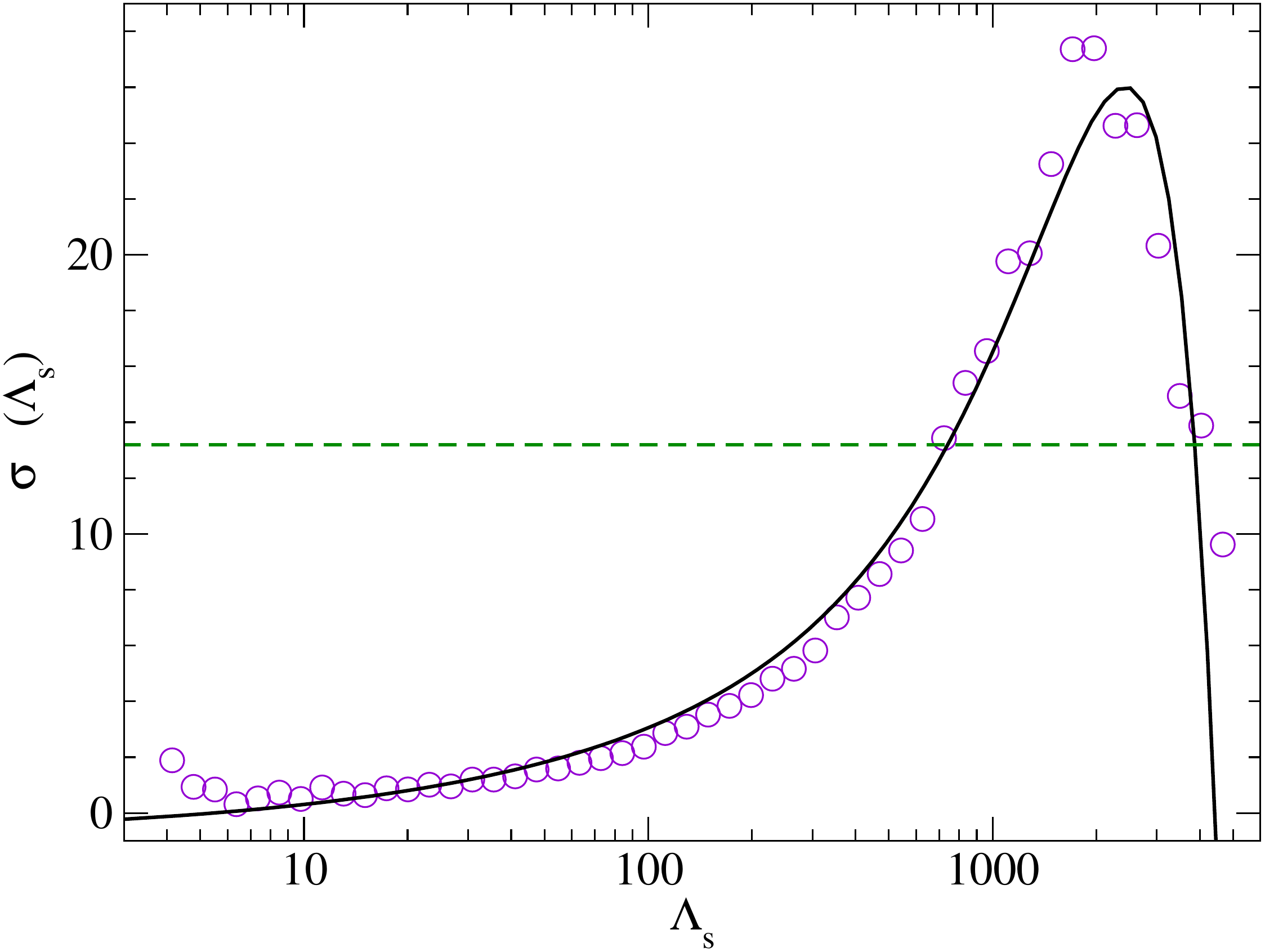}
\put(7,118){\tiny\rotatebox{90}{$\mathcal{R}$}}
\end{overpic}
\end{center}
\caption[$\lambda_0$ residual standard deviations]{Standard deviations of the $\lambda_0$ residuals $\mathcal{R}$ binned in $q$ (top) and $\Lambda_s$ (bottom), highlighting the different error weights across the entire $(q,\Lambda_s)$ parameter space.
In this figure, the violet circles indicate the standard deviation of each bin in $q$-space ($\Lambda_s$-space), while the solid black lines represent the best fit given by Eqs.~\eqref{eq:margFit3}-\eqref{eq:margFit4}.
The uncertainty is maximal for $q\sim0.5$ and $\Lambda_s\sim2000$, while it becomes minimal for both low and high values of $q$ and $\Lambda_s$.
Also shown in the figure is the 90th percentile of the un-binned residuals seen in Fig.~\ref{fig:residuals}, taken to be the overall systematic uncertainty introduced by using binary Love relations.
}
\label{fig:qLsResiduals}
\end{figure}

\subsection{Future observations}\label{sec:futureObservations}

We now estimate the feasibility of using the improved EoS-insensitive relations in future GW observations of the coalescence of binary NSs. We estimate the statistical accuracy to which parameters can be extracted through the simple Fisher analysis~\cite{Finn:Fisher,Cutler:Fisher} method described in detail in Chapter~\ref{chap:ParameterEstimation}, assuming sufficiently high signal-to-noise ratio and Gaussian noise~\cite{Cutler:Fisher,Berti:Fisher,Poisson:Fisher}. 

We consider here 6 different interferometer designs as described in Chapter~\ref{chap:ParameterEstimation}: aLIGO O2~\cite{aLIGO}, aLIGO at design sensitivity~\cite{aLIGO}, A\texttt{+}~\cite{Ap_Voyager_CE}, Voyager~\cite{Ap_Voyager_CE}, CE~\cite{ET}, ET-D~\cite{Ap_Voyager_CE}, in order to compare the statistical uncertainties accrued on parameter extraction using future upgraded LIGO detectors, as well as third generation detectors.
We here consider the PhenomD (IMRD) waveform template~\cite{PhenomDI,PhenomDII} modified by a 6PN tidal correction~\cite{Wade:tidalCorrections} (IMRD + 6PN) as described in Chapter~\ref{chap:ParameterEstimation}, as well as the same PhenomD template modified instead with a NRTidal correction~\cite{Samajdar:NRTidal} (IMRD + NRTidal). Considering two template models will allow us to estimate systematic uncertainties due to mismodeling of the GW signal. 

We begin by testing our Fisher analysis (with an IMRD + 6PN waveform injection) against a simulated event identical to GW170817 with O2 detector sensitivity~\cite{aLIGO}. Because only 1 event was detected, we use $\tilde\Lambda$ and $\delta\tilde\Lambda$ as the tidal parameters for comparison purposes. Further, we scale the luminosity distance such that the signal-to-noise-ratio ($SNR \equiv \rho$) is fixed to $\rho=32.4$, as found in GW170817. We used fiducial template parameter values of $\mathcal{M}=1.22\text{ M}_\odot$ for the chirp mass, $\eta=0.249$ for the symmetric mass ratio, $\tilde\Lambda=395$ (Corresponding to GW170817), and $0$ for the remaining parameters. We also assume low spin priors $|\chi| \leq 0.05$, as well as $\tilde{\Lambda} \leq 3000$ and $|\delta \tilde{\Lambda}| \leq 500$~\cite{Wade:2014vqa}. The resulting $90\%$ credible region of the posterior distribution on $\tilde{\Lambda}$ has a range of $\pm 276.99$, which is in close agreement to that found by LIGO~\cite{TheLIGOScientific:2017qsa,Abbott2018}

\begin{figure}
\begin{center} 
\includegraphics[width=0.7\columnwidth]{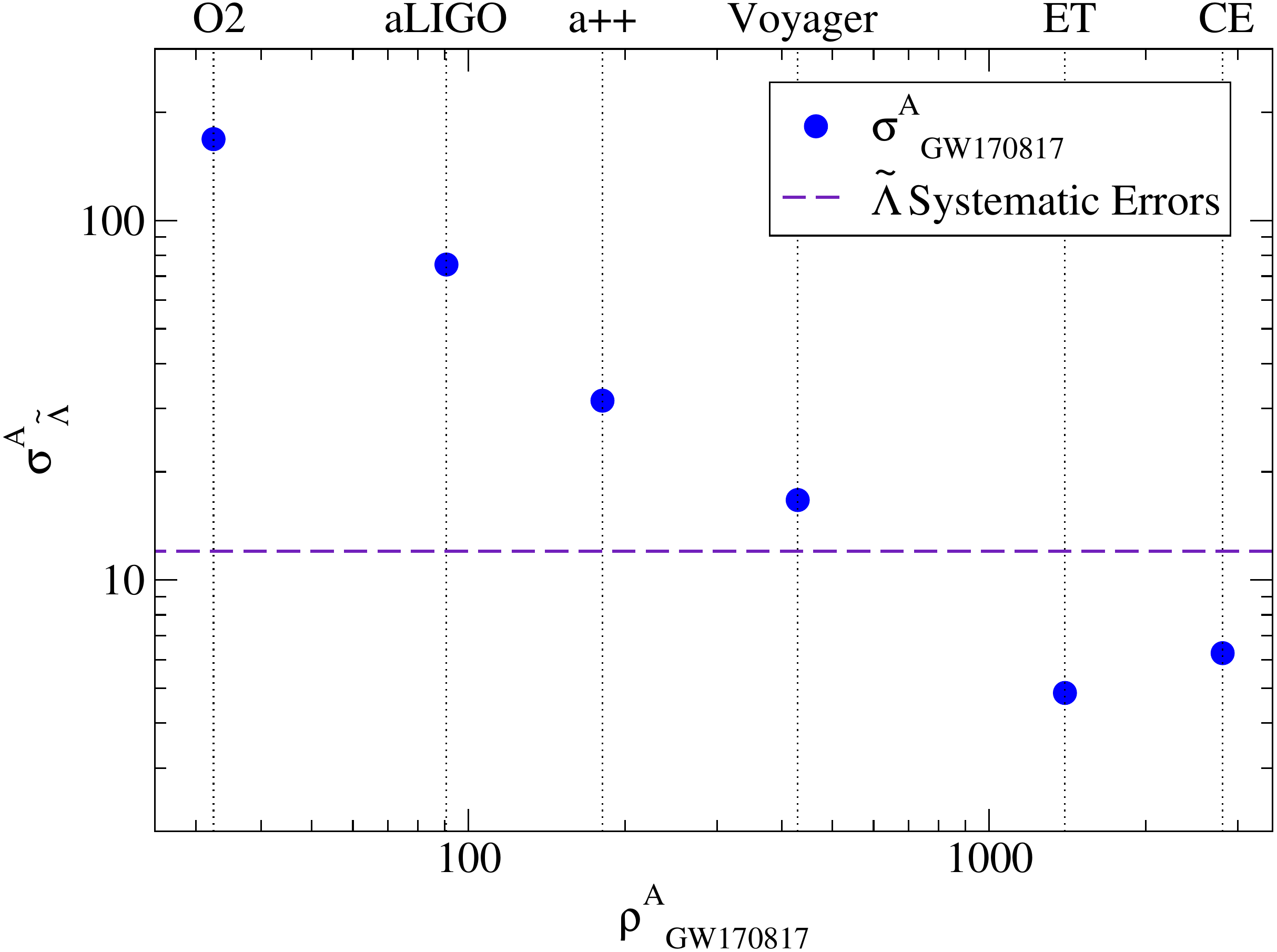}
\end{center}
\caption[Single-event uncertainties on $\tilde\Lambda$]{Estimated statistical uncertainty $\sigma^A_{\text{GW170817}}$ in the extraction of $\tilde\Lambda$ from a single GW170817-like event as if observed with aLIGO at design sensitivity, A\texttt{+}, Voyager, CE, and ET-D, plotted as a function of the signal-to-noise-ratio $\rho^A_{\text{GW170817}}$ that those detectors would measure for such an event. For comparison, we also plot the systematic uncertainty on $\tilde\Lambda$ due to the use of the binary Love relations. The statistical and the systematic errors become comparable for Voyager-class detectors or better.
}
\label{fig:singleFisherLt}
\end{figure} 

Next, we consider events similar to GW170817 detected on upgraded detectors and future detectors, as well as the combined statistical uncertainties of $N_A$ events detected over a 1 year observation with each detector. The latter is calculated by integrating the local binary NS merger rate over redshifts up to the horizon redshift of each detector. 
Similar to before, we used fiducial template parameter values of $\mathcal{M}=1.22\text{ M}_\odot$ for the chirp mass, $\eta=0.249$ for the symmetric mass ratio, $\lambda_0=150$, $\lambda_1=-213$ (corresponding to the GW170817), and $0$ for the remaining parameters.
Also as before, for each Fisher calculation we assume low spin priors $|\chi| \leq 0.05$, as well as $0 \leq \lambda_0 \leq 3207$ and $-4490 \leq \lambda_1 \leq 0$~\cite{delPozzo:TaylorTidal}\footnote{These are converted from the dimensional forms found in~\cite{delPozzo:TaylorTidal} into their corresponding dimensionless forms.}.
The process we use to compute single and combined statistical uncertainties for each detector sensitivity $S_n^A(f)$ is detailed in Sec.~\ref{intro:Combined}.
From this, we determine if and when the statistical uncertainties associated with the parameter extraction of $\lambda_0$ drop below the systematic EoS variation errors from using the binary Love relations.

\begin{table*}
\centering
\resizebox{\linewidth}{!}{%
\begin{tabular}{|c|@{\extracolsep{4pt}}C{1.7cm}@{\extracolsep{0pt}}|C{1.7cm}|@{\extracolsep{4pt}}C{1.7cm}@{\extracolsep{-2pt}}|C{1.7cm}|@{\extracolsep{-2pt}}C{1.7cm}@{\extracolsep{2pt}}|C{1.7cm}|@{\extracolsep{0pt}}C{1.7cm}@{\extracolsep{0pt}}|C{1.7cm}|}
\cline{1-1}\cline{2-3}\cline{4-9}
    \multicolumn{1}{|c|}{\bfseries Detectors (A)} & \multicolumn{2}{|c|}{\bfseries GW170817} & \multicolumn{6}{|c|}{\bfseries Multiple events} \\
\cline{1-1}\cline{2-3}\cline{4-9}
\noalign{\smallskip}
\cline{2-3}\cline{4-6}\cline{7-9}
\multicolumn{1}{c}{} & \multicolumn{1}{|c|}{} & \multicolumn{1}{c|}{} & \multicolumn{3}{|c|}{} & \multicolumn{3}{|c|}{}
\\[-1em]
\multicolumn{1}{c}{}  &  \multicolumn{1}{|c|}{\multirow{2}{*}{$\rho^A_{\text{GW170817}}$}}  &  \multirow{ 2}{*}{$\sigma^A_{\text{GW170817}}$}  &  \multicolumn{3}{|c|}{$N_A$}  &  \multicolumn{3}{|c|}{$\sigma^A_N$}  \\
\cline{4-6}\cline{7-9}
\multicolumn{1}{c}{}  &  \multicolumn{1}{|c|}{}  &  \multicolumn{1}{c|}{}  &  \multicolumn{1}{|c|}{Low}  &  \multicolumn{1}{c|}{Central} &  \multicolumn{1}{c|}{High}  & \multicolumn{1}{|c|}{Low}  &  \multicolumn{1}{c|}{Central} &  \multicolumn{1}{c|}{High}\\
\cline{2-3}\cline{4-6}\cline{7-9}
\noalign{\smallskip}
\noalign{\smallskip}
\cline{1-1}\cline{2-3}\cline{4-6}\cline{7-9}
&\multicolumn{1}{|c|}{}&&\multicolumn{1}{|c|}{}&&&\multicolumn{1}{|c|}{}&&\\[-1em]
 O2  & \multicolumn{1}{|c|}{$\sci{3.2}{1}$}  & $\sci{1.7}{2}$ &  \multicolumn{1}{|c|}{--} & -- & -- & \multicolumn{1}{|c|}{--} & -- & --\\
\cline{1-1}\cline{2-3}\cline{4-6}\cline{7-9}
&\multicolumn{1}{|c|}{}&&\multicolumn{1}{|c|}{}&&&\multicolumn{1}{|c|}{}&&\\[-1em]
 aLIGO  & \multicolumn{1}{|c|}{$\sci{9.1}{1}$}  & $\sci{1.1}{2}$ &  \multicolumn{1}{|c|}{$\sci{2.0}{1}$} & $\sci{9.8}{1}$ & $\sci{3.0}{2}$ & \multicolumn{1}{|c|}{$\sci{1.8}{2}$} & $\sci{8.3}{1}$ & $\sci{4.7}{1}$\\
\cline{1-1}\cline{2-3}\cline{4-6}\cline{7-9}
&\multicolumn{1}{|c|}{}&&\multicolumn{1}{|c|}{}&&&\multicolumn{1}{|c|}{}&&\\[-1em]
 A\texttt{+}  & \multicolumn{1}{|c|}{$\sci{1.8}{2}$}  & $\sci{4.6}{1}$ &  \multicolumn{1}{|c|}{$\sci{1.6}{2}$} & $\sci{7.9}{2}$ & $\sci{2.4}{3}$ & \multicolumn{1}{|c|}{$\sci{5.9}{1}$} & $\sci{2.5}{1}$ & $\sci{1.4}{1}$\\
\cline{1-1}\cline{2-3}\cline{4-6}\cline{7-9}

\multicolumn{1}{|c|}{} & \multicolumn{1}{|c|}{} & \multicolumn{1}{c|}{} & \multicolumn{1}{|c|}{} & \multicolumn{1}{c|}{} & \multicolumn{1}{c|}{} & \multicolumn{1}{|c|}{} & \multicolumn{1}{c|}{} & \multicolumn{1}{c|}{}
\\[-1em]
&\multicolumn{1}{|c|}{}&&\multicolumn{1}{|c|}{}&&&\multicolumn{1}{|c|}{}&&\\[-1em]
 Voyager  & \multicolumn{1}{|c|}{$\sci{4.3}{2}$}  & $\sci{2.5}{1}$ &  \multicolumn{1}{|c|}{$\sci{2.2}{3}$} & $\sci{1.1}{4}$ & $\sci{3.2}{4}$ & \multicolumn{1}{|c|}{$\sci{2.1}{1}$} & $\sci{9.6}{0}$ & $\sci{5.3}{0}$\\
\cline{1-1}\cline{2-3}\cline{4-6}\cline{7-9}

\multicolumn{1}{|c|}{} & \multicolumn{1}{|c|}{} & \multicolumn{1}{c|}{} & \multicolumn{1}{|c|}{} & \multicolumn{1}{c|}{} & \multicolumn{1}{c|}{} & \multicolumn{1}{|c|}{} & \multicolumn{1}{c|}{} & \multicolumn{1}{c|}{}
\\[-1em]
&\multicolumn{1}{|c|}{}&&\multicolumn{1}{|c|}{}&&&\multicolumn{1}{|c|}{}&&\\[-1em]
 ET-D  & \multicolumn{1}{|c|}{$\sci{1.4}{3}$}  & $\sci{6.9}{0}$ &  \multicolumn{1}{|c|}{$\sci{7.2}{4}$} & $\sci{3.4}{5}$ & $\sci{1.1}{6}$ & \multicolumn{1}{|c|}{$\sci{3.8}{0}$} & $\sci{1.7}{0}$ & $\sci{9.6}{-1}$\\
\cline{1-1}\cline{2-3}\cline{4-6}\cline{7-9}

\multicolumn{1}{|c|}{} & \multicolumn{1}{|c|}{} & \multicolumn{1}{c|}{} & \multicolumn{1}{|c|}{} & \multicolumn{1}{c|}{} & \multicolumn{1}{c|}{} & \multicolumn{1}{|c|}{} & \multicolumn{1}{c|}{} & \multicolumn{1}{c|}{}
\\[-1em]
&\multicolumn{1}{|c|}{}&&\multicolumn{1}{|c|}{}&&&\multicolumn{1}{|c|}{}&&\\[-1em]
 CE  & \multicolumn{1}{|c|}{$\sci{2.8}{3}$}  & $\sci{7.7}{0}$ &  \multicolumn{1}{|c|}{$\sci{3.0}{5}$} & $\sci{1.4}{6}$ & $\sci{4.4}{6}$ & \multicolumn{1}{|c|}{$\sci{3.7}{0}$} & $\sci{1.7}{0}$ & $\sci{9.0}{-1}$\\
\cline{1-1}\cline{2-3}\cline{4-6}\cline{7-9}
\end{tabular}
}
\caption[$\lambda_0$ extraction uncertainty and SNR from multiple events]{
Approximate signal-to-noise ratio $\rho^A_{\text{GW170817}}$ and ($1\sigma$) statistical uncertainty on the extraction of $\lambda_{0}$ had a single event like GW170817 been observed by future interferometer $A$ and had interferometer $A$ observed $N_{A}$ events in a 1 year observation, using aLIGO, A\texttt{+}, Voyager, CE, and ET. The number of events $N_{A}$, and the combined statistical uncertainty depends on the binary NS merger detection rate, and thus we include results assuming an upper, a central and a lower limit on this rate. The statistical uncertainties on $\lambda_0$ becomes comparable with the systematic uncertainty (set to be $P_{90}=13.19$) from using the improved binary Love relations with detectors of Voyager-class or better.
}\label{tab:variances}
\end{table*}

Figures~\ref{fig:singleFisherLt} and~\ref{fig:stackedFisher}, and Table~\ref{tab:variances} summarize our results graphically using an IMRD + 6PN template. In Fig.~\ref{fig:singleFisherLt}, one can see the statistical accuracy to which $\tilde{\Lambda}$ can be estimated as a function of $\rho^A_{\text{GW170817}}$ for single GW170817-like events, i.e.~the signal-to-noise ratio that future detector $A$ would have measured for a GW170817 event. The systematic uncertainty in $\tilde{\Lambda}$ (the horizontal dashed line) becomes comparable to the statistical uncertainty for Voyager-class detectors or better. Figure~\ref{fig:stackedFisher} presents similar results but for the parameter $\lambda_{0}$, for which one can combine posteriors and obtained a reduced combined statistical error. This figure also shows that the statistical and systematic uncertainties become comparable for Voyager-class detectors or better. The careful reader will notice that Figs.~\ref{fig:stackedFisher} and~\ref{fig:singleFisherLt} show that the single-event statistical uncertainties using CE is higher than using ET despite the latter having a larger SNR for a GW170817-like event. This is because the 3-detector geometry of ET enables it to have higher sensitivity than CE at frequencies above 300Hz, which is precisely where the tidal deformabilities are encoded, but ET has lower sensitivity at lower frequencies, where a lot of the signal-to-noise ratio accumulates.


Is the statistical error calculated here robust to mismodeling systematics in the template? We repeat the analysis above using an IMRD + NRTidal template model  and find results consistent with those presented above. Indeed, Fig.~\ref{fig:stackedFisher} shows the individual and combined statistical error on $\lambda_{0}$ when using this template model (maroon circles and dashed region). The accuracy to which $\lambda_{0}$ can be measured with this template model is systematically better than when using the IMRD+6PN model (by roughly a factor of two). This implies that more accurate template models will indeed be required in the third-generation detector era, as previously pointed out in~\cite{Samajdar:NRTidal}. However, our conclusion that for Voyager-class detectors or better the statistical and binary Love systematic uncertainties become comparable seems robust.   


\section{Conclusions}
\label{sec:conclusion}

The recent GW observation of a binary NS coalescence, GW170817, placed constraints on the supranuclear matter EoS for NSs. We used this observation to generate a restricted set of spectral EoSs that agree with it, in order to reduce the uncertainties upon the extraction of tidal parameters from future GW events. Previous work by Yagi and Yunes~\cite{Yagi:2013bca,Yagi:ILQ,Yagi:2015pkc,Yagi:binLove} had found EoS-insensitive relations between symmetric and antisymmetric combinations of NS tidal deformabilities, which aid in the extraction of said tidal parameters. We here found that the GW170817-constrained set of EoSs are more EoS-insensitive by a factor of $\sim 60$\% for stars with mass ratios of $0.75$ relative to previous work. Similarly, we find an increase in EoS-insensitivity in the C-Love and I-Love-Q  relations by factors of $\sim 75$\% and $\sim 50$\% respectively. The former further allowed us to improve the R-Love relation leading to uncertainties below $400$ meters in the entire parameter space. We also studied the EoS-insensitive relations of hybrid stars and found that for the most part the relations remain insensitive for isolated stars, albeit with slightly higher EoS variability. The binary Love relations, however, do not satisfy the same EoS-insensitive relations as in the case of purely hadronic star, when the binary contains at least one hybrid star. 

The second half of this chapter focused on when the improvements in the EoS-insensitive relations would become necessary in future detectors. Current detectors are not yet sensitive enough that the systematic uncertainties in the EoS-insensitive relations make a large difference. However, we did find that for Voyager-class detectors or better, the systematic uncertainties in the EoS-insensitive relations due to EoS variability become comparable to statistical uncertainties in the estimation of $\tilde{\Lambda}$ and $\lambda_{0}$. We also considered the effect of waveform mismodeling and found that the above conclusion remains robust, but that more accurate waveform models will be necessary to take full advantage of the improved sensitivity of future detectors. 

Future work on this subject could entail an investigation into the improvement of alternative EoS-insensitive relations, such as the multipole Love relations between various $\ell$-th order electric, magnetic, and shape tidal deformabilities, as discussed in~\cite{Yagi:Multipole}. Lackey \textit{et al.}~\cite{Lackey:Surrogate, Lackey:EOB} presented surrogate models of non-spinning effective-one-body waveforms with the use of universal relations. By reducing the number of waveform model parameters, surrogate models aid in the extraction of NS observables from GW detections. The improvement in the multipole Love relations can then be used to increase the accuracy of such surrogate models. Another possible avenue for future research includes a more comprehensive analysis into the intricacies of new hybrid star binary Love relations.

\addtocontents{toc}{\protect\end{adjustwidth}}


\part{Testing general relativity}\label{part:TestingGR}
\addtocontents{toc}{\protect\begin{adjustwidth}{0.5cm}{0cm}}


\chapter{Parameterized tests}\label{chap:Parameterized}
\begin{adjustwidth}{\parindent}{\parindent}
\begin{spacing}{1.5}
Gravitational waves from extreme gravity events such as the coalescence of two black holes in a binary system fill our observable universe, bearing with them the underlying theory of gravity driving their process.
10 binary black hole merger events observed so far can be used to test Einstein's theory of general relativity, which has otherwise been proven to agree with observations from several sources in the weak- or static-field regimes.
We here demonstrate use of the so-called parameterized tests of general relativity where we introduce generic non-Einsteinian corrections to the waveform, which can easily be mapped to parameters in known example theories beyond general relativity.
We find that multi-band observations can be crucial in probing several modified theories of gravity, including those with gravitational parity violation.
We then study prospects on constraining scalar tensor theories with mixed black hole - neutron star systems.
Finally, we investigate the effects of correcting the general relativity PhenomD template waveform with the ensuing interactions during both the inspiral and the remnant black hole quasinormal mode portions of the binary black hole coalescence event.
In particular, we study the detectability of EdGB effects by including corrections to both the inspiral and ringdown portions, as well as those to the mass and spin of remnant black holes.
\end{spacing}
\end{adjustwidth}
\newpage


\section{Introduction}\label{sec:intro}

Einstein's theory of General Relativity (GR), which elegantly relates the geometries of spacetime to the manifestation of gravity, has remained at its post as the prevailing theory of gravity for over 100 years\footnote{This chapter is based on the following papers: Z. Carson, B. Seymour and K. Yagi, \textit{Future Prospects for Probing Scalar-Tensor Theories with Gravitational Waves from Mixed Binaries}, Class. Quant. Grav. \textbf{37}, 065008 (2020), Z. Carson and K. Yagi, \textit{Probing string-inspired Gravity with the inspiral and ringdown of gravitational waves}, Phys. Rev. D \textbf{101}, 104030 (2020), Z. Carson and K. Yagi, \textit{Parameterized and inspiral-merger-ringdown consistency tests of gravity with multi-band gravitational wave observations}, Phys. Rev. D \textbf{101}, 044047 (2020), and Z. Carson and K. Yagi, \textit{Parameterized and Consistency Tests of Gravity with Gravitational Waves: Current and Future}, Proceedings, Recent Progress in Relativistic Astrophysics: Shanghai, China, Vol. 17(1) (2019)}.
Throughout this era, GR has been subject to a plethora of tests in search for minute deviations which may point to alternative theories of gravity.
As pointed out by Popper~\cite{popper}, scientific theories such as GR can never be entirely proven, however alternative theories may be constrained.
When subject to observations on the solar-system scale where gravity is weak and approximately static, such as photon-deflection, Shapiro time-delay, perihelion advance of Mercury, and the Nordvedt effect~\cite{Will_SolarSystemTest}, 
GR has passed the tests with flying colors.
Observations concerning the strong-field, static systems of binary pulsar systems~\cite{Stairs_BinaryPulsarTest,Wex_BinaryPulsarTest} similarly proved GR to be entirely consistent.
On the large-scale side, cosmological observations~\cite{Ferreira_CosmologyTest,Clifton_CosmologyTest,Joyce_CosmologyTest,Koyama_CosmologyTest,Salvatelli_CosmologyTest} have also proven Einstein to be correct.
See also Ref.~\cite{Ishak:2018his} for a review on testing general relativity on cosmological scales.

On September 14, 2015, the LIGO detectors in Hanford and Livingston chirped with activity as they, for the first time ever, observed the iconic GW signal from the explosive coalescence of two BHs 1.4 billion lightyears away.
Aptly named GW150914~\cite{GW150914} by the LVC, this historic detection has ushered in an entirely new era of observational astrophysics, finally allowing us to probe the extreme gravity regime of spacetime~\cite{Abbott_IMRcon2,Yunes_ModifiedPhysics,Berti_ModifiedReviewSmall}, where the fields are strong, non-linear, and highly dynamical.
GWs such as these carry multitudes of information across the universe regarding the local spacetime properties of the event, including clues highlighting the underlying theory of gravity driving the show.
For the past 100 years, Einstein's theory of GR has remained at its post as the prevailing theory of gravity, despite GW150914 and the following 10 events~\cite{GW_Catalogue} all being found to be consistent with his theory~\cite{TheLIGOScientific:2016pea,Abbott_IMRcon,Monitor:2017mdv,Abbott:2018lct}.
Even though the marvel of modern engineering that is the current LVC infrastructure~\cite{aLIGO} might not yet be sensitive enough to expose the subtle signs of a theory beyond GR, the next generation of ground- and space-based GW detectors~\cite{Ap_Voyager_CE,ET,LISA,B-DECIGO,DECIGO,TianQin} promise hefty sensitivity improvements across the GW frequency spectrum.
This may yet prove to finally be enough to study the traces of new hidden theory of gravity describing our universe. 

For the past 100 years, GR has been put under the microscope, with countless observations and tests performed in a wide variety of spacetime environments, all ultimately finding agreement with Einstein's famous theory.
Observations on the solar-system scale where gravity is weak and approximately static~\cite{Will_SolarSystemTest}, or the strong-field, static observations of binary pulsar systems~\cite{Stairs_BinaryPulsarTest,Wex_BinaryPulsarTest}, even cosmological observations~\cite{Ferreira_CosmologyTest,Clifton_CosmologyTest,Joyce_CosmologyTest,Koyama_CosmologyTest,Salvatelli_CosmologyTest}, and extreme-gravity observations of GWs~\cite{Abbott_IMRcon2,Yunes_ModifiedPhysics,Berti_ModifiedReviewSmall,TheLIGOScientific:2016pea,Abbott_IMRcon,Monitor:2017mdv,Abbott:2018lct}, have all ultimately found results remarkably consistent with the predictions of GR.

Even with the substantial list of past observational success, we must continue to test GR.
While this theory still explains all of our gravitational observations, there still remains several open questions which could potentially be explained by alternative theories of gravity.
To give a few examples, the accelerated expansion of the universe due to dark energy~\cite{Jain:2010ka,Salvatelli:2016mgy,Koyama:2015vza,Joyce:2014kja}, the inconsistent galactic rotation curves due to dark matter~\cite{Famaey:2011kh,Milgrom:DarkMatter,Milgrom:2008rv,Clifton:2011jh,Joyce:2014kja}, the matter/anti-matter asymmetry in the current universe~\cite{Clifton:2011jh,Famaey:2011kh}, the inflationary period of the early universe~\cite{Joyce:2014kja,Clifton:2011jh,Famaey:2011kh,Koyama:2015vza}, or even the question of unifying GR and quantum mechanics~\cite{Clifton:2011jh,Joyce:2014kja,Famaey:2011kh,Milgrom:2008rv,Jain:2010ka,Koyama:2015vza} all remain open to this day.
Several modified theories of gravity have been proposed to date, many of which have been found to explain some of the open questions remaining.
Similar to the historical Newtonian description of gravity, these advanced theories could potentially reduce to GR in weak-gravity environment, and activate in the un-probed extreme-gravity spacetimes.
For this reason, binary BH inspirals with immense gravitational fields and velocities reaching $\sim50\%$ the speed of light could very well prove to be vital in probing such a theory's untested side.

With such a large array of proposed modified gravity alternatives, how does one go about determining which one most accurately describes nature?
As always, we must rule them out one at a time with experimental observations.
For example, weak-field observations of the solar system and binary pulsar systems have placed very stringent constraints on several scalar-tensor theories~\cite{Jordan1959,Brans1961,Damour1993_2,Damour1993,Damour1996,Damour1992}, as well as the spontaneous scalarization of neutron stars~\cite{Sampson2014,Anderson2016}.
There are some remaining theories that have not yet been strongly constrained -- for example, theories with gravitational parity violation~\cite{Jackiw:2003pm,Yunes_dcs,Alexander_cs} which may not be activated in the weak-curvature systems currently studied.
In this chapter, we study various modified theories of gravity, all of which affect the gravitational waveform with different dependence on the relative velocities of binary constituents, or gravitational wave frequencies.

We currently live in a very exciting era of gravitational wave astronomy.  
With great success in both the past and on-going observing runs, many new ground-based GW interferometers are planned: several upgrades to the current LIGO infrastructure (Advanced LIGO, LIGO A\texttt{+}, LIGO Voyager)~\cite{Ap_Voyager_CE}, as well as new third-generation detectors like Cosmic Explorer (CE)~\cite{Ap_Voyager_CE} and the Einstein Telescope (ET)~\cite{ET}, each with improved sensitivity in the $1-10^4$ Hz range.
While such detectors have been designed with incredible sensitivities that are able to observe millions of events per year, with signal-to-noise ratios (SNRs) on the order of $10^3$~\cite{Zack:URrelations}, they can not probe the sub-unity frequency bands dominated by compact binary early inspirals, supermassive BH binaries, white-dwarf binaries, etc.
Space-based detectors such as LISA~\cite{LISA}, TianQin~\cite{TianQin}, B-DECIGO~\cite{B-DECIGO}, and DECIGO~\cite{DECIGO} on the other hand, have long Mm- to Gm-scale arms which allow them to accurately probe the low-frequency $10^{-4}-1$ Hz portion of the GW spectrum.
While ground-based detectors are proficient at observing the late, high-frequency, high-velocity, merger-ringdown portion of stellar-mass binary BH gravitational waveforms, space-based detectors can probe the early, low-frequency, low velocity inspiral portion.

Shortly after the observation of GW150914, Sesana~\cite{Sesana:2016ljz} pointed out that joint \textit{multi-band} detections of GW150914-like events\footnote{Multi-band GW observations are also possible for more massive binary BHs~\cite{AmaroSeoane:2009ui,Cutler:2019krq} and binary neutron stars~\cite{Isoyama:2018rjb}.} could be made using both LISA and ground-based detectors, with multi-band detection rates on the order of $\mathcal{O}(1)$~\cite{Gerosa:2019dbe,Sesana:2016ljz}.
Such events would first be observed in their early inspiral stage by space-based telescopes, until leaving the space-band at $1$ Hz for LISA or TianQin for several months before entering the ground-band to eventually merge at $\sim300$ Hz.
The early detections by space-based interferometers could give alert to EM telescopes~\cite{Sesana:2016ljz} for follow-up observations, as well as ground-based detectors, allowing for potential sensitivity optimizations which could be used to improve upon tests of GR~\cite{Tso:2018pdv}.
Similarly, ground-based observations will allow one to revisit sub-threshold space-based data, effectively lowering the detection threshold SNR from 15 to 9~\cite{Wong:2018uwb,Moore:2019pke}, and enhance the overall number of detections~\cite{Moore:2019pke,Cutler:2019krq}.
Additionally, multi-band GW observations can improve upon the measurement accuracy of many binary parameters, specifically the masses and sky positions~\cite{Nair:2015bga,Nair:2018bxj,Vitale:2016rfr,Cutler:2019krq}. 

In this chapter, we begin with a demonstration of the improvements one can gain with multi-band observations for the parameterized tests of GR~\cite{Nair:2015bga,Barausse:2016eii,Vitale:2016rfr,Gnocchi:2019jzp}.
We study a variety of example theories of gravity, including
Einstein-dilaton Gauss-Bonnet (EdGB) gravity, dynamical Chern-Simons (dCS) gravity, scalar-tensor theories, noncommutative gravities, theories with time-varying $G$, time-varying BH mass or modified dispersion relations.

Following this, we focus on the present and future implications on constraining non-GR theories with an additional massless scalar field, known as scalar tensor theories (STTs).
This has been studied for binary pulsar systems~\cite{Wex_BinaryPulsarTest,Anderson:2019eay}, pulsar-white dwarf (WD) systems~\cite{Freire:2012mg,Wex_BinaryPulsarTest,Shao:2017gwu,Anderson:2019eay}, and triple pulsar-WD-WD systems~\cite{Berti_ModifiedReviewLarge,Archibald:2018oxs}.
See also Refs.~\cite{Bonilla:2019mbm,DAgostino:2019hvh} for constraints on STTs from both GWs and pulsar timing measurements.
Here, we consider the present and future constraints obtained from the GW detections of mixed BH-NS coalescences.
In STTs, compact objects acquire scalar charges that source the scalar field. A scalar force acts between two scalarized objects, giving rise to a fifth force which depends on the internal structure of the massive objects and violates the strong equivalence principle (SEP). Binaries consisting of scalarized astrophysical objects further emit scalar dipole radiation (on top of gravitational quadrupolar radiation in GR), causing the binaries to evolve faster.  

Finally, we focus our attention on a particular string-inspired STT known as Einstein-dilaton Gauss-Bonnet (EdGB) gravity, where a dilaton scalar field is coupled to a quadratic curvature term in the action~\cite{Kanti_EdGB,Maeda:2009uy,Sotiriou:2013qea,Yagi:2015oca}, with coupling parameter $\alpha_\EdGB$.
With this new interaction in hand, BHs can become \textit{scalarized}~\cite{Campbell:1991kz,Yunes:2011we,Takahiro,Sotiriou:2014pfa,Berti_ModifiedReviewSmall,Prabhu:2018aun,Bakopoulos:2018nui,Antoniou:2017hxj,Antoniou:2017acq} (similar to conducting spheres becoming electrically charged), and a new fifth force interaction can be experienced between two such objects in a binary orbit.
Similar to analagous interactions found in nature (i.e. EM dipole radiation), such binary systems would decay faster than proposed by GR through additional scalar dipole radiation.

The current observational constraint found on the EdGB coupling parameter to date has been set to $\sqrt{\alpha_\EdGB}\lesssim2$ km~\cite{Kanti_EdGB,Pani_EdGB,Yagi_EdGB,Blazquez-Salcedo:2017txk,Witek:2018dmd,Nair_dCSMap,Yamada:2019zrb,Tahura:2019dgr}. Previous work on constraining EdGB gravity with GWs from BBH mergers mainly focused on looking at the correction in the inspiral due to the scalar dipole emission~\cite{Yunes_ModifiedPhysics,Nair_dCSMap,Yamada:2019zrb,Tahura:2019dgr}. Black hole quasinormal modes (QNMs) can also be used to probe this theory~\cite{Blazquez-Salcedo:2017txk}, while Ref.~\cite{Witek:2018dmd} estimated a rough bound on the theory from the dephasing due to the scalar field radiation computed via NR simulations.
Additionally, see Ref.~\cite{Barausse:2016eii} where the authors found constraints on dipole emission with space-based detector LISA, as well as multiband observations. 
See also a recent analysis by Ref.~\cite{Okounkova:2020rqw}, where the first NR model of an EdGB merger-ringdown waveform was presented, finding a coupling parameter constraint of $\sqrt{\alpha_\EdGB}\leq11$ km.

In this part of the chapter, we probe EdGB gravity with GWs from BBH mergers by including both inspiral and ringdown corrections.
The former correction is computed using the commonly used ppE formalism~\cite{Yunes:2009ke}, in which generic amplitude and phase modifications are introduced into the inspiral GR waveform, and the mapping to EdGB is known~\cite{Yagi:2011xp,Yunes_ModifiedPhysics,Tahura:2019dgr}.
The latter corrections are computed with the EdGB corrections to the individual QNM ringing frequency and damping time found in Ref.~\cite{Blazquez-Salcedo:2016enn} (see also~\cite{Blazquez-Salcedo:2017txk,Blazquez-Salcedo:2016yka,Blazquez-Salcedo:2018pxo}). Moreover, we take into account EdGB corrections to the final mass and spin of the remnant BH as a function of the initial masses and spins, which can be estimated from corrections to the orbital energy and angular momentum found in Ref.~\cite{Ayzenberg:2014aka}.
We then directly estimate the measurement accuracy of $\alpha_\EdGB$ using corrections to the full waveform, rather than only the inspiral ones as usually considered.
In previous analyses, BHs were assumed to be slowly-rotating and corrections were derived up to quadratic order in spin. 
Here, we estimate how much higher-order corrections in spin may affect bounds on $\alpha_\EdGB$ by deriving corrections to quartic order in spin as well.


\section{Modified theories of gravity}\label{sec:theory}

In this section we describe the several modified theories of gravity considered in this chapter, which can be thought of as breaking or deforming the fundamental pillars of GR.
First among them is the \textit{strong equivalence principle} (SEP) pillar~\cite{Will_SEP}, which states that the trajectories of free-falling and self-gravitating bodies are independent of their internal structure.
Second, the \textit{Lorentz invariance} (LI) pillar tells us that there is no preferred direction in our universe.
And last, the \textit{four dimensional spacetime} (4D) pillar, which conveys that the universal spacetime is composed of only four dimensions: 3 spatial and 1 temporal.
Finally, we consider the principle of \textit{massless gravitons} ($m_g$) as a result of GR, which describes gravity as being mediated by massless bosons traveling at the speed of light.
Theories which violate these fundamental pillars of GR can be broadly cataloged into two groups:
\begin{itemize}
\item \textbf{Modifications to GW generation mechanisms}: These modifications to GR alter how GWs are formed, and are active only during the coalescence event, with non-zero time derivatives of the source multipole moments.
Because of this, these theories depend only on the local properties of the source, such as the masses and spins.
\item \textbf{Modifications to GW propagation mechanisms}: These modifications alter the speed or dispersion relations of GWs themselves, and are only active  during their travel between their source and Earth.
Because of this, such theories depend on global properties such as the luminosity distance $D_L$ to the event.
\end{itemize} 
In the following subsections, we provide a brief description of the modified theories of gravity considered in this investigation, together with the mapping between the ppE parameter $\beta$ and the theoretical parameters.
We point the reader towards the more comprehensive analyses of Refs.~\cite{Berti_ModifiedReviewLarge,Berti_ModifiedReviewSmall,Yunes_ModifiedPhysics} for more complete descriptions of each theory.


\subsection{Scalar-tensor theories and Damour Esposito-Far$\acute{e}$se gravity}
Scalar-tensor theories which violate the SEP include a coupling into the Einstein Hilbert action, where the Ricci scalar $R$ is multiplied by some function of the scalar field $\phi$.
If such a scalar field is time-dependent with a growth rate of $\dot\phi$ (for example, from a cosmological background~\cite{Jacobson_STcosmo,Horbatsch_STcosmo,Berti_STcosmo}), BHs will accumulate scalar charges which accelerate the inspiral.
The mapping between $\beta_\ST$ and $\dot\phi$ is given by~\cite{Horbatsch_STcosmo,Jacobson_STcosmo}
\begin{equation}
\beta_\ppE^{(\ST)}=-\frac{5}{1792}\dot{\phi}^2\eta^{2/5}(m_1 s_1^\ST-m_2s_2^\ST)^2,
\end{equation}
where the dimensionless BH scalar charges $s_i^\ST$ are given by $s_i^\ST\equiv(1+\sqrt{1-\chi_i^2})/2$~\cite{Horbatsch_STcosmo}.
The ppE exponent is $b = -7$, thus the correction enters in the waveform at $-1$PN order.
We also note here that the small coupling approximation $\dot \phi \, m_i \ll 1$ must be upheld for meaningful constraints to be extracted, else the coupling terms in the action are not sufficiently small compared to the Einstein-Hilbert action.
The current most stringent constraint on $\dot\phi$ is $10^{-6}$ s$^{-1}$~\cite{Horbatsch_STcosmo} obtained from the orbital decay of a supermassive BH binary OJ287.

In STTs, compact objects acquire scalar charges that source the scalar field. A scalar force acts between two scalarized objects, giving rise to a fifth force which depends on the internal structure of the massive objects and violates the strong equivalence principle (SEP). Binaries consisting of scalarized astrophysical objects further emit scalar dipole radiation (on top of gravitational quadrupolar radiation in GR), causing the binaries to evolve faster. 

Such radiation becomes larger when the difference between the scalar charges of the binary constituents become larger, and thus a mixed binary consisting of one black hole and one neutron star system is ideal for probing such theories~\cite{Berti:spaceFreq,Takahiro,Sagunski:2017nzb,Huang:2018pbu}. 
Specifically, the increased mass difference $m_1^2-m_2^2$ and small total mass $M$ of a BH/NS system will minimize the allowable constraints on $\sqrt{\alpha_\EdGB}\sim M$ (for non-GR coupling parameter $\alpha_\EdGB$), while the offset acquired from the increased SNR of the alternative scenario of a large-mass binary BH system\footnote{In many cases, low-mass binary BH systems and BH/NS systems with slowly rotating BHs may be indistinguishable. However, in the Einstein-dilaton Gauss-Bonnet theory of gravity, the dipole-radiation slightly decreases  when comparing a BH/NS system with the equivalent binary BH system with the NS replaced by a slowly rotating BH but the effect is insignificant, and thus the constraints are not significantly affected.} only improves constraints by a small factor of $\sim \text{SNR}^{-1/4}$, while the SNR only itself increases by $\sim M^{5/6}$. Moreover, a smaller binary system has a lower relative velocity for a fixed frequency, which leads to an enhanced dipole radiation.
It is then extremely advantageous to decrease the total mass of the system rather than maximize it.
Such sources are particularly interesting and extremely timely to consider as two of the candidates in the O3 run by the LIGO/Virgo Collaborations, S190426c and S190814bv, are likely to be the merger of a black hole and a neutron star, if they are of astrophysical origin~\cite{gracedb,gracedb2}\footnote{We note that the LVC categorizes BH/NS candidate events as $m_1>5\text{ M}_\odot$ and $m_2<3\text{ M}_\odot$~\cite{classification}, and therefore O3 candidates S190426c and S190814bv potentially could be binary BH mergers. In addition, S190426c currently has a 58\% possibility of being terrestrial noise.}.
Because BHs have vanishing scalar charges in the quasi-Brans-Dicke theory of gravity considered in this analysis, the presence of a NS is required to place constraints on such a theory.
Therefore, it is of vital importance that, not only the events be of astrophysical origin, but they must with high confidence also be a BH/NS system rather than a binary BH system.

A particularly interesting class of theories within STTs is the quasi-Brans-Dicke theory. The former was introduced first by Damour and Esposito-Far$\acute{e}$se (DEF)~\cite{Damour:1992we,Damour:1996ke} which induces a non-linear growth of the scalar charges onto neutron stars called \textit{spontaneous scalarization}~\cite{Damour:1992we,Damour:1993hw,Damour:1996ke}, while black holes remain hairless as in GR. The latter is motivated from string theory and the dilaton scalar field is coupled with a quadratic curvature term (Gauss-Bonnet invariant) in the gravitational action~\cite{Kanti_EdGB,Maeda:2009uy}. In this theory, black holes have non-vanishing scalar charges~\cite{Campbell:1991kz,Yunes:2011we,Yagi_EdGBmap,Sotiriou:2014pfa} while neutron stars do not if the scalar field coupling is linear~\cite{Yagi_EdGBmap,Yagi:2015oca}.
In this chapter we consider the single BH-NS detections with future GW detectors, as well as the multi-band detections between both space- and ground-based detectors~\cite{Barausse:2016eii,Carson_multiBandPRL,Nair:2015bga,Nair:2018bxj}, and finally the combination of multiple observations~\cite{Takahiro,Abadie:2010cf} made on future detectors with expanded horizons.

In quasi-Brans-Dicke theories of gravity, matter fields couple to the scalar field $\varphi$ through the effective metric $A^2(\varphi)g_{\mu\nu}$~\cite{Shao:2017gwu,Damour:1993hw,Damour:1996ke,Anderson:2019hio}.
One can then define the gradient and curvature of the conformal potential $\ln A(\varphi)$ to be $\alpha(\varphi) \equiv d\ln{A(\varphi)}/d\varphi$, and $\beta(\varphi) \equiv d\alpha/d\varphi$.
In particular, we focus on the DEF model~\cite{Damour:1996ke,Damour:1992we}\footnote{See App.~\ref{app:theoryCompare} for a comparison with the similar Mendes-Ortiz (MO)~\cite{Mendes:2016fby} model.}, where the coupling function can be written in one of its simplest forms as $A(\varphi)=\exp{(\beta_0 \varphi^2/2)}$.
Such a theory can be completely characterized by the two weak-field parameters $\alpha_0=\alpha(\varphi_0)=\beta_0\varphi_0$ and $\beta_0=\beta(\varphi_0)$, where $\varphi_0= \alpha_0/\beta_0$ is the asymptotic value of the scalar field $\varphi$ at infinity.

Similarly in the strong-field case, NSs with mass $m_A$ couple to the scalar field with an effective coupling $\alpha_A=\partial \ln m_A/\partial\phi_0$, known as the (dimensionless) \textit{scalar charge} \footnote{Scalar charges depend on the NSs underlying equation-of-state (EoS). In this analysis we assume the APR4 EoS, consistent with the binary NS observation GW170817~\cite{LIGO:posterior,TheLIGOScientific:2017qsa}. See App.~\ref{app:EosSpinCompare} for a comparison between results found with different EoSs.} (the scalar charge for BHs is 0~\cite{Damour:1992we}).
Such scalar charges induce scalar dipole radiation in a compact binary, which enters at $-1$PN order relative to the leading GR quadrupole radiation and makes the binary evolve faster.
Following Ref.~\cite{Damour:1998jk,Shao:2017gwu,Littenberg:2018xxx}, one can derive the corresponding ppE correction to the waveform to be
\begin{equation}\label{eq:ppeBeta}
\beta_\ppE^{(\DEF)}=-\frac{5\eta^{2/5}(\Delta\alpha)^2}{7168}, \hspace{5mm} n=-1,
\end{equation}
where $\eta\equiv m_1m_2/M^2$ is the symmetric mass ratio, and $\Delta\alpha\equiv(\alpha_1-\alpha_2)$ is the difference in scalar charges between orbiting compact objects.
Additionally, see Sec.~\ref{sec:highorder} for a discussion and comparison on the inclusion of higher-order PN corrections to the waveform phase, as well as to the amplitude.
See also Ref.~\cite{Zhao:2019suc} for constraints from GW170817, and predictions for future binary NS detections.


\subsection{EdGB gravity}\label{sec:edgbCorrections}
Another interesting class of STTs is the EdGB theory of gravity.
In this particular theory, the ``dilaton" scalar field $\varphi$ is coupled to a quadratic curvature term in the action.
Correspondingly, the Einstein-Hilbert action is modified by the additional coupling term and the scalar field kinetic term~\cite{Kanti_EdGB,Maeda:2009uy,Sotiriou:2013qea}
\begin{equation}\label{eq:EdGB}
S_\EdGB=\int d^4x \sqrt{-g} \left[f(\varphi) \mathcal{R}^2_{\GB} - \frac{1}{2} \nabla_\mu \varphi \nabla^\mu \varphi \right]\,,
\end{equation}
where $g$ is the determinant of the metric $g_{\mu\nu}$ and $\mathcal{R}_{\GB}$ is the curvature-dependent Gauss-Bonnet invariant given by
\begin{equation}
\mathcal{R}^2_\GB\equiv R_{abcd}R^{abcd}-4R_{ab}R^{ab}+R^2.
\end{equation}
$f(\varphi)$ is a function of $\varphi$ and some types of string theory effectively reduces to the correction in Eq.~\eqref{eq:EdGB} with an exponential coupling between the scalar field and $\mathcal{R}^2_\GB$. If one expands such a function about a fiducial value of the scalar field $\varphi_0$, the leading constant term does not contribute to the field equations since $\mathcal{R}^2_\GB$ is a topological invariant. Thus, the leading effect arises from a linear coupling, and in this chapter, we consider\footnote{Additionally, refer to~\cite{Bakopoulos:2018nui,Antoniou:2017hxj,Antoniou:2017acq} for more general couplings.}
\begin{equation}\label{eq:linear-coupling}
f(\varphi) = \alpha_\EdGB \varphi\,,
\end{equation}
where $\alpha_\EdGB$ is the coupling parameter of EdGB gravity.

In scalar-tensor theories of gravity including EdGB gravity, compact objects can accumulate scalar monopole charges, which in turn source a scalar field.
This effect is naturally analogous to the classical effects of electric/mass/color charges sourcing the electric/gravitational/strong fields.
Pairs of such scalarized objects will then give rise to a new ``fifth force" interaction between them, altering their ensuing trajectories.
This effect is dependent on the internal structures of the compact objects, therefore violating the strong equivalence principle, one of the fundamental pillars of GR.
While two such compact objects orbiting each other in a binary system will decay under the emission of gravitational radiation (as predicted by GR)\footnote{Gravitational radiation is also modified from GR in EdGB gravity, though such an effect enters at higher order than the scalar dipole radiation in the binary evolution.}, the new scalar interaction will additionally induce scalar dipole radiation.
This effect will of course accelerate the coalescence process more than the predictions of GR estimate.

Scalar charges in EdGB gravity with a linear coupling as in Eq.~\eqref{eq:linear-coupling} only anchor to BHs~\cite{Campbell:1991kz,Yunes:2011we,Takahiro,Sotiriou:2014pfa,Berti_ModifiedReviewSmall,Prabhu:2018aun}, and not to other objects such as neutron stars~\cite{Yagi:2011xp,Yagi:2015oca}.
Such scalar charges $s$ depend on the BH's spin, mass, and the EdGB coupling parameter, and have been found to be~\cite{Berti_ModifiedReviewSmall,Prabhu:2018aun}
\begin{equation}
s_i^{(\EdGB)}=2\frac{\sqrt{1-\chi_i^2}-1+\chi_i^2}{\chi_i^2}\frac{\alpha_\EdGB}{M}.
\end{equation}

EdGB gravity may be treated as an effective field theory only if the correction $S_\EdGB$ to the action is much smaller than the Einstein-Hilbert action $S_\EH$.
Such an assumption allows one to neglect the higher-order curvature terms of order $\mathcal{O}(R^3)$.
This approximation is known as the \textit{small coupling approximation}, and enforces the requirement that
\begin{equation}
\zeta_\EdGB \equiv \frac{16\pi\alpha_\EdGB^2}{M^4} \ll 1,
\end{equation}
for binaries with total mass $M\equiv m_1+m_2$.
If this inequality fails to be upheld, constraints on $\alpha_\EdGB$ are deemed to be invalid, as the assumption $S_\EdGB \ll S_\EH$ no longer holds.
In addition, beyond the small-coupling approximation corresponds to large couplings between the scalar field and the curvature, which have largely been ruled out with observations.
Typically, constraints on the EdGB coupling parameter are presented for the quantity $\sqrt{\alpha_\EdGB}$, which has units of length (commonly in km).
Current constraints on this quantity have been found to be $10^7$ km from solar system observations~\cite{Amendola_EdGB}, and $\mathcal{O}(1\mathrm{km})$ from theoretical considerations, and observations of BH low-mass X-ray binaries, neutron stars, and GWs~\cite{Kanti_EdGB,Pani_EdGB,Yagi_EdGB,Witek:2018dmd,Nair_dCSMap,Yamada:2019zrb,Tahura:2019dgr}.
We urge caution that the constraint $\sqrt{\alpha_\EdGB} \lesssim 2$ found in Refs.~\cite{Nair_dCSMap,Yamada:2019zrb} take into account certain approximations which warrant such results to be a rough estimate.
For example, the authors of Ref.~\cite{Nair_dCSMap} used posterior samples provided by the LIGO/Virgo Collaborations though data more finely-sampled around the GR value seem to be necessary to derive a more reliable posterior distribution on $\sqrt{\alpha_\EdGB}$.
In addition, the authors utilize $-1$PN constraints obtained by the LIGO-Virgo analysis of the binary black hole signals, where it is warned that they cannot be interpreted as dipole radiation constraints, due to higher-order non-negligible terms from dipole radiation.
Also, Ref.~\cite{Yamada:2019zrb} utilized a simplified grid-search method rather than a full stochastic sampling procedure.

Now let us find the various corrections to the gravitational waveform under an EdGB prescription of gravity.
This includes corrections to the inspiral portion of the waveform, to the remnant BH's QNMs, and finally to the remnant BH's mass and spin predictions. 
We point out a recent analysis in Ref.~\cite{Okounkova:2020rqw}, in which a NR binary black hole merger-ringdown waveform in EdGB gravity was presented for the first time.

\subsubsection{Inspiral}\label{sec:inspiral}
In our analysis, we consider the commonly-used ppE formalism~\cite{Yunes:2009ke}\footnote{ppE phase corrections have a one-to-one correspondence to the inspiral corrections in the generalized IMRPhenom formalism~\cite{Abbott_IMRcon2} used by LVC~\cite{Yunes_ModifiedPhysics}.} to enact EdGB corrections to the inspiral gravitational waveform.
See Chapter~\ref{chap:GWs} for a summary of the ppE formalism, which allows one to modify the phase and amplitude of the GR waveform with generic parameterized corrections, in a theory-agnostic way.
In this part of the chapter, we solely consider corrections to the GR inspiral waveform derived from the EdGB theory of gravity.
Such a theory affects the waveform amplitude at $-1$PN order ($a=-2$) with magnitude~\cite{Tahura_GdotMap}
\begin{equation}\label{eq:alpha_ppE}
\alpha_\ppE^{(\EdGB)}=-\frac{5}{192}\zeta\frac{(m_1^2 \tilde{s}^{(\EdGB)}_2-m_2^2\tilde{s}^{(\EdGB)}_1)^2}{M^4\eta^{18/5}},
\end{equation}
with $\tilde s^{(\EdGB)}_i = s^{(\EdGB)}_i m_i/\alpha_\EdGB$.
Similarly, the waveform phase is modified at $-1$PN order ($b=-7$) with magnitude~\cite{Yagi_EdGBmap}
\begin{equation}\label{eq:beta_ppE}
\beta_\ppE^{(\EdGB)}=-\frac{5}{7168}\zeta\frac{(m_1^2 \tilde{s}^{(\EdGB)}_2-m_2^2\tilde{s}^{(\EdGB)}_1)^2}{M^4\eta^{18/5}}.
\end{equation}
To have consistency with other EdGB corrections to be explained later, we only keep up to quadratic order in BH spins.
However, in Sec.~\ref{sec:4thOrdSpin} we consider EdGB corrections to the waveform up to $\mathcal{O}(\chi^4)$, comparing it to those found here.
For the remainder of this chapter, corrections labeled ``inspiral" correspond to the addition of both phase and amplitude corrections to the GR inspiral waveform.
We note that Tahura \textit{et al.}~\cite{Tahura:2019dgr} showed that corrections to the GR amplitude is not as important as those in the phase, but we keep the former as well in this chapter for completeness.

\subsubsection{Ringdown}\label{sec:mergerRingdown}
While the ppE formalism described above allows us to include EdGB corrections to the inspiral description of the waveform, we can additionally model corrections to the ringdown waveform.
As the orbits of the inspiraling BHs decay under the emission of gravitational radiation, they eventually become close enough to each other to enter plunging orbits, where a common horizon is formed as they merge together.
The remnant BH then relaxes down to its final state via the radiation of QNMs~\cite{Berti:2009kk}.
QNMs can be described by just two parameters: the ringdown frequency $f_\RD$ and the damping frequency $f_\text{damp}$~\cite{Berti:2005ys,Berti:2009kk}.
We refer the readers to Ref.~\cite{Maselli:2019mjd} where similar corrections were made, and constraints with multiple GW events were quantified.
See also Refs.~\cite{McManus:2019ulj,Cardoso:2019mqo} where a general formalism to map ringdown corrections directly to specific theories of gravity was developed.
$f_\RD$ and $f_\text{damp}$ are described by the remnant BH's mass and spin $M_f$ and $\chi_f$ (from the BH no-hair theorem), which in turn only depend on $m_1$, $m_2$, $\chi_1$, and $\chi_2$ of the original BH binary system obtained through NR simulations~\cite{PhenomDII}.

However, within the EdGB viewpoint of gravity, the QNMs additionally depend upon the EdGB coupling parameter $\zeta_\EdGB$.
In this analysis, we attempt to model corrections to the ringdown and damping frequencies $f_\RD$ and $f_\text{damp}$ up to first order in $\zeta_\EdGB$, like so
\begin{align}
f_{\RD} &=f_{\RD,\GR}+\zeta_\EdGB f_{\RD,\zeta} + \mathcal{O}(\zeta_\EdGB^2),\label{eq:fRD}\\
f_{\text{damp}} &=f_{\text{damp},\GR}+\zeta_\EdGB f_{\text{damp},\zeta} + \mathcal{O}(\zeta_\EdGB^2),\label{eq:fdamp}
\end{align}
where $f_{\RD,\GR}$ and $f_{\text{damp},\GR}$ are the GR QNM frequency predictions~\cite{PhenomDI,PhenomDII}, and $f_{\RD,\zeta}$ and $f_{\text{damp},\zeta}$ are the first order EdGB corrections. 
To derive such QNM frequency corrections to first order in $\zeta_\EdGB$, we use the results in Ref.~\cite{Blazquez-Salcedo:2016enn} to compute the complex QNM frequency up to quadratic order in spin $\chi_f$ of the remnant BH\footnote{Reference~\cite{Blazquez-Salcedo:2016enn} follows a slightly different EdGB notation than considered here, beginning with the coupling parameter $\alpha_\EdGB$ in the action as well as their definition of $\zeta_\EdGB'$. The quantities can be mapped to our definitions by letting $\zeta_\EdGB' \rightarrow 4 \sqrt{\zeta_\EdGB}$.}.
We take note that this assumption was tested for accuracy, where it was found that by taking EdGB corrections to the waveform up to $\mathcal{O}(\chi^4)$ from Ref.~\cite{Maselli:2019mjd}, the resulting Fisher-estimated constraints only varied by $\sim1.5\%$ from the $\mathcal{O}(\chi^2)$ case, well within the accuracies of this analysis.
We consider the leading order, $\ell = m = 2$ axial and polar QNMs. 
As discussed in Ref.~\cite{Blazquez-Salcedo:2016enn}, the QNM non-spinning components  have been computed for both of these modes.
The spinning components of the axial modes were then obtained by adopting the null geodesic correspondence\footnote{See Refs.~\cite{Silva:2019scu,Glampedakis:2017dvb,Glampedakis:2019dqh} where the null geodesic correspondence was used to estimate corrections for rotating BHs.}~\cite{Yang:2012he} since such modes do not couple to the scalar field perturbation. 
The spinning components of the polar modes in EdGB gravity is currently unknown, though based on the claim in~\cite{Blazquez-Salcedo:2016enn}, we assume the polar modes have the same spin dependence as the axial modes to carry out a rough estimate on the latter.

Now let us compute the ringdown and damping frequencies.
Such QNM frequencies can be written as
\begin{align}
2 \pi M_f f_{\RD,\zeta} &=  \left(\mathcal{R}_{\text{ns},0}+\mathcal{R}_{\text{ns},1}\zeta_\EdGB \right)\left(1+\mathcal{R}_{\text{s},0}+\mathcal{R}_{\text{s},1}\zeta_\EdGB\right) + \mathcal{O}(\zeta_\EdGB^2), \\
2 \pi M_f f_{\mathrm{damp},\zeta} &=\left(\mathcal{I}_{\text{ns},0}+\mathcal{I}_{\text{ns},1}\zeta_\EdGB \right)\left(1+\mathcal{I}_{\text{s},0}+\mathcal{I}_{\text{s},1}\zeta_\EdGB\right) + \mathcal{O}(\zeta_\EdGB^2), 
\end{align}
where $\mathcal{R}_{\text{ns/s},i}$ and $\mathcal{I}_{\text{ns/s},i}$ $(i=0,1)$ are arbitrary non-spinning (ns) and spinning (s) coefficients. 
The above expressions are expanded to linear order in $\zeta$ once again in the following analysis.
The non-spinning components are tabulated in Table~\ref{tab:QNMs}, while the spinning components are given by
\begin{align}
\mathcal{R}_{\text{s},0} &=0.3849\chi_f+0.2038\chi_f^2+ \mathcal{O}(\chi_f^3),\\ 
 \mathcal{R}_{\text{s},1} &=0.5216\chi_f+0.4224\chi_f^2+ \mathcal{O}(\chi_f^3),\\ 
 \mathcal{I}_{\text{s},0} &=-0.0741\chi_f^2+ \mathcal{O}(\chi_f^3),\\ 
 \mathcal{I}_{\text{s},1} &=-0.0944\chi_f-0.1056\chi_f^2+ \mathcal{O}(\chi_f^3).
\end{align}
Finally, we find the EdGB corrections to the ringdown and damping frequencies as
\begin{align}
f_{\RD,\zeta}&= \frac{\mathcal{R}_{\text{ns},1}(1+\mathcal{R}_{\text{s},0})+\mathcal{R}_{\text{ns},0} \mathcal{R}_{\text{s},1}}{2\pi M_f},\label{eq:fRD}\\
f_{\text{damp},\zeta} &= \frac{\mathcal{I}_{\text{ns},1}(1+ \mathcal{I}_{\text{s},0}) + \mathcal{I}_{\text{ns},0} \mathcal{I}_{\text{s},1}}{2\pi M_f}.
\label{eq:fdamp}
\end{align}
which can be simplified to
\begin{align}
f_{\RD,\zeta}&= \frac{a_0(1+a_1\chi_f+a_2\chi_f^2)}{2\pi M_f} + \mathcal{O}(\chi_f^3),\label{eq:fRD}\\
f_{\text{damp},\zeta} &= \frac{b_0(1+b_1\chi_f+b_2\chi_f^2)}{2\pi M_f} + \mathcal{O}(\chi_f^3),
\label{eq:fdamp}
\end{align}
where $a_i$ and $b_i$ are presented in Table~\ref{tab:Mfchif}.
In this analysis, we include EdGB corrections into the merger-ringdown waveform up to linear order in $\zeta_\EdGB$ and quadratic in $\chi_f$ using the above prescription, with corrections like so labeled as ``axial/polar QNMs".
See Sec.~\ref{sec:4thOrdSpin} for a demonstration of the inclusion of spin effects into the remnant BH QNMs, where we include corrections up to $\mathcal{O}(\chi^4)$, and also remove all spin effects.
In particular, we find that such spin effects only affect the constraints $\sqrt{\alpha_\EdGB}$ by at most $1.6\%$.

\renewcommand{\arraystretch}{1.2}
\begin{table}
\centering
\addvbuffer[12pt 8pt]{\begin{tabular}{c c c c}
$\mathcal{R}_{\text{ns},0}$ &$\mathcal{R}_{\text{ns},1}$ &$\mathcal{I}_{\text{ns},0}$&$\mathcal{I}_{\text{ns},1}$ \\
\hline
\hline
$0.3737$&$-0.1874$&$-0.08896$&$-0.0622$
\end{tabular}}
\caption[Remnant BH quasinormal modes in EdGB gravity]{
Non-spinning parameters $\mathcal{R}_{\text{ns},i}$ and $\mathcal{I}_{\text{ns},i}$~\cite{Blazquez-Salcedo:2016enn} required to reconstruct the EdGB corrections to the ringdown and damping frequencies $f_{\RD,\zeta}$ and $f_{\text{damp},\zeta}$ as found in Eqs.~\eqref{eq:fdamp}.
}\label{tab:QNMs}
\end{table}

\renewcommand{\arraystretch}{2}
\begin{table}
\centering
\resizebox{.8\textwidth}{!}{%
\addvbuffer[12pt 8pt]{\begin{tabular}{c c c}
$a_0$&$a_1$&$a_2$\\
\hline
$-0.1874$ & $-0.6552$ & $-0.6385$\\
\hline
$b_0$&$b_1$&$b_2$\\
\hline
$-0.0622$ & $-0.1350$ & $-0.2251$\\
\hline
\hline
$c_0$&$c_1$&$c_2$\\
\hline
$\frac{43740-2233\sqrt{2}\eta^2}{262440\eta}$ & $\frac{50659\sqrt{3}\eta^2-116640\sqrt{6}}{12(2233\sqrt{2}\eta^2-43740)}$ & $\frac{1361569247\sqrt{2}\eta^2-1285956000}{264600(2233\sqrt{2}\eta^2-43740)}$\\
$d_0$&$d_1$&$d_2$\\
\hline
$\frac{13571}{29160 \sqrt{3}}$&$\frac{75371}{40713}\sqrt{\frac{2}{3}}$&$\frac{58180627}{149620275}$\\
\end{tabular}}
}
\caption[Coefficients for EdGB corrections to the gravitational waveform]{
Coefficients $a_i$ $b_i$, $c_i$, and $d_i$ required for the reconstruction of the EdGB corrections to the remnant BH QNM ringdown and damping frequencies $f_\RD$, $f_\text{damp}$, as well as the mass and spin $M_{f,\zeta}$ and $\chi_{f,\zeta}$ as found in Eqs.~\eqref{eq:fRD},~\eqref{eq:fdamp},~\eqref{eq:Mf} and~\eqref{eq:chif} respectively.
}\label{tab:Mfchif}
\end{table}

\subsubsection{Remnant BH mass and spin}\label{sec:MfChif}
In addition to the direct waveform modifications displayed in the preceding sections, a post-merger remnant BH in EdGB gravity will settle down into a non-GR final mass and spin configuration, due to the increased levels of energy and angular momentum radiation.
This effect will also indirectly modify the gravitational waveform.
In GR, the final spin angular momentum of the post-merger BH can be roughly approximated to be the sum of the spin angular momentum of the initial BHs and the orbital angular momentum $L_z$ of a particle with mass $\mu=m_1 m_2/M$ orbiting about the remnant BH at the radius of the ISCO, $r_\ISCO$~\cite{Barausse:2009uz}.
More specifically, the full expression for a spin-aligned system is found to be~\cite{Barausse:2009uz,Yunes_ModifiedPhysics}
\begin{equation}
\mu L_z(M,\chi_f,r_\ISCO)=M(M \chi_f-a_s-\delta_m a_a)
\end{equation}
where $a_{s,a}\equiv (m_1\chi_1 \pm m_2\chi_2)/2$ are the symmetric/anti-symmetric combinations of spins, and $\delta_m\equiv (m_1-m_2)/M$ is the weighted mass difference, and  $L_z$ is the specific orbital angular momentum.
Similarly, the final mass of the remnant BH $M_f$ can be expressed in relation to the specific orbital energy $E_\text{orb}$ 
of a particle with mass $\mu$ orbiting at $r_\ISCO$ as
\begin{align}\label{eq:M_f}
\mu \left[1-E_\text{orb}(M_f,\chi_f,r_\ISCO)\right]=M-M_f
\end{align}
Here $1-E_b\equiv E_\text{orb}$ is equivalent to the binding energy of the particle. 

We here make an assumption that the above GR picture also holds in EdGB gravity and derive corrections to $M_f$ and $\chi_f$. To do so, we take into account the EdGB corrections to $E_\text{orb}$, $L_z$ and $r_\ISCO$. Unfortunately, these expressions are not known to all orders in the BH spin. Thus, we use the expressions valid to quadratic order in spin presented in~\cite{Ayzenberg:2014aka}. In addition, there is a scalar interaction between two scalarized BHs, and thus Eq.~\eqref{eq:M_f} needs to be modified to
\begin{equation}
\mu \left[1-E_\text{orb}(M_f,\chi_f,r_\ISCO)-E_\text{scalar}(\mu,M,\chi_f,r_\ISCO,\zeta_\EdGB) \right]=M-M_f.
\end{equation}
Here~\cite{Stein:2013wza},
\begin{equation}
E_\text{scalar}(\mu,M,\chi_f,r_\ISCO,\zeta_\EdGB)=\frac{\zeta_\EdGB}{\eta ^2}\left(1-\frac{\chi_f^2}{4}\right) \frac{M}{r},
\end{equation}
corresponds to the specific scalar interaction energy between the particle (with mass $\mu$ and zero spin) and the central BH (with mass $M_f$ and spin $\chi_f$).
See Sec.~\ref{sec:4thOrdSpin} for an investigation into higher-order spin effects up to $\mathcal{O}(\chi^4)$ for each EdGB correction considered here.

Similar to the merger-ringdown corrections to the QNM ringing and damping frequencies, we consider corrections to the remnant BH mass and spin to linear order in $\zeta_\EdGB$ and quadratic (and also quartic) in $\chi_f$.
The complete expressions for $M_f$ and $\chi_f$ can then be written as
\begin{align}
M_f &=M_{f,\GR}+\zeta_\EdGB M_{f,\zeta} + \mathcal{O}(\zeta_\EdGB^2),\label{eq:Mf} \\
\chi_f &=\chi_{f,\GR}+\zeta_\EdGB \chi_{f,\zeta} + \mathcal{O}(\zeta_\EdGB^2),\label{eq:chif}
\end{align}
where $M_{f,\GR}$ and $\chi_{f,\GR}$ are the GR predictions of the final mass and spin from the NR fits of Ref.~\cite{PhenomDII}, and $M_{f,\zeta}$ and $\chi_{f,\zeta}$ are the resulting EdGB corrections at first order in $\zeta_\EdGB$.
Having the above expressions for EdGB energy and angular momentum in hand, one can estimate the EdGB corrections to these quantities as
\begin{align}
M_{f,\zeta}&= M c_0 \left( 1+c_1 \chi_f+c_2\chi_f^2 \right)  + \mathcal{O}\left(\chi_f^3\right), \label{eq:Mf}\\
\chi_{f,\zeta} &= -d_0\eta \left(1 + d_1 \chi_{f,\GR} + d_2 \chi_{f,\GR}^2\right) + \mathcal{O}\left(\chi_{f,\GR}^3\right),\label{eq:chif}
\end{align}
where $c_i$ and $d_i$ are presented in Table~\ref{tab:Mfchif}.
Observe that the above expressions themselves depend on the remnant BH spin in GR ($\chi_{f,\GR}$), found in Ref.~\cite{PhenomDII}.
We also note that the above expression for $M_f$ depends on the solution for $\chi_f$. 
$M_f$ is then expanded once again after $\chi_f$ has been substituted in to quadratic order in spin.


Finally, in this part of the chapter we investigate the effect of each type of EdGB correction present in the template waveform: inspiral and ringdown effects.
In particular, we consider the following five cases in which we perform a Fisher analysis:
\begin{enumerate}
\item \textit{Inspiral:} EdGB corrections only in the inspiral waveform.
\item \textit{Axial QNMs:} EdGB corrections only in the ringdown waveform for the case of purely axial QNMs.
\item \textit{Polar QNMs:} Same as 2 but with polar QNMs.
\item \textit{Inspiral+Axial QNMs:} Combination of 1 and 2, with corrections to both the inspiral and ringdown portions.
\item \textit{Inspiral+Polar QNMs:} Same as 4 but for polar QNMs.
\end{enumerate}
We include remnant BH mass and spin corrections within only the latter four cases listed above.
Within each of the above listed cases, we compare the results from each detector and event considered.


\subsection{dCS gravity}
Similar to EdGB gravity, dCS gravity is a SEP-violating effective field theory which modifies the Einstein-Hilbert action with a quadratic curvature term called the Pontryagin density, which violates parity, and is non-minimally coupled to a scalar field~\cite{Jackiw:2003pm,Alexander_cs}.
Scalar dipole charge is accumulated on the BHs, inducing scalar quadrupole radiation which in turn accelerates the inspiral.
The magnitude of the correction is proportional to the dCS coupling parameter $\alpha_\dCS$. 
The mapping between $\beta_\ppE^{(\dCS)}$ and $\alpha_\dCS$ can be written as~\cite{Nair_dCSMap}
\begin{align}
 \nonumber \beta_\ppE^{(\dCS)}=&-\frac{5}{8192}\frac{16\pi\alpha_\dCS^2}{M^4\eta^{14/5}}\frac{ m_1 \left(s_2^{(\dCS)}\right)^2-m_2 \left(s_1^{(\dCS)}\right)^2}{M^2}\\
 &+\frac{15075}{114688}\frac{16\pi\alpha_\dCS^2}{M^4\eta^{14/5}}\left( \frac{m_1^2 \chi_1^2 +m_2^2 \chi_2^2}{M^2}-\frac{305}{201}\eta \chi_1 \chi_2  \right),
\end{align}
where the dimensionless BH scalar charge can be written as $s_i^{(\dCS)}=(2+2\chi_i^4-2\sqrt{1-\chi_i^2}-\chi_i^2 \lbrack 3-2\sqrt{1-\chi_i^2} \rbrack)/2\chi_i^3$~\cite{Yagi:2012vf}. The ppE exponent is $b = -1$, which corresponds to a $+2$PN correction.
We also note that once again the small coupling approximation $\zeta_\dCS \equiv 16\pi\alpha_\dCS^2/M^4 \ll 1$ must be satisfied in order to place meaningful constraints on $\sqrt{\alpha_\dCS}$.
Current constraints on $\sqrt{\alpha_\dCS}$ are obtained from solar system and table-top experiments as $10^8$ km~\cite{AliHaimoud_dCS,Yagi_dCS}.

\subsection{Noncommutative gravity (NC)}
Noncommutative theories of gravity~\cite{Harikumar:2006xf} have been proposed to quantize the spacetime coordinates, which have been promoted to operators~\cite{Snyder:QuantizedST} $\hat{x}^\mu$, in order to eliminate the quantum field theory ultraviolet divergences.
Such theories have the ultimate goal of unifying the theories of GR and quantum mechanics.
The spacetime operators as such, satisfy the familiar canonical commutation relations
\begin{equation}
\lbrack \hat{x}^\mu , \hat{x}^\nu \rbrack = i \theta^{\mu\nu},
\end{equation}
where $\theta^{\mu\nu}$ quantifies the ``fuzziness" of spacetime coordinates, similar to the reduced Planck's constant $\hbar$ in quantum mechanics.
Within this non-commutating formalism, we strive to constrain the scale of quantum spacetime.
A useful parameter to do so normalizes the magnitude of $\theta^{\mu\nu}$ to the Planck length and time scales $l_p$ and $t_p$: $\Lambda^2 \equiv \theta^{0i}\theta_{0i}/l_p^2t_p^2$.
The Lorentz-violating effects from noncommutative gravity enters the gravitational waveform at $+2$PN order ($b=-1$), and has the ppE phase correction given by~\cite{Tahura_GdotMap}
\begin{equation}
\beta_\ppE^{(\NC)}=-\frac{75}{256}\eta^{-4/5}(2\eta-1)\Lambda^2.
\end{equation}
The current constraints on the scale of quantum spacetime $\sqrt{\Lambda}$ come from the GW observation of GW150914, found to be $\sqrt{\Lambda}<3.5$~\cite{Kobakhidze:2016cqh}, which is on the order of the Planck scale.

\subsection{Time-varying $G$ theories}
The gravitational constant $G$ may vary with time at a rate of $\dot G$, producing an anomalous acceleration of the binary system. 
In this SEP-violating theory, the mapping between $\beta_\ppE^{(\dot G)}$ and $\dot G$ is given by~\cite{Yunes_GdotMap,Tahura_GdotMap}
\begin{align}
 \beta_\ppE^{(\dot G)}=&-\frac{25}{851968}\eta^{3/5}\dot{G}_C\lbrack 11M+3(s_1^{(\dot G)}+s_2^{(\dot G)})M-41(m_1s_1^{(\dot G)}+m_2s_2^{(\dot G)}) \rbrack,
\end{align}
with $b= - 13$ ($-4$PN order).
Here, the sensitivities are given by $s_i^{(\dot G)}\equiv -\frac{G_C}{\delta G_C}\frac{\delta m_i}{m_i}$.
The current strongest constraint on $|\dot G|$ is $(0.1-1)\times 10^{-12}$ yr$^{-1}$~\cite{Bambi_Gdot,Copi_Gdot,Manchester_Gdot,Konopliv_Gdot}.

\subsection{Time-varying BH mass theories}
Some (4D-pillar-violating) modified theories of gravity as well as astrophysical processes predict time-variation in the BH mass, $\dot{m}_A$.
Many of the string-inspired models suggest that our four-dimensional brane spacetime is embedded in larger dimensional bulks~\cite{ArkaniHamed:1998rs,ArkaniHamed:1998nn,Randall:1999ee,Randall:Braneworld,Berti_ModifiedReviewSmall}.
One example is the RS-II~\cite{Randall:Braneworld} ``braneworld" model by Randall and Sundrum, in which BHs may evaporate \textit{classically}~\cite{Emparan:2002px,Tanaka:2002rb}\footnote{This scenario is now in question given the construction of brane-localized \textit{static} BH solutions~\cite{Figueras:2011gd,Abdolrahimi:2012qi}.}. 
The evaporation rate is proportional to the size $l$ of the extra dimension, which has previously been constrained to $(10-10^3)$ $\mu$m~\cite{Adelberger:2006dh,Johannsen_ED,Johannsen_ED2,Psaltis_ED,Gnedin_ED}.
Such a modification to the BH mass enters the waveform at $-4$PN order ($b=-13$), and $\beta_\ppE^{(\dot{M})}$ can be mapped to $\dot{M}$ via~\cite{Yagi_EDmap}
\begin{equation}
\beta_\ppE^{(\dot{M})}=\frac{25}{851968} \dot{M}\frac{3-26\eta+34\eta^2}{\eta^{2/5}(1-2\eta)}.
\end{equation}
The evaporation rate of the binary system $\dot M$ can be written as a function of $l$~\cite{Emparan_Mdot,Berti_ModifiedReviewSmall} in the RS-II model, or mapped to any other desired model.
Alternatively, BH mass losses can be explained by cosmological effects such as the accretion of dark (or ``phantom") energy~\cite{Babichev:2014lda,Babichev:2005py,Babichev:2004yx}.
For comparison purposes, we compute the astrophysical Eddington mass accretion rate $\dot{M}_{\text{Edd}}$ at which the BH radiates the Eddington luminosity $L_{\text{Edd}}$.
For a GW150914-like binary BH, it is found to be $\dot{M}_{\text{Edd}} = 1.4\times10^{-6}\text{ M}_\odot/\text{yr}$.

\subsection{Dynamical graviton mass}
The ``dynamical massive graviton" theory~\cite{Zhang:2017jze} (violating the massless graviton pillar of GR) models the graviton's mass to be smaller than all current constraints in weak gravity regions (see Sec.~\ref{sec:MDR} below), while becoming much larger in dynamical, strong-field spacetimes such as in the presence of binary BH mergers.
As such, this theory enters the gravitational waveform as a generation modification, rather than the usual propagation mechanism.
Here, we offer a new ppE correction to the gravitational waveform via the fractional discrepancy between the observed and predicted decay rates of the binary system's period $\dot{P}$ found in Ref.~\cite{Finn:2001qi}. 
In particular, we focus on a class of massive gravity theories that correctly reduces to GR in the limit $m_g \to 0$~\cite{deRham:2016nuf} by abandoning Lorentz invariance~\cite{Finn:2001qi}.
We found that such an effect enters the waveform at $-3$PN order ($b=-11$), with the correction given by
\begin{equation}
\beta_\ppE^{(m_g)}=\frac{25}{19712}\frac{\mathcal{M}^2}{\hbar^2F(e)}m_g^2,
\end{equation}
where $F(e)$ is a function of the eccentricity (Eq. (4.11) of Ref.~\cite{Finn:2001qi}), taken to be $1$ for our analysis (corresponding to quasi-circular binaries).
Current constraints on the dynamical graviton mass have been found to be $5.2\times10^{-21}$ eV from binary pulsar observations~\cite{Miao:2019nhf}, and $\sim 10^{-14}$ eV from GW measurements~\cite{Chung:2018dxe}.

As we discuss in the next section, the mass of the graviton also changes the propagation of GWs. However, the amount of the graviton mass can be different between (i) in the vicinity of a BH and (ii) in the region where GWs propagate from a source to us. Thus, we treat these effects separately in this chapter. We consider these effects one at a time, though one could introduce two different graviton masses, like the dynamical and propagating graviton mass, and measure these two graviton masses. However, the dynamical graviton mass that gives rise to non-GR corrections at the level of the GW generation introduces modifications to the waveform phase at $-3PN$, while those from modifications to the GW propagation enters at 1PN order. Since these two PN orders are well-separated, the amount of correlation is small, and thus we expect the bound presented here gives us a good estimate on each effect.

\subsection{Modified dispersion relations (MDR)}\label{sec:MDR}

Now let us discuss the modified dispersion of GWs.
Modifications to the propagation of GWs activate during their transport between the binary coalescence source and Earth.
As such, these modifications typically violate the LI pillar of gravity as well as the massless graviton, and describe corrections to the frequency dispersion of GWs, which in turn modifies the propagation speed of GWs.
These modifications depend primarily on the distance between the binary and Earth.

In general, the dispersion relation for GWs with modified theories of gravity takes the following form~\cite{Mirshekari_MDR}:
\begin{equation}
E^2=p^2+\mathbb{A} \, p^{a_\MDR},
\end{equation}
where $E$ and $p$ are the graviton's energy and momentum, $a_\MDR$ is related to the PN order via $n=1+\frac{3}{2}a_\MDR$, and $\mathbb{A}$ corresponds to the strength of the dispersion. 
The mapping between $(\beta_\ppE^{(\MDR)},b)$ and $(\mathbb{A},a_\MDR)$ is given by~\cite{Mirshekari_MDR}
\begin{eqnarray}
\beta_\ppE^{(\MDR)}&=&\frac{\pi^{2-a_\MDR}}{1-a_\MDR}\frac{D_{a}}{\lambda_{\mathbb{A}}^{2-a_\MDR}}\frac{\mathcal{M}^{1-a_\MDR}}{(1+z)^{1-a_\MDR}}, \\
b&=& 3 (a_\MDR -1).\label{eq:MDR}
\end{eqnarray}
Here, $z$ is the redshift, $\lambda_{\mathbb{A}}\equiv h \mathbb{A}^{1/(a_\MDR-2)}$ is similar to the Compton wavelength with Plank's constant $h$, and the effective distance $D_a$ is given by~\cite{Mirshekari_MDR,Yunes_ModifiedPhysics}
\begin{eqnarray}
D_{a}&=&\frac{z}{H_0\sqrt{\Omega_M+\Omega_{\Lambda}}} \left \lbrack 1-\frac{z}{4} \left(  \frac{3 \Omega_M}{\Omega_M+\Omega_{\Lambda}}+2a_\MDR \right)  \right\rbrack + \mathcal{O}(z^3),
\end{eqnarray}
where $H_0=67.9$ km sec$^{-1}$ Mpc$^{-1}$ is the local Hubble constant, and $\Omega_M=0.303$, and $\Omega_{\Lambda}=0.697$ are the energy densities of matter and dark energy~\cite{Aghanim:2018eyx}.

In this chapter, we mainly investigate bounds on the specific case of the \textit{massive graviton}~\cite{deRham_mg,Hinterbichler_mg,Rubakov_mg,Will_mg} (propagation), where $\mathbb{A}=m_g^2$ and $a_\MDR=0$.
The current constraints on the graviton mass have been found to be $6\times10^{-24}$ eV~\cite{Will:2018gku,Abbott_IMRcon} from solar-system constraints (Yukawa-like corrections to the binding energy and Kepler's law), $5\times10^{-23}$ eV~\cite{Abbott_IMRcon} from the combination of GW signals from the LVC catalog (GW propagation modifications), and $\sim 10^{-14}$ eV~\cite{Chung:2018dxe} or $5 \times 10^{-21}$ eV~\cite{Miao:2019nhf}  from GW150914 and binary pulsars respectively (GW generation modifications). 
Stronger bounds have been obtained from cosmological observations (see e.g.~\cite{deRham:2016nuf,Desai:2017dwg,Gupta:2018hgm}).

Additionally, we offer general constraints on $\mathbb{A}$ in Sec.~\ref{sec:amplitude}, applicable to many alternative theories of gravity with modified dispersion relations.
Some examples of these include~\cite{Mirshekari_MDR,Yunes_ModifiedPhysics} 
\begin{itemize}
\item \textit{Double special relativity}~\cite{Magueijo_dsr,AmelinoCamelia_dsr,AmelinoCamelia_dsr2,AmelinoCamelia_dsr3} with $\mathbb{A}=\eta_{\text{dsrt}}$ and $a_\MDR=3$; 
\item \textit{Extra-dimensional theories}~\cite{Sefiedgar_edt} with $\mathbb{A}=-\alpha_{\text{edt}}$ and $a_\MDR=4$;
\item \textit{Ho\v rava-Lifshitz Gravity}~\cite{Vacaru_horava,blas_horava,Horava,Horava_2} with $\mathbb{A}=\kappa_{\text{hl}}^4\mu_{\text{hl}}^2/16$ and $a_\MDR=4$;
\item \textit{Multifractional Spacetime Theory}~\cite{Calcagni_mf,Calcagni_mf2,Calcagni_mf3,Calcagni_mf4} with $\mathbb{A}=2E_*^{2-a_\MDR}/(3-a_\MDR)$ and $a_\MDR=2-3$.
\end{itemize}

\section{Detectability of GW150914-like events}\label{sec:SNRs}

In the following chapter, we primarily consider only GW150914-like~\cite{GW150914} events, with masses $(m_1,m_2)=(35.8\text{ M}_\odot,29.1\text{ M}_\odot)$ and spins $(\chi_1,\chi_2)=(0.15,0)$.
Such spins are taken to be non-vanishing so that the spin-dependent BH scalar charges are non-zero in dCS gravity, yet still small enough to be consistent with the LVC's measurement of the effective spin~\cite{GW150914}.
The luminosity distance is scaled such that an SNR of $\rho_{\text{O2}}=25.1$ would be achieved on the sensitivity for the LIGO/Virgo's 2nd Observing Run (O2)~\cite{aLIGO}. 
We also note that we assume the initial LISA detection of GW150914-like events to take place exactly four years prior to their merger, corresponding to the expected lifetime of the LISA mission.
Such an assumption is considered for its validity in further detail in upcoming Sec.~\ref{sec:paramResults}.

In this section, we discuss the feasibility of detecting GW150914-like events using the space-based GW interferometer LISA\footnote{We found that space-based detector TianQin observes very similar, yet slightly louder ($\rho=10.7$ for GW150914-like events) results to that of LISA. 
Additionally, DECIGO and B-DECIGO can detect strong GW150914-like signals with SNRs of $10^2-10^4$.}.
As described in Refs.~\cite{Wong:2018uwb}, the standard threshold SNR of $\rho_\text{th} \sim 8$ can be reduced to $\rho_\text{th} \sim 4-5$ for LISA by revisiting sub-threshold events in prior LISA data with information from high-SNR events in the ground-based bands of e.g.~CE. 
Moore \textit{et al.}~\cite{Moore:2019pke} later pointed out that a template-based search for LISA requires a much larger SNR threshold of $\rho_\text{th} \sim 15$, which can be further reduced to $\rho_\text{th} \sim 9$ in combination with ground-based detectors.
However, such an estimation may be pessimistic, as non-template-based approaches may bring this threshold down further.

To demonstrate how well such events can be detected in either case, Fig.~\ref{fig:SNRs} displays the region in the $(m_1,m_2)$ parameter space where SNRs exceeds the threshold value of $\rho_\text{th} = 5$ or 9 for both CE and LISA.
Observe how in both cases, GW150914-like events with $(m_1,m_2)=(35.8\text{ M}_\odot,29.1\text{ M}_\odot)$ fall within the multiband detectability region defined by both $\rho>5$ or $\rho>9$.
For LISA observations of GW150914-like events, the SNR of $\rho=9.3$ only marginally falls within the larger threshold of $9$, while the CE observation well exceeds both thresholds by $\sim2$ orders-of-magnitude.
We note that for the following analysis, we consider nearly non-rotating GW150914-like events that satisfy such detectability criteria.
We refer to the discussion by Jani \textit{et al.}~\cite{Jani:2019ffg} for a more in-depth analysis into the multi-band detection between third-generation detectors and LISA.

\begin{figure}
\centering
\includegraphics[width=0.7\columnwidth]{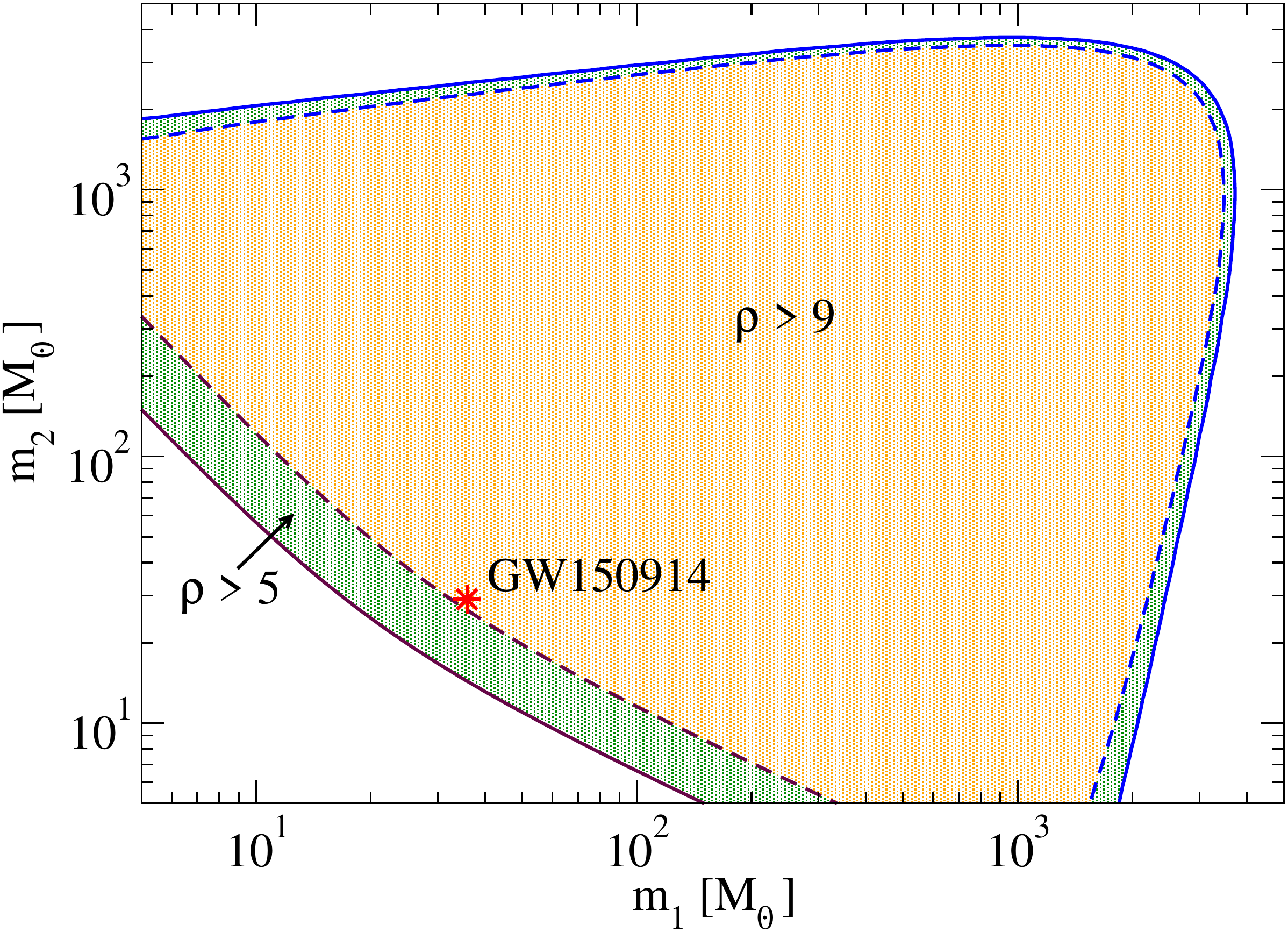}
\caption[Multi-band detectability]{Multi-band detectability region as a function of the constituent BH masses $m_1$ and $m_2$.
This region is formed by SNRs in agreement with the condition $\rho>\rho_\text{th}$ for $\rho_\text{th}=5$ (optimistic) and $\rho_\text{th}>9$~\cite{Wong:2018uwb,Moore:2019pke} (pessimistic) for events detected by both the ground-based detector CE, and space-based detector LISA. 
Such SNRs have been computed with the assumption of non-spinning BHs at luminosity distances of $410$ Mpc.
The upper-right edge (blue) of the region corresponds to CE's $\rho_\text{th}$ contour, while the lower-left edge (maroon) is formed by LISA's contour. 
The SNRs are computed following Eq.~\eqref{eq:SNR}.
Additionally shown as a red star is the event GW150914 with $(m_1,m_2)=(35.8\text{ M}_\odot,29.1\text{ M}_\odot)$.
Observe how GW150914-like events marginally fall within LISA's larger observational SNR threshold of $\rho_{\text{th}}=9$.
Alternatively, such events exceed both ground-based SNR thresholds by more than 2 orders-of-magnitude.
}\label{fig:SNRs}
\end{figure}

\section{Results}\label{sec:paramResults}
In this section we describe the resulting constraints on modified theories of gravity using the parameterized tests of GR.
We begin with those found from multiband observations, followed up by those obtained from mixed GW binaries.
Finally, we present constraints on the EdGB theory of gravity with waveform corrections to the inspiral, merger-ringdown, and remnant BH parameters present.

\subsection{Multiband observations of GWs}

In this subsection, we present our findings for constraints on several different modified theories of gravity with multiband observations between ground- and space-based detectors.
We first show results for bounds on GW generation mechanisms, followed by those on GW propagation mechanisms.
All such constraints are summarized in \hypertarget{link:theories}{Tab.~\ref{tab:theories}}.

\subsubsection{GW generation mechanisms: phase corrections}

\begin{figure*}
\centering
\includegraphics[width=\columnwidth]{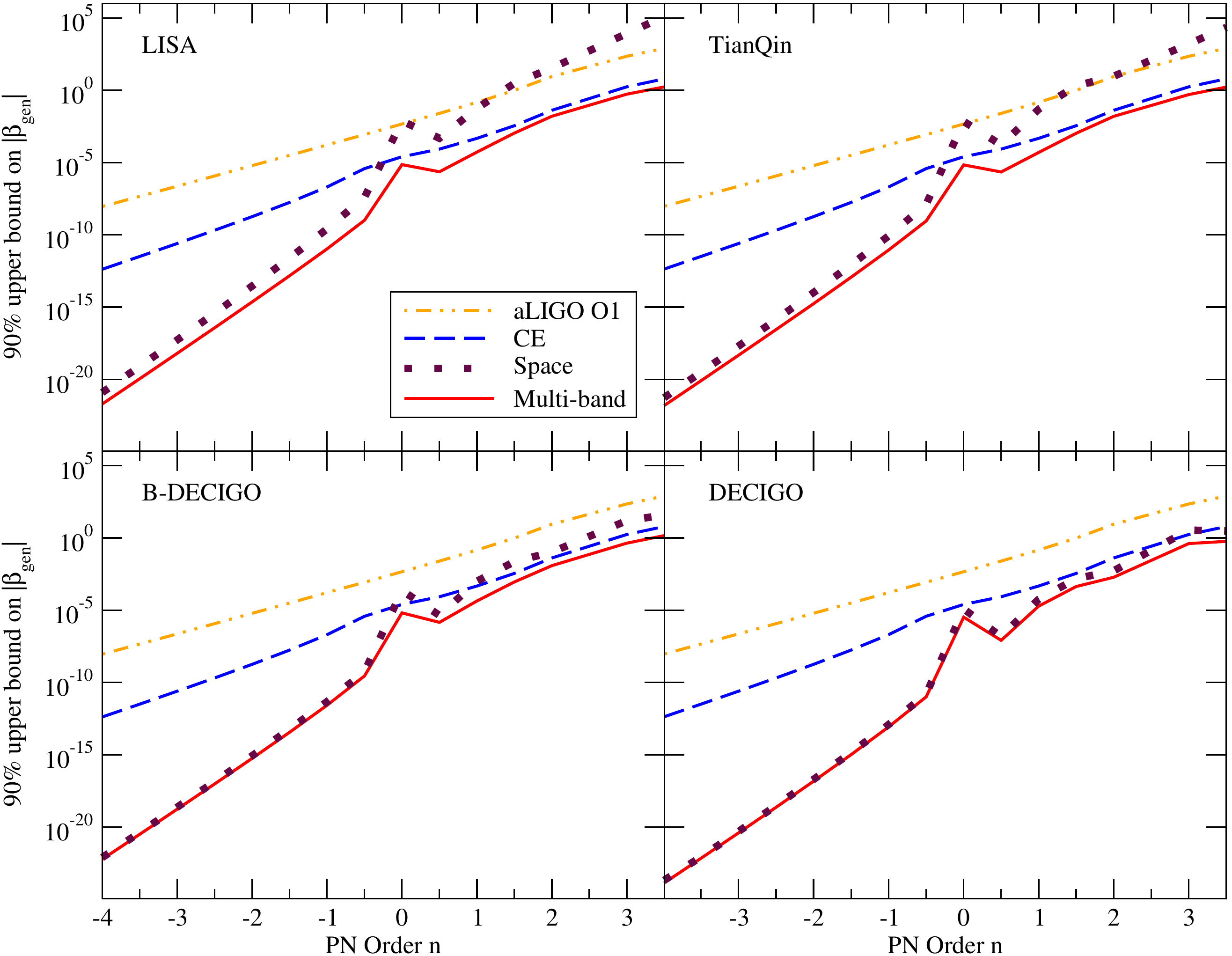}
\caption[Constraints on ppE phase parameter for generation effects]{
Constraints on the generalized non-GR phase parameter for generation effects $|\beta_{\text{gen}}|$ as a function of PN order for GW150914-like events observed on various space- and ground-based detectors individually.
Observe how space-based detectors are most proficient at probing effects that enter at negative-PN orders, with ground-based detectors more suited to probing positive-PN effects.
The combination of the two (multi-band) results in improved bounds across all PN orders.
}\label{fig:betaBoundsGen}
\end{figure*}

Figure~\ref{fig:betaBoundsGen} presents 90\% upper credible level bounds on the generalized non-GR phase parameter $\beta_\mathrm{gen}$ for generation mechanisms\footnote{While certain theories of gravity correspond to different signs of the ppE parameter $\beta$, in this analysis we only constrain the modulus of the parameter to remain as generic and theory-agnostic as possible. In future, more directed analyses with a Bayesian approach, priors of $\beta<0$ or $\beta>0$ can be imposed to improve constraints.}, and we show bounds on the amplitude parameter $\alpha_\mathrm{gen}$ in Sec.~\ref{sec:amplitude}.
We consider CE, LISA, TianQin, B-DECIGO, DECIGO, and the combinations of each space-based detector with CE.
Bounds are obtained for each half-integer PN order between $-4$PN and +3.5PN, with the exception of +2.5PN which observes complete degeneracies with the coalescence phase $\phi_c$.
Additionally, the bound from O1 (LVC's 1st observing run) is taken from Ref.~\cite{Yunes_ModifiedPhysics} for comparison.

\begin{figure*}
\centering
\includegraphics[width=.48\columnwidth]{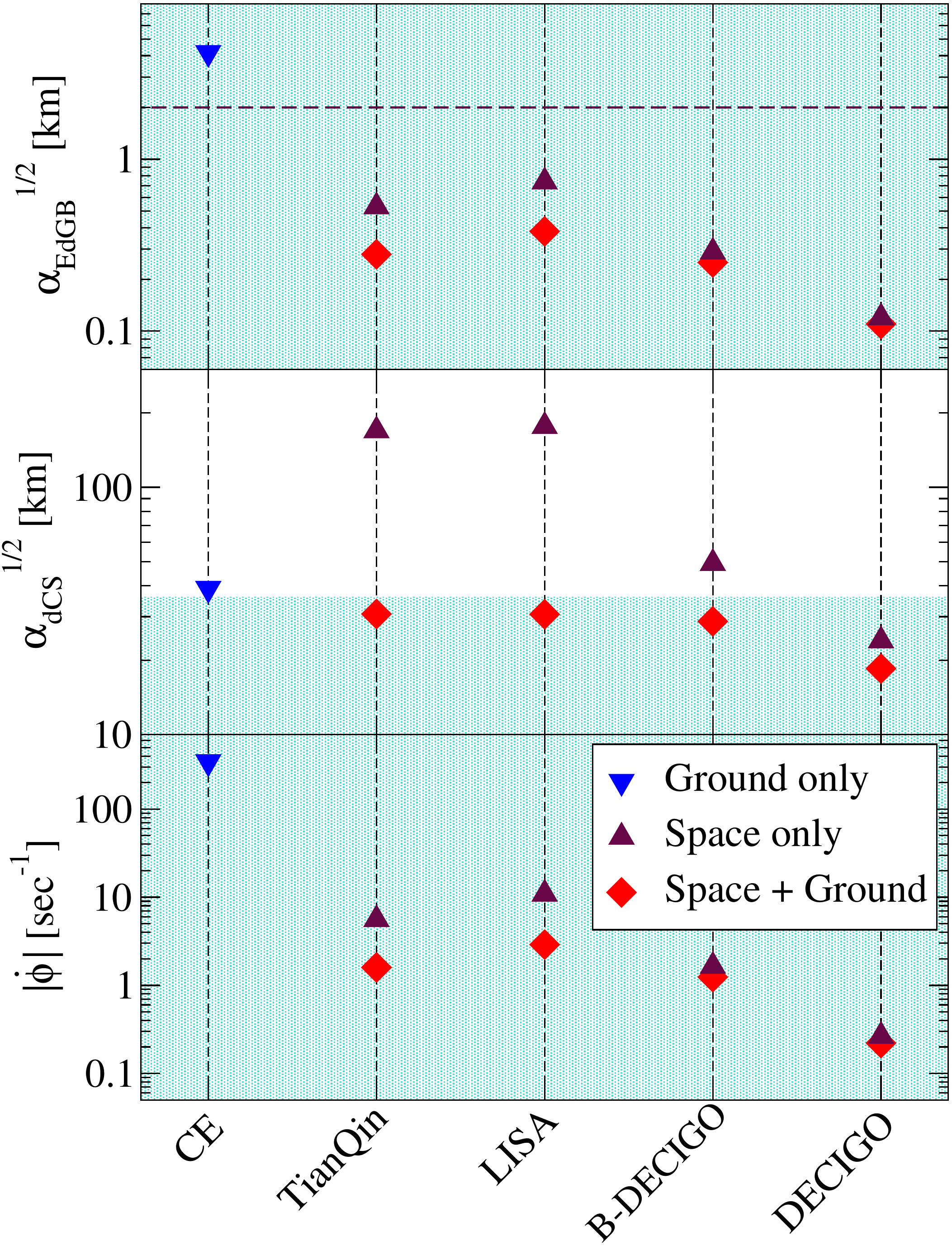}
\includegraphics[width=.48\columnwidth]{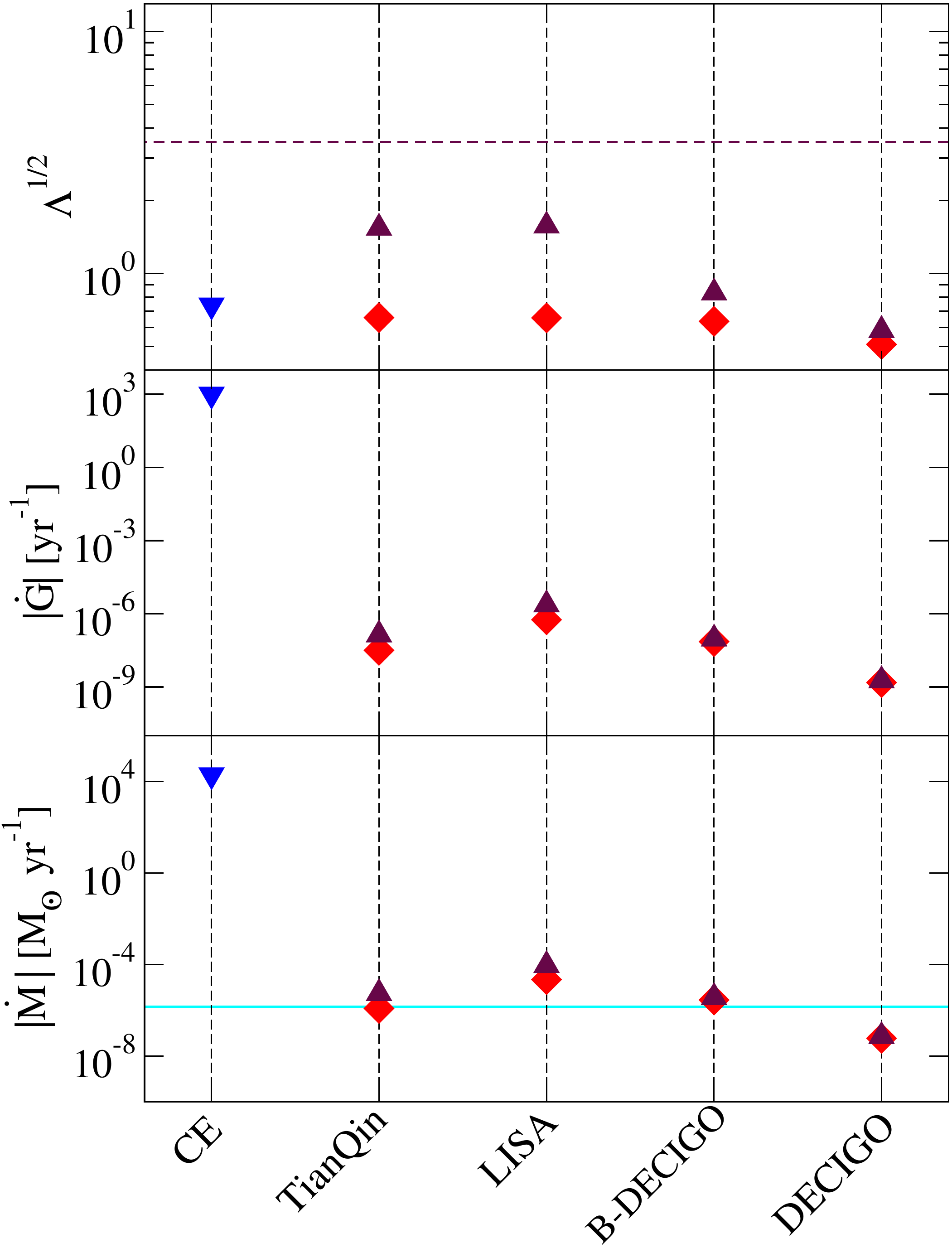}
\caption[Constraints on modified theories of gravity]{
90\% upper-bound credible level constraints on the theoretical parameters representative of 6 of the modified theories of gravity considered in this analysis for GW150914-like events.
Bounds are presented for: EdGB gravity ($\sqrt{\alpha_\EdGB}$, $-1$PN order), dCS gravity ($\sqrt{\alpha_\dCS}$, +2PN order), scalar tensor theories ($\dot{\phi}$, $-1$PN order), noncommutative gravity ($\sqrt{\Lambda}$, +2PN order), varying-G theories ($\dot{G}$, $-4$PN order), and BH mass-varying theories ($\dot{M}$, $-4$PN order), and are additionally tabulated in Table~\ref{tab:theories}.
The blue shaded regions correspond to the region such that the small coupling approximations $\zeta_\EdGB \ll 1$, $\zeta_\dCS \ll 1$, and $m\dot{\phi} \ll 1$ are valid (the definition of the dimensionless coupling constants $\zeta$ can be found in Sec.~\ref{sec:theory}), and the dashed maroon lines correspond to the current bounds in the literature, also tabulated in Table~\ref{tab:theories}.
The cyan line in the bottom right panel corresponds to the Eddington accretion rate for GW150914-like events.
}\label{fig:modifiedBounds}
\end{figure*}

We can make the following observations from the figure. 
First, notice that non-GR effects entering the gravitational waveform at negative-PN orders can be constrained most stringently by space-based detectors, while positive-PN effects are most proficiently constrained by ground-based detectors.
The obvious exception being DECIGO, which bridges the gap between the two frequency bands and provides the strongest bounds at both positive and negative PN orders.
Second, when one considers multi-band observations by combining both space- and ground-based detectors, we see large improvements of up to a factor of 40 across all PN orders.

Here we briefly discuss the effect of LISA's mission lifetime on observations of the theory-agnostic parameter $\beta$.
In particular, for the above calculations we assumed the best-case scenario in which all four years of LISA's mission can be used to observe the same GW signal from a GW150914-like event.
For comparison, we instead consider an estimate of the same effect (at $-4$PN order for the largest effect possible) given that only the last three, two, or one years of LISA's lifetime will be able to observe the GW150914-like signal.
We find that the resulting constraints on $\beta$ are weakened by ratios of $1.8$, $4.3$, and $16$ respectively, compared to the best-case four-year scenario.
Thus we conclude that such effects change our results on the order of an order of magnitude for the worst-case scenario of only one-year observation by LISA. 
Additionally, we find such weakened observations to have SNRs of $9.2$, $8.2$, $6.7$, and $5$ for four-, three-, two-, and one-year observations respectively.
All such SNRs still remain within the multi-band detectability region found in Fig.~\ref{fig:SNRs} for the best-case threshold SNR.

Now we consider the cases in which LISA can observe the GW signal more than four years prior to the coalescence.
In particular, we consider the two new scenarios in which (i) LISA observes the early inspiral signal from six years to two years prior to merger before going offline, and then CE observes the merger two years later, and (ii) LISA observes the early inspiral signal from ten years to six years prior to merger before going offline, and then CE observes the merger 6 years later.
The above to cases have LISA SNRs of 8.8 and 7.7 respectively, each above the best-case SNR threshold.
Relative to the scenario considered in this chapter in which LISA begins observing four years prior to the merger, we find such constraints to be \textit{strengthened} by factors of 1.7, and 3.1 respectively at $-4$PN order.
Such constraints have been improved because the $-4$PN order correction is the best-case scenario for observing the earlier inspiral.
In the worst-case scenario of $3.5$PN order corrections, we find constraints on $\beta$ to \textit{weaken} by factors of 1.04, and 1.12 respectively.
Such changes are insignificant to our analysis, and for almost all PN orders give way to improved ppE constraints, making our estimates conservative.

Now that constraints on the agnostic non-GR parameter $\beta$ have been obtained, we apply them to the specific theories of gravity reviewed in Sec.~\ref{sec:theory} to constrain their theoretical parameters using various single-band and multi-band observations.
Such bounds are obtained by simply selecting the constraints on $|\beta_{\text{gen}}|$ corresponding to the PN order associated with the desired modified theory of gravity, and finally mapping them to the theoretical parameters with the ppE expressions found in Sec.~\ref{sec:theory}.
When obtaining constraints on theory-specific parameters from the theory-agnostic ppE parameter $\beta$, we assume a fiducial value of $\beta=0$ corresponding to GR. 
The resulting root-mean-square variance on $\beta$ describes the statistical variance $\beta$ is allowed to observe within the detector noise.
For this reason, under consideration of propagation of uncertainties when transforming $\sigma_\beta$ to $\sigma_\epsilon$ for some theory-specific parameter $\epsilon$, all terms containing measurement errors on intrinsic template parameter vanish due to their proportionality with $\beta\rightarrow0$.
See Ref.~\cite{Perkins:2018tir} for a more in-depth discussion on this topic.

\begin{figure}
\centering
\includegraphics[width=0.6\columnwidth]{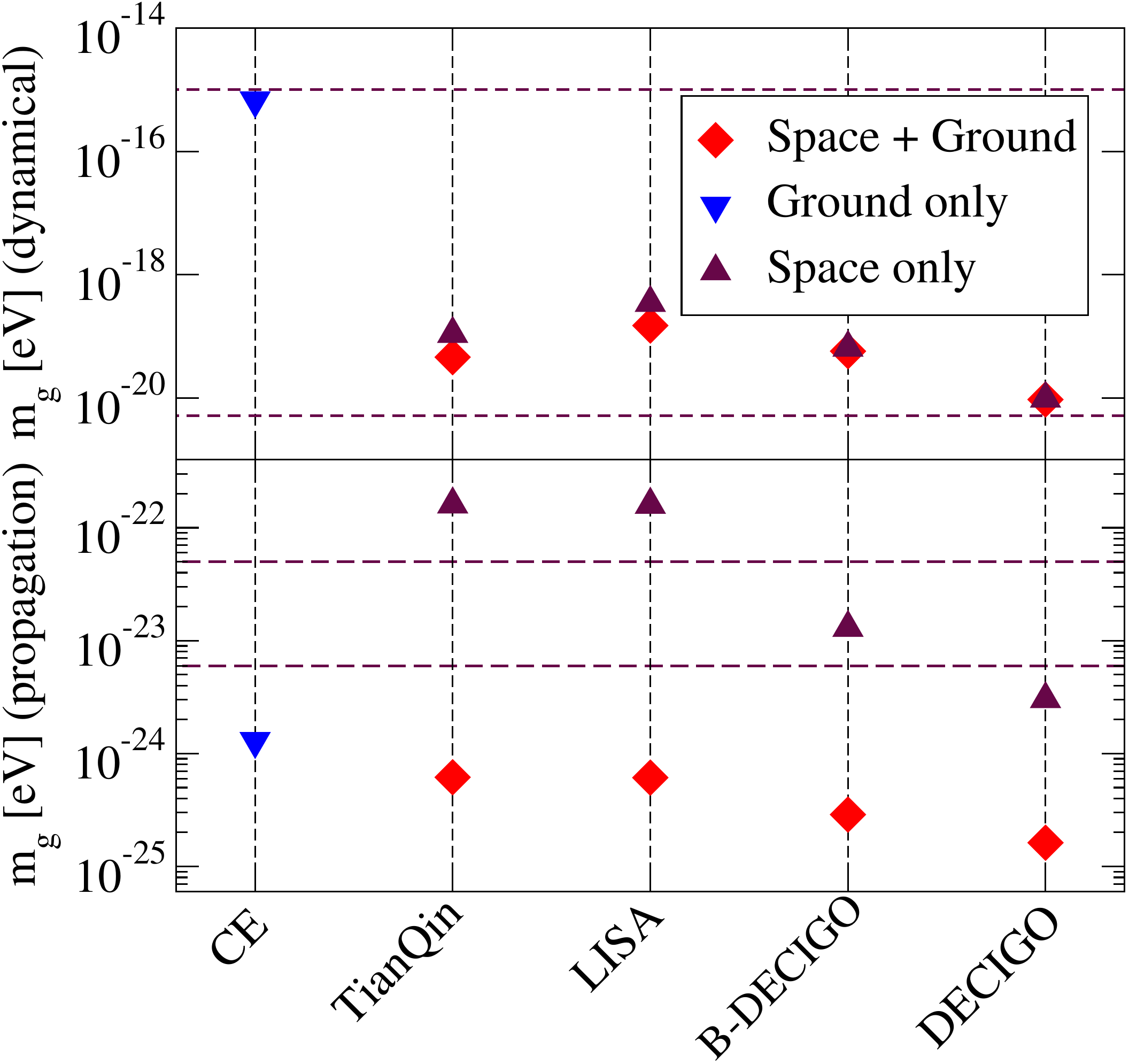}
\caption[Graviton mass constraints]{Similar to Fig.~\ref{fig:modifiedBounds}, but for constraints on the graviton mass in the dynamical, generation mechanism regime (top), and in the weak-field, propagation mechanism regime (bottom).
}\label{fig:gravitonMassBounds}
\end{figure}

Figure~\ref{fig:modifiedBounds} and the top panel of Fig.~\ref{fig:gravitonMassBounds} display the 90\% upper credible level limit on the associated parameters for theories that modify the generation of GWs: EdGB, dCS, scalar-tensor, noncommutative, varying-$G$, varying-$M$, and massive graviton theories with dynamical effects for both single- and multi-band detections of GW150914-like events.
Additionally, the constraints are tabulated in Table~\ref{tab:theories} for each detector analyzed, along with the current strongest bounds from the literature. We summarize below our findings for each theory.

\begin{itemize}

\item \textit{EdGB, scalar-tensor:}  
In both theories, corrections to the waveform were derived within the small-coupling approximations, in which non-GR corrections are assumed to be always smaller than the GR contribution. Bounds on the theoretical constants (square-root of the coupling constant $\sqrt{\alpha_\EdGB}$ for EdGB gravity and the time-variation of the scalar field $\dot \phi$ for scalar-tensor theories) both satisfy the small-coupling approximations for every detection scenario, however only $\sqrt{\alpha_\EdGB}$ can improve upon the current strongest bound of 2--6 km~\cite{Kanti_EdGB,Pani_EdGB,Yagi_EdGB,Nair_dCSMap,Yamada:2019zrb,Tahura:2019dgr}, for both space-based and multi-band detections.

\item \textit{dCS:} Similar to EdGB and scalar-tensor theories, corrections to the waveform have been derived within the small-coupling approximation. 
Constraints placed on the parity-violation constant $\sqrt{\alpha_\dCS}$ with CE, LISA, TianQin, and B-DECIGO fall short of the small-coupling approximation, and thus are not reliable for GW150914-like events. One can place valid constraints only when multi-band detections are made, improving upon the current constraint of $10^8$ km~\cite{AliHaimoud_dCS,Yagi_dCS} by $\sim$7 orders-of-magnitude.

\item \textit{noncommutative:} Bounds on the noncommutative parameter $\sqrt{\Lambda}$ can slightly be improved upon the current most stringent bound of 3.5~\cite{Kobakhidze:2016cqh} with multi-band observations.

\item \textit{varying-$M$:} Constraints on the time-variation of the total black hole mass $\dot M$ (motivated not only by astrophysical gas accretion, but also by a classical evaporation in a braneworld scenario~\cite{Emparan:2002px,Tanaka:2002rb} or dark energy accretion~\cite{Babichev:2014lda,Babichev:2005py,Babichev:2004yx}) are below that of the Eddington accretion rate for BHs in GW150914-like events for B-DECIGO(+CE), DECIGO(+CE), and LISA+CE multi-band detections.
See also a recent analysis~\cite{Caputo:2020irr} in which the impact of gas accretion on the orbital evolution of BH binaries, and thus the GW emission.

\item \textit{varying-$G$:} Space and multi-band observations can improve significantly over ground-based ones, though the former bounds on the time-variation in $G$ are still much weaker than other existing bounds~\cite{Bambi_Gdot,Copi_Gdot,Manchester_Gdot,Konopliv_Gdot}.\footnote{Space-based bounds can be comparable with current strongest bound for GW observations of supermassive BH binaries~\cite{Yunes_GdotMap}.}

\item \textit{massive graviton (dynamical):} CE bounds (from modifications in the inspiral) on the mass of the graviton are comparable to GW150914-bounds from ringdown~\cite{Chung:2018dxe} while bounds from space-based detectors can be comparable to those from binary pulsars~\cite{Miao:2019nhf}.

\end{itemize}

Finally, we comment that, similar to the constraints on $\beta$, theories that modify GR at negative-PN orders (EdGB, scalar-tensor, varying-$G$, varying-$M$, and dynamical massive graviton) are more strongly constrained by space-based detectors, while positive-PN theories (dCS and noncommutative) observe stronger bounds with ground-based detectors.

\subsubsection{GW propagation mechanisms: phase corrections}

\begin{figure*}
\centering
\includegraphics[width=\columnwidth]{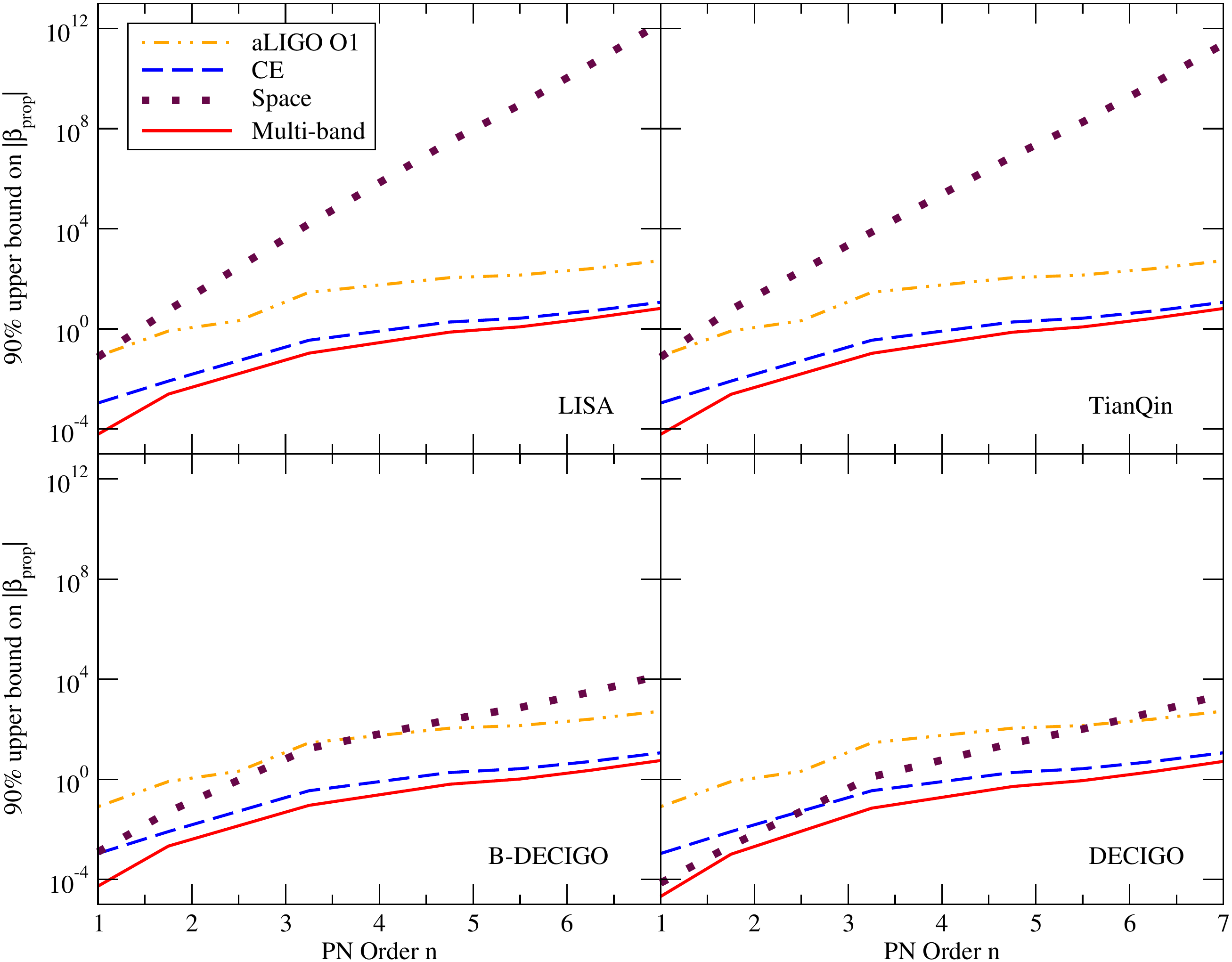}
\caption[Constraints on ppE phase parameter for propagation effects]{
Similar to Fig.~\ref{fig:betaBoundsGen}, but for propagation effects $\beta_{\text{prop}}$ entering the GW phase.
Observe how (with the exception of DECIGO), CE gives stronger constraints on $\beta_{\text{prop}}$ than space-based detectors, while the multi-band case can be even stronger by up to an order-of-magnitude.
}\label{fig:betaBoundsProp}
\end{figure*}

We next move on to studying bounds on the ppE phase parameter $\beta_{\text{prop}}$ from the GW propagation mechanisms (bounds on the ppE amplitude parameter $\alpha_{\text{prop}}$ are shown in Sec.~\ref{sec:amplitude}). 
Figure~\ref{fig:betaBoundsProp} presents bounds on $\beta_{\text{prop}}$ against each PN order at which the correction enters. 
We only show bounds on positive PN orders because all of the example theories discussed in Sec.~\ref{sec:MDR} show such a feature. 
Observe that bounds placed with CE dominate those by space-based detectors, with little improvement via multi-band observations. 
These bounds on $\beta_{\text{prop}}$ can easily be mapped to those on the magnitude $\mathbb{A}$ of the correction to the graviton dispersion relation using Eq.~\eqref{eq:MDR}, as shown in Fig.~\ref{fig:Abounds} in Sec.~\ref{sec:amplitude}.

While these constraints may be used to compute bounds on a variety of propagation mechanism non-GR effects, we here focus on the case of massive gravitons with $\mathbb{A}=m_g^2$ and $a_\MDR=0$, now in the weak-field regime.
The bottom panel of Fig.~\ref{fig:gravitonMassBounds} displays such bounds for each detector considered.
We observe that CE places the strongest constraints out of all single-band observations\footnote{The bound becomes much stronger for observing supermassive BH binaries with space-based detectors~\cite{Will_mg,Berti:Fisher,Yagi:2009zm,Berti:2011jz,Cornish:2011ys,Keppel:2010qu,Chamberlain:2017fjl}.}.
When combined to make multi-band detections, we see an improvement on the graviton mass bound, with more than an order-of-magnitude reduction from the current solar system bound of $6\times10^{-24}$ eV~\cite{Will:2018gku}.

We finally consider a comparison of different space-based detectors' ability to test GR.
In particular, we compare constraints on coupling parameters found in a selected few modified theories of gravity investigated with each space-based GW detector: LISA, TianQin, B-DECIGO, and DECIGO.
For EdGB gravity, we find constraints on $\sqrt{\alpha_\EdGB}$ to be $0.7$ km, $0.6$ km, $0.3$ km, and $0.1$ km respectively.
In dCS gravity, we find constraints on $\sqrt{\alpha_\dCS}$ to be respectively $169$ km, $176$ km, $49$ km, and $24$ km.
Finally, for the propagation effect of massive gravitons, we find constraints of $1.6\times10^{-22}$ eV, $1.6\times10^{-22}$ eV, $1.3\times10^{-23}$ eV, and $1.6\times10^{-24}$ eV respectively.
We see that in general, space-based detector DECIGO forms the strongest constraints on all theories of gravity by nearly an order of magnitude in some cases, while B-DECIGO places similar, yet slightly weaker bounds.
When comparing similar space-based detectors LISA and TianQin, we see comparable constraints that differ by less than $\sim10\%$.
In particular, we see that TianQin can place slightly stronger constraints for theories that enter at higher PN orders, and vice-versa for LISA.

\subsubsection{GW amplitude and dispersion relation corrections}\label{sec:amplitude}

In this section, we present constraints on the ppE amplitude parameter $\alpha$, as well as on corrections to the graviton dispersion relation.
Figures~\ref{fig:alphaBoundsGen} and~\ref{fig:alphaBoundsProp} display the 90\% credible level upper limits on $|\alpha_{\text{gen}}|$ and $|\alpha_{\text{prop}}|$ for modified theories which affect GW generation, and propagation effects respectively.
Observe that the multi-band results simply follow bounds from space- (ground-)based detectors for corrections entering at negative (positive) PN orders, and do not have much improvement from single-band results.

Such constraints can be mapped to the desired coupling parameters of many modified theories of gravity~\cite{Tahura_GdotMap}.
Figure~\ref{fig:Abounds} presents constraints on the generalized dispersion relation correction, $\mathbb{A}$.
Such bounds can be further applied to modified theories of gravity which predict modified dispersion relations, as discussed in the previous section.
Observe that the multiband bounds are very similar to those from CE, consistent with Fig.~\ref{fig:alphaBoundsProp}.

\begin{figure}
\centering
\includegraphics[width=\columnwidth]{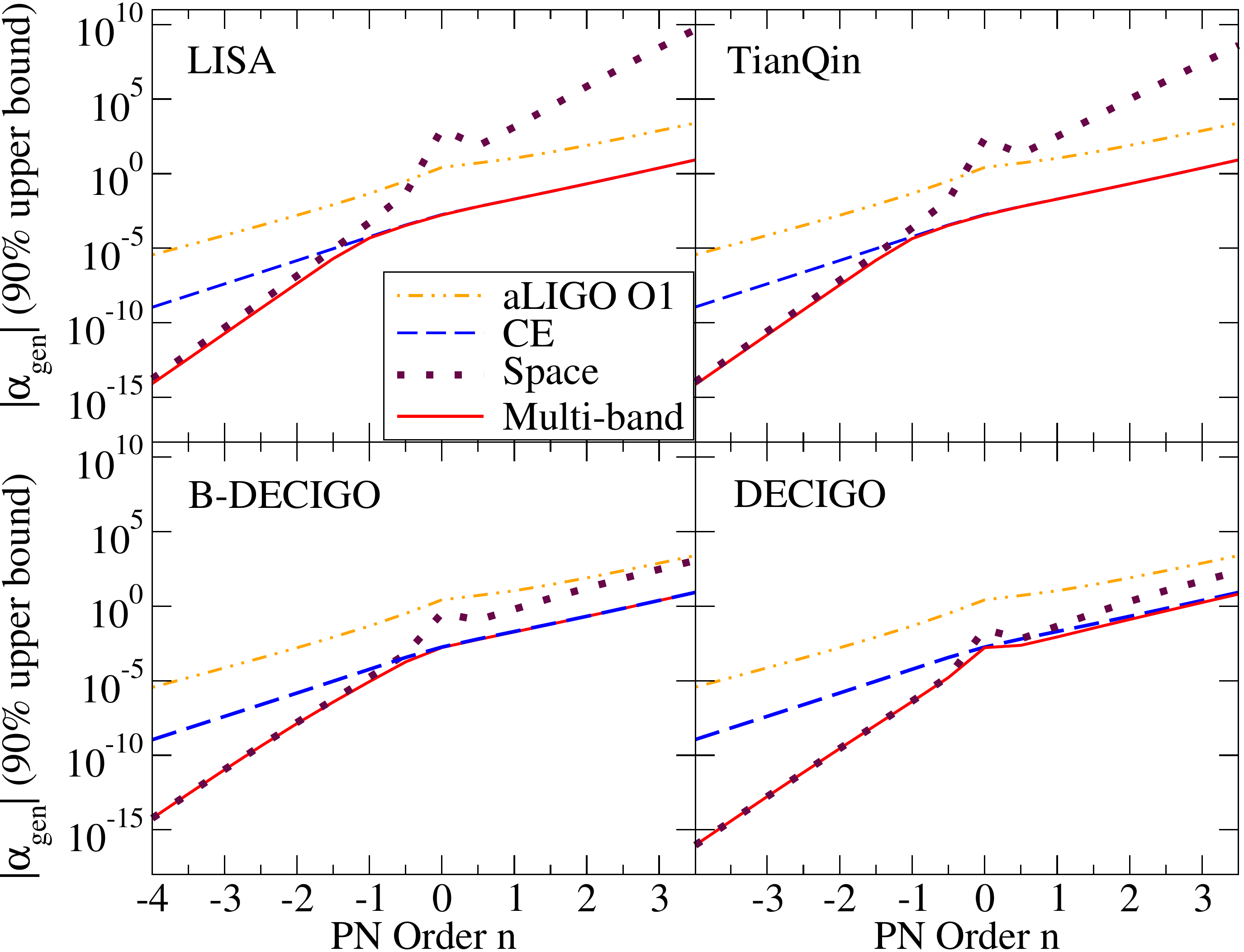}
\caption[Constraints on ppE amplitude parameter for generation effects]{
Similar to Fig.~\ref{fig:betaBoundsGen}, but for generation effects $\alpha_{\text{gen}}$ entering the GW amplitude.
Observe how for the case of amplitude corrections, multi-band observations do not provide for much improvement over space-based detectors for negative-PN orders, and ground-based detectors for positive-PN orders.
}\label{fig:alphaBoundsGen}
\end{figure}

\begin{figure}
\centering
\includegraphics[width=\columnwidth]{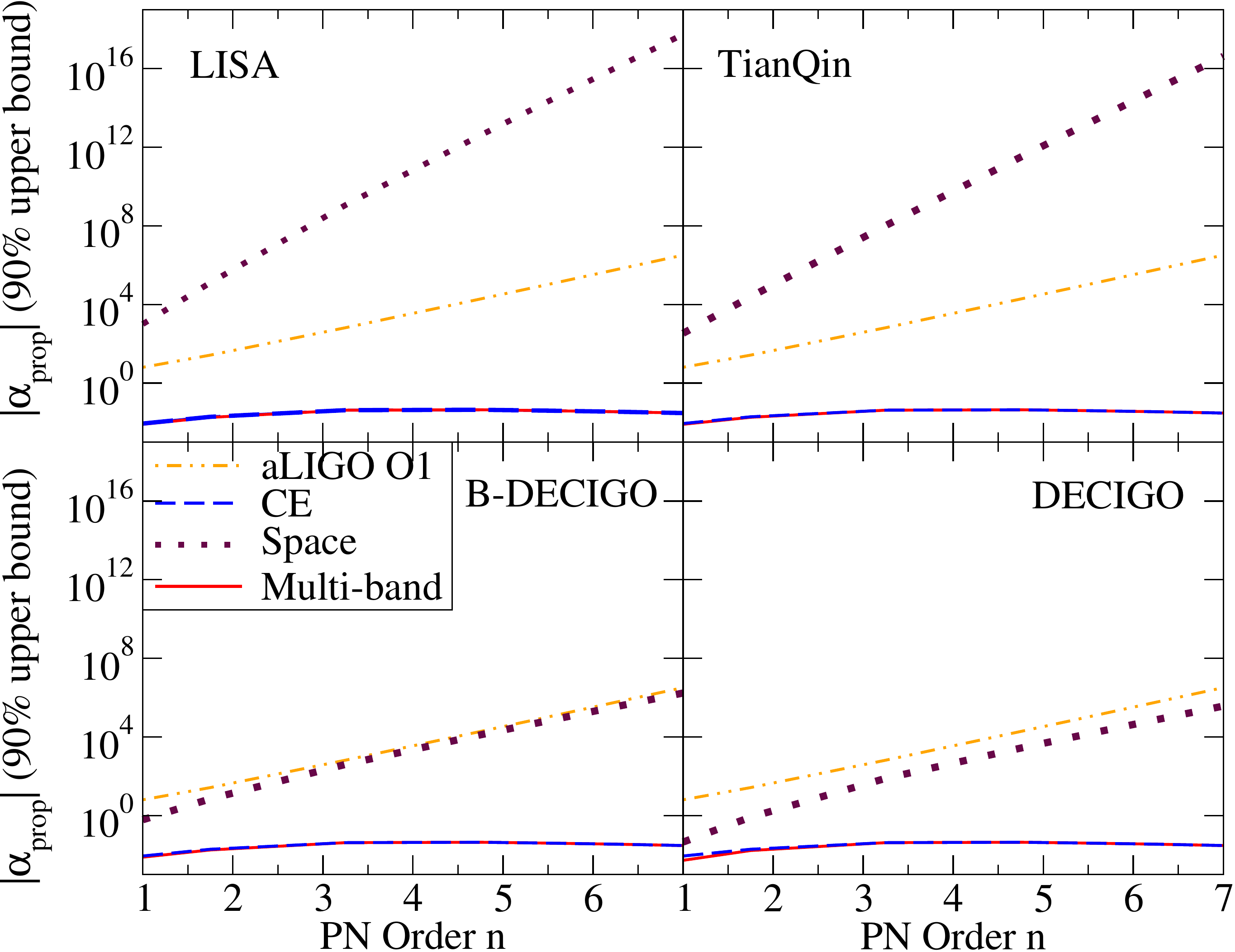}
\caption[Constraints on ppE amplitude parameter for propagation effects]{
Similar to Fig.~\ref{fig:betaBoundsGen}, but for propagation effects $\alpha_{\text{prop}}$ entering the GW amplitude.
Observe how for the case of amplitude corrections, multi-band observations do not provide for much improvement over the constraints provided by CE.
}\label{fig:alphaBoundsProp}
\end{figure}

\begin{figure}
\centering
\includegraphics[width=\columnwidth]{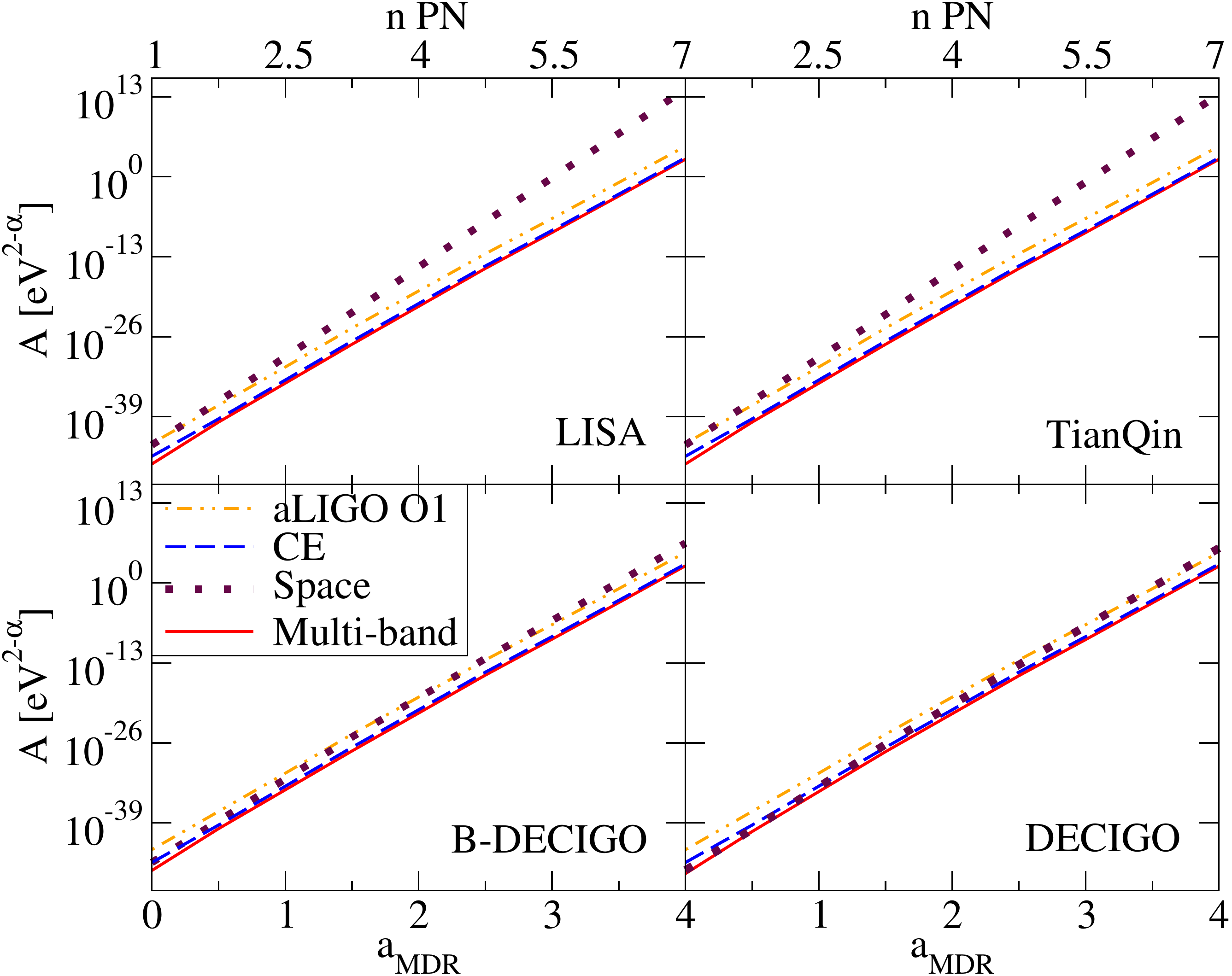}
\caption[Constraints on graviton dispersion relation]{
90\% upper-bound constraints on the magnitude of the correction to the graviton dispersion relation $\mathbb{A}$ as a function of $a_\MDR$, found in Eq.~\eqref{eq:MDR}.
Such bounds are computed from the ppE parameter for propagation mechanisms $\beta_{\text{prop}}$ for GW150914-like events on various space- and ground-based detectors, and the combination of the two.
Constraints on modifications to GR can be found by selecting a value of $\alpha$ and mapping $\mathbb{A}$ to its corresponding form, e.g. for massive gravitons with $a_\MDR=0$ and $\mathbb{A}=m_g^2$, with bounds displayed in Table~\ref{tab:theories} and Fig.~\ref{fig:gravitonMassBounds}.
}\label{fig:Abounds}
\end{figure}


\subsection{Constraining STTs with mixed binaries}\label{sec:STTs}
In this section we consider constraints obtained from mixed BH-NS binaries on the quasi-Brans-Dicke theory DEF, as well as the EdGB theories of gravity.
Finally, we conclude with a summary of the approximations made in this analysis, and their effects on the results.

\subsubsection{Quasi-Brans-Dicke theories}\label{sec:bransdicke}
Let us first focus on the quasi-Brans-Dicke theory, DEF.
We begin with current considerations possibly obtained from the O3 BH-NS merger candidates S190426c and S190814bv~\cite{gracedb}.
This is followed up by a foray into the future of GW astronomy, with constraints obtained by third-generation detectors, multi-band measurements, and multi-event combinations.

\begin{figure}
\begin{center}
\includegraphics[width=.7\columnwidth]{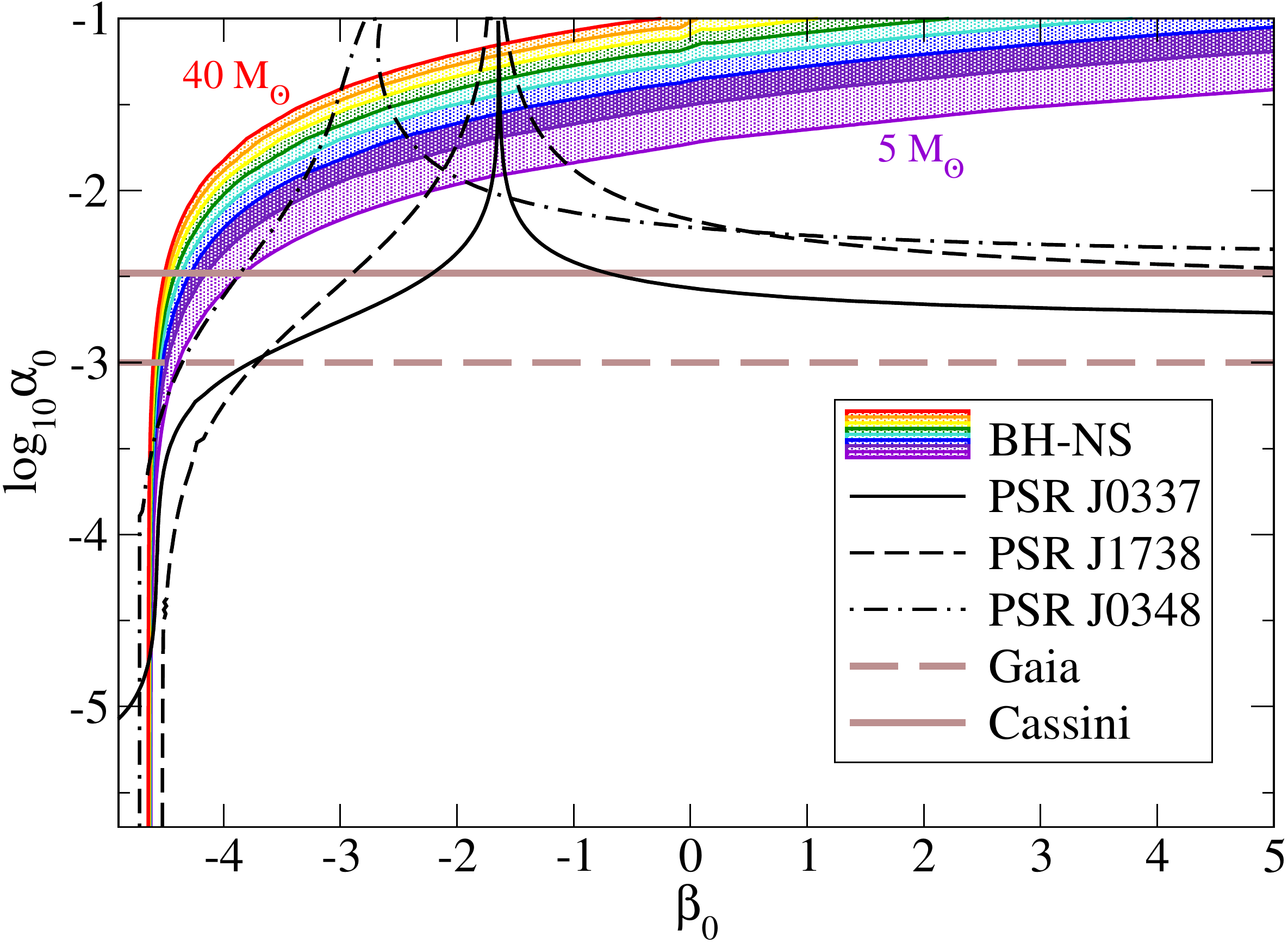}
\caption[Current DEF constraints]{Estimated $68$\% confidence interval bounds on the DEF quasi-Brans-Dicke modified theory of gravity with an assumed EoS of APR4, detected on the LIGO O3 detector with a SNR of 10 that is close to a detection threshold SNR and thus the bounds serve as conservative.
Such bounds are presented for a BH-NS system with $m_\NS=1.4\text{ M}_\odot$ and $m_\BH$ varying between $5\text{ M}_\odot$ and $40\text{ M}_\odot$ (with iterations of $5\text{ M}_\odot$).
The solid, dashed, and dash-dotted black curves correspond to constraints placed by the pulsar triple system PSR J0337+1715~\cite{Ransom:2014xla,Archibald:2018oxs}, and the pulsar-WD systems PSR J1738+0333~\cite{Kilic:2014yxa} and PSR J0348+0432~\cite{Antoniadis:2013pzd}, respectively.
The solid and dashed brown horizontal lines correspond to constraints by the existing Cassini spacecraft~\cite{Bertotti:2003rm} and those predicted by Gaia~\cite{mignard_2009}.
Such bounds are computed via $\alpha_0^2=\frac{|1-\gamma|}{2-|1-\gamma|}$ for parameterized post-Newtonian parameter $\gamma$ (see Eq.~(18) of Ref.~\cite{Anderson:2019eay}).
Take note that the Cassini constraints converted here to $\alpha_0$ were obtained with a few assumptions that make them applicable as an order-of-magnitude estimation.
}\label{fig:aligoDEF}
\end{center}
\end{figure} 

We first discuss the present considerations of DEF constraints using GW and pulsar timing observations.
Figure~\ref{fig:aligoDEF} presents the estimated constraints in the DEF theory parameter $\alpha_0-\beta_0$ plane for the various observations considered in this analysis.
Observe that the combination of Cassini and pulsar timing measurements from PSR J0337 and PSR J1738 places the strongest constraints on DEF gravity. Moreover, even if the O3 candidates S190426c and S190814bv were BH-NS merger events~\cite{gracedb}, they struggle to place competitive bounds on DEF theory. Thus, this motivates why we must consider future bounds on DEF from GW measurements.

We conclude with an expedition into the future of GW astronomy.
We consider the BH-NS system described previously, with fixed BH and NS masses of $10\text{ M}_\odot$ and $1.4\text{ M}_\odot$, respectively.
We assume detections on the future GW interferometers A\texttt{+}, Voyager, CE, ET, B-DECIGO, and DECIGO, and following the spirit of Ref.~\cite{Takahiro} we combine the bounds on $\Delta\alpha$ from $N$ BH-NS detections falling within the horizon of each detector over one observing year, as described in Chapter~\ref{chap:ParameterEstimation}.
Further, we consider the multi-band observations~\cite{Carson_multiBandPRD,Carson_multiBandPRL,Gnocchi:2019jzp} of such binaries between both ground-based detector ET and space-based detectors DECIGO/B-DECIGO.
Unlike the multi-band case with space-based detector LISA, in which the possibly large SNR threshold of $\sim9$~\cite{Moore:2019pke} would prevent one from obtaining event rates larger than $\mathcal{O}(1)$, multi-band detections between ET and DECIGO/B-DECIGO will instead be limited by the ET event rate because of the large SNRs obtained by the space-based telescopes (this reasoning primarily applies to DECIGO, with event rates approximately equivalent or greater than those on ET and CE, rather than B-DECIGO with significantly smaller rates).
Such event rates are still significantly large at $\sim (500-500,000)$ (see Table~\ref{tab:eventRates}).

\begin{figure}
\begin{center}
\includegraphics[width=\textwidth]{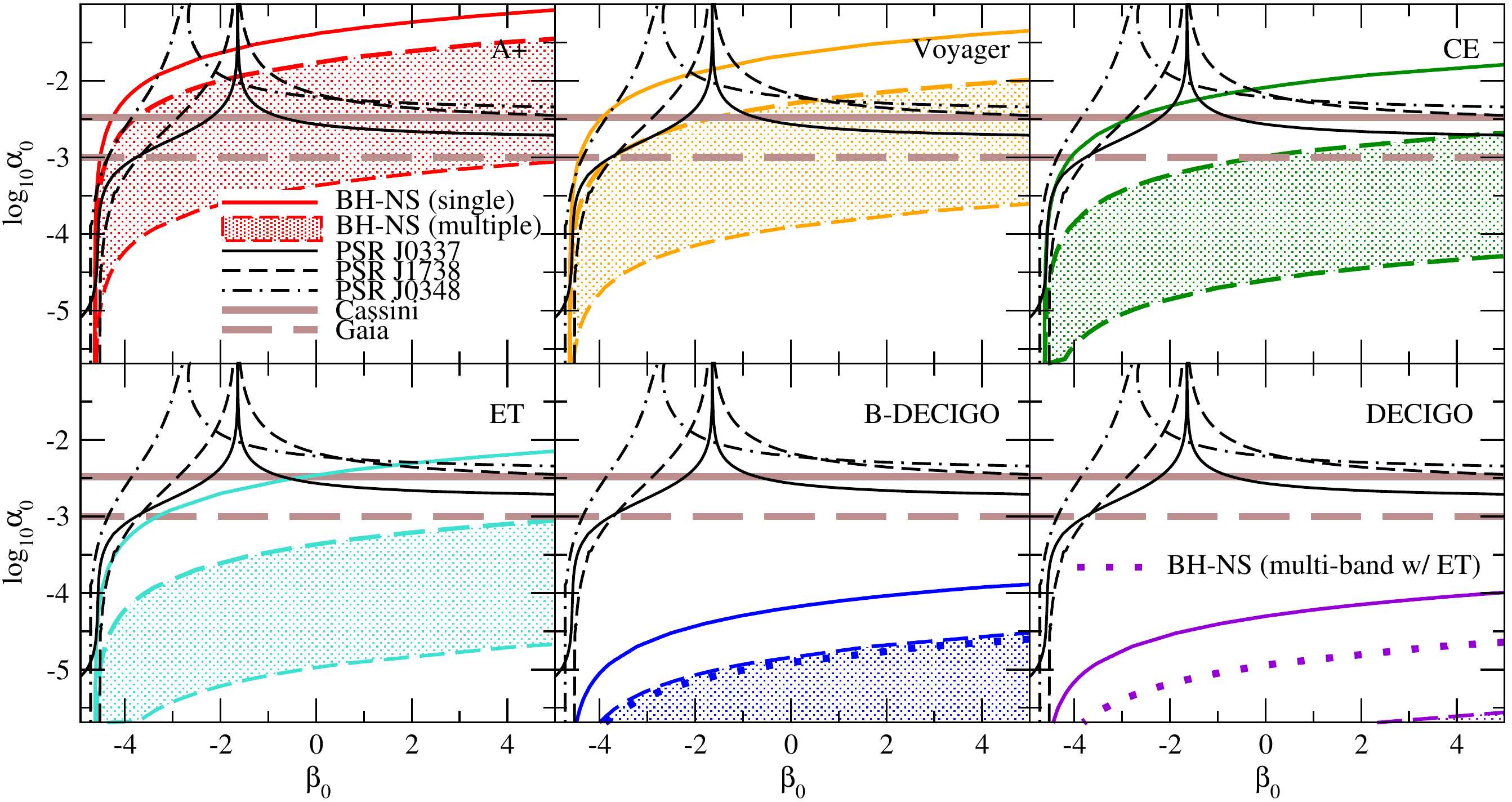}
\caption[Future DEF constraints]{Predicted $68$\% confidence interval constraints on the DEF quasi-Brans-Dicke modified theory of gravity with an assumed EoS of APR4, for $10\text{ M}_\odot-1.4\text{ M}_\odot$ BH-NS merger events at 1 Gpc detected by A\texttt{+}, Voyager, CE, ET, B-DECIGO, and DECIGO, with SNRs of 8.5, 21, 143, 71, 24, and 600 respectively.
In each panel, the solid colored lines represent bounds for single events, while the shaded region between dashed colored lines represent the combined constraints from multiple events, from the pessimistic to optimistic coalescence rates.
Additionally, the dotted lines represent the bounds from the multi-band observations between ET and B-DECIGO/DECIGO.
The brown horizontal lines and solid/dashed/dot-dashed black curves are the same as Fig.~\ref{fig:aligoDEF}.
}\label{fig:stackedDEF}
\end{center}
\end{figure}

Figure~\ref{fig:stackedDEF} presents the bounds in the DEF theory placed for the above-mentioned procedures.
Observe how all of the current constraints can be improved upon with the optimistic number of detections on the A\texttt{+} detector, while CE and ET begin to approach the same point with only the pessimistic number of detections.
Further, all predicted bounds placed with DECIGO/B-DECIGO (single-event, multiple-event, and multi-band) improve the current constraints by several orders of magnitude.
Of course, existing bounds from solar system experiments and binary pulsar observations will also improve in future. For example, bounds on $\alpha_0$ from Gaia will improve those from Cassini by a factor of a few~\cite{mignard_2009}, while the bounds from the pulsar triple system PSR J0337 will improve by a factor of $\sim 10$ with SKA~\cite{Berti_ModifiedReviewLarge}. Future GW bounds with 3rd generation detectors (ET/CE) and space-based detectors (B-DECIGO/DECIGO) are likely to be even stronger than them.
We also note that the bounds for Brans-Dicke theory with $\beta_0=0$ for ET and DECIGO are consistent with those in~\cite{Zhang:2017sym,Takahiro}.

\subsubsection{Einstein-dilaton Gauss-Bonnet gravity}\label{sec:EdGB} 
Now let us consider the EdGB theory of gravity.
Similar to before, we consider current constraints on the coupling parameter $\sqrt{\alpha_\EdGB}$, as well as future constraints from third-generation detectors, multi-band observations, and multi-event combinations.

\begin{figure}
\begin{center}
\includegraphics[width=.7\columnwidth]{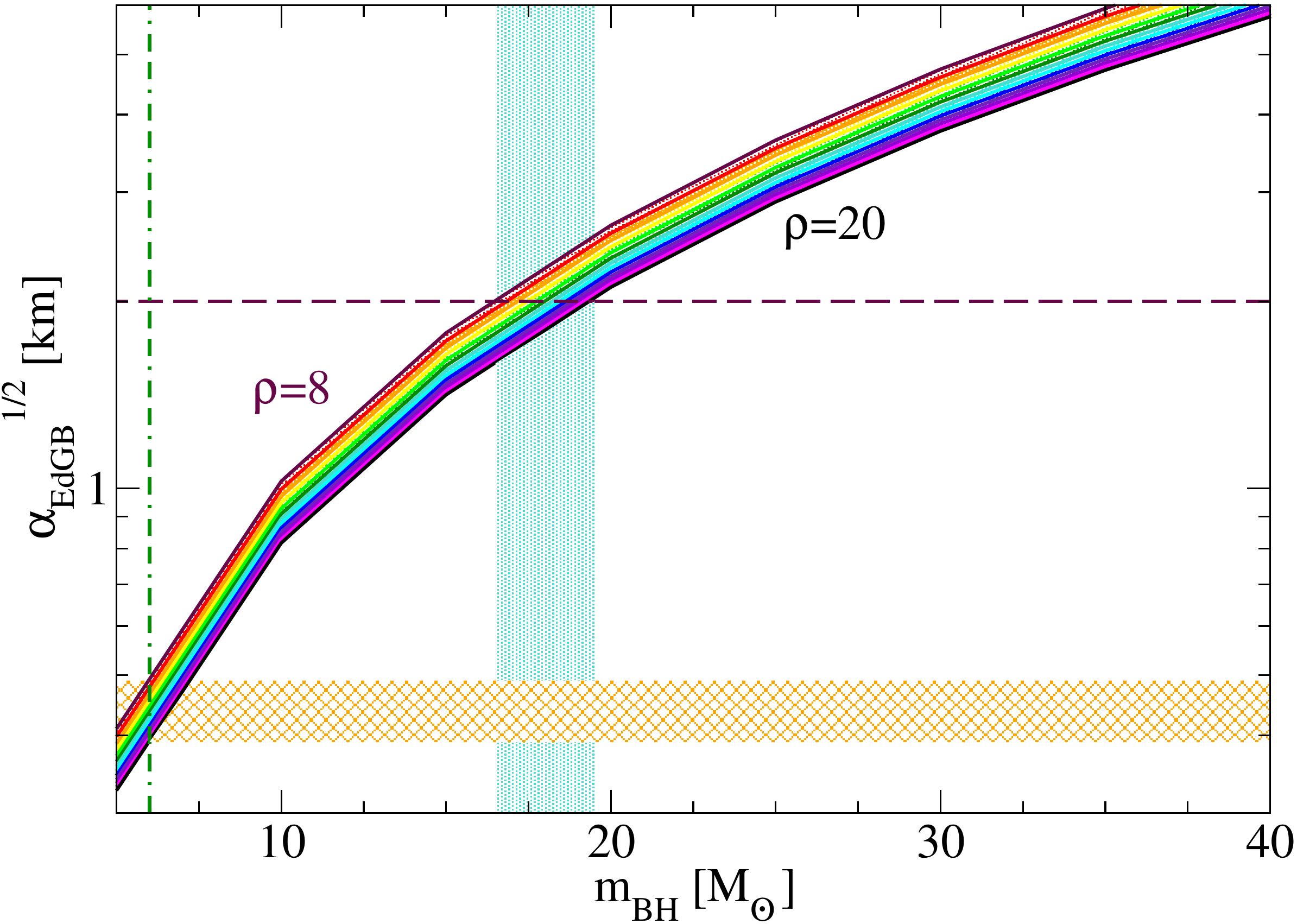}
\caption[Current EdGB constraints]{Projected $68$\% confidence interval bounds on the EdGB coupling parameter $\sqrt{\alpha_\EdGB}$ as a function of the black hole mass $m_\BH$ merging with a $1.4\text{ M}_\odot$ NS.
Such constraints are presented for event SNRs ranging from $\rho=8$ to 20 with iterations of 1.
Observe that the strongest constraint in the literature~\cite{Yagi_EdGB,Nair_dCSMap,Yamada:2019zrb} can be improved upon for events with $m_\BH<16.5\text{ M}_\odot$, with the intersection displayed by the vertical shaded turquoise region.
Observe also that the BH/NS candidate S190426c (with a 58\% probability of terrestrial origin rather than astrophysical) with a likely BH mass of $\sim6\text{ M}_\odot$~\cite{Lattimer:2019qdc} can place a constraint of $\sqrt{\alpha_\EdGB}<\lbrack0.4-0.5\rbrack$ km, indicated by the shaded orange region, which is stronger than the current bound by a factor of 4--5.
}\label{fig:aligoEDGB}
\end{center}
\end{figure}

We begin by discussing the current observational constraints on $\sqrt{\alpha_\EdGB}$, had a BH-NS coalescence been observed by the current iteration of LIGO interferometers.
Figure~\ref{fig:aligoEDGB} projects the prospective constraints on $\sqrt{\alpha_\EdGB}$ for BH-NS binaries with $m_\NS=1.4\text{ M}_\odot$ as a function of $m_\BH$ for detection SNRs ranging between 8 and 20 on the aLIGO O3 detector.
Observe how for BHs with mass less than $16.5\text{ M}_\odot$ ($19.5\text{ M}_\odot$), the current constraint in the literature of $2$ km can be improved upon for events with SNR $=8$ (20).
Thus, if S190426c or S190814bv are NS-BH merger events with sufficiently low-mass BHs, such events would place a bound in EdGB gravity that is stronger than the existing bounds.
Reference~\cite{Lattimer:2019qdc} estimated the properties of S190426c from the probability of the system being in specific categories, such as BH/NS. 
In particular, the BH mass is estimated as $\sim 6M_\odot$. 
If S190426c was indeed a BH/NS system (58\% probability of terrestrial origin) and if this mass estimate was correct, we can place strong constraints on $\sqrt{\alpha_\EdGB}$ of $0.4$ ($0.5$) km for a 20 (8) SNR event -- a factor of 4--5 improvement from the current observational constraint.

We follow this up with a discussion of future constraints placed on EdGB gravity.
Similar to the previous section, we estimate the constraints placed on $\sqrt{\alpha_\EdGB}$ from a $10\text{ M}_\odot-1.4\text{ M}_\odot$ BH-NS merger event detected on each detector, in the single-event, multiple-event, and multi-band cases.
Figure~\ref{fig:stackedEDGB} displays the corresponding bounds for each scenario.
Observe that the single-event rates can place constraints between $0.02-1$ km, all stronger than the current bound of $2$ km.
Further, we see that the multi-band constraints do not offer much improvement from the single-band case, while the combined event bounds can reach down to $\sim 10^{-5}$ km with DECIGO, improving the current bounds by up to five orders-of-magnitude. 
These bounds with DECIGO are consistent with the rough estimate presented in~\cite{Yagi_EdGB} and a recent analysis of~\cite{Gnocchi:2019jzp} for binary black holes with single events.

\begin{figure}
\begin{center}
\includegraphics[width=.7\columnwidth]{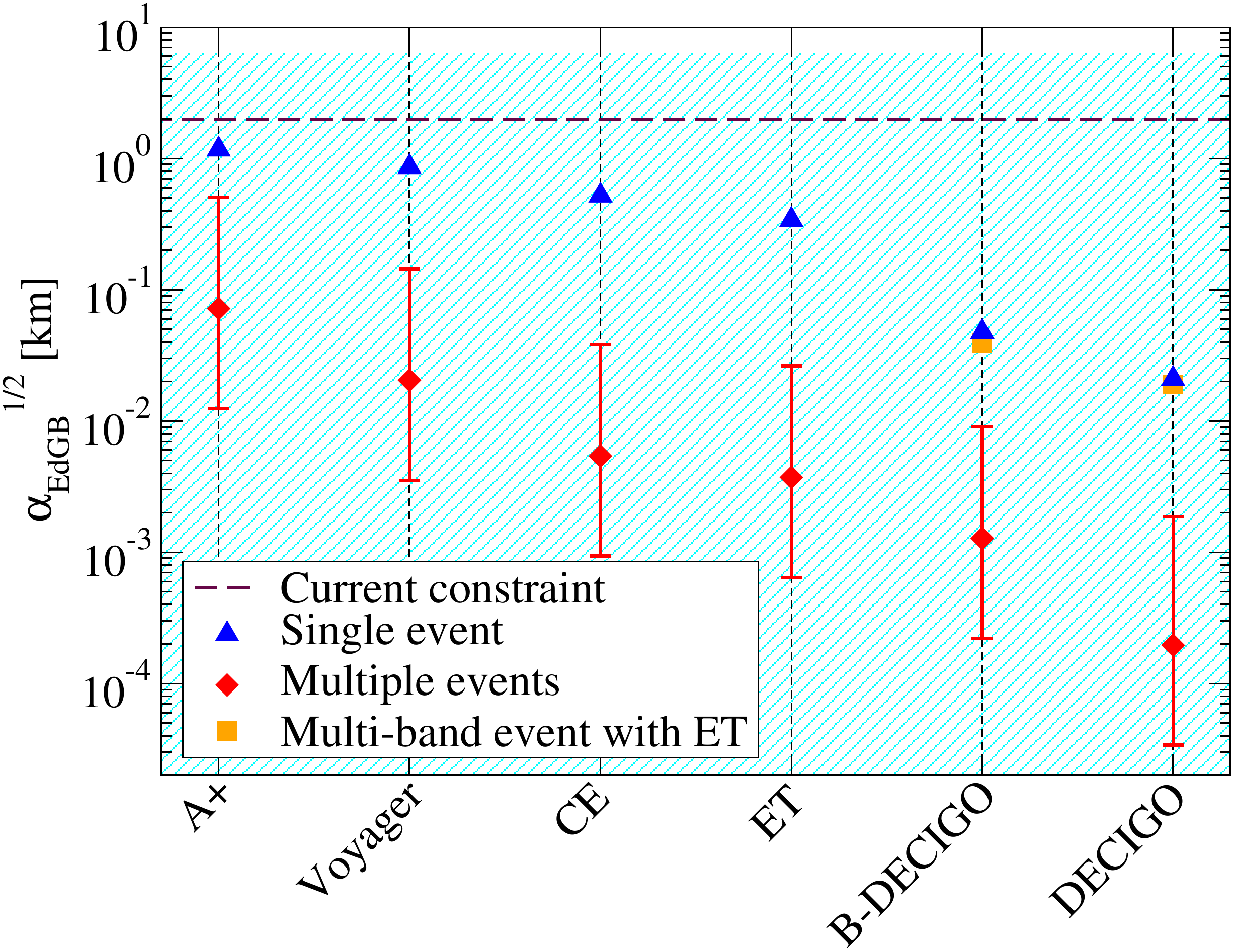}
\caption[Future EdGB constraints]{
Estimated $68$\% confidence interval constraints on the EdGB coupling parameter $\sqrt{\alpha_\EdGB}$ for a $10\text{ M}_\odot-1.4\text{ M}_\odot$ BH-NS merger event as observed on each detector.
The blue triangles represent single-event detections, while the red error bars correspond to the combined constraints from multiple events, with the upper, central, and lower bounds corresponding to the optimistic, ``realistic", and pessimistic number of detections~\cite{Abadie:2010cf}.
The orange squares give the multi-band result in conjunction with ET, and the shaded cyan region is where the small coupling approximation is valid.
Finally, the horizontal dashed line corresponds to the current most stringent result~\cite{Yagi_EdGB,Nair_dCSMap,Yamada:2019zrb}.
}\label{fig:stackedEDGB}
\end{center}
\end{figure}

\subsubsection{Validity of approximations}\label{sec:approximations}
In this section we explore the simplifying approximations made in the above analysis, and their effects on the presented results.
We begin with a discussion on the number of higher-order PN corrections added to the gravitational waveform.
In the main analysis, only the leading order $-1$PN correction term was taken into account in the gravitational waveform, and here we discuss the effect of including corrections up to $0.5$PN order. 
Next we consider the effect of rotating BHs, rather than the static ones considered in the analysis.
Finally we conclude with an alternative method to combine multiple events with varying BH and NS masses, rather than the fixed masses considered in the main analysis.

\begin{itemize}
\item \textit{Higher-order STT corrections to the waveform}\label{sec:highorder}

Let us begin with a discussion of the higher-order PN corrections present in the gravitational waveform.
We start by taking into account the $l=m=2$ dipolar contributions to the Fourier domain phase found in Eq.~(81c) of Ref.~\cite{Sennett:2016klh}\footnote{Note the non-dipolar phase corrections are not relative to the $-1$PN phase correction, proportional to $\Delta\alpha$, so were not included in this approximation for simplicity.} up to $0.5$PN order. 
We make the approximation $\alpha_\text{A} \ll 1$ such that only terms proportional to $\Delta\alpha^2$ (and thus also proportional to $\beta_\ppE^{(\DEF)}$) remain, making it possible to keep correlations minimal by allowing only one non-GR parameter to remain in the waveform, $\beta_\ppE^{(\DEF)}$\footnote{Note that the addition of new parameters to the gravitational waveform template may act to weaken the obtained constraints on $\Delta\alpha$ due to increased correlations between the parameters. A detailed analysis on the magnitude of this effect is beyond the scope of this analysis. The error due to linearization of the dipole radiation is expected to be minimal because $\beta_\ppE^{(\DEF)} \ll (\Delta\alpha)^2$.}.
The resulting DEF corrections to the gravitational waveform up to $0.5$PN order are given by
\begin{equation}
  \delta\psi=\beta_\ppE^{(\DEF)} u^{-7} \left(1+\frac{2623+2640\eta}{4280} u^{2}-6\pi u^{3}\right)\,,
\end{equation}
where $\beta_\ppE^{(\DEF)}$ is the $-1$PN ppE parameter given in Eq.~\eqref{eq:ppeBeta}.
We now include these additional corrections into the gravitational waveform and recompute bounds on the DEF theory for an SNR 10 event with a BH mass of $10\text{ M}_\odot$ on the LIGO O3 detector.
We find constraints on $\Delta\alpha$ to be $\sim 0.104$ with leading-order $-1$PN corrections included, and $\sim 0.105$ (negligible difference on a plot such as Fig.~\ref{fig:aligoDEF}) with higher-order corrections to $0.5$PN included.
Such results have a $\sim 0.4$\% difference, and are likely to be negligible in the presented analysis. 
A similar result was found in~\cite{Yunes:2016jcc} for Brans-Dicke theory.
We also bring to attention Ref.~\cite{Tahura:2019dgr}, where it was discovered that the additional presence of non-GR amplitude corrections to the waveform only differs from those in phase by $\sim4$\% - another negligible difference in our analysis.

\item \textit{Rotating BHs}\label{sec:rotating}
Now let us discuss the effect of considering rotating BHs in our analysis, rather than the static ones considered previously. 
To perform this simple comparison, we compute constraints on $\Delta\alpha$ in the DEF theory of gravity, and $\sqrt{\alpha_\EdGB}$ in the EdGB theory of gravity for a SNR 10 event with a BH mass of $10\text{ M}_\odot$ on the LIGO O3 detector, with the assumptions of BH spins $\chi_\BH=(0,0.5,1)$.
The resulting constraints were found to be $\Delta\alpha < (0.105,0.104,0.103)$ and $\sqrt{\alpha_\EdGB} < (1.80\text{ km},1.79\text{ km},1.78\text{ km})$ respectively, with only a maximal difference of $\sim1.5$\% found between them in either case.
Thus, we conclude that the effect of rotating BHs in our analysis is sufficiently negligible.

We also point out that the inclusion of spin precession does not play a crucial role in the order-of-magnitude constraint of $-1$PN effects. 
This is demonstrated in Table~IV of Ref.~\cite{Yagi:2009zm}, where it was shown that spin precession strengthens constraints on the Brans-Dicke parameter by $\sim43$\%, due to small correlations between the $-1$PN ppE parameter and the spins entering at $1.5$PN.
Such an effect is magnified for increasing mass-ratio systems, much larger than those considered in this analysis, as they tend to increase the effects of precession.

\item \textit{BH/NS mass populations}\label{sec:varymass}
In this section we model an appropriate BH-NS mass distribution function and implement it into the procedure used to combine the uncertainty in non-GR parameters from $N$ events, now with variational masses.
To do this, we modify the expression given in Eq.~\eqref{eq:stacking} by injecting a mass-distribution function $f(m_\BH,m_\NS)$ like so
\begin{align}
   \nonumber \sigma_{\theta^a}^{-2}=\Delta T\int\limits^{40\text{ M}_\odot}_{5\text{ M}_\odot}\int\limits^{2\text{ M}_\odot}_{1\text{ M}_\odot}\int\limits_0^{z_h(m_\NS, m_\BH)}&4\pi \lbrack a_0 r(z) \rbrack^2 \mathcal{R} R(z) \frac{d\tau}{dz} \sigma_{\theta^a}^{-2}(z,m_\BH,m_\NS)\\
     &\times f(m_\BH,m_\NS) dz dm_\NS dm_\BH,
\end{align}
where $\sigma_{\theta^a}(z,m_\BH,m_\NS)$ and $z_h(m_\NS,m_\BH)$ are now interpolated functions that also depend on the individual binary masses. For simplicity, we assume that $f(m_\BH,m_\NS)$ for BH/NS is simply given by a product of the individual mass distributions $f_\BH(m_\BH)$ and $f_\NS(m_\NS)$ as $f(m_\BH,m_\NS)=\mathcal{C}f_\BH(m_\BH)f_\NS(m_\NS)$, where the constant $\mathcal{C}$ is determined by 
normalizing the function to be unity when being integrated over $m_\BH$ and $m_\NS$. We use $f_\BH(m_\BH)$ as the mass distribution of primary black holes in stellar-mass BH binaries derived by the LIGO/Virgo Collaborations~\cite{LIGOScientific:2018jsj}, while we adopt $f_\NS(m_\NS)$ as a Gaussian distribution used e.g. in~\cite{Taylor:2011fs}. Namely, we have
\begin{equation}
    f_\BH(m_\BH) \propto\left\{\begin{array}{ll}{\left(\frac{m_\BH}{\text{M}_\odot}\right)^{-\alpha} } & {\text { if } m_{\min }  \leq m_\BH \leq m_{\max }} \\ {0} & {\text { otherwise }}\end{array}\right\},\hspace{5mm}f_\NS(m_\NS) \propto\mathcal{N}(\mu_\NS,\sigma_\NS).
\end{equation}
The relevant parameters $\alpha$, $m_\text{min}$, $m_\text{max}$, $\mu_\NS$, and $\sigma_\NS$ have been fit to be $0.4$, $5\text{ M}_\odot$, $41.6\text{ M}_\odot$, $1.34\text{ M}_\odot$, and $0.06\text{ M}_\odot$ respectively.
We perform a grid search with 8 redshift values between 0 and 8, 10 NS masses between $1\text{ M}_\odot$ and $2\text{ M}_\odot$ and 10 BH masses between $5\text{ M}_\odot$ and $40\text{ M}_\odot$ to compute $\sigma_{\sqrt{\alpha_\EdGB}}(z,m_\BH,m_\NS)$ for 800 mass/redshift samples, which is then interpolated.

We perform this example computation for the EdGB theory of gravity on the CE detector with the ``realistic" number of events ($730,000$) and compare the resulting constraint on $\sqrt{\alpha_\EdGB}$ to the case of fixed-mass binaries presented in the main analysis.
Under the described circumstances, we find a constraint on $\sqrt{\alpha_\EdGB}$ of $0.003$ km with the new variational mass model. 
Compared to the static-mass model result of $0.004$ km, we find that the two methods agree to within 25\%.
Interestingly, we find the new constraints to be stronger than the old ones as the new analysis includes BH masses lower than $10M_
\odot$ considered originally, and thus the results displayed in the main text can be presented as a conservative estimate.
Because the relationship between $\alpha_0$ and $\beta_0$ in DEF theory of gravity themselves depend on the constituent masses, this method can not be used to reliably compute bounds in the $(\alpha_0,\beta_0)$ plane.
However, we expect to find similar results to the EdGB case (where $\alpha_\EdGB^2$ doesn't depend on the masses). This is verified by instead estimating constraints on $\Delta\alpha$ only, which agrees to the static-mass model to within 25\% as well.
\end{itemize}


\subsection{Probing EdGB gravity with full-waveform corrections}\label{sec:QNMs}

Now let us discuss constraints on the EdGB theory of gravity obtained with corrections to the gravitational waveform in both the inspiral, merger-ringdown, and remnant BH predictions.
For a comparison between constraints found in this section and with the inspiral-merger-ringdown consistency tests found in Chapter~\ref{chap:IMRconsistency}, see Tab.~\ref{tab:alphas}.
We perform the parameterized test of GR with EdGB corrections to the waveform to $\mathcal{O}(\chi^2)$ in BH spin, followed up by a discussion and demonstration of corrections to $\mathcal{O}(\chi^4)$.

\subsubsection{$\mathcal{O}(\chi^2)$ corrections to BH spin}\label{sec:2ndOrdSpin}

We start by performing the analysis with EdGB corrections to $\mathcal{O}(\chi^2)$ in BH spin.
Figure~\ref{fig:Param_alpha2} presents the upper bounds on $\sqrt{\alpha_\EdGB}$ from the most massive binary BH event detected to date, GW170729, with various EdGB corrections considered. Observe first that when only the inspiral correction is considered, the bound is beyond the validity of the small coupling approximation, while those become valid once we consider corrections to QNMs. This shows the importance of the latter when constraining EdGB gravity with large mass binaries for which the contribution of the ringdown is larger.

\begin{figure}[!htbp]
\begin{center}
\includegraphics[width=.7\columnwidth]{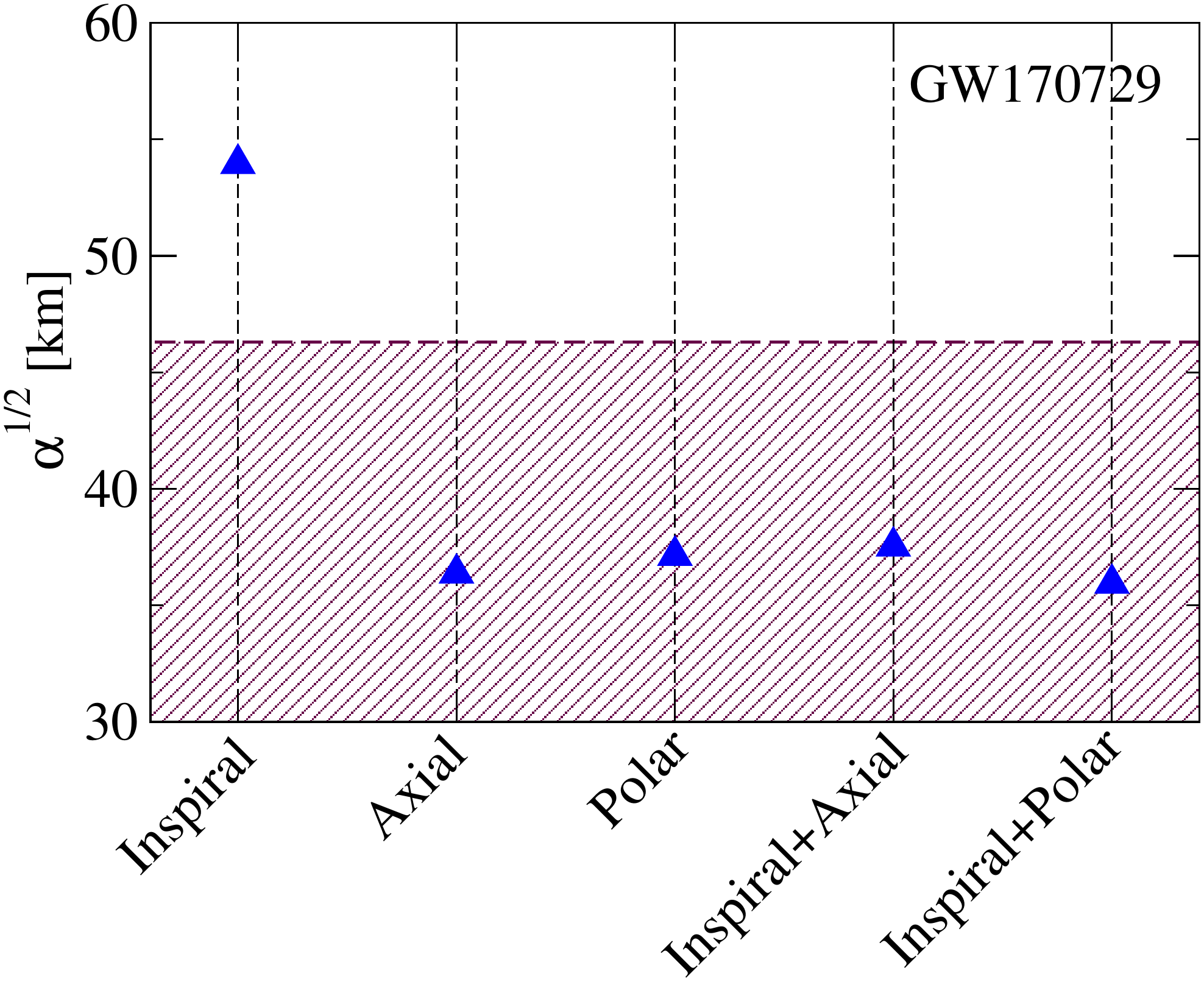}
\caption[GW170729 constraints on $\sqrt{\alpha_\EdGB}$ with different waveform corrections]{
90\%-credible upper bounds on the EdGB parameter $\sqrt{\alpha_\EdGB}$ for the most massive binary BH event detected to date: GW170729.
Such constraints are obtained via a Fisher analysis with the parameterized tests of GR.
Such bounds are organized into six categories (represented by the columns in each panel) of EdGB corrections introduced to the GR waveform as discussed in Sec.~\ref{sec:edgbCorrections}: GR, inspiral, axial QNMs, polar QNMs, inspiral+axial QNMs, and inspiral+polar QNMs.
Observe the importance of including non-GR effects in the merger-ringdown waveform for massive events, as the small-coupling approximation (valid only in the shaded region) becomes invalid otherwise.
}\label{fig:Param_alpha2}
\end{center}
\end{figure}

Figure ~\ref{fig:Param_alpha1} displays the upper bounds on $\sqrt{\alpha_\EdGB}$ observed on O1/O2 runs for each GW event considered in this analysis.
We observe several things in regards to this.
Firstly, the smaller total mass events correspond to stronger constraints.
This is because the expressions in Eqs.~\eqref{eq:alpha_ppE}-\eqref{eq:beta_ppE} minimize $\sqrt{\alpha_\EdGB}$ for both minimal mass ratio \textit{and}, more notably, the individual mass.
Second, we observe that the type of EdGB corrections to the waveform does not strongly affect the two more massive events (GW150914, and GW170729), while the two lighter events (GW170608 and GW151226) observe a deterioration in constraining EdGB gravity when only including the axial/polar QNMs. This is because the fraction of the ringdown portion in the observed waveform becomes larger for larger-mass binaries, and hence QNM corrections become more important for these binaries.
Third, we observe that for more massive events such as GW170729, the inclusion of only inspiral EdGB effects results in an invalid constraint due to violation of the small coupling approximation, as already shown in Fig.~\ref{fig:Param_alpha2}. 
Similar conclusions are made later in Chapter~\ref{chap:IMRconsistency} for massive events, in which the merger-ringdown portion of the gravitational waveform began to make significant contributions to the inspiral-merger-ringdown consistency test, compared to the inspiral portion.

\begin{figure}[!htbp]
\begin{center}
\includegraphics[width=.7\columnwidth]{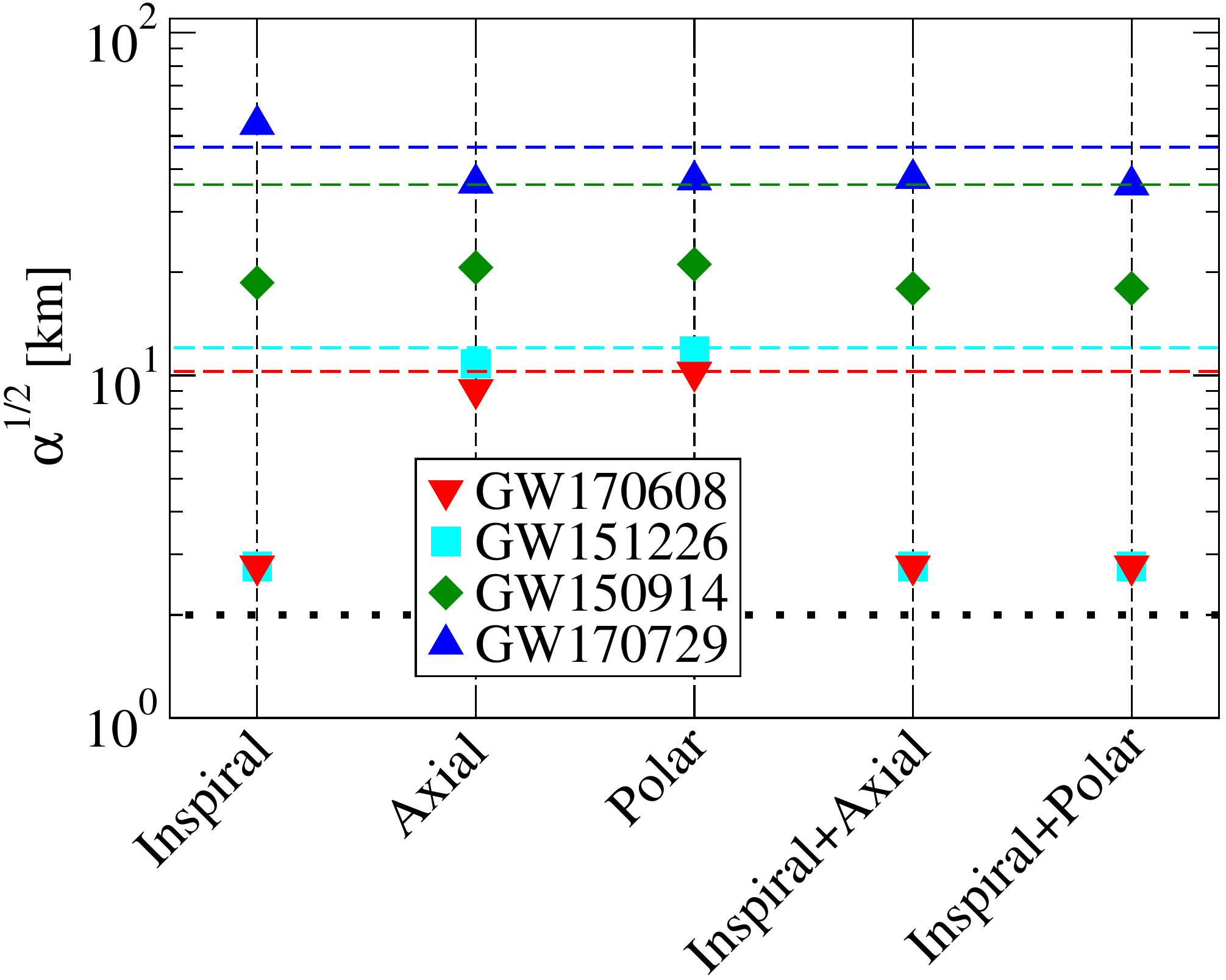}
\caption[Constraints on $\sqrt{\alpha_\EdGB}$ with different waveform corrections]{
Same as Fig.~\ref{fig:Param_alpha2} but for various GW events GW170608, GW151226, GW150914, and GW170729 (in order of increasing mass).
The dashed horizontal lines represent the small coupling approximation $\zeta_\EdGB \ll 1$ for events of the same color, representing invalid constraints when placed above the corresponding line.
The dotted black horizontal line corresponds to the current constraint of $\sqrt{\alpha_\EdGB}\leq 2$.
}\label{fig:Param_alpha1}
\end{center}
\end{figure}

Next we consider the future detectability of EdGB effects in the waveform.
Figure~\ref{fig:Param_alpha3} displays the possible upper bound on $\sqrt{\alpha_\EdGB}$ observed by CE, LISA, and the multiband observation between the two for GW150914-like events, which indeed lie in the multiband detectability region displayed in Sec.~\ref{sec:SNRs}.
We note that EdGB effects with only axial/polar QNM corrections can not be probed by LISA (thus multiband observations give the same result as CE detections alone) due to its cutoff frequency of $1$ Hz.
We observe that LISA observations alone can improve the ability to probe EdGB gravity by roughly one order of magnitude from CE observations alone, with little difference made by the addition of axial/polar QNM corrections. 
Multiband observations further improve the bound by about a factor of two. Notice also that the LISA and multiband bounds are stronger than current bounds~\cite{Kanti_EdGB,Pani_EdGB,Yagi_EdGB,Witek:2018dmd,Nair_dCSMap,Yamada:2019zrb,Tahura:2019dgr}.
See also Table~\ref{tab:alphas} for a comparison between the $\sqrt{\alpha_\EdGB}$ constraints found in this chapter with the parameterized tests, and the ones found in Chapter~\ref{chap:IMRconsistency} with the inspiral-merger-ringdown consistency tests.
In particular, the latter analysis utilized the same EdGB corrections to the gravitational waveform used here, and then tested the consistency between the inspiral and merger-ringdown signals for varying values of $\alpha_\EdGB$.
We find that the bounds from the two analyses are comparable to each other.

\begin{figure}[htb]
\begin{center}
\includegraphics[width=.7\columnwidth]{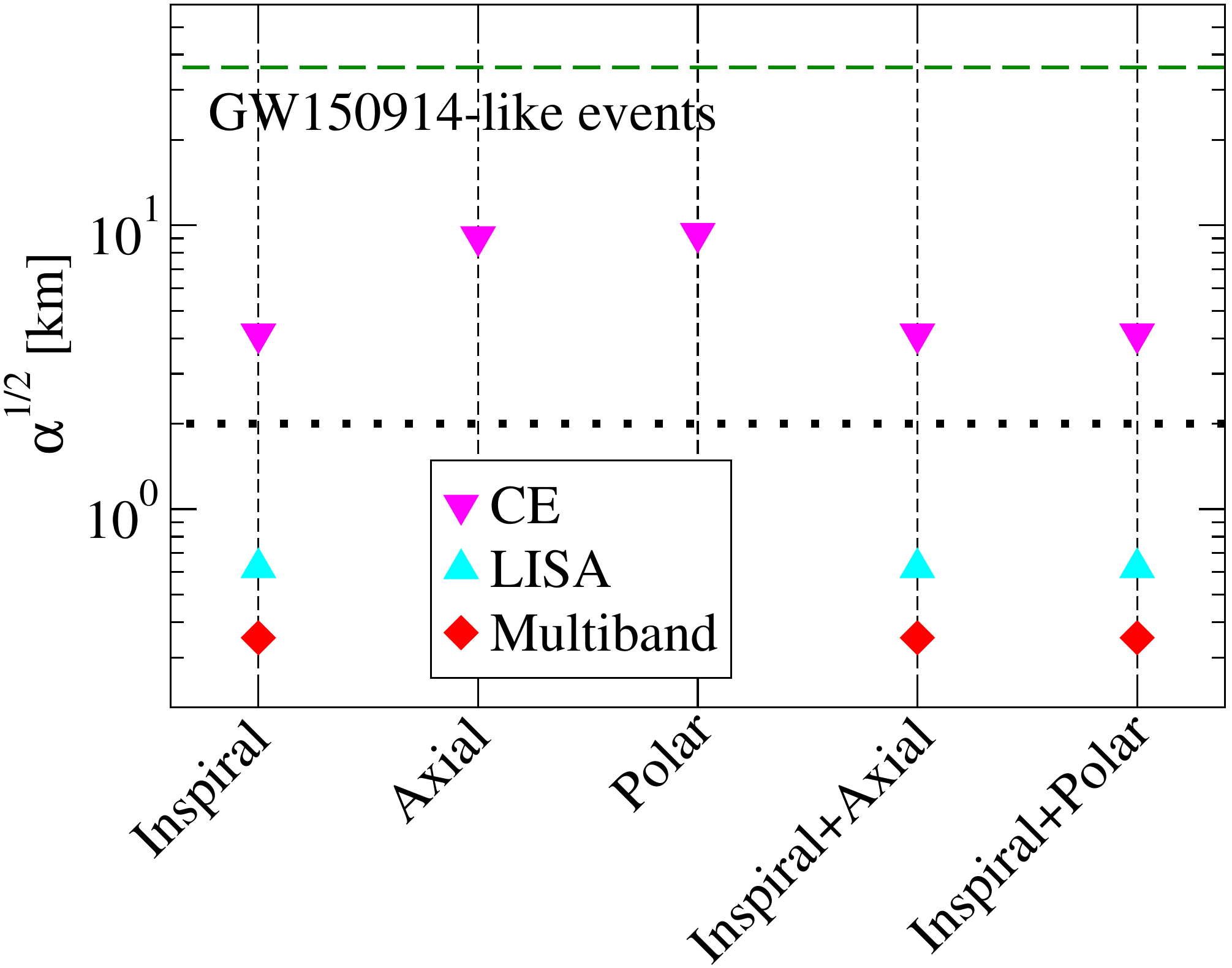}
\caption[Future constraints on $\sqrt{\alpha_\EdGB}$ with different waveform corrections]{
Same as Fig.~\ref{fig:Param_alpha1} but for future GW150914-like events detected by the ground-based detector CE, space-based detector LISA, and the multi-band observation between the two.
We note that no space-based or multiband bounds appear in the axial/polar QNMs columns, because space-based detectors such as LISA can not observe the merger-ringdown effects for GW150914-like events occurring at high frequencies.
}\label{fig:Param_alpha3}
\end{center}
\end{figure}

\renewcommand{\arraystretch}{1.2}
\begin{table}
\centering
\addvbuffer[12pt 8pt]{\begin{tabular}{c c c}
\multirow{2}{*}{Detector} & $\sqrt{\alpha_\EdGB}$ [km] & $\sqrt{\alpha_\EdGB}$ [km] \\
 & (Param.) &  (IMR consist. Chapter~\ref{chap:IMRconsistency})\\
\hline
aLIGO & $17$  & $15$  \\
CE & $5$  & $8$ \\
LISA & $0.6$  & -- \\
Multiband & $0.3$  & $0.2$  \\
\end{tabular}}
\caption[EdGB constraints from the IMR consistency and parameterized tests]{Comparison between the current and future upper bound on $\sqrt{\alpha_\EdGB}$ obtained in this chapter with the parameterized tests (Param.) and with the inspiral-merger-ringdown consistency tests (IMR. consist.) discussed in Chapter~\ref{chap:IMRconsistency}.
Such constraints were formed from GW150914-like events, with the both inspiral and axial QNM EdGB effects included in the waveform template.
Observe how constraints obtained from both tests produce comparable results on the detectability of EdGB effects in the GW signal. 
}\label{tab:alphas}
\end{table}

\subsubsection{$\mathcal{O}(\chi^4)$ corrections to BH spin}\label{sec:4thOrdSpin}
In this section we compute EdGB corrections to the gravitational waveform up to quartic order in BH spin, to check the validity of the slow-rotation approximation to quadratic order in spin used in the previous subsection.
We begin by expanding the expressions for the inspiral dipole radiation and QNM corrections already computed in Sec.~\ref{sec:theory} to quartic order in BH spin.
Next we compute corrections to $r_\ISCO$,  $E_\text{orb}$ and $L_z$ via the EdGB spacetime metric $g_{\alpha\beta}^\EdGB$ found in Ref.~\cite{Maselli:2015tta}, where they computed each element up to 5th order in BH spin $\chi$.
The orbital energy and angular momentum can be obtained from $g_{\alpha\beta}^\EdGB$ by simultaneously solving the equations $V_\text{eff}(r)=0$ and $\frac{d}{dr}V_\text{eff}(r)=0$ for $E_\text{orb}$ and $L_z$ with effective potential given by
\begin{equation}\label{eq:Veff}
V_\text{eff}(r)=\frac{E_\text{orb}^2g_{\phi\phi}^\EdGB+2E_\text{orb}L_zg_{t\phi}^\EdGB+L_z^2g_{tt}^\EdGB}{(g_{t\phi}^\EdGB)^2-g_{tt}^\EdGB g_{\phi\phi}^\EdGB}-1
\end{equation}
Finally, the location of the ISCO is given by  $\frac{d}{dr}E_\text{orb}(r_\ISCO)=0$.

\renewcommand{\arraystretch}{1.2}
\begin{table}
\centering
\addvbuffer[12pt 8pt]{\begin{tabular}{c c c c}
\multirow{2}{*}{GW Event} & $\sqrt{\alpha_\EdGB}$ $(\chi^2)$ & $\sqrt{\alpha_\EdGB}$ $(\chi^4)$ &frac. diff.  \\
  & [km] & [km] & [$\%$]\\
\hline
GW170608~\cite{GW170608} &2.29 & 2.28 & 0.4 \\
GW151226~\cite{GW151226} & 2.76 & 2.75 & 1.1 \\
GW150914~\cite{GW150914} & 17.16 & 17.15 & 0.1 \\
GW170729~\cite{GW170729} & 28.71 & 28.29 & 1.5 \\
\end{tabular}}
\caption[$\sqrt{\alpha_\EdGB}$ constraints with $\mathcal{O}(\chi^2)$ and $\mathcal{O}(\chi^4)$ waveform corrections]{Constraints on $\sqrt{\alpha_\EdGB}$ obtained with EdGB corrections to the waveform up to  quadratic order in BH spin (2nd column), and quartic order in BH spin (3rd column). The last column shows the fractional difference between the two.
We observe that such results agree to within $1.5\%$ in all cases, with the the largest difference appearing for the most massive event GW170729.
}\label{tab:4thOrderSpin}
\end{table}

With the above corrections to the entire gravitational waveform to quadratic order in spin, we estimate constraints on EdGB parameter $\sqrt{\alpha_\EdGB}$.
In particular, we compute constraints on $\sqrt{\alpha_\EdGB}$ for each GW event considered in this analysis: GW150914, GW151226, GW170608, and GW170729 as detected on the O2 detector, with non-zero fiducial BH spins.
We compare these results with those of the main analysis, with corrections to only quadratic order in spin.
Table~\ref{tab:4thOrderSpin} presents a comparison between constraints on $\sqrt{\alpha_\EdGB}$ obtained from (i) waveforms with corrections to quadratic order in BH spin, and (ii) to quartic order in BH spin.
We find that such results agree with each other to between $0.2\%$ and $1.5\%$, with the latter resulting from the massive BHs in GW170729, in which spin effects become more important as it seems to have the largest final spin out of the 4 GW events considered here.
Therefore we conclude that the effect of higher-order spin corrections to the gravitational waveform has up to a $\sim1.5\%$ effect on our predictions, which validates our order-of-magnitude estimation presented in this chapter including up to quadratic order.

Finally, we consider the effect of including spin effects into the remnant BH QNMs.
For example, in the dCS theory of gravity, all of the ingredients required to correct the full waveform considered here are available, with the exception of the QNM spin corrections.
Here, we remove all EdGB spin effects to the remnant QNM corrections and compute constraints on $\sqrt{\alpha_\EdGB}$.
We find the constraint to be $27.58\text{ km}$ for GW170729 with which the contribution of the ringdown is most significant out of the 4 GW events considered. Such a constraint agrees very well with those tabulated in Table~\ref{tab:4thOrderSpin} for spin corrections to both $\mathcal{O}(\chi^2)$ and $\mathcal{O}(\chi^4)$, with the largest difference being with the most massive event again.
Therefore, we conclude that spin effects in the remnant BH QNMs make only a negligible impact on constraints on $\sqrt{\alpha_\EdGB}$. 


\section{Conclusions}\label{sec:conclusion}
In this chapter, we have highlighted the power in found in the parameterized tests of GR.
This was applied to three different cases: the multi-band observations between ground- and space-based GW detectors on the constraint of various modified theories of gravity; the constraint of STTs such as DEF and EdGB under the observation of mixed BH-NS binaries; and finally probes of EdGB gravity with waveform corrections to both the inspiral and ringdown of the gravitational waveform.

We began by performing parameterized tests of GR by considering generalized modifications to the GW phase, finding that multi-band observations can provide constraints reaching up to 40 times stronger than their single-band counterparts.
Such constraints were applied to the specific cases of EdGB, dCS, scalar-tensor, noncommutative, varying-$G$, varying-$M$, and massive graviton theories of gravity, resulting in constraints on the theories' associated parameters.
In particular, we find that constraints placed on the EdGB, dCS, noncommutative, and massive graviton non-GR effects show improvement upon the current constraints (by up to seven orders-of-magnitude for dCS gravity) found in the literature when making use of multi-band detections.

Next we demonstrated the present and future considerations on constraining STTs which violate the SEP with mixed BH-NS binaries.
We considered both the DEF and EdGB theories, which predict massless scalar fields $\varphi$ which couple to matter and alter the consequent trajectories of gravitating bodies.
We investigate constraints placed on these theories' coupling parameter spaces for the possible detection of BH-NS coalescences, both on the current iteration of LIGO interferometers, and with future GW detectors both on the ground and in space.
In the DEF theory, we find that if such an event (such as the possible candidates S190426c or S190814bv in the O3 run) were to be observed with the present GW detection capabilities, competing bounds to those from pulsar timing observations can be presented.
In EdGB theory, we find that with BH masses less than $19.5\text{ M}_\odot$, improvements to the current constraint on the coupling parameter $\sqrt{\alpha_\EdGB}<2$ km can be made to the order of $\mathcal{O}(0.1)$ km.
Such events detected on future GW detectors (single-event, multi-band observations, and multiple-event stacking) have been demonstrated to improve upon the current bounds by several orders of magnitude in many cases.

Finally, we have modeled the resulting EdGB effects throughout various parts of the gravitational waveform, including the inspiral, the characteristic ringdown QNMs, and finally, to the final mass and spin properties of the remnant BH.
With these new tools in hand, we offer predictions on the future detectability of such EdGB effects present in the gravitational waveform. 

We studied the detectability of EdGB effects in an observed GW signal by introducing various combinations of EdGB modifications to the inspiral and merger-ringdown portions of the waveform.
In particular, we discovered that for more massive events such as GW170729, the EdGB merger-ringdown contributions begin to hold high significance.
When only the inspiral corrections to the waveform (as is typically considered) were applied, the small-coupling approximation $\zeta_\EdGB \ll 1$ failed to be upheld. 
Only upon the inclusion of the merger-ringdown corrections does this quantity become satisfied, allowing for valid constraints on $\sqrt{\alpha_\EdGB}$.
We found that future space-based and multiband observations can place bounds that are stronger than current bounds on EdGB gravity.
We also found that the constraints on $\sqrt{\alpha_\EdGB}$ found here with parameterized tests agree well with those found in Chapter~\ref{chap:IMRconsistency} with the inspiral-merger-ringdown consistency tests.
Here we additionally considered the effects at higher order in spin to justify the use of slow-rotation approximation, finding that that such higher-order corrections only change our results up to a maximum of $1.5\%$.
We additionally investigate the effect of spin corrections to the remnant BH QNMs, finding that their inclusion has a negligible impact on parameter estimation.

Although we have taken into account the known EdGB corrections to the waveform to date as much as possible, there are some other modifications that have been left out. 
Below, we list some of the caveats in our analysis presented in this chapter:
\begin{itemize}
\item We only include leading-order post-Newtonian terms in the waveform, while a more advanced analysis could include higher order corrections. Though such corrections do not seem to be important in certain scalar-tensor theories~\cite{Yunes_ModifiedPhysics} (see App. B).
\item In GR, axial and polar QNMs are identical (isospectral), while such isospectrality is broken in EdGB~\cite{Blazquez-Salcedo:2016enn}. Thus, the ringdown portion of the waveform may be more complicated than that for GR.
\item Our estimate for the mass and spin of the BH remnant in EdGB gravity is based on the picture verified in GR, though this needs to be justified once NR simulations of binary black hole mergers are available in such a theory~\cite{Witek:2018dmd}.
\item We did not include corrections during the merger phase of the waveform. Again, it is likely that one needs to wait for NR simulations to realize how the corrections enter in this phase.
\end{itemize}
Having said this, we believe our calculations should be valid as an order of magnitude estimate. One reason to support this point is because corrections to the waveform enter linearly in $\zeta_\EdGB \propto \alpha_\EdGB^2$. Thus, even if our estimates are off by an order of magnitude in $\zeta_\EdGB$, bounds on $\sqrt{\alpha_\EdGB}$ are affected only by a factor of $\sim 10^{1/4} \sim 2$.

Faults such as the ones listed above can be remedied by the full construction of an EdGB (or any non-GR theory) waveform.
Work in this direction is already in progress such as Ref.~\cite{Witek:2018dmd}, where the scalar field dynamics during binary BH mergers have been expressed in EdGB gravity.
Very recently, the EdGB correction to the merger-ringdown waveform from a binary black hole has been computed~\cite{Okounkova:2020rqw}\footnote{See also Refs.~\cite{Okounkova:2017yby,Okounkova:2019dfo,Okounkova:2019zjf} for similar works in dCS gravity.}. 
In the future, we plan to compare such numerical-relativity waveforms with the simple analytic model presented here to quantify the validity of the latter.

For the purposes and scope of this investigation, the Fisher analysis has been used to predict valid order-of-magnitude constraints on the EdGB theory of gravity.
As thoroughly discussed in Ref.~\cite{Yunes_ModifiedPhysics}, for large enough SNR the results approximate well a Bayesian analysis. In the former reference, the Fisher-estimated non-GR parameter $\beta_\ppE$ in the inspiral agreed with its Bayesian counterpart to within $\sim40\%$ at $-1$PN for GW150914-like events with an SNR of 25. This corresponds to only a $\sim10\%$ difference on the coupling parameter $\sqrt{\alpha_\EdGB}$ in EdGB gravity. Regarding the latter, the 90\% credible contours in the final mass-spin plane obtained with Fisher and Bayesian analyses agreed with an error of 20\% for GW150914. Such agreements only strengthen considerably for the future detectors considered in this analysis.

In the above investigation, we utilized an approximate Fisher analysis based approach to predict posterior probability distributions on BH source parameters by assuming fiducial values given by the median values reported by the LVC.
A more comprehensive analysis would instead make use of the maximal likelihood values of such source parameters obtained directly from posterior probability distributions.
For highly skewed posterior probability distributions, the difference between the two values could potentially be significant.
Albeit, we find this point to be beyond the scope of this analysis, which is provided as a first step approximation to test GR with order-of-magnitude estimates on source parameters.
We leave this point, as well as a full Bayesian analysis to future work.

The analysis described about can be improved in numerous ways.
One example of such improvements would be to consider a full Bayesian analyses rather than the Fisher analysis used here.
Additionally, one can simulate the multi-band event rates discussed in Refs.~\cite{Sesana:2016ljz,Cutler:2019krq,Gerosa:2019dbe} to combine the signals and further reduce the systematic errors residing in our bounds on non-GR theoretical parameters.
Finally, we can use the results in Sec.~\ref{sec:amplitude} to consider alternative theories of gravity which modify the GW amplitude rather than phase~\cite{Alexander:2007kv,Yunes:2008bu,Yunes:2010yf,Yagi:2017zhb}.
Further, more accurate BH-NS population simulations other than those found in Ref.~\cite{Abadie:2010cf} may be utilized in future analyses, together with different masses for different events.
Finally, one could consider a more comprehensive list of STTs to study, rather than the select few examples investigated here: DEF, MO (see App.~\ref{app:theoryCompare} for a comparison between the two), and EdGB.

In the preceding chapter, we considered inspiral-ringdown waveform modifications from the EdGB theory of gravity as one given example. 
Future analyses could, given all the necessary ingredients described above, repeat the entire investigation using any given modified theory of gravity.
By simply knowing the leading PN corrections to the inspiral portion (known for most modified theories of gravity~\cite{Tahura_GdotMap}), corrections to the specific orbital energy $E_\text{orb}$ and angular momentum $L_z$, (known for theories such as dCS gravity~\cite{Yagi:2012ya}), and corrections to the QNMs (for dCS gravity, these are only known for non-spinning BHs~\cite{Yunes:2007ss,Cardoso:2009pk,Molina:2010fb}), the simple ``patchwork" analysis presented in this chapter could be revisited, without the need for a full non-GR waveform.


\newpage
\chapter{Inspiral-merger-ringdown consistency tests}\label{chap:IMRconsistency}
\begin{adjustwidth}{\parindent}{\parindent}
\begin{spacing}{1.5}
The extreme-gravity collisions between black holes allow us to probe the underlying theory of gravity in the extreme gravity sector of spacetime, where the fields are strong, non-linear, and dynamical.
To probe this theory of gravity, we apply the theory agnostic inspiral-merger-ringdown consistency test where one derives the mass and spin of a merger remnant from the inspiral and merger-ringdown independently assuming general relativity is correct, and then check their consistency.
One interesting future possibility is to detect gravitational waves from GW150914-like stellar-mass black hole binaries with both ground-based and space-based detectors using \textit{multi-band} gravitational-wave observations.
We apply a predictive forecast of the theory-agnostic inspiral-merger-ringdown consistency test to an example theory beyond general relativity for the first time, for future gravitational wave observations.  
Here we focus on the string-inspired Einstein-dilaton Gauss-Bonnet gravity and modify the inspiral, ringdown, and remnant black hole properties of the gravitational waveform. 
The formalism developed here can easily be applied to other theories.
\end{spacing}
\end{adjustwidth}
\newpage


\section{Introduction}\label{sec:intro}

The historic observation of GWs from the merger of two BHs by the LVC~\cite{GW150914} has ushered in the birth of a new era of astrophysics, for the first time probing the extreme gravity regime where spacetime is strong, non-linear, and dynamical\footnote{This chapter is based on the following papers: Z. Carson and K. Yagi, \textit{Probing string-inspired gravity with the inspiral-merger-ringdown consistency tests of gravitational waves}, Class. Quant. Grav. \textbf{37}, 215007 (2020), Z. Carson and K. Yagi, \textit{Parameterized and inspiral-merger-ringdown consistency tests of gravity with multi-band gravitational wave observations}, Phys. Rev. D \textbf{101}, 044047 (2020), Z. Carson and K. Yagi, \textit{Multi-band gravitational wave tests of general relativity}, Class. Quant. Grav. Letters \textbf{37}, 02LT01 (2019), and Z. Carson and K. Yagi, \textit{Parameterized and Consistency Tests of Gravity with Gravitational Waves: Current and Future}, Proceedings, Recent Progress in Relativistic Astrophysics: Shanghai, China, Vol. 17(1) (2019)}.
GWs such as these carry with them multitudes of information; not only regarding the sources' astrophysical properties, but also about the underlying theory of gravity driving the process.
However, this first event, as well as the following 10~\cite{GW_Catalogue}, have failed to detect any significant deviations from the predictions of GR~\cite{Abbott_IMRcon}, the prevailing theory of gravity for the past century~\cite{Will_SEP}.
While the current LVC infrastructure~\cite{TheVirgo:2014hva,TheLIGOScientific:2014jea} is a marvel of modern engineering, it may not yet be enough to uncover the elusive traces of a modified theory of gravity.
The next generation of GW detectors~\cite{advancedLIGO,Ap_Voyager_CE,ET,LISA,B-DECIGO,DECIGO,TianQin}, on the other hand, promise improvements on the order of $100$ times the sensitivity, as well as new sensitivity in the mHz regime.
Will this be enough to pull back the curtain on the hidden theories of gravity running the show?

Throughout the last century, countless tests and observations of GR have been performed~\cite{Will_SolarSystemTest,Stairs_BinaryPulsarTest,Wex_BinaryPulsarTest,Ferreira_CosmologyTest,Clifton_CosmologyTest,Joyce_CosmologyTest,Koyama_CosmologyTest,Salvatelli_CosmologyTest,Berti_ModifiedReviewLarge,Abbott_IMRcon,Yunes_ModifiedPhysics,Ishak:2018his}, all finding agreement with Einstein's theory in a variety of environments.
However, even with such success, GR still needs to be tested.
While it explains a majority of our observations, there yet remain several unanswered questions which could be explained by new theories of gravity.
For example, ``dark energy" and the accelerated expansion of the universe~\cite{Jain:2010ka,Salvatelli:2016mgy,Koyama:2015vza,Joyce:2014kja}, ``dark matter" and its influence on galactic rotation curves~\cite{Famaey:2011kh,Milgrom:DarkMatter,Milgrom:2008rv,Clifton:2011jh,Joyce:2014kja}\footnote{such rotation curves as well as other observations can be well explained by dark matter particle models, as well as certain modified theories of gravity, although the former typically gives stronger agreement with various observations.}, and more~\cite{Clifton:2011jh,Famaey:2011kh,Joyce:2014kja,Koyama:2015vza,Milgrom:2008rv,Jain:2010ka} all remain open to this day.
To date, a plethora of modified theories of gravity have been proposed to explain some of the open questions listed above.

In this chapter, we demonstrate the power of testing the consistency between the inspiral and merger-ringdown GW signals~\cite{Ghosh_IMRcon,Ghosh_IMRcon2,Abbott_IMRcon,Abbott_IMRcon2}.
Known as the inspiral-merger-ringdown (IMR) consistency tests of GR, we describe how one can, given independent measurements of (i) only the inspiral signal, and (ii) only the merger-ringdown signal, test the theory of GR assumed in a template waveform.
We begin with a demonstration of the power of multi-band observations of such signals between ground- and space-based detectors.
We then forecast current and future constraints on the EdGB theory of gravity as described thoroughly in Chapter~\ref{chap:Parameterized} with the IMR consistency test, for the first time applying such a test to a specific theory of gravity.
We consider EdGB corrections to not only the inspiral properties of a binary BH coalescence~\cite{Takahiro}, but also to the characteristic QNMs~\cite{Blazquez-Salcedo:2016enn} and final properties of the post-merger BH~\cite{Ayzenberg:2014aka} as discussed in Chapter~\ref{chap:Parameterized}.
To the best of our knowledge, the IMR consistency test has been put into context for an example modified theory of gravity for the first time, and can indeed be applied to other alternative theories of gravity, given the required ingredients.

While the following analysis is not entirely robust, it is presented as a new alternative route to obtain order-of-magnitude estimates (or better in most scenarios) without the significant time concerns required with full NR solutions, which do not yet exist for most alternative theories of gravity\footnote{Additionally, such NR simulations face challenges such as a lack of numerically stable formulations which prevent them from being simulated with currently-known methods.}.
In particular, in the following analysis we only consider the leading-order post-Newtonian corrections to the waveform, we utilize a predictive Fisher analysis rather than a comprehensive Bayesian one, we assume the QNMs are isospectral as they are in GR, and we neglect merger corrections to the merger-ringdown and only include the QNM ringdown corrections.
Such approximations lead the analysis to being less-robust, however it offers a new method to forecast estimated constraints on any given modified theory of gravity by taking into account additional pieces of information available to make the gravitational waveform closer to completion with a minimal degree of effort and computational time.
Following the Fisher analysis techniques found in Chapter~\ref{chap:ParameterEstimation}, we estimate the two-dimensional posterior probability distributions of remnant BH parameters $M_f$ and $\chi_f$ from both the inspiral and merger-ringdown signals.


\section{IMR consistency tests of GR}\label{sec:IMRmethod}

While two GW150914-like stellar-mass BHs in a binary system inspiral together via GW radiation, space-based GW interferometers can effectively probe the inspiral portion of the waveform, occurring at low frequencies.
Once the separation distance between the bodies become close enough, they fall into a plunging orbit until finally they merge, forming a common horizon which will settle down via radiation of quasi-normal modes~\cite{Chandrasekhar_QNM,Vishveshwara_QNM} - a high-frequency merger-ringdown signal which is best observed by ground-based GW detectors.
The remnant BH with mass $M_f=M_f(m_1,m_2,\chi_1,\chi_2)$ and spin $\chi_f=\chi_f(m_1,m_2,\chi_1,\chi_2)$ (provided by NR fits in Refs.~\cite{PhenomDII}) can then be entirely described by the same two parameters, in accordance with the BH no-hair theorems.

Using only the inspiral portion of the signal ($f<f_\ISCO=(6^{3/2}\pi M)^{-1}$~\cite{Abbott_IMRcon2}), the final mass and spin of the remnant BH can be uniquely estimated using predictions of the initial mass and spin parameters $m_1$, $m_2$, $\chi_1$, and $\chi_2$, while having no information about the merger-ringdown portion.
The opposite is also true: the final mass and spin may be predicted from the merger-ringdown portion of the signal ($f>f_\ISCO=(6^{3/2}\pi M)^{-1}$) with no accompanying information about the inspiral.
We utilize the Fisher analysis method from Chapter~\ref{chap:ParameterEstimation} to predict a four-dimensional probability distribution between $m_1$, $m_2$, $\chi_1$, and $\chi_2$, which can be transformed into the two-dimensional probability distribution between $M_f$ and $\chi_f$ using fits from numerical-relativity simulations.
Assuming the SNR is sufficiently large\footnote{GW150914 was observed with total SNR of 25.1, which is assumed throughout the analysis.} for both the inspiral and merger-ringdown waveforms, the estimates of $(M_f^\II,\chi_f^\II)$ should agree with those of $(M_f^\MR,\chi_f^\MR)$ within the statistical errors, assuming that GR is the correct theory of gravity.
This test, known as the \textit{IMR consistency test}~\cite{Ghosh_IMRcon,Ghosh_IMRcon2,Abbott_IMRcon,Abbott_IMRcon2}, enables one to detect emergent modified theories of gravity, manifesting themselves as a difference between the remnant BH parameters $(M_f,\chi_f)$, as computed from the inspiral, and merger-ringdown portions of the waveform individually.
Such a test can be performed by computing the two-dimensional posterior probability distributions $P_\II(M_f,\chi_f)$ and $P_\MR(M_f,\chi_f)$ from each section of the waveform.
The overlap of such distributions can determine how well GR describes the observed signal.

All detected GW signals to date have been found to be consistent with GR~\cite{Abbott_IMRcon,Abbott_IMRcon2,Ghosh_IMRcon,Ghosh_IMRcon2,Ghosh_2017}.
Reference~\cite{Ghosh_IMRcon2} phenomenologically introduced a non-GR correction at second post-Newtonian order in the gravitational wave energy flux and studied the IMR consistency test, though the reference did not include corrections to the QNM ringdown spectrum for simplicity.

While most similar tests are performed through a Bayesian statistical analysis~\cite{Abbott_IMRcon,Abbott_IMRcon2,Ghosh_IMRcon,Ghosh_IMRcon2,Ghosh_2017}, here we offer a new method using the Fisher analysis techniques described in Chapter~\ref{chap:ParameterEstimation} that is computationally less expensive. 
Namely, for each of the inspiral and merger-ringdown portions of the waveform, we first derive posterior distributions of parameters 
\begin{equation}
\theta^a_\GR=(\ln \mathcal{A}, \phi_c, t_c, m_1,m_2, \chi_1, \chi_2),
\end{equation} 
using the Fisher analysis method. 
Next, we marginalize over the first three parameters to find the posterior distributions for $(m_1,m_2,\chi_1,\chi_2)$.
Marginalization over a given parameter is typically accomplished by integration over the full range of values, or in the case of multi-variate Gaussian distributions by simply removing the corresponding row and column from the covariance matrix $\Sigma_{ij}\equiv\Gamma_{ij}^{-1}$.
Finally, using the Jacobian transformation matrix and the NR fits provided in Ref.~\cite{PhenomDII}, the two-dimensional Gaussian probability distributions $P_\II(M_f,\chi_f)$ and $P_\MR(M_f,\chi_f)$ are constructed.

What are the other limitations of the Fisher analysis? 
Below, we will only use the GR gravitational waveform, which corresponds to injecting the GR waveform and also recovering it with the GR waveform. 
Such a method does not allow us to estimate the systematic errors, and thus the final distribution is always centered around the true GR value, which is not the case in a real analysis~\cite{Abbott_IMRcon2,Abbott_IMRcon}. 
Moreover, the posterior distribution from the Fisher method is always Gaussian, and thus a 90\% credible contour in a two-dimensional parameter space is always given by an ellipse, which is also not true in reality.

However, what a Fisher analysis can accurately describe is the \textit{size} and  direction of correlation of the posterior distributions for $(\epsilon,\sigma)$, which are of high value when predicting the future resolving power from the GR value of $(0,0)$.
Throughout the first half of this chapter, we consider the \textit{area} of the 90\% confidence region as a metric of the discriminatory power one can gain upon use of this test with ground-based, space-based, and multi-band detections.
Such information may be used to gain valuable insight on how well future observations can discern GR effects from non-GR effects.

In the second half of this chapter, we present a method to estimate the systematic uncertainties present in the IMR consistency test given a arbitrary theory of gravity beyond-GR.
In particular, by following the analysis of Ref.~\cite{Cutler:2007mi}, one can estimate the ``theoretical'', or systematic errors present in the extraction of template parameters $\theta^a$ due to mismodeling present in the template waveform. 
Specifically, one can estimate systematic errors on $\theta^a$ by assuming use of the GR template, while EdGB gravity is in fact the correct theory in nature.
The theoretical errors can be computed as
\begin{equation}
\Delta_\text{th}\theta^a \approx \Sigma^{ab} \left( \lbrack \Delta A+iA_\GR\Delta\Psi \rbrack e^{i\Psi_\GR} \Big| \partial_b \tilde{h}_\GR \right),
\end{equation}
where $\Sigma^{ab} = (\Gamma^{-1})^{ab}$ is the covariance matrix, a summation over $b$ is implied, and $\Delta A \equiv A_\GR-A_\EdGB$ and $\Delta\Psi \equiv \Psi_\GR-\Psi_\EdGB$ are the differences between the amplitude and phase in GR and EdGB gravity.
We note that the above expression for the systematic errors is most accurate when the difference between the GR and non-GR signals are small.
Thus for large enough values of EdGB coupling parameter the above approximation will become less accurate.
However, in the main analysis presented here, we impose the small coupling approximation which ensures small magnitudes of $\sqrt{\alpha_\EdGB}$.
The probability distributions in $(\Delta M_f/\bar{M}_f,\Delta \chi_f/\bar{\chi}_f)$ have included both statistical errors ($\sqrt{\bm{\Sigma}_{\I,\MR}}$) which determine their size, and systematic errors ($\bm{\Delta_\text{th}X}_{\I,\MR}$) which determine their offset from the GR predictions.

For the injection waveform, we use the EdGB waveform by modifying the IMRPhenomD GR waveform in three ways. The first modification is in the inspiral portion, where we add the EdGB leading post-Newtonian correction as in Eq.~\eqref{eq:alpha_ppE}. The second modification is in the ringdown portion, where we modify the QNM ringdown and damping frequencies as in Eq.~\eqref{eq:fdamp}. The third modification is in the estimate of the final mass and spin, which is given in Eqs.~\eqref{eq:Mf} and~\eqref{eq:chif}.
We utilize fiducial values such that $\eta$ and $\chi_s$ correspond to the initial parameters of the GW event in question, $M_f$ and $\chi_f$ correspond to those predicted by Eqs.~\eqref{eq:Mf} and~\eqref{eq:chif}, and $\phi_c=t_c=0$.
Finally, we allow the fiducial value of $\zeta$ to vary slowly as we proceed with the IMR consistency test with different magnitudes of EdGB coupling.

The IMR consistency test can be performed with the following prescription. 
First, we generate the two-dimensional posterior probability distributions $P_{\I,\MR}(M_f,\chi_f)$ in the $M_f-\chi_f$ plane from each portion of the waveform described above.
Such posterior distributions are described as a two-dimensional Gaussian probability distribution function with root-mean-square errors estimated via a Fisher-based analysis, as described below.
Combined with the \textit{theoretical} (systematic) uncertainty ``shifts" $\bm{\Delta_\text{th}X}_{\I,\MR}\equiv(\Delta_\text{th} M_f,\Delta_\text{th}\chi_f)$ (in the first half of this chapter, we assume systematic errors to be vanishing, in order to focus on the \textit{size} of the probability distributions for future observations), the final probability distributions in the $M_f-\chi_f$ plane are taken to be Gaussian
\begin{align}
\nonumber P_{\I,\MR}\equiv \frac{1}{2\pi\sqrt{|\bm{\Sigma}_{\I,\MR}|}}\exp \Bigg\lbrack -\frac{1}{2} &\left(  \bm{X} - \bm{X}^\GR_{\I,\MR} -\bm{\Delta_\text{th}X}_{\I,\MR} \right)^\text{T} \\
& \times \bm{\Sigma}_{\I,\MR}^{-1}\left( \bm{X} - \bm{X}^\GR_{\I,\MR} -\bm{\Delta_\text{th}X}_{\I,\MR} \right) \Bigg\rbrack,\label{eq:pdf}
\end{align}
where $\bm{\Sigma}_{\I,\MR}$ represents the covariance matrix, $\bm{X}\equiv(M_f,\chi_f)$ contains the final state variables, and $\bm{X}^\GR_{\I,\MR}$ contains their GR predictions from the inspiral and merger-ringdown portions respectively.
The agreement between the two above distributions is typically measured by transforming to the new parameters $\epsilon$ and $\sigma$, describing the departures $\Delta M_f$ and $\Delta \chi_f$ from the GR predictions of final mass and spin from the inspiral and merger-ringdown, normalized by the averages between the two $\bar{M}_f$ and $\bar{\chi}_f$~\cite{Ghosh_2017}
\begin{align}\label{eq:epsilon}
\epsilon &\equiv \frac{\Delta M_f}{\bar{M}_f} \equiv 2 \frac{M_f^\II-M_f^\MR}{M_f^\II+M_f^\MR},\\
\sigma &\equiv \frac{\Delta \chi_f}{\bar{\chi}_f} \equiv 2 \frac{\chi_f^\II-\chi_f^\MR}{\chi_f^\II+\chi_f^\MR}.\label{eq:sigma}
\end{align}
The probability distributions $P_\II(M_f,\chi_f)$ and $P_\MR(M_f,\chi_f)$ can be transformed to $P(\epsilon,\sigma)$ by following the Appendix of Ref.~\cite{Ghosh_2017}, resulting in the final expression given by:
\begin{align}\label{eq:transform}
\nonumber P(\epsilon,\sigma)=\int\limits^1_0 \int\limits^{\infty}_0 P_\II \left( \left\lbrack 1+\frac{\epsilon}{2} \right\rbrack \bar{M}_f , \left\lbrack 1+\frac{\sigma}{2} \right\rbrack \bar{\chi}_f \right)&\times P_\MR \left( \left\lbrack 1-\frac{\epsilon}{2} \right\rbrack \bar{M}_f , \left\lbrack 1-\frac{\sigma}{2} \right\rbrack \bar{\chi}_f \right) \\
&\times\bar{M}_f \bar{\chi}_f d\bar{M}_f d\bar{\chi}_f.
\end{align}
Finally, the consistency of the posterior probability distribution with the GR value of $(\epsilon,\sigma)|_\GR\equiv(0,0)$ will determine the agreement of the signal with GR.
Any statistically significant deviations from the GR value may uncover evidence of modified theories of gravity present in any portion of the GW signal.

\begin{figure}[htb]
\begin{center}
\includegraphics[width=0.7\textwidth]{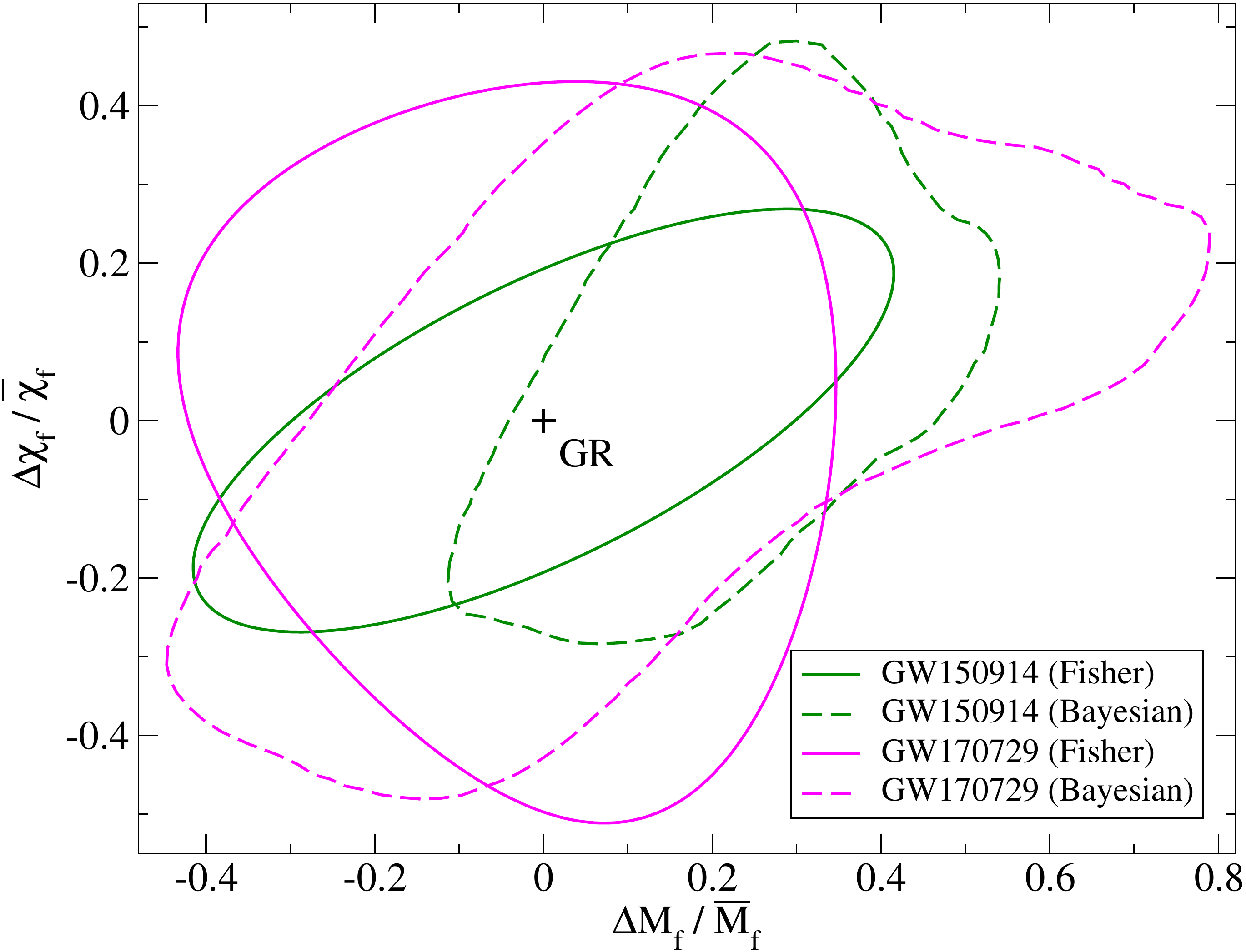}
\caption[Fisher vs. Bayesian comparison for the IMR consistency test]{
Comparison between the transformed posterior probability distributions in the IMR consistency test for both the Fisher analysis method (solid) used in this chapter, and the Bayesian one (dashed) done by the LVC in~\cite{Abbott_IMRcon}.
We display the results for both GW events GW150914 (green) and GW170729 (magenta) considered in this chapter.
We observe that in both cases, the total enclosed areas of the probability distributions agree between the Fisher and Bayesian analyses to within 10\% accuracy, confirming that the former can capture some qualitative features of the latter and thus is reliable as an order-of-magnitude estimate, at least for the magnitude of statistical uncertainties.
We do note that in the case of GW170729 the Fisher analysis distribution does not quite represent the \textit{correlations} observed in the more comprehensive Bayesian analysis: something we expect to improve with future high-SNR observations.
}\label{fig:FisherBayesian}
\end{center}
\end{figure}

See Fig.~\ref{fig:FisherBayesian} for a comparison between the Fisher analysis method considered in this chapter, and the Bayesian done by the LVC in~\cite{Abbott_IMRcon}.
We note that for both GW150914 and GW170729, the total enclosed areas of the Fisher and Bayesian probability distributions agree with each other to within 10\% accuracy.
This confirms the validity of the Fisher analysis method considered in this chapter as a qualitative estimate, something which we expect to improve considerably for future observations with increased SNRs as we also consider here.
We do note, however, that while the total areas agree well (indicative of the total statistical uncertainties), the direction of correlations for the case of GW170729 do not agree particularly well.
We expect this to improve as well for future high-SNR events.

In the second half of the chapter, we apply this method to test the specific example theory of EdGB gravity as follows. 
We choose the template waveform to be the IMRPhenomD waveform in GR, while we inject a signal in EdGB gravity by implementing the EdGB corrections to the inspiral, ringdown and final mass/spin of the IMRPhenomD waveform given by Eqs.~\eqref{eq:alpha_ppE},~\eqref{eq:beta_ppE},~\eqref{eq:fdamp},~\eqref{eq:Mf},~\eqref{eq:chif}.
We increase the fiducial value of $\zeta$ from 0 ($\bm{\Delta_\text{th}X}_{\I,\MR}=0$), until finally the GR prediction of $(\Delta M_f/\bar{M}_f,\Delta \chi_f/\bar{\chi}_f)|_\GR=(0,0)$ falls outside of the 90\% confidence region (i.e. the systematic uncertainties are larger than the statistical errors).
This indicates the magnitude of $\zeta$ required to observe non-GR effects in the waveform.
We note that in the following presented analysis, the explicit role of the theoretical error found in~\cite{Cutler:2007mi} is different than that used in Ref.~\cite{Cutler:2007mi} and a similar multi-band analysis paper~\cite{Glampedakis:2019dqh}.
In both of the above references, the authors describe such theoretical error as a source of theoretical uncertainty due to mismodeling of the waveform.
On the other hand, in this analysis the theoretical errors are used to simulate the shift that best-fit parameters $M_f$ and $\chi_f$ would experience given EdGB corrections were present in the true signal while the GR waveform has been used for the data analysis.
Such shifts in best-fit parameters are then directly compared to the parameter covariances found with the Fisher information matrix.
We believe this is the first analysis where the IMR consistency tests have been applied to a concrete non-GR theory, where both inspiral and ringdown corrections are consistently included.


\section{Results}\label{sec:results}
In this section we present the results from the IMR consistency tests of GR.
We begin with the consideration of a multiband IMR consistency test, where we analyze the area of the transformed posterior probability distributions as a metric of future detectability on various detection schema.
We conclude with a probe of EdGB gravity with the IMR consistency test by injecting EdGB effects into the waveform inspiral, merger-ringdown, and remnant properties to compute the resulting systematic errors, increasing the magnitude of non-GR effects until the test is failed.

\subsection{Multiband IMR consistency tests}

Figure~\ref{fig:IMRDconsistency} displays the 90\% confidence regions of the remnant mass and spin predictions from the inspiral, merger-ringdown, and full waveforms as detected on LIGO O1, in comparison with the Bayesian results of Ref.~\cite{Abbott_IMRcon}.
Here we see good agreement between the probability distributions, in both the direction of correlation, and the area of the 90\% confidence regions -- the latter agreeing to within 10\% for all contours considered.
We remind the reader that the agreement between the inspiral and merger-ringdown probability distributions indicates the degree of consistency with GR.

\begin{figure}
\centering
\includegraphics[width=0.7\columnwidth]{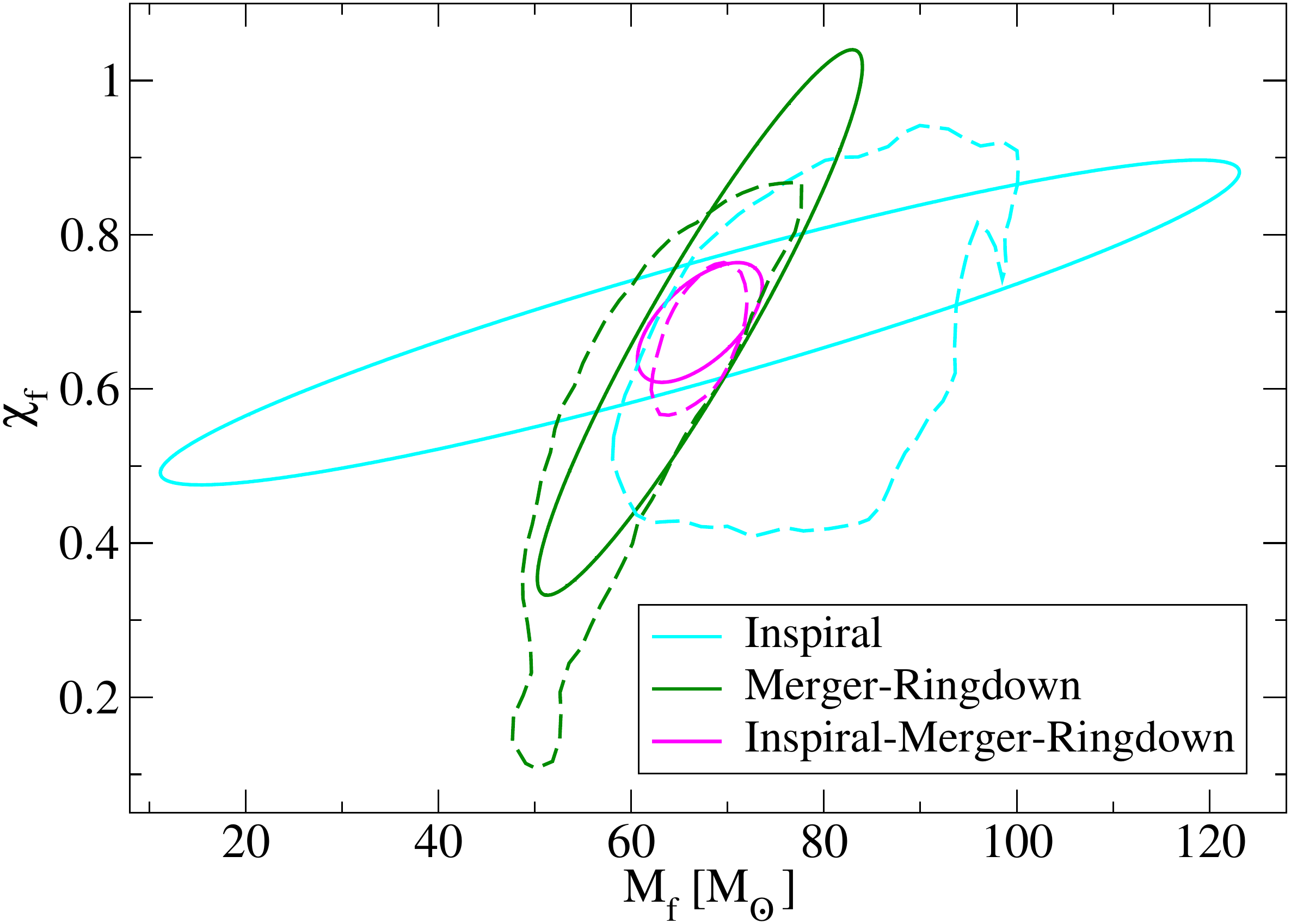}
\caption[IMR consistency test for GW150914-like events]{90\% credible region contours of the inspiral, merger-ringdown, and complete waveform posterior distributions in the $M_f-\chi_f$ plane, for GW150914-like events observed on the LIGO O1 detector.
We present both the Fisher analysis results (solid) discussed here and the Bayesian results of Ref.~\cite{Abbott_IMRcon} (dashed) for comparison.
We observe good agreement between the two analyses in both the direction of correlation, and in the overall areas, which agree to within 10\% for all three distributions.
}\label{fig:IMRDconsistency}
\end{figure}

Next, we follow Ref.~\cite{Ghosh_2017} to transform the individual inspiral and merger-ringdown probability distributions into the joint probability distribution between new parameters $(\epsilon,\sigma)$ via Eq.~\eqref{eq:transform}.
These quantities determine the remnant mass and spin (predictions assuming GR) discrepancies $\Delta M_f$ and $\Delta \chi_f$ between the inspiral and merger-ringdown waveforms, normalized by the averages between the two $\bar{M}_f$ and $\bar{\chi}_f$.
\hypertarget{link:IMRDconsistencyTransformed}{Figure~\ref{fig:IMRDconsistencyTransformed}} displays the estimated 90\% credible regions in the $\epsilon-\sigma$ plane for GW150914-like events observed on the following detectors: LIGO O1 (Fisher and Bayesian\footnote{Such Bayesian results are extracted from the \texttt{IMRPhenomPv2} results of Ref.~\cite{Abbott_IMRcon}. Similar results were found with the non-precessing \texttt{SEOBNRv4} model presented there.}~\cite{Abbott_IMRcon}), CE, and the multi-band observations between CE and TianQin, LISA, B-DECIGO, and DECIGO\footnote{As the merger-ringdown portion of the signal begins beyond the observing capacity of all space-based detectors for GW150914-like events, the IMR consistency test may not be performed solely by space-based detectors for such events. However, Ref.~\cite{Hughes:2004vw} showed that supermassive BH binaries are compatible with these observations.}.
The consistency of such distributions with the GR value of $(\sigma,\epsilon)|_\GR=(0,0)$ gives insight into how well the entire waveform agrees with GR, while any statistically significant deviations may indicate non-GR effects present within the signal.


Now we quantify the resolving power gained for each single-band and multi-band observation, describing how effectively one can discriminate between GR and non-GR effects.
To do this, we compute and compare the areas of the 90\% confidence regions as a metric towards this resolution.
Figure~\ref{fig:IMRDareas} presents the ratio of such areas for the LIGO O1 (Fisher and Bayesian~\cite{Abbott_IMRcon}) detector relative to CE, and to the multi-band observations with CE and TianQin, LISA, B-DECIGO, and DECIGO.
Here we observe three important features.
First, the results obtained here for LIGO O1 agree very well (within 10\%) with the Bayesian analysis of Ref.~\cite{Abbott_IMRcon}, showing good validity of our Fisher-estimated analysis.
Second, we observe almost a three-order-of-magnitude improvement upon the use of CE from the results of LIGO O1.
Third, we see additional gains in resolving power by a factor of 7-10 upon the use of multi-band observations.

\begin{figure}
\centering
\includegraphics[width=0.7\columnwidth]{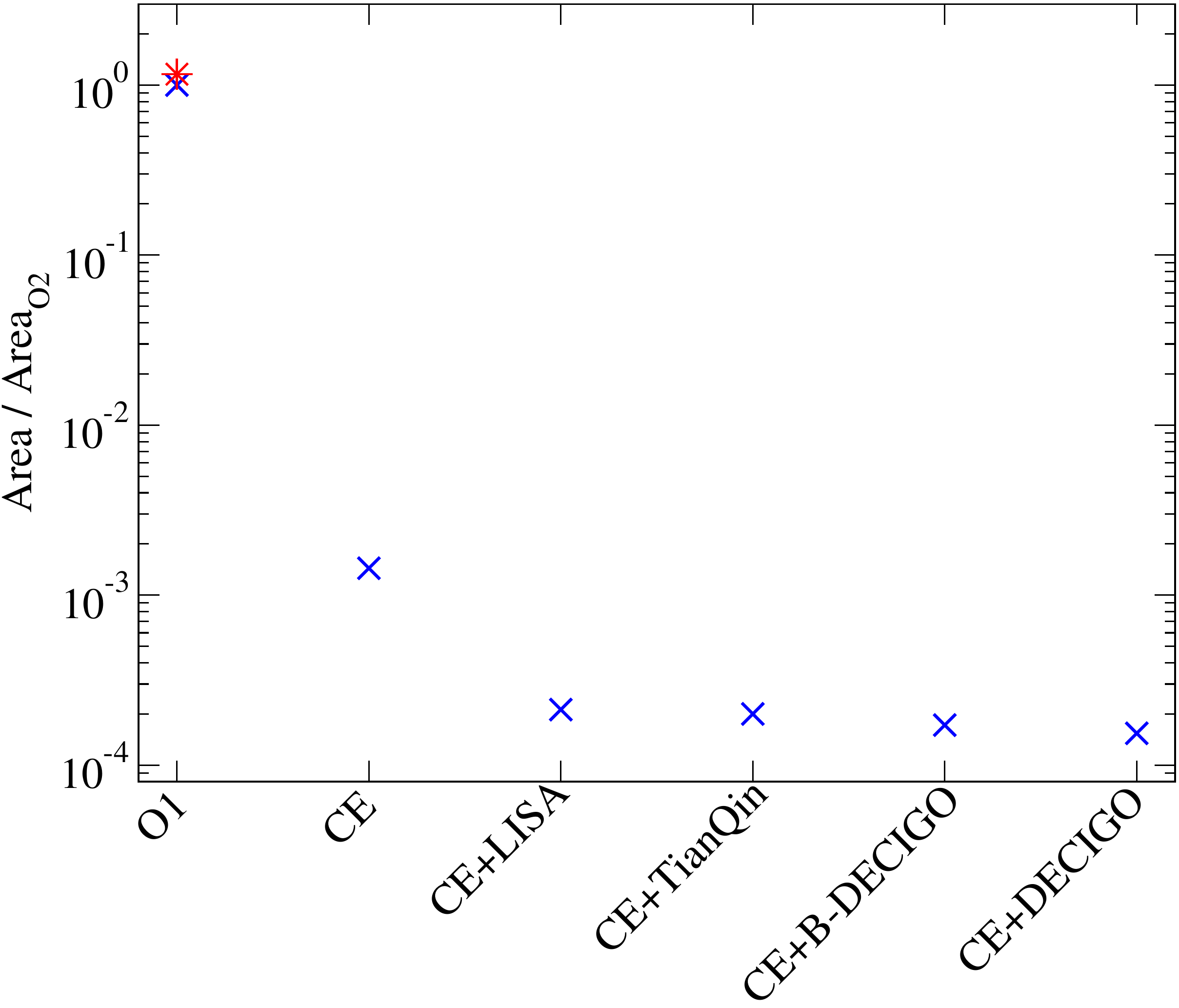}
\caption[IMR consistency test contour areas]{Ratios of the areas of the 90\% credible regions relative to that found with the LIGO O1 detector (Fisher result) shown in Fig.~\ref{fig:IMRDconsistency} for GW150914-like events, obtained from a Fisher analysis (blue cross).
We report good agreement within 10\% between the LIGO O1 Fisher and Bayesian~\cite{Abbott_IMRcon} (red star) results.
We also observe up to three orders-of-magnitude of improvement from the results of LIGO O1 to CE, and a further improvement of 7-10 times upon the use of multi-band observations.
}\label{fig:IMRDareas}
\end{figure}

\subsection{Probing EdGB gravity with the IMR consistency test}\label{sec:IMRconsistencyTest}
Now let us discuss the resulting detectability of EdGB effects using the IMR consistency tests of GR, using the process outlined in Sec.~\ref{sec:IMRmethod}.
This is done by injecting varying magnitudes of EdGB effects into the waveform until the IMR consistency test is failed.

\begin{figure*}[htb]
\begin{center}
\includegraphics[width=.46\textwidth]{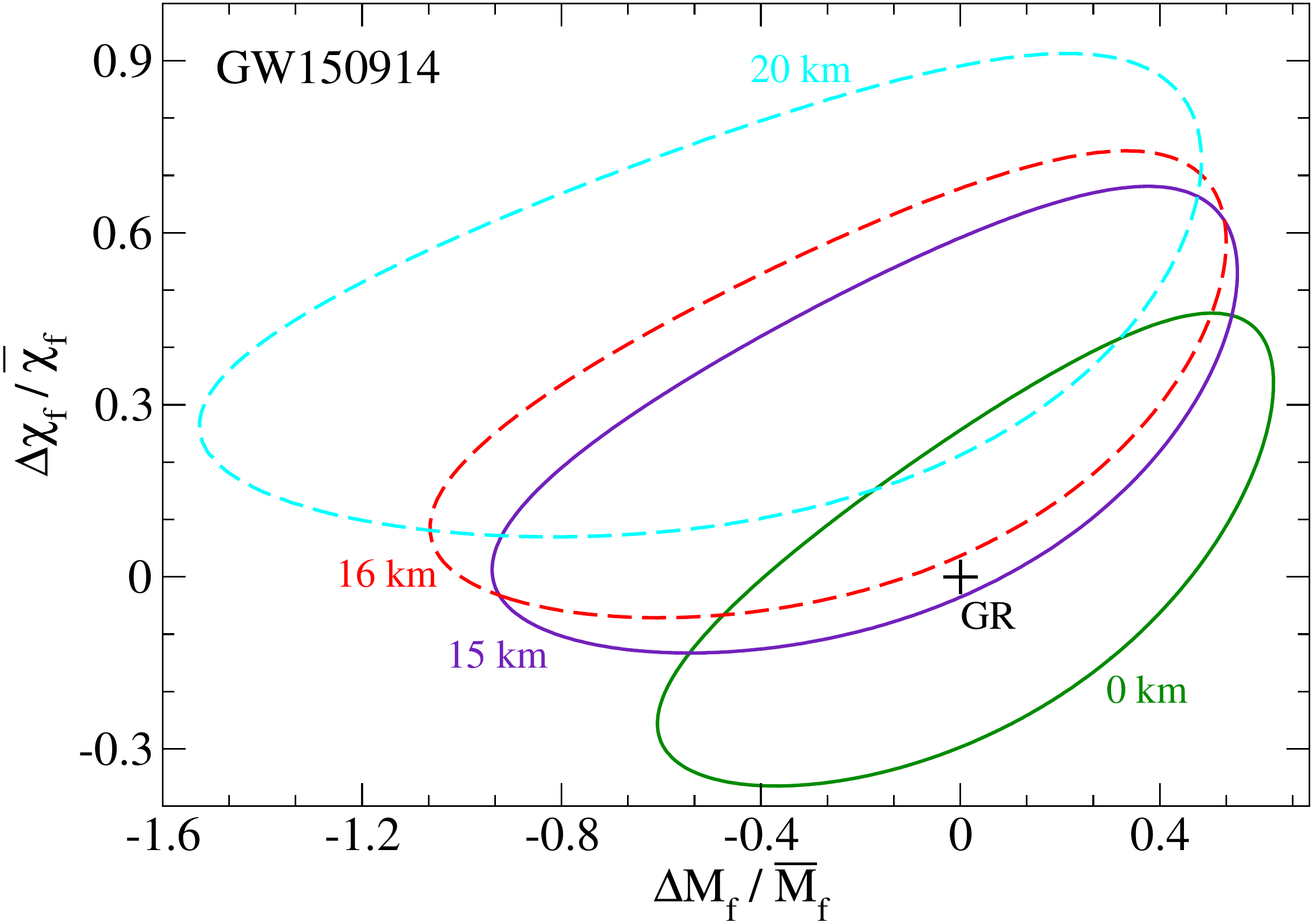}
\includegraphics[width=.46\textwidth]{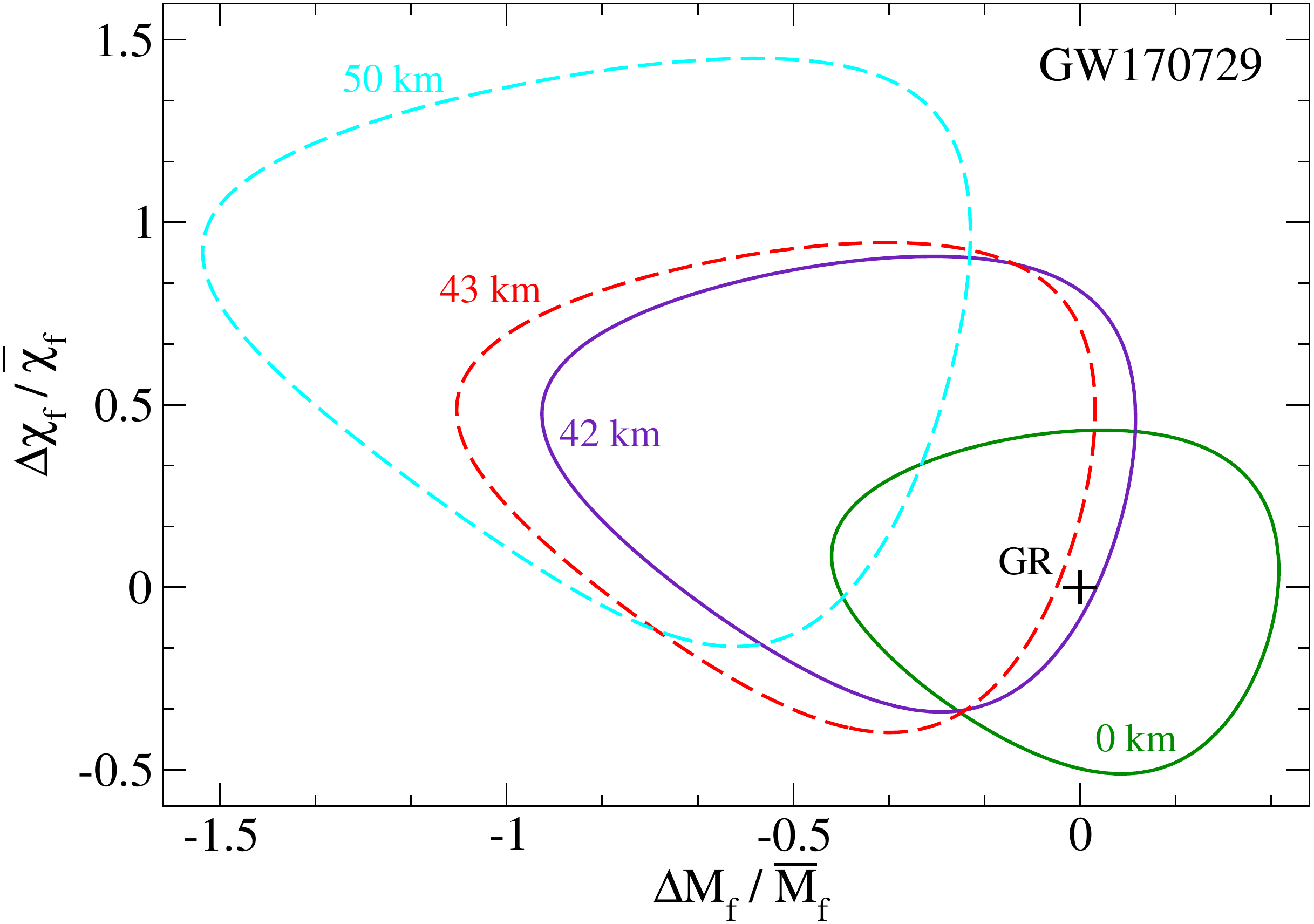}
\caption[IMR consistency test on the O2 detector]{
(left) IMR consistency test performed under EdGB gravity for GW150914 with a corresponding waveform generated via IMRPhenomD with the O2 detector.
Displayed is the 90\% confidence region of the transformed probability distribution in the $\Delta M_f/\bar{M}_f-\Delta \chi_f/\bar{\chi}_f$ plane, with the GR value of $(0,0)$.
The analysis is repeated for various fiducial values of $\sqrt{\alpha_\EdGB}$.
(right) Same as the left panel but for the more massive GW event GW170729 with $(M_f,\chi_f)_\GR=(80.3\text{ M}_\odot,0.81)$.
}\label{fig:IMRD_O2}
\end{center}
\end{figure*}

Let us first discuss the current prospects of observing EdGB effects upon the detection of binary BH merger events by the LIGO O2~\cite{aLIGO} detector.
The left panel of Fig.~\ref{fig:IMRD_O2} present the results of the test for GW150914 with $\sqrt{\alpha_\EdGB}=(0\text{ km},15\text{ km},16\text{ km}, 20\text{ km})$.
Such a waveform was generated with the PhenomD model assuming BH masses and spins of $(m_1,m_2,\chi_1,\chi_2)=(38.9 M_\odot,31.6M_\odot,0.32,-0.44)$, with a luminosity distance scaled to a signal-to-noise (SNR) ratio of $25.1$.
The above masses and spins were obtained from the median values of each distribution as reported in~\cite{GW150914}, and the alignment of the spins were chosen to be in agreement with the median value of effective spin $\chi_\text{eff}$ as reported by the same reference.
We observe that, at the 90\% confidence interval, EdGB effects can be observed for $\sqrt{\alpha_\EdGB}>15$ km, much larger than the current constraint of $2$ km~\cite{Kanti_EdGB,Pani_EdGB,Yagi_EdGB,Nair_dCSMap,Yamada:2019zrb,Tahura:2019dgr} with GW observations from GW150914, using ppE corrections to the waveform.
Therefore, we confirm that the current LVC infrastructure is unable to detect EdGB effects based on the existing observational constraints of $\sqrt{\alpha_\EdGB}<2$ km.
Similarly, we repeat the process for the more massive event GW170729 in the right panel of Fig.~\ref{fig:IMRD_O2}, observing how contributions from the merger-ringdown signal are much more significant in this scenario, with large uncertainties now present in the inspiral signal instead, resulting in EdGB detectability of $\sqrt{\alpha_\EdGB}>42$ km.
This waveform was generated with BH masses and spins of $(m_1,m_2,\chi_1,\chi_2)=(50.6M_\odot,34.3M_\odot,0.60,-0.57)$ with an SNR of $10.8$.
Similar to before, the masses and spins were chosen from the median values reported in~\cite{GW170729}, with the spin alignments chosen to be in agreement with $\chi_\text{eff}$.

\begin{figure}[htb]
\begin{center}
\includegraphics[width=.45\textwidth]{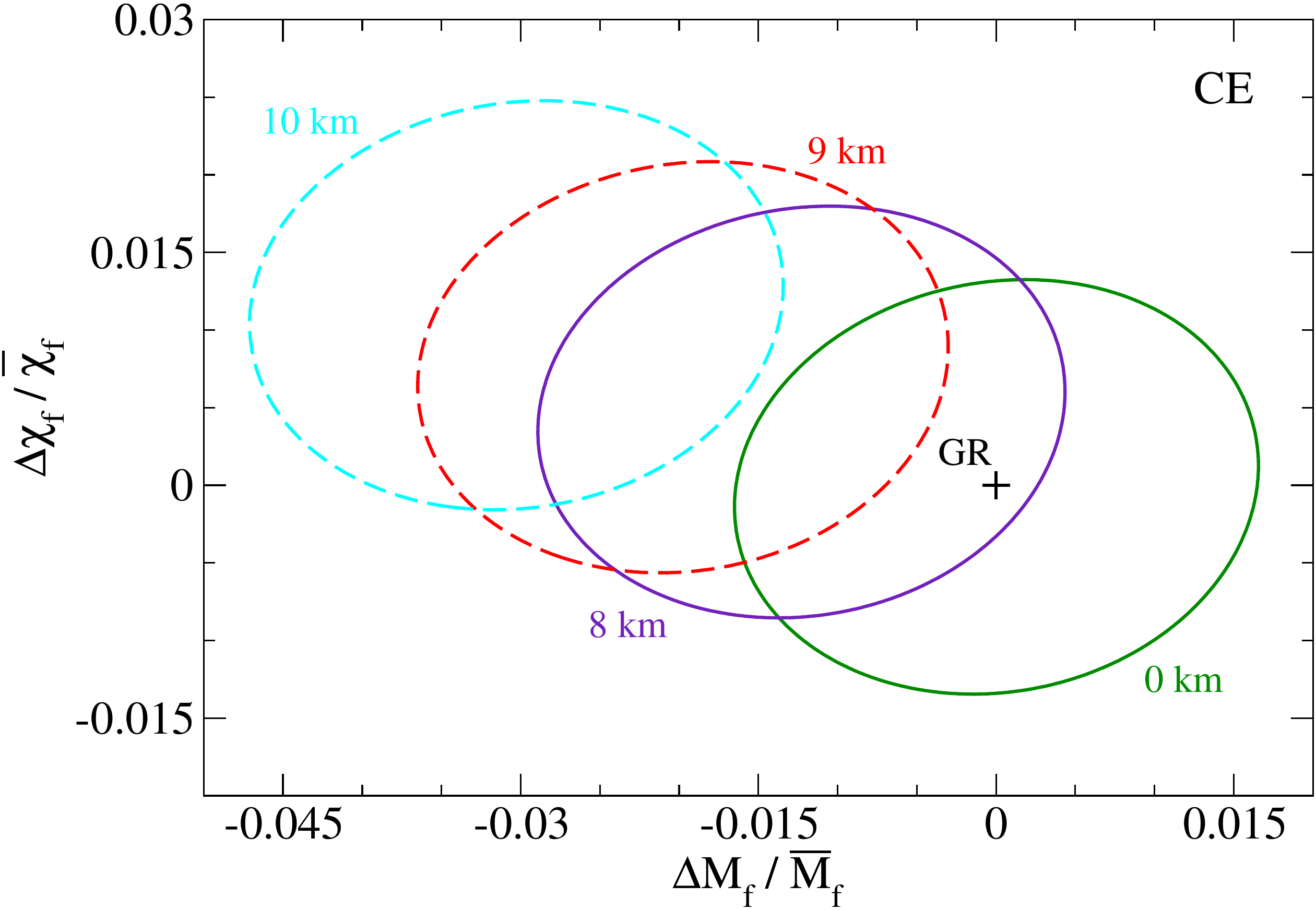}
\includegraphics[width=.45\textwidth]{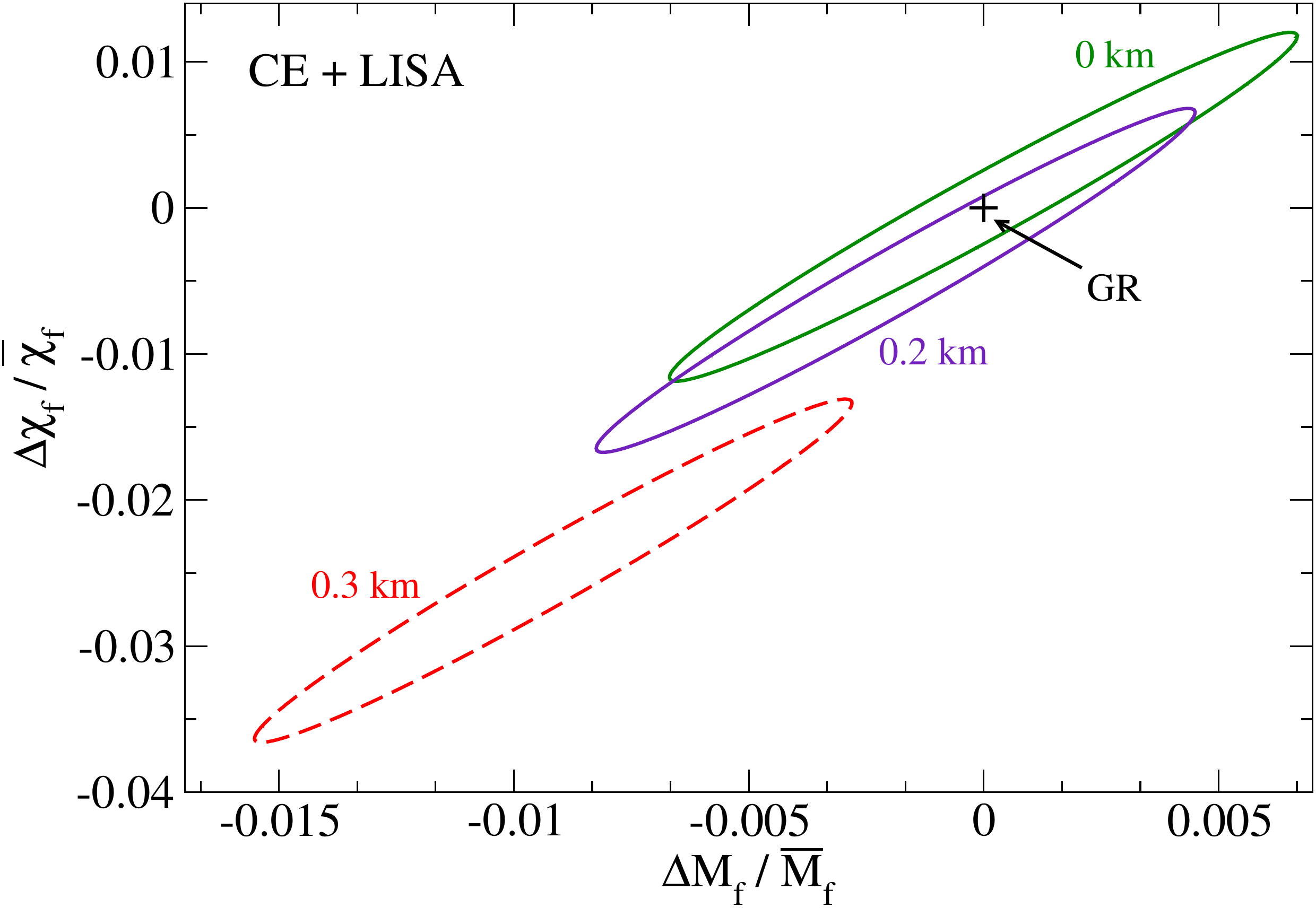}
\caption[The future of IMR consistency tests]{
Similar to Fig.~\ref{fig:IMRD_O2} but with the CE detector (left), and the multi-band observation between LISA and CE (right).
}\label{fig:IMRD_future}
\end{center}
\end{figure}

Next we consider the future prospects of observing such effects in the waveform with third-generation ground-based detectors.
The left panel of Fig.~\ref{fig:IMRD_future} shows the resulting probability distributions in $(\Delta M_f/\bar{M}_f,\Delta \chi_f/\bar{\chi}_f)$ found with the Cosmic Explorer~\cite{Ap_Voyager_CE} (CE) observations of GW150914-like events, with $\sqrt{\alpha_\EdGB}=(0\text{ km},8\text{ km},9\text{ km},10\text{ km})$.
We see that with CE, EdGB effects can be determined to a 90\% confidence interval for $\sqrt{\alpha_\EdGB}>8$ km, still above the current constraint of $\sqrt{\alpha_\EdGB}<2$ km.

By noting that a majority of both the statistical (size of the contours) and systematic uncertainties (shift of the contour centers) come from the inspiral signal, we consider a multiband observation by combining CE with the space-based detector LISA~\cite{LISA} to further probe the inspiral event.
The right panel of Fig.~\ref{fig:IMRD_future} shows the resulting probability distributions in $(\Delta M_f/\bar{M}_f,\Delta \chi_f/\bar{\chi}_f)$ observed by multiband observations with $\sqrt{\alpha_\EdGB}=(0\text{ km},0.2\text{ km},0.3\text{ km})$.
Here we see that multiband detections can probe EdGB effects with magnitudes of $\sqrt{\alpha_\EdGB}>0.2$ km, an order-of-magnitude smaller than the current constraint of $2$ km.
Thus, if non-GR effects such as EdGB are indeed present in nature with $0.2\text{ km}<\sqrt{\alpha_\EdGB}<2$ km, multiband detections between CE and LISA can uncover them to the 90\% confidence interval. On the other hand, if one does not find deviations from GR, one would be able to place bounds on EdGB gravity that are stronger than current bounds by an order of magnitude.
The projected bounds with future detectors presented here using IMR consistency tests are comparable to those found with the parameterized tests discussed in Chapter~\ref{chap:Parameterized}, with the differences between the two tabulated in Tab.~\ref{tab:alphas}.

In the above analysis we have shown the considerable increase one might gain by introducing milli-Hz era GW detectors such as LISA (or e.g. TianQin~\cite{TianQin}) to the ground-based observations with third-generation detector CE.
We note that such constraints can also be expected to improve considerably upon the additional observation from deci-Hz detectors such as (B-)DECIGO~\cite{B-DECIGO,DECIGO}.
As found in Ref.~\cite{Carson_multiBandPRD} by the same authors, multi-band observations with CE+DECIGO considerably outperformed those with any other space-based detector such as LISA or TianQin.
In particular, Tab.~I of~\cite{Carson_multiBandPRD} shows such bounds to improve anywhere from a factor of two to two orders-of-magnitude when introducing multi-band observations with DECIGO.
For the specific theory of EdGB, constraints on $\sqrt{\alpha_\EdGB}$ were shown to improve by nearly a factor of three.


\section{Conclusions}\label{sec:conclusion}
In this chapter we investigated the consistency between the inspiral and merger-ringdown portions of the gravitational waveform, in the so called IMR consistency test.
We applied this test to future multiband observations of GWs to analyze the size of the remnant BH mass and spin posterior probability distributions, indicitave of the future resolving power for non-GR effects.
We then apply such tests to the specific example of EdGB gravity, where full waveform corrections were injected allowing us to find the magnitude of coupling paramter $\alpha_\EdGB$ required to be detectable in future GW observations.

We began by demonstrating the resolving power gained upon the use of multi-band observations, finding up to an order-of-magnitude improvement relative to the single-band detections made by ground-based detectors alone.
Such an improvement gives way to the enhanced opportunity to shed light on even the most minuscule deviations from GR in the extreme gravity regime.

We then consider the example of EdGB theory of gravity, which is a proposed scalar-tensor theory of gravity with curvature coupling to the dilaton scalar field.
This string-inspired theory predicts the scalarization of BHs~\cite{Campbell:1991kz,Yunes:2011we,Takahiro,Sotiriou:2014pfa}, calling forth ``fifth force" interactions between orbiting BHs in a binary system and giving rise to scalar dipole radiation that predicts an increased rate of inspiral between them.
In this chapter, we have modeled the resulting EdGB effects throughout various parts of the gravitational waveform, including the inspiral, the characteristic ringdown QNMs, and finally, to the final mass and spin properties of the remnant BH.
With these new tools in hand, we offer predictions on the future detectability of such EdGB effects present in the gravitational waveform. 

In particular, we chose EdGB gravity as an example non-GR theory to study the power of IMR consistency tests, though the formalism that we developed here can easily be applied to other theories if all of the ingredients are available. 
For example, dynamical Chern-Simons gravity is another theory beyond GR that breaks parity and is motivated from string theory, loop quantum gravity and is an effective field theory for inflation~\cite{Jackiw:2003pm,Alexander_cs,Nishizawa:2018srh}. 
Leading post-Newtonian corrections to the inspiral waveform have been derived in~\cite{Yagi:2012vf,Nair_dCSMap,Tahura:2019dgr}, while the scalar interaction energy and corrections to the specific orbital energy, angular momentum and the location of the ISCO have been computed in~\cite{Stein:2013wza,Yagi_dCS,Yagi:2012vf}. The QNM ringdown spectrum for non-spinning BHs in such a theory has been studied in~\cite{Yunes:2007ss,Cardoso:2009pk,Molina:2010fb}. Therefore, once the spin corrections to the BH ringdown in this theory is available, one can repeat the analysis here to investigate how accurately one can probe dynamical Chern-Simons gravity with the IMR consistency test.

We studied the detectability of EdGB effects emergent in an observed GW signal by introducing EdGB modifications to the inspiral, merger-ringdown, and remnant BH properties of the waveform.
In particular, for a given GW detector we increased the magnitude of $\alpha_\EdGB$ present in the systematic uncertainties until they overcame the statistical errors.
Once this happens, the inspiral and merger-ringdown signals are no longer consistent and we can claim the presence of  non-GR effects in the waveform.
In particular, we find this magnitude to be one-order-of-magnitude stronger than the current constraint of $\sqrt{\alpha_\EdGB}<2$ km when utilizing the multiband observation between ground- and space-based detectors CE and LISA.

If a non-GR effect is observed, how can one potentially infer whether it originated from one non-GR theory or another? Given that the inspiral-merger-ringdown tests discussed here were originally designed to test the consistency of GR, a different test would be more appropriate to address the above question. For example, one could directly try to measure the leading corrections to the inspiral and ringdown frequency independently and check for the consistency between the two quantities within a given non-GR theory. In the case of EdGB gravity, one can eliminate $\zeta$ from the two to find such a relation, which is unique to the theory.

While this work demonstrated the large gains one can make on testing GR upon the use of multi-band observations and the IMR consistency test, the analysis can be improved in numerous ways.
One example of such improvements would be to consider a full Bayesian analyses rather than the Fisher analysis used here -- although it was found that our results agree well with their Bayesian counterparts.
Additionally, one can simulate the multi-band event rates discussed in Refs.~\cite{Sesana:2016ljz,Cutler:2019krq,Gerosa:2019dbe} to combine the signals and further reduce the systematic errors residing in our bounds on non-GR theoretical parameters.
Finally, we can use the results in Sec.~\ref{sec:amplitude} to consider alternative theories of gravity which modify the GW amplitude rather than the phase~\cite{Alexander:2007kv,Yunes:2008bu,Yunes:2010yf,Yagi:2017zhb}.


\newpage
\chapter{Testing beyond-Kerr spacetimes}\label{chap:beyondKerr}
\begin{adjustwidth}{\parindent}{\parindent}
\begin{spacing}{1.5}
Recently the Event Horizon Telescope Collaboration, with very-long baseline interferometric observations, resolved structure about the center of the supermassive black hole M87$^*$, paving the way for a test of the spacetime surrounding black holes.
Generic, parameterized spacetimes beyond Kerr allow one to arbitrarily test the no-hair theorem for deviations from the Kerr result with no prior theoretical knowledge or motivation.
In this chapter, we present such a new general, stationary, axisymmetric and asymptotically flat black hole solution with separable geodesic equations, in which five free non-linear functions parameterically deviate from the Kerr result, allowing one to transform to many alternative black hole solutions present in the literature.
We then derive analytic expressions for the Keplerian and epicyclic frequencies, the orbital energy and angular momentum, the location of the innermost stable orbit of circular equatorial particle orbits, and the image of the photon rings in the new spacetime, which correspond to the boundary of the black hole shadow image taken by the Event Horizon Telescope.
We then focus on two singly-parameterized spacetime metrics and derive parameterized corrections to the waveform inspiral, ringdown, and remnant properties.
We predict the beyond-Kerr parameter magnitudes required in an observed gravitational wave signal to be statistically inconsistent with general relativity.
\end{spacing}
\end{adjustwidth}
\newpage


\section{Introduction}\label{sec:intro}
The no-hair theorem tells us that the spacetime surrounding isolated BHs is uniquely described by the Kerr metric\footnote{This chapter is based on the following papers: Z. Carson and K. Yagi, \textit{A parameterized black hole metric preserving Kerr symmetries}, Phys. Rev. D \textbf{101}, 084030 (2020) and Z. Carson and K. Yagi, \textit{Probing beyond-Kerr spacetimes with inspiral-ringdown corrections to gravitational waves}, Phys. Rev. D \textbf{101}, 084050 (2020)}.
This famous metric is asymptotically flat, stationary, axisymmetric, and is parameterized by only two BH parameters: the mass ($M$), and the spin ($a$).
Further, within this metric exists an event horizon masking the true coordinate singularity found within.
Supposing the no-hair theorem holds true, we can expect all BH observations, i.e. of photon orbits about them, to agree with those as described by the Kerr metric.
To date, there has been no sufficient evidence that points to otherwise~\cite{Menou:1997ys,Narayan:1997xv}, however only now have our capabilities advanced enough to accurately probe the near-field structure of these mysterious objects.
In the near future, we may have the opportunity to resolve effects that go beyond the standard Kerr model.

To date, several tests of the no-hair theorem have been enacted and proposed, as reviewed in~\cite{Yunes:2013dva,Gair:2012nm,Johannsen:2011mt,Bambi:2015ldr}.
Such tests include observations of pulsar-BH binaries~\cite{Liu_2012,Pfahl_2004,Wex_1999}, the orbits of supermassive BH (SMBH) stellar companions~\cite{Sadeghian:2011ub,Merritt:2009ex,Will_2008}, the EM accretion flows of SMBHs (continuum spectrum, iron lines, quasi-periodic oscillations, etc.)~\cite{Johannsen:2011dh,Bambi:2013sha,Bambi:2012at,Johannsen:2012ng,Bambi:2011jq,Bambi:2012ku,Johannsen:2010ru,Johannsen:2010xs,Psaltis_2011,Johannsen_2016,Krawczynski_2012,PhysRevD.83.103003,Johannsen_2010,Maselli:2014fca,Maselli:2017kic}, the quasinormal ringdown modes of a post-coalescence perturbed remnant BH~\cite{Berti:2007zu,Dreyer:2003bv,PhysRevD.73.064030}, and even the GW observations of extreme mass-ratio inspirals of super-massive and stellar-mass BHs~\cite{Vigeland:2011ji,Gair:2011ym,Apostolatos:2009vu,Gair:2007kr,Collins:2004ex,Glampedakis:2005cf,PhysRevD.84.064016,PhysRevD.81.024030,PhysRevD.78.102002,PhysRevD.69.082005,Barack:2006pq,Johannsen:2011mt,Mandel_2014,PhysRevD.77.064022,PhysRevD.56.1845,PhysRevD.52.5707,Isi:2019aib}.
No significant deviations from the Kerr result have been detected so far. 

Recent developments in the very-long baseline interferometric (VLBI) array in the Event Horizon Telescope (EHT) have given us the unique opportunity to probe the spacetimes of SMBHs.
With the sole purpose of imaging SMBHs, the EHT and VLBI span the entire globe with an an array of millimeter and sub-millimeter instruments, effectively creating an Earth-sized telescope~\cite{doeleman2009imaging}.
Currently operating with 8 telescopes, the EHT achieved the impressive feat of resolving the lensed photon orbits about Messier 87's central SMBH M87$^*$, with an angular resolution of $\sim50\mu$as~\cite{1435171,1435174,1435175,1435177,1435168}.
Even with these extraordinary results, strong deviations from the no-hair theorem have not been detected.
However, the EHT will continue to develop with the addition of new facilities, with a planned resolution increase of $\sim40\%$ over the next 3-5 years.
Along with the addition of new SMBH targets, this boost in resolution and image fidelity will further provide us with the ability to probe these extreme spacetimes.

While the Kerr BH metric, developed under the solutions to Einstein's theory of general relativity (GR), has had unprecedented success in describing our BH observations, we must continue to test the no-hair theorem.
In particular, to this day we have left unanswered several important observational questions regarding the nature of the universe, which could potentially be described by a new theory of gravity.
For example, the elusive ``dark energy'' and ``dark matter'' accelerating the expansion of our universe, and the rotation of our galaxies~\cite{Jain:2010ka,Salvatelli:2016mgy,Koyama:2015vza,Joyce:2014kja,Famaey:2011kh,Milgrom:DarkMatter,Milgrom:2008rv,Clifton:2011jh,Joyce:2014kja}, or the early universe's rapid inflationary period~\cite{Joyce:2014kja,Clifton:2011jh,Famaey:2011kh,Koyama:2015vza} and our current universe's extreme matter/anti-matter asymmetry~\cite{Clifton:2011jh,Famaey:2011kh}, or even the issue of unifying GR and quantum mechanics~\cite{Clifton:2011jh,Joyce:2014kja,Famaey:2011kh,Milgrom:2008rv,Jain:2010ka,Koyama:2015vza}, all remain unanswered.
Such questions could potentially be explained by a new theory of gravity, which would certainly exhibit itself within the extreme-gravity BH spacetimes currently described by the Kerr metric.

To test the current theory of gravity, one needs to first develop a new spacetime metric as a solution to modified field equations.
In GR, this spacetime metric is described by the Kerr result $g_{\alpha\beta}^\K$ for rotating BHs.
In order to perform the tests in an efficient, theory-agnostic way in beyond GR, one would presumably introduce parameterized deviations from the Kerr metric which, when vanishing, reproduces the Kerr result again.
In this scenario, one or more non-Kerr deviation parameters could be observationally constrained in a model-independent way that requires no prior theoretical knowledge.
One could then map such bounds to those on theoretical parameters in specific non-Kerr theories.
To date, several such metrics have been developed~\cite{Collins:2004ex,Glampedakis:2005cf,Vigeland:2011ji,PhysRevD.81.024030,Johannsen:2011dh,Manko,Konoplya:2020hyk}, each of which are stationary, axisymmetric, asymptotically flat, and contain one or more parameters deviating from the Kerr metric. 
For example, Johannsen developed a more general parameterized metric with separable geodesic equations in~\cite{Johannsen:2015pca}, followed up by an even more general metric preserving the same symmetries found in~\cite{Papadopoulos:2018nvd}.
Several of the above parameterized metrics can then be mapped to many known BH solutions found in the literature~\cite{Papadopoulos:2018nvd,Randall:1999ee,Aliev:2005bi,Jai-akson:2017ldo,Ding:2019mal,Kerr-Sen,Kanti_EdGB,Maeda:2009uy,Sotiriou:2014pfa,Ayzenberg:2014aka,Jackiw:2003pm,Yagi_dCS,Yunes_dcs,Bardeen,Kumar:2019uwi,Pani:2011gy,Kumar:2020hgm}, with popular transformations for the latter two metrics tabulated in~\cite{Johannsen:2015mdd,Carson_BumpyPhotonRings}.

In this chapter, we focus on testing several beyond-Kerr spacetimes with both GW and EM observations.
For the former, we consider singly-parameterized spacetime metrics introduced elsewhere.
We then follow the IMR consistency test process outlined in Chapter~\ref{chap:IMRconsistency} with the Fisher analysis techniques described in Chapter~\ref{chap:ParameterEstimation} to identify the magnitude of non-Kerr parameters required to be considered as a sufficient non-Kerr observation.
For the latter test, we build a new general spacetime following the works of~\cite{Johannsen:2015pca}.
In this deformed spacetime which preserves Kerr symmetries, we introduce 6 free non-linear functions that deviate away from the Kerr metric.
We then consider the effect of such parameters on several astrophysical phenomena.

When testing such parameterized spacetimes with GWs, we focus our attentions on two parameterized metrics. The first one was derived by Johannsen and Psaltis (JP)~\cite{Johannsen:2011dh}, which has a single deviation parameter $\epsilon_3$. This spacetime is an example of the more general metric found in~\cite{Johannsen:2015pca}. 
The second one is motivated by Johannsen in~\cite{Johannsen:2015pca}, where a new deviation parameter $\beta$\footnote{$\beta$ is introduced in $\Delta = r^2 - 2 M r + a^2$ as $\Delta \to \bar \Delta = \Delta + \beta$, where $M$ and $a$ characterize the mass and spin of a BH while $r$ is the radial coordinate.} was introduced into the more general metric in~\cite{Johannsen:2015pca}.
Here, we remove all non-Kerr deviation parameters with the exception of $\beta$ to form the singly-parameterized ``modified-$\Delta$'' (mod.~$\Delta$, or ``MD'' in superscripts/subscripts) metric.
Because these single-parameter spacetimes have been obtained from the more general beyond-Kerr metrics which can be mapped to several known BH solutions, they make ideal candidates for testing GR in a simple way.

With a model-independent beyond-Kerr metric in hand, one next needs to find the modifications to the gravitational waveform imparted under the new spacetime.
As accomplished previously in Chapters~\ref{chap:Parameterized} and~\ref{chap:IMRconsistency} for the Einstein-dilaton Gauss-Bonnet theory of gravity~\cite{Kanti_EdGB,Maeda:2009uy,Sotiriou:2013qea,Yagi:2015oca}, one can obtain analytic expressions for various corrections to the gravitational waveform in an alternative theory of gravity.
In this chapter, given a arbitrary spacetime metric $g_{\alpha\beta}^\X$, we show how one can obtain correctional expressions to the GW inspiral, ringdown quasinormal modes (QNMs), and the remnant BH's mass and spin.
When inserted into the standard GR gravitational waveform, these singly-parameterized corrections can be used to test future incoming signals for deviations from GR.
See also Ref.~\cite{Suvorov:2019qow} where it was detailed how one can test beyond-GR theories of gravity, even for Kerr BHs~\cite{Psaltis:2007cw} if one considers their perturbations..

Finally, with a generalized beyond-Kerr metric and its resulting corrections to the gravitational waveform template, one needs to test the observed GW signals for deviations present within.
Specifically, we focus our attention on the so-called inspiral-merger-ringdown (IMR) consistency tests of GR~\cite{Ghosh_IMRcon,Ghosh_IMRcon2,Abbott_IMRcon,Abbott_IMRcon2} described thoroughly in Chapter~\ref{chap:IMRconsistency}. 
In this application, one tests the consistency between the inspiral and merger-ringdown GW signals to predict the possibility of emergent non-Kerr effects present in the observed signal.
In particular, we estimate (with the Fisher analaysis techniques~\cite{Poisson:Fisher,Berti:Fisher,Yagi:2009zm} described in Chapter~\ref{chap:ParameterEstimation}) the final BH's mass and spin individually from the inspiral signal, and then from the merger-ringdown signal.
If both predictions show significant disagreement from each other, one can conclude with evidence of non-Kerr effects present in the observed signal (provided systematic errors are under control).

In addition to the IMR consistency tests, we also test the gravitational waveform in a parameterized way.
Indicated as the parameterized tests of GR throughout the following chapter, we begin by introducing corrections to the waveform inspiral, merger-ringdown, and to the remnant BH properties.
All such corrections are parameterized by the single JP or mod.~$\Delta$ parameters $\epsilon_3$ and $\beta$ which allows for a convenient test.
We assume the waveform is described by GR ($\epsilon_3=0$ or  $\beta=0$ ) and estimate the resulting root-mean-square uncertainties on the non-GR parameters.
Such variations then describe the ``wiggle room'' such non-GR parameters have to still remain consistent within the GW detector's noise, and can be taken as an upper-bound constraint.

We refer readers to related works on testing beyond-Kerr spacetimes with GWs.
Reference~\cite{Gair:2011ym,Moore:2017lxy} constructs an approximate, multipolar gravitational waveform suitable for extreme-mass-ratio-inspirals (EMRIs) detectable by space-based detector LISA for inspiral using the analytic kludge method from a beyond-Kerr ``bumpy'' spacetime, which can be use to test GR with GW signals by placing constraints on the deviations, as was considered for future LISA observations.
Reference~\cite{Chua:2018yng} considered similar EMRI analytic kludge waveforms and performed a Bayesian model selection analysis for distinguishing Kerr and beyond-Kerr models.
Additionally, Refs.~\cite{Barack:2006pq,Glampedakis:2005cf} considered quadrupole corrections to the GR Kerr analytic kludge waveforms for EMRIs in a bumpy spacetime to consider the accuracy with which LISA could constrain such deformations.
Reference~\cite{Xin:2018urr} considered the JP metric considered in this chapter to build a parameterized EMRI waveform and test it with future space-based observations.
Even more recently Ref.~\cite{Cardenas-Avendano:2019zxd} considered a singly-parameterized beyond-Schwarzschild (non-spinning BHs) metric and derive corrections to the inspiral waveform to place constraints on previous LVC detections. 
See also Ref.~\cite{Maselli:2019mjd} where similar corrections to the QNMs were made, and constraints with future observations of multiple GW events were quantified.

The analysis presented in this chapter differs from the above ones in at least a few ways. 
For example, we not only consider different beyond-Kerr spacetimes than the ones considered above (except for~\cite{Xin:2018urr}), but we additionally find corrections to the ringdown waveform and also to the remnant BH properties, all up to quadratic order in BH rotation.
Additionally, while all of the above analyses focus on waveforms suitable for EMRIs detectable by LISA, in our analysis, we find corrections to the commonly-used IMRPhenomD gravitational waveform which is more suited to comparable-mass systems (this waveform has been calibrated for mass-ratios up to $1:18$, significantly smaller than that for EMRIs of $\sim 1:10^5$).

With this prescription, we present for the first time a recipe for one to quickly estimate corrections to the inspiral, ringdown, and remnant BH properties given only an arbitrary spacetime metric $g_{\alpha\beta}^\X$.
We exemplify this for both the JP and mod.~$\Delta$ spacetime metrics $g_{\alpha\beta}^\JP$ and $g_{\alpha\beta}^\MD$, which are parameterized beyond-Kerr by the single parameters $\epsilon_3$ and $\beta$, deviating from the Kerr metric $g_{\alpha\beta}^\K$.
We follow this up with a demonstration of the power of these corrections by performing the IMR consistency test to predict the magnitudes of $\epsilon_3$ and $\beta$ required for one to observe statistically significant deviations from the Kerr result.
With current generation GW detectors, we find comparable constraints on the JP deviation parameter $\epsilon_3$ to those from x-ray observations of BH accretion disks~\cite{Kong:2014wha,Bambi:2015ldr}, found to be loosely $\epsilon_3\lessapprox5$.
With future space-based and ground-based GW observatories we find constraints a few orders of magnitude stronger.
We find that such results from the IMR consistency tests are mostly comparable to those from the parameterized tests.
In particular, we find that the extreme-mass-ratio-inspirals observable by future space-based detector LISA~\cite{LISA} can probe such effects by three orders-of-magnitude stronger than the current constraints found in the literature.

We then consider future tests of beyond-Kerr spacetimes with EM observations rather than the GW ones discussed above.
In order to sufficiently test these spacetimes for the possibility of non-Kerr effects, we must first model a ``beyond-Kerr'' spacetime in a generic way~\cite{Collins:2004ex,Glampedakis:2005cf,Vigeland:2011ji,PhysRevD.81.024030,Johannsen:2011dh,Manko,Cardoso:2014rha,Rezzolla:2014mua,Konoplya:2016jvv,Konoplya:2020hyk}.
Ideally, each metric element should parametrically deviate from the Kerr metric separately, in such a way that the Kerr spacetime is obtained when all deviations vanish.
In this chapter, we restrict ourselves to BHs preserving the symmetries of Kerr BHs, namely asymptotically flat, stationary, axisymmetric, and with separable geodesic equations.
The latter condition avoids chaotic particle orbits, and equates to the existence of a fourth constant of motion, in addition to the proper mass, the energy and angular momentum,: the so-called ``Carter constant''~\cite{PhysRevD.78.102002,Carter}.
The metric considered here is a more broad example of the general class of metrics presented in~\cite{Konoplya:2018arm} which admit separable Klein Gordon equations, and is reduced to the latter for certain assumptions on the beyond-Kerr functions presented here.
Several such metrics have been derived in the literature~\cite{Vigeland:2011ji,Johannsen:2015pca,Konoplya:2018arm}, each with one or more parametric deviations which reduce to Kerr when vanishing.
See also Ref.~\cite{Chen:2019jbs} where the authors obtained several separable spacetimes from the Newman-Janis algorithm, and present a Venn diagram showing the relationship among such metrics and others found in the literature.

In 2015, Johannsen designed a Kerr-like BH solution to the Einstein field equations which is stationary, axisymmetric, asymptotically flat, and contains four constants of motion, and an event horizon~\cite{Johannsen:2015pca}.
His metric depends non-linearly on four free functions which parametrically deviate from the Kerr solution, and is general enough that it can be mapped to several other known BH solutions~\cite{Johannsen:2015pca}.
This is revisited by Papadopoulos and Kokkotas~\cite{Papadopoulos:2018nvd} and Yagi \textit{et al}~\cite{Yagi:2012ya}, in which new metrics preserving Kerr symmetries (yet no constraints such as asymptotic flatness) were presented following~\cite{Benenti}.
Following this, Johannsen constrained several of the free-functions from weak-field solar system observations~\cite{Williams:2004qba}, and proceeded to derive expressions for several spacetime properties including the orbital energy and angular momentum, the Keplerian and epicyclic frequencies, and the location of the ISCO.
Further, the same author derived expressions for the photon orbits about the parameterized beyond-Kerr BH in~\cite{Johannsen:2015qca}.
In this chapter, we aim to extend this important work by introducing a fifth free function and investigating the validity of the constraints placed from solar-system observations.

In this chapter, we follow the work of Johannsen~\cite{Johannsen:2015pca} and design a more generic, stationary and axisymmetric, asymptotically flat Kerr-like BH metric with separable structure.
The components of the inverse metric for a generic spacetime with separability structure were derived in~\cite{Benenti}. 
Such a metric has been used to construct a generic beyond-Kerr (inverse) metric with separable structure in~\cite{Yagi:2012ya} (Appendix B) and also recently in~\cite{Papadopoulos:2018nvd}, which contains five arbitrary functions of $r$ and five arbitrary functions of $\theta$.

We construct the new metric as follows. We first introduce the most generic deviation into the contravariant Kerr metric in such a way that the Hamilton-Jacobi separability condition is preserved.
The new metric is then simplified by imposing the constraint of asymptotic flatness at null infinity (radial infinity along a fixed retarded time). 
Further, we impose constraints consistent with the weak-field solar system tests as Johannsen did in~\cite{Johannsen:2015pca}, however we note that such constraints may not be explicitly valid in the strong-gravity regions surrounding BHs.

We follow this up by exploring several properties of the new spacetime.
We first locate the positions of the event horizon, Killing horizon, and ergosphere, finding that the former two reduce to the Kerr horizons, and the latter depends on just one of the 5 free functions found in the metric.
We additionally explore the spheroidicity conditions found in Ref.~\cite{Glampedakis:2018blj}, where we find the $\theta$-independent functions to admit Kerr-like spherical photon orbits.
Following this, we investigate the orbital properties of circular equatorial particle orbits, finding analytic expressions for the orbital energy and angular momentum, the Keplerian and epicyclic frequencies, and also the location of the ISCO.
We next derive analytic expressions for the photon rings as can be observed by e.g. the EHT, and present plots of the viewing plane as seen by a distant observer at null infinity for several parameterizations of the metric.
We then demonstrate the effect each parameterization has on each of the above BH properties, and also investigate the presence of naked singularities emergent for certain parameterizations.
Finally, we produce the required mappings that relate the new metric to seven other BH solutions found in the literature~\cite{Papadopoulos:2018nvd,Randall:1999ee,Aliev:2005bi,Jai-akson:2017ldo,Ding:2019mal,Kerr-Sen,Kanti_EdGB,Maeda:2009uy,Sotiriou:2014pfa,Ayzenberg:2014aka,Jackiw:2003pm,Yagi_dCS,Yunes_dcs,Bardeen,Kumar:2019uwi,Pani:2011gy,Kumar:2020hgm}.


\section{Corrections to gravitational waveforms}\label{sec:waveform}
In this section, we describe how modified BH solutions affect the gravitational waveform. In particular, we consider corrections to the inspiral, ringdown (through QNMs) and the final mass and spin of the remnant BH.
In particular, this general recipe is used to calculate corrections to the gravitational waveform for the two singly-parameterized JP and mod.~$\Delta$ spacetimes considered in this chapter in the following section.

\subsection{Inspiral}

Among the many corrections to the gravitational waveform described in Sec.~\ref{sec:bumpy}, we consider the ppE formalism~\cite{Yunes:2009ke} once again for corrections to the inspiral phase and amplitude, as described in Chapter~\ref{chap:GWs}.
Let us first describe how to compute the ppE phase and amplitude parameters for a given metric, following and slightly modifying App.~A of~\cite{Tahura_GdotMap}. 
The calculation below is similar to that in~\cite{Cardenas-Avendano:2019zxd}, but has been extended for a more generic correction in the metric.
In particular, for the two example metrics that we consider in this chapter, the dominant modifications to the binary evolution comes from the correction to the $(t,t)$ component of the metric. 
First, we make an assumption that such a metric component is given by
\begin{equation}
g_{tt} = -1 + \frac{2m}{r}\left( 1 + A \frac{m^p}{r^p} \right) + \mathcal{O}\left(\frac{m^2}{r^2}\right),
\end{equation}
where $m$ is the mass of an isolated BH, and the parameters (A,p) characterize the leading correction to the potential. Then, the reduced effective potential of a binary becomes
\begin{equation}\label{eq:V_eff}
V_\mathrm{eff} = - \frac{M}{r}\left( 1 + A \frac{M^p}{r^p} \right) + \frac{L_z^2}{2 \mu^2 r^2},
\end{equation}
where $\mu$ is the reduced mass, and $L_z$ is the $z$-component of the orbital angular momentum. 
Taking the radial derivative of $V_\mathrm{eff}$ with respect to $r$, equating it with 0 and setting $L_z = \mu r^2 \Omega$ with the orbital angular velocity $\Omega$, one finds the modified Kepler's law as
\begin{equation}
\Omega^2 = \frac{M}{r^3}\left[ 1 +  (p+1) A \frac{M^p}{r^p} \right].
\end{equation}
This equation can be inverted to yield
\begin{equation}\label{eq:r}
r = \left( \frac{M}{\Omega^2} \right)^{1/3}\left(1 +  \frac{p+1}{3} A \; v^{2p} \right),
\end{equation}
where  $v = (M \Omega)^{1/3}$ is the relative velocity and we only keep to leading correction in $A$. We substitute this back into Eq.~\eqref{eq:V_eff} and find the binding energy as
\begin{equation}
E_b = - \frac{1}{2} \eta^{-2/5}u^2 \left[1 - \frac{2(2 p-1)}{3} A \; v^{2p} \right].
\end{equation}
Next, we look at corrections to the GW luminosity. To take into account such dissipative corrections, one needs a specific theory. Thus, we neglect such effects in this chapter and assume that the the GW luminosity is given by the one in GR:
\begin{equation}
\mathcal{L}_\GW = \frac{32}{5} \pi^6 \mu^2 r^4 f^6. 
\end{equation}
This luminosity acquires a conservative correction from that in Kepler's law as
\begin{equation}
\mathcal{L}_\GW = \frac{32}{5}  \eta^2 v^{10} \left[1 + \frac{4(p+1)}{3} A \; v^{2p} \right].
\end{equation}

Having these ingredients at hand, we are now ready to compute the ppE parameters. We first look at the frequency evolution of the binary, given by
\begin{eqnarray}
\label{eq:fdot}
\dot f = \frac{df}{dE_b} \frac{dE_b}{dt} &=& - \frac{df}{dE_b} \mathcal{L}_\GW = \frac{96}{5 \pi \mathcal{M}^2} u^{11} \left( 1 + \gamma_{\dot f} u^{2p}\right),
\end{eqnarray}
where $f = \Omega/\pi$ and 
\begin{equation}
\gamma_{\dot f} = \frac{2}{3}  (p+1) (2 p+1) A.
\end{equation}
Equation~(20) of~\cite{Tahura_GdotMap} gives one the ppE parameters in the phase as\footnote{When we substitute $A=-4 a_1$ and $p=2$ where $a_1$ is the non-Kerr parameter used in~\cite{Cardenas-Avendano:2019zxd}, one finds $\beta_\ppE$ in agreement with that in~\cite{Cardenas-Avendano:2019zxd} modulo a minus sign that originates from the different convention used for the phase $\Psi$.}
\begin{eqnarray}
\beta_\ppE &=& - \frac{15}{16(2p-8)(2p-5)} \gamma_{\dot f} = -\frac{5  (p+1) (2 p+1)}{8 (2 p-8) (2 p-5)} A, \\
b_\ppE &=& 2p -5. 
\end{eqnarray}
On the other hand, the amplitude correction can be obtained from Eqs.~\eqref{eq:r} and~\eqref{eq:fdot} and the fact that the amplitude is proportional to $r^2/\sqrt{\dot f}$:
\begin{eqnarray}
\alpha_\ppE &=&  -\frac{1}{3}  (p+1) (2 p-1) A, \\
a_\ppE &=& = 2p.
\end{eqnarray}
Notice that $\alpha_\ppE$ and $\beta_\ppE$ are related to each other as
\begin{equation}
\alpha_\ppE = \frac{16  (p-4) (2 p-5) (2 p-1)}{15 (2 p+1)} \beta_\ppE.
\end{equation}
Both corrections enter at $p$th PN order relative to the leading contribution in GR (or Kerr).
These expressions are generic and can be applied to any beyond-Kerr metrics, as long as the dominant correction to the metric comes from the correction to the Newtonian potential.

\subsection{Ringdown}

We next explain how to derive modifications to the ringdown portion of the waveform.
Following in the footstpdf of the \textit{post-Kerr} formalism developed in Ref.~\cite{Silva:2019scu,Glampedakis:2017dvb,Glampedakis:2019dqh}, we estimate the QNM ringdown and damping frequencies $\omega_\R$ and $\omega_\I$ in the eikonal limit.
In this limit, $\omega_\R$ and $\omega_\I$ are associated with the light ring's angular frequency $\Omega_0$ and the Lyapunov exponent $\gamma_0$ (corresponding to the divergence rate of photon orbits grazing the light ring) at the light ring's radius $r_0$ as
\begin{eqnarray}
\omega_\R &=& 2 \Omega_0 =  2 (\Omega_\K + \delta \Omega_0),  \\
\omega_\I &=&  - \frac{1}{2}|\gamma_0| =  - \frac{1}{2}|\gamma_\K + \delta \gamma_0|.
\end{eqnarray}
Here 
\begin{equation}
\Omega_\K = \pm \frac{m^{1/2}}{r_\K^{3/2} \pm a m^{1/2}}
\end{equation}
is the angular frequency of the Kerr light ring\footnote{The upper (lower) sign corresponds to prograde (retrograde) orbit.} at 
\begin{equation}
r_\K = 2m \left\{ 1+\cos\left[ \frac{2}{3} \cos^{-1} \left(\mp \frac{a}{m} \right)\right] \right\},
\end{equation}
while 
\begin{align} 
\delta \Omega_{0} &=\mp\left(\frac{m}{r_{\K}}\right)^{1 / 2}\left[h_{\varphi \varphi} \pm\left(\frac{r_{\K}}{m}\right)^{1 / 2}\left(r_{\K}+3 m\right) h_{t \varphi}\right. \nonumber \\ 
&\left.+\left(3 r_{\K}^{2}+a^{2}\right) h_{t t}\right] /\left[\left(r_{\K}-m\right)\left(3 r_{\K}^{2}+a^{2}\right)\right] 
\end{align}
is the correction to $\Omega_\K$ with $h_{\mu\nu}$ representing the metric deviation away from Kerr. On the other hand,
\begin{equation}
\gamma_\K = 2 \sqrt{3 m} \frac{\Delta_{\K} \Omega_{\K}}{r_{\K}^{3 / 2}\left(r_{\K}-m\right)}
\end{equation}
is the Lyapunov exponent for Kerr with $\Delta_\K = r_\K^2 - 2 M r_\K + a^2$, while $\delta \gamma_0$ is the non-Kerr correction given in Eq.~(18) of~\cite{Glampedakis:2017dvb}.
See Refs.~\cite{McManus:2019ulj,Cardoso:2019mqo} where a general formalism to map ringdown corrections similar to the ones explained above directly to specific theories of gravity was developed.

\subsection{Final mass and spin}

Finally, we discuss modifications to the remnant BH's mass and spin, $M_f$ and $\chi_f$.
In GR, one can approximately estimate such parameters from the initial masses and spins via the specific energy $E$ and specific orbital angular momentum $L_z$ (found in e.g. in Eqs.~\eqref{eq:energy} and ~\eqref{eq:angMom} or~\cite{Johannsen:2015pca}) of a particle of mass $\mu=m_1m_2/M$ orbiting the remnant BH at the ISCO.
This corresponds to solving the equations~\cite{Buonanno:2007sv,Barausse:2009uz} 
\begin{align}
\nonumber \mu [1-E(M_f,\chi_f,r_\ISCO)]&=M-M_f,\\
\mu L_z(M_f,\chi_f,r_\ISCO)&=M(M_f \chi_f-a_s-\delta_m a_a),
\end{align}
as described previously in Chapter~\ref{chap:Parameterized}.
Since $E(M_f,\chi_f,r_\ISCO)$ is dimensionless in the geometric units, the $M_f$ dependence cancels and it only depends on $\chi_f$. 
Given the difference between $M$ and $M_f$ is small, we can approximate $M_f \approx M$ in the second equation.
We estimate corrections to $M_f$ and $\chi_f$ in the beyond-Kerr metrics assuming this picture still holds.
The orbital energy and angular momentum are obtained such that the expressions $\bar V_\text{eff}=0$ and $\bar V_\text{eff}'=0$ are simultaneously satisfied for effective potential $\bar V_\mathrm{eff}$ given in Eq.~\eqref{eq:Ueff}~\cite{Johannsen:2015pca}.
Corrections to the ISCO radius are further obtained by solving the expression $E'(r_\ISCO)=0$ where $E$ is the specific energy of a particle orbiting around the BH given by Eq.~\eqref{eq:energy}~\cite{Johannsen:2015pca}.
Combining these, one can find expressions for the corrections to the remnant black hole's mass $(\delta M_f)$ and spin $(\delta\chi_f)$ as
\begin{equation}
M_f =M_f^\K+\delta M_f, \quad
\chi_f=\chi_f^\K+ \delta\chi_f, 
\end{equation}
where $M_f^\K$ and $\chi_f^\K$  are the results for Kerr, which we take to be the ones in~\cite{PhenomDII}.
See App.~\ref{app:arbitraryCorrections} for the general expressions for $\delta M_f$ and $\delta\chi_f^\K$ in an arbitrary spacetime with metric $g_{\alpha\beta}^\X=g_{\alpha\beta}^\K+\zeta h_{\alpha\beta}^\X$.


\section{Beyond Kerr spacetimes}\label{sec:bumpy}
In this section we discuss the three beyond Kerr spacetimes considered in this analysis.
The first two which were derived elsewhere are the Johannsen-Psaltis (JP) metric introduced in Ref.~\cite{Johannsen:2011dh} and a modified version of Johannsen's metric in Refs.~\cite{Johannsen:2015mdd,Johannsen:2015pca}, denoted as the modified $\Delta$, or the mod.~$\Delta$ metric.
Finally, we build a new spacetime metric which preserves the Kerr symmetries, and is axisymmetric, stationary, and asymptotically flat.
For the former two spacetimes, we begin with an introduction to each spacetime, followed by the theoretical framework developed in the current analysis used to calculate the various non-Kerr corrections to the binary system present in each spacetime.
Such corrections include those to: the waveform phase and amplitude, Kepler’s law, the orbital energy and angular momentum, the radiated flux, the QNM’s ringing and damping frequencies, the ISCO radius, and the remnant BH mass and spin predictions. 
These corrections may then be used to modify the gravitational waveform as sourced in the given bumpy-Kerr spacetimes.
For the latter spacetime, we first derive it following in the footstpdf of Johannsen in Ref.~\cite{Johannsen:2015pca}.
We then show that the event and Killing horizons in such a spacetime coincides with those of a Kerr BH, and finally we derive the locations of the ergosphere.

\subsection{Johannsen-Psaltis metric}\label{sec:JP}
We begin our discussion on the JP metric, introduced by Johannsen and Psaltis in Ref.~\cite{Johannsen:2011dh}.
In this article, the authors begin with the Kerr metric $g_{\mu\nu}^\K$, with metric elements in Boyer-Lindquist coordinates given by
\begin{align}
\nonumber g_{tt}^\K&=-\left( 1-\frac{2mr}{\Sigma} \right) = - \frac{\Delta-a^2 \sin^2\theta}{\Sigma},\\ 
\nonumber g_{rr}^\K&=\frac{\Sigma}{\Delta},\\
\nonumber g_{\theta\theta}^\K&=\Sigma,\\
\nonumber g_{\phi\phi}^\K&=\left( r^2+a^2+\frac{2ma^2r\sin^2\theta}{\Sigma} \right)\sin^2\theta\nonumber = \frac{ [(r^2+a^2)^2-a^2\Delta \sin^2\theta]\sin^2\theta}{\Sigma}, \nonumber \\
g_{t\phi}^\K&=-\frac{2mar\sin^2\theta}{\Sigma} = - \frac{a (r^2+a^2-\Delta)\sin^2\theta}{\Sigma}, 
\label{eq:Kerr_metric}
\end{align}
with
\begin{align}
\nonumber \Sigma&\equiv r^2+a^2\cos^2\theta,\\
\Delta&\equiv r^2-2mr+a^2,
\end{align}
where $(r,\theta,\phi)$ are the radial, polar, and azimuthal coordinates, and $m$, $a$ are the BH's mass and spin.
Generalized parametric deviations $h(r,\theta)$ of the form
\begin{equation}
h(r,\theta)=\sum\limits_{k=0}^{\infty}\left(\epsilon_{2k}+\epsilon_{2k+1}\frac{mr}{\Sigma}\right) \left( \frac{m^2}{\Sigma} \right)^k
\end{equation}
for some non-Kerr deviation parameters $\epsilon_k$ were introduced into each metric element as seen in Eq.~(44) of Ref.~\cite{Johannsen:2011dh}.
By further applying the constraint of asymptotically flat spacetime at null infinity, as well as observational constraints on the parameterized post-Newtonian (PPN) framework~\cite{Will:2005va}, the deviation function $h(r,\theta)$ was reduced to a single non-Kerr parameter $\epsilon_3$
\begin{equation}\label{eq:JPparameter}
h(r,\theta)=\epsilon_3\frac{m^3r}{\Sigma^2}.
\end{equation}
Assuming that deviations from Kerr are small and keeping only up to linear order in $\epsilon_3$, the resulting JP metric $g_{\mu\nu}^\JP$ can be written as
\begin{align}
\nonumber g_{tt}^\JP&=-\left(1-\frac{2mr}{\Sigma} \right)-\epsilon_3\frac{m^3 (r-2 m)}{r^4},\\ 
\nonumber g_{rr}^\JP&=\frac{\Sigma}{\Delta}+\epsilon_3\frac{m^3(r-2m)}{\Delta ^2},\\
\nonumber g_{\theta\theta}^\JP&=\Sigma,\\
\nonumber g_{\phi\phi}^\JP&=\left( r^2+a^2+\frac{2ma^2r\sin^2\theta}{\Sigma} \right)\sin^2\theta+\epsilon_3\frac{a^2m^3(r+2m)}{r^4},\\
 g_{t\phi}^\JP&=-\frac{2mar\sin^2\theta}{\Sigma}-\epsilon_3\frac{2am^4}{r^4}.\label{eq:JP}
\end{align}
With this choice of $h(r,\theta)$, the JP metric now allows one to probe strong-field gravity to any order of spin in a parameterized way.
Observe how in the limit of $\epsilon_3\rightarrow0$, we recover the original Kerr metric for a spinning BH.
See Refs.~\cite{Kong:2014wha,Bambi:2015ldr} for constraints on the JP deviation parameter $\epsilon_3$ from BH accretion disk thermal spectra, found to be loosely $\epsilon_3\lessapprox5$.

We next identify the dominant contribution to the binary evolution. For a particle orbiting around a BH, the angular velocity $\Omega$ is determined from the radial derivative of $g_{tt}$, $g_{t\phi}$ and $g_{\phi\phi}$ (see Eq.~\eqref{eq:KeplerianFrequency} found in e.g. Ref.~\cite{Johannsen:2015pca}). When we expand the JP metric components about $r =\infty$, one finds that the leading correction to $\partial_r g_{tt}$, $\partial_r g_{t\phi}$ and $\partial_r g_{\phi\phi}$ enters at $\mathcal{O}(m^2/r^2)$, $\mathcal{O}(m^3/r^3)$ and $\mathcal{O}(m^5/r^5)$ relative to the leading Kerr contribution respectively. Thus, the dominant correction comes from $g_{tt}$ and we find 
\begin{equation}
A_\JP = - \frac{\epsilon_3}{2}, \quad p_\JP = 2. 
\end{equation}

Now let us lay the groundwork for the JP modifications to the gravitational waveform by applying the results presented in Sec.~\ref{sec:waveform}. 
First, the ppE parameters are given by
\begin{align}
\nonumber \beta_\ppE^\JP&=\frac{75 \epsilon_3}{64 \eta ^{4/5}}, \hspace{5mm} b_\ppE^\JP=-1,\\ \label{eq:JPppe}
\alpha\ppE^\JP&=\frac{3 \epsilon_3}{2 \eta ^{4/5}}, \hspace{5mm} a_\ppE^\JP=4,
\end{align}
and the corrections enter at 2PN order. 
This is of the same order as the correction for the beyond-Kerr metric proposed in Ref.~\cite{Rezzolla:2014mua}, as found in Ref.~\cite{Cardenas-Avendano:2019zxd}.
Next, the QNM corrections in a JP spacetime to first order in JP deviation parameter, and quadratic in spin are given by
\begin{align}
\nonumber \omega_\R^\JP &= \omega_\R^\K+ \epsilon_3\left( \frac{1}{81\sqrt{3}M}+\frac{10}{729M}\chi+\frac{47}{1458\sqrt{3}M}\chi^2 \right),\\
\omega_\I^\JP &= \omega_\I^\K- \epsilon_3\left( \frac{1}{486M}\chi+\frac{16}{2187\sqrt{3}M}\chi^2 \right),
\end{align}
for unitless spin parameter $\chi\equiv a / M$, and Kerr QNM frequencies $\omega_{\R,\I}^\K$. 
It is interesting to note that $\omega_\I$ does not acquire corrections if the BH is non-spinning.
Finally, corrections to the final mass and spin are given by
\begin{align}
\delta M_f^\JP&=-\epsilon_3\frac{ \mu}{139968} \left\lbrack864 \delta\chi_f^\JP \left(5 \sqrt{2} \chi_f^\K+3 \sqrt{3}\right)+545 \sqrt{2} \left( \chi_f^\K \right)^2+324 \sqrt{3} \chi_f^\K+216 \sqrt{2}\right\rbrack\\
\nonumber\delta \chi_f^\JP &= -\epsilon_3 \frac{1}{384 \sqrt{3} \kappa  \mu} \Big\lbrack420 \mu  M \chi _a \delta _m+420 \mu  \chi _a \lambda-152 \sqrt{2} \kappa  \mu +2416 \sqrt{3} \mu ^2\\
&+420 \mu  \delta _m \lambda \chi _s+945 \sqrt{3} M^2-315 \kappa  M+1086 \sqrt{6} \mu  M+420 \mu  M \chi _s\Big\rbrack,
\end{align}\\
which is valid to linear order in $\epsilon_3$ and to quadratic order in the final spin, and
\begin{align}
\nonumber \kappa&\equiv \Big\lbrack 8 \sqrt{3} \mu  \chi _a \left(M \delta _m+\lambda\right)+8 \sqrt{3} \mu  \chi _s \left(\delta _m \lambda+M\right)+3 \left(40 \mu ^2+9 M^2+12 \sqrt{2} \mu  M\right)\Big\rbrack.\\
\lambda&\equiv\sqrt{M (M-4\mu )}
\end{align}
This is derived from $r_\ISCO$, which, to linear order in JP deviation and quadratic in BH spin, is given by
\begin{equation}
r_\ISCO^\JP=r_\ISCO^\K\left\lbrack 1-\epsilon_3\left(\frac{1}{27} + \frac{37}{324\sqrt{6}}\chi +\frac{1229}{23328}\chi^2 \right) \right\rbrack,
\end{equation}
with Kerr result $r_\ISCO^\K$~\cite{wald_1984}.

\subsection{Modified $\Delta$ metric}\label{sec:MD}
Now let us discuss the newly constructed mod.~$\Delta$ metric, following in the footstpdf of Johannsen in Refs.~\cite{Johannsen:2015mdd,Johannsen:2015pca}.
We begin in Ref.~\cite{Johannsen:2015pca}, in which 4 free functions $A_1(r)$, $A_2(r)$, $A_5(r)$, and $f(r)$ are introduced to the Kerr spacetime,  parameterically describing deviations from GR, as shown in Eq.~(51) of~\cite{Johannsen:2015pca}.
Such a metric is found to be stationary, axisymmetric, asymptotically flat, admits freely-rotating BHs, reduces to the Kerr metric for $A_1(r)=A_2(r)=A_5(r)=1$, and $f(r)=0$, and posess a fourth constant of motion, a Carter-like constant~\cite{Carter:1968rr}.
This symmetry, as in the Kerr metric, gives rise to separable, non-chaotic geodesic equations for particle motion.

Following this, in Ref.~\cite{Johannsen:2015mdd}, Johannsen further modified the obtained spacetime metric  by introducing a pure-deviation $\beta$ from the Kerr metric, by substituting
\begin{equation}
\label{eq:Delta_replacement}
\Delta\rightarrow \bar{\Delta}\equiv\Delta+\beta m^2
\end{equation}
into the metric found in Eq.~(51) of Ref.~\cite{Johannsen:2015pca}.
We further equate all other free functions to their Kerr values, $A_1(r)=A_2(r)=A_5(r)=1$, and $f(r)=0$, resulting in the modified-$\Delta$ metric $g_{\mu\nu}^\MD$ with elements given by\footnote{This metric can also be obtained by applying the replacement in Eq.~\eqref{eq:Delta_replacement} to Eq.~\eqref{eq:Kerr_metric}.}
\begin{align}
\nonumber g_{tt}^\MD&=-\left( 1-\frac{2mr}{\Sigma} \right)-\beta \frac{m^2}{\Sigma},\\ 
\nonumber g_{rr}^\MD&=\frac{\Sigma}{\Delta}-\beta\frac{m^2\Sigma}{\Delta^2},\\
\nonumber g_{\theta\theta}^\MD&=\Sigma,\\
\nonumber g_{\phi\phi}^\MD&=\left( r^2+a^2+\frac{2ma^2r\sin^2\theta}{\Sigma} \right)\sin^2\theta\\
\nonumber& \hspace{5mm} -\beta\frac{a^2m^2\sin ^4\theta}{\Sigma},\\
g_{t\phi}^\MD&=-\frac{2mar\sin^2\theta}{\Sigma}+\beta\frac{a m^2  \sin ^2\theta}{\Sigma},\label{eq:MD}
\end{align}
where we assume that the deviation from Kerr is small and we keep only to linear order in $\beta$.
This spacetime is entirely parameterized by the single, pure-deviation parameter $\beta$, reduces to the Kerr metric for $\beta=0$, and is useful as it can be mapped to BH solutions other than Kerr.
Such metrics include the Kerr-Newman metric for charged BHs~\cite{delaCruzDombriz:2009et}, the RS-II braneworld BHs~\cite{Aliev:2005bi}, and those in the modified gravity (MOG)~\cite{Moffat:2014aja}

Now let us consider the various corrections to the gravitational waveform present in the mod.~$\Delta$ spacetime. Just like in the case of the JP metric, the leading correction comes from $g_{tt}^\MD$ and 
\begin{equation}
A_\MD = - \frac{\beta}{2}, \quad p_\MD = 1,
\end{equation}
which means that the correction enters at 1PN order.
First, the ppE parameters entering in the inspiral waveform are given by
\begin{align}
\nonumber \beta_\ppE^\MD&=\frac{5\beta}{48\eta^{2/5}}, \hspace{5mm} b_\ppE^\MD=-3,\\
\alpha_\ppE^\MD&=-\frac{\beta}{3\eta^{2/5}}, \hspace{5mm} a_\ppE^\MD=2.
\end{align}
Next, the ringdown frequencies are modified as
\begin{align}
\nonumber \omega_\R^\MD &= \omega_\R^\K+ \beta \left( \frac{1}{9\sqrt{3}M}+\frac{2}{27M}\chi+\frac{61}{486\sqrt{3}M}\chi^2 \right),\\
\omega_\I^\MD &= \omega_\I^\K+\beta \left( \frac{1}{108\sqrt{3}M}-\frac{1}{243M}\chi-\frac{11}{729\sqrt{3}M}\chi^2 \right).
\end{align}
Finally, the corrections to the final mass and spin are given by 
\begin{align}
\delta M_f^\MD&=-\beta\frac{\mu}{7776}\left\lbrack48 \delta\chi_f^\MD \left(5 \sqrt{2} \chi_f^\K+3 \sqrt{3}\right)+119 \sqrt{2} \left(\chi_f^\K\right)^2+84 \sqrt{3}\chi_f^\K+72 \sqrt{2}\right\rbrack,\\
\nonumber\delta\chi_f^\MD&=-\beta\frac{1}{256 \sqrt{3} \kappa  \mu}\Big\lbrack876 \mu  M \chi _a \delta _m+876 \mu  \chi _a \lambda -286 \sqrt{2} \kappa  \mu +5288 \sqrt{3} \mu ^2\\
&+876 \mu  \delta _m \lambda \chi _s+1971 \sqrt{3} M^2-657 \kappa  M+2172 \sqrt{6} \mu  M+876 \mu  M \chi _s\Big\rbrack,
\end{align}
where we used the ISCO radius expression of
\begin{equation}
r_\ISCO^\MD=r_\ISCO^\K \left\lbrack 1-\beta \left( \frac{1}{4} + \frac{1}{2\sqrt{6}}\chi + \frac{77}{432}\chi^2 \right) \right\rbrack.
\end{equation}


\subsection{A new metric preserving Kerr symmetries}\label{sec:CYst}
In this section, we present a new general spacetime metric with parameterized deviations beyond GR.
This spacetime preserves all of the Kerr symmetries, and is axisymmetric, stationary, and asymptotically flat.
We also show that the event and Killing horizons in such a spacetime coincides with those of a Kerr BH, and then we derive the locations of the ergosphere.
Finally, we compute the spheroidicity conditions of Ref.~\cite{Glampedakis:2018blj}, showing them to be independent of $\theta$, thus admitting Kerr-like spherical photon orbits.

\subsubsection{The metric}\label{sec:CYmetric}
Here we compute the new spacetime metric as used throughout this analysis.
We obtain this metric by following and modifying the analysis thoroughly done by Johannsen in Ref.~\cite{Johannsen:2015pca}.
There, a regular parameterized BH solution was created to be stationary, axisymmetric, asymptotically flat, and separable.
The latter property comes forth from the existence of a fourth constant of motion, the Carter-like constant~\cite{Carter}.
The metric presented here, while very similar to Johannsen's, is more general and admits an additional deviation function from the Kerr metric.
In addition, we allow several parameters assumed to vanish by Johannsen via weak-field PPN constraints, to remain intact.

We begin with the Kerr metric for a rotating BH.
This well-known spacetime has a line element given by
\begin{align}
\nonumber &ds_\K^2=-\left( 1-\frac{2mr}{\Sigma} \right)dt^2+
\Sigma d\theta^2 -\frac{4mar\sin^2\theta}{\Sigma}dtd\phi\\
 &+\frac{\Sigma}{\Delta} dr^2+\left( r^2+a^2+\frac{2ma^2r\sin^2\theta}{\Sigma} \right)\sin^2\theta d\phi^2.
\end{align}
Similar to Ref.~\cite{Johannsen:2015pca}, we introduce scalar deviation functions $f(r)$, $g(\theta)$, $A_1(r)$, $A_2(r)$, $A_3(\theta)$, $A_4(\theta)$, and $A_6(\theta)$ into the contravariant Kerr metric, as well as new functions $\bar A_{0}(\theta)$ and $A_{0}(r)$
\begin{align}
\nonumber &g^{\alpha\beta}\frac{\partial}{\partial x^\alpha}\frac{\partial}{\partial x^\beta}=\\
\nonumber &-\frac{1}{\Delta\tilde\Sigma} \Bigg\lbrack (r^2+a^2)^{2}A_1(r)^{2}\left(\frac{\partial}{\partial t}\right)^{2}+a^{2} A_2(r)^{2}\left(\frac{\partial}{\partial \phi}\right)^{2}+ 2a(r^2+a^2)A_{0}(r) \frac{\partial}{\partial t}\frac{\partial}{\partial \phi}\Bigg\rbrack\\
&+\frac{1}{\tilde\Sigma\sin^2\theta} \Bigg\lbrack A_3(\theta)^{2}\left(\frac{\partial}{\partial\phi}\right)^{2}+a^{2} \sin^{4}\theta A_4(\theta)^{2}\left(\frac{\partial}{\partial t}\right)^{2}+ 2 a \sin^2\theta \bar A_{0}(\theta)\frac{\partial}{\partial t}\frac{\partial}{\partial \phi}  \Bigg\rbrack \nonumber \\
&+\frac{\Delta}{\tilde\Sigma}A_5(r)\left( \frac{\partial}{\partial r} \right)^2+\frac{1}{\tilde\Sigma}A_6(r)\left( \frac{\partial}{\partial\theta} \right)^2,\label{eq:CYcontravariant}
\end{align}
with 
\begin{equation}
\tilde\Sigma\equiv\Sigma+f(r)+g(\theta).
\end{equation}
Observe how this expression is similar to Eq.~(10) of Ref.~\cite{Johannsen:2015pca}, however the two additional scalar functions $A_0(r)$ and $\bar A_0(\theta)$ introduce more generality into the function.
One recovers the metric in~\cite{Johannsen:2015pca} by setting $A_0 = A_1 A_2$ and $\bar A_0 = A_3 A_4$, while it reduces to the Kerr metric in the limit $A_i \to 1$ and $f \to 0$, $g \to 0$.
Such modifications guarantee the resulting Hamilton-Jacobi equations remain seperable, and a fourth constant of motion appears as thoroughly described in~\cite{Johannsen:2015pca}.
Additionally, following Ref.~\cite{Konoplya:2018arm} we find that our metric is a more broad example of the general class of metrics that admit separable Klein Gordon equations.
As was the case for the metric presented in~\cite{Johannsen:2015pca}, we find that with the additional assumption of $f(r)=(r^2+a^2)\left(\frac{A_1(r)}{A_2(r)}-1\right)$, our metric also reduces to one that allows for the separability of the Klein Gordon equations.

Next we define functional forms of our scalar deviation functions, and apply various constraints.
We expand the radial functions as a power series in $m/r$, $g(\theta)$ as a Legendre expansion, and ignore the remaining polar functions for now~\cite{Johannsen:2015pca}:
\begin{align}
A_i(r)&\equiv\sum\limits_{n=0}^\infty\alpha_{in}\left( \frac{m}{r} \right)^n, \hspace{5mm} (i=0,1,2,5),\\
f(r)&\equiv r^2\sum\limits_{n=0}^\infty\epsilon_n \left( \frac{m}{r} \right)^n\\
g(\theta)&\equiv M^2\sum\limits_{n=0}^\infty\gamma_{n} P_n(\cos\theta),
\end{align}
with Legendre polynomials $P_n(\cos\theta)$.
We note that the Legendre expansion of $g(\theta)$ differs from that presented in~\cite{Johannsen:2015pca}, where the author utilized a trigonometric expansion in powers of $\sin\theta$ and $\cos\theta$.
The Legendre expansion given here is a more natural choice given an axisymmetric spacetime metric, and gives unique choices on parameters $\gamma_n$, where degeneracies occur in the the trigonometric expansion utilized previously.

We begin constraining the deviation parameters by imposing the condition of asymptotic flatness~\cite{heusler_1996,wald_1984}.
This corresponds to imposing that our metric line element must limit to 
\begin{equation}
ds_\infty^2=-\left( 1-\frac{2m}{r} \right)dt^2-\frac{4ma}{r}\sin^2\theta dtd\phi+dr^2+r^2d\Omega^2,
\end{equation}
at spatial infinity $r\to \infty$ for $d\Omega\equiv d\theta^2+\sin^2\theta d\phi^2$.
Doing so reveals the need to constrain $\bar A_0(\theta)=A_3(\theta)=A_4(\theta)=A_6(\theta)=1$, as well as $\alpha_{00}=\alpha_{10}=\alpha_{20}=\alpha_{50}=1$ and $\epsilon_0=0$.
The asymptotic behavior of $g_{tt}$ and $g_{t\phi}$ become
\begin{align}
g_{tt} &= -1 + (2 + 2 \alpha_{11} -\epsilon_1)\frac{m}{r} + \mathcal{O}\left( \frac{m^2}{r^2} \right)\,, \\
g_{t\phi} &= -(2+\alpha_{01}) a  \frac{m}{r} \sin^2\theta + \mathcal{O}\left( \frac{m^2}{r^2} \right)\,.
\label{eq:g_tph_asympt}
\end{align}
Thus, we can rescale $m$ and $a$ to further set $\alpha_{01} = 0$ and $\alpha_{11} = \epsilon_{1}/2$ without loss of generality. 

The final covariant form of the metric for a BH with mass $m$ and spin $a$ is given by
\begin{align}
\allowdisplaybreaks
\nonumber g_{tt}&=- \frac{\tilde\Sigma \left(\Delta -a^2 A_2^2 \sin ^2\theta \right)}{\tilde \rho^4},\\
\nonumber g_{rr}&=\frac{\tilde\Sigma}{A_5\Delta},\\
\nonumber g_{\theta\theta}&=\tilde\Sigma,\\
\nonumber g_{\phi\phi}&=\frac{\tilde{\Sigma } \sin^2\theta \left[\left(a^2+r^2\right)^2A_1^2 -a^2 \Delta  \sin ^2\theta \right]}{\tilde \rho^4},\\
g_{t\phi}&=-\frac{a \tilde{\Sigma } \sin^2\theta  \left[ \left(a^2+r^2\right)A_0 - \Delta\right]}{\tilde \rho^4}, \label{eq:CYmetric}
\end{align}
with
\begin{align}
\nonumber\tilde \rho^4 \equiv & \left[ \left(a^2+r^2\right)A_1 -a^2 A_2   \sin ^2\theta \right]^2 \\
&+ a^2 \left(a^2+r^2\right)(A_0-A_1A_2) \sin ^2\theta \left\lbrack\frac{a^2+r^2}{\Delta} (A_0+A_1 A_2)-2  \right\rbrack,
\end{align}
and
\begin{align}
\label{eq:Sigma_tilde}
\tilde\Sigma&\equiv\Sigma+f(r)+g(\theta),\\
\label{eq:Sigma}
\Sigma&\equiv r^2+a^2\cos^2\theta,\\
\label{eq:Delta}
\Delta&\equiv r^2+a^2-2Mr.
\end{align}
The arbitrary functions are expanded about spatial infinity as
\begin{align}
A_i(r)&\equiv 1+ \sum\limits_{n=1}^\infty\alpha_{in}\left(\frac{m}{r}\right)^n, \quad (i=0,1,2,5),\\
f(r)&\equiv r^2\sum\limits_{n=1}^\infty \epsilon_n\left(\frac{m}{r}\right)^n,\\
g(\theta)&\equiv m^2\sum\limits^\infty_{n=0}\gamma_n P_n(\cos\theta),
\end{align}
for Legendre polynomials $P_n(\cos\theta)$.
Here, the parameters $\alpha_{in}$, $\epsilon_n$ and $\gamma_n$ control the amount of deviation from Kerr. We can set $\alpha_{01} = 0$ and $\alpha_{11} = \epsilon_{1}/2$ without loss of generality by rescaling $m$ and $a$. One can further impose $\epsilon_1 =\epsilon_2= \alpha_{51}=\alpha_{12}=g(\theta)=0$ to satisfy solar system bounds, though such weak-field constraints may not apply to spacetime outside of a BH. The difference from Johannsen's metric in~\cite{Johannsen:2015pca} is that we have introduced a new radial function $A_0$ that enters in $g_{tt}$, $g_{t\phi}$ and $g_{\phi\phi}$.
We believe this new metric is the most general stationary, axisymmetric, asymptotically flat, and separable spacetime.

Next we consider imposing constraints obtained from the parameterized-post-Newtonian (ppN) framework~\cite{Will:2005va}\footnote{Apart from $\gamma_\ppN$ and $\beta_\ppN$ considered here, one could in principle consider other PPN parameters, including the one in~\cite{Alexander:2007zg} which is associated with the Lense-Thirring precession. We leave this possibility for future work.}.
This is done by further imposing that the metric for a non-spinning object must reduce to the line element given by
\begin{align}
\nonumber ds^2_\ppN=&-\left[1-\frac{2m}{r}+2(\beta_\ppN-\gamma_\ppN)\frac{m^2}{r^2}  + \mathcal{O}\left( \frac{m^3}{r^3} \right) \right]dt^2\\
&+\left[ 1+2\gamma_\ppN \frac{m}{r}  + \mathcal{O}\left( \frac{m^2}{r^2} \right)\right]dr^2 +r^2 \left[ 1 + \mathcal{O}\left( \frac{m^2}{r^2} \right)\right]d\Omega,
\end{align}
for ppN parameters $\gamma_\ppN$ and $\beta_\ppN$, while the asymptotic behavior of the new metric is given by
\begin{align}
g_{tt} &= -1 + 2\frac{m}{r} + \left[\frac{m^2}{4}  (8\alpha_{12} + \epsilon_1^2 - 4 \epsilon_2)-g \right]\frac{m^2}{r^2} + \mathcal{O}\left( \frac{m^3}{r^3} \right)\,, \\
g_{rr} &= 1 + \left(2 - \alpha_{51} + \epsilon_1 \right)\frac{m}{r}+ \mathcal{O}\left( \frac{m^2}{r^2} \right)\,, \\
g_{\theta\theta} &= r^2 \left[1 + \epsilon_1\frac{m}{r}+ \mathcal{O}\left( \frac{m^2}{r^2} \right) \right]\,, \\
g_{\phi\phi} &= r^2 \sin^2\theta \left[1 + \epsilon_1\frac{m}{r}+ \mathcal{O}\left( \frac{m^2}{r^2} \right) \right]\,.
\end{align}
Given the strong observational constraints of $\beta_\ppN$~\cite{Williams:2004qba} and $\gamma_\ppN$~\cite{Bertotti:2003rm} from solar system experiments, one can further impose the conditions $\epsilon_1 = 0$ (which automatically sets $\alpha_{11}=0$), $\alpha_{51} = 0$ and $2\epsilon_2-2\alpha_{12}-g(\theta)/M^2=0$. The simplest choice of the last condition is $\epsilon_2=\alpha_{12}=g(\theta)=0$~\cite{Johannsen:2015pca}, which is what we consider in the main part of this chapter.

Notice, however, that because Birkhoff's theorem is not guaranteed to \textbf{hold} in theories beyond GR, such ppN constraints obtained in the weak-field environment of the local solar system may not necessarily apply to the strong-gravity conditions present near the BHs considered here. 
This is indeed the case for BHs in e.g. Einstein-dilaton Gauss-Bonnet gravity, in which the BH exterior spacetime is different from that for stars due to the presence (absence) of the BH (stellar) scalar charge~\cite{Campbell:1991kz,Yunes:2011we,Yagi:2011xp,Sotiriou:2014pfa,Yagi:2015oca,Berti_ModifiedReviewSmall,Prabhu:2018aun}.
Thus, the presented constraints on $\epsilon_1$, $\alpha_{51}$ $\epsilon_2$, $\alpha_{12}$, and $g(\theta)$ may not necessarily hold, and App.~\ref{app:lowerOrder} provides a description of the effects of including such parameters in the metric.


\subsubsection{Location of the event horizon, Killing horizon, and ergosphere}\label{sec:horizons}

In this section we describe the locations of the event horizon, Killing horizon, and ergosphere in the new spacetime.
In particular, we note that the locations of each of these appear identically to those as presented in Ref.~\cite{Johannsen:2015pca}, thus we refer the reader there for a thorough description of each.

We begin by briefly describing the event horizon in both the new spacetime presented here, and Johannsen's spacetime.
The angular function $r_\EH\equiv H(\theta)$ is a solution to
\begin{equation}
g^{rr}+g^{\theta\theta}\left( \frac{dH}{d\theta} \right)^2=0,
\end{equation}
which results in the solution
\begin{equation}
\Delta A_5(H)+\left( \frac{dH}{d\theta} \right)^2=0.
\end{equation}
One solution is either the Kerr result ($\Delta=0$, $dH/d\theta = 0$)
\begin{equation}
r_\EH=m+\sqrt{m^2-a^2},
\end{equation}
or a solution to $A_5(r)=0$ with $dH/d\theta = 0$. 
The latter set is realized if $A_5$ diverges when $\Delta = 0$, which is the case for some example BH solutions that can be mapped to the metric presented in this paper. 
These solutions are the same as those for Johannsen's spacetime because $A_0(r)$ does not appear in either $g^{rr}$ or $g^{\theta\theta}$.
We see here that this expression only depends on the non-Kerr deviation function $A_5(r)$.

Next we find the location of the Killing horizon, which is located at the solution of $(g_{t\phi})^2-g_{tt} g_{\phi\phi}=0$.
Such an expression reduces down to $\Delta=0$, coinciding with the event horizon, in both the spacetimes considered here and by Johannsen.

Finally, the ergosphere exists at the roots of $g_{tt}=0$.
Because $A_0(r)$ only appears in the denominator of $g_{tt}$, the new solution is identical to that of Johannsen's, reducing to
\begin{equation}
\Delta=a^2A_2(r)^2\sin^2\theta
\end{equation}
in both cases, which is displayed in Fig.~1 of Ref.~\cite{Johannsen:2015pca}.
We see that this expression only depends on the non-Kerr deviation function $A_2(r)$. 


\subsubsection{The spheroidicity condition}\label{sec:spheroid}
In this section we compute the spheroidicity condition as detailed in the analysis of Ref.~\cite{Glampedakis:2018blj} for the new general metric presented in this chapter.
In the above investigation, the authors found the most general form of the ``spheroidicity condition''.
Such a condition defines non-equatorial circular orbits confined on a spheroidal shell described by $r_0(\theta)$.
In particular, we compute the spheroidicity condition found in Eq.~(14) of~\cite{Glampedakis:2018blj} as a function of $r_0(\theta)$ for the general metric presented here in Eq.~\eqref{eq:CYmetric}, with only the leading order terms of each beyond-Kerr function present.
We find the resulting condition to be
\begin{align}
\nonumber 0 &= 3 \alpha _{13} a^6 M^3 \left(\alpha _{13} M^3+r_0^3\right)-a^5 b M^2 r_0 \left(5 \alpha _{13} \alpha _{22} M^3+3 \alpha _{13} M r_0^2+2 \alpha _{22} r_0^3\right)\\
\nonumber & -a^4 r_0 \Big(M^4 \left(8 \alpha _{13} r_0^3-2 \alpha _{22}^2 b^2 r_0\right) -2 \alpha _{22} b^2 M^2 r_0^3+7 \alpha _{13}^2 M^7-8 \alpha _{13}^2 M^6 r_0-7 \alpha _{13} M^3 r_0^4\\
\nonumber&+M r_0^6+r_0^7\Big)+2 a^3 b M r_0^2 \Big(6 \alpha _{13} \alpha _{22} M^5-5 \alpha _{13} \alpha _{22} M^4 r_0+4 \alpha _{13} M^3 r_0^2+3 \left(\alpha _{22}-\alpha _{13}\right) M^2 r_0^3\\
\nonumber&-2 \alpha _{22} M r_0^4+r_0^5\Big)-a^2 r_0^3 \Big(b^2 \big(5 \alpha _{22}^2 M^5-3 \alpha _{22}^2 M^4 r_0+6 \alpha _{22} M^3 r_0^2-4 \alpha _{22} M^2 r_0^3+ M r_0^4\\
\nonumber& -r_0^5\big)+10 \alpha _{13}^2 M^7-7 \alpha _{13}^2 M^6 r_0+8 \alpha _{13} M^4 r_0^3-5 \alpha _{13} M^3 r_0^4-2 M r_0^6+2 r_0^7\Big)\\
\nonumber&-a b M r_0^4 \Big(-8 \alpha _{13} \alpha _{22} M^5+5 \alpha _{13} \alpha _{22} M^4 r_0-4 \alpha _{13} M^3 r_0^2+\left(3 \alpha _{13}-2 \alpha _{22}\right) M^2 r_0^3+2 \alpha _{22} M r_0^4\\
&+2 r_0^5\Big)-r_0^5 \Big(3 \alpha _{13}^2 M^7-2 \alpha _{13}^2 M^6 r_0-\alpha _{13} M^3 r_0^4-3 M r_0^6+r_0^7\Big)
\end{align}
for orbital impact parameter $b$.
Similar to Johannsen's metric, we find the above condition to be independent of $\theta$, and we conclude that this metric admits $r_0=\text{const.}$ Kerr-like spherical photon orbits.


\section{Testing beyond-Kerr spacetimes}\label{sec:testingBeyondKerr}
In this section we consider tests of the beyond-Kerr spacetimes developed above.
In particular, for the singly-parameterized JP and mod.~$\Delta$ metrics derived elsewhere we consider tests by way of GWs.
Specifically, we use the IMR consistency tests introduced in Chapter~\ref{chap:IMRconsistency} with the Fisher analysis techniques found in Chapter~\ref{chap:ParameterEstimation} to estimate the magnitudes of $\epsilon_3$ and $\beta$ required to fail the test.
For the new metric preserving Kerr symmetries presented here, we consider EM tests.
We find corrections to several astrophysical phenomena including the Keplerian and epicyclic frequencies of orbiting particles, the orbital energy and angular momentum of particle orbits, the location of the ISCO, and finally the photon orbits.

\subsection{GW tests of beyond-Kerr spacetimes}\label{sec:GWtests}
In this section we test the singly-parameterized JP and modified~$\Delta$ spacetimes by way of the IMR consistency tests and the parameterized tests of GR.
See Tab.~\ref{tab:results} for a summary of all estimated constraints in both the JP and mod.~$\Delta$ spacetimes obtained from both (i) the IMR consistency tests of GR, and (ii) re-computed with the parameterized tests of GR for comparison.

\subsubsection{IMR consistency test in the JP spacetime}\label{sec:IMRJP}

\begin{figure*}[!htbp]
\begin{center}
\includegraphics[width=.45\textwidth]{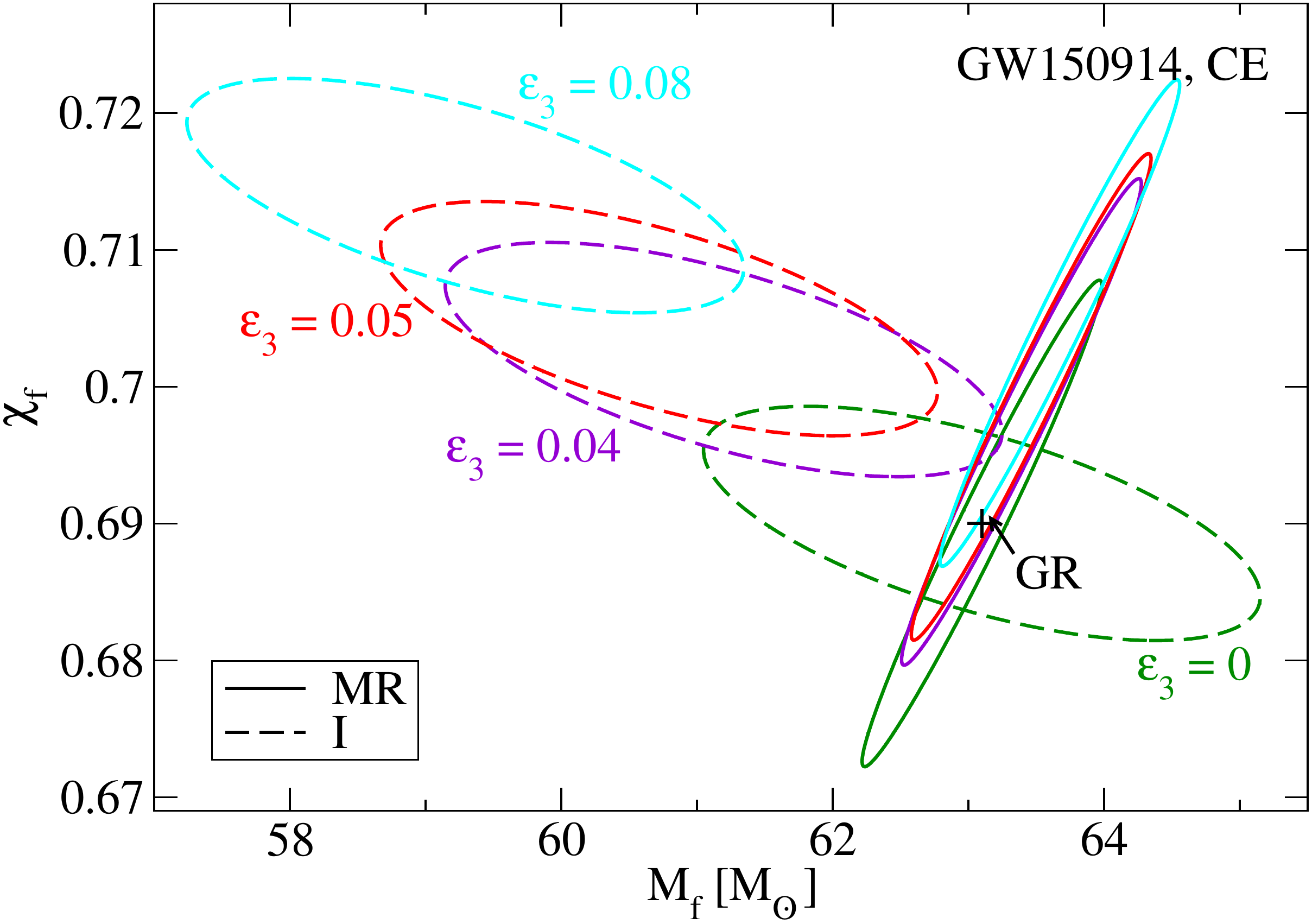}
\includegraphics[width=.45\textwidth]{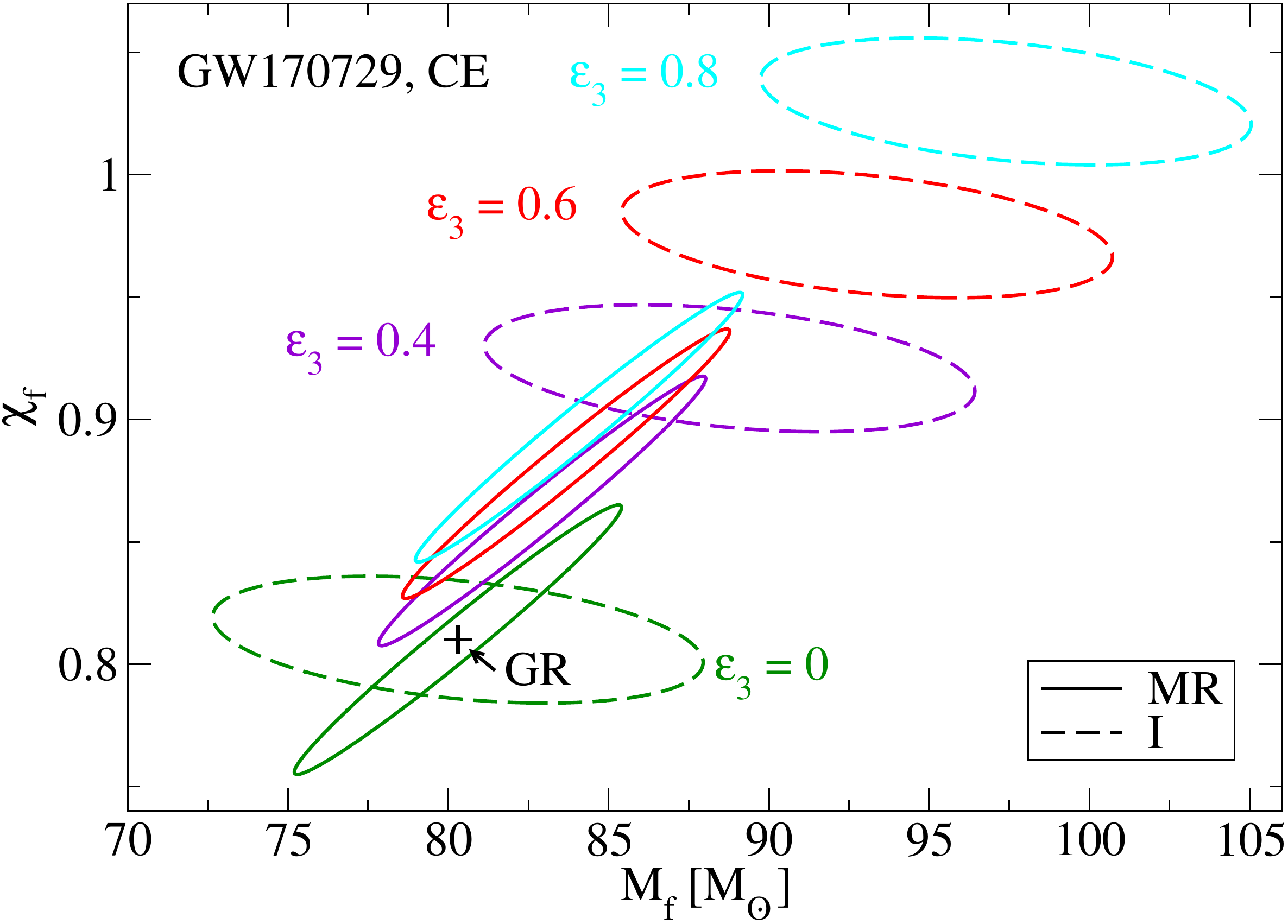}
\includegraphics[width=.45\textwidth]{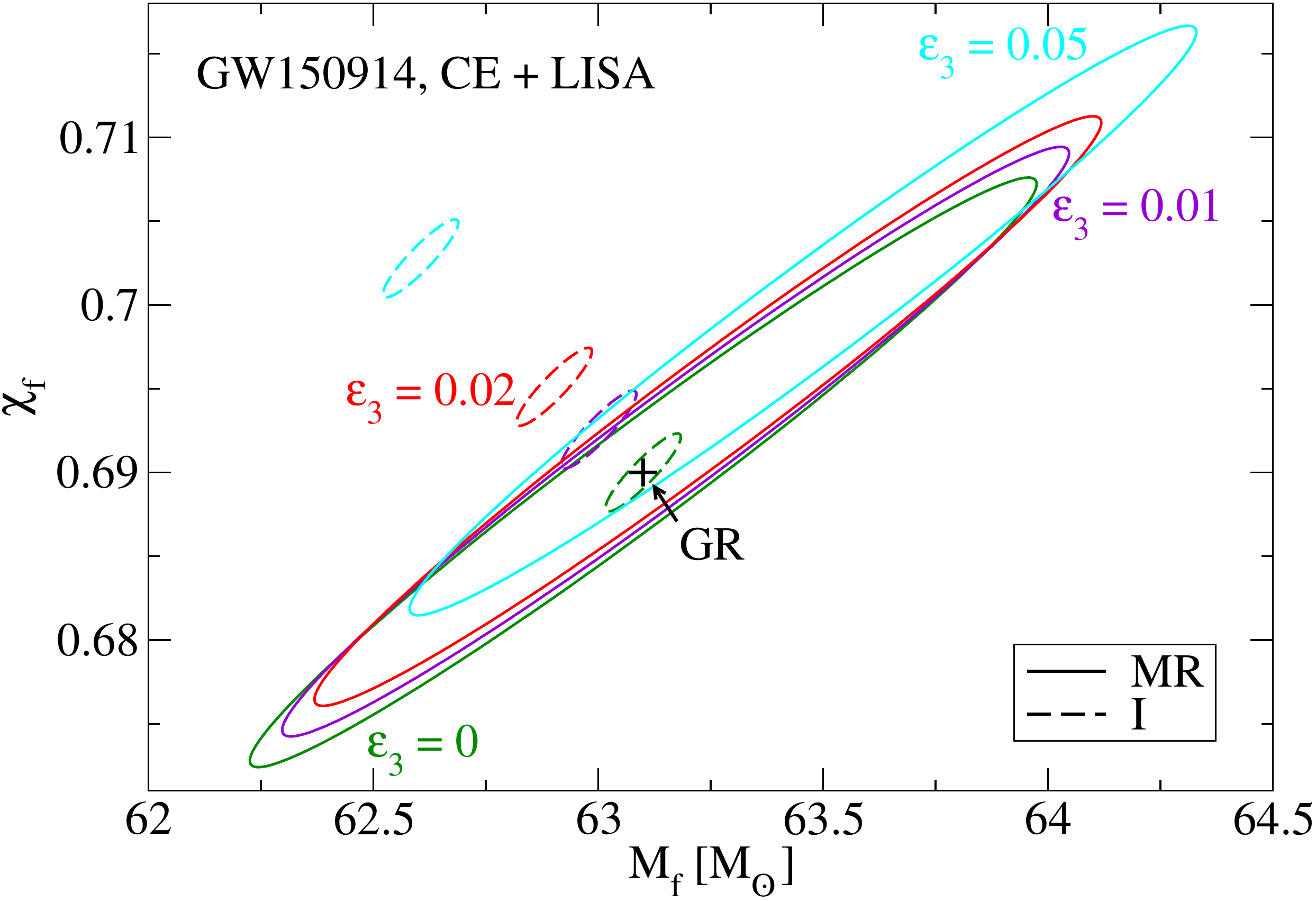}
\includegraphics[width=.45\textwidth]{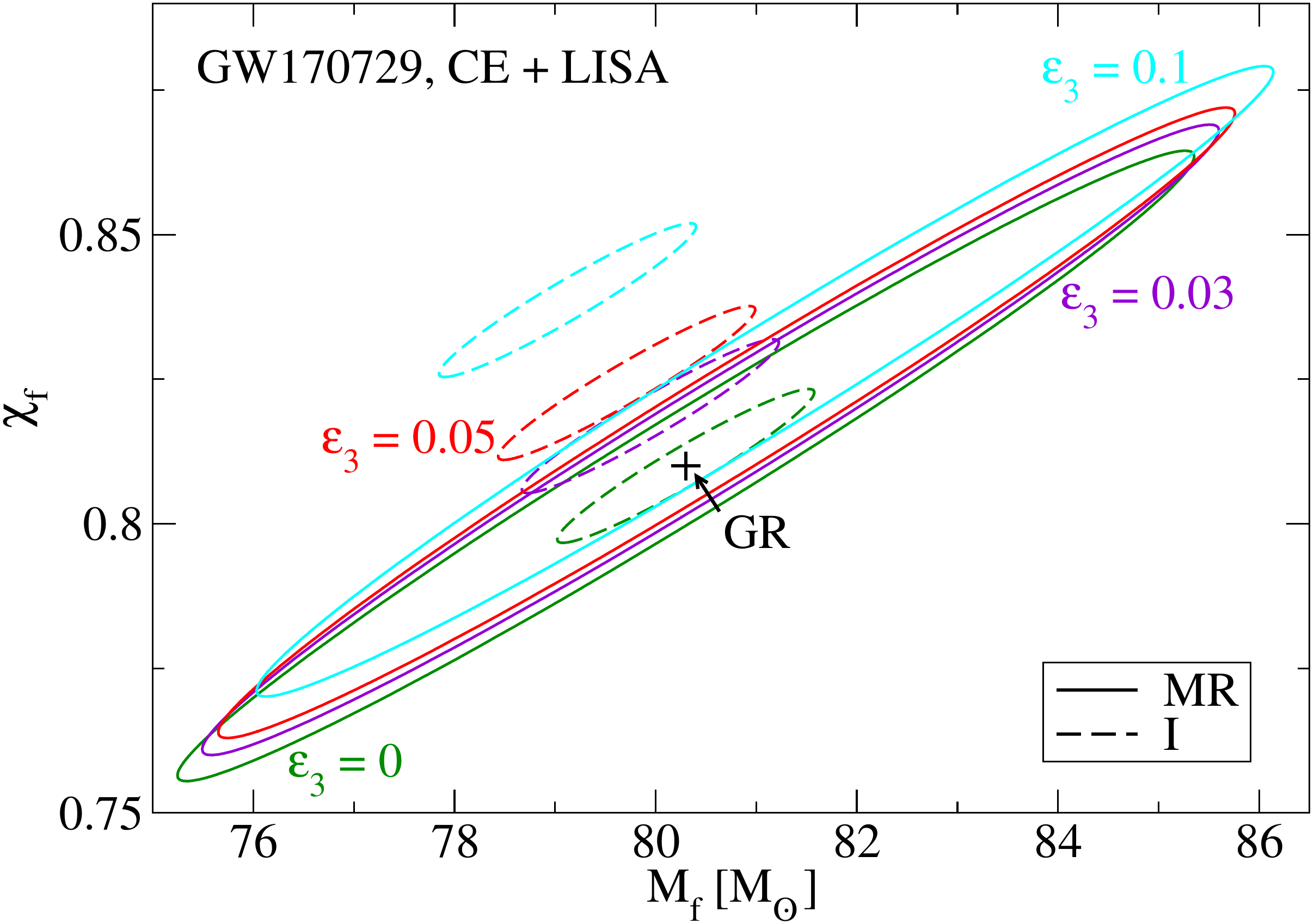}
\caption[IMR consistency tests in a JP spacetime]{IMR consistency test for the ``golden event'' GW150914 (left) and the massive event GW170729 (right) in the JP spacetime using the CE detector (top), and through the multiband observation between CE and LISA (bottom).
In particular, in each panel we plot the 90\% confidence regions in the $(M_f,\chi_f)$ plane as observed from (i) only the inspiral (I) signal, and (ii) only the merger-ringdown (MR) signal, for consecutively increasing values of the JP deviation parameter $\epsilon_3$.
Only when such probability distributions begin to disagree with each other can one decisively admit there may be evidence of beyond-Kerr spacetimes present.
}\label{fig:imrJP}
\end{center}
\end{figure*}

\begin{figure}[!htbp]
\begin{center}
\includegraphics[width=.7\columnwidth]{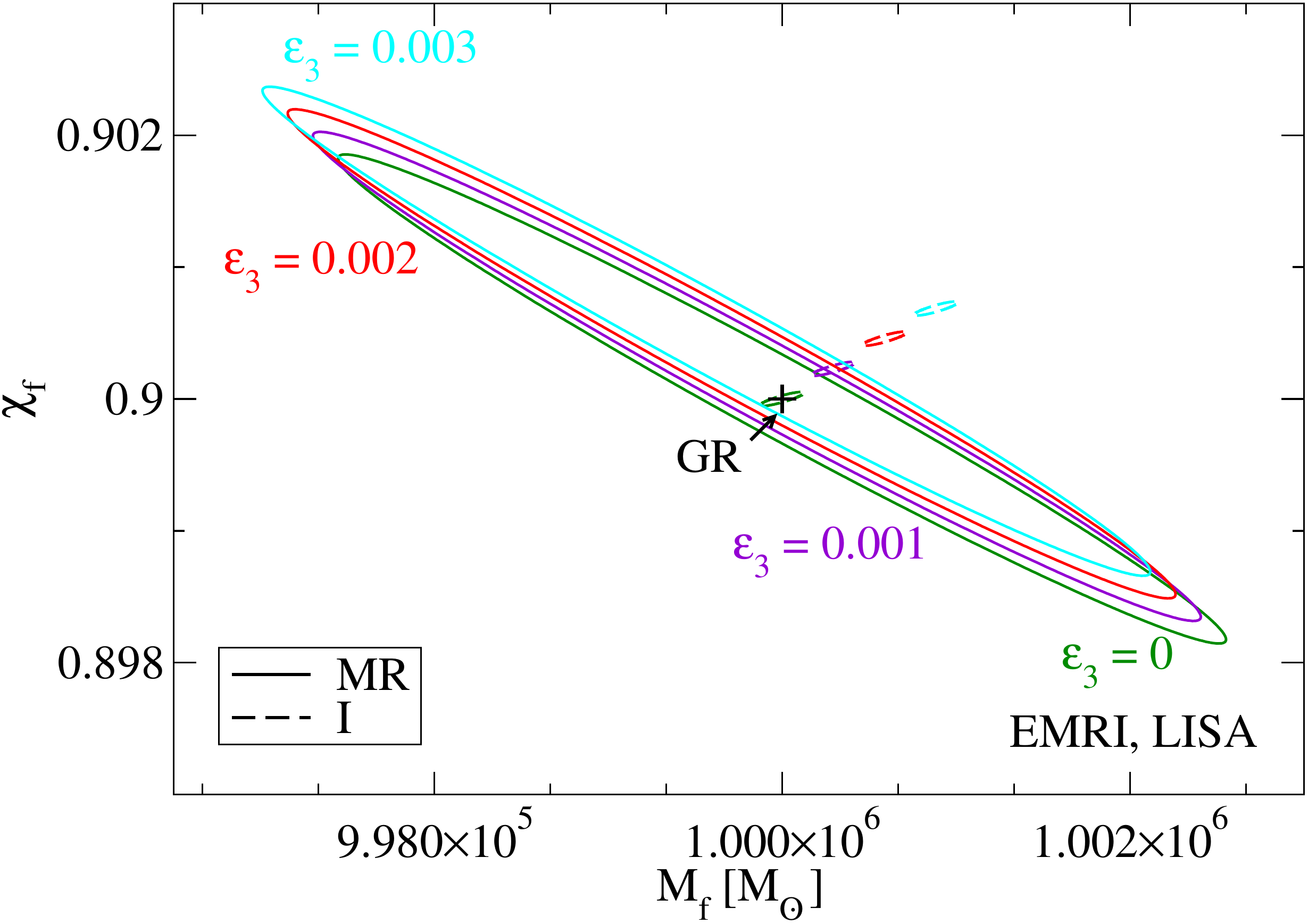}
\includegraphics[width=.7\columnwidth]{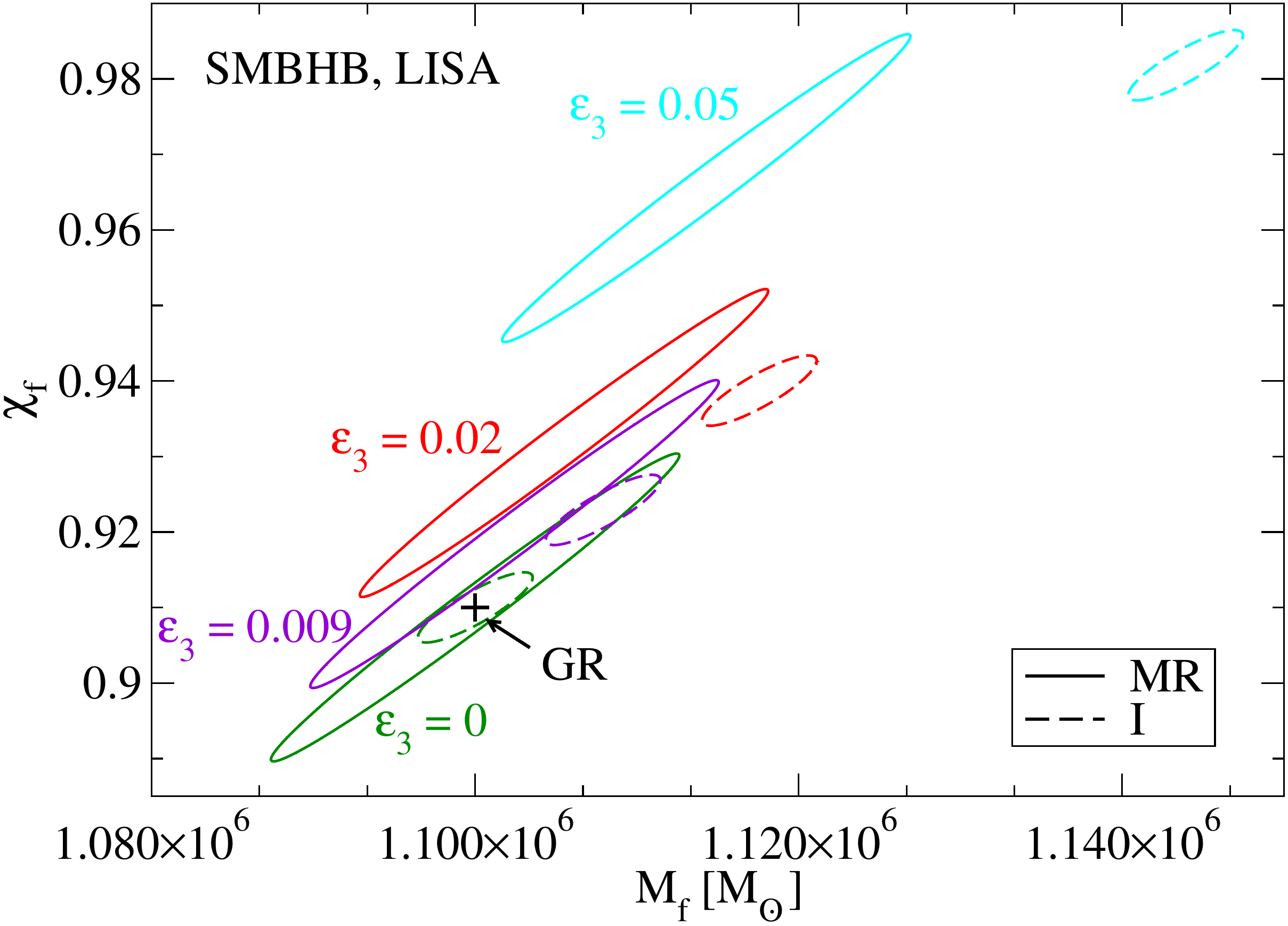}
\caption[EMRI and SMBHB IMR consistency tests in a JP spacetime]{Similar to Fig.~\ref{fig:imrJP}, but for the EMRI and SMBHB GW events.
}\label{fig:imrJPmassive}
\end{center}
\end{figure}

Let us begin by performing the IMR consistency test in the JP spacetime, to predict how well one can observe deviations from GR.
By following the procedure outlined in Chapter~\ref{chap:IMRconsistency}, we perform the IMR consistency test for several consecutively increasing values of the JP deviation parameter $\epsilon_3$ until the inspiral and merger-ringdown 90\% confidence interval probability distributions begin to disagree.
Only then can one provide evidence of non-Kerr behaviors in the gravitational signal.

We start with an investigation into the GW events already detected on the aLIGO O2 detector, namely GW150914 and GW170729.
We perform the IMR consistency test for several values of $\epsilon_3$ injected into the gravitational waveform with the aLIGO O2 GW detector.
We find that for GW150914-like (GW170729-like) events, when $\epsilon_3\approx7$ ($\epsilon_3\approx10$) the systematic uncertainties begin to overtake the statistical errors, and the I and MR contours begin to disagree. 
Such constraints on $\epsilon_3$ are on the same order of magnitude as those from x-ray observations presented in~\cite{Kong:2014wha,Bambi:2015ldr}.
On the other hand, for GW170729-like events we observe that, due to the large BH masses, there is large statistical and systematic uncertainty in the inspiral determination of remnant properties.
No such large errors are present in the merger-ringdown determination, but, similar to the GW150914-like case, we see the two contours begin to disagree at $\epsilon_3\approx10$.
However, they fail to satisfy the small-deviation assumption made in the derivation of ppE parameters, thus the resulting constraints are less valid than the following ones presented for future GW detectors.
As a result of this we do not present the resulting contours in this chapter, however the constraints are still tabulated in \hypertarget{link:results}{Tab.~\ref{tab:results}} for reference.

We next focus our attention on future observations of the same GW events GW150914 and GW170729.
We now consider such events as detected by the future CE detector, as well as the increased observation from the multiband observation between ground- and space-based detectors CE and LISA.
Figure~\ref{fig:imrJP} displays the results of the IMR consistency test in such cases.
For the CE case, we find that when $\epsilon_3\approx0.05$ and $\epsilon_3\approx0.6$, we can begin to distinguish the inspiral and merger-ringdown signals for GW150914- and GW170729-like events respectively.
For the multiband case, we find that while the inspiral has significantly smaller statistical uncertainties than the merger ringdown (due to the low-frequency space-based observations by LISA), its systematic uncertainties are much larger.
This allows one to constrain deviation parameters to $\epsilon_3\approx0.02$ and $\epsilon_3\approx0.05$ for GW150914- and GW170729-like events respectively.
Such constraints are about two-orders-of-magnitude stronger than existing bounds presented by~\cite{Kong:2014wha,Bambi:2015ldr}.

Finally, we consider the more extreme events detectable in the low frequency bands by LISA: EMRIs\footnote{Such EMRIs are not valid in the NR fits presented in the PhenomD waveform, which have been calibrated to NR simulations with mass ratios of only up to 1:18. To take this into account in the parameterized tests, we remove all NR fits from the gravitational waveform, and cut-off all frequency integrations before the merger-ringdown, at $f_\ISCO$. Namely, we use the TaylorF2 waveform in GR up to 3.5PN order included in the phase and introduce the ppE corrections to account for the inspiral corrections. Thus, such estimates are more conservative than the other ones presented in this chapter, as we only consider the inspiral portion of the waveform.} and SMBHBs.
Figure~\ref{fig:imrJPmassive} presents the resulting IMR consistency test for such two events involving massive BHs.
When considering EMRI systems, we find that the inspiral signal is very deterministic for the remnant BH properties, with such contours orders of magnitude smaller than their merger-ringdown counterparts.
With a majority of systematic uncertainties present in the inspiral signal, we find that we can constrain $\epsilon_2\approx6\times10^{-3}$ -- a few orders-of-magnitude stronger than those found in~\cite{Kong:2014wha,Bambi:2015ldr}.
These constraints are much stronger because in high mass-ratio inspirals, the quadrupole radiation is smaller, thus the orbit decays slower and the number of GW cycles is greatly increased compared to equal-mass systems, so the non-Kerr effect is significantly enhanced.
This can be seen by the factor of $\eta^{-4/5}$ present in the ppE phase parameter in Eq.~\eqref{eq:JPppe}, which is very large for large mass-ratio systems ($\sim 10^{4}$ for EMRIs, and only $\sim 3$ for i.e. GW150914).
However, as noted previously, we point out that such results are not as reliable due to the IMRPhenomD NR fits only being calibrated up to mass ratios of 1:18.
Finally, we see that for SMBHB events detected by LISA, we can constrain $\epsilon_3\approx0.02$, significantly weaker than those from EMRIs, and similar to those found by future GW170729 and GW150914 observations.

Finally, for comparison we perform a parameterized test of GR for the deviation parameter $\epsilon_3$.
To do so, we include $\epsilon_3$ into the template waveform with fiducial value of $0$, and perform a Fisher analysis to estimate root-mean-square uncertainties on $\epsilon_3$.
Such results are displayed in Tab.~\ref{tab:results} in comparison to all of the constraints found via the IMR consistency test as presented here.
We find that they give comparable bounds on $\epsilon_3$ for each case considered, even for EMRIs.
In this case, the IMRD consistency test is less valid as mentioned above, while in the parameterized test we used the TaylorF2 waveform with the ppE correction and stopped all integrations before the merger-ringdown.

\subsubsection{IMR consistency test in the modified $\Delta$ spacetime}\label{sec:IMRDMD}

\begin{figure*}
\begin{center}
\includegraphics[width=.45\textwidth]{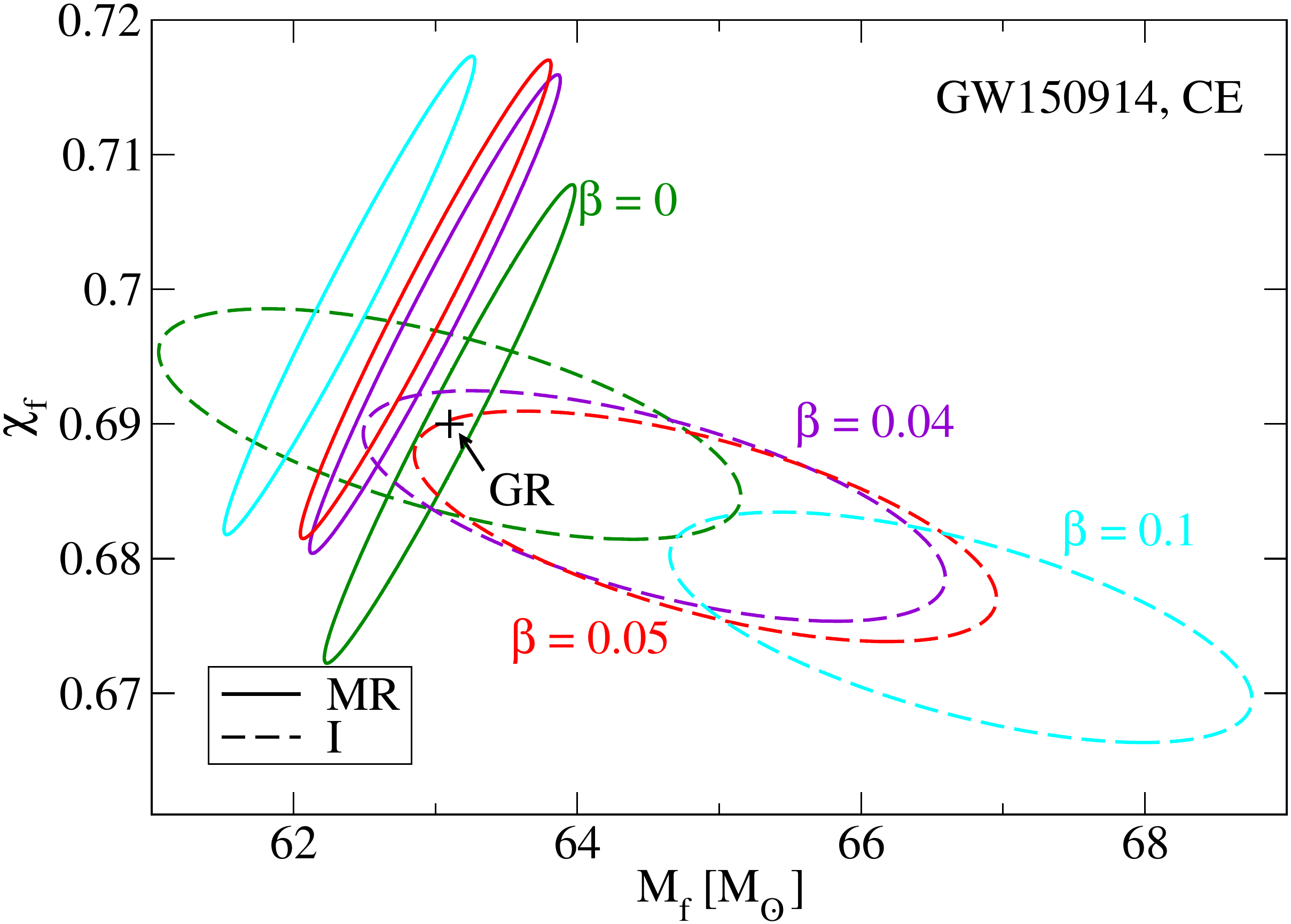}
\includegraphics[width=.45\textwidth]{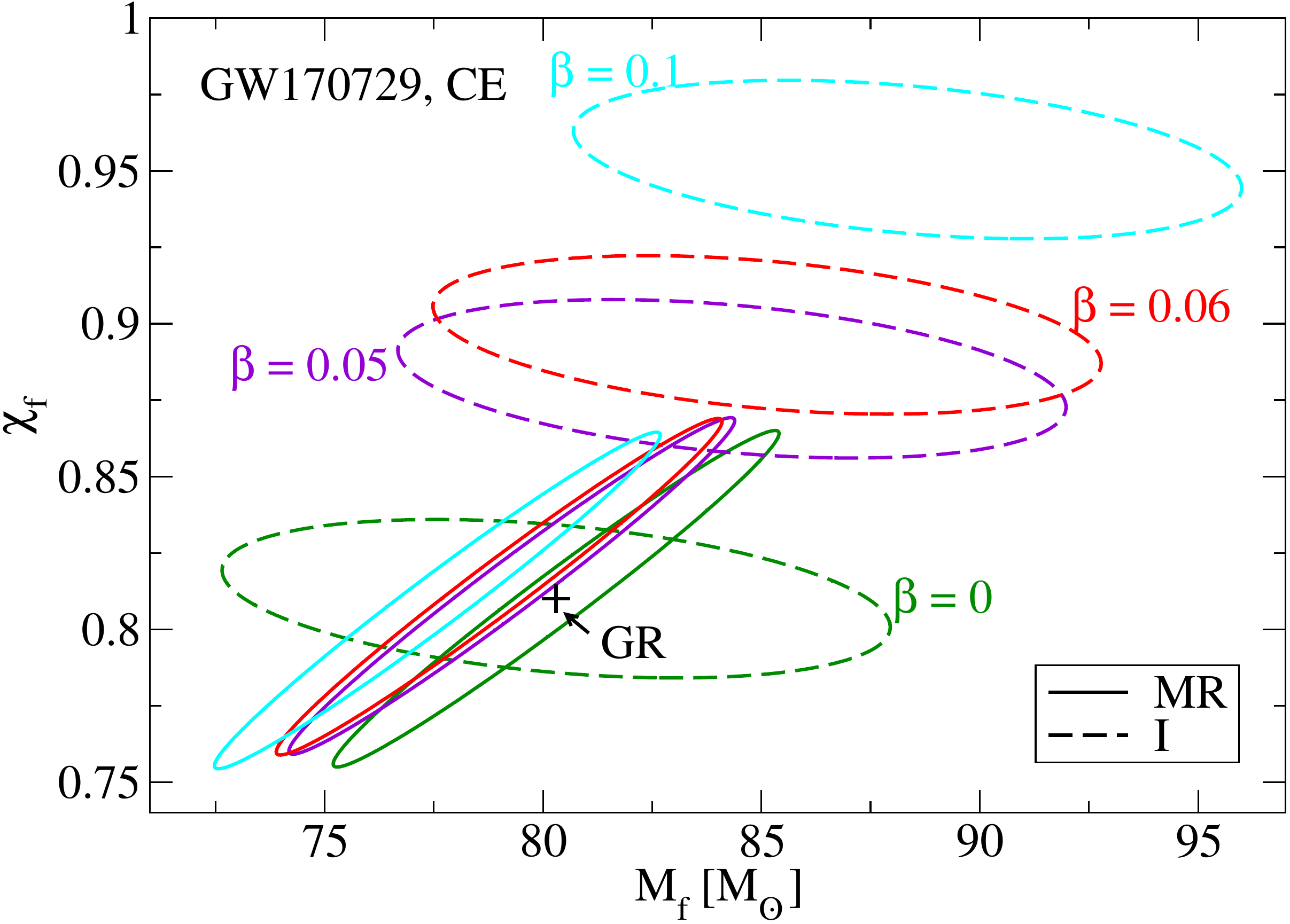}
\includegraphics[width=.45\textwidth]{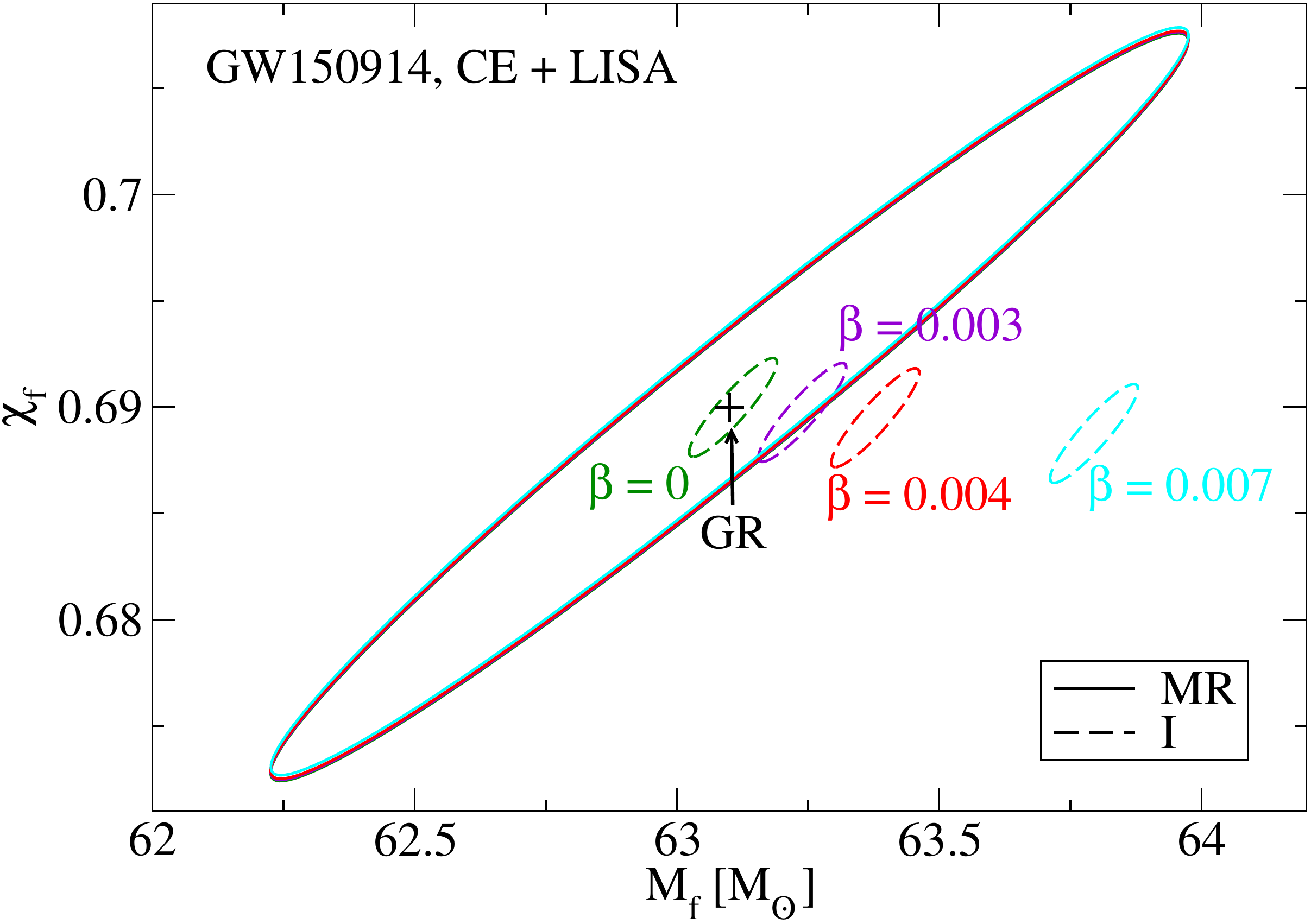}
\includegraphics[width=.45\textwidth]{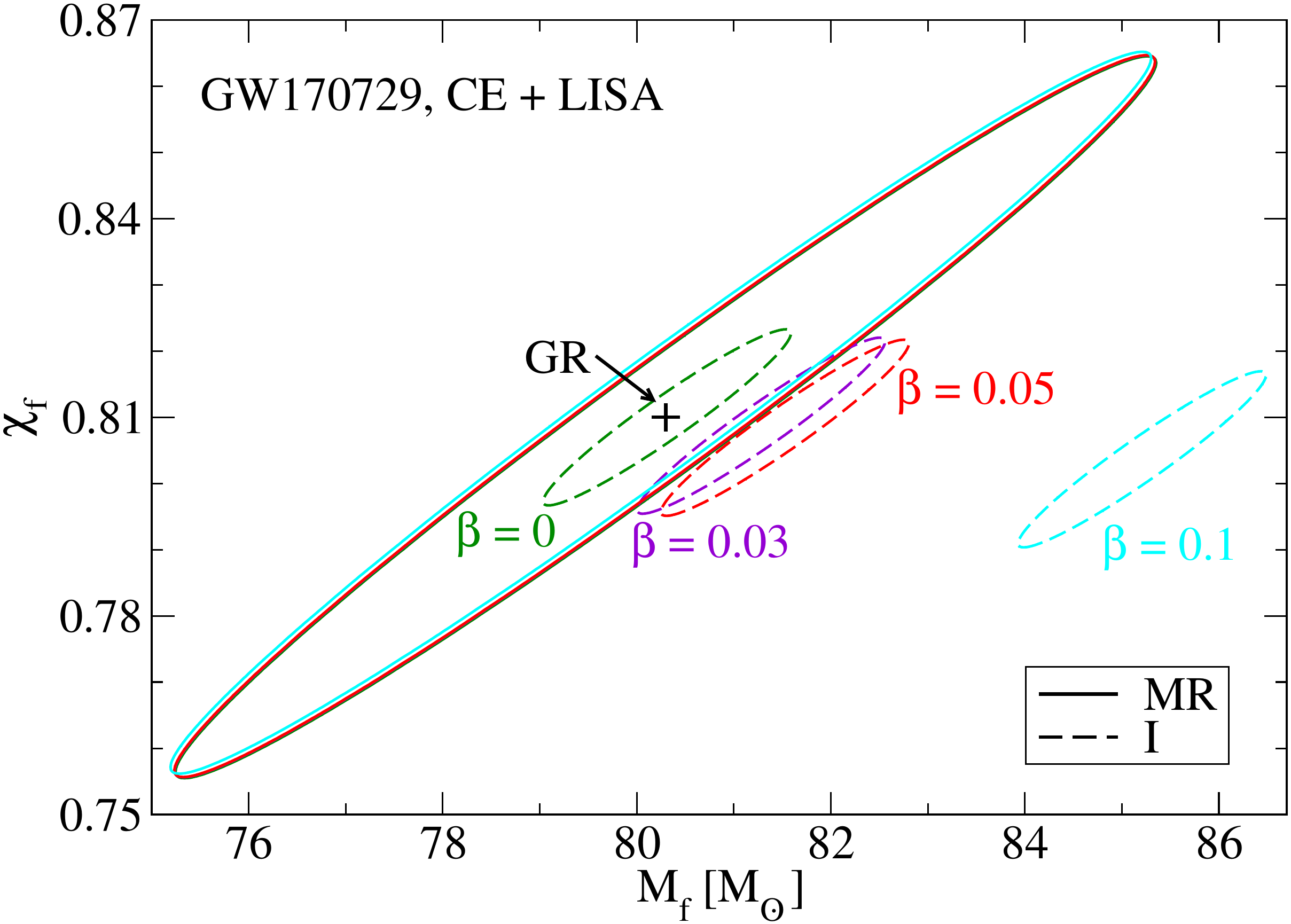}
\caption[IMR consistency tests in a mod.~$\Delta$ spacetime]{
Same as Fig.~\ref{fig:imrJP} but under the mod.~$\Delta$ spacetime instead, with $\beta$ being the beyond-Kerr deviation parameter.
}\label{fig:imrKB}
\end{center}
\end{figure*}

\begin{figure}
\begin{center}
\includegraphics[width=.7\columnwidth]{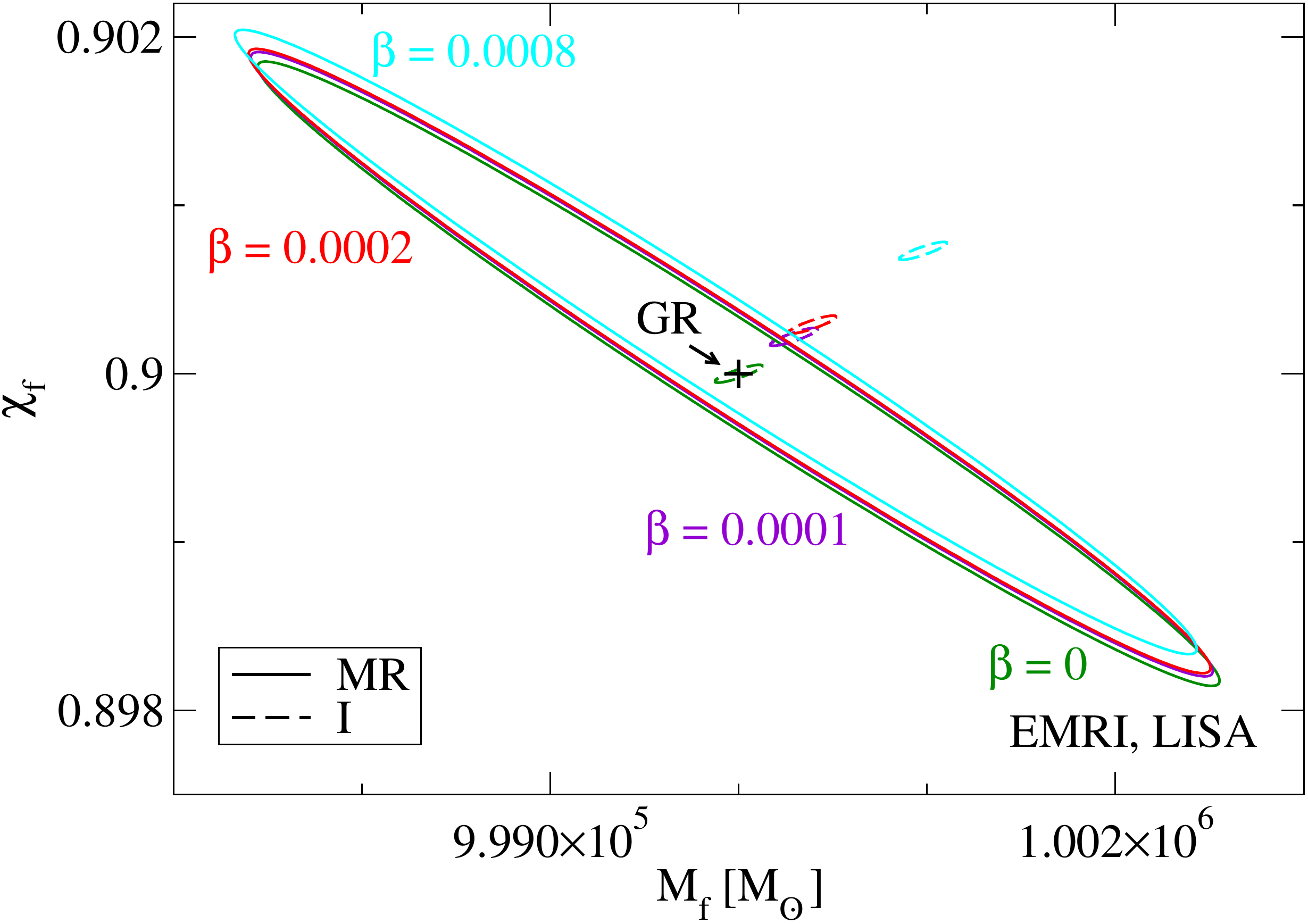}
\includegraphics[width=.7\columnwidth]{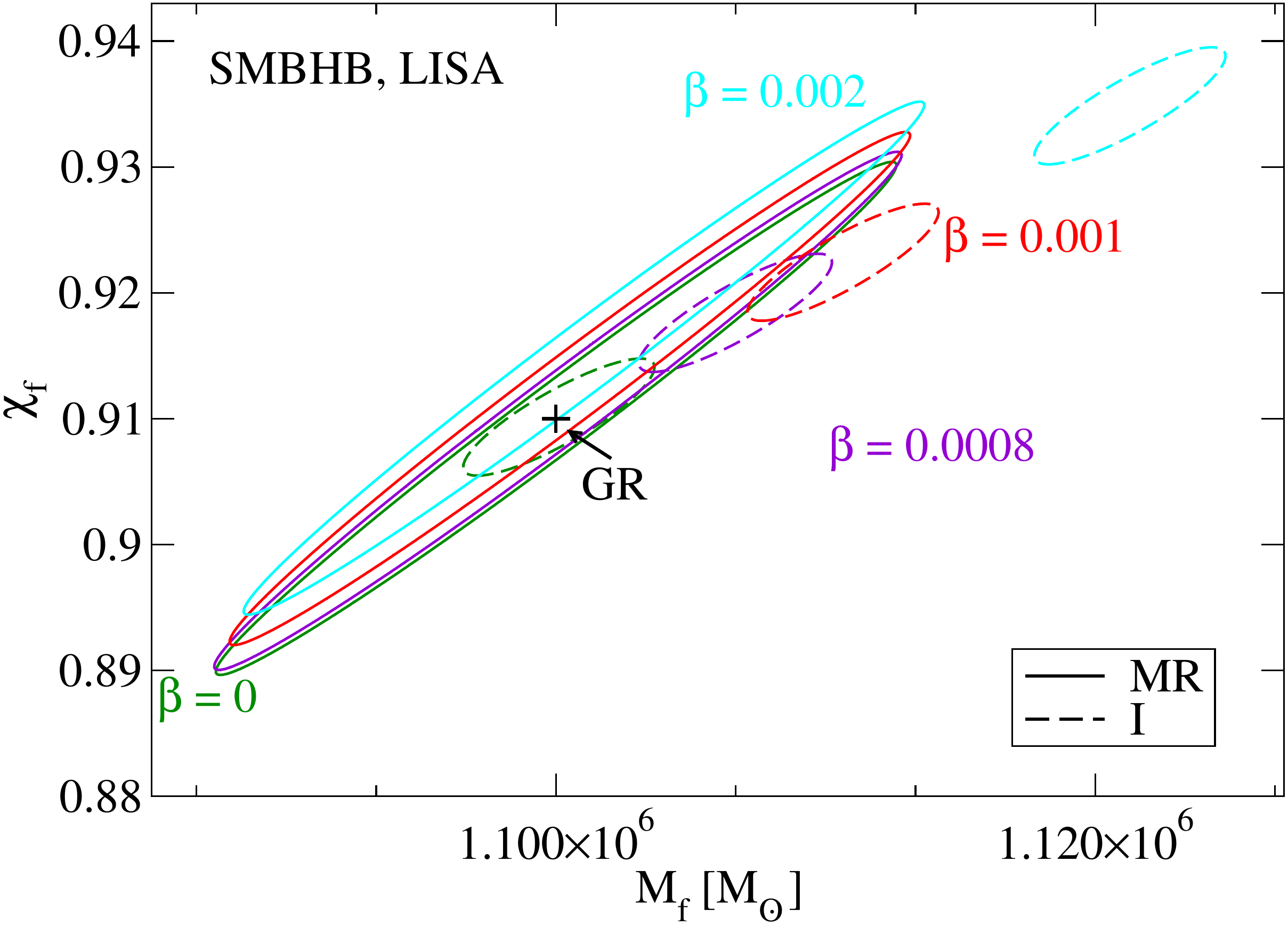}
\caption[EMRI and SMBHB IMR consistency tests in a mod.~$\Delta$ spacetime]{
Same as Fig.~\ref{fig:imrJPmassive} but within the mod.~$\Delta$ spacetime with deviation parameter $\beta$.
}\label{fig:imrKBmassive}
\end{center}
\end{figure}

Now we repeat the analysis performed in Sec.~\ref{sec:results} in the mod.~$\Delta$ spacetime.
Because the results here are very similar to those found in the preceding section, we only outline a brief overview here.

We begin by performing the IMR consistency test on GW150914-like and GW170729-like events observed on both CE, as well as with the multi-band observation between CE and LISA.
Figure~\ref{fig:imrKB} presents the resulting 90\% credible error ellipse in the $(M_f,\chi_f)$ plane for each case.
Similar to above in the JP spacetime, we observe that for O2, we can detect non-Kerr effects on GW150914-like events for $\beta\approx2$, and significantly higher at $\beta\approx14$  for GW170729-like events, due to the large inspiral uncertainty resulting from the large BH masses.
Such results are still less reliable than the following ones due to the large deviations, and we do not present the resulting contours, however the constraints are still tabulated in Tab.~\ref{tab:results}.
Following this, we see that when observed on future detector CE, GW150914-like events can resolve non-Kerr effects at a significantly smaller $\beta\approx0.05$, and a very similar value of $\beta\approx0.06$ for GW170729-like events.
Finally, we observe constraints of $\beta\approx5\times10^{-3}$ (GW150914-like) and $\beta\approx0.05$ (GW170729-like) for the multiband observations between CE and LISA.
The former strong constraint is a result of the small inspiral statistical uncertainties and large systematic uncertainties.

Following this, we repeat the IMR consistency test for LISA observations of EMRIs and SMBHBs.
Heeding the warning discussed previously in Sec.~\ref{sec:IMRJP} about the validity of EMRIs in the PhenomD waveform, we once again present these results in Fig.~\ref{fig:imrKBmassive}.
Once again, the inspiral statistical uncertainty on EMRI observations is minuscule, resulting in the strong constraint of $\beta\approx2\times10^{-4}$.
Finally, we observe inconsistencies between the inspiral and merger-ringdown signals in a SMBHB event at $\beta\approx10^{-3}$.
Finally, we note that in the mod.~$\Delta$ spacetime, typically the direction of systematic uncertainties in the $(M_f,\chi_f)$ plane are opposite to those in the JP spacetime. 
We found that this is primarily due to the different PN orders at which each spacetime alters the inspiral waveform at ($2$PN order in JP, $1$PN order in mod.~$\Delta$).
This effect is dominant among the corrections provided in this analysis, and serves to shift the direction of systematic uncertainties present in each spacetime.

In addition, we perform a set of parameterized tests of GR for each case considered here for comparison to the ones found with the IMR consistency test.
As in the JP case, we find that such bounds are comparable for each case discussed in this section.
As discussed in Sec.~\ref{sec:IMRJP}, the IMR consistency test is less valid due to the invalid use of the IMRPhenomD waveform, while such pieces were removed for the parameterized test.
We refer the reader to Table~\ref{tab:results} for a comprehensive display of all results found in this section.

\subsubsection{The effects of ringdown and remnant BH corrections to the waveform}\label{sec:comparison}
In this section, we investigate the effects of including ringdown and remnant BH corrections into the waveform.
In other words, how much does this change our results if only the inspiral corrections were included as is commonly done in parameterized tests?

\renewcommand{\arraystretch}{1.2}
\begin{table}
        \centering
        \begin{tabular}{c | c c}
        & Inspiral-corr. only & all corrections \\ 
        \hline
        GW150914 (CE) & $0.0586$ & $0.0509$ \\
        GW170729 (CE) & $0.464$ & $0.406$ \\
        SMBHB (LISA) & $0.0101$ & $0.0098$ \\
        \end{tabular}
        \caption[Comparison of inspiral vs. full waveform corrections]{Comparison between constraints on JP parameter $\epsilon_3$ when (left) only inspiral corrections to the waveform are included, and (right) when all of the inspiral, ringdown, and remnant BH property corrections are included for parameterized tests.
        Constraints for the GW150914- and GW170729-like events are assumed to be made with the third-generation detector CE for demonstration purposes, while the SMBH ones are assumed to have been observed with space-based detector LISA. Observe that additional corrections do not give significant contribution on bounding beyond-Kerr spacetimes with parameterized tests.
        }\label{tab:comparison}
\end{table}

We begin by performing parameterized tests in two separate cases: (i) with only ppE inspiral corrections present within the whole inspiral-merger-ringdown waveform, and (ii) with inspiral, ringdown, and remnant BH property corrections present in the waveform, as was done in the main analysis.
For demonstration purposes, we choose the third-generation detector CE observing GW150914-like events, GW170729-like events, and then space-based detector LISA observing a SMBHB event as considered in the main analysis.
See Table~\ref{tab:comparison} for a summary of obtained results in each case.
We see that for the smaller-mass events GW150914 and GW170729, the two cases differ by up to $\sim15$\%.
As expected, the large-mass SMBHB event observed by LISA only differs by $\sim3$\% due to the small-frequency window available to space-based detectors, where the inspiral corrections make the largest difference.
We conclude that such additional corrections to the ringdown and remnant BH properties in the waveform do not have significant contribution on constraining beyond-Kerr spacetimes with parameterized tests.

In fact, a similar feature can be seen for the IMR consistency tests. Since the systematic error in the merger-ringdown portion is typically smaller than that of the inspiral, even if we do not include corrections to the ringdown and final BH's mass and spin, we would still find bounds that are comparable to those presented in Table~\ref{tab:results}. 
These findings give us supporting evidence that in many cases, the dominant contribution comes from the corrections to the inspiral, as considered e.g. in~\cite{Yunes_ModifiedPhysics}.


\subsection{Astrophysical implications in the new metric}\label{sec:astrophysical}
In this section we present the various astrophysical implications emergent under a BH described by the new metric preserving Kerr symmetries presented in this analysis.
Specifically, we derive expressions for the various astrophysical observables one might consider about a such a BH.
Such properties include the Keplerian and epicyclic frequencies of orbiting particles, the orbital energy and angular momentum of particle orbits, the location of the ISCO, and finally the photon orbits.


\subsubsection{Keplerian and epicyclic frequencies}\label{sec:frequencies}

Now let us describe the computation of the Keplerian and epicyclic frequencies $\nu_\phi$, $\nu_r$, and $\nu_\theta$.
The former frequency describes a particle's motion in the polar direction as observed at null infinity, while the latter two describe the motion in the azimuthal and radial directions for perturbed orbits.

We begin by finding the Keplerian frequency $\nu_\phi=\Omega_\phi/2\pi $.
We start with the definition of $\Omega_\phi\equiv\dot{\phi}/\dot{t}$, which can be determined from the geodesic equations
\begin{equation}
\frac{d^2x^\alpha}{d\tau^2}=-\Gamma^\alpha_{\beta\gamma}\frac{dx^\beta}{d\tau}\frac{dx^\alpha}{d\tau},
\end{equation}
with Christoffel symbols $\Gamma^\alpha_{\beta\gamma}$ and proper time $\tau$.
Following Ref.~\cite{Johannsen:2015pca}, axi-symmetry and reflection symmetry of particles on circular equatorial orbits allow us to reduce this relation to
\begin{equation}
\partial_r  g_{rr} \dot{t}^2+2\partial_r g_{t\phi}\dot{t}\dot{\phi}+\partial_r g_{\phi\phi}\dot{\phi}^2=0.
\end{equation}
The above expression allows us to express the Keplerian and frequency as
\begin{equation}
\Omega_\phi=\frac{-\partial_r g_{t\phi}\pm\sqrt{(\partial_r g_{t\phi})^2-\partial_r g_{tt}\partial_r g_{\phi\phi}}}{\partial_r g_{\phi\phi}}.\label{eq:KeplerianFrequency}
\end{equation}
We see that this expression only depends on the non-Kerr deviation functions $A_1(r)$, $A_2(r)$, $A_0(4)$, and $f(r)$.

Next we obtain expressions for the vertical and radial epicyclic frequencies describing the radial and polar motion of orbiting particles with mass $\mu$.
Following the derivation presented in Sec.~4 and~5 of~\cite{Abramowicz:2004tm}, the general epicyclic frequencies  observed with respect to the proper time of a comoving observer in the $X$-direction are given by
\begin{equation}
\omega_\X=\sqrt{\frac{\partial^2 V_\text{eff}}{\partial X^2}},
\end{equation} 
for effective potential
\begin{equation}
V_\text{eff}=-\frac{1}{2}(g^{tt} E^2-2g^{t\phi} E L_z+g^{\phi\phi} L_z^2+\mu^2).\label{eq:Ueff}
\end{equation}
The resulting radial and vertical epicyclic frequencies observed at null infinity are found to be
\begin{align}
\Omega_r&=\sqrt{\frac{(g_{tt}+\Omega_\phi g_{t\phi})^2}{2g_{rr}}\left( \partial^2_r g^{tt} - 2 L_z \partial^2_r g^{t\phi}+L_z^2 \partial^2_r g^{\phi\phi} \right)},\\
\Omega_\theta&=\sqrt{\frac{(g_{tt}+\Omega_\phi g_{t\phi})^2}{2g_{\theta\theta}}\left( \partial^2_\theta g^{tt} - 2 L_z \partial^2_\theta g^{t\phi}+L_z^2 \partial^2_\theta g^{\phi\phi} \right)}.
\end{align}
We see such epicyclic frequencies depend on the non-Kerr deviation functions $A_1(r)$, $A_2(r)$, $A_0(r)$, $A_5(r)$, and $f(r)$, while the vertical frequencies depend on all but $A_5(r)$.

Finally, we plot the Keplerian and epicyclic frequencies $\nu_\X\equiv\Omega_\X/2\pi$ for various combinations of lower-order deviation paramters.
Here take note that certain combinations of deviation parameters produce naked singularities outside of the BH event horizon, as discussed further in App.~\ref{app:nakedSingularities}.
Such exotic singularities originate from disallowed combinations of parameters $A_1(r)$, $A_2(r)$, and $A_0(r)$, due to their unique relationships as seen in Eq.~\eqref{eq:CYcontravariant}.
Namely, we find that if $\alpha_{13}\ne0$ or $\alpha_{22}\ne0$, $\alpha_{02}$ must additionally be non-vanishing and of the same sign, else the photon orbit energies and angular momentum become discontinuous and negative, and photon orbits become open, letting photons escape to null infinity as discussed in~\cite{Hioki:2009na,Papnoi:2014aaa}.
Here we vary only the lowest-order non-vanishing parameters present in the given expressions $\alpha_{13}$, $\alpha_{22}$, $\alpha_{02}$, and $\epsilon_3$ for the Keplerian and vertical epicyclic frequencies, and also $\alpha_{52}$ for the radial epicyclic frequency.
In each case, all other non-Kerr deviation parameters that are not specifically mentioned are set to be 0.
For a further analysis on the further-lower-order parameters assumed to vanish here, see App.~\ref{app:lowerOrder}.
In Fig.~\ref{fig:nuPhi} we plot the Keplerian frequencies $\nu_\phi$, while in Figs.~\ref{fig:nuTheta} and~\ref{fig:nuR} the vertical and radial epicyclic frequencies $\nu_\theta$ and $\nu_r$ are plotted for various non-vanishing parameters.
We observe that, in general, the parameters $\epsilon_3$ and $\alpha_{52}$ introduce very little change into the frequencies $\nu_\X$, while combinations of $\alpha_{02}$ and $\alpha_{13}$, $\alpha_{22}$ have the power to significantly alter the ensuing trajectories.
Observe how the frequencies (especially epicyclic ones) can deviate significantly from the Kerr case when varying the new parameter $\alpha_{02}$ introduced for the first time in this chapter.


\subsubsection{Energy and angular momentum}\label{sec:EandLz}

In this section, we present the expressions for the orbital energy $E_\text{orb}$ and angular momentum $L_z$ of a particle orbiting a BH described by the new metric.
We begin with the effective potential $V_\text{eff}$ given in Eq.~\eqref{eq:Ueff}.
In the equatorial plane, circular orbits obey the expressions $V_\text{eff}(r)=\frac{d V_\text{eff}(r)}{dr}=0$.
When combined with the Keplerian frequency in Eq.~\eqref{eq:KeplerianFrequency}, and the 
Keplerian frequency written in terms of constants of motion
\begin{equation}
\Omega_\phi=\frac{p_\phi}{p_t}=-\frac{g_{t\phi} E_\text{orb}+g_{tt} L_z}{g_{\phi\phi} E_\text{orb}+g_{t\phi} L_z},
\end{equation}
we obtain expressions for the energy and angular momentum of a particle orbiting our central BH
\begin{align}
E_\text{orb}&=-\mu\frac{g_{tt}+g_{t\phi}\Omega_\phi}{\sqrt{-g_{tt}-2g_{t\phi}\Omega_\phi-g_{\phi\phi}\Omega_\phi^2}},\label{eq:energy}\\
L_z&=-\mu\frac{g_{t\phi}+g_{\phi\phi}\Omega_\phi}{\sqrt{-g_{tt}-2g_{t\phi}\Omega_\phi-g_{\phi\phi}\Omega_\phi^2}}.\label{eq:angMom}
\end{align}
We see that these expressions only depend on the non-Kerr deviation functions $A_1(r)$, $A_2(r)$, $A_0(4)$, and $f(r)$.

\begin{figure*}[htb]
\begin{center}
\includegraphics[width=.45\textwidth]{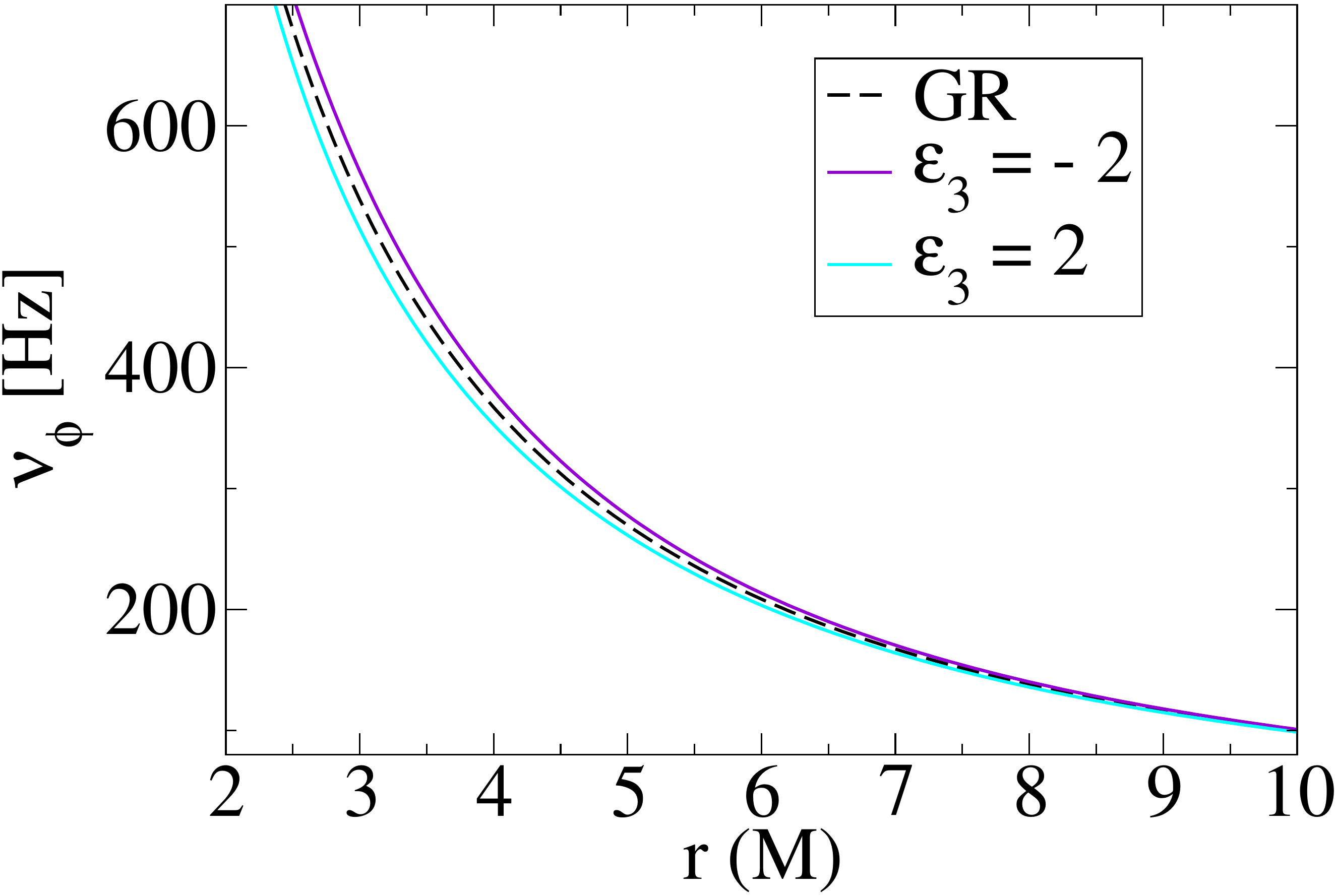}
\includegraphics[width=.45\textwidth]{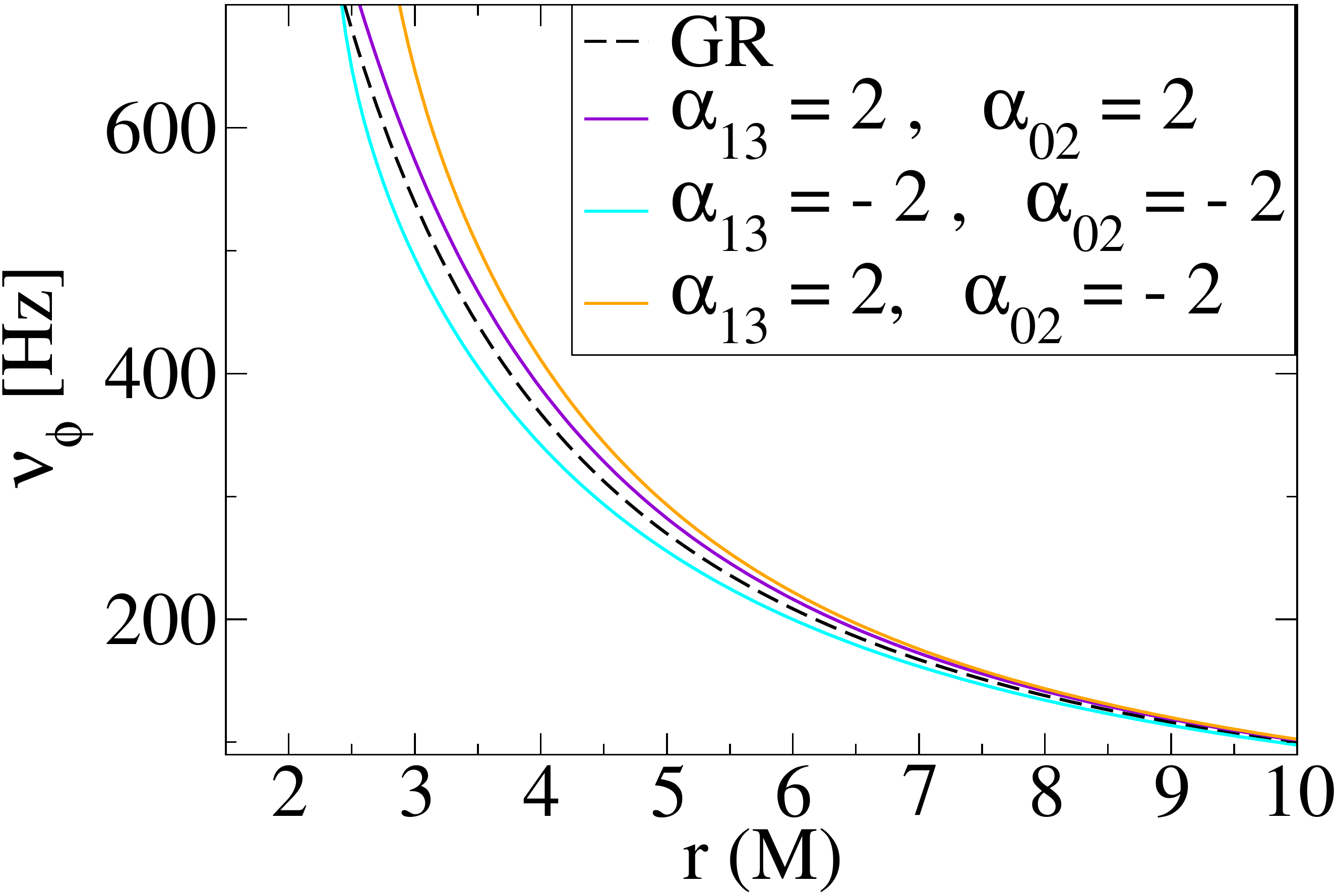}
\includegraphics[width=.45\textwidth]{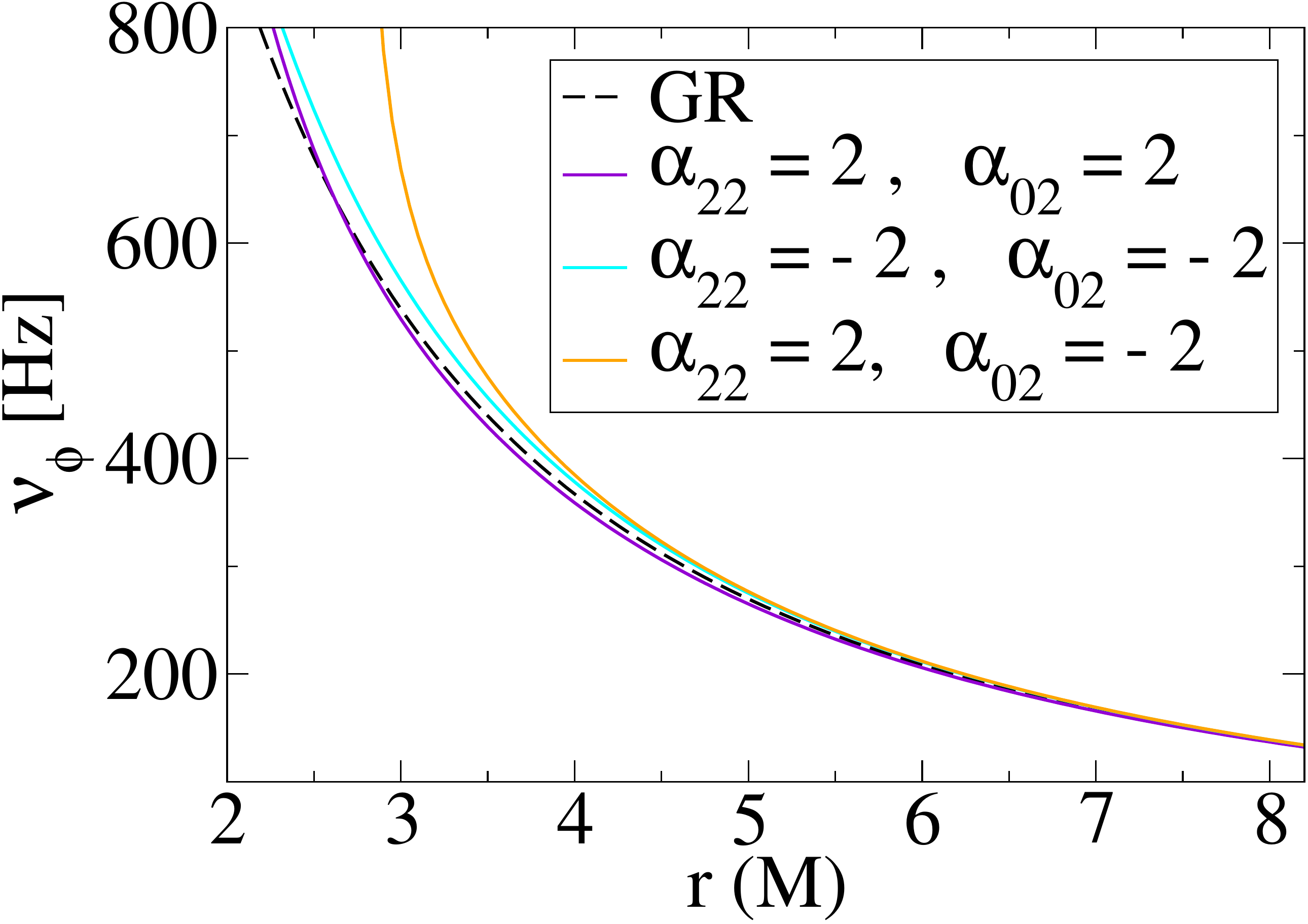}
\caption[Keplerian frequencies]{Keplerian frequencies of a particle orbiting a $10\text{ M}_\odot$ BH with $a=0.8M$ on circular equatorial orbits for various values of the non-Kerr deviation parameters, while setting all remaining parameters to 0.
The frequencies are plotted for varying the $f(r)$ lowest-order parameter $\epsilon_3$ (left), varying the $A_1(r)$ and $A_0(r)$ lowest-order parameters $\alpha_{13}$ and $\alpha_{02}$ (center), and varying the $A_2(r)$ and $A_0(r)$ lowest-order parameters $\alpha_{22}$ and $\alpha_{02}$ (right).
Several cases with $\alpha_{22}=0$ while $\alpha_{02} \ne 0$ or $\alpha_{12}=0$ while $\alpha_{02} \ne 0$, or vice versa, produce BHs with naked singularities which are not shown here. The left-most plot agrees with that in Fig.~3 of~\cite{Johannsen:2015pca}.
}\label{fig:nuPhi}
\end{center}
\end{figure*}

\begin{figure*}[htb]
\begin{center}
\includegraphics[width=.45\textwidth]{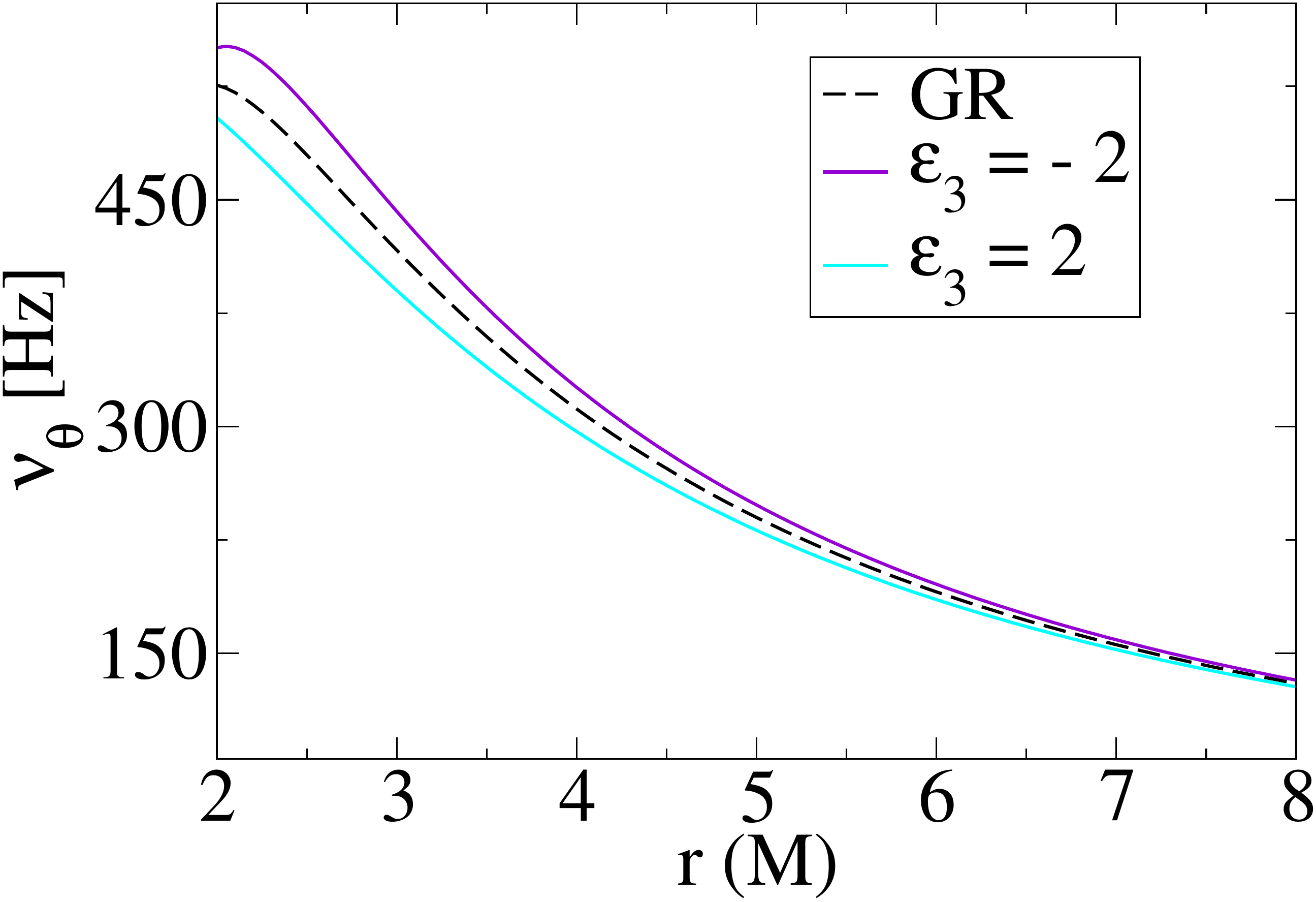}
\includegraphics[width=.45\textwidth]{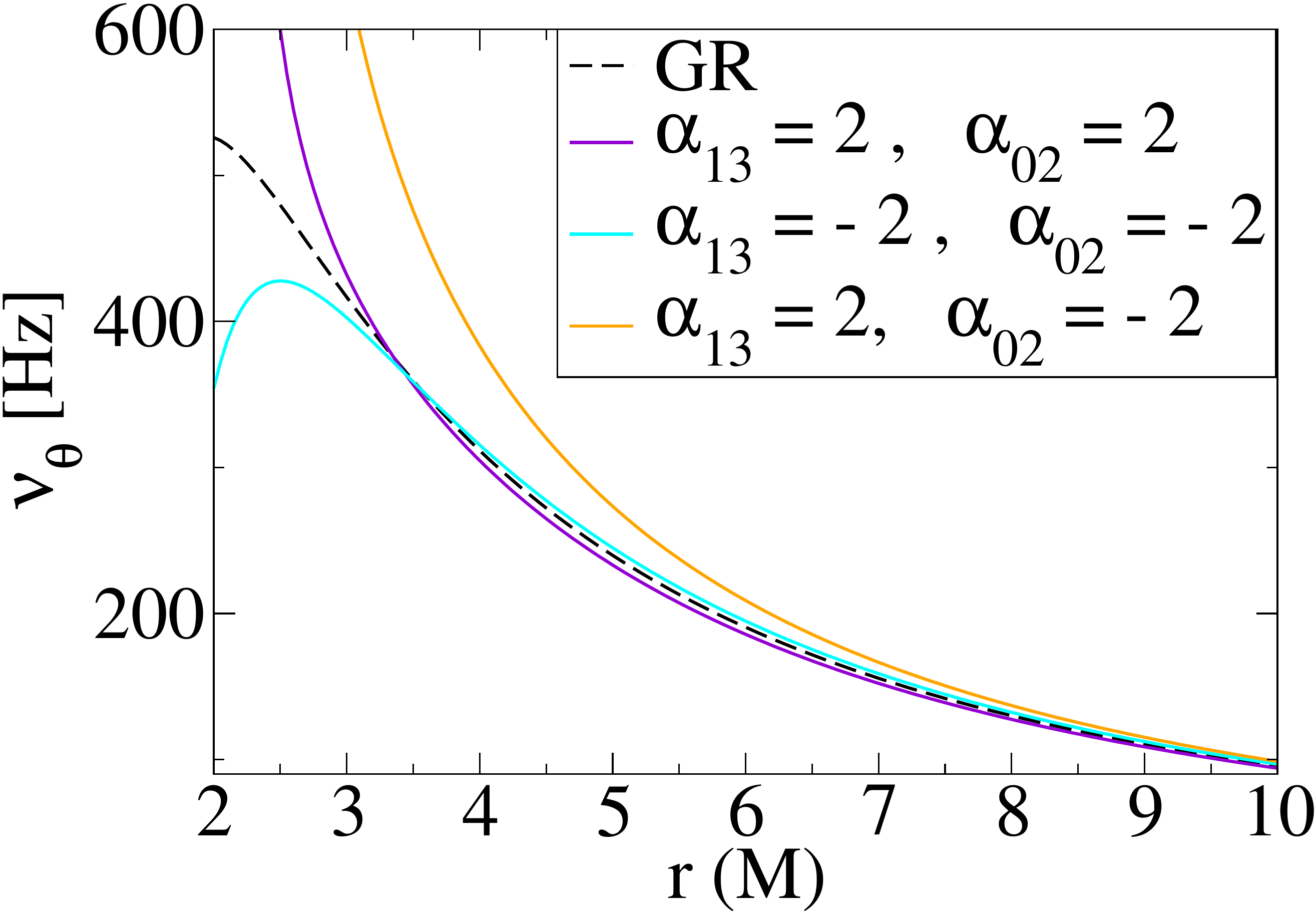}
\includegraphics[width=.45\textwidth]{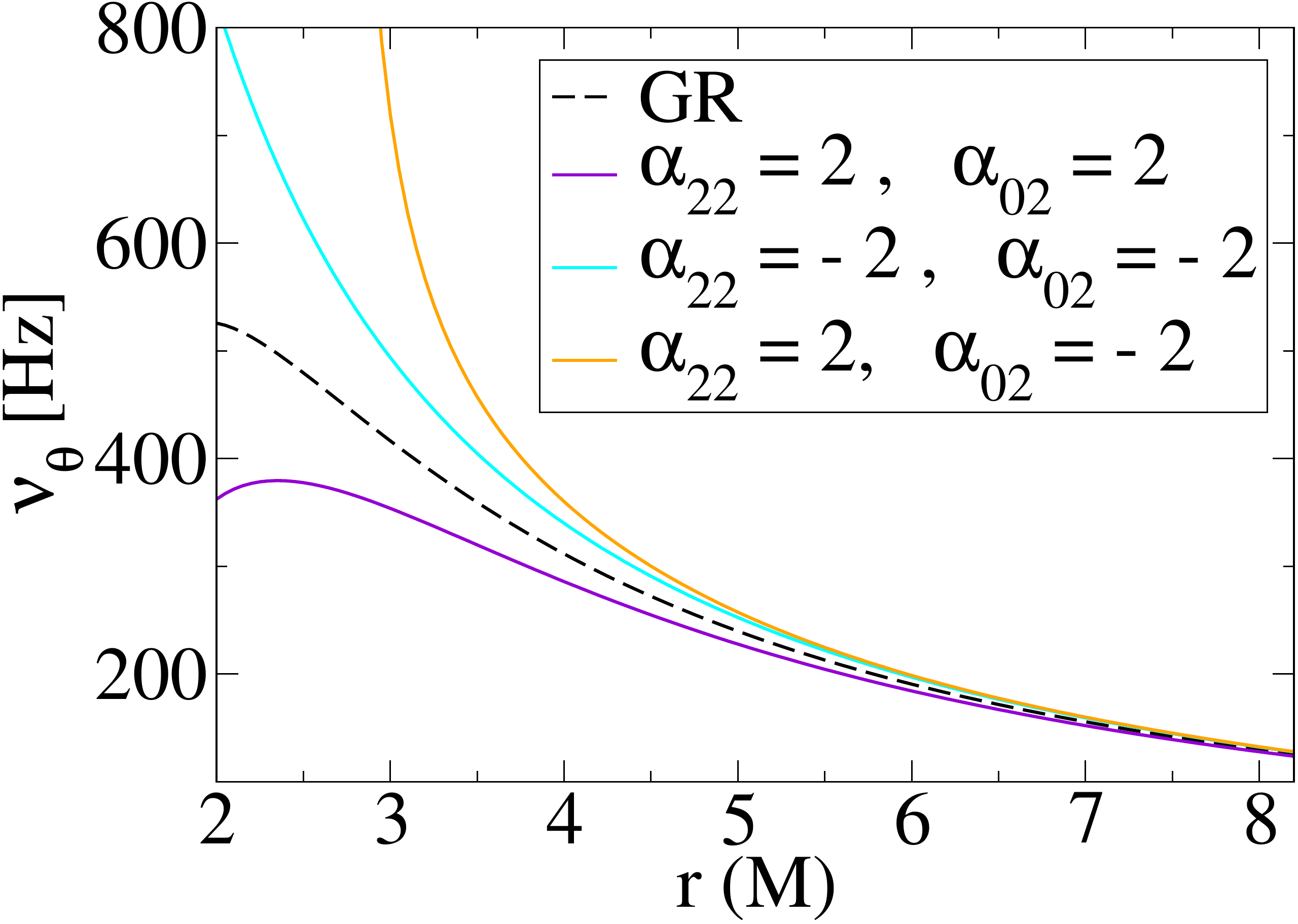}
\caption[Vertical epicyclic frequencies]{Similar to Fig.~\ref{fig:nuPhi} but for the vertical epicyclic frequency $\nu_{\theta}$.
}\label{fig:nuTheta}
\end{center}
\end{figure*}

\begin{figure*}[htb]
\begin{center}
\includegraphics[width=.45\textwidth]{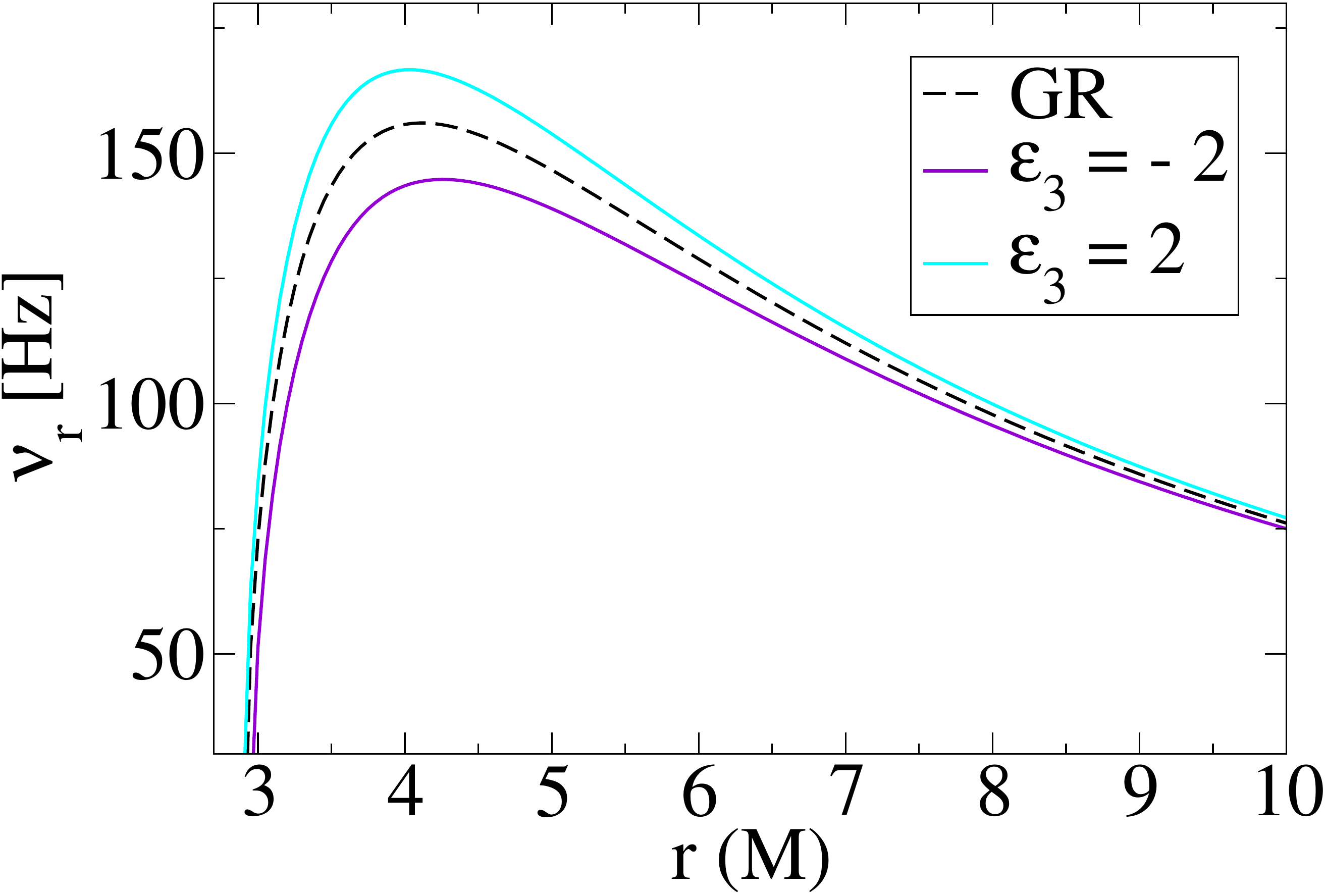}
\includegraphics[width=.45\textwidth]{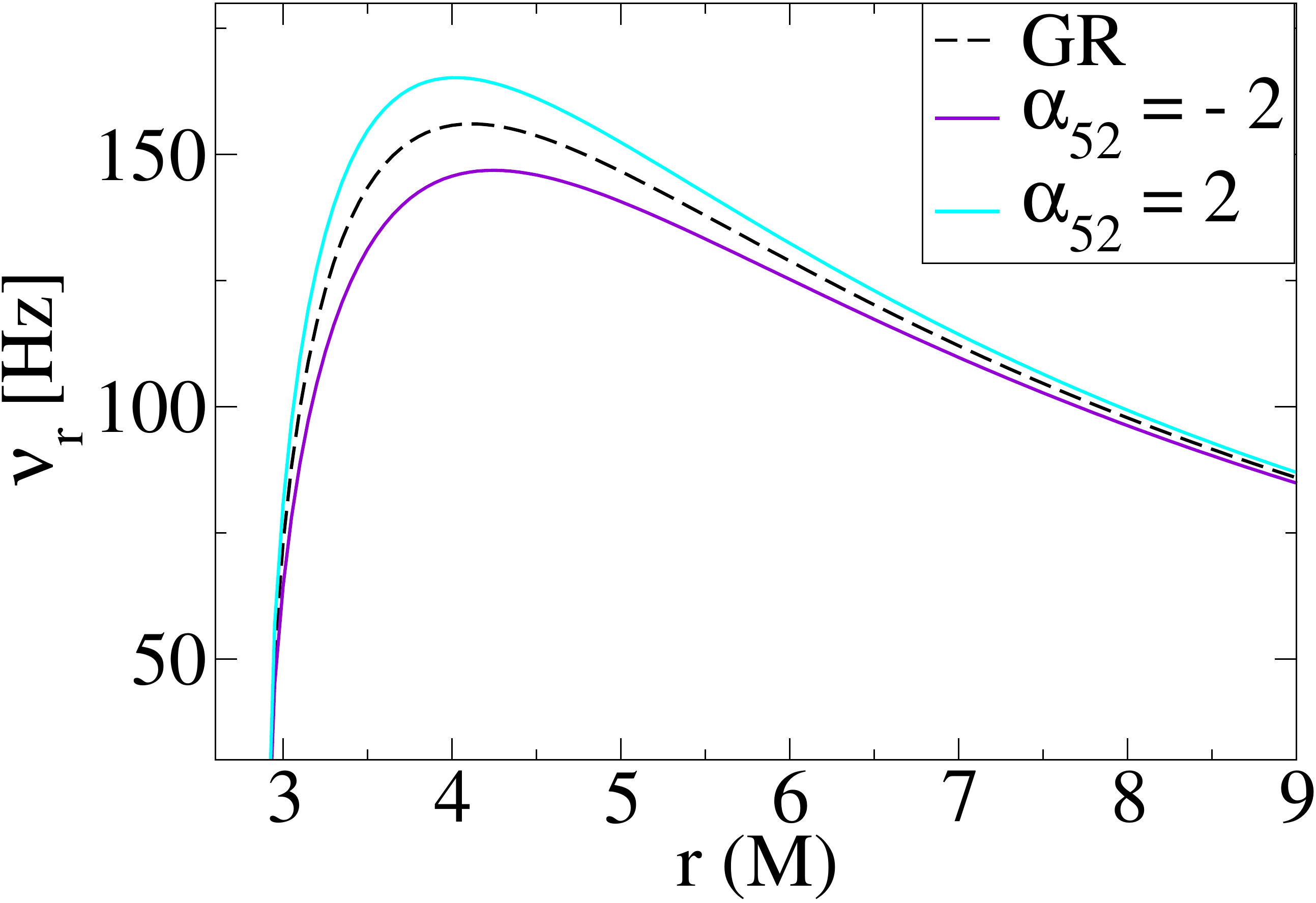}\\
\includegraphics[width=.45\textwidth]{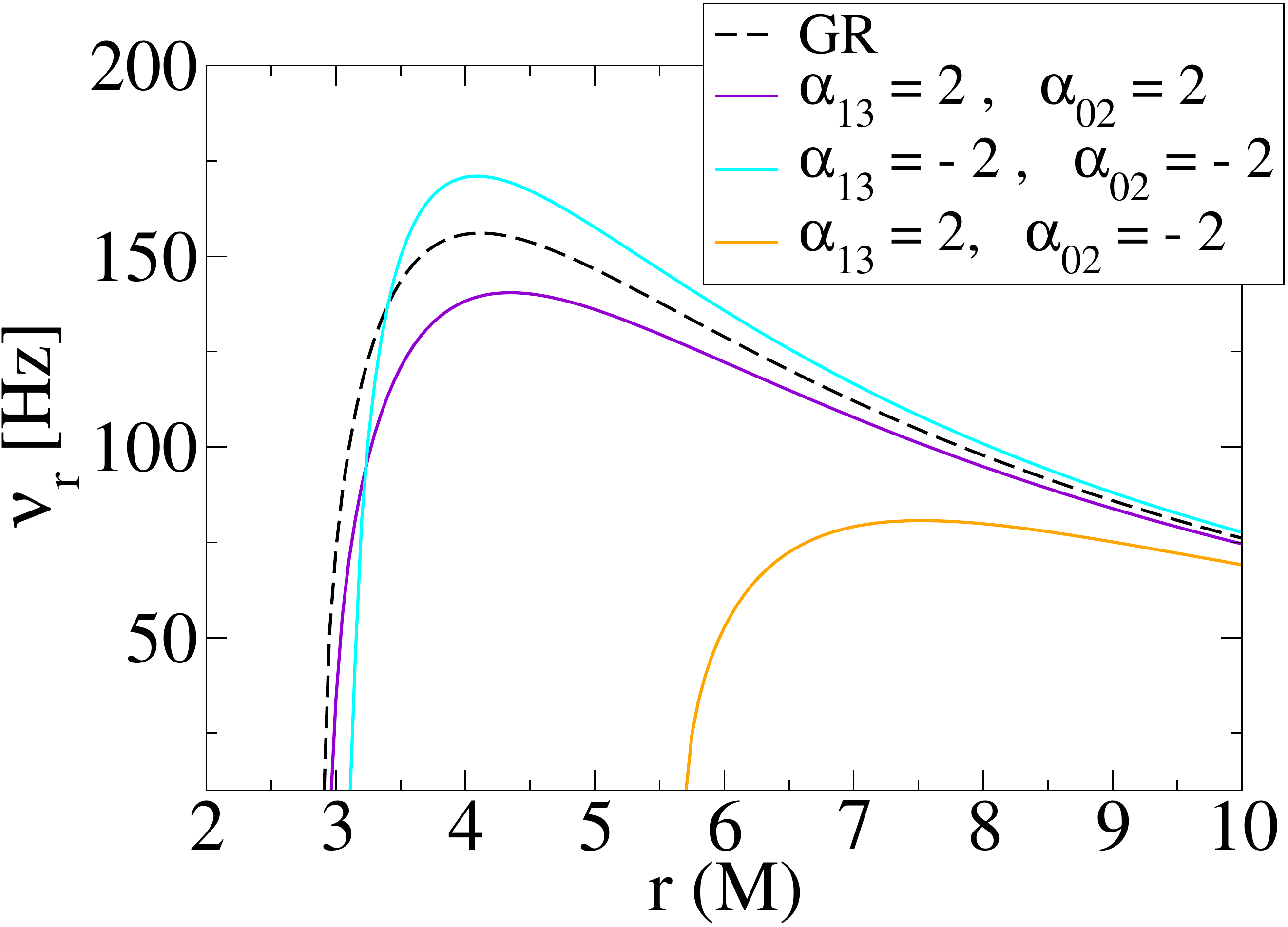}
\includegraphics[width=.45\textwidth]{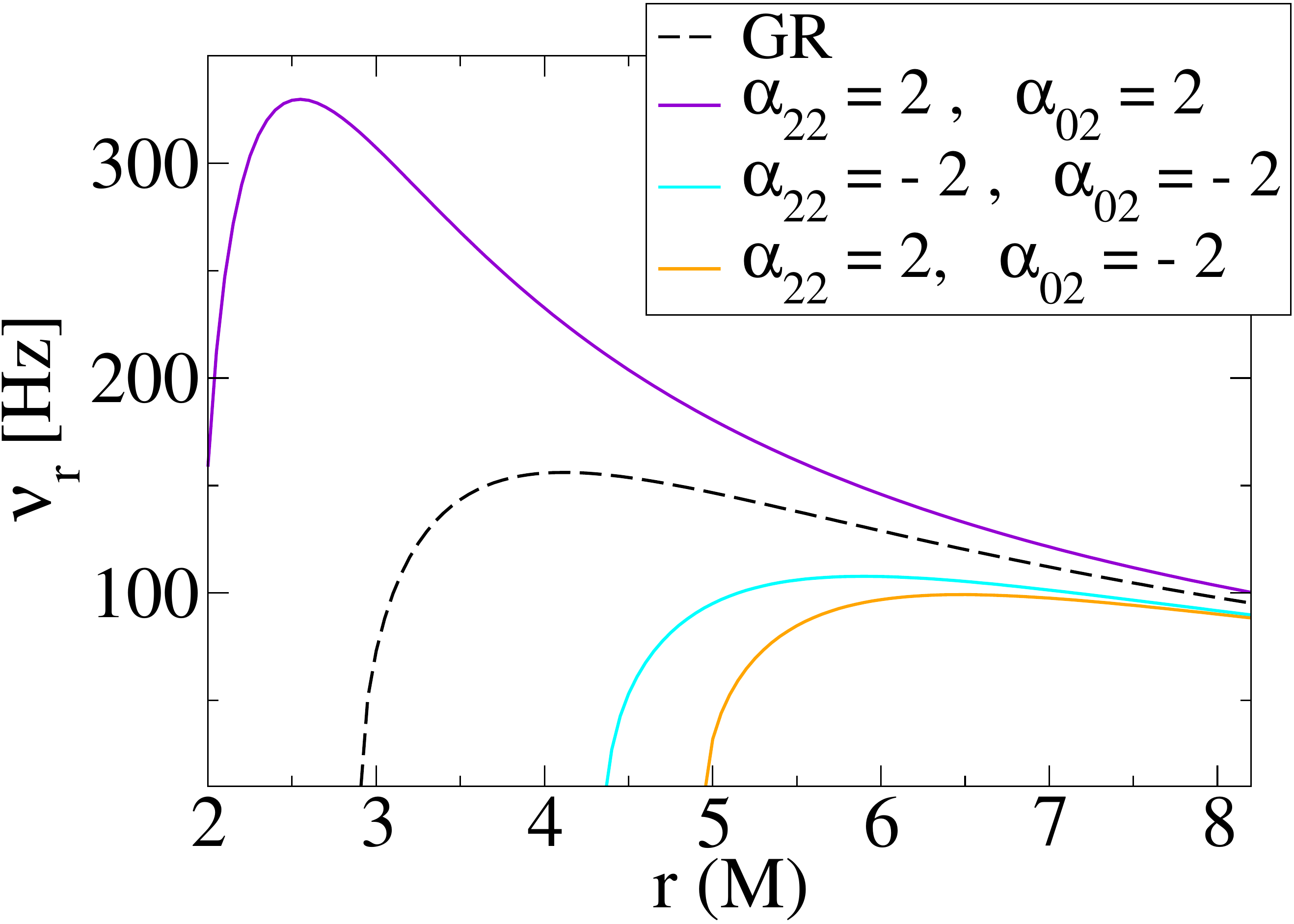}
\caption[Radial epicyclic frequencies]{
Similar to Fig.~\ref{fig:nuPhi} but for the radial epicyclic frequency $\nu_r$. 
Additionally plotted here (top right) is the dependence on the lowest-order parameter $\alpha_{52}$ of the deviation function $A_5(r)$ appearing only in the $g_{rr}$ metric element.
}\label{fig:nuR}
\end{center}
\end{figure*}

\begin{figure*}[htb]
\begin{center}
\includegraphics[width=.45\textwidth]{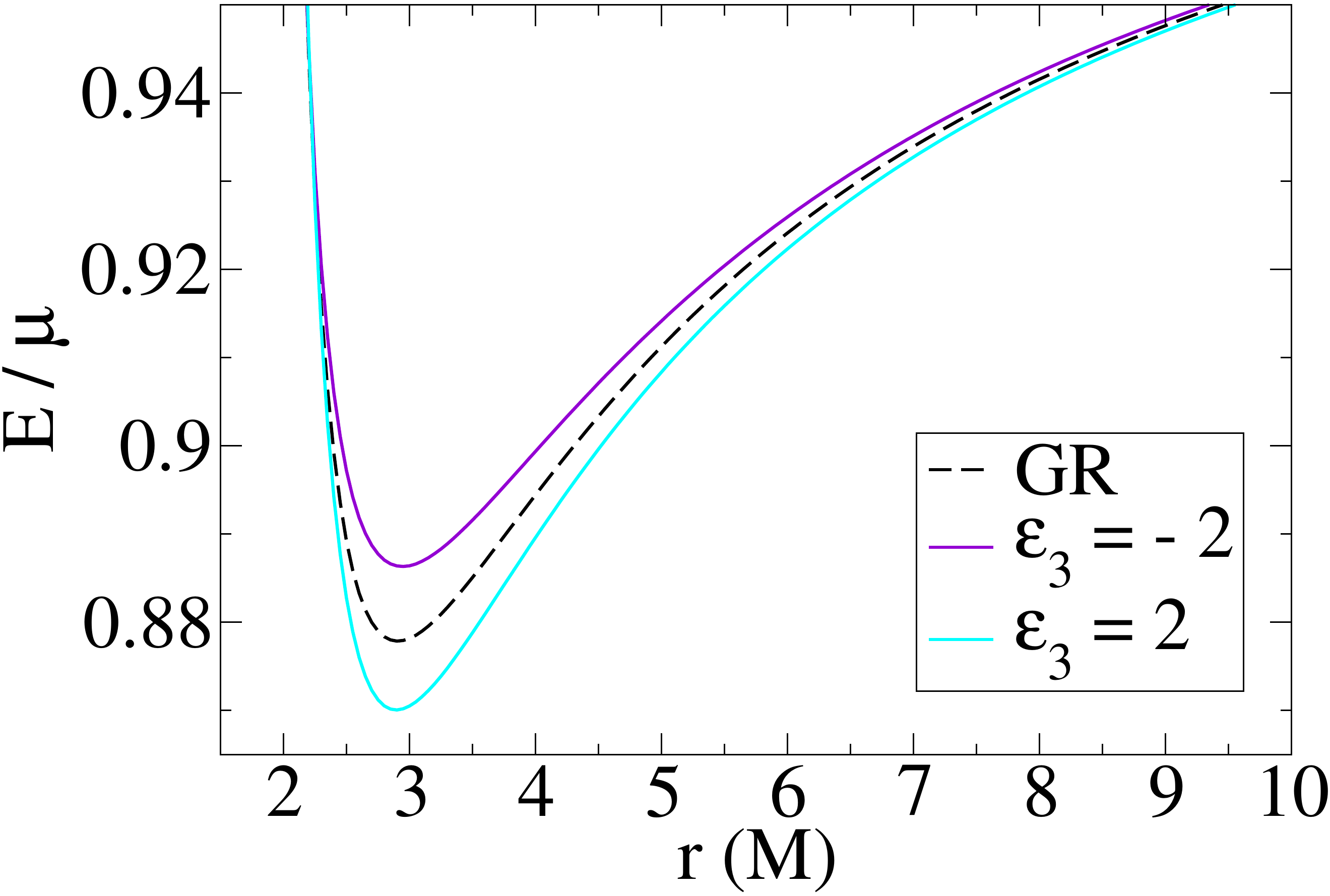}
\includegraphics[width=.45\textwidth]{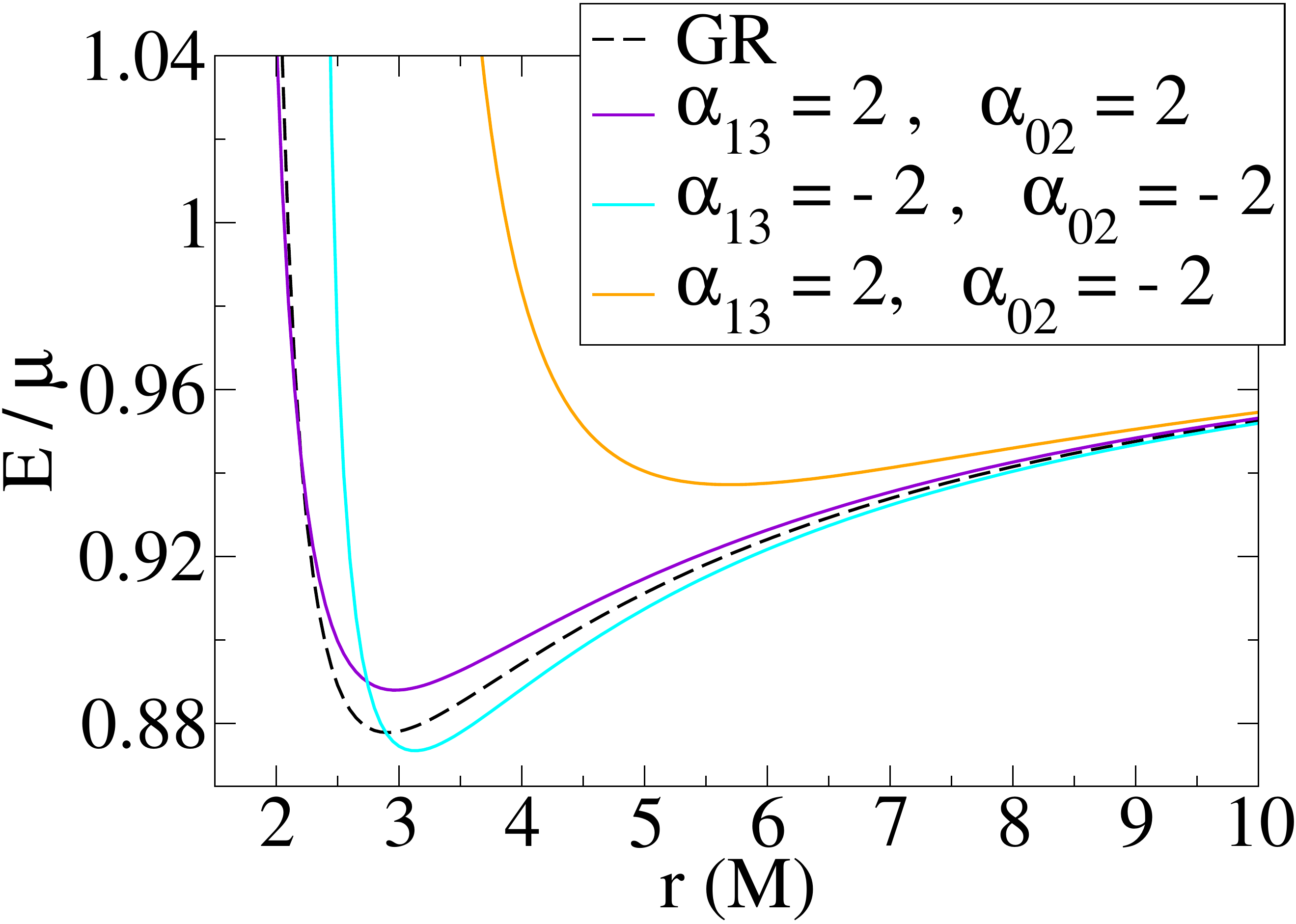}
\includegraphics[width=.45\textwidth]{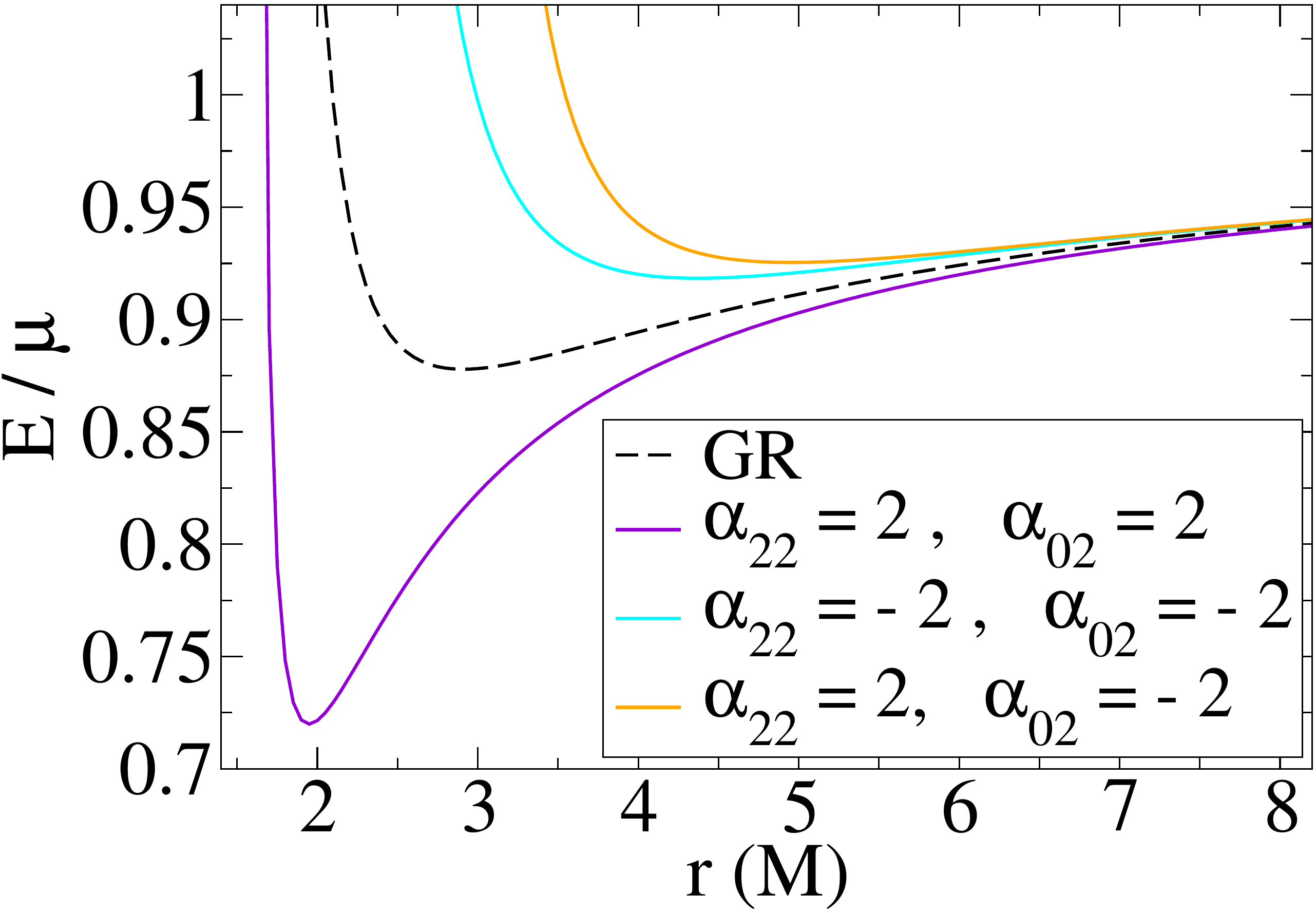}
\caption[Orbital energy]{Similar to Fig.~\ref{fig:nuPhi} but for the specific orbital energy $E_\text{orb}/\mu$.
The left-most plot agrees with that in Fig.~4 of~\cite{Johannsen:2015pca}.
We note that the non-Kerr deviation parameters, especially the new one $A_0(r)$ introduced in this chapter significantly impact the orbital energy of particles. 
}\label{fig:energy}
\end{center}
\end{figure*}

\begin{figure*}[htb]
\begin{center}
\includegraphics[width=.45\textwidth]{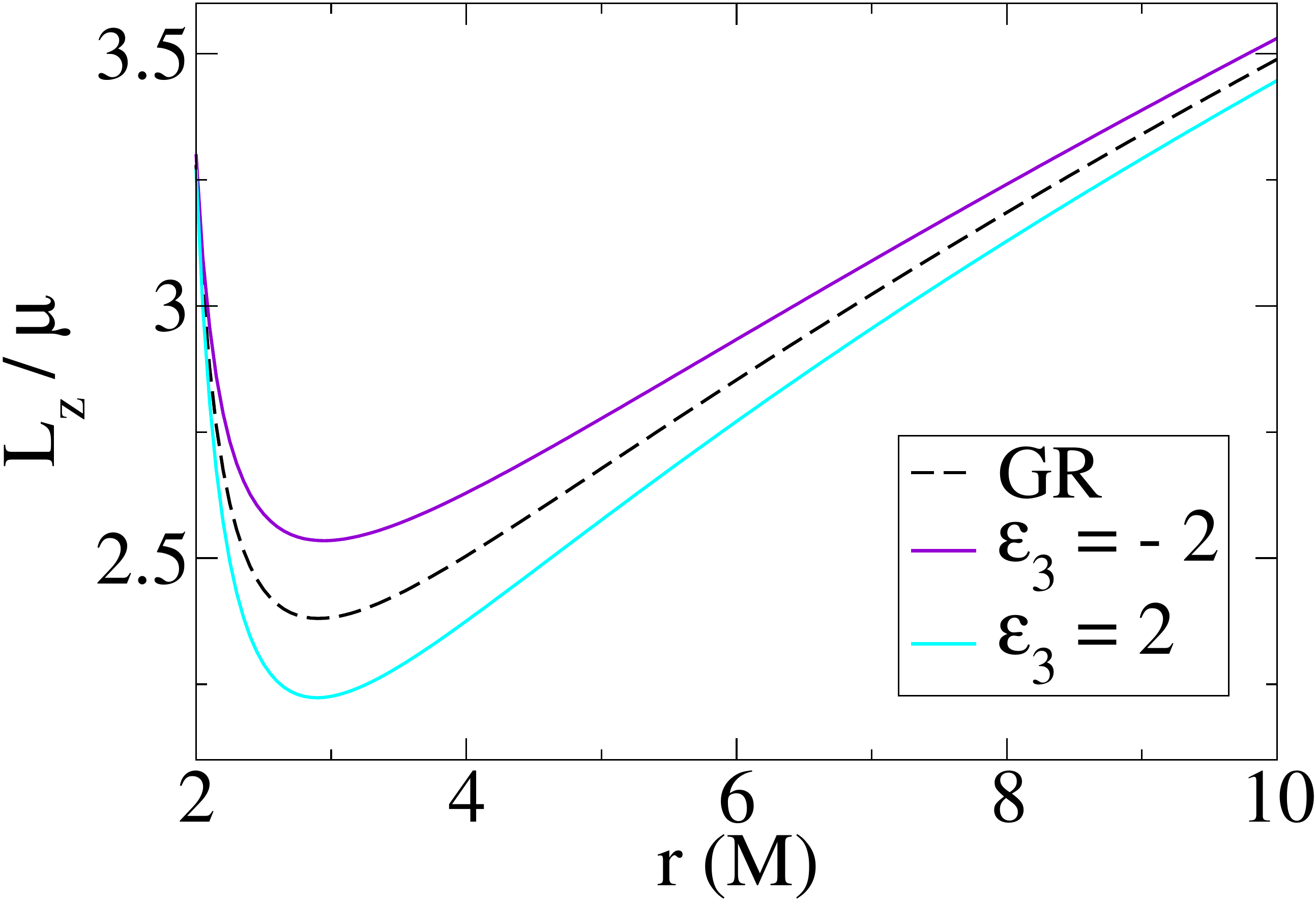}
\includegraphics[width=.45\textwidth]{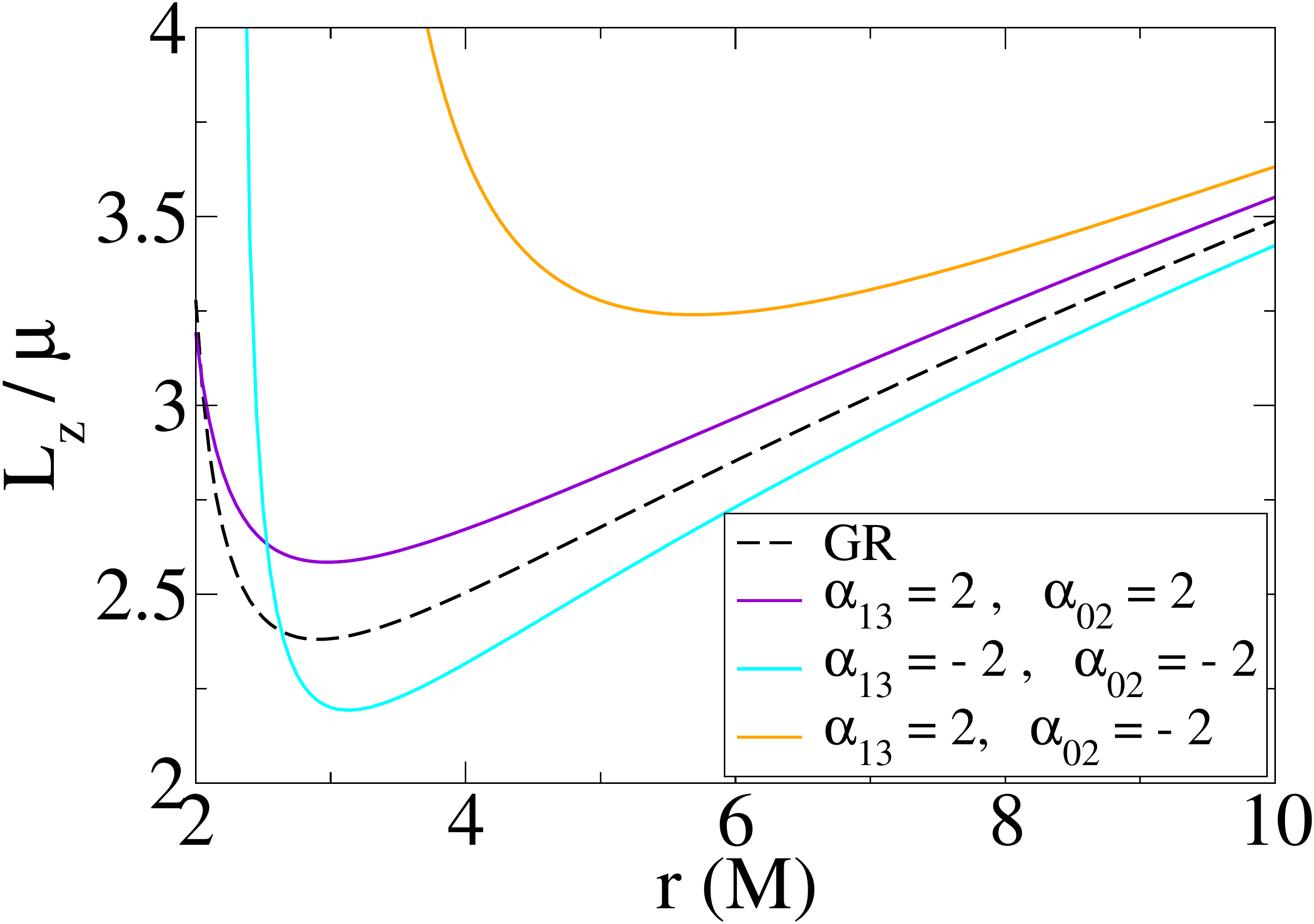}
\includegraphics[width=.45\textwidth]{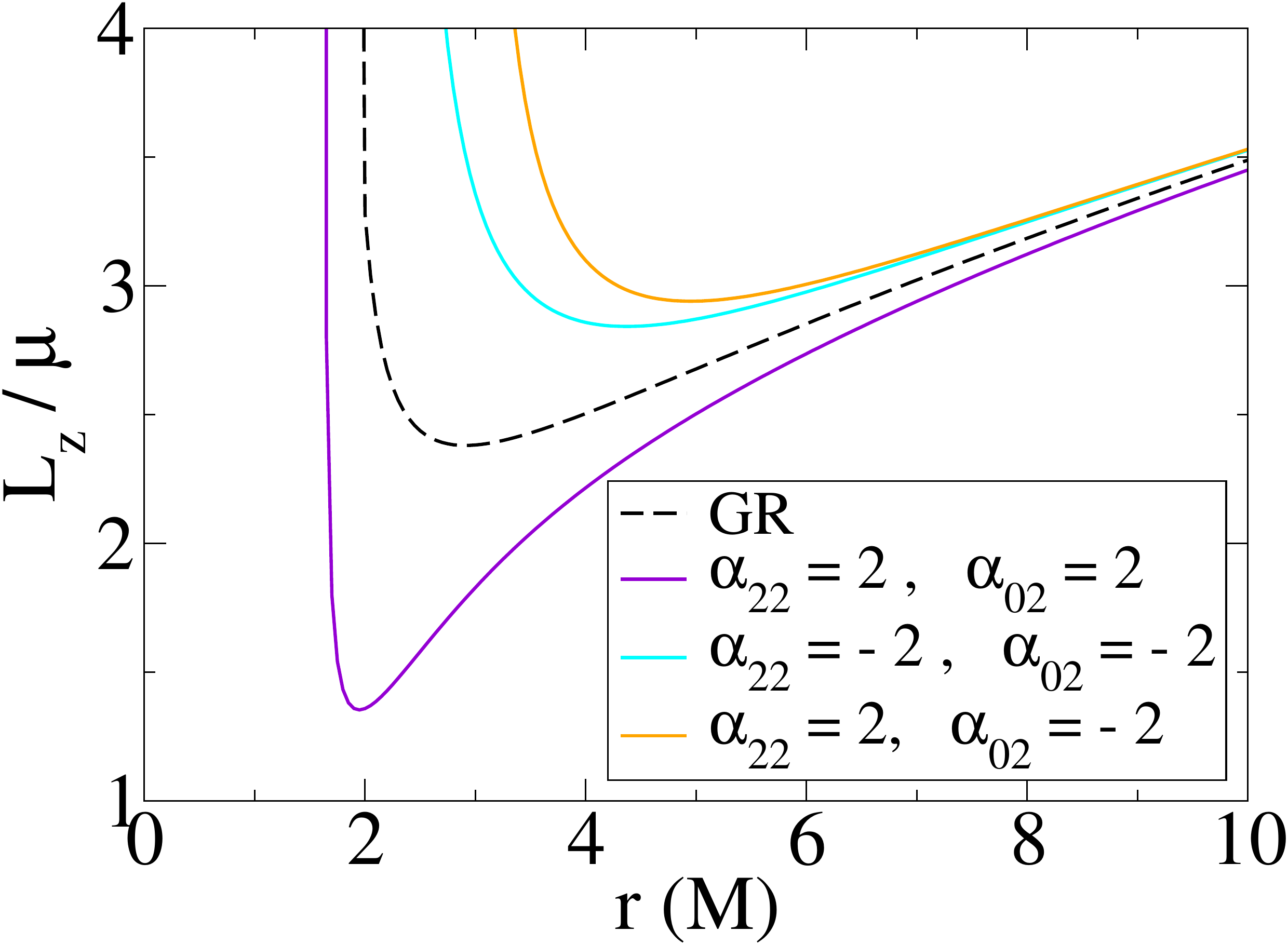}
\caption[Orbital angular momentum]{Same as Fig.~\ref{fig:energy} but for the orbital angular momentum $L_z$. The left-most plot agrees with that in Fig.~5 of~\cite{Johannsen:2015pca}.
}\label{fig:momentum}
\end{center}
\end{figure*}

\begin{figure*}[!htbp]
\begin{center}
\includegraphics[width=.45\textwidth]{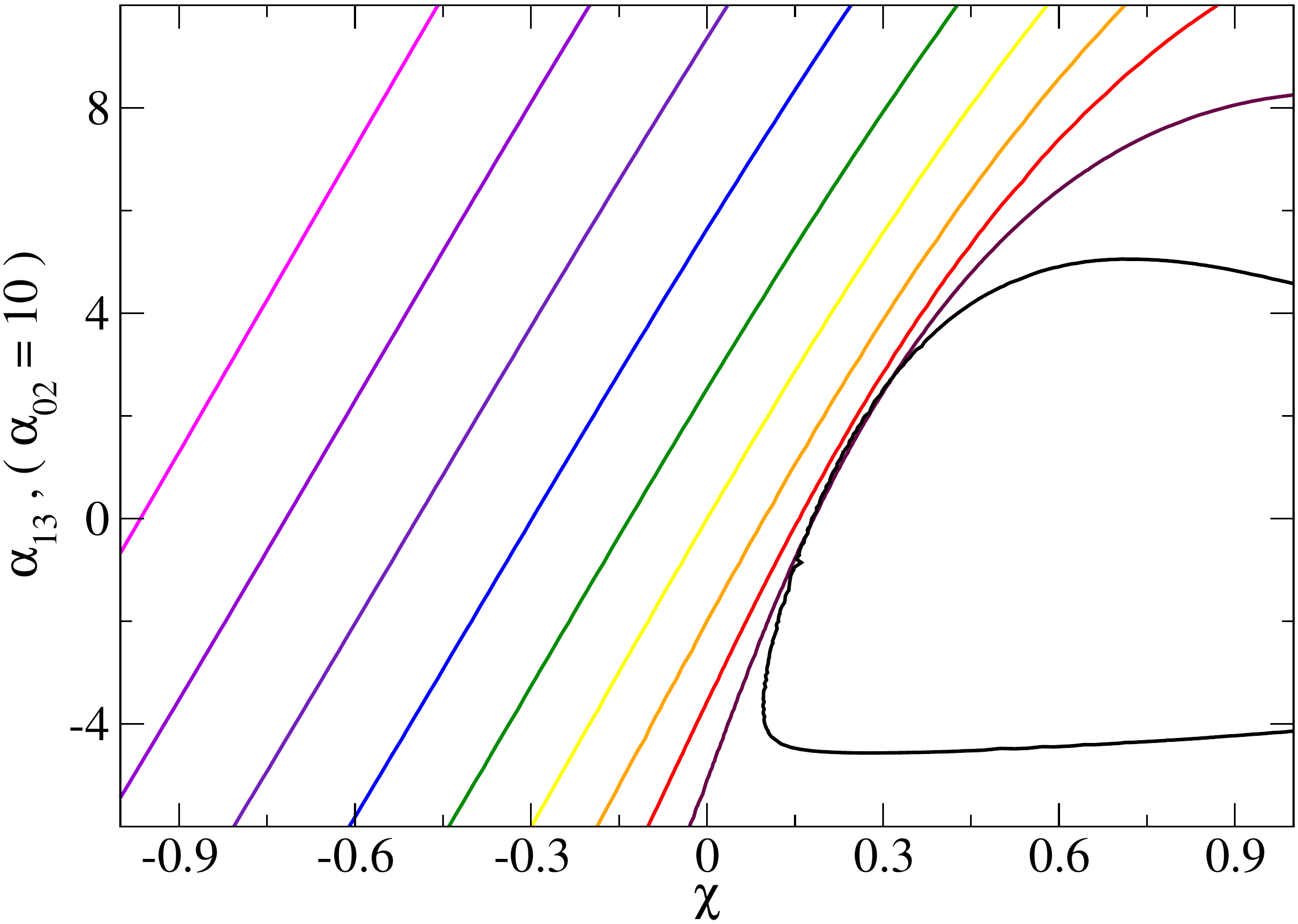}
\includegraphics[width=.45\textwidth]{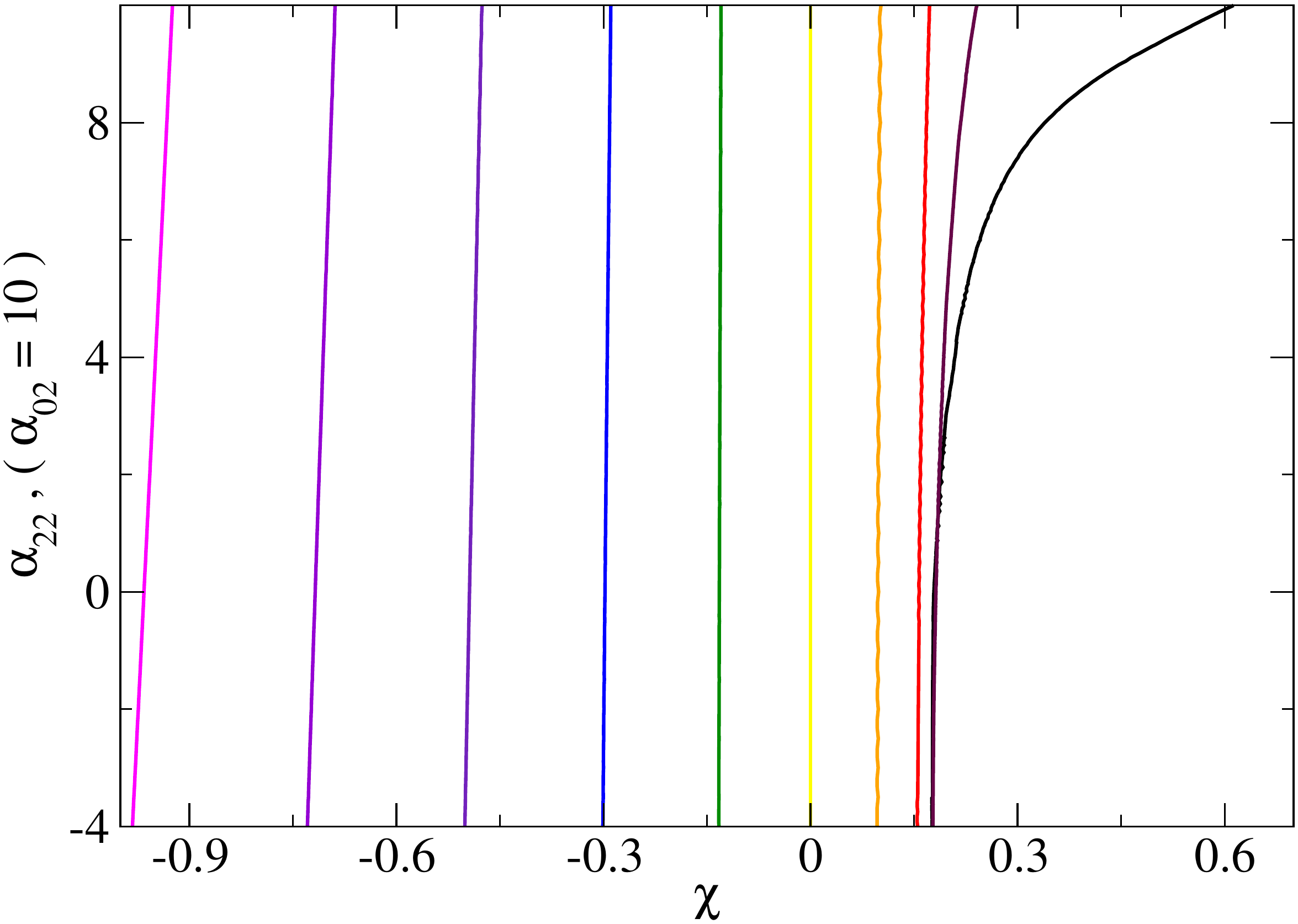}\\
\includegraphics[width=.45\textwidth]{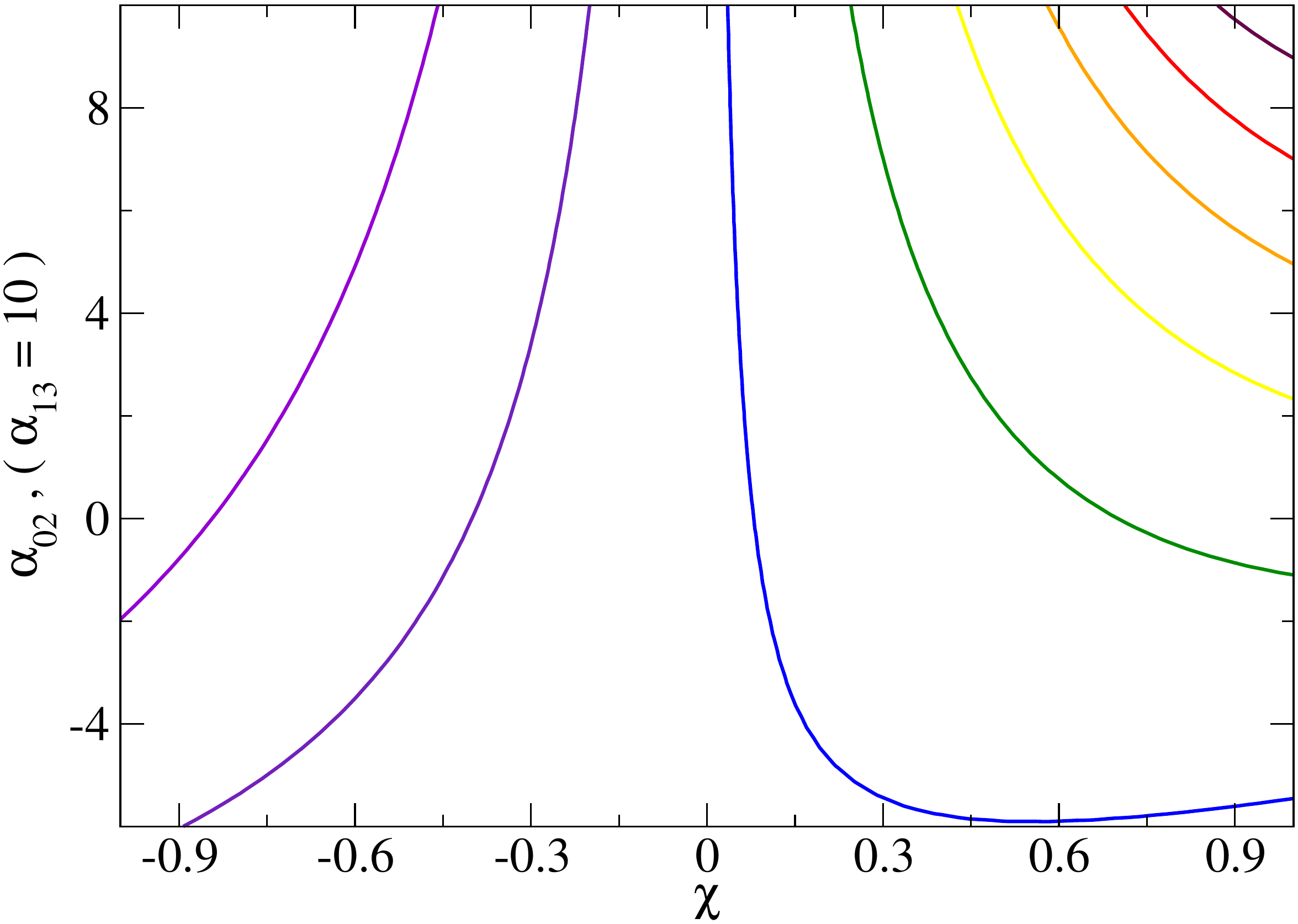}
\includegraphics[width=.45\textwidth]{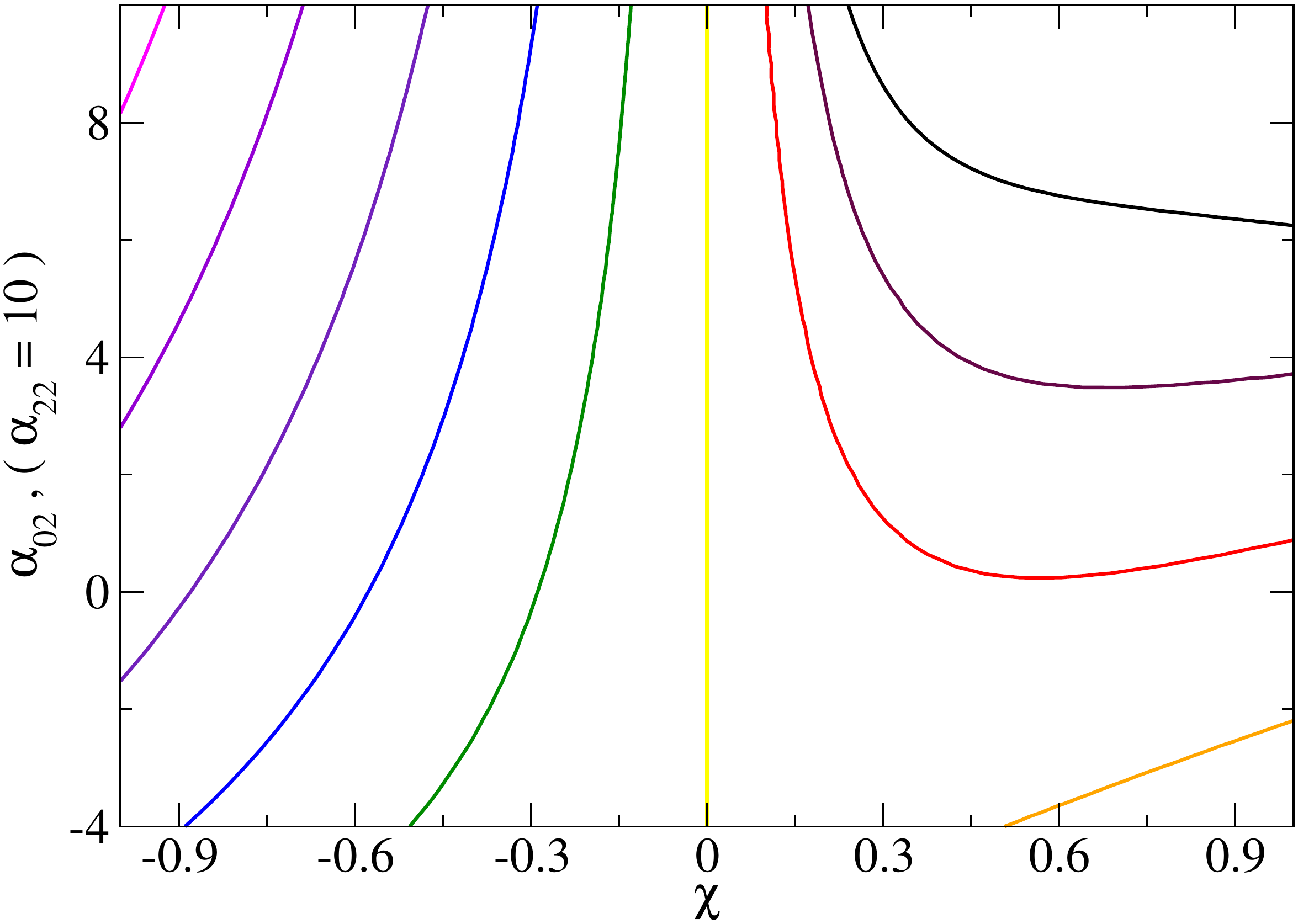}\\
\includegraphics[width=.45\textwidth]{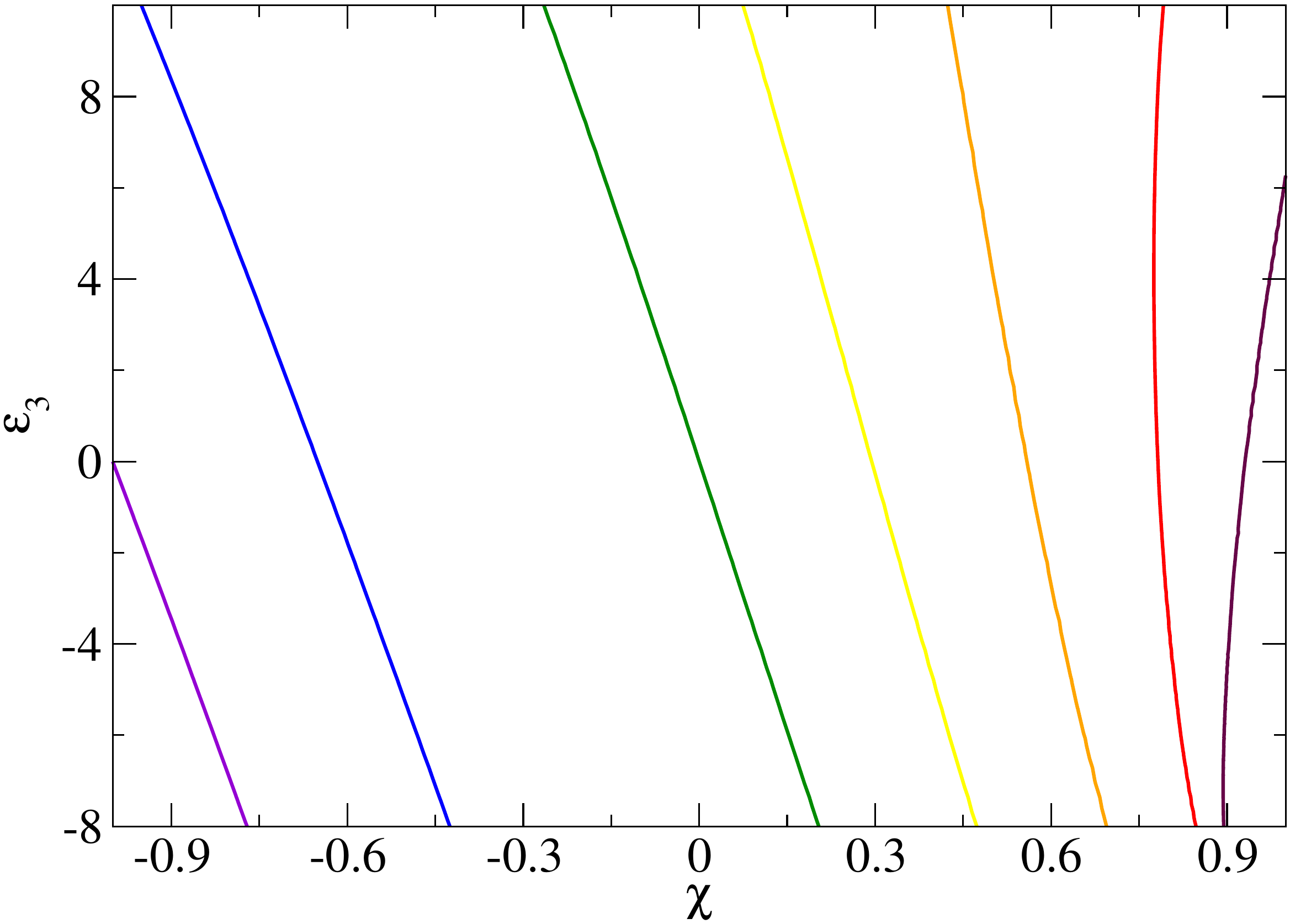}
\caption[Contours of constant ISCO radius]{Constant $r_\ISCO$ contours displaying their dependence on the unitless BH spin $\chi$ and non-Kerr deviation parameters $\epsilon_3$, $\alpha_{02}$, $\alpha_{13}$, and $\alpha_{22}$.
Such contours are presented (from right to left) for $r_\ISCO$ values of $2M$ (black), $3M$ (maroon), $4M$ (red), $5M$ (orange), $6M$ (yellow), $7M$ (green), $8M$ (blue), $9M$ (indigo), $10M$ (violet), and $11M$ (magenta). 
When varying the parameters $\alpha_{13}$, or $\alpha_{22}$, we fix $\alpha_{02}=10$, and vice versa, to avoid the presence of naked singularities.
}\label{fig:rISCO}
\end{center}
\end{figure*}

Now let us plot the resulting energy $E_\text{orb}/\mu$ and angular momentum $L_z/\mu$ for various deviation parameters present in the expressions obtained above.
Here we vary only the lowest-order non-vanishing parameters present in the given expressions: $\alpha_{13}$, $\alpha_{22}$, $\alpha_{02}$, and $\epsilon_3$.
For a further analysis on lower-order parameters assumed to vanish here, see App.~\ref{app:lowerOrder}.
In Figs.~\ref{fig:energy} and~\ref{fig:momentum} we plot the energy and angular momentum as a function of radius for a particle of mass $\mu$ on a circular orbit for several combinations of non-Kerr deviation parameters that produce BHs without naked singularities, where the energies and angular momenta become discontinuous and non-positive.
In each case, all non-Kerr parameters that are not specifically mentioned are set to be 0.
We see that in general, non-Kerr parameters (including the new parameters introduced in this manuscript, $\alpha_{0n}$) have a significant impact on the energy and angular momentum of orbiting particles, and also the ISCO radius (minimum point of the energy curves) that we will describe in more detail in the next section.


\subsubsection{Innermost stable circular orbits}\label{sec:ISCO}

In this section, we compute the location of the ISCO.
In particular, the ISCO occurs at the minimum stable point of the orbital energy $E_\text{orb}$ of a particle with a circular orbit, or 
\begin{equation}\label{eq:risco}
\frac{dE_\text{orb}}{dr}\Big|_{r=r_\ISCO}=0.
\end{equation} 
Because the dependence of this solution is very complicated in terms of the lower-order deviation parameters, we here plot contours of constant $r_\ISCO$ for varying unitless BH spins $\chi$, and magnitude of deviation parameters.
Fig.~\ref{fig:rISCO} does just this for 5 different classes of non-vanishing deviation parameters, taking note that the parameters $\alpha_{13}$ or $\alpha_{22}$ can not be the sole non-vanishing parameter unless $\alpha_{02}$ is also non-vanishing, else naked singularities appear as discussed previously and in App.~\ref{app:nakedSingularities}.
Thus, to vary $\alpha_{13}$ or $\alpha_{22}$, we fix $\alpha_{02}=10$ and vice versa, in order to check the $r_\ISCO$ dependence on individual non-Kerr parameters.

Now we discuss the ISCO dependence on the lower-order non-Kerr parameters $\epsilon_3$, $\alpha_{02}$, $\alpha_{13}$, and $\alpha_{22}$ as seen in Fig.~\ref{fig:rISCO}.
When varying the parameter $\epsilon_3$, we see that for $\chi<0.8$ the ISCO is mildly dependent on non-Kerr perturbations.
When varying $\alpha_{22}$ we observe that $r_\ISCO$ stays almost constant for any given value of $\alpha_{22}$ except for very large spin BHs.
As for $\alpha_{13}$, we see that the location of the ISCO depends very strongly on the non-Kerr parameter.
Finally we observe that for BHs with non-vanishing spin, the dependence of $r_\ISCO$ on $\alpha_{02}$ becomes increasingly stronger for increasingly larger BH spins $\chi$.


\subsubsection{Photon rings}\label{sec:shadows}
In this section, we obtain solutions describing the orbit of a photon about a BH described by the new metric with various non-vanishing deviation parameters.
Following Refs.~\cite{Johannsen:2015pca,Johannsen:2015qca}, we begin with the Hamilton-Jacobi function
\begin{equation}
S\equiv - \frac{1}{2}\mu\tau-E_\text{orb} t+L_z\phi+S_r(r)+S_\theta(\theta)
\end{equation}  
for particle mass $\mu$, proper time $\tau$, orbital energy $E_\text{orb}$, angular momentum $L_z$ (these are the first two constants of motion), and generalized radial and polar functions $S_r(r)$ and $S_\theta(\theta)$.
We compute the Hamilton-Jacobi equations
\begin{equation}
-\frac{\partial S}{\partial\tau}=\frac{1}{2}g^{\alpha\beta}\frac{\partial S}{\partial x^\alpha}\frac{\partial S}{\partial x^\beta},
\end{equation}
to obtain
\begin{align}
\nonumber&-\mu ^2\left(a^2 \cos ^2(\theta )+f(r)+g(\theta)+r^2\right)=\frac{1}{\Delta}\Bigg\lbrack-a^4 A_1^2 E_\text{orb}^2+2 a^3 A_0 E_\text{orb} L_z\\
\nonumber&-2 a^2 A_1^2 E_\text{orb}^2 r^2-a^2 A_2^2 L_z^2+a^2 \Delta  E_\text{orb}^2 \sin ^2\theta +2 a A_0 E_\text{orb} L_z r^2-2 a \Delta  E_\text{orb} L_z\\
&+\Delta  L_z^2 \csc ^2\theta +\Delta  \left(\frac{\partial S_\theta}{\partial \theta}\right)^2-A_1^2 E_\text{orb}^2 r^4+A_5 \Delta ^2 \left(\frac{\partial S_r}{\partial r}\right)^2\Bigg\rbrack.
\end{align}

Next we separate the Hamilton-Jacobi equations, using the separation constant
\begin{align}
\nonumber C=&-\mu ^2 -\left(f(r)+r^2\right)-\frac{1}{\Delta}\Big\lbrack-a^4 A_1^2 E_\text{orb}^2+2 a^3 A_0 E_\text{orb} L_z-2 a^2 A_1^2 E_\text{orb}^2 r^2\\
&-a^2 A_2^2 L_z^2+2 a A_0 E_\text{orb} L_z r^2-A_1^2 E_\text{orb}^2 r^4+A_5 \Delta ^2 \left( \frac{\partial S_r}{\partial r} \right)^2\Big\rbrack,\\
 C=& a^2 E_\text{orb}^2 \sin^2\theta+\mu ^2 (g(\theta)+a^2\cos ^2\theta)-2 a E_\text{orb} L_z+L_z^2 \csc^2\theta+\left(\frac{\partial S_\theta}{\partial \theta} \right)^2.
\end{align}
We then define the Carter-like constant of motion $Q\equiv C-(L_z-a E_\text{orb})^2$ which gives us a solution for $S_r(r)$ (and $S_\theta(\theta)$, not displayed here)
\begin{align}
& \hspace{5cm} S_r(r)=\pm\int dr\frac{1}{\Delta}\sqrt{\frac{R(r)}{A_5(r)}},\\
\nonumber & R(r)\equiv a^4 A_1^2 E_\text{orb}^2-2 a^3 A_0E_\text{orb} L_z+2 a^2 A_1^2 E_\text{orb}^2 r^2+a^2 A_2^2 L_z^2-a^2 \Delta  E_\text{orb}^2\\
&\hspace{1cm}-2 a A_0 E_\text{orb} L_z r^2+2 a \Delta  E_\text{orb} L_z+A_1^2 E_\text{orb}^2 r^4-\Delta  f(r) \mu ^2-\Delta  L_z^2-\Delta  Q-\Delta  \mu ^2 r^2,
\end{align}
where the different signs represent particles with prograde and retrograde motion.
This proves that the new metric presented here indeed has a separable structure, and thus generalizing Johannsen's~\cite{Johannsen:2015pca}.

Finally, we compute the generalized momenta $p_\alpha$ given by 
\begin{equation}
p_\alpha=\frac{\partial S}{\partial x^\alpha}.
\end{equation}
In particular, we focus on the radial momenta, given in covariant and contravariant form as
\begin{align}
p_r=\pm \frac{1}{\Delta}\sqrt{\frac{R(r)}{A_5(r)}},\\
p^r=\pm \frac{A_5(r)R(r)}{\tilde\Sigma}.\label{eq:momentum}
\end{align}
Following Ref.~\cite{Johannsen:2015qca}, the impact parameters $x'$ and $y'$~\cite{Bardeen_1973} describing the image plane from an observer's point of view at infinity with an inclination angle $i$ can be found to be
\begin{equation}
x'=-\frac{\xi}{\sin i}, \hspace{4mm} y'=\pm\sqrt{\eta+a^2\cos^2 i-\xi^2\cot^2i}.\label{eq:photonRings}
\end{equation}
In the above expression, the new invariant parameters $\xi\equiv L_z/E_\text{orb}$ and $\eta\equiv Q/E^2$ have been constructed entirely out of constants of motion.

Now we describe the solutions $(\xi,\eta)$ which describe the photon rings about the central BH.
Such new constants of motion are conserved along the null geodesics, and, because they are the same everywhere, can be simply solved for along the special case of a circular-orbit for simplicity.
Here, the radial photon momentum $p^r$ found in Eq.~\eqref{eq:momentum} as well as its radial derivative much vanish.
Because $\tilde\Sigma$ and $A_5(r)$ are both non-negative, this results in the system of equations
\begin{equation}\label{eq:SolveEqs}
R(r)=0, \hspace{5mm} \frac{dR(r)}{dr}=0,
\end{equation}
with the full re-parameterized expression for $R(r)$ for an orbiting photon ($\mu=0$) given by
\begin{align}
\nonumber R(r)=&a^4 A_1(r)^2-2 a^3 A_0(r) \xi +2 a^2 A_1(r)^2 r^2 +a^2 A_2(r)^2 \xi ^2-a^2 \Delta -2 a A_0(r) \xi  r^2\\
&+2 a \Delta  \xi+ A_1(r)^2 r^4-\Delta  \eta -\Delta  \xi ^2.
\end{align}
With this, one can easily simultaneously solve Eqs.~\eqref{eq:SolveEqs} for $\xi$ and $\eta$ to give parameterized expressions for the photon orbits in Eqs.~\eqref{eq:photonRings}.
Because such results are quite lengthy we do not show them here, however in the Kerr limit, they are found to correctly reduce to the GR expressions found in~\cite{Bardeen}
\begin{align}
\xi&=-\frac{r^2(r-3m)+a^2(r+m)}{a(r-m)}\\
\eta&=\frac{r^3\lbrack 4a^2m-r(r-3m)^2 \rbrack}{a^2(r-m)^2}.
\end{align} 
We also note that the expressions for $\xi$ and $\eta$ only depend on the non-Kerr deviation functions $A_1(r)$, $A_2(r)$, and $A_0(r)$.

Now we compute the image of the photon rings about a BH described by the new metric.
In particular, we focus our attention on only the lowest order parameters $\alpha_{13}$, $\alpha_{22}$, and $\alpha_{02}$, as was done in~\cite{Johannsen:2015pca,Johannsen:2015qca}.
However, we refer the reader to App.~\ref{app:lowerOrder} for an analysis of the inclusion of lower-order parameter $\alpha_{12}$, which was assumed to vanish in the ppN framework.
There we also consider photon rings in the EdGB theory of gravity, where we investigate the validity of the $\frac{1}{r}$ expansion in the EdGB metric.
Here we vary each parameter $\alpha_{13}$, $\alpha_{22}$, and $\alpha_{02}$, while avoiding the cases with emergent naked singularities (See App.~\ref{app:nakedSingularities}), else closed photon orbits do not appear and the photons escape to null infinity as discussed in~\cite{Hioki:2009na,Papnoi:2014aaa}.
This means that if $\alpha_{13} \ne 0$ or $\alpha_{22} \ne 0$, then $\alpha_{02}$ must also be non-vanishing and of the same sign.

Figure~\ref{fig:BHshadowsInclinationSpin} shows the image of the photon ring as it depends on the BH's spin ($\chi$), and the observers inclination angle ($i$) for the case of all deviation parameters vanishing (Kerr) for simplicity.
We observe that for a fixed inclination angle, increasing the BHs spin serves to increasingly displace and deform the photon's orbit.
Similarly, for a fixed BH spin, the inclination dilutes the displacement and deformation gained from the rotating BH for all but the highest angles. 
Such displacement and deformation as a function of inclination only reaches the maximum value allowable by the spin, with none present for a static non-rotating BH.

\begin{figure}[htb]
\begin{center}
\includegraphics[width=.45\textwidth]{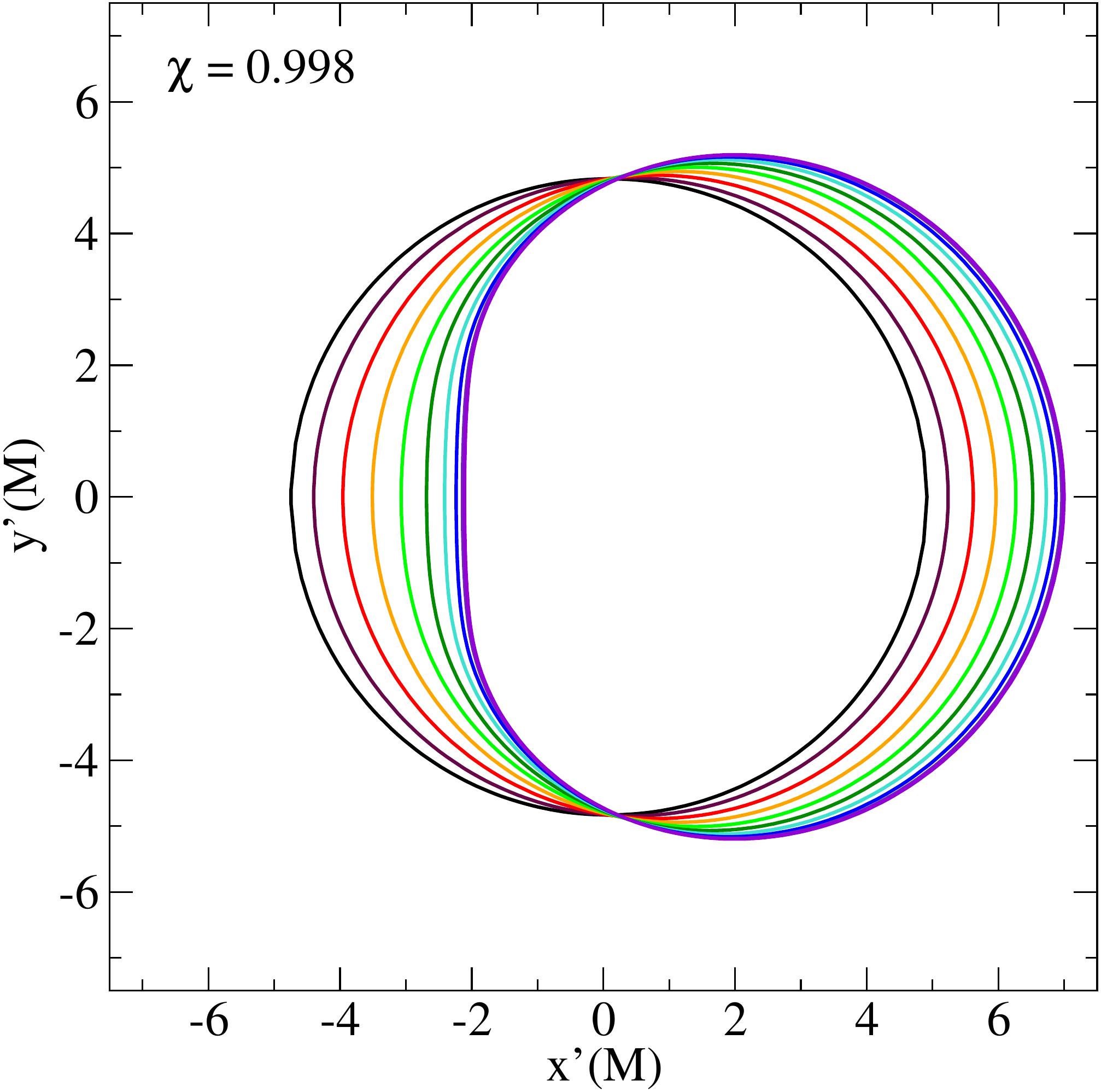}
\includegraphics[width=.45\textwidth]{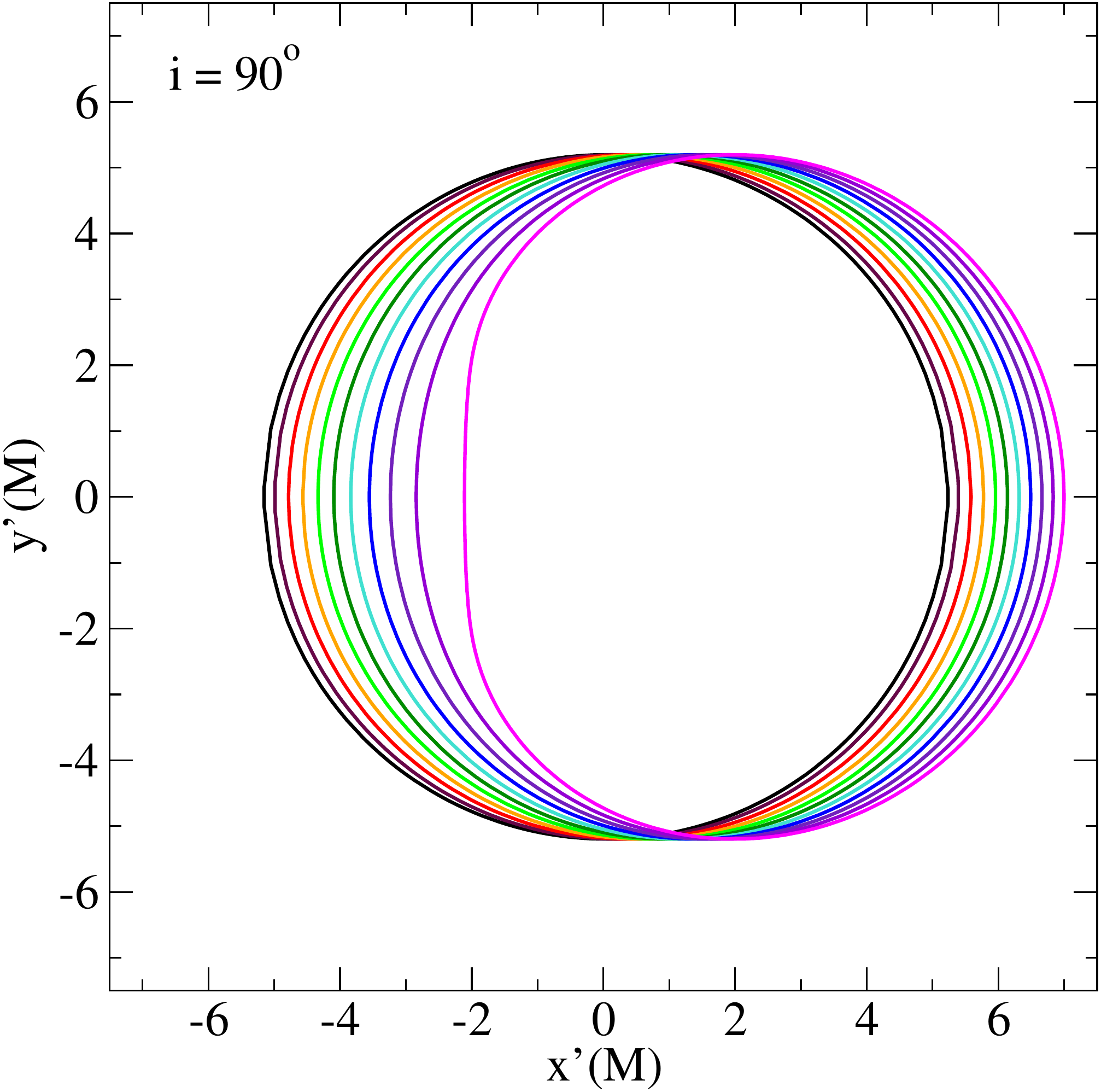}
\caption[Photon orbits about a Kerr black hole]{Images of photon rings of a Kerr BH seen by an observer at infinity (all deviation parameters vanishing in the new metric) for increasing degrees of inclination at a fixed BH spin (top), and increasing BH spins at a fixed inclination (bottom).
For the former, we fix the BH spin at the extreme case of $\chi=0.998$ for demonstration purposes, and increase the inclination angle going left to right from $i=0^\circ$ to $i=90^\circ$ in intervals of $10^\circ$.
For the latter, we fix the inclination angle at the extreme case of $i=90^\circ$, and increase the BH spin going left to right from $\chi=0$ to $\chi=0.998$ in intervals of $0.1$.
}\label{fig:BHshadowsInclinationSpin}
\end{center}
\end{figure}


Finally, we compute the images of the closed photon rings about a BH for several non-vanishing deviation parameters in \hypertarget{link:BHshadowsNonGR}{Fig.~\ref{fig:BHshadowsNonGR}}.
Specifically, for highly-rotating BHs ($\chi=0.998$), moderately-rotating BHs ($\chi=0.5$), and slowly-rotating BHs ($\chi=0.002$), we generate the photon orbits for different non-vanishing values of $\alpha_{13}$, $\alpha_{22}$, and $\alpha_{02}$.
We observe that the effect of increasing $\alpha_{13}$ and $\alpha_{02}$ acts to increase the image size, and negative values of each parameter works to deform the image.
The latter becomes less apparent as the spin decreases, while the former still holds true for even low BH spins.
Next we see that non-vanishing values of $\alpha_{22}$ and $\alpha_{02}$ only marginally affect the image size, but highly deform the orbits for fast-rotating BHs.
In this case (and not in the case of non-vanishing $\alpha_{13}$ and $\alpha_{02}$), we see that positive values of the parameters work to deform the image inwards, while negative values distort outwards.
We conclude with the remark that, especially for highly-rotating and/or highly-inclined observations, that BHs with deviations from Kerr are highly distinguishable from the exact Kerr result.
This is because the deviation parameters $\alpha_{22}$ and $\alpha_{02}$ (corresponding to free functions $A_2(r)$ and $A_0(r)$) are associated with modifications to the angular portions ($\phi$-components) of the contravariant metric in Eq.~\eqref{eq:CYcontravariant}.


\section{Transformation of the new metric to other spacetimes}\label{sec:mappings}
In this section we present the maps that take one from the new metric presented here to several other deformed spacetimes present in the literature.
In particular, we focus on the following theories and spacetimes:

\begin{enumerate}

\item the separable deformed spacetime: 

\begin{itemize}

\item spacetime in Papadopoulos and Kokkotas~\cite{Papadopoulos:2018nvd};

\item parameters: generic deviation parameters $\mathcal{A}_i$ and $\mathcal{B}_i$ for $i=1\dots5$;

\end{itemize}

\item the string-inspired RS2 Braneworld~\cite{Randall:1999ee}:

\begin{itemize}

\item spacetime in~\cite{Aliev:2005bi};

\item parameters: the tidal charge $\beta$;

\end{itemize}

\item the heterotic string theory:

\begin{itemize}

\item spacetime in Kerr \& Sen~\cite{Kerr-Sen};

\item parameters: deviation parameter $b$ related to the magnetic dipole moment;

\end{itemize}

\item Einstein-dilaton-Gauss-Bonnet (EdGB) gravity~\cite{Kanti_EdGB,Maeda:2009uy,Sotiriou:2014pfa}: 

\begin{itemize}

\item spacetime in~\cite{Pani:2011gy,Ayzenberg:2014aka};

\item parameters: the dimensionless coupling constant $\zeta_\EdGB$;

\end{itemize}

\item dynamical Chern-Simons (dCS) gravity~\cite{Jackiw:2003pm,Alexander_cs} 

\begin{itemize}

\item spacetime in~\cite{Yunes_dcs,Yagi_dCS,Pani:2011gy};

\item parameters: the dimensionless coupling constant $\zeta_\dCS$;

\end{itemize}

\item quantum-corrected regular BHs:

\begin{itemize}

\item spacetime in Bardeen~\cite{Bardeen};

\item parameters: deviation $g$ controlling the amount of BH ``regularity'';

\end{itemize}

\item  the Kalb-Ramond BH with Kalb-Ramond parameter $s=1$ and $s=2$:

\begin{itemize}

\item spacetime in~\cite{Kumar:2020hgm};

\item parameters: the Kalb-Ramond Lorentz-violating parameter $\Gamma$.

\end{itemize}

\end{enumerate}

Now let us provide a brief overview of the procedure used to find such mappings between the new spacetime presented here and the ones (X) listed above.
Using a computer-algebra software, this is done by first equating each metric element $g_{\alpha\beta}=g_{\alpha\beta}^\X$ to solve for the the functions $A_0(r)$, $A_1(r)$, $A_2(r)$, $A_5(r)$, and $f(r)$ found in the new metric as a function of the GR deformation parameters present in metric X.
Such functions are then expanded in powers of $\frac{m}{r}$ about $r=\infty$ to obtain the mappings between non-Kerr parameters $\alpha_{0n}$, $\alpha_{1n}$, $\alpha_{2n}$,  $\alpha_{5n}$, and $\epsilon_n$ and non-Kerr parameters in metric $X$.
We present all such mappings in Tab.~\ref{tab:maps}, for the three lowest-order non-vanishing parameters from each free function.
We note that here, we do not take into account the ppN bounds mentioned in Ref.~\cite{Johannsen:2015pca}, and instead allow all lower-order parameters to enter.

\renewcommand{\arraystretch}{1.2}
\begin{table}[!htbp]
\centering
\resizebox{.8\textwidth}{!}{%
\begin{tabular}{|c| l|}
\hline
BH spacetime & Deviation parameters\\
\hline
\hline
\multirow{5}{*}{separable deformed spacetime~\cite{Papadopoulos:2018nvd}} & $f(r)=A_1^\PK(r)-r^2$, \\
& $A_1(r)=\frac{\sqrt{-A_5^\PK(r)\Delta}}{a^2+r^2}$, \\
& $A_2(r)=\frac{\sqrt{-A_3^\PK(r)\Delta}}{a}$, \\
& $A_5(r)=\frac{A_2^\PK(r)}{\Delta}$, \\
& $A_0(r)=-\frac{A_4^\PK(r)\Delta}{a(a^2+r^2)}$ \\
\hline
\multirow{4}{*}{RS2 Braneworld~\cite{Randall:1999ee}} & $\alpha_{10}=1$, $\alpha_{12}=-\frac{\beta}{2m^2}$, $\alpha_{13}=-\frac{\beta}{m^2}$, \dots, \\
& $\alpha_{20}=1$, $\alpha_{22}=-\frac{\beta}{2m^2}$, $\alpha_{23}=-\frac{\beta}{m^2}$, \dots, \\
& $\alpha_{50}=1$, $\alpha_{52}=\frac{\beta}{m^2}$, $\alpha_{53}=\frac{2\beta}{m^2}$, \dots, \\
& $\alpha_{00}=1$, $\alpha_{02}=-\frac{\beta}{m^2}$, $\alpha_{03}=-\frac{2\beta}{m^2}$, \dots \\
\hline
\multirow{5}{*}{heterotic string (Kerr-Sen)~\cite{Kerr-Sen}} & $\epsilon_1=\frac{2b}{m}$, \dots, \\
& $\alpha_{10}=1$, $\alpha_{11}=\frac{b}{m}$, $\alpha_{12}=-\frac{b^2+4bm}{2m^2}$, \dots, \\
& $\alpha_{20}=1$, $\alpha_{21}=-\frac{b}{m}$, $\alpha_{22}=\frac{3b^2-4bm}{2m^2}$, \dots, \\
& $\alpha_{50}=1$, $\alpha_{51}=\frac{2b}{m}$, $\alpha_{52}=\frac{4b}{m}$, \dots, \\
& $\alpha_{00}=1$, $\alpha_{02}=-\frac{4b}{m}$, $\alpha_{03}=\frac{8b(b-m)}{m^2}$, \dots \\
\hline
\multirow{4}{*}{EdGB gravity~\cite{Ayzenberg:2014aka}} & $\alpha_{10}=1$, $\alpha_{13}=-\frac{\zeta}{6}$, $\alpha_{14}=-\frac{14\zeta}{3}$, \dots, \\
& $\alpha_{20}=1$, $\alpha_{23}=-\frac{13\zeta}{30}$, $\alpha_{24}=-\frac{16\zeta}{3}$, \dots, \\
& $\alpha_{50}=1$, $\alpha_{52}=\zeta$, $\alpha_{53}=3\zeta$, \dots, \\
& $\alpha_{00}=1$, $\alpha_{03}=-\frac{3\zeta}{5}$, $\alpha_{04}=-10\zeta$, \dots \\
\hline
\multirow{4}{*}{dCS gravity~\cite{Yagi_dCS,Yunes_dcs}} & $\alpha_{10}=1$,\\
& $\alpha_{20}=1$, $\alpha_{24}=-\frac{5\zeta}{8}$, $\alpha_{25}=-\frac{15\zeta}{14}$, \dots, \\
& $\alpha_{50}=1$, \\
& $\alpha_{00}=1$, $\alpha_{04}=-\frac{5\zeta}{8}$, $\alpha_{05}=-\frac{15\zeta}{14}$, \dots \\
\hline
\multirow{4}{*}{quantum-corrected (Bardeen)~\cite{Bardeen}} & $\alpha_{10}=1$, $\alpha_{13}=-\frac{3g^2}{2m^2}$, $\alpha_{14}=-\frac{3g^2}{m^2}$, \dots,\\
& $\alpha_{20}=1$, $\alpha_{23}=-\frac{3g^2}{2m^2}$, $\alpha_{24}=-\frac{3g^2}{m^2}$, \dots, \\
& $\alpha_{50}=1$, $\alpha_{53}=\frac{3g^2}{m^2}$, $\alpha_{54}=\frac{6g^2}{m^2}$, \dots, \\
& $\alpha_{00}=1$, $\alpha_{03}=-\frac{3g^2}{m^2}$, $\alpha_{04}=-\frac{6g^2}{m^2}$, \dots \\
\hline
\multirow{8}{*}{Kalb-Ramond~\cite{Kumar:2020hgm}} \multirow{4}{*}{($s=1$)} & $\alpha_{10}=1$, $\alpha_{12}=-\frac{\Gamma}{2m^2}$, $\alpha_{13}=-\frac{\Gamma}{m^2}$, \dots, \\
& $\alpha_{20}=1$, $\alpha_{22}=-\frac{\Gamma}{2m^2}$, $\alpha_{23}=-\frac{\Gamma}{m^2}$, \dots, \\
& $\alpha_{50}=1$, $\alpha_{52}=\frac{\Gamma}{m^2}$, $\alpha_{53}=\frac{2\Gamma}{m^2}$, \dots, \\
& $\alpha_{00}=1$, $\alpha_{02}=-\frac{\Gamma}{m^2}$, $\alpha_{03}=-\frac{2\Gamma}{m^2}$, \dots \\[-0.01em]
\cline{2-2}
\multirow{4}{*}{\color{white}Kalb-Ramond~\lbrack00\rbrack \color{black}($s=2$)} & $\alpha_{10}=1$, $\alpha_{11}=-\frac{\Gamma}{2m}$, $\alpha_{12}=\frac{\Gamma(3\Gamma-8m)}{8m^2}$, \dots,\\
& $\alpha_{20}=1$, $\alpha_{21}=-\frac{\Gamma}{2m}$, $\alpha_{22}=\frac{\Gamma(3\Gamma-8m)}{8m^2}$, \dots, \\
& $\alpha_{50}=1$, $\alpha_{51}=\frac{\Gamma}{m}$, $\alpha_{52}=\frac{2\Gamma}{m}$, \dots, \\
& $\alpha_{00}=1$, $\alpha_{01}=-\frac{\Gamma}{m}$, $\alpha_{02}=\frac{\Gamma(\Gamma-2m)}{m^2}$, \dots \\[-0.01em]
\hline
\end{tabular}
}
\caption[Transformations from the new metric to other spacetimes]{Mappings from the new metric presented in this chapter to several other BH solutions in related works.
For the separable deformed spacetime, we show only the transformation between the arbitrary functions defined here and in~\cite{Papadopoulos:2018nvd} ($A_i^\PK(r)$).
The mappings to EdGB and dCS gravity are only valid up to linear order in BH spin and first order in deformation parameters $\zeta$, and the mapping to Bardeen is only valid to quartic order in deviation parameter $g$.
Parameters which are missing correspond to those that are vanishing in the series expansion.
}\label{tab:maps}
\end{table}

We now discuss the results presented in Table~\ref{tab:maps}.
We first note that in the separable deformed spacetime presented by Papadopoulos and Kokkotas, we only find the transformation from the arbitrary functions presented here $A_i(r)$, $f(r)$ to the ones ($A_i^\PK(r)$) found in~\cite{Papadopoulos:2018nvd}.
Additionally, Ref.~\cite{Johannsen:2015pca} states that the RS2 Braneworld metric could not be related to Johannsen's metric, and in Ref.~\cite{Johannsen:2015qca} the author found a mapping by introducing a new non-Kerr parameter $\beta$ such that $\Delta \to \Delta+\beta$.
However, we have found that both Johannsen's metric and the metric presented here could be mapped to RS2 Braneworld as shown in Table~\ref{tab:maps}, with some difficulty.
In both the EdGB and dCS theories of gravity, the mapping is only valid up to first order in spin $\chi$ and coupling parameter $\zeta$.
We note that all the mappings presented in Table~\ref{tab:maps} for the new metric can also be mapped to Johannsen's.

We finish this section by noting the high versatility of this new metric, with the ability to map to many  BH solutions found in the literature. Having said this, we point out there are many BH solutions that cannot be mapped to the new metric found here. Such metrics include BHs in 
Einstein-scalar gravity~\cite{Bogush:2020lkp} and Einstein-Maxwell dilaton theory~\cite{Jai-akson:2017ldo}, the Bumblebee metric~\cite{Ding:2019mal}, and slowly-rotating BHs in EdGB and dCS gravity~\cite{Yagi_dCS,Ayzenberg:2014aka}.
These metrics do not contain separable geodesic equations, thus no Carter-like constant is present and the geodesic equations may become chaotic~\cite{Cardenas-Avendano:2018ocb}.
Such fundamental differences manifest themselves as an inability to transform to Johannsen's metric or the metric presented here, and can be seen by the appearance of angular functions within the mapping functions of $f(r)$ and $A_i(r)$.


\section{Conclusions}\label{sec:conclusion}
The no-hair theorem tells us that isolated BHs create\sout{s} a spacetime described by the famous Kerr metric.
In this metric, we can predict the shape and size of photon rings seen by a far-away observer that depend only on the central BH's mass and spin.
While several tests to date have confirmed this hypothesis~\cite{Yunes:2013dva,Gair:2012nm,Liu_2012,Pfahl_2004,Wex_1999,Sadeghian:2011ub,Merritt:2009ex,Will_2008,Johannsen:2011dh,Bambi:2013sha,Bambi:2012at,Johannsen:2012ng,Bambi:2011jq,Bambi:2012ku,Johannsen:2010ru,Johannsen:2010xs,Psaltis_2011,Johannsen_2016,Krawczynski_2012,PhysRevD.83.103003,Johannsen_2010,Berti:2007zu,Dreyer:2003bv,PhysRevD.73.064030,Vigeland:2011ji,Apostolatos:2009vu,Gair:2007kr,Collins:2004ex,Glampedakis:2005cf,PhysRevD.84.064016,PhysRevD.81.024030,PhysRevD.78.102002,PhysRevD.69.082005,Barack:2006pq,Johannsen:2011mt,Mandel_2014,PhysRevD.77.064022,PhysRevD.56.1845,PhysRevD.52.5707,Isi:2019aib}, what if small deviations from the Kerr metric yet exist in nature?
That is one question the EHT with the VLBI aim to answer, by using an effective earth-sized telescope to accurately map photon rings about SMBHs located at the center of galaxies.

In this chapter, we have extended the important analysis done by Johannsen in~\cite{Johannsen:2015pca} to design a more general Kerr-like BH solution.
This new metric can be interpreted as the most general stationary, axisymmetric, and asymptotically flat spacetime we can create with intact separable geodesic equations.
Such a new metric is parameterized non-linearly by 5 free functions $A_i(r)$ for $i=0,1,2,5$ and $f(r)$ which deviate from the Kerr metric, and reproduce the exact Kerr metric when vanishing.
This general metric can be mapped to a large range of BH solutions found in the literature, as demonstrated for seven different cases~\cite{Papadopoulos:2018nvd,Randall:1999ee,Aliev:2005bi,Jai-akson:2017ldo,Ding:2019mal,Kerr-Sen,Kanti_EdGB,Maeda:2009uy,Sotiriou:2014pfa,Ayzenberg:2014aka,Jackiw:2003pm,Yagi_dCS,Yunes_dcs,Bardeen,Kumar:2019uwi,Pani:2011gy,Kumar:2020hgm}. 
The metric has been shown to produce an event horizon and Killing horizon coexistent with the Kerr one. 
Finally, we look at the spheroidicity conditions found in Ref.~\cite{Glampedakis:2018blj}, finding a $\theta$-independent function which admits a Kerr-like spherical photon orbits.

Now that we have a new, general metric in hand, we proceeded to calculate several properties of the ensuing spacetime.
In particular, we focused on circular equatorial particle orbits and found analytic expressions for the orbital energy and angular momentum of such a particle, along with its Keplerian and epicyclic frequencies for perturbed radial and vertical orbits.
We plot these quantities for several different parameterizations of the new metric for comparison against the Kerr result to show the effect of the parameterized deviations.
We then compute the location of the ISCO, once again comparing the results for several parameterizations against the Kerr result.

We finally shift our attention to the orbits of photons about BHs described by the new metric presented here.
By following the analysis of Johannsen in~\cite{Johannsen:2015qca}, we derive analytic expressions for thin photon orbit solutions, called ``photon rings''.
The images of such orbits are observable by, e.g. the EHT, and are extremely timely due to the recent image of the lensed photon orbits about the SMBH M87$^*$~\cite{1435171,1435174,1435175,1435177,1435168}, with future resolution and fidelity improvements imminent.
We then compare the photon rings about BHs with several different parameterizations against the Kerr result.
We find that, especially for highly-rotating BHs and/or highly inclined observation angles, the non-Kerr photon rings indeed distinguish themselves prominently against the standard Kerr result.

Future work in this topic includes constraining the new metric found in this chapter with current and future observations. For example, one can use observations of X-ray continuum spectrum and iron line emissions from accretion disks around BHs to constrain some of the parameters, as already done in~\cite{Choudhury:2018zmf,Tripathi:2018lhx,Tripathi:2019bya,John:2019rhj,Tripathi:2019fms}.
Another way to constrain the metric is to use future gravitational-wave observations. 
For example, extreme-mass ratio inspirals can probe accurately spacetime around BHs~\cite{Vigeland:2011ji,Gair:2011ym,Apostolatos:2009vu,Gair:2007kr,Collins:2004ex,Glampedakis:2005cf,PhysRevD.84.064016,PhysRevD.81.024030,PhysRevD.78.102002,PhysRevD.69.082005,Barack:2006pq,Johannsen:2011mt,Mandel_2014,PhysRevD.77.064022,PhysRevD.56.1845,PhysRevD.52.5707,Isi:2019aib}. Other possibilities include pulsars orbiting around BHs~\cite{Liu_2012,Liu:2014uka}, stars orbiting around the center of Sgr A$^*$~\cite{Sadeghian:2011ub,Merritt:2009ex,Will_2008} and low-mass X-ray binaries with BHs~\cite{Yagi_EdGB}.
One could additionally repeat the analysis done in~\cite{Cardenas-Avendano:2019zxd} with the new metric presented here, where the authors compared current and future gravitational wave and x-ray constraints on deformed spacetime metric parameters.

In addition, future work on the presented topic includes a detailed investigation into a finite stress-energy tensor that sources the beyond-Kerr nature presented here, if such a metric corresponds to a non-vacuum spacetime.
In particular, one could compute Einstein's Equations with the new metric considered in this chapter assuming GR, and attribute the purely beyond-Kerr components to an additional stress-energy tensor.
While the key purpose of this investigation was to determine an arbitrary, theory agnostic spacetime model, a source term for such effects is interesting to study, and e.g. determine if the energy conditions are satisfied.
We found that the stress-energy tensor and energy conditions required for the beyond-Kerr corrections presented here are lengthy and complicated given the number of arbitrary beyond-Kerr functions, and don't provide any immediately meaningful observations. 
We leave a further analysis on this point for future work.

Additionally, one could introduce a stress-energy tensor corresponding to the accretion disk found outside of e.g. M87$^*$ in the EHT observations, find a black hole solution with such an accretion disk perturbatively and see if such a solution can be mapped to the beyond-Kerr metric presented in this chapter (or with any other beyond-Kerr spacetimes).
Such future work could be used to probe properties of the disk with the arbitrary free parameters utilized in this chapter.

Alternatively, we consider testing singly-parameterized beyond-Kerr spacetimes with GWs and the inspiral-merger-ringdown consistency test as done before in Chapter~\ref{chap:IMRconsistency}.
Parameterized BH solutions to modified Einstein's field equations allow us to test the extreme-gravity regime of GR in a model-independent way.
From here, parameterized corrections to the gravitational waveform for inspiraling BHs can be predicted.
Once one has these tools in hand, future GW signals can be tested against the beyond-Kerr metric by (i) IMR consistency tests (comparing the consistency between the inspiral and merger-ringdown portions of the signal) and (ii) parameterized tests.

In this chapter, we presented the necessary recipe required to estimate corrections to the inspiral, ringdown, and remnant BH properties of the gravitational waveform given an arbitrary spacetime metric $g_{\alpha\beta}^\X$, and then test future GW signals against this template with the above two tests.
In particular, we extend the work of Chapter~\ref{chap:IMRconsistency} where this was done for the specific example of the EdGB theory of gravity.
We first derive corrections in a generic way without specifying the beyond-Kerr spacetime. 
As an application, we focused on the JP metric introduced by Johannsen and Psaltis in Ref.~\cite{Johannsen:2011dh} and the modified $\Delta$ metric, modified from Johannsen's metric in Refs.~\cite{Johannsen:2015mdd,Johannsen:2015pca}.
Each spacetime metric considered here is singly-parameterized beyond the Kerr metric with parameters $\epsilon_3$ and $\beta$.
Such spacetimes can then be mapped to BH solutions other than Kerr found in the literature.

With the arbitrary JP and mod.~$\Delta$ metrics in hand, we next estimate corrections to the gravitational waveform for inspiraling BHs immersed in a JP or mod.~$\Delta$ spacetime.
Specifically, we found corrections to the GW amplitude and phase in the inspiral, the ringdown and damping QNM frequencies, the orbital energy and angular momentum of a particle about the BH, and finally the remnant BH's mass and spin.
Each of the above-listed corrections are parameterized by the single parameters $\epsilon_3$ and $\beta$ in the JP and mod.~$\Delta$ spacetimes respectively, and can be accordingly added into the gravitational waveform template.
With this modified-gravity waveform template, we next test GW signals for their feasibility within such a spacetime.

We then perform the IMR consistency test to predict the magnitude of $\epsilon_3$ ($\beta$) required to differentiate between Kerr and JP (mod.~$\Delta$) GW signals.
Within this test, we compute statistical uncertainties on the remnant BH mass and spin parameters from the inspiral and merger-ringdown signals independently, using a Fisher analysis.
We next estimated the systematic uncertainties in each measurement representing the waveform mismodeling uncertainty present by using a GR template with Kerr BHs, and yet observing a GW signal with a given magnitude of $\epsilon_3$ ($\beta$) present within.
We then increase the magnitude of $\epsilon_3$ ($\beta$) until the inspiral and merger-ringdown estimates of remnant BH properties begin to disagree to a statistically significant level.
Only at this point can we reliably claim the observed GW signal indeed has emergent  JP (mod.~$\Delta$) effects present within.
We also computed bounds on $\epsilon_3$ ($\beta$) using the parameterized test and compared them with those from the IMR consistency test.

We now discuss our findings.
We performed the IMR consistency test in each considered spacetime metric for the current-generation aLIGO O2 detector, the third-generation CE detector, the future space-based detector LISA, and finally the multi-band observation between the latter two.
As summarized in Table~\ref{tab:results}, we first found that observations by the O2 detector can detect JP (mod.~$\Delta$) deviations from the GR waveform for magnitudes of $\epsilon_3$ ($\beta$) on the order of unity, in agreement with current constraints.
For future GW detectors CE and LISA, we found that constraints about two orders-of-magnitude stronger can be claimed.
Finally, for the observation of EMRIs by the space-based detector LISA, we found that constraints three orders-of magnitude stronger can be made.
Such strong constraints occur because EMRI BH systems radiate GWs less compared to comparable-mass systems with the same total mass, thus increasing the amount of time JP (mod.-$\Delta$) effects are observed for, which results in a factor of $\eta^{-4/5}\sim10^{4}$ for EMRI systems in the ppE correction to the inspiral waveform.

In this analysis, several assumptions were made that somewhat weaken our results. 
In particular, we have assumed the following caveats:
\begin{itemize}
    \item We only included corrections to the ringdown phase of the waveform, neglecting those to the merger. 
    \item We only consider conservative corrections to the inspiral waveform, rather than dissipative ones. The resulting presented bounds are therefore conservative in nature. Once dissipative effects are additionally included, constraints may become stronger.
    \item We only included corrections to the leading-order PN terms in the waveform, and also to quadratic order in spin, and first order in beyond-Kerr parameters $\zeta$.
    \item We assumed that the QNMs are isospectral between axial and polar modes, something that may not be entirely true in beyond-Kerr spacetimes. 
    \item We estimated the BH final mass and spin following the result that holds for Kerr BH binary mergers in GR, which may not be true in beyond-Kerr spacetimes. 
\end{itemize}
One needs to specify a theory of gravity to overcome most of the points raised above, which goes beyond the scope of probing beyond-Kerr spacetime in a generic, model-independent way with GWs.
We present this article as a new method to quickly and easily estimate various corrections in the full waveform from an arbitrary beyond-Kerr metric, to obtain order-of-magnitude parameter constraints.
Future analyses could improve upon this work for more valid, yet significantly slower and computationally expensive results.
Specifically, repeated calculations with the more-robust Bayesian parameter-estimation analysis could be performed.
One could also study higher PN order-corrections and higher spin corrections beyond $\mathcal{O}(\epsilon_3,\beta,\chi^2)$ to the gravitational waveform. 
Another avenue for future work includes studying beyond-Kerr spacetimes other than those considered here.

\addtocontents{toc}{\protect\end{adjustwidth}}



\phantomsection
\newpage
\part*{Appendix}
\newpage 
\clearpage
\phantomsection
\addcontentsline{toc}{part}{Appendix}

\addtocontents{toc}{\protect\begin{adjustwidth}{0.5cm}{0cm}}

\newpage 
\clearpage
\phantomsection

\appendix

\chapter{$\tilde \Lambda$ versus $\Lambda_{1.4}$}\label{app:LTL14}

Malik \textit{et al}.~\cite{Malik2018} first studied correlations between nuclear and tidal parameters for individual NSs\footnote{This appendix provides supplementary material for Chapter~\ref{chap:ConstrainingNuclearParameters}.}. Given that the tidal parameter measured from GW observations is $\tilde \Lambda$, corresponding to the mass-weighted average of two tidal parameters in a binary, the authors of Ref.~\cite{Malik2018} assumed the masses of the two NSs in GW170817 to be $m_1 = 1.4M_\odot$ and $m_2 = 1.33 M_\odot$. Next they studied correlations between $\tilde \Lambda$ in such a binary and $\Lambda_{1.4}$, representing the tidal deformability for an individual NS with a mass of $1.4M_\odot$. 

The above assumption can be dangerous because the individual mass measurements of GW170817 are not very accurate. Although the chirp mass has been measured with great accuracy as $\mathcal{M}=(m_1m_2)^{3/5} (m_1 + m_2)^{-1/5}=1.188^{+0.004}_{-0.002} \text{ M}_{\odot}$, the mass ratio varies as $q = m_2/m_1 \in \lbrack 0.73 , 1.00 \rbrack$~\cite{Abbott2017}. 

The top panel of Fig.~\ref{fig:LTofq_L14} presents the $\tilde \Lambda$--$\Lambda_{1.4}$ correlation for various $q$ within the above range with the chirp mass fixed to $\mathcal{M} = 1.188M_\odot$, while the bottom panel shows the absolute fractional difference from the linear fit. Observe that a strong correlation exists between $\tilde \Lambda$ and $\Lambda_{1.4}$ for any $q$. The maximum fractional error for this case is $\sim 5$\%, with a correlation coefficient of $C=0.998$. 
On the other hand, once we include the hybrid EoSs discussed in more detail in Appendix~\ref{app:hybrid}, one clearly sees a large deviation from the correlation with other EoSs, with the fractional difference reaching up to 60\%.
\setlength{\belowcaptionskip}{-10pt}
\begin{figure}
\begin{center} 
\includegraphics[width=0.7\columnwidth]{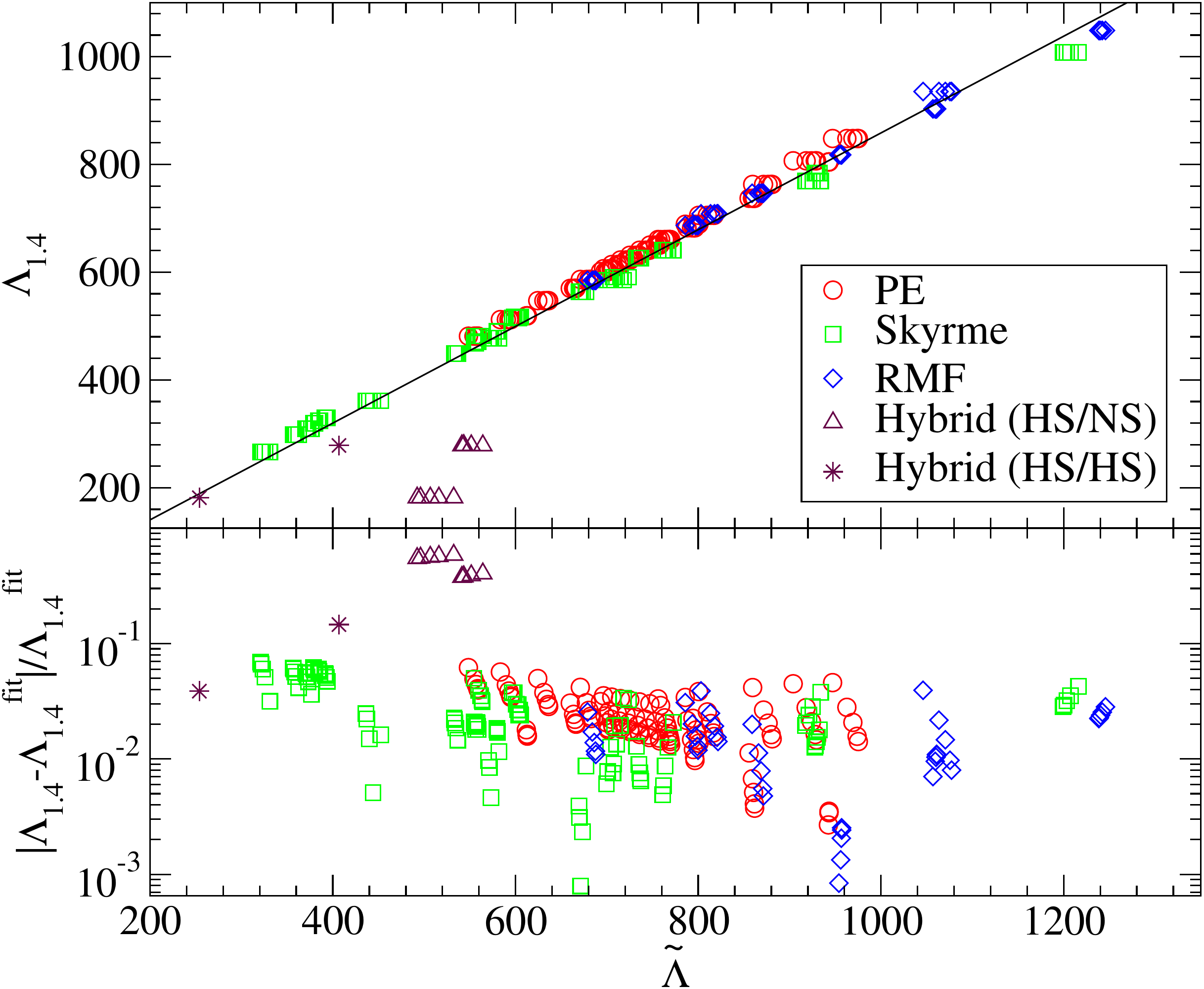}
\end{center}
\caption[Correlations between $\tilde\Lambda$ and $\Lambda_{1.4}$]{(Top) Correlation between mass-weighted average tidal deformability $\tilde{\Lambda}$ and $\Lambda_{1.4}$ (individual tidal deformability at the mass of $1.4 \text{ M}_{\odot}$) for various EoSs each evaluated at mass ratios $q=0.73$, $0.80$, $0.87$, $0.93$ and $1.00$. The chirp mass is fixed to be the measured value of $\mathcal{M} = 1.188M_\odot$. (Bottom) Fractional difference from the fit for each EoS. Notice how the HS EoSs interrupt the universality between the two parameters by up to 60\% (5\% maximal percent difference in the absence of hybrid EoSs). 
}
\label{fig:LTofq_L14}
\end{figure} 
\setlength{\belowcaptionskip}{0pt}

The behavior in Fig.~\ref{fig:LTofq_L14} can be understood from Fig.~\ref{fig:LTofq}, where we show $\tilde \Lambda$ against $q$ with $\mathcal{M}$ fixed to the measured value for GW170817. If we do not consider hybrid EoSs, $\tilde \Lambda$ is insensitive to $q$~\cite{Radice2018,Burgio2018}, which is the origin of the strong correlation in the $\tilde \Lambda$--$\Lambda_{1.4}$ relation. On the other hand, for hybrid EoSs considered here, GW170817 can be either HS/HS or HS/NS when the mass ratio is close to unity\footnote{We note that hybrid EoSs considered in~\cite{Paschalidis2018} admit either NS/NS or HS/NS for GW170817.}. 
Thus, one finds a significant drop in $\tilde \Lambda$ as one increases $q$~\cite{Burgio2018}, which changes the $\tilde \Lambda$--$\Lambda_{1.4}$ relation drastically.
\begin{figure}
\begin{center} 
\includegraphics[width=0.7\columnwidth]{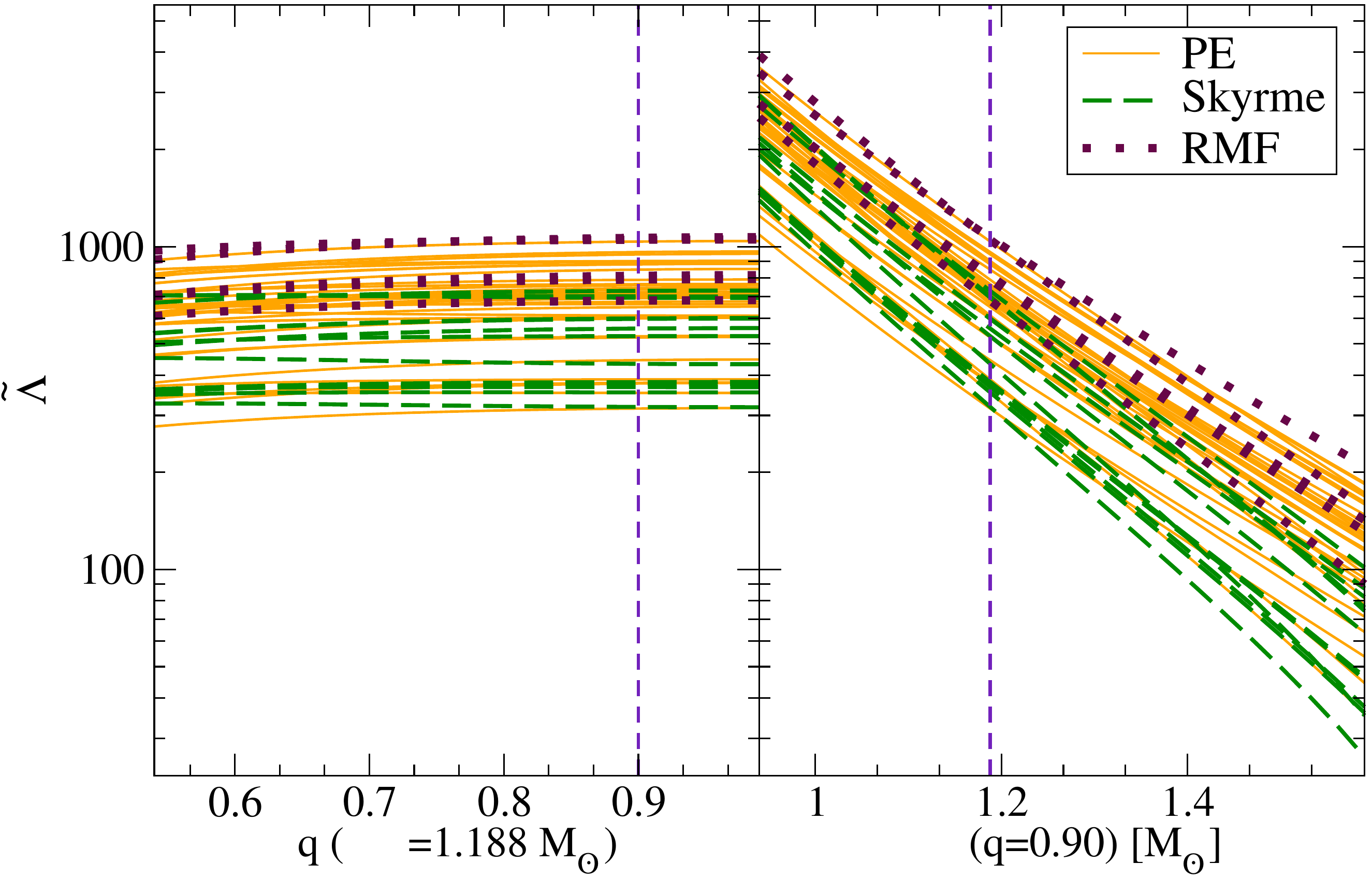}
\end{center}
\caption[$\tilde\Lambda$ as a function of mass ratio]{$\tilde{\Lambda}$ for a representative set of EoSs as a function of mass ratio $q$ in GW170817's observed range of $q \in \lbrack 0.73 , 1.00 \rbrack$ with the chirp mass fixed to $\mathcal M = 1.188M_\odot$. Notice how $\tilde{\Lambda}$ only varies slightly in this region of interest for Skyrme, RMF, and PEs. Hybrid EoSs on the other hand admit two different configurations for GW170817, HS/NS (solid maroon) and HS/HS (dashed maroon), with the former giving a significant variation in $\tilde{\Lambda}$.
For demonstration purposes, the black vertical line corresponds to mass ratio $q=0.995$, where it can be seen that two different binary configurations emerge, discussed in more detail in Appendix~\ref{app:hybrid}.
}
\label{fig:LTofq}
\end{figure} 

\newpage
\chapter{Repeated analysis without PEs}\label{app:sansHybrid}

\begin{figure}
\begin{center} 
\includegraphics[width=0.7\columnwidth]{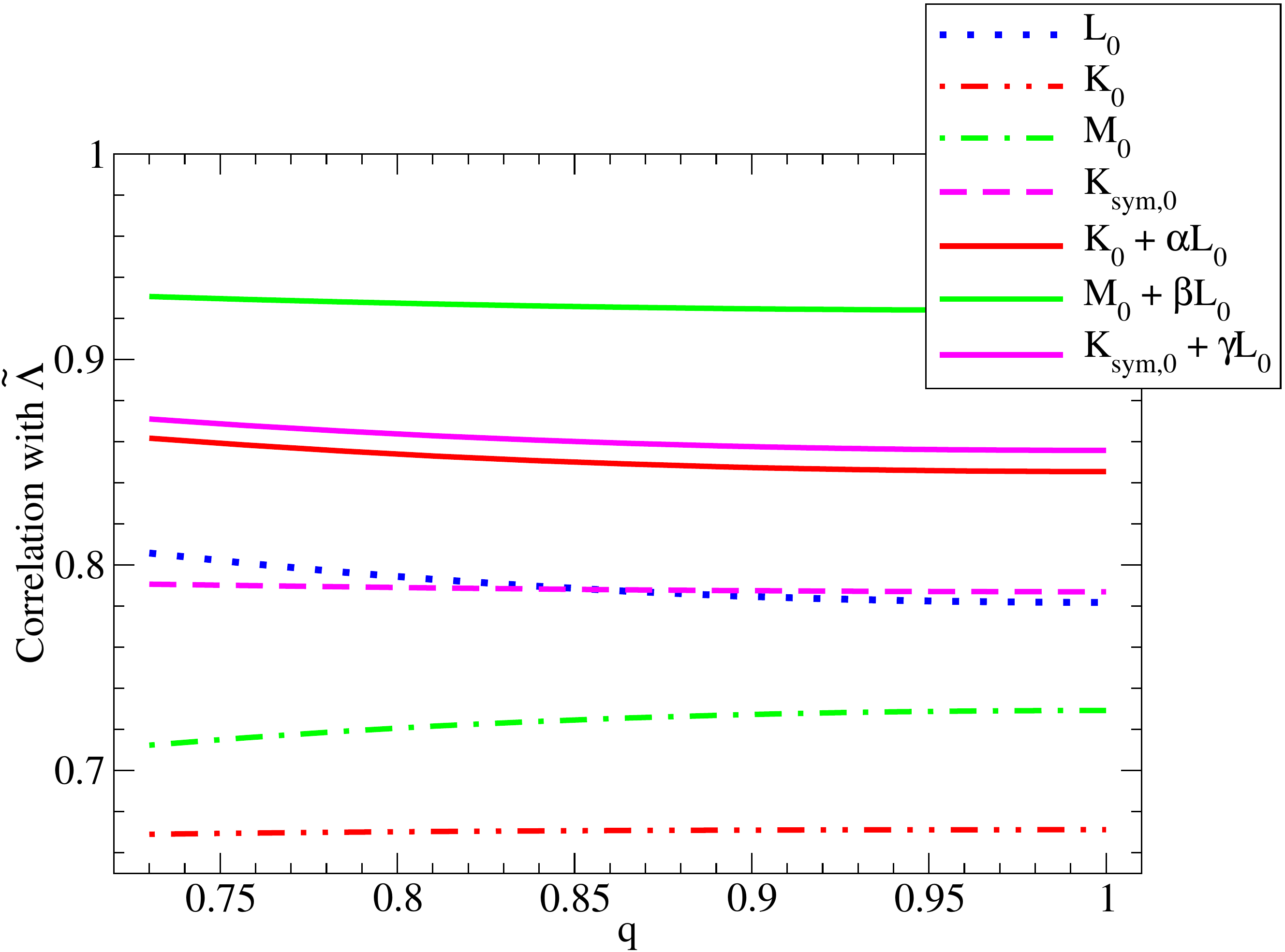}
\end{center}
\caption[Correlations with nuclear parameters sans PEs]{
Similar to Fig.~\ref{fig:correlation} upon the removal of PEs. Observe that the correlations for linear combinations involving lower order parameters improve by up to 55\%, while linear combinations with high order parameter $K_{\text{sym},0}$ shows slightly diminished, yet comparable, correlations. Observe also that the correlations are insensitive to $q$.
}
\label{fig:alamCorrelations}
\end{figure} 

\begin{table*}
\centering
\resizebox{\linewidth}{!}{%
\begin{tabular}{|| C{3.4cm} || C{5.6cm} | C{5.6cm}||@{}m{0pt}@{}} 
 \hline
 \diagbox[width=3.8cm]{\hspace*{0pt} \raisebox{2pt}{$L_0$ \small{[MeV]}}}{\raisebox{-5pt}{$\tilde{\Lambda}$} \hspace*{0.5cm}} & 70-720~\cite{Abbott2018} & 279--822~\cite{Coughlin:2018fis} \\
\hline
\hline
 40--62~\cite{Lattimer2013,Lattimer2014,Tews2017} & \makecell{\\ \Krange{161}{309}  \\ \\ \Mrange{1506}{3506} \\ \\ \Ksymrange{-327}{140} \\ \\}  & \makecell{\\ \Krange{182}{324} \\ \\ \Mrange{1851}{3723} \\ \\ \Ksymrange{-246}{190} \\ \\ } \\
 \hline
 30--86~\cite{Oertel2017} & \makecell{\\ \Krange{134}{320}  \\ \\ \Mrange{1131}{3662} \\ \\ \Ksymrange{-394}{168} \\ \\} & \makecell{\\ \Krange{155}{335} \\  \\ \Mrange{1476}{3880} \\  \\ \Ksymrange{-313}{218}\\ \\ }\\
 \hline
\end{tabular}
}
\caption[Nuclear constraints sans PEs]{
Similar to Table~\ref{tab:Constraints}, when excluding PEs, and only considering the ``first method" of computing nuclear parameter constraints.
Observe how the bounds upon removal of PEs show drastic improvement - showing closer agreement with Ref.~\cite{Malik2018} (in addition to uncertainty from EoS variation), and also highlighting the effect of utilizing a large set of additional EoSs.
The exception is high order nuclear parameter $K_{\text{sym},0}$ - showing weakened constraints due to the inclusion of uncertainty in $L_0$.
}\label{tab:alamConstraints} 
\end{table*}

In this appendix, we study the effect of PEs on nuclear parameter bounds by re-analyzing them without including such EoSs\footnote{This appendix provides supplementary material for Chapter~\ref{chap:ConstrainingNuclearParameters}.}.
This way,  we can directly compare our results with those in Ref.~\cite{Malik2018} which did not include these additional EoSs.
Figure~\ref{fig:alamCorrelations} once again presents correlations between $\tilde{\Lambda}$ and 
linear combinations of nuclear parameters as a function of  mass ratio. 
Here, for comparison purposes we choose $\alpha=1.10$, $\beta=15.62$, and $\gamma=2.81$ such that correlations become maximum, as was done in Ref.~\cite{Malik2018}.
Observe that correlations with $\tilde{\Lambda}$ remain almost constant throughout the entire region of allowable mass ratios.
In addition, note how correlations for linear combination involving $K_0$ and $M_0$ are increased by up to  $55\%$ from Fig.~\ref{fig:correlation} which includes PEs, while linear combinations with higher order nuclear parameter $K_{\text{sym},0}$ interestingly shows a small decrease in correlation, yet remains comparable.
This is revealing of the flexible nature of the $K_{\text{sym},0}$ nuclear parameter.

We now derive constraints on nuclear parameters without PEs. 
Following the procedure outlined in Sec.~\ref{sec:constraints}, new bounds on $K_0$, $M_0$, and $K_{\text{sym},0}$ are calculated for a central mass ratio of $q=0.87$, and summarized in Table~\ref{tab:alamConstraints}. Comparing this with Table~\ref{tab:Constraints}, one sees that the additional PEs significantly weaken estimated constraints for low order nuclear parameters $K_0$ and $M_0$, and interestingly, improve them for high-order nuclear parameter $K_\mathrm{sym,0}$.
Here we find results somewhat agreeable to what was found in Malik et al~\cite{Malik2018}, however enlarged due to the addition of EoS variation uncertainties. 

\newpage
\chapter{Hybrid quark-hadron stars}\label{app:hybrid}
In this appendix, we investigate the use of an additional valid class of EoS: hybrid quark-hadron stars based on Ref.~\cite{Paschalidis2018}\footnote{This appendix provides supplementary material for Chapter~\ref{chap:ConstrainingNuclearParameters}.}.
Here, the low-density nucleonic matter region of PEs transition into a high-density quark matter phase in a given transitional energy density region $\epsilon_1 \leq \epsilon \leq \epsilon_2$. 
For our purposes, we consider Set I quark matter EoSs, where the pressure following transition is given by~\cite{Alford:2017qgh} (see also~\cite{Montana:2018bkb, 1971SvA....15..347S,Zdunik:2012dj,Alford:2013aca}):
\begin{equation}
P(\epsilon) =    \left\{
\begin{array}{ll}
      P_{\text{tr}} & (\epsilon_1  \leq \epsilon \leq \epsilon_2) \\
      P_{\text{tr}}+c_{\text{s}}^2(\epsilon-\epsilon_2) & (\epsilon > \epsilon_2)
\end{array} 
\right. 
\end{equation}
with $c_s$ being the constant speed of sound in the quark matter, $\epsilon_1$ and $\epsilon_2$ characterizing the energy density ``jump" $\epsilon_2 - \epsilon_1 \equiv \epsilon_1 j$, and $P_{\text{tr}}$ representing the transition pressure, such that the low density hadronic matter's energy density equals $\epsilon_1$. 
In this paper, we adopt the ACS-II parameterization in~\cite{Paschalidis2018} as $P_{\text{tr}} = 1.7 \times 10^{35}$dyn/cm$^2$, $\epsilon_2 = 8.34 \times 10^{14}$g/cm$^3$ and $c_s^2=0.8$ with $j=0.8$ or 1. 
 
As we show in Fig.~\ref{fig:LTofq}, strong phase transitions in the star admit a secondary stable HS configuration (denoted HS/HS).
HSs  evaluate to a reduction in tidal deformability $\tilde{\Lambda}$ from their NS-branch counterparts, thus altering universal relations accordingly.
Here, we examine how this additional possibility of binary HSs and the choice of fiducial nuclear matter EoS impacts correlations between $\tilde{\Lambda}$ and nuclear parameters.

Figure~\ref{fig:fiducialVariation} investigates this phenomena by choosing 3 different fiducial nuclear matter EoSs with soft ($\tilde{\Lambda}\approx 465$), intermediate ($\tilde{\Lambda} \approx 800$), and stiff ($\tilde{\Lambda} \approx 1045$) representative values of tidal deformability for $\mathcal{M} = 1.188M_\odot$ and $q=0.995$. 
Next, HS EoSs are formulated, and new universal relations are derived - including both stellar configurations at high values of $q$, as can clearly be seen by the dashed vertical line in Fig.~\ref{fig:LTofq}.
Observe how the choice of fiducial nuclear matter EoS impacts the universal relations differently depending on which combination of nuclear parameters is used.
For example, use of the stiff fiducial EoS compared to the intermediate one results in a small decrease in correlation for $K_0 + \alpha L_0$, a negligible decrease for $M_0 + \beta L_0$, and a large decrease for $K_{\text{sym},0} + \alpha L_0$.
Alternatively, choice of the soft fiducial EoS results in medium decreases in correlation for $K_0 + \alpha L_0$ and $M_0 + \alpha L_0$, and an \textit{increase} in correlation for $K_{\text{sym},0} + \alpha L_0$.

In conclusion, we find that the use of valid hybrid quark-hadron star EoSs in universal relations can influence universality in unexpected ways. Thus, the bounds derived in Table~\ref{tab:Constraints} are strictly valid only for NSs, and they are subject to change once one includes the possibility for HSs.
Refer also to Ref.~\cite{Montana:2018bkb} for a more detailed analysis of hybrid star EoSs in conjunction with GW170817.

\vspace*{8cm}
\begin{figure*}[!t]
\begin{center} 
\includegraphics[width=\columnwidth]{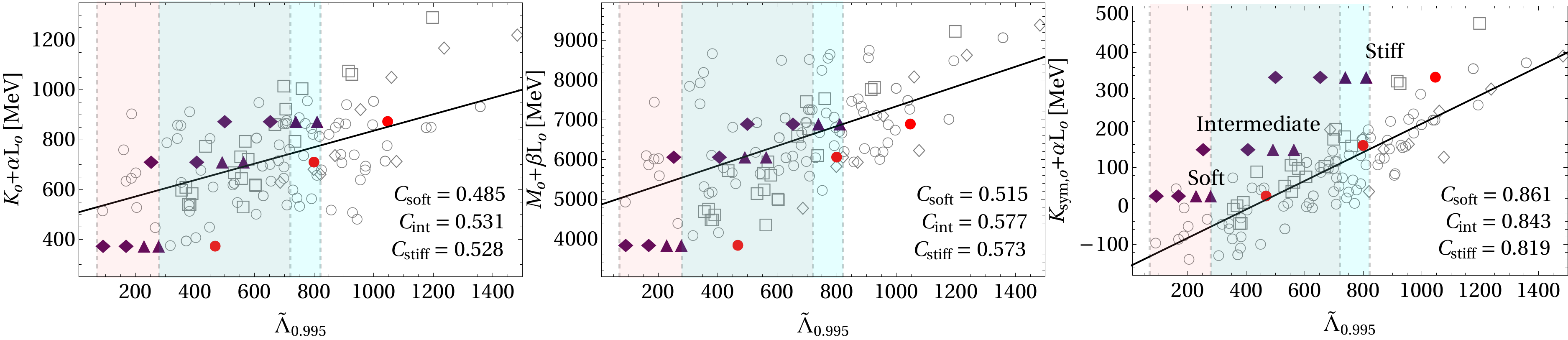}
\end{center}
\caption[Nuclear correltions with hybrid stars]{
Scatter plots demonstrating variation in correlation at $q=0.995$ (with $\mathcal{M} = 1.188M_\odot$) when introducing HS EoSs based on 3 different fiducial nuclear matter EoSs with $\tilde \Lambda \approx 465$, $\tilde \Lambda \approx 800$, $\tilde \Lambda \approx 1045$, represented by filled red circles. 
These correspond to soft, intermediate and stiff fiducial EoSs respectively. 
These are followed by two different star configurations of hybrid HS/NS values (purple triangle), and HS/HS (purple diamond) with a reduction in $\tilde{\Lambda}$, as is shown in Fig.~\ref{fig:LTofq} for two different HS EoSs, corresponding to the $j=0.8$ and $j=1.0$ configurations.
As demonstrated in Fig.~\ref{fig:LTofq}, $q=0.995$ clearly admits both HS/NS and HS/HS binary configurations.
Shown in gray as reference are the PE, Skyrme, and RMF EoSs, irrelevant to this investigation.
Displayed in the bottom right corner is the correlation between nuclear parameter combinations and $\tilde{\Lambda}$ when imposing soft, intermediate, and stiff fiducial EoSs in the generation of HS structure.
Notice how the choice of fiducial EoS alters correlations between $\tilde{\Lambda}$ and combinations of nuclear parameters differently.
This indicates that potential HS EoSs could impact nuclear bounds significantly.
}
\label{fig:fiducialVariation}
\end{figure*} 


\newpage
\chapter{Original versus restricted sets of EoS}
\label{app:EoS-comparison}

\begin{figure}
\begin{center} 
\includegraphics[width=0.7\columnwidth]{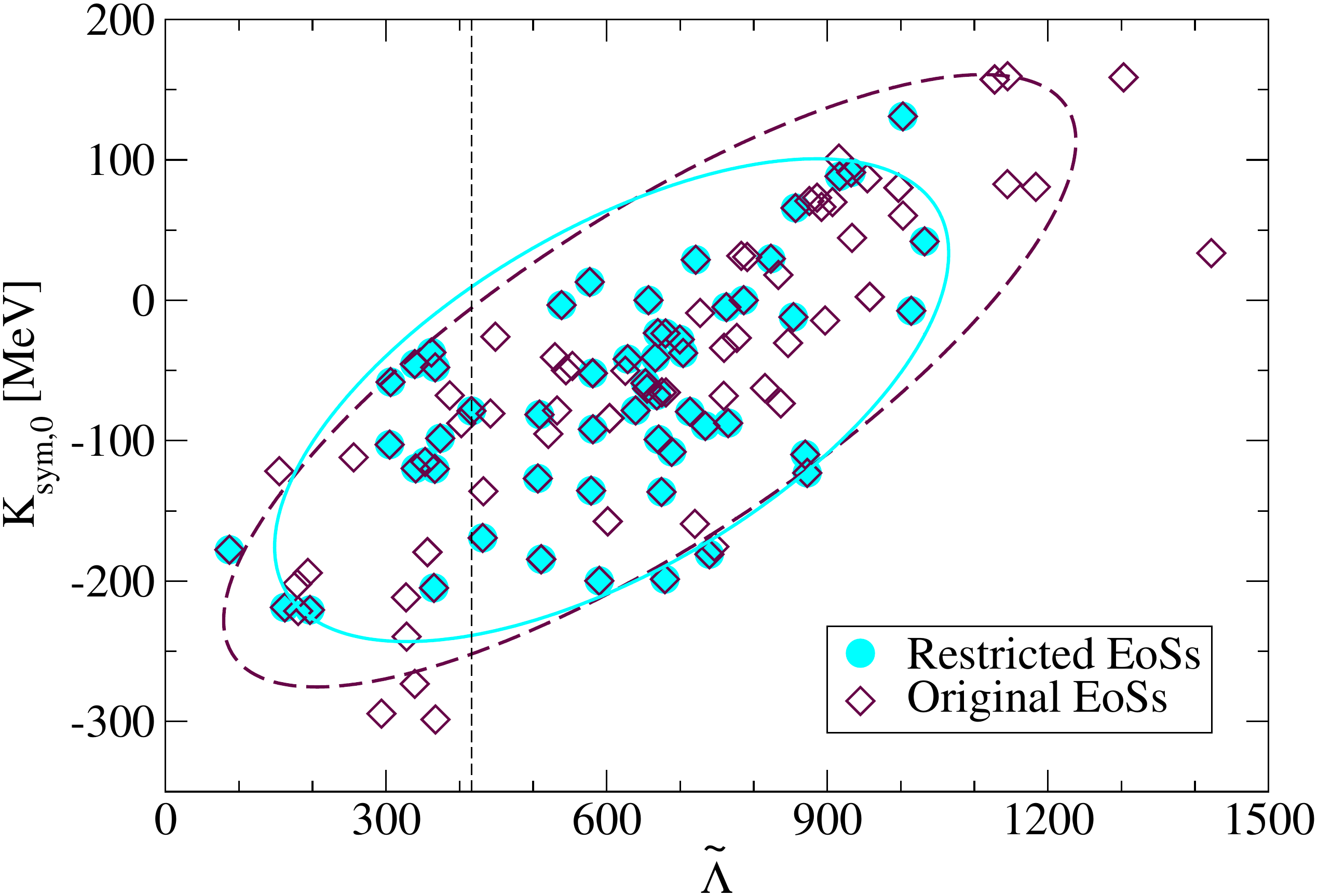}
\end{center}
\caption[Restricted and un-restricted EoS $K_{\text{sym},0}-\tilde\Lambda$ PDFs]{Comparison of the two-dimensional $K_{\text{sym},0}-\tilde\Lambda$ correlations when using (i) the reduced set of EoSs taking into account the nuclear parameter correlations found in Ref.~\cite{Tews2017} (filled circle) and (ii) the original set of EoSs used in Ref.~\cite{Zack:nuclearConstraints} (open diamond).
Depicted by their respective 90\% confidence ellipses, we observe that the reduced set of EoSs shrinks the probability distribution in the direction of correlation, though such a set does not appreciably change the width.
The systematic errors, computed to be the 90\% width of the $K_{\text{sym},0}$ probability distribution evaluated at the fiducial value of $\tilde\Lambda$ (depicted by the vertical dashed line), are seen to be both $\sim104$ MeV, independent of which set of EoSs are used.
However, the overall errors \textit{are} observed to be reduced for the reduced set.
}
\label{fig:eosCompare0}
\end{figure} 

In this appendix, we show how the restriction of EoSs described in Sec.~\ref{sec:eos} used in Chapter~\ref{chap:ConstrainingNuclearParameters} impacts our observations, as compared to the original, unrestricted set of EoSs\footnote{This appendix provides supplementary material for Chapter~\ref{chap:ConstrainingNuclearParameters}.}.
Figure~\ref{fig:eosCompare0} shows a comparison between the two-dimensional probability distributions $P(K_{\text{sym},0},\tilde\Lambda)$ resulting from each set of EoSs.
We see that while the restriction to EoSs does indeed shrink the 90\% confidence intervals in the direction of correlation, the widths are approximately equal at the fiducial value of $\tilde\Lambda$ (where the systematic errors are analyzed).
This shows that while using a subset of EoSs may reduce the overall error\footnote{We indeed observed large reductions in the overall errors found in Fig.~\ref{fig:OverallVsSystematic} when using the restricted EoSs rather than the original ones, while the level of systematics stayed constant at $\sim104$ MeV.}, the level of systematic errors will remain mostly fixed.
Our result is consistent, for example, with Ref.~\cite{Carriere2003}, which found that the correlation between $L_0$ and the radius of a $1.4\text{ M}_\odot$ neutron star was weak because of the contribution of the high-density component of the EoS.

\newpage
\chapter{Multiplicative combinations of nuclear parameters}\label{app:multiplicative}

In this appendix, we discuss the feasibility of using multiplicative combinations of nuclear parameters, such as $K_0 L_0^\eta$, rather than the linear combinations such as $K_0+\alpha L_0$ considered in Refs.~\cite{Alam2016,Malik2018} and the main analysis of Chapter~\ref{chap:ConstrainingNuclearParameters}\footnote{This appendix provides supplementary material for Chapter~\ref{chap:ConstrainingNuclearParameters}.}.
Here we consider the following multiplicative combinations for comparison purposes: $K_0L_0^{\eta}$, $M_0L_0^{\mu}$, and $K_{\text{sym},0}L_0^{\nu}$, where coefficients $\eta$, $\mu$, and $\nu$ are similarly chosen to achieve maximal correlation. Such multiplicative combinations are similar to those considered in~\cite{Sotani:2013dga,Silva:2016myw}.

Figure~\ref{fig:CorrOfMmult} presents the correlations between $\tilde\Lambda$ and all 6 multiplicative and linear combinations of nuclear parameters considered in this analysis.
While the two classes of nuclear parameter combinations produce very similar correlations with $\tilde\Lambda$, we observe that the linear cases slightly outperform the multiplicative cases for nearly all values of chirp mass.
Similarly, repeating the analysis\footnote{Because the two-dimensional probability distribution is now between $\tilde\Lambda$ and $K_{\text{sym},0}L_0^{\nu}$, an additional marginalization over $L_0$: $\int^{\infty}_{-\infty}P(K_{\text{sym},0}L_0^\nu)P(L_0)dL_0$ must be performed ($P(L_0)$ is an additional prior distribution on $L_0$ given by Refs.~\cite{Lattimer2013,Lattimer2014,Tews2017,Oertel2017}) in order to extract the posterior distribution on $K_{\text{sym},0}$.} found in Sec.~\ref{sec:chirpmass} returns constraints on $K_{\text{sym},0}$ to be slightly worse than that considered in the the main analysis, due to the additional inclusion of uncertainties from nuclear parameter $L_0$. 
One arrives at a similar conclusion if one uses a linear combination with $\gamma \neq 0$~\cite{Zack:nuclearConstraints}.

\begin{figure}
\begin{center} 
\begin{overpic}[width=0.7\columnwidth]{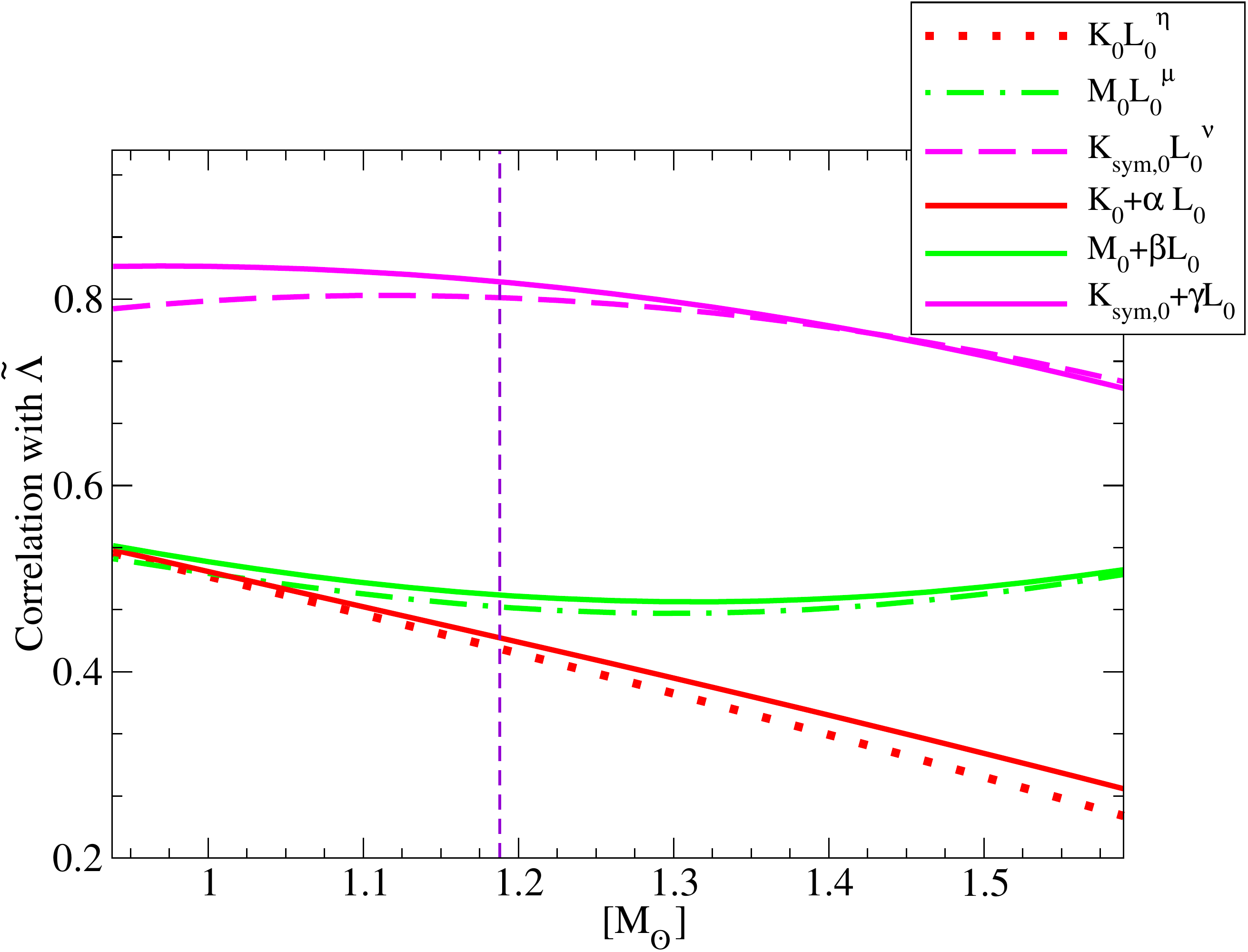}
\put(130,3){\small$\mathcal{M}$}
\end{overpic}
\end{center}
\caption[Correlations for multipliplicative combinations of nuclear parameters]{
Similar to Fig.~\ref{fig:CorrOfMmult}, but for the comparison between multiplicative and linear combinations of nuclear parameters: $K_0L_0^{\eta}$, $M_0L_0^{\mu}$, $K_{\text{sym},0}L_0^{\nu}$, $K_0+\alpha L_0$, $M_0+\beta L_0$, and $K_{\text{sym},0}+\gamma L_0$.
Here, parameters $\eta$, $\mu$, $\nu$, $\alpha$, $\beta$, and $\gamma$ are chosen such that the correlations with $\tilde\Lambda$ are maximal at each value of chirp mass.
Observe how both the linear and multiplicative combinations of nuclear parameters produce similar correlations with $\tilde\Lambda$, though the former outperforms the latter marginally for nearly all values of chirp mass.
}
\label{fig:CorrOfMmult}
\end{figure} 

From this evidence, we conclude with the remarks that the multiplicative combinations of nuclear parameters offer nothing new in terms of enhanced constraints on nuclear parameters.
The multiplicative combinations of nuclear parameters slightly under-perform their linear combination counterparts in terms of correlations with $\tilde\Lambda$.

\newpage
\chapter[Example computation of $K_{\text{sym},0}$ PDF]{Example computation of the $K_{\text{sym},0}$ posterior distribution}\label{app:posteriorExample}

In this appendix, we demonstrate the process of computing the posterior distribution on $K_{\text{sym},0}$ (used in Sec.~\ref{sec:chirpmass}) for one value of chirp mass, $\mathcal{M}=1.188\text{ M}_{\odot}$, corresponding to GW170817 on detector O2\footnote{This appendix provides supplementary material for Chapter~\ref{chap:ConstrainingNuclearParameters}.}.
This case corresponds to the large dot in Fig.~\ref{fig:OverallVsSystematic}.
Referring to Figs.~\ref{fig:meanLt} and~\ref{fig:sigmaLt}, we observe that the mean and root-mean-square $\tilde\Lambda$ values for O2 detector sensitivity at $\mathcal{M}=1.188\text{ M}_{\odot}$ are given by $\mu_{\tilde\Lambda}=430.8$ and $\sigma_{\text{O2}}=172.5$, respectively.
This results in a prior distribution on $\tilde\Lambda$ shown in Fig.~\ref{fig:LTofqPrior}, given by
\begin{equation}
\label{eq:PO2}
P_{\text{O2}}(\tilde\Lambda)= \frac{1}{\sqrt{2\pi(172.5)^2}} e^{-(\tilde\Lambda-430.8)^2/2(172.5)^2}.
\end{equation}
We additionally show the true posterior distribution on $\tilde\Lambda$ derived in Ref.~\cite{LIGO:posterior}, which was used as a prior in our original analysis found in Ref.~\cite{Zack:nuclearConstraints} for comparison purposes.

\begin{figure}
\begin{center} 
\includegraphics[width=0.7\columnwidth]{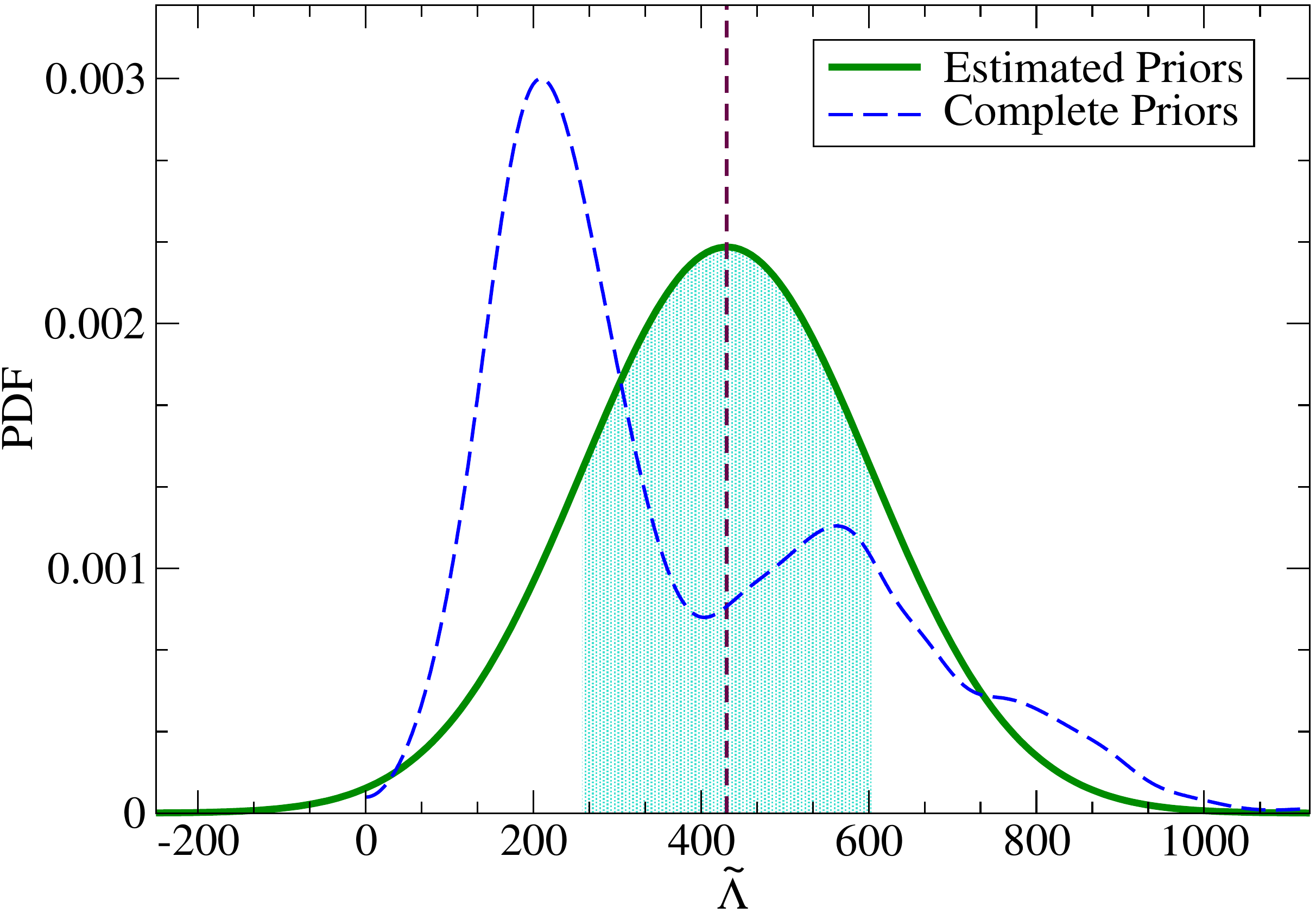}
\end{center}
\caption[Gaussian vs. actual $\tilde\Lambda$ prior PDFs]{
Prior distribution (solid green) on $\tilde\Lambda$ for O2 detector sensitivity with a chirp mass of $\mathcal{M}=1.188\text{ M}_{\odot}$.
This distribution, used to compute posteriors on $K_{\text{sym},0}$, is generated by assuming $\tilde\Lambda$ follows a Gaussian distribution in Eq.~\eqref{eq:PO2} with mean $\mu_{\tilde\Lambda}=430.8$ (dashed vertical line), and root-mean-square $\sigma_{\text{O2}}=172.5$ (cyan shaded region).
These are computed from the GW170817-constrained EoSs from Ref.~\cite{Zack:nuclearConstraints}, and from a simple Fisher analysis respectively.
Additionally shown in the figure is the posterior distribution on $\tilde\Lambda$ derived by the LIGO and Virgo Collaborations in Ref.~\cite{LIGO:posterior} (dashed blue).
}
\label{fig:LTofqPrior}
\end{figure} 

Following along with Sec.~\ref{sec:futureSingle}, we generate the two-dimensional probability distribution between $K_{\text{sym},0}$ and $\tilde\Lambda$, given by Eq.~\eqref{eq:2dPDF}.
We find the covariance matrix and the mean vector to be
\begin{equation}
\mathbf{\Sigma}=\begin{pmatrix} 45610 & 10410 \\ 10410 & 6418 \end{pmatrix}, \quad
\bm{\mu}=\begin{pmatrix} 606.7 \\ -71.16 \end{pmatrix}\,,
\end{equation}
for $\bm x = (\tilde \Lambda, K_{\mathrm{sym,0}})$
This results in the two-dimensional probability distribution between $K_{\text{sym},0}$ and $\tilde\Lambda$ shown in Fig.~\ref{fig:probDistExample}.
The systematic error on $K_{\text{sym},0}$ is then computed by evaluating the 90\% confidence interval width of the distribution in the $K_{\text{sym},0}$-dimension at $\mu_{\tilde\Lambda}=430.8$, corresponding to the mean of the prior distribution in $\tilde\Lambda$.
The resulting (one-sided 90\% confidence level) systematic errors for this case are found to be $\sigma_{\text{sys}}=104.6\text{ MeV}$.

\begin{figure}
\begin{center} 
\begin{overpic}[width=0.7\columnwidth]{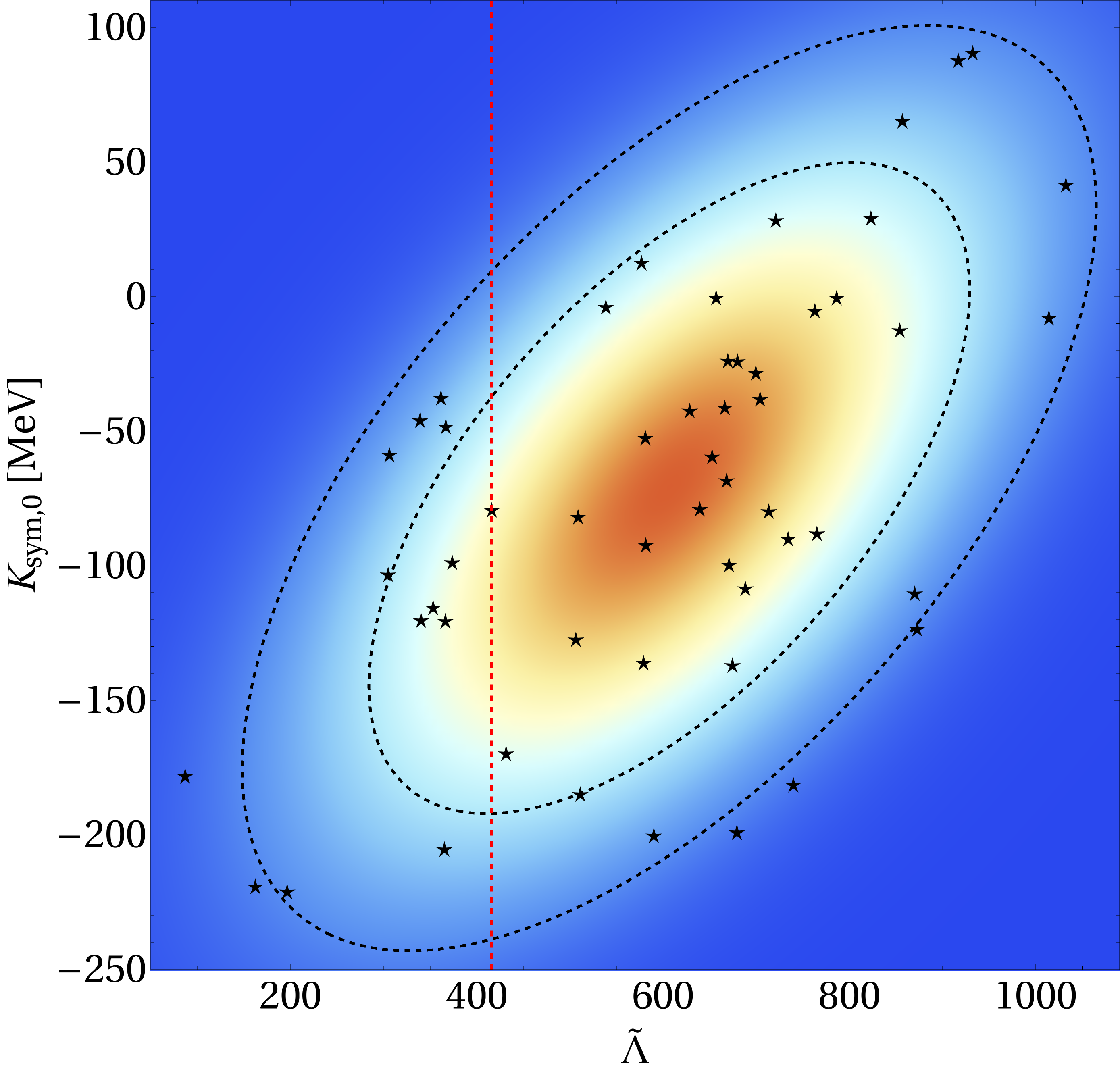}
\put(175,240){\large $68\%$}
\put(270,140){\large $90\%$}
\end{overpic}
\end{center}
\caption[Example two-dimensional PDF between $K_{\text{sym},0}$ and $\tilde\Lambda$]{
Two-dimensional normalized probability distribution between $K_{\text{sym},0}$ and $\tilde\Lambda$ as given by Eq.~\eqref{eq:2dPDF}, with the 68\% and 90\% confidence regions highlighted in black.
Overlaid on the distribution is the set of 58 data points corresponding to the various EoS models used in the analysis.
The vertical dashed line represents the mean of the prior $\tilde\Lambda$ distribution, at which the (one-side 90\% confidence interval) systematic error in $K_{\text{sym},0}$ is computed to be $\sigma_{\text{sys}}=104.6\text{ MeV}$.
}
\label{fig:probDistExample}
\end{figure} 

Next, we obtain the one-dimensional conditional probability distributions on $K_\text{sym,0}$ given tidal deformability observations of $\tilde\Lambda_\text{obs}$.
By following Eq.~\eqref{eq:conditional}, this is simply given by
\begin{equation}
P(K_\text{sym,0}|\tilde\Lambda_\text{obs})= \frac{\text{Exp}\left\lbrack-\frac{\left(K_\text{sym,0}-(0.228(\tilde\Lambda_\text{obs}-606.7)-71.16)\right)^2}{2(63.59)^2}\right\rbrack}{\sqrt{2\pi(63.59)^2}}.
\end{equation}
Finally, the posterior distribution on $K_{\text{sym},0}$ can be computed by combining the one-dimensional conditional probability distributions with the prior distribution on $\tilde\Lambda$, and then integrating over all observations of $\tilde\Lambda$:
\begin{equation}
P(K_{\text{sym},0})=\int\limits^{\infty}_{-\infty}P(K_{\text{sym},0}|\tilde{\Lambda})P_{\text{O2}}(\tilde\Lambda)d\tilde\Lambda.
\end{equation}
Figure~\ref{fig:KsymPosterior} finally displays the resulting posterior distribution on $K_{\text{sym},0}$, with a mean value of $-115^{+75}_{-73} \text{ MeV}$, giving a 90\% confidence interval of $-227\text{ MeV}\leq K_{\text{sym},0}\leq 7\text{ MeV}$ (or a one-sided $90\%$ confidence interval of $117$ MeV corresponding to the maroon dot in Fig.~\ref{fig:OverallVsSystematic}).
Comparing this to the resulting posterior distribution (also shown in Fig.~\ref{fig:KsymPosterior}) using the LVC priors on $\tilde\Lambda$ giving a 90\% confidence interval of $-285\text{ MeV}\leq K_{\text{sym},0}\leq 7\text{ MeV}$, we find that this approximation of $\tilde\Lambda$ priors slightly \textit{underestimates} the errors in $K_{\text{sym},0}$, but otherwise works quite well.
We also see that this approximation of $\tilde\Lambda$ priors skews the distribution less so than the LVC posterior, making it more normally distributed. 
We also note that here, we utilize a restricted set of EoSs, resulting in a slightly more accurate posterior distribution on $K_{\text{sym},0}$.

\begin{figure}
\begin{center} 
\includegraphics[width=0.7\columnwidth]{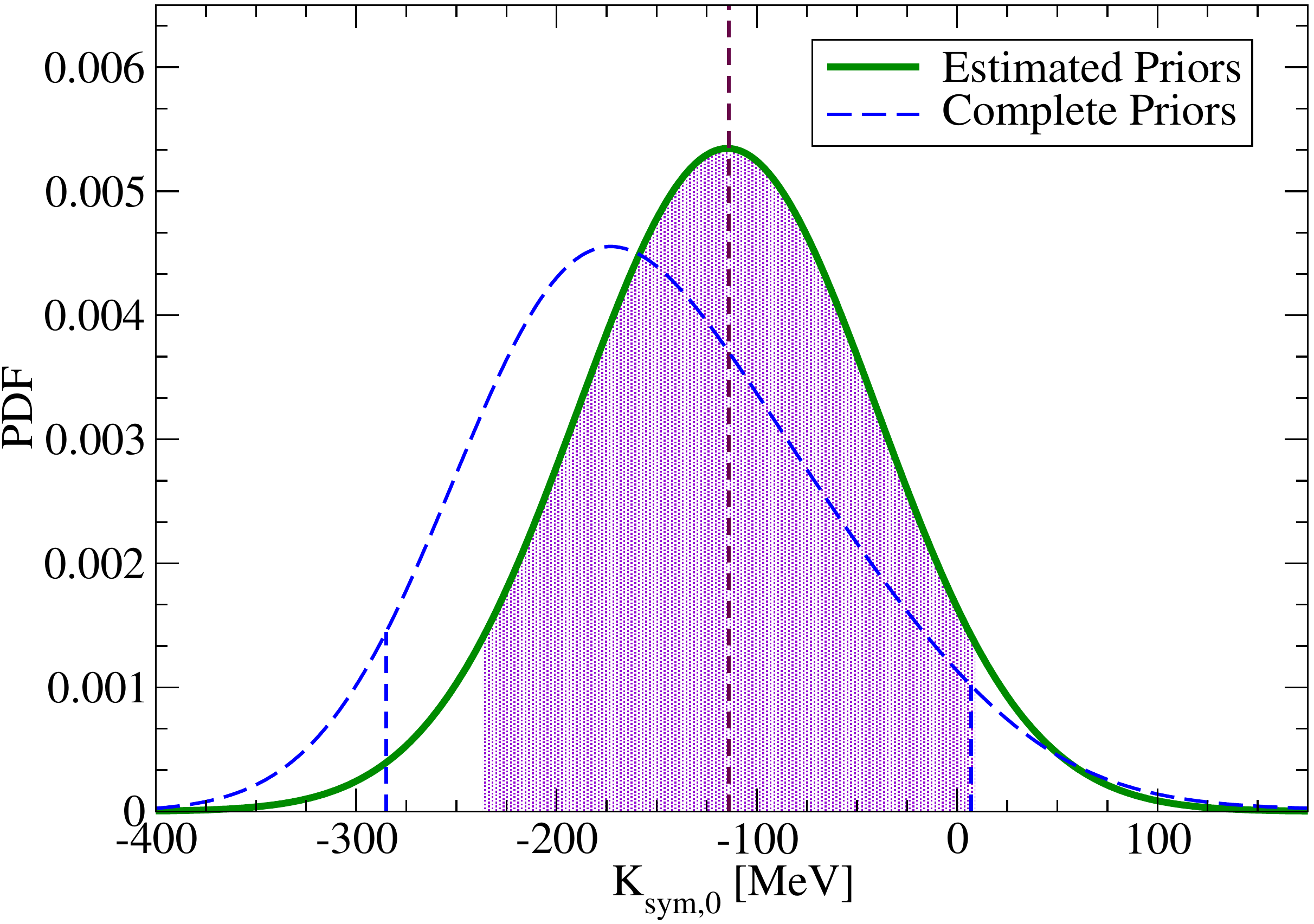}
\end{center}
\caption[Example $K_{\text{sym},0}$ posterior PDF]{
Resulting posterior distribution on $K_{\text{sym},0}$ (solid green), displaying a mean of $-97.63\text{ MeV}$ (dashed magenta) and a one-sided 90\% confidence interval of $116.0\text{ MeV}$ (shaded magenta).
Additionally shown (dashed blue) are the results found when using the full set of 121 EoSs and the full prior distribution in $\tilde\Lambda$, rather than the Gaussian estimation and restricted set of EoSs used here.
Observe that the approximation of Gaussian $\tilde\Lambda$ prior distributions slightly underestimates the uncertainties in $K_{\text{sym},0}$.
}
\label{fig:KsymPosterior}
\end{figure} 




\newpage
\chapter{DEF and MO theory comparison}\label{app:theoryCompare}
In this appendix, we compare results between the DEF~\cite{Damour_1992,Damour_1993} and MO~\cite{Mendes:2016fby} quasi-Brans-Dicke theories of gravity\footnote{This appendix provides supplementary material for Chapter~\ref{chap:Parameterized}.}.
The latter theory is defined by the coupling $\alpha(\varphi)=\text{tanh}(\sqrt{3}\beta_0\varphi)/\sqrt{3}$, while the former relies upon only the first term in the above expansion about $\varphi_0$, namely $\alpha(\varphi)=\beta_0 \varphi$.
Figure~\ref{fig:theoryCompare} compares the results for the PSR J0337~\cite{Ransom:2014xla,Archibald:2018oxs} system from the SEP-violation test, assuming an APR4 EoS\footnote{The BH-NS system constraints were found to be indistinguishable between theories, due to their lack of the horn structure present only in pulsar-WD binaries.}.
Observe how the ``horn" structure\footnote{The horns arise in pulsar-WD systems because certain values of $\alpha_0$ and $\beta_0$ suppresses the dipole term  and deteriorating the constraints~\cite{Anderson:2019eay}.} is more pronounced in MO theory, and the drop-off in the lower left region is shifted.
Otherwise, the two theories predict nearly-identical values. This finding is consistent with that in~\cite{Anderson:2019hio} for the orbital decay rate measurement of pulsar-WD binaries.

\begin{figure}
\begin{center}
\includegraphics[width=.7\columnwidth]{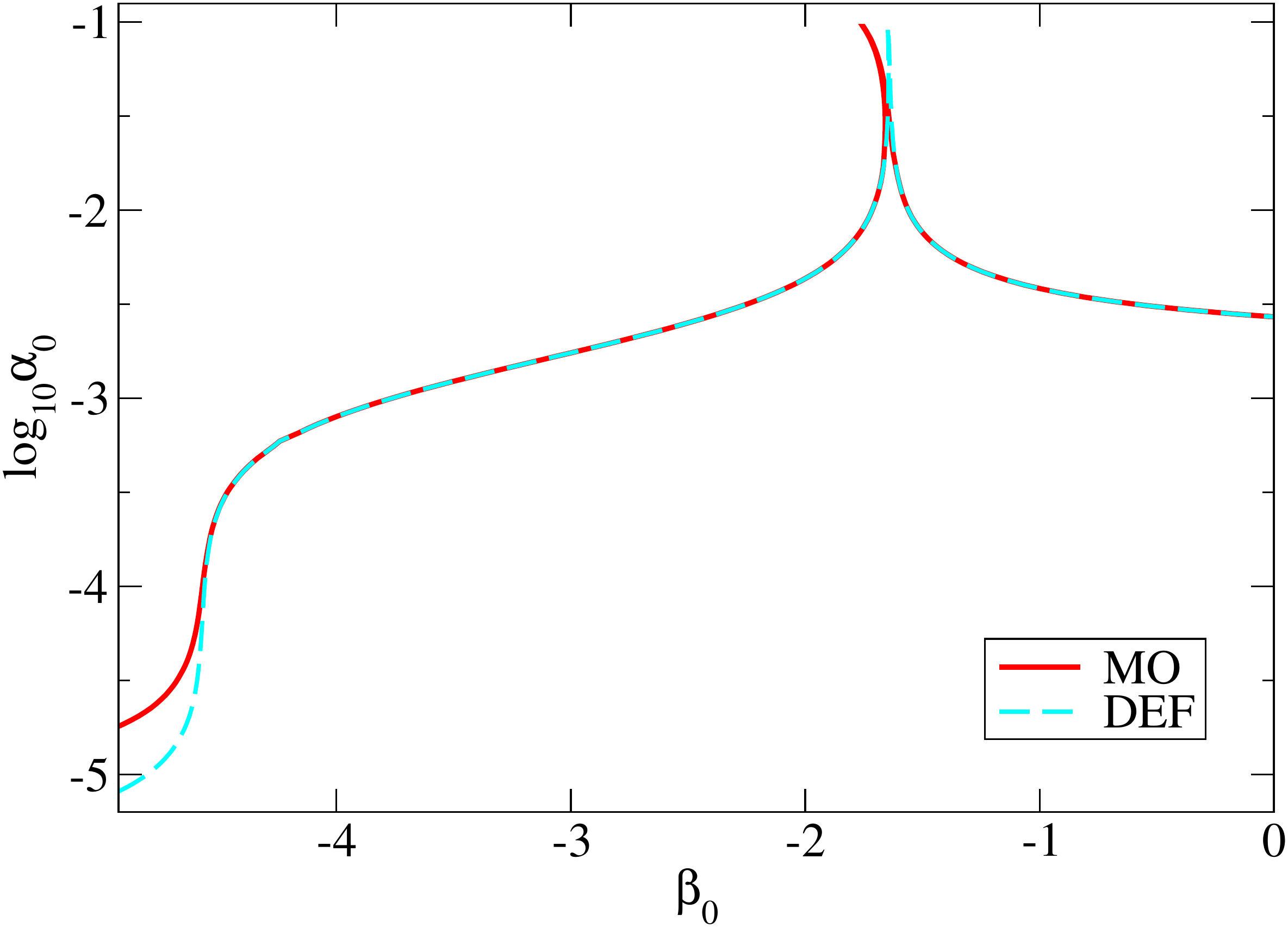}
\caption[DEF vs. MO theory comparison]{
Comparison between the PSR J0337~\cite{Ransom:2014xla,Archibald:2018oxs} constraints formed in the $\alpha_0-\beta_0$ plane for the two quasi-Brans-Dicke theories: DEF and MO.
Bounds formed from GW constraints of BH-NS binaries were found to be indistinguishable from one another in each theory, and have been excluded from this figure.
Observe how for a majority of the contours each theory predicts identical constraints.
The two obvious exceptions being the tilt of the horn, and the drop-off at the lower-left region of the parameter space. 
}\label{fig:theoryCompare}
\end{center}
\end{figure}


\chapter[EoS comparison and spin effects]{Equation of state comparison and the effects of spin in the waveform}\label{app:EosSpinCompare}
In this appendix, we present a comparison between the assumption of different NS EoSs, as well as an investigation into spin effects in the gravitational waveform\footnote{This appendix provides supplementary material for Chapter~\ref{chap:Parameterized}.}.
Figure~\ref{fig:eosCompare} compares bounds in the $\alpha_0-\beta_0$ plane for $10\text{ M}_\odot-1.4\text{ M}_\odot$ BH-NS system detected by ET, assuming three different EoSs: WFF1~\cite{WFF1,WFF1dat,Read2009}, APR4~\cite{APR4,Read2009}, and MPA1~\cite{MPA1,Read2009}.
Such EoSs were chosen to be consistent with the GW observation of binary NSs, GW170817~\cite{LIGO:posterior,TheLIGOScientific:2017qsa,LIGOScientific:2019eut}, and with increasing degrees of stiffness.
We observe that, while the constraints do not differ much, the softer EoSs produce stronger bounds for small values of $\beta_0\lessapprox-3$, while the stiffer EoSs give stronger results for large values of $\beta_0\gtrapprox-3$.
Thus, for consistency, we present results in the main text for the APR4 EoS.

\begin{figure}
\begin{center}
\includegraphics[width=.7\columnwidth]{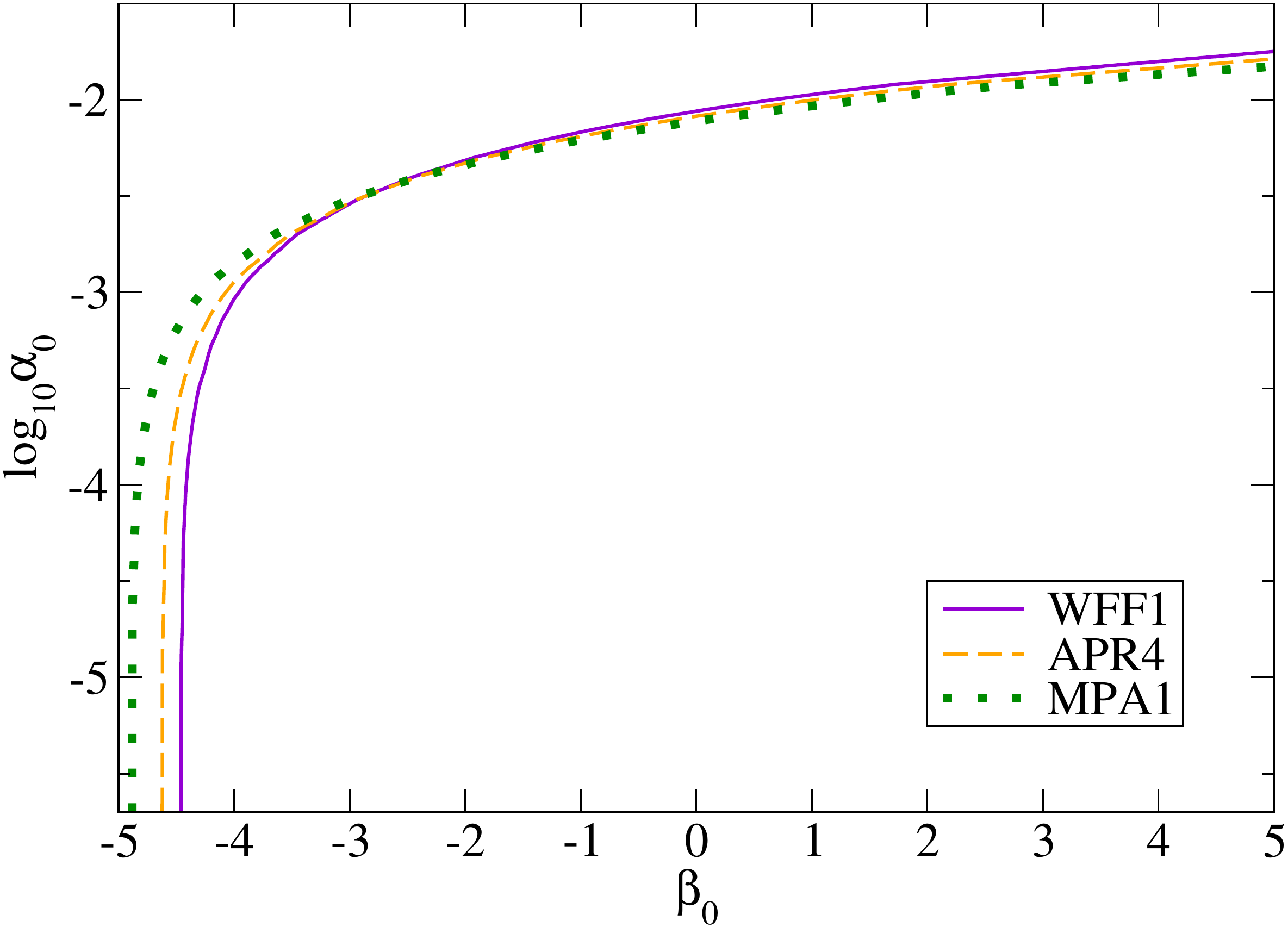}
\caption[DEF equation of state comparison]{Comparison between quasi-Brans-Dicke constraints formed in the $\alpha_0-\beta_0$ plane assuming WFF1, APR4, and MPA1 NS EoSs, all compatible with the observation of GW170817~\cite{LIGO:posterior}.
These constraints were computed assuming a $10\text{ M}_\odot-1.4\text{ M}_\odot$ BH-NS system at 1 Gpc detected by ET.
We observe that softer EoSs give stronger bounds for smaller values of $\beta_0\lessapprox-3$, while stiffer EoSs give stronger constraints for larger values of $\beta_0\gtrapprox-3$.
}\label{fig:eosCompare}
\end{center}
\end{figure}

We now consider the advisement of including spin effects in the gravitational waveform when computing constraints on quasi-Brans-Dicke theories.
Such bounds were computed for binary NS systems found in Ref.~\cite{Shao:2017gwu} with a waveform template not including any spin effects.
In our analysis, we utilize the PhenomD~\cite{PhenomDII,PhenomDI} gravitational waveform which does indeed include spin effects.
Figure~\ref{fig:spinCompare} compares the constraints formed from the ET observation of a $10\text{ M}_\odot-1.4\text{ M}_\odot$ BH-NS system, both with and without spin effects included in the PhenomD waveform.
We see that the latter under-estimates bounds on $\Delta\alpha$ by a factor of 2, indicating the necessity to include spin effects in the waveform.
These discrepancies arise from the correlations between spin, and the other parameters in the waveform, in particular the non-GR parameter, which ultimately increases the uncertainties in parameter estimation.

\begin{figure}
\begin{center}
\includegraphics[width=.7\columnwidth]{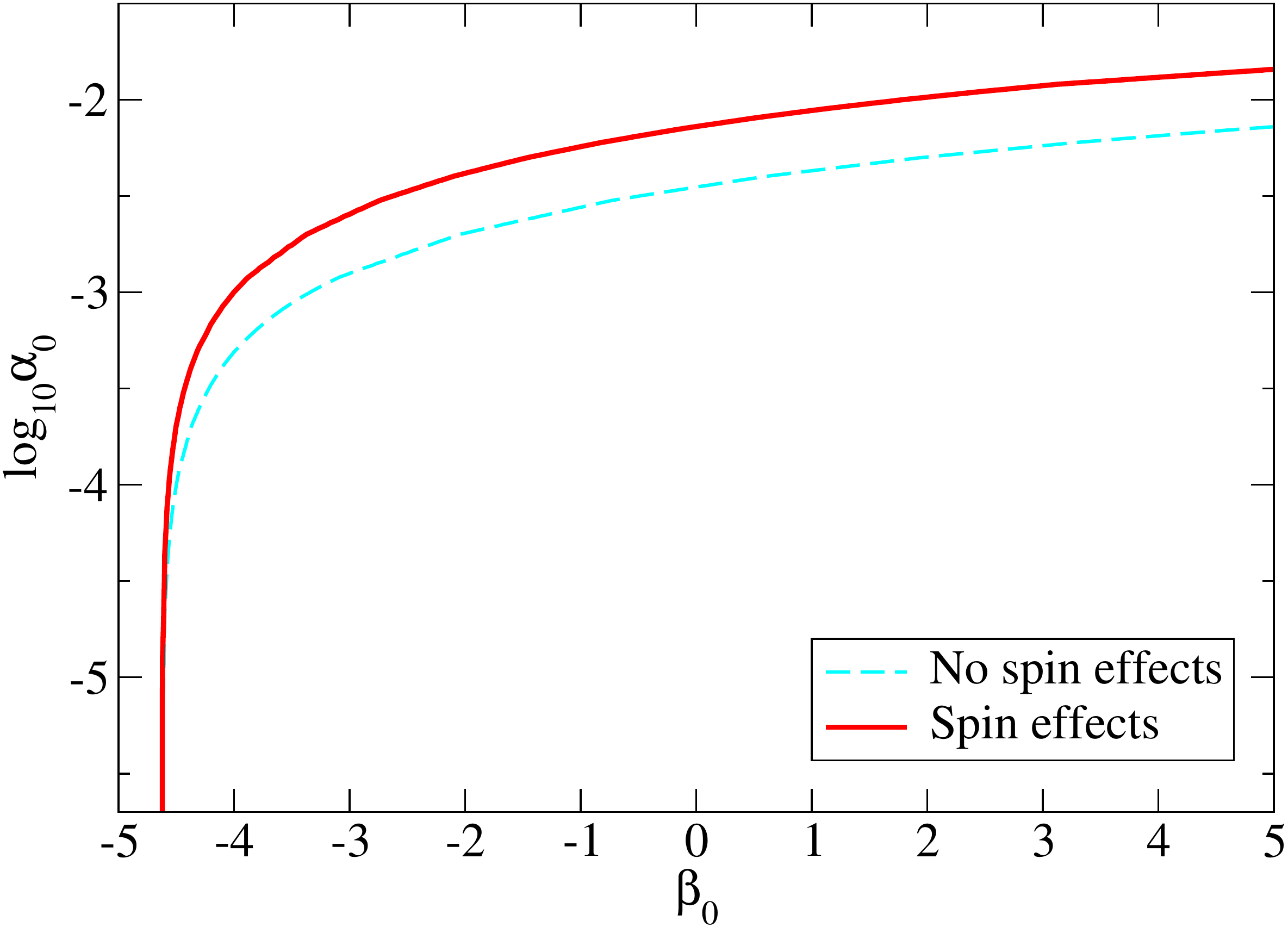}
\caption[DEF spin-effects comparison]{Comparison between the quasi-Brans-Dicke constraints formed in the $\alpha_0-\beta_0$ plane with and without including spin effects in the PhenomD gravitational waveform.
The latter, which produces much stronger constraints, can be seen here to under-estimate $\Delta\alpha$ by a factor of $\sim2$.
The constraints displayed here were computed assuming a $10\text{ M}_\odot-1.4\text{ M}_\odot$ BH-NS system at 1 Gpc detected by ET.
}\label{fig:spinCompare}
\end{center}
\end{figure}



\newpage

\chapter{Arbitrary remnant BH mass and spin corrections}\label{app:arbitraryCorrections}
In this appendix, we display the lengthy corrections to the remnant BH mass and spin given in an arbitrary spacetime metric $g_{\alpha\beta}^\X=g_{\alpha\beta}^\K+\zeta h_{\alpha\beta}^\X$ for general deviation parameter $\zeta$ and perturbation metric $h_{\alpha\beta}^\X$. The perturbation metric is further expanded up to quadratic order in BH spin as $h_{\alpha\beta}^\X=h_{\alpha\beta,0}+h_{\alpha\beta,1}\chi_f+h_{\alpha\beta,2}\chi_f^2$.
In the following expressions, all of the metric components are to be evaluated at $r=6M$.

\begin{gather}
\nonumber\delta M_f^\X=\frac{\mu}{23328 M^2}  \Bigg\lbrack2 \chi _f^\K \Big(62208 \sqrt{3} M^4 h_{\text{tt},0}''+15552 \sqrt{3} M^3 h_{\text{tt},0}'-23328 \sqrt{2} M^3 h_{\text{tt},1}'\\
\nonumber+10368 \sqrt{2} M^3 h_{\text{t$\phi $},0}''+432 \sqrt{3} M^2 h_{\text{tt},0}-3888 \sqrt{2} M^2 h_{\text{tt},1}-2376 \sqrt{2} M^2 h_{\text{t$\phi $},0}'-2592 \sqrt{3} M^2 h_{\text{t$\phi $},1}'\\
\nonumber+288 \sqrt{3} M^2 h_{\phi \phi ,0}''+432 \sqrt{2} M h_{\text{t$\phi $},0}+216 \sqrt{3} M h_{\text{t$\phi $},1}-204 \sqrt{3} M h_{\phi \phi ,0}'-108 \sqrt{2} M h_{\phi \phi ,1}'\\
\nonumber+55 \sqrt{3} h_{\phi \phi ,0}+36 \sqrt{2} h_{\phi \phi ,1}\Big)+\left(\chi _f^\K\right)^2 \Big(-248832 \sqrt{2} M^5 h_{\text{tt},0}'''-64800 \sqrt{2} M^4 h_{\text{tt},0}''\\
\nonumber+124416 \sqrt{3} M^4 h_{\text{tt},1}''-27648 \sqrt{3} M^4 h_{\text{t$\phi $},0}'''-1296 \sqrt{2} M^3 h_{\text{tt},0}'+31104 \sqrt{3} M^3 h_{\text{tt},1}'\\
\nonumber-46656 \sqrt{2} M^3 h_{\text{tt},2}'+12384 \sqrt{3} M^3 h_{\text{t$\phi $},0}''+20736 \sqrt{2} M^3 h_{\text{t$\phi $},1}''-1152 \sqrt{2} M^3 h_{\phi \phi ,0}'''\\
\nonumber+1332 \sqrt{2} M^2 h_{\text{tt},0}+864 \sqrt{3} M^2 h_{\text{tt},1}-7776 \sqrt{2} M^2 h_{\text{tt},2}-4344 \sqrt{3} M^2 h_{\text{t$\phi $},0}'\\
\nonumber-4752 \sqrt{2} M^2 h_{\text{t$\phi $},1}'-5184 \sqrt{3} M^2 h_{\text{t$\phi $},2}'+1332 \sqrt{2} M^2 h_{\phi \phi ,0}''+576 \sqrt{3} M^2 h_{\phi \phi ,1}''\\
\nonumber+880 \sqrt{3} M h_{\text{t$\phi $},0}+864 \sqrt{2} M h_{\text{t$\phi $},1}+432 \sqrt{3} M h_{\text{t$\phi $},2}-741 \sqrt{2} M h_{\phi \phi ,0}'-408 \sqrt{3} M h_{\phi \phi ,1}'\\
\nonumber-216 \sqrt{2} M h_{\phi \phi ,2}'+183 \sqrt{2} h_{\phi \phi ,0}+110 \sqrt{3} h_{\phi \phi ,1}+72 \sqrt{2} h_{\phi \phi ,2}\Big)-72 \sqrt{2} \Big(648 M^3 h_{\text{tt},0}'\\
\nonumber+108 M^2 h_{\text{tt},0}+36 \sqrt{6} M^2 h_{\text{t$\phi $},0}'-3 \sqrt{6} M h_{\text{t$\phi $},0}+3 M h_{\phi \phi ,0}'-h_{\phi \phi ,0}\Big)\\
-720 \sqrt{2} M^2 \chi _f^\K \delta \chi _f^\X-432 \sqrt{3} M^2 \delta \chi _f^\X\Bigg\rbrack
\end{gather}
\begin{gather}
\nonumber\delta\chi_f^\X=\frac{1}{6912}  \Bigg\{746496 \sqrt{6} h_{\text{tt},0}'' M^2+171072 \sqrt{6} h_{\text{tt},0}' M-559872 h_{\text{tt},1}' M+248832 h_{\text{t$\varphi $},0}'' M\\
\nonumber+2592 \sqrt{6} h_{\text{tt},0}+46656 h_{\text{tt},1}+\frac{648 \sqrt{6} h_{\varphi \varphi ,0}}{M^2}+\frac{1512 h_{\varphi \varphi ,1}}{M^2}-62208 h_{\text{t$\varphi $},0}'-31104 \sqrt{6} h_{\text{t$\varphi $},1}'\\
\nonumber+3456 \sqrt{6} h_{\varphi \varphi ,0}''+\frac{1}{M^2 \mu }\Bigg\lbrack\sqrt{3} \left(-9 M-6 \sqrt{2} \mu +\kappa  \sqrt{3}\right)\Big(-746496 h_{\text{tt},0}''' M^5-349920 h_{\text{tt},0}'' M^4\\
\nonumber+186624 \sqrt{6} h_{\text{tt},1}'' M^4-41472 \sqrt{6} h_{\text{t$\varphi $},0}''' M^4-42768 h_{\text{tt},0}' M^3+42768 \sqrt{6} h_{\text{tt},1}' M^3\\
\nonumber-139968 h_{\text{tt},2}' M^3+9936 \sqrt{6} h_{\text{t$\varphi $},0}'' M^3+62208 h_{\text{t$\varphi $},1}'' M^3-3456 h_{\varphi \varphi ,0}''' M^3+2106 h_{\text{tt},0} M^2\\
\nonumber+648 \sqrt{6} h_{\text{tt},1} M^2+11664 h_{\text{tt},2} M^2-4320 \sqrt{6} h_{\text{t$\varphi $},0}' M^2-15552 h_{\text{t$\varphi $},1}' M^2-7776 \sqrt{6} h_{\text{t$\varphi $},2}' M^2\\
\nonumber+3276 h_{\varphi \varphi ,0}'' M^2+864 \sqrt{6} h_{\varphi \varphi ,1}'' M^2+852 \sqrt{6} h_{\text{t$\varphi $},0} M+2376 h_{\text{t$\varphi $},1} M+2592 \sqrt{6} h_{\text{t$\varphi $},2} M\\
\nonumber-1677 h_{\varphi \varphi ,0}' M-630 \sqrt{6} h_{\varphi \varphi ,1}' M-648 h_{\varphi \varphi ,2}' M+389 h_{\varphi \varphi ,0}+162 \sqrt{6} h_{\varphi \varphi ,1}+378 h_{\varphi \varphi ,2}\Big)\Bigg\rbrack\\
\nonumber-\frac{12}{\kappa M^2} \Bigg\lbrack 48 \mu  \left(2592 h_{\text{tt},0}' M^3-216 h_{\text{tt},0} M^2+144 \sqrt{6} h_{\text{t$\varphi $},0}' M^2-48 \sqrt{6} h_{\text{t$\varphi $},0} M+12 h_{\varphi \varphi ,0}' M-7 h_{\varphi \varphi ,0}\right)\\
\nonumber+18 \left(3 M+2 \mu  \sqrt{2}\right) \Big(10368 \sqrt{2} h_{\text{tt},0}'' M^4+2376 \sqrt{2} h_{\text{tt},0}' M^3-2592 \sqrt{3} h_{\text{tt},1}' M^3+1152 \sqrt{3} h_{\text{t$\varphi $},0}'' M^3\\
\nonumber+36 \sqrt{2} h_{\text{tt},0} M^2+216 \sqrt{3} h_{\text{tt},1} M^2-288 \sqrt{3} h_{\text{t$\varphi $},0}' M^2-432 \sqrt{2} h_{\text{t$\varphi $},1}' M^2+48 \sqrt{2} h_{\varphi \varphi ,0}'' M^2\\
\nonumber+44 \sqrt{3} h_{\text{t$\varphi $},0} M+144 \sqrt{2} h_{\text{t$\varphi $},1} M-35 \sqrt{2} h_{\varphi \varphi ,0}' M-12 \sqrt{3} h_{\varphi \varphi ,1}' M+9 \sqrt{2} h_{\varphi \varphi ,0}+7 \sqrt{3} h_{\varphi \varphi ,1}\Big)\\
\nonumber+\left(4 \mu  \sqrt{3}+(M \delta_m+\lambda ) \chi _a+(M+ \delta_m \lambda ) \chi _s\right) \Big(-746496 \sqrt{3} h_{\text{tt},0}''' M^5-349920 \sqrt{3} h_{\text{tt},0}'' M^4\\
\nonumber+559872 \sqrt{2} h_{\text{tt},1}'' M^4-124416 \sqrt{2} h_{\text{t$\varphi $},0}''' M^4-42768 \sqrt{3} h_{\text{tt},0}' M^3+128304 \sqrt{2} h_{\text{tt},1}' M^3\\
\nonumber-139968 \sqrt{3} h_{\text{tt},2}' M^3+29808 \sqrt{2} h_{\text{t$\varphi $},0}'' M^3+62208 \sqrt{3} h_{\text{t$\varphi $},1}'' M^3-3456 \sqrt{3} h_{\varphi \varphi ,0}''' M^3\\
\nonumber+2106 \sqrt{3} h_{\text{tt},0} M^2+1944 \sqrt{2} h_{\text{tt},1} M^2+11664 \sqrt{3} h_{\text{tt},2} M^2-12960 \sqrt{2} h_{\text{t$\varphi $},0}' M^2-15552 \sqrt{3} h_{\text{t$\varphi $},1}' M^2\\
\nonumber-23328 \sqrt{2} h_{\text{t$\varphi $},2}' M^2+3276 \sqrt{3} h_{\varphi \varphi ,0}'' M^2+2592 \sqrt{2} h_{\varphi \varphi ,1}'' M^2+2556 \sqrt{2} h_{\text{t$\varphi $},0} M+2376 \sqrt{3} h_{\text{t$\varphi $},1} M\\
\nonumber+7776 \sqrt{2} h_{\text{t$\varphi $},2} M-1677 \sqrt{3} h_{\varphi \varphi ,0}' M-1890 \sqrt{2} h_{\varphi \varphi ,1}' M-648 \sqrt{3} h_{\varphi \varphi ,2}' M+389 \sqrt{3} h_{\varphi \varphi ,0}\\
\nonumber+486 \sqrt{2} h_{\varphi \varphi ,1}+378 \sqrt{3} h_{\varphi \varphi ,2}\Big)\Bigg\rbrack+\frac{1}{M}\left(9504 h_{\text{t$\varphi $},0}+10368 \sqrt{6} h_{\text{t$\varphi $},1}-2520 \sqrt{6} h_{\varphi \varphi ,0}'-2592 h_{\varphi \varphi ,1}'\right)\Bigg\}\\
\end{gather}



\newpage
\chapter{The effects of lower-order parameters in the metric}\label{app:lowerOrder}
In this appendix we investigate the effects of lower-order parameters on the direct observables in the new spacetime: the Keplerian and epicyclic frequencies, and the photon rings\footnote{This appendix provides supplementary material for Chapter~\ref{chap:beyondKerr}.}.
Such lower-order parameters were originally assumed to be vanishing by Johannsen in Ref.~\cite{Johannsen:2015pca} due to strong constraints on the ppN parameters~\cite{Williams:2004qba}.
However, such constraints were obtained via observations of the local, weak-field solar system and may not hold true in the strong-gravity  regimes present near BHs, where the spacetime is not guaranteed to even be similar to that surrounding a star.
Thus, in this appendix we revive these neglected parameters:
\begin{itemize}
\item the first order parameter $\epsilon_1$ from the function $f(r)$\footnote{We note that to avoid a rescaling of the observable BH mass $M$, when $\epsilon_1\neq0$, we must set $\alpha_{11}=\epsilon_1/2$.}; 
\item the lowest-order parameter $\alpha_{12}$ from the $A_1(r)$ function; 
\item the lowest-order parameter $\alpha_{51}$ from the function $A_5(r)$; 
\item the first- and second-order parameters $\gamma_1$ and $\gamma_2$ from the function $g(\theta)$\footnote{We additionally consider the second-order parameter $\gamma_{2}$ corresponding to terms of $\cos^2 \theta$. This is because in all example metrics considered, typically $\cos^2\theta$ enters at first order in $g(\theta)$, as is the case of the Kerr-Sen metric.}.
\end{itemize} 

\begin{figure*}[htb]
\begin{center}
\includegraphics[width=.43\columnwidth]{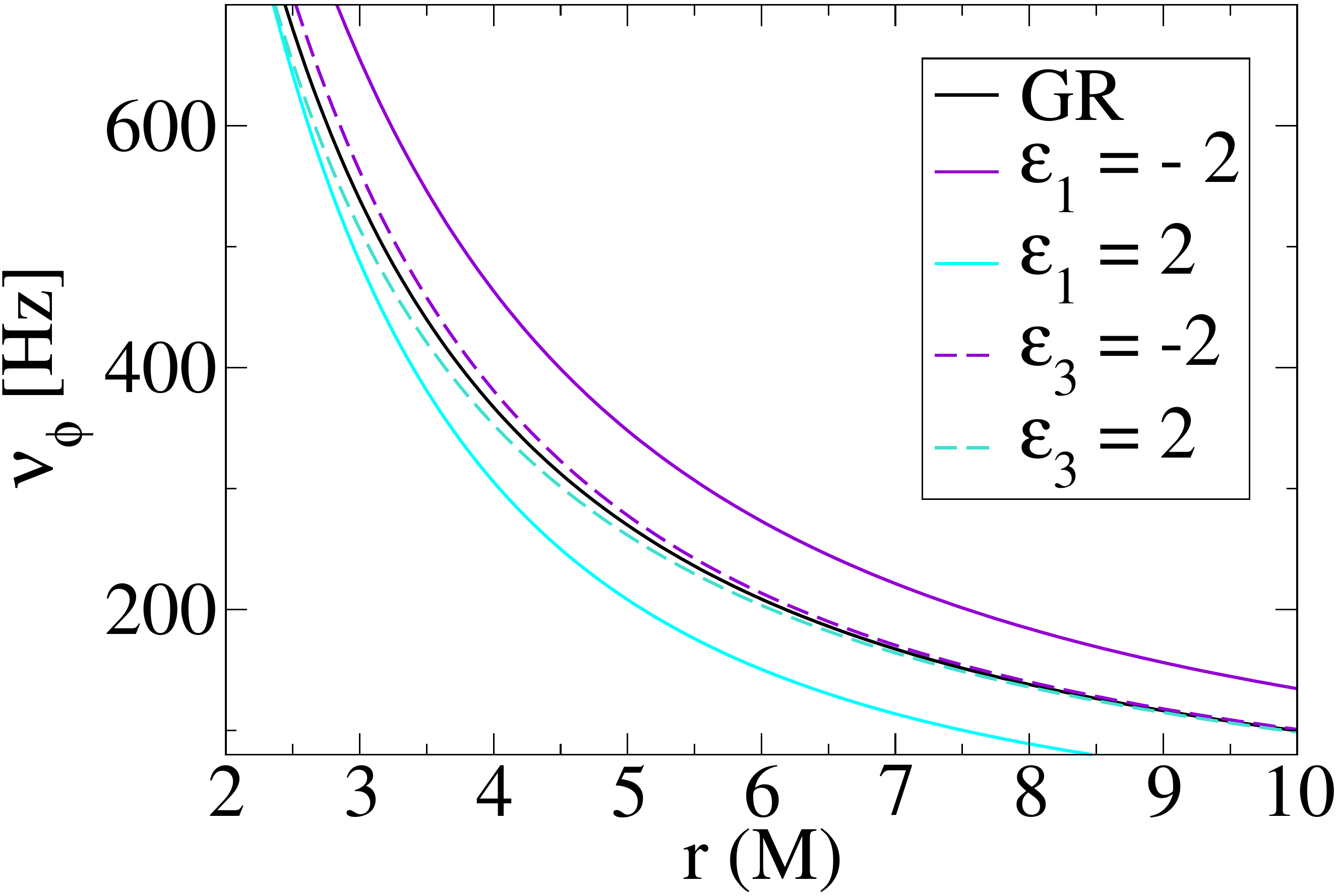}
\includegraphics[width=.43\columnwidth]{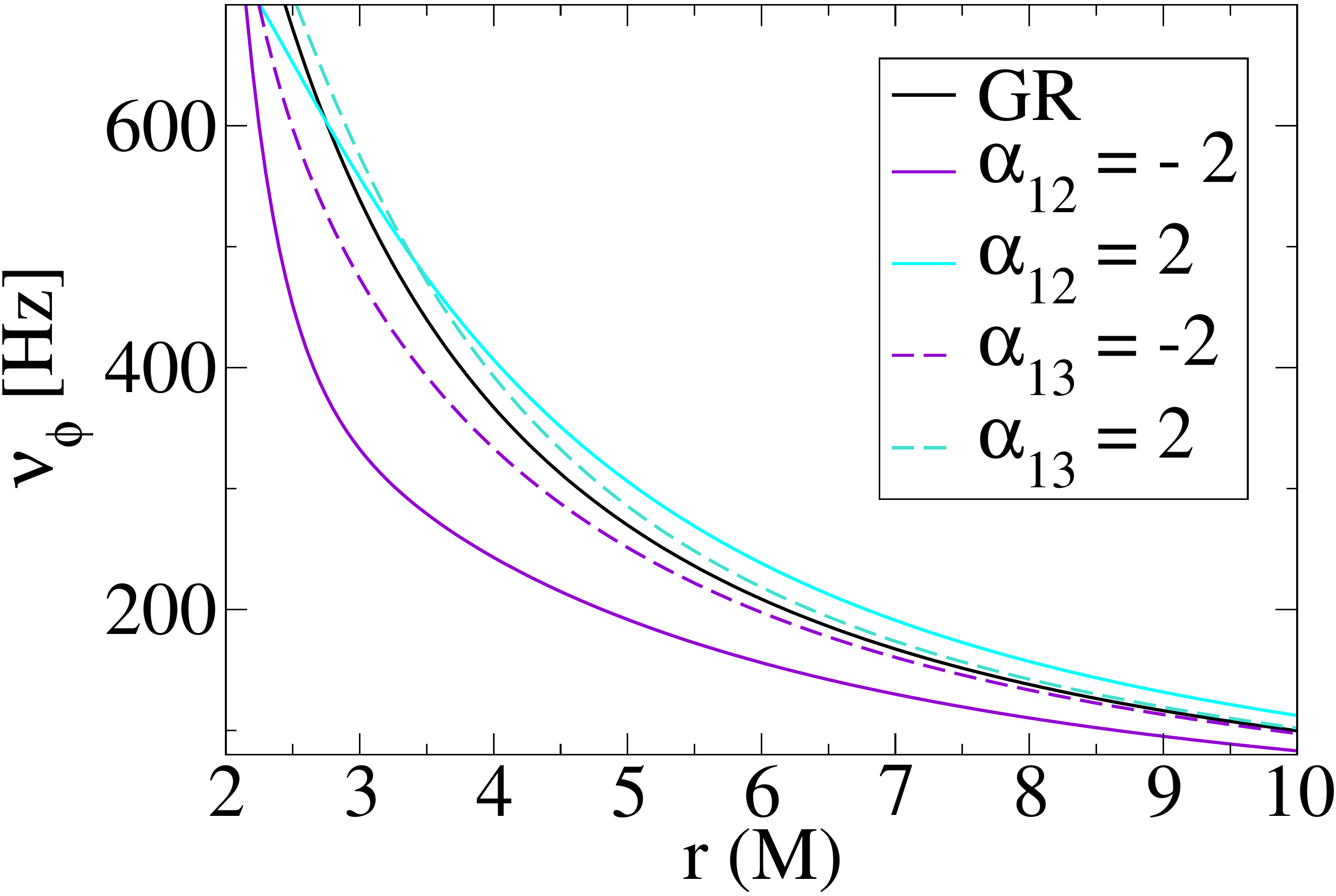}
\includegraphics[width=.43\columnwidth]{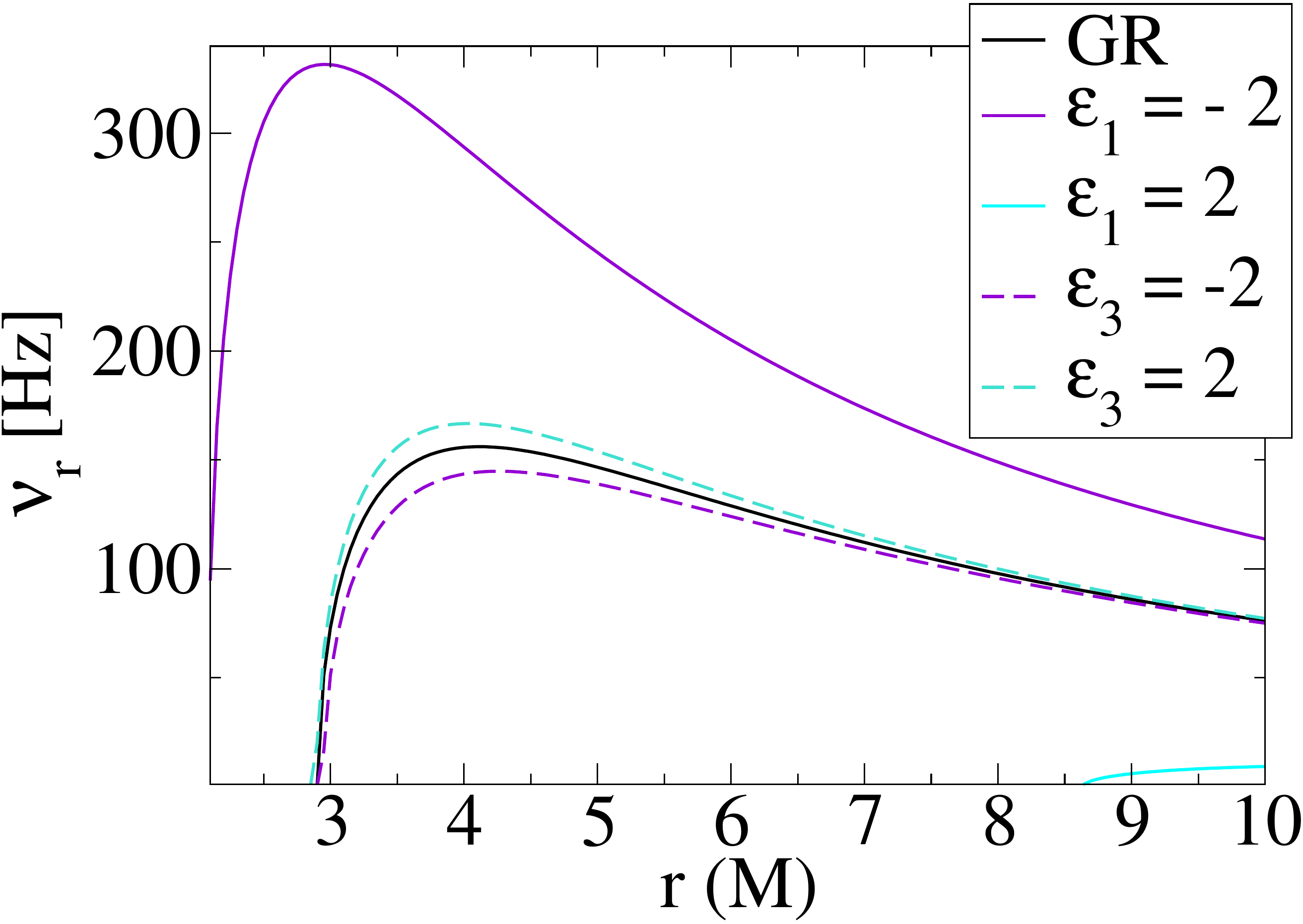}
\includegraphics[width=.43\columnwidth]{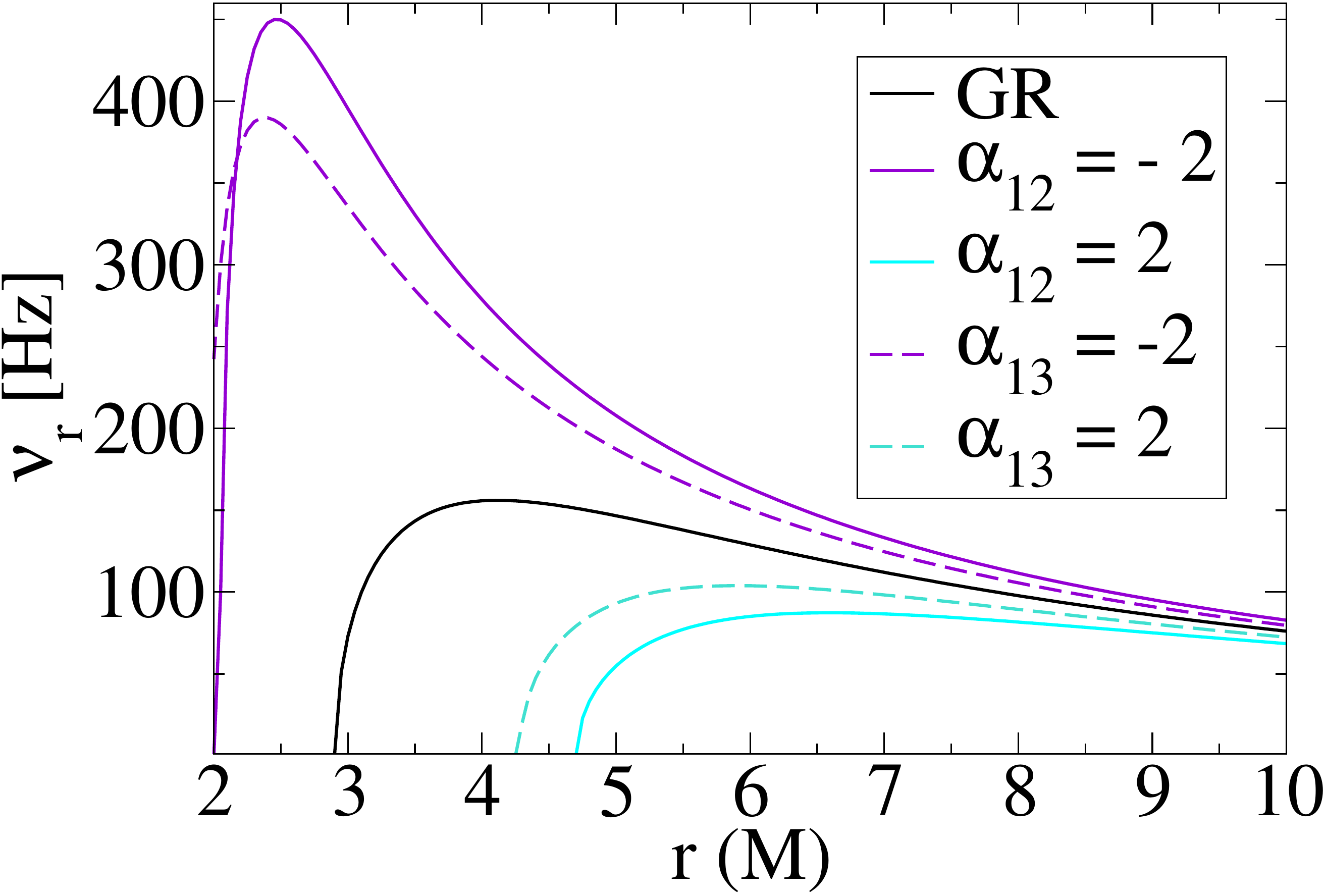}
\includegraphics[width=.43\columnwidth]{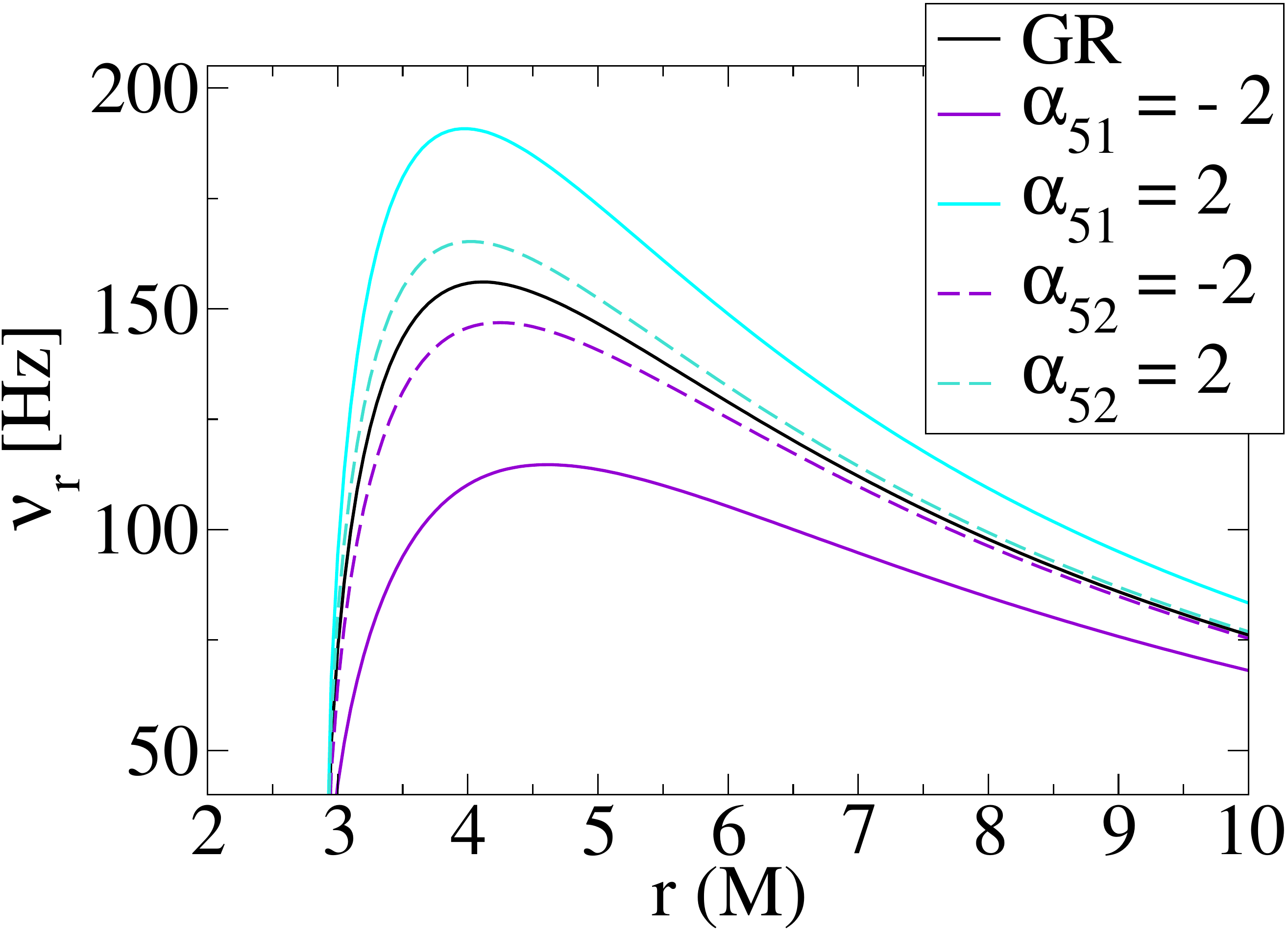}
\includegraphics[width=.43\columnwidth]{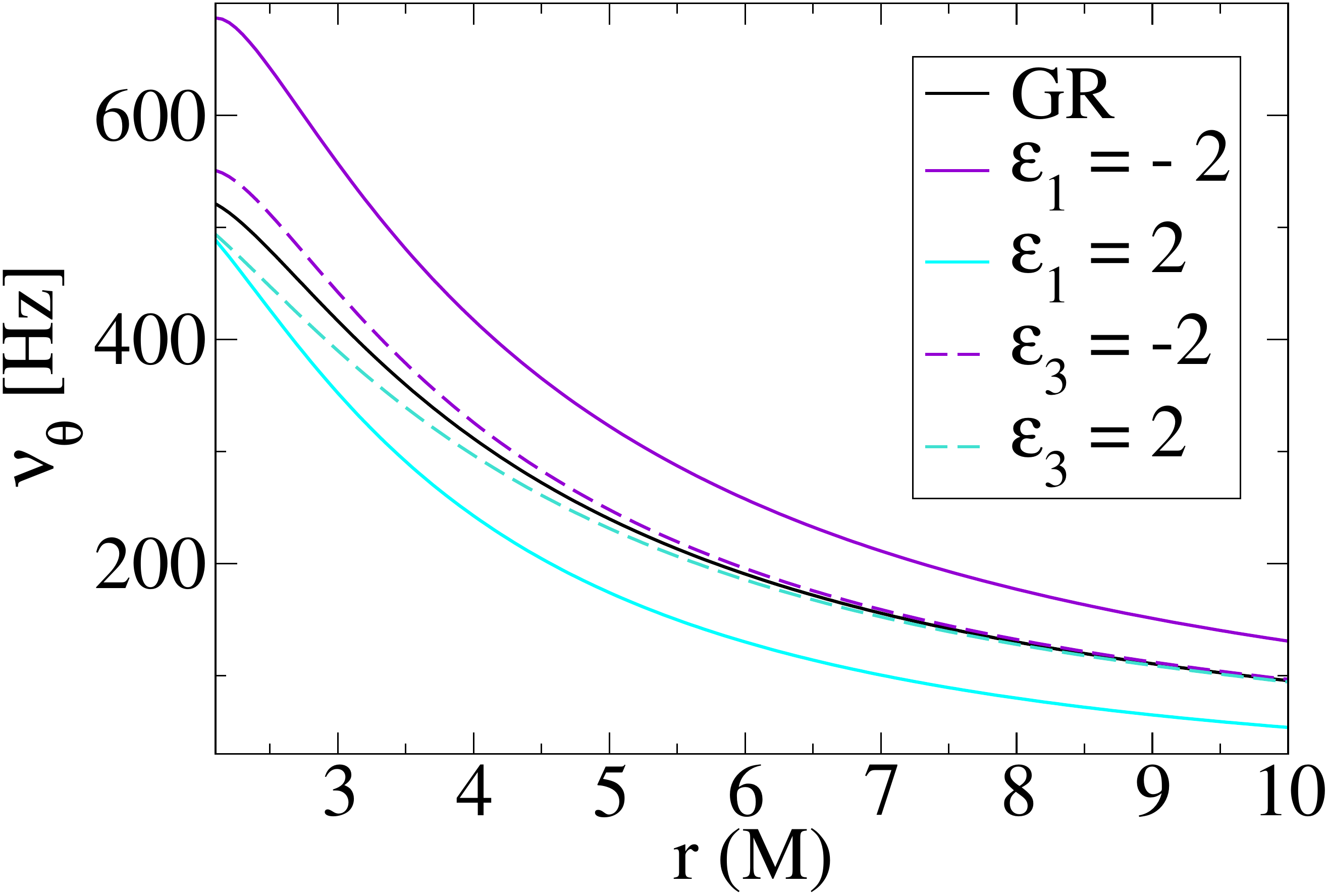}
\includegraphics[width=.43\columnwidth]{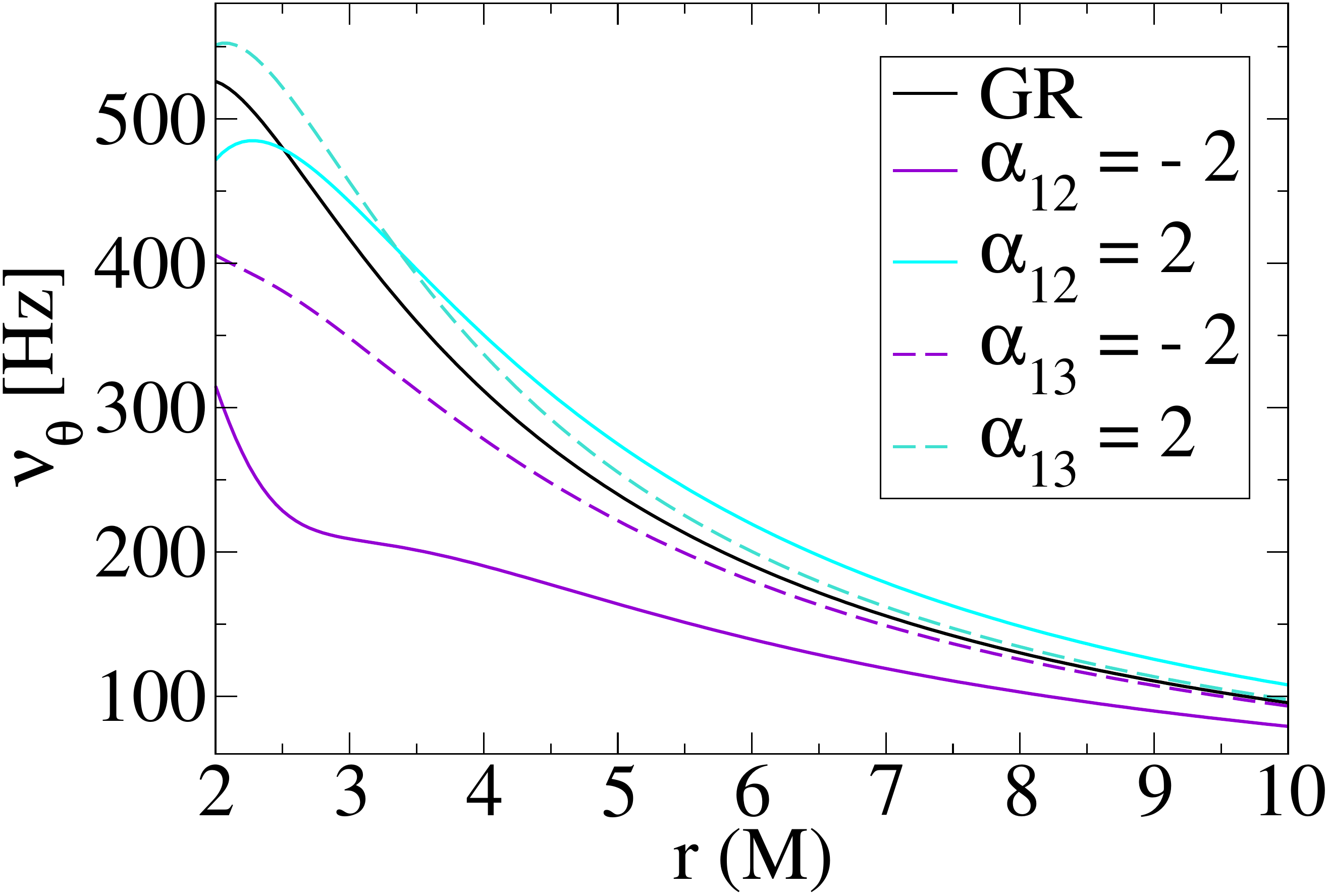}
\includegraphics[width=.43\columnwidth]{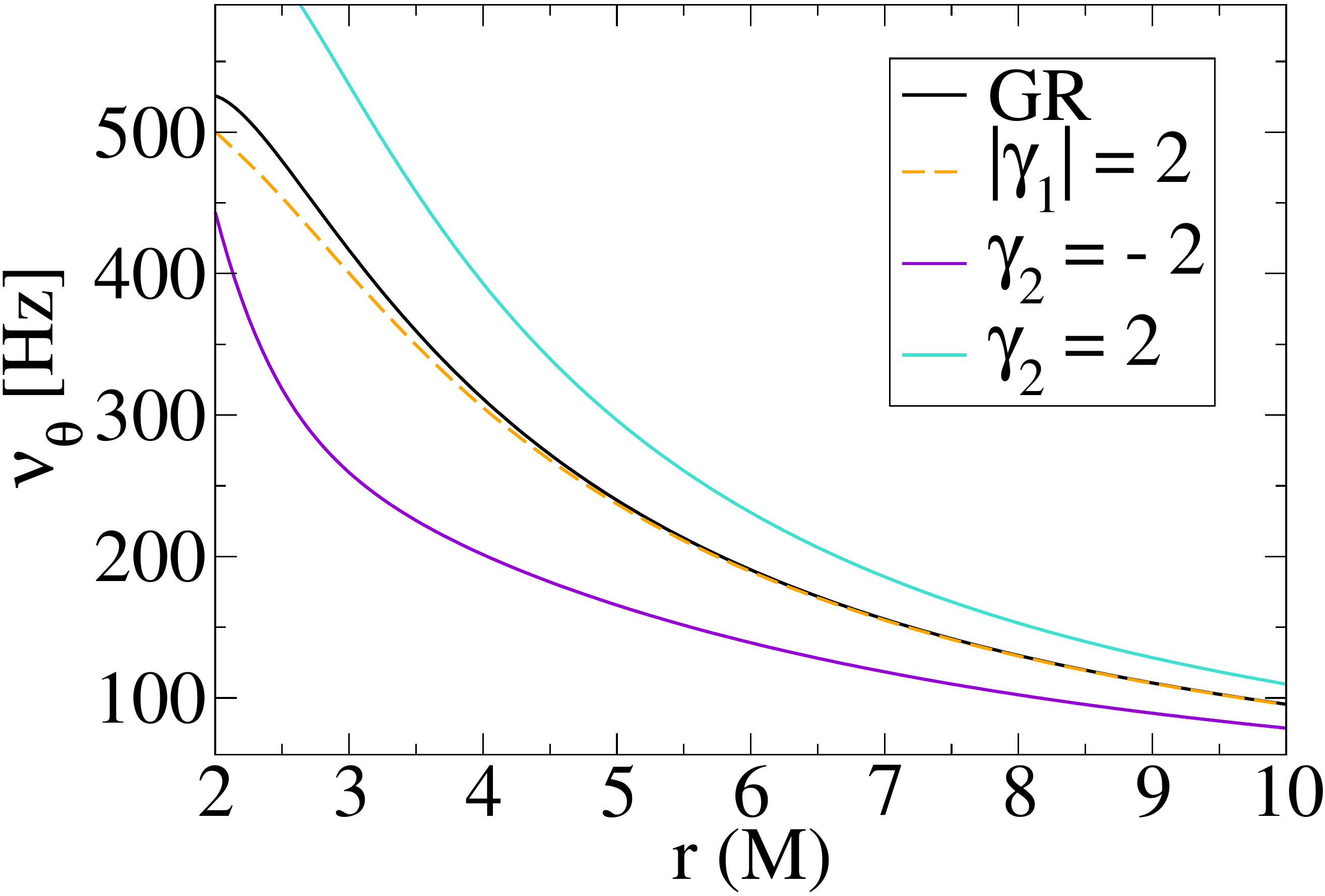}
\caption[Orbital frequencies with lower-order non-vanishing parameters]{Keplerian (top row), radial epicyclic (middle row), and vertical epicyclic frequencies about a BH with the lower order parameters $\epsilon_1$, $\alpha_{12}$, $\alpha_{51}$, $\gamma_1$, and $\gamma_2$ that were presumed to vanish in Ref.~\cite{Johannsen:2015pca} due to strong constraints on the ppN parameters~\cite{Williams:2004qba}.
Also shown for comparison in each case are the high-order parameters $\epsilon_3$, $\alpha_{13}$, and $\alpha_{52}$ that were used in the main analysis and~\cite{Johannsen:2015pca}.
We note that when $\epsilon_1\ne0$, we have set $\alpha_{11}=\epsilon/2$ to prevent a rescaling of the BH mass.
}\label{fig:lowerOrder_nu}
\end{center}
\end{figure*}

We begin our investigation on the Keplerian and epicyclic frequencies' dependence on such lower-order parameters.
In the case of the former, we note that of the parameters we focus on in this appendix, only $\epsilon_1$ and $\alpha_{12}$ enter the expression (which is dependent on $g_{tt}$, $g_{\phi\phi}$, and $g_{t\phi}$) for equatorial orbits (where $\gamma_1 P_1(\cos\theta)$ vanishes entirely).
Similarly, the radial epicyclic frequencies depend only on $\epsilon_1$, $\alpha_{12}$, and now $\alpha_{51}$, due to the $g_{rr}$ dependence.
Finally, the vertical epicyclic frequencies depend on $\epsilon_1$, $\alpha_{12}$, and now $\gamma_1$ and $\gamma_2$, as a result of the $\partial_\theta$ derivatives.
In Fig.~\ref{fig:lowerOrder_nu}, we compare each of these orbital frequencies for two different cases: (i) only including the lower order parameters ($\epsilon_1$, $\alpha_{12}$, $\alpha_{51}$, $\gamma_1$ and $\gamma_2$) that were assumed to vanish in the ppN framework, and (ii) only including the next-higher-order parameters already used in the main analysis ($\epsilon_3$, $\alpha_{13}$, $\alpha_{52}$).
We observe that for every case the inclusion of the neglected lower-order parameters makes quite a large difference on the observables $\nu_\phi$, $\nu_r$, and $\nu_\theta$ as compared to the higher-order ones.
In particular, the radial epicyclic frequencies are significantly impacted upon the inclusion of $\epsilon_1$, $\alpha_{12}$, and $\alpha_{51}$.
This indicates that, if the parameters $\epsilon_1$, $\alpha_{12}$, $\alpha_{51}$, $\gamma_1$ and $\gamma_2$ are indeed non-vanishing and the ppN constraints applied to BHs are invalid, such lower-order parameters must be included for accuracy.

\begin{figure}[htb]
\begin{center}
\includegraphics[width=.7\columnwidth]{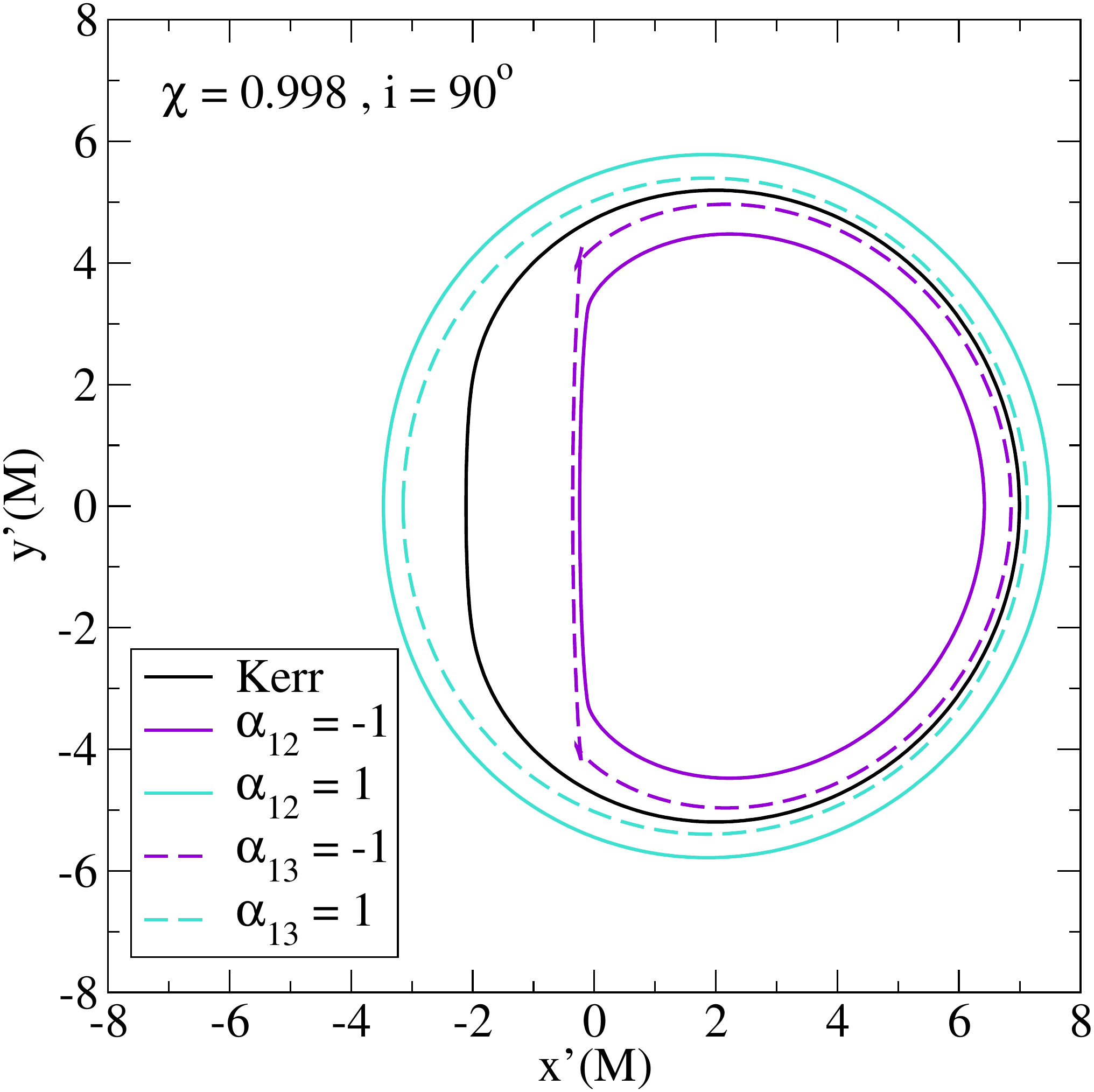}
\caption[Photon orbits with lower-order non-vanishing parameters]{
Comparison between the image of the photon rings about a BH when including lower order parameter $\alpha_{12}$ (presumed to vanish in Ref.~\cite{Johannsen:2015pca} due to strong constraints on the ppN parameters~\cite{Williams:2004qba}) and when instead including the next-order parameter $\alpha_{13}$.
Observe the difference made in the observable photon orbit when including the lower-order parameter that may be non-vanishing if the solar-system ppN constraints are invalid.
}\label{fig:lowerOrder_shadows}
\end{center}
\end{figure}

Next we discuss another observable -- the BH photon orbits.
We find that, of the lower-order parameters considered here, such expressions only depend on $\alpha_{12}$.
This is a result of $g(\theta)$ not entering the separated radial equations, $A_5(r)$ canceling out on each side of $R(r)=\frac{d}{dr}R(r)=0$, and a vanishing $\mu f(r)$ for photon orbits with $\mu=0$.
In Fig.~\ref{fig:lowerOrder_shadows} we plot the ensuing photon rings with (i) only $\alpha_{12}$ included, and (ii) only $\alpha_{13}$ included.
We observe that, if such lower-order parameters were non-vanishing in the ppN framework, their inclusion would make a sizable impact on the size (but not the shape) of the photon rings.

\begin{figure}[htb]
\begin{center}
\includegraphics[width=.7\columnwidth]{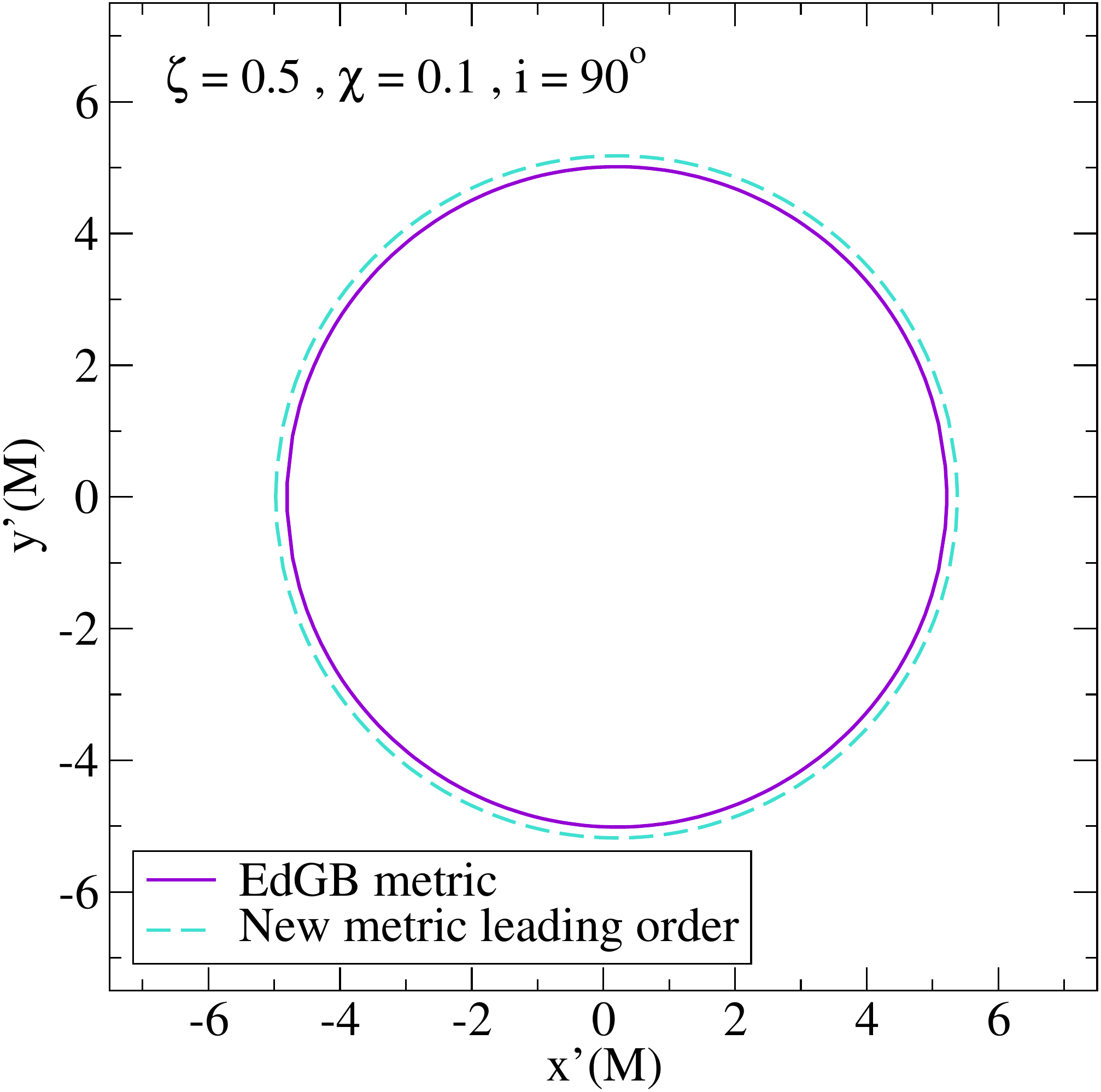}
\caption[Photon orbits in EdGB gravity]{
Comparison between photon orbits about a BH computed with two different methods: (i) using the new metric leading-order mapping to EdGB found in Tab.~\ref{tab:maps} and (ii) using the full EdGB metric to first order in spin.
Here, the BH spin or EdGB coupling parameter $\zeta$ can not be too large or else the small-coupling approximation or small-spin approximations begin to break down and the photon orbits become nonsensical.
Obsesrve how the two photon orbits agree quite well, indicating the validity of the leading-order expansion in $\frac{1}{r}$.
}\label{fig:lowerOrder_EdGB_shadows}
\end{center}
\end{figure}

Finally we provide a brief analysis on the validy of including leading-order parameters in the computation of photon rings for the example of EdGB gravity.
To do this, we begin by computing the photon ring solutions in the EdGB theory of gravity, to first order in spin as described in Ref.~\cite{Ayzenberg:2014aka} (exact in the $\frac{1}{r}$ expansion).
Next we take the new metric and map to the EdGB theory of gravity as described in Tab.~\ref{tab:maps} using only the leading-order parameters (to leading order in the $\frac{1}{r}$ expansion).
Figure~\ref{fig:lowerOrder_EdGB_shadows} compares the photon orbit paths for these two cases for a EdGB BH with coupling parameter $\zeta=0.5$, spin $\chi=0.1$, and observer inclination $i=90^\circ$\footnote{We note that, due to the small-coupling and slow-rotation approximations used in the EdGB metric, neither $\zeta$ nor $\chi$ can be too large, else the approximations break down and the photon ring results become unreliable.}.
We observe that the two photon orbit solutions agree quite well, giving some indication to the validity of using only the leading-order terms in the $\frac{1}{r}$ expansion.

\newpage
\chapter{Naked singularities in the new spacetime}\label{app:nakedSingularities}
In this appendix we briefly discuss the emergent naked singularities for certain sections of the new metric's parameter space\footnote{This appendix provides supplementary material for Chapter~\ref{chap:beyondKerr}.}.
As also shown in e.g. Refs.~\cite{Hioki:2009na,Papnoi:2014aaa}, when a naked singularity is present outside of the BH event horizon $r_\EH$, closed photon orbits no longer exist and the photons escape to radial infinity.
Fig.~\ref{fig:nakedShadows} demonstrates this phenomena for several cases of BHs with naked singularities, such as $\alpha_{13}=2$, $\alpha_{22}=2$, $\alpha_{02}=2$, $\alpha_{13}=-\alpha_{02}=2$, or $\alpha_{22}=-\alpha_{02}=2$, with all other deviation parameters vanishing in each case.
Observe how for each case, a photon ``arc'' appears, and the orbit is not closed.
We similarly find in each such scenario that the other spacetime properties such as the orbital energy and angular momentum become discontinuous, negative, and complex.
We find that these naked singularities appear when any of $\alpha_{13}$, $\alpha_{22}$, or $\alpha_{02}$ appear as the sole non-vanishing parameter of the three, while the former two must appear alongside $\alpha_{02}$ to avoid such singularities.
We also find that when non-vanishing, the parameters $(\alpha_{13},\alpha_{02})$ or $(\alpha_{22},\alpha_{02})$ must share the same sign, else naked singularities appear.

\begin{figure}[htb]
\begin{center}
\includegraphics[width=.7\columnwidth]{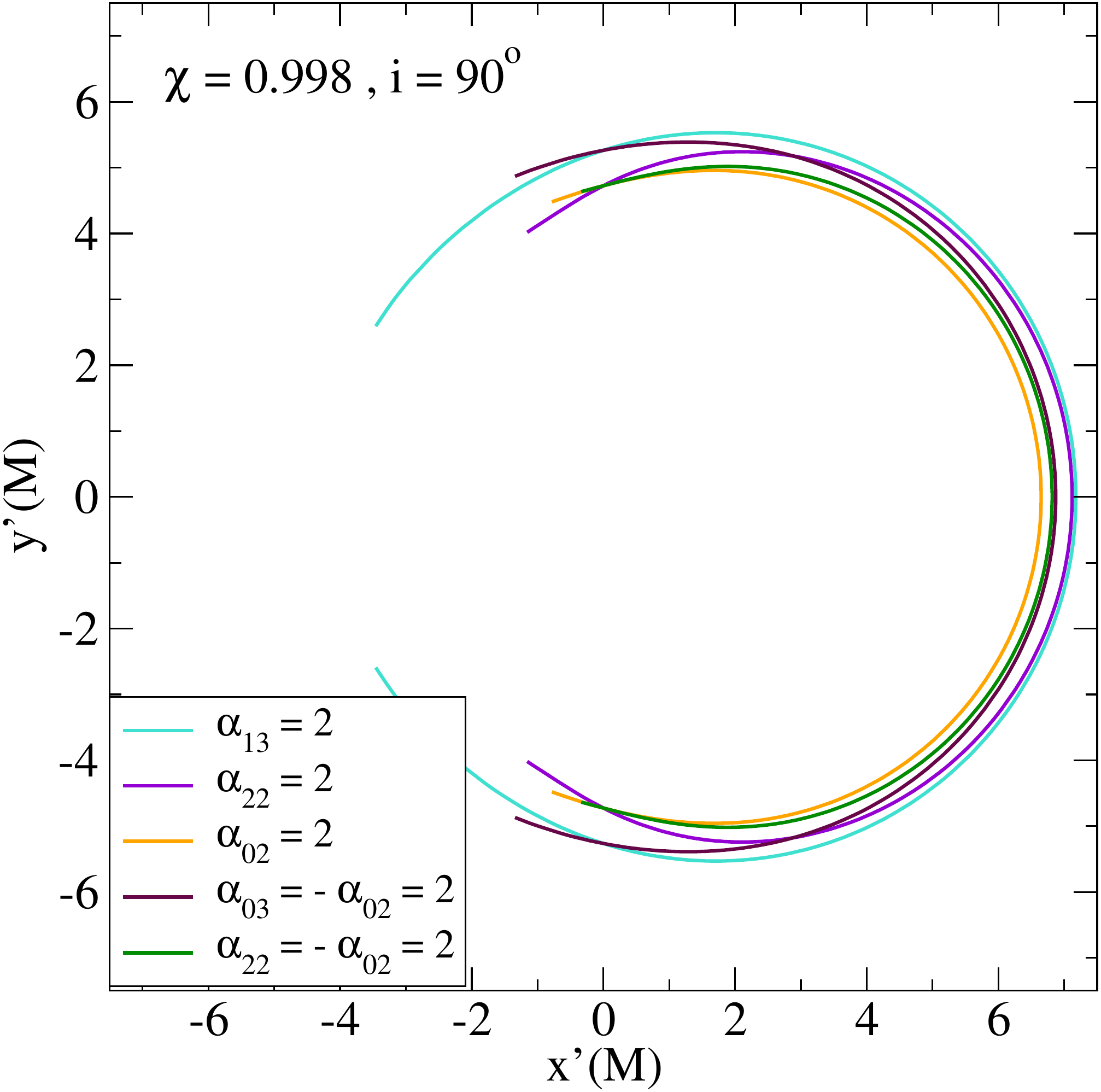}
\caption[Open photon orbits about BHs with naked singularities]{Open photon orbits (``arcs'') about a BH with photons escaping to radial infinity.
Such orbits appear for certain parameterizations of the new metric similar to the ones presented here: $\alpha_{13}=2$, $\alpha_{22}=2$, $\alpha_{02}=2$, $\alpha_{13}=-\alpha_{02}=2$, and $\alpha_{22}=-\alpha_{02}=2$, with all other deviation parameters set to 0.
}\label{fig:nakedShadows}
\end{center}
\end{figure}

Next we demonstrate that such open photon orbits are indeed indicative of emergent naked singularities in the new spacetime.
This is done by examining the nature of the Kretschmann invariant $K$ in both of the spacetimes that do and do not exhibit naked-singularity symptoms.
The Kretschmann invariant is given by
\begin{equation}
K=R_{\alpha\beta\gamma\delta}R^{\alpha\beta\gamma\delta}
\end{equation}
for Riemann curvature tensor given by
\begin{equation}
R^\alpha_{\beta\gamma\delta}=\Gamma^\alpha_{\beta\delta,\gamma}-\Gamma^\alpha_{\beta\gamma,\delta}+\Gamma^\mu_{\beta\delta}\Gamma^\alpha{\mu\gamma}+\Gamma^\mu_{\beta\gamma}\Gamma^\alpha{\mu\delta}
\end{equation}
with Christoffel symbols given by
\begin{equation}
\Gamma_{\alpha\beta\gamma}=\frac{1}{2}(g_{\alpha\beta,\gamma}+g_{\alpha\gamma,\beta}-g_{\beta\gamma,\alpha}).
\end{equation}
The scalar quantity $K$ is gauge invariant, and thus a divergence of $K$ is a sign for the presence of a true singularity.
For demonstration purposes, we pick a highly-rotating BH with $\chi=0.998$, for the two cases of $\alpha_{13}=2$, and $\alpha_{13}=\alpha_{02}=2$, where the former exhibits naked singularity behaviors, and the latter does not.
The event horizon for the latter case is located at $r_\EH=m+\sqrt{m^2-a^2}$, which reduces to $1.06$m for our given BH rotation.

\begin{figure}[htb]
\begin{center}
\includegraphics[width=.7\columnwidth]{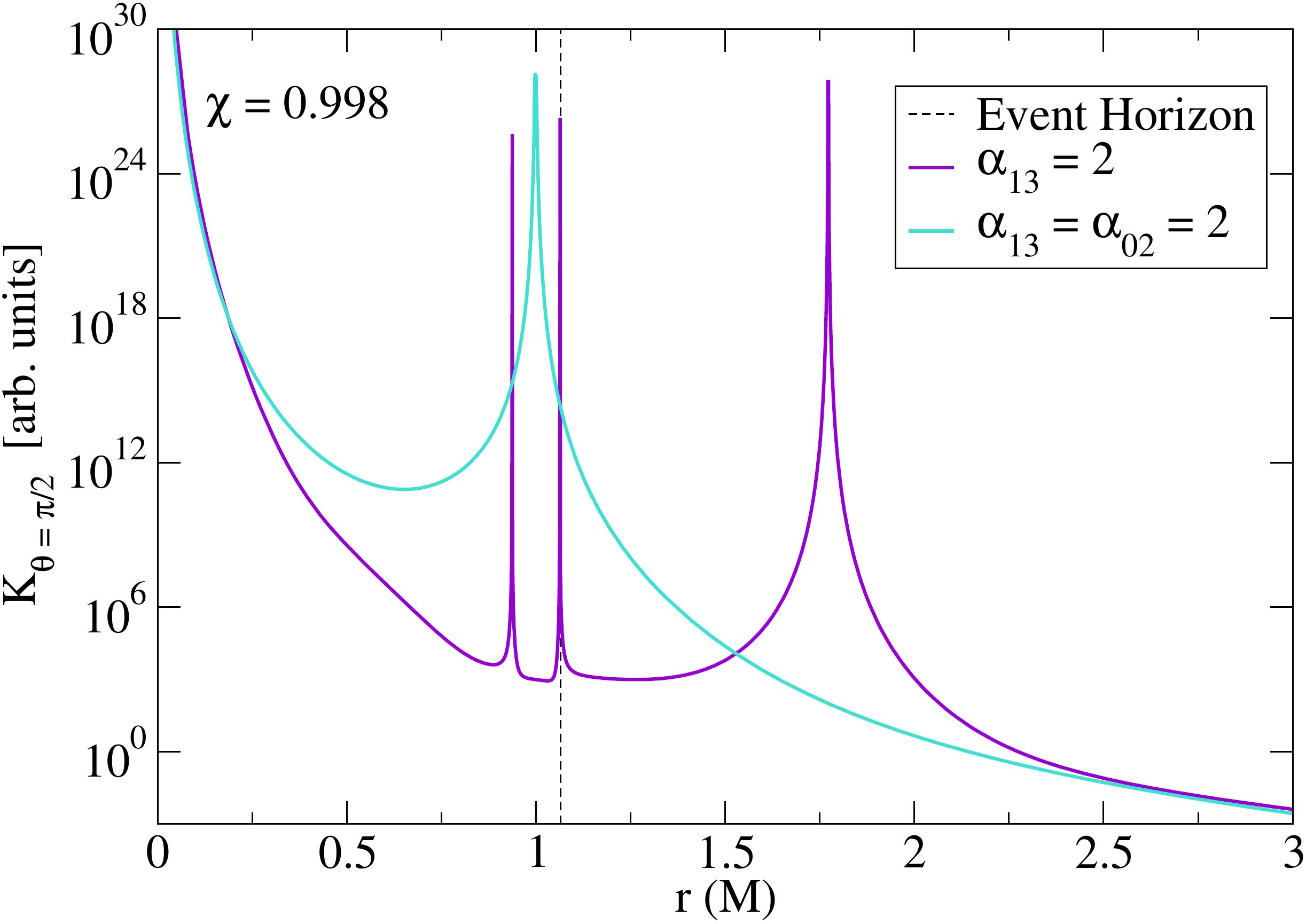}
\caption[Kretschmann invariant for a BH with naked singularities]{Scalar Kretschmann invariant $K$ on the equatorial plane plotted in arbitrary units as a function of radius from the central BH described by the new metric.
This is plotted for two cases: (i) BH with non-vanishing parameter $\alpha_{13}=2$, and a (ii) BH with non-vanishing parameters $\alpha_{13}=\alpha_{02}=2$.
Also shown by the dashed vertical line is the event horizon of $r_\EH=m+\sqrt{m^2-a^2}=1.07$m for case (ii).
In case (i) we see, as predicted, that there exists a naked singularity at $r=1.77$m, well outside of the ``would-be'' event horizon.
Additionally, in this case there also exists a singularity at $r=0.93$m, and interestingly, on the ``would-be'' event horizon at $r=1.07$m.
In case (ii), we observe a singularity at $r=a=0.998$m, behind the event horizon as usual, confirming our predictions.
}\label{fig:invariant}
\end{center}
\end{figure}

Finally, we compute the Kretschmann invariant along the equatorial plane.
The results are shown in Fig.~\ref{fig:invariant} for both cases.
We observe that for the closed-photon orbit case of $\alpha_{13}=\alpha_{02}=2$, there exists a singularity at $r=a=0.998$m as usual, well behind the event horizon of $r_\EH=1.07$m.
Alternatively, for the open-orbit case of sole-non-vanishing parameter $\alpha_{13}=2$, we observe several interesting features.
First, we see a singularity behind the ``would-be'' event horizon at $r=0.93$m as one would expect.
Next, there exists a singularity directly on the ``would-be'' event horizon at $r=1.07$m, and finally we see a singularity well beyond the ``would-be'' event horizon at $r=1.77$m.
This confirms our suspicion of the existence of naked singularities, thus for the remainder of this analysis we avoid parameterizations that create such anomalies.

\addtocontents{toc}{\protect\end{adjustwidth}}


\newpage 
\clearpage
\phantomsection
\addcontentsline{toc}{part}{Bibliography}
\bibliography{../bibliography/Zack.bib}


\newpage
\thispagestyle{plain}
\section*{\Huge Vita}
\indent
\indent
\addcontentsline{toc}{part}{Vita} 
\begin{spacing}{1.2}
Zack Carson was born in Salt Lake City, Utah on February 3, 1992. 
He obtained a B.~S. in physics and also in mathematics with a minor in astronomy from the University of Utah in 2014. 
Since then, he has been a Ph.~D. student at the University of Virginia. 
His honors include the University of Virginia distinguished graduate teaching award for STEM fields, a departmental fellowship, and a second place departmental poster contest winner.
He has presented invited talks at the Southeastern Section of the American Physical Society (SESAPS) meeting and at Bharat Kumar's nuclear theory group webinar series, as well as several contributed talks elsewhere.
His publications are presented below:
\begin{enumerate}
\item Z. Carson and K. Yagi, \textit{Probing Einstein-dilaton Gauss-Bonnet Gravity with the inspiral and ringdown of gravitational waves}, Phys. Rev. D \textbf{101}, 104030 (2020)
\item Z. Carson and K. Yagi, \textit{Parameterized and inspiral-merger-ringdown consistency tests of gravity with multi-band gravitational wave observations}, Phys. Rev. D \textbf{101}, 044047 (2020)
\item Z. Carson and K. Yagi, \textit{Asymptotically flat, parameterized black hole metric preserving Kerr symmetries}, Phys. Rev. D \textbf{101}, 084030 (2020)
\item Z. Carson and K. Yagi, \textit{Probing beyond-Kerr spacetimes with inspiral-ringdown corrections to gravitational waves}, Phys. Rev. D \textbf{101}, 084050 (2020)
\item Z. Carson, B. Seymour and K. Yagi, \textit{Future Prospects for Probing Scalar-Tensor Theories with Gravitational Waves from Mixed Binaries}, Class. Quant. Grav. \textbf{37}, 065008 (2020)
\item Z. Carson and K. Yagi, \textit{Probing string-inspired gravity with the inspiral-merger-ringdown consistency tests of gravitational waves}, Class. Quant. Grav. \textbf{37}, 215007 (2020)
\item Z. Carson and K. Yagi, \textit{Multi-band gravitational wave tests of general relativity}, Class. Quant. Grav. Letters \textbf{37}, 02LT01 (2019)
\item Z. Carson and K. Yagi, \textit{Parameterized and Consistency Tests of Gravity with Gravitational Waves: Current and Future}, Proceedings, Recent Progress in Relativistic Astrophysics: Shanghai, China, Vol. 17(1) (2019)
\item S. Tahura, Z. Carson and K. Yagi, \textit{Testing Gravity with Gravitational Waves from Binary Black Hole Mergers: Contributions from Amplitude Corrections}, Phys. Rev. D \textbf{100}, 104001 (2019)
\item Z. Carson, A. W. Steiner and K. Yagi, \textit{Future Prospects for Constraining Nuclear Matter Parameters with Gravitational Waves}, Phys. Rev. D \textbf{100}, 023012 (2019)
\item Z. Carson, K. Chatziioannou, C. Haster, K. Yagi, and N. Yunes \textit{Equation-of-state insensitive relations after GW170817}, Phys. Rev. D \textbf{99} 083016 (2019)
\item Z. Carson, A. W. Steiner and K. Yagi, \textit{Constraining nuclear matter parameters with GW170817}, Phys. Rev. D \textbf{99} 043010 (2019)
\item J. Zimmerman, Z. Carson, K. Schumacher, A. W. Steiner and K. Yagi, \textit{Measuring Nuclear Matter Parameters with NICER and LIGO/Virgo}, submitted to Phys. Rev. Letters (2020)
\item J. Zimmerman, Z. Carson, K. Schumacher, A. W. Steiner and K. Yagi, \textit{Bounds on Nuclear Matter Parameters by Combining Measurements of Radius and Tidal Deformabilty}, in preparation for Phys. Rev. D (2020)
\item J. Zang, Z. Yang, X. Xie, M. Ren, Y. Shen, Z. Carson, \textit{et al}, \textit{High quantum efficiency uni-traveling-carrier photodiode}, IEEE Photonics Technology Letters, \textbf{29} (3), 302-305 (2017)
\item X. Zhu, Z. Carson, R. Alexander and O. Pfister \textit{Leveraging qumode scalability: high squeezing and entanglement from redistributed multitudinous-mode squeezing}, In preparation (2020)
\end{enumerate}
\end{spacing}
\thispagestyle{plain}

\end{document}